\documentclass[a4paper,arxiv,twosidenoshift]{doc}

\begin{document}

\docinfo{The Dilatation Operator of N=4 Super Yang-Mills Theory and Integrability}
        {Niklas Beisert}
        {Dissertation}
        {Gauge Theory, AdS/CFT, String Theory, Spin Chains, Integrable Models, Conformal Field Theory, QFT}

\docphysrept[\frontmatter]{\begin{frontmatter}}
\docphysrept{\input{secprtit}}
\dochumboldt{\include{sechutit}}
\docprint{\thispagestyle{empty}

\begin{center}\parindent0pt

\nonumtoc{chapter}{}{Title}

\begingroup\Huge\renewcommand{\bfdefault}{bx}\bfseries\mathversion{bold}
The Dilatation Operator\\
of $\superN=4$ Super Yang-Mills Theory\\
and Integrability\par
\vspace{1cm}
\endgroup

\begingroup
\docphysrept[%
Dissertation\\
Eingereicht an der \href{http://www.hu-berlin.de}{Humboldt-Universit\"at zu Berlin}
\\
\"Uberarbeitete Fassung\par
]{}
\vspace{1.5cm}
\endgroup

\includegraphics[width=8cm]{sectitle.logo.eps}
\vspace{1.5cm}

\begingroup\LARGE\sffamily
Niklas Beisert\par
\vspace{0.5cm}
\endgroup

\begingroup
\large
\emph{22.~Oktober 2005}\par\bigskip

\texttt{\arxivno{hep-th/0407277}}\\
\texttt{AEI-2004-057}\par
\endgroup

\vfill

\includegraphics[width=5cm]{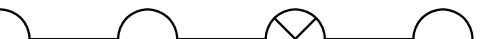}
\vfill

\dochumboldt[%
\textit{Max-Planck-Institut f\"ur Gravitationsphysik\\
\href{http://www.aei.mpg.de}{Albert-Einstein-Institut}\\
Am M\"uhlenberg 1, 14476 Potsdam, Deutschland}
\vspace{1.5cm}
]{}

\docprint{
\parbox{2.5cm}{\centering
\href{http://www.hu-berlin.de}{$\color{white}\left|\color{black}
\parbox{2cm}{\centering\includegraphics[width=2cm]{sectitle.hu.eps}}
\color{white}\right|\color{black}$}}\qquad\qquad
\parbox{2.5cm}{\centering
\href{http://www.aei.mpg.de}{$\color{white}\left|\color{black}
\parbox{1.5cm}{\centering\includegraphics[width=1.5cm]{sectitle.aei.eps}}
\color{white}\right|\color{black}$}}
}

\end{center}

\clearpage\thispagestyle{empty}\cleardoublepage

}
\docarxiv{\thispagestyle{empty}

\begin{center}\parindent0pt

\nonumtoc{chapter}{}{Title}

\begingroup\Huge\renewcommand{\bfdefault}{bx}\bfseries\mathversion{bold}
The Dilatation Operator\\
of $\superN=4$ Super Yang-Mills Theory\\
and Integrability\par
\vspace{1cm}
\endgroup

\begingroup
\docphysrept[%
Dissertation\\
Eingereicht an der \href{http://www.hu-berlin.de}{Humboldt-Universit\"at zu Berlin}
\\
\"Uberarbeitete Fassung\par
]{}
\vspace{1.5cm}
\endgroup

\includegraphics[width=8cm]{sectitle.logo.eps}
\vspace{1.5cm}

\begingroup\LARGE\sffamily
Niklas Beisert\par
\vspace{0.5cm}
\endgroup

\begingroup
\large
\emph{22.~Oktober 2005}\par\bigskip

\texttt{\arxivno{hep-th/0407277}}\\
\texttt{AEI-2004-057}\par
\endgroup

\vfill

\includegraphics[width=5cm]{sectitle.dynkin.eps}
\vfill

\dochumboldt[%
\textit{Max-Planck-Institut f\"ur Gravitationsphysik\\
\href{http://www.aei.mpg.de}{Albert-Einstein-Institut}\\
Am M\"uhlenberg 1, 14476 Potsdam, Deutschland}
\vspace{1.5cm}
]{}

\docprint{
\parbox{2.5cm}{\centering
\href{http://www.hu-berlin.de}{$\color{white}\left|\color{black}
\parbox{2cm}{\centering\includegraphics[width=2cm]{sectitle.hu.eps}}
\color{white}\right|\color{black}$}}\qquad\qquad
\parbox{2.5cm}{\centering
\href{http://www.aei.mpg.de}{$\color{white}\left|\color{black}
\parbox{1.5cm}{\centering\includegraphics[width=1.5cm]{sectitle.aei.eps}}
\color{white}\right|\color{black}$}}
}

\end{center}

\clearpage\thispagestyle{empty}\cleardoublepage

}
\docphysrept[\begin{docnophysrept}

\chapternonum{Zusammenfassung}

\begin{center}
\Large\mathversion{bold}\bfseries
Der Dilatationsoperator der\\
$\superN=4$ Super Yang-Mills Theorie\\
und Integrabilit\"at\par
\vspace{1cm}
\end{center}
Der Dilatationsoperator mi{\ss}t Skalendimensionen 
von lokalen Operatoren in einer konformen Feldtheorie.
In dieser \docphysrept[Dissertation]{Arbeit} betrachten wir ihn am Beispiel
der maximal supersymmetrischen Eichtheorie in
vier Raumzeit-Dimensionen.
Wir entwicken und erweitern Techniken um den Dilatationsoperator
abzuleiten, 
zu untersuchen 
und anzuwenden.
Diese Werkzeuge sind ideal geeignet um Pr\"azisionstests
der dynamischen AdS/CFT-Vermutung anzustellen.
Insbesondere wurden er im Zusammenhang mit 
Stringtheorie 
auf dem \emph{plane-waves} Hintergrund
(ebenfrontige planare Wellen) 
und dem Thema \emph{spinning strings}
erfolgreich angewendet.

Wir konstruieren den Dilatationsoperator 
ausschlie{\ss}lich mittels algebraischer Methoden:
Indem wir die Symmetriealgebra und strukturelle
Eigenschaften von Feynman-Dia\-gram\-men ausn\"utzen,
k\"onnen wir aufwendige, feldtheoretische Berechnungen 
auf h\"oheren Schleifen umgehen.
Auf diese Weise erhalten wir den kompletten
ein-schleifen Di\-la\-ta\-tions\-ope\-ra\-tor und
die planare drei-schleifen Deformation in einem
interessanten Untersektor.
Diese Resultate erlauben es uns auf das Thema
Integrabilit\"at in vier-dimensionalen planaren Eichtheorien
einzugehen:
Wir beweisen, da{\ss} der komplette 
Di\-la\-ta\-tions\-ope\-ra\-tor auf einer Schleife integrabel ist,
und pr\"asentieren den dazugeh\"origen Bethe-Ansatz.
Weiterhin argumentieren wir, da{\ss} die Integrabilit\"at 
sich bis drei Schleifen und dar\"uber hinaus fortsetzt.
Unter der Annahme der Integrabilit\"at konstruieren
wir schlie{\ss}lich ein neuartiges Spinketten-Modell
auf f\"unf Schleifen und schlagen einen Bethe-Ansatz vor,
der sogar auf beliebig vielen Schleifen g\"ultig sein mag!\par
Wir veranschaulichen den Nutzen unserer Methoden
in zahlreichen Beispielen und stellen zwei 
wichtige Anwendungen im Rahmen der AdS/CFT-Korrespondenz vor: 
Wir leiten aus dem Dilatationsoperator den Hamiltonoperator
der \emph{plane-wave} String-Feld\-theorie her 
und berechnen damit die Energieverschiebung auf dem Torus.
Weiterhin wenden wir den Bethe-Ansatz an,
um Skalendimensionen von Operatoren mit gro{\ss}en 
Quantenzahlen zu finden.
Der Vergleich mit der Energie von 
\emph{spinning strings} Konfigurationen
zeigt eine erstaunliche \"Ubereinstimmung.

{\def\cleardoublepage{\clearpage}\chapternonum{Abstract}}

The dilatation generator measures the scaling dimensions 
of local operators in a conformal field theory.
In this thesis we consider the example of maximally
supersymmetric gauge theory in four dimensions
and develop and extend techniques to derive, investigate and apply 
the dilatation operator. 
These tools are perfectly suited for precision 
tests of the dynamical AdS/CFT conjecture. 
In particular, they have been successfully applied 
in the context of strings on plane waves and spinning strings.

We construct the dilatation operator
by purely algebraic means:
Relying on the symmetry algebra and structural properties 
of Feynman diagrams we are able to bypass involved, 
higher-loop field theory computations.
In this way we obtain the complete
one-loop dilatation operator and the
planar, three-loop deformation in an interesting subsector.
These results allow us to address the issue
of integrability within a planar four-dimensional gauge theory:
We prove that the complete dilatation generator is integrable 
at one-loop and present the corresponding Bethe ansatz.
We furthermore argue that integrability extends to three-loops and beyond.
Assuming that it holds indeed, we finally construct a
novel spin chain model at five-loops and propose
a Bethe ansatz which might be valid at arbitrary loop-order!

We illustrate the use of our technology
in several examples and also present two
key applications for the AdS/CFT correspondence:
We derive the plane-waves string field theory Hamiltonian
from the dilatation operator and compute the
energy shift on the torus.
Furthermore, we use the Bethe ansatz to find
scaling dimensions of operators with 
large quantum numbers. A comparison to 
the energy of spinning strings shows an
intricate functional agreement.

\end{docnophysrept}

\begin{docdophysrept}

\begin{abstract}
In this work we review recent progress in 
four-dimensional
conformal quantum field theories 
and scaling dimensions of local operators. 
Here we consider the example of maximally
supersymmetric gauge theory and present techniques
to derive, investigate and apply the dilatation operator
which measures the scaling dimensions.

  We construct the dilatation operator by purely algebraic means: Relying on the
symmetry algebra and structural properties of Feynman diagrams we are able to
bypass involved, higher-loop field theory computations. In this way we obtain
the complete one-loop dilatation operator and the planar, three-loop deformation
in an interesting subsector. These results allow us to address the issue of
integrability within a planar four-dimensional gauge theory: We prove that the
complete dilatation generator is integrable at one-loop and present the
corresponding Bethe ansatz. We furthermore argue that integrability extends to
three-loops and beyond. Assuming that it holds indeed, we finally construct a
novel spin chain model at five-loops and propose a Bethe ansatz which might be
valid at arbitrary loop-order!
\end{abstract}


\end{docdophysrept}

\finishchapter 
]{}
\docphysrept{\end{frontmatter}}
\tableofcontents

\finishchapter

\docphysrept[\mainmatter]{}
\chapternonum{Introduction}


\begin{docnophysrept}

Probably the two most important advances 
in the deeper understanding of our world
in terms of theoretical physics 
were made at the beginning of the
twentieth century: 
The \emph{theory of general relativity} and \emph{quantum mechanics}. 
On the one hand, Einstein's theory of general relativity (GR) has replaced 
Newton's theory of gravity and, in its original form, 
is still the most accurate theory 
to describe forces between massive bodies. 
It brought about a major change in the
notion of space and time. Both were unified
into spacetime which, in addition, is
curved by the masses that propagate on it. 
Two of the most important conceptual improvements
of GR are \emph{symmetry} and \emph{locality}.
The symmetry of GR is called diffeomorphism-invariance
and allows to label points of spacetime in 
an arbitrary way, the equations of GR do not depend on this. 
Furthermore, GR is a local field theory, there is no 
action at a distance, but instead,
just like in Maxwell's Electrodynamics, 
forces are mediated by a field.

On the other hand, there is Quantum Mechanics.
It proposes a completely new notion of particles and forces,
both of which should be considered as two manifestations 
of the same object. 
It also departed from a deterministic weltanschauung;
a measurement is inevitably probabilistic 
and moreover must be considered as an action which influences
the outcome of future measurements.
Many aspects of quantum mechanics seem odd at first and second sight 
and truly make sense only in a quantum field theory (QFT).
QFT introduced the notion of particle creation and annihilation,
an essential element for local interactions.
The price that has to be paid 
are spurious divergencies due to particles
being created and annihilated at the same place and instant.
It required some effort to understand, \emph{regularise}
and \emph{renormalise} the divergencies in order to 
obtain finite, physical results.
 
The first fully consistent physical QFT was 
Quantum Electrodynamics (QED), the quantum counterpart 
of Electrodynamics. 
A guiding principle in the construction of 
QED is, again, \emph{symmetry}.
Here, the symmetry is given by gauge transformations;
they allow to change some
unphysical degrees of freedom of the theory
by an arbitrary amount. 
Consequently, QED is termed a \emph{gauge theory}
and, in particular, it has an Abelian $\grU(1)$ gauge group.
Amplitudes in QED can be expanded in a coupling constant $g$
related to the fundamental charge of an electron.
This perturbative treatment leads to 
Feynman diagrams which describe interactions in a rather
intuitive fashion.
Besides electromagnetism, two other interactions
between particles have been observed in 
particle accelerators: the weak and the strong interactions. 
Let us discuss the second kind. 
The strong nuclear force is responsible for the binding of
nucleons to nuclei, which would otherwise disperse due to
their electromagnetic charge. 
One of the earlier candidates for a description of these
interactions was a \emph{string theory}, a theory of string-like extended 
objects instead of point-like particles. 
It explained some qualitative aspects of particle (excitation) spectra correctly;
yet, soon it was found that it embodies some 
insurmountable theoretical as well as phenomenological shortcomings
and interest in it declined.

In the meantime, 
an alternative description of strong (and weak)
interactions had emerged. Like QED, it is based on a gauge theory, 
the so-called Quantum Chromodynamics (QCD). 
The gauge group for strong interactions is $\grSU(3)$, 
which may, for instance, be inferred indirectly from the spectrum of hadrons.
Here, symmetry is important for several reasons. 
First of all, and even more so for QCD than for QED, symmetry
is essential for the theoretical consistency of the model. 
Furthermore, the particular gauge group of QCD leads to a 
feature called asymptotic freedom/confinement. 
It implies that QCD is effectively weak at very short distances, 
but becomes infinitely strong at larger dimensions
(on the scale of nucleons). On a qualitative level,
this may be understood as follows: The attraction/repulsion between two charges 
is mediated by flux lines. As opposed to QED, 
in QCD flux lines attract each other
and will form a small tube stretching between the charges. 
The tube effectively behaves like a string with tension and 
binds the particles irrespective of their distance. 
This explains why it is not possible to observe
an individual charged particle and leads us to confinement, 
which allows only uncharged particles to propagate freely.

A peculiarity of generic gauge theories with gauge group $\grU(N)$, 
which we will make heavy use of, 
was observed by 't~Hooft \cite{'tHooft:1974jz}:
He derived a relationship between the topological structure 
of a Feynman graph and its $N$-dependence.
When $1/N$ is interpreted as a coupling constant, he observed that
the perturbative expansion in $1/N$ is very similar in nature to 
the perturbative genus expansion in a generic interacting 
string theory (string field theory).

Not only due to their mathematical beauty, 
the theory of General Relativity and Quantum Mechanics/QFT 
have become the foundations of modern physics, 
but mainly because of the
accuracy to which they describe the world.
On the one hand, gravity is a very weak force and it requires a large
amount of matter to feel its effects. 
Consequently, GR describes the world at very large scales.
For instance, GR was first confirmed when the 
aberration of light near the perimeter of the sun was investigated.
On the other hand, the remaining three forces described by 
QFT's are incomparably stronger.
Therefore quantum field theories chiefly describe the microcosm. 
In particular, the standard model of particle physics, 
the union of the above three gauge theories, has led to 
some non-trivial predictions which have been confirmed with 
unprecedented accuracy, e.g.~the electric moment of the electron
and muon.

One of the major open problems of theoretical physics 
is to understand what happens when an enormous amount
of matter is concentrated on a very small region of space. 
For example, this situation arises at 
the singularity of a black hole or shortly after
the big bang.
To describe such a situation, 
we would need to combine General Relativity with the 
concepts of quantum field theory and consider
\emph{quantum gravity} (QG). 
Despite the better part of a century of research, 
such a unification correctly describing our world 
has not yet been found.
The main obstacle for the direct construction of
a quantum theory of gravity are the divergencies
mentioned above, which cannot be renormalised in this case 
and render the quantum theory meaningless.

Currently, the most favoured theory for a consistent quantisation
of gravity is \emph{superstring theory}. 
It is a refinement of the (bosonic) string theory found in
connection with strong interactions and 
involves an additional symmetry which 
relates fermions and bosons, namely \emph{supersymmetry}.
Supersymmetry makes string theory very appealing to theorists:
It overcomes several of the shortcomings of bosonic string theory
and restricts the form such that there are only five types of 
string theories (IIA,IIB,I,HO,HE), which were, moreover, argued to be
equivalent via duality. This is a very good starting point for a 
\emph{theory of everything}, given that string theory
not only naturally incorporates gravity, but also 
gauge theories, the type of theory on which the standard model is based.

With the advent of superstring theory, supersymmetry has
been applied to field theories as well, giving rise
to beautiful structures. 
One important aspect is that many of the divergencies 
observed in ordinary QFT's are absent in supersymmetric ones.
Indeed, this is the case for 
the unique, maximally supersymmetric gauge theory 
in four spacetime dimensions,
\emph{$\superN=4$ super Yang-Mills theory} 
($\superN=4$ SYM) \cite{Gliozzi:1977qd,Brink:1977bc}. 
This remarkable feature 
\cite{Sohnius:1981sn,Mandelstam:1983cb,Howe:1984sr,Brink:1983pd}
allows the theory to be conformally invariant,
even at the quantum level!
Conformal symmetry is a very constraining 
property in field theory. Most importantly,
two-point and three-point correlation functions 
are completely determined by the scaling dimensions
and structure constants of the involved local operators.
For instance, the two-point function of a scalar operator $\Op$ of
dimension $D$ must be of the form 
\[\label{eq:Intro.TwoPoint}
\bigvev{\Op(x)\,\Op(y)}=\frac{M}{|x-y|^{2D}}\,,
\]
where $M$ is an unphysical normalisation constant.
In two dimensions, conformal symmetry is even more
powerful, it makes a theory 
mathematically quite tractable 
and leads to a number of exciting phenomena such as integrability.
Consequently, it plays a major role 
in the world-sheet description of string theory 
and was thoroughly investigated.
In four dimensions, however, conformal invariance 
appeared to be more of a shortcoming at first sight: 
It makes the model incompatible
with particle phenomenology, which might be the reason why 
$\superN=4$ SYM was abandoned soon after its discovery.

New interest in this theory was triggered by 
the \emph{AdS/CFT correspondence}.
Inspired by the studies of string/string dualities and D-branes,
Maldacena conjectured that IIB string theory 
on the curved background%
\footnote{This manifold consists of the five-sphere and 
the five-dimensional anti-de Sitter spacetime,
which is an equivalent of hyperbolic space but
with Minkowski signature.}
$AdS_5\times S^5$
should be equivalent to $\superN=4$ SYM 
\cite{Maldacena:1998re,Gubser:1998bc,Witten:1998qj}
(see \cite{Aharony:1999ti,D'Hoker:2002aw} for comprehensive 
reviews of the subject)
and thus substantiated the gauge/string duality proposed
earlier by 't~Hooft.
The correspondence is supported by the well-known fact
that the symmetry groups of both theories, $\grPSU(2,2|4)$, match.
Consequently, the representation theory of the 
\emph{superconformal algebra} $\alPSU(2,2|4)$ \cite{Dobrev:1985qv}
was investigated more closely \cite{Andrianopoli:1998ut,Andrianopoli:1999vr},
and numerous non-renormalisation theorems were derived
(see e.g.~\cite{Lee:1998bx}).
In addition, some unexpected non-renormalisation theorems, 
which do not follow from $\alPSU(2,2|4)$ representation theory,
were found \cite{Arutyunov:2000ku}. 
Once thought to be somewhat boring,
it gradually became clear that conformal $\superN=4$ gauge theory 
is an extremely rich and non-trivial theory with many hidden secrets;
eventually, the correspondence has helped in formulating 
the right questions to discover some of them.

\end{docnophysrept}


\begin{docdophysrept}
Conformal quantum field theories have been fascinating theoretical
physicists for a long time. In two dimensions they are arguably
the most important class of field theories: Mathematically, 
they are quite tractable once one puts to full use the fact that
the conformal algebra has an infinite number of
generators. In consequence many exact results on the spectrum
of operators and the structure of correlation functions may be
derived. Apart form their pivotal importance to string theory,
they are physically of great value due to their relationship
with critical phenomena and integrable models of statistical
mechanics. For example, in many cases the representation theory
of 2D conformal field theories fixes the scaling dimensions of
local operators, which in turn are often related to critical exponents
of experimentally relevant systems of solid state physics.

In four dimensions conformal symmetry was long believed to play only a
minor role. The algebra has only finitely many generators, the
QFT's relevant to particle physics are certainly not conformal
and the only, trivial, example seemed to be free, massless field theory.
However, after the discovery of supersymmetry, it has become clear
that supersymmetric gauge theories can be exactly conformally
invariant on the quantum level! 
In many cases the phase diagram
of such gauge theories containes conformal points or regions,
quite analogously to phase diagrams in two dimensions. 
There is therefore obvious theoretical interest in increasing our understanding
of these phases. In particular, the 4D gauge theory with the maximum
possible number $\superN=4$ of rigid supersymmetries, discovered in 1976 
\cite{Gliozzi:1977qd,Brink:1977bc}, has a superconformal phase
\cite{Sohnius:1981sn,Howe:1984sr,Brink:1983pd}.

A few years earlier, 't~Hooft made a connection between
gauge theories with gauge group $\grU(N)$ and 
the world-sheet theory of strings \cite{'tHooft:1974jz}.
He derived a relationship between the two-dimensional
topological structure of a Feynman graph and its $N$-dependence.
When, for large $N$, the quantity $1/N$ is interpreted as a coupling constant, 
he found that the perturbative expansion in $1/N$ is very similar in nature to 
the perturbative genus expansion in a generic interacting 
string theory. Although a precise relationship to string theory 
remained unclear for a long time, large $N$ gauge theories and 
the classification into planar and non-planar graphs have
deepened our understanding of the strong interactions and 
quantum field theory in general.

The gauge/string duality proposed by 't~Hooft 
and conformal $\superN=4$ Super Yang-Mills theory (SYM) 
were combined within the \emph{AdS/CFT correspondence}.
Inspired by the studies of string/string dualities and D-branes,
Maldacena conjectured that IIB string theory 
on the curved background%
\footnote{This manifold consists of the five-sphere and 
the five-dimensional anti-de Sitter spacetime,
which is an equivalent of hyperbolic space but
with Minkowski signature.}
$AdS_5\times S^5$
should be equivalent to $\superN=4$ SYM 
\cite{Maldacena:1998re,Gubser:1998bc,Witten:1998qj}
(see \cite{Aharony:1999ti,D'Hoker:2002aw} for comprehensive 
reviews of the subject).
The correspondence is supported by the well-known fact
that the symmetry groups of both theories, $\grPSU(2,2|4)$, match.
Consequently, the representation theory of the 
\emph{superconformal algebra} $\alPSU(2,2|4)$ \cite{Dobrev:1985qv}
was investigated more closely \cite{Andrianopoli:1998ut,Andrianopoli:1999vr},
and numerous non-renormalisation theorems were derived
(see e.g.~\cite{Lee:1998bx}).
In addition, some unexpected non-renormalisation theorems, 
which do not follow from $\alPSU(2,2|4)$ representation theory,
were found \cite{Arutyunov:2000ku}. 
Once thought to be somewhat boring,
it gradually became clear that conformal $\superN=4$ gauge theory 
is an extremely rich and non-trivial theory with many hidden secrets;
eventually, the correspondence has helped in formulating 
the right questions to discover some of them.

\end{docdophysrept}


Yet, the conjecture goes beyond kinematics 
and claims the full dynamical agreement of both theories. 
For example, it predicts that the spectrum of scaling dimensions $D$ in
the conformal gauge theory should coincide with the 
spectrum of energies $E$ of string states
\[\label{eq:Intro.AdSSpectrum}
\set{D}=\set{E}.
\]
\begin{docdophysrept}%
The scaling dimension of a local operator 
determines the behaviour of the operator under scale transformations.
Scaling dimensions appear most prominently within two-point
functions of local operators whose structure is completely fixed
by conformal symmetry.
For instance, the two-point function of a scalar operator $\Op$ of
dimension $D$ must be of the form 
\[\label{eq:Intro.TwoPoint}
\bigvev{\Op(x)\,\Op(y)}=\frac{M}{|x-y|^{2D}}\,,
\]
where $M$ is an unphysical normalisation constant.%
\end{docdophysrept}%
Unfortunately, like many dualities, 
Maldacena's conjecture is of the strong/weak type: 
The weak coupling regime of gauge theory maps to the
strong coupling (i.e.~tensionless) regime in string theory
and vice versa.
The precise correspondence is given by
\[\label{eq:Intro.AdSCoupling}
\gym^2 N=\lambda=\frac{R^4}{\alpha^{\prime\,2}}\,,
\qquad
\frac{1}{N}=\frac{4\pi g\indup{s}}{\lambda}\,,
\]
where $\gym$ is the Yang-Mills coupling constant 
and $\alpha'$ is the inverse string tension.%
\footnote{The actions are inversely related
to these constants, $S\indup{YM}\sim 1/\gym^2$ 
and $S\indup{string}\sim 1/\alpha'$.
Therefore, quantum effects are suppressed at small
$\gym$ and small $\alpha'$ in the respective theories.}
Furthermore, $N$ is the rank of the $\grU(N)$ gauge group 
of Yang-Mills theory, $\lambda$ is the effective 't~Hooft 
coupling constant in the large $N$ limit, 
$g\indup{s}$ is the topological expansion
parameter in string theory and $R$ is the radius of the 
$AdS_5\times S^5$ background.
%
%
It is not known how to fully 
access the strong coupling regime in either 
theory, let alone how to rigorously quantise string theory 
on the curved background.
Therefore, the first tests of the AdS/CFT correspondence
were restricted to the infinite tension regime of string theory
which is approximated by supergravity
and corresponds to the strong coupling regime on the gauge theory side.
Gauge theory instanton calculations of four-point functions
of operators which are protected by supersymmetry
were shown to agree with the supergravity results see
e.g.~\cite{Bianchi:1998nk}. 

Despite a growing number of confirmations of the conjecture
in sectors protected by symmetry, 
the fundamental problem of a strong/weak duality remained. 
For example, the AdS/CFT correspondence predicts that
the scaling dimensions $D$ of generic, unprotected operators 
in gauge theory should scale as 
\[\label{eq:Intro.AdSGenericEnergy}
D\sim \lambda^{1/4}\]
for large $\lambda$, but how could this conjecture be tested?
It was Berenstein, Maldacena and Nastase (BMN) who proposed
a limit where this generic formula may be evaded \cite{Berenstein:2002jq}:
In addition to a large $\lambda$, consider local operators
with a large charge $J$ on $S^5$, 
whose scaling dimension $D$ is separated from the charge $J$
by a finite amount only. 
More explicitly, the limit proposed by BMN is 
\[\label{eq:Intro.BMNLimit}
\lambda,J\longrightarrow \infty\qquad
\mbox{with}\qquad
\lambda'=\frac{\lambda}{J^2}\quad
\mbox{and} \quad
D-J
\quad \mbox{finite}.
\]
In this limit, the $AdS_5\times S^5$ background 
effectively reduces to a so-called \emph{plane-wave} background
\cite{Blau:2001ne,Blau:2002dy}
on which the spectrum of string modes 
can be found exactly and 
the theory can be quantised \cite{Metsaev:2001bj,Metsaev:2002re}.
Remarkably, the light-cone energy $E\indups{LC}$ of a
string-mode excitation
\[\label{eq:Intro.BMNEnergy}
E\indups{LC}=\sqrt{1+\lambda' n^2}=1+\half \lambda' n^2+\ldots
\]
has a perturbative expansion 
at a small effective coupling constant $\lambda'$.
As the light-cone energy corresponds to the combination $D-J$ 
in gauge theory, suddenly the possibility of a quantitative comparison for 
unprotected states had emerged!
Indeed, BMN were able to show the agreement 
at first order in $\lambda'$ for a set of operators.
Their seminal article \cite{Berenstein:2002jq} 
has sparked a long list of further investigations
and we would like to refer the reader to 
\cite{Pankiewicz:2003pg,Plefka:2003nb,Kristjansen:2003uy,Sadri:2003pr,Russo:2004kr} 
for reviews.
Let us only comment on one direction of research:
In its original form, the BMN limit was proposed only 
for non-interacting strings and gauge theory in the planar limit.
Soon after the BMN proposal, it was demonstrated that also
non-planar corrections can be taken into 
account in gauge theory \cite{Kristjansen:2002bb,Constable:2002hw},
they correspond to energy shifts due to string interactions
\cite{Spradlin:2002ar}. 
In gauge theory, the effective genus counting parameter in the 
so-called \emph{double-scaling limit} is $g_2=J^2/N$.
The first order correction in $\lambda'$ and $g_2^2$ was computed
in \cite{Beisert:2002bb,Constable:2002vq} 
and was argued to agree with string theory 
\cite{Roiban:2002xr,Pearson:2002zs}.
This is yet another confirmation of the AdS/CFT correspondence,
but for the first time within an interacting string theory!

In the study of the BMN correspondence, the attention
has been shifted away from lower dimensional operators
to operators with a large number of constituent fields
\cite{Kristjansen:2002bb,Constable:2002hw,Beisert:2002bb,Constable:2002vq,Gross:2002su}.
There, the complications are mostly of a combinatorial nature.
It was therefore desirable to develop efficient methods 
to determine anomalous dimensions without having to deal
with artefacts of the regularisation procedure.
This was done in various papers, on the planar 
\cite{Berenstein:2002jq,Gross:2002su,Parnachev:2002kk,Beisert:2002tn,Klose:2003tw} 
and non-planar level 
\cite{Kristjansen:2002bb,Constable:2002hw,Beisert:2002bb,Constable:2002vq,Eynard:2002df,Gursoy:2002yy,Gursoy:2002fj},
extending earlier work on protected half-BPS 
\cite{D'Hoker:1998tz,Penati:1999ba,Penati:2000zv,Penati:2001sv}
and quarter-BPS operators \cite{Ryzhov:2001bp}.
In \cite{Beisert:2002ff} it was realised, 
following important insights in
\cite{Gross:2002mh,Janik:2002bd}, 
that these well-established
techniques can be considerably simplified and extended
by considering the \emph{Dilatation Operator}.
The dilatation operator $\algD$ is one of the generators
of the conformal algebra and it measures the scaling
dimension $D$ of a local operator%
\footnote{To avoid confusion, we will later speak of `states' instead
of local operators.}
$\Op$ 
\[\label{eq:Intro.DilEigen}
\algD\,\Op = D\,\Op.
\]
In general, there are many states and finding the scaling dimension 
is an eigenvalue problem which requires to resolve the mixing of states.
Once the dilatation operator has been constructed, 
it will generate the matrix of scaling dimensions
for any set of local operators of a conformal field theory
in a purely algebraic way
(in \appref{app:Example} we present an 
introductory example of how to apply the dilatation operator).
What is more, scaling dimensions can be obtained 
exactly for all gauge groups and, 
in particular, for the group $\grU(N)$ with finite $N$ \cite{Beisert:2003tq}.
Even two or higher-loop calculations of anomalous dimensions,
which are generically plagued by multiple divergencies,
are turned into a combinatorial exercise!
Using the dilatation operator techniques, 
many of the earlier case-by-case studies 
of anomalous dimensions 
\cite{Penati:2001sv,Anselmi:1997mq,Anselmi:1998ms,Bianchi:1999ge,Bianchi:2000hn,Arutyunov:2001mh,Kotikov:2000pm,Kotikov:2001sc,Dolan:2001tt,Bianchi:2002rw,Arutyunov:2002rs}
were easily confirmed \cite{Beisert:2003tq}.
They furthermore enabled a remarkable all-genus comparison 
between BMN gauge theory and
plane-wave string theory \cite{Spradlin:2003bw}.
The subject of this dissertation is the 
construction and investigation of the dilatation operator
in $\superN=4$ SYM, 
a conformal quantum field theory, in perturbation theory. 
\bigskip

Classical scaling dimensions of states are easily found by
counting the constituent fields weighted by their respective 
scaling dimensions. It is just as straightforward to construct the 
classical dilatation operator to perform this counting. 
Scaling dimensions in a field theory generally 
receive quantum corrections, $D=D(g)$ and consequently the dilatation operator
must receive radiative corrections $\algD=\algD(g)$, too.
In the path integral framework there will be no natural way to obtain 
quantum corrections to the dilatation operator; we will
have to derive them from correlators, for example from two-point functions. 
Now what is the benefit in considering the dilatation operator 
if a conventional calculation uses two-point correlators as well?
There are two major advantages: Firstly, the dilatation generator is computed once 
and for all, while a two-point function will have to be evaluated for each 
pair of states (unless one makes use of some effective vertex 
e.g.~\cite{Beisert:2002bb,Constable:2002vq}).
Secondly, the dilatation operator computes only the scaling dimension $D(g)$.
The two-point function also includes a contribution $M(g)$ 
from the normalisation of states.
These two quantities will have to be disentangled before the 
scaling dimension can be read off from the two-point
function \eqref{eq:Intro.TwoPoint}.
Here, a complicating issue is that in general the normalisation coefficient 
$M(g)$ obtained in field theory is divergent.

A radiative correction to the dilatation operator in the context
of $\superN=4$ SYM has first 
been computed in \cite{Minahan:2002ve,Beisert:2002ff}.%
\footnote{Note that the correction is precisely given by 
the effective vertices found earlier in \cite{Constable:2002hw,Beisert:2002bb}.}
This one-loop correction was restricted to the sector of states composed
from the six scalar fields of the theory only, the so-called $\alSO(6)$ subsector,
on which the one-loop dilatation operator closes.
\medskip

However, there is nothing special about the scalar fields, 
except maybe their conceptual simplicity. 
Generic local operators can as well consist of fermions or 
gauge fields (in the guise of a field strength). 
What is more, we can also apply an
arbitrary number of (covariant) derivatives to the
basic constituent fields. 
In principle, one could now compute the one-loop dilatation operator 
for all fields 
(we shall denote a generic field with derivatives 
by the symbol $\fldW$). 
This is feasible, but certainly much more involved
than the calculations for the $\alSO(6)$ subsector
due to infinitely many types of fields $\fldW$ and 
a complicated structure of spacetime indices in the
expected conformal two-point function, 
see e.g.~\cite{Gursoy:2002yy,Klose:2003tw,Chu:2003ji,Georgiou:2004ty}.

In \cite{Beisert:2003tq} a different approach 
to obtain contributions to the dilatation generator 
has been proposed: Just as in field
theory, all contributing diagrams to a two-point function are written down.
The most complicated part of their computation is to evaluate
the spacetime integrals due to 
vertices of the Feynman diagram.
Nevertheless, the structural result of the integrals is known; it is 
some power of the distance $|x-y|^a$ of the local operators 
multiplied to some function $f(\epsilon)$ of the regulator.%
\footnote{For integrals with open spacetime indices the
result is a linear combination of such terms
with spacetime indices on $(x-y)_\mu$ or $\eta_{\mu\nu}$.}
The power $a$ can be inferred by matching dimensions, 
but the function $f(\epsilon)$ is a genuine result of the integral. 
The crucial idea is not to compute the
function, but to assume the most general singular 
behaviour when the regulator is removed, 
e.g.~$f(\epsilon)=c_{-1}/\epsilon+c_0+c_1\epsilon+\ldots$\,\,.
This allows to write down the contributions to the dilatation operator
in terms of the unknown coefficients $c_k$. Now one can
investigate the structure of the dilatation generator
to simplify and combine the contributions.
Usually, it turns out that there are only a few independent coefficients
which actually contribute to anomalous dimensions.
The proposed trick is to make use of known results or other constraints
to determine these coefficients. 

To derive the complete one-loop dilatation operator, it is useful to 
consider its symmetry. A common practice in physics is to derive some
result only for one component of a multiplet of objects; symmetry
will then ensure that the result applies to all components 
of the same multiplet. The same simplification can be applied to
the one-loop dilatation operator: It was shown in 
\cite{Beisert:2003jj} that superconformal symmetry considerably reduces 
the number of independent coefficients to just a single
infinite sequence. This sequence was subsequently
evaluated in field theory. Furthermore, it was conjectured
that this last step might be unnecessary and making
full use the symmetry algebra would \emph{constrain the complete
one-loop dilatation operator uniquely} up to an overall
constant (the coupling constant).
This is indeed the case as we shall prove in this work.
Put differently, superconformal symmetry and some basic facts from field
theory (i.e.~the generic structure of a one-loop contribution)
completely determine all two-point functions at the one-loop level!
To outline the form of the dilatation operator, let us just note
that the radiative correction acts on two fields at a time. 
The contribution $\algD_{12}$ from
a pair of fields depends on their `total spin'%
\footnote{The total spin is a quantity of the representation theory of the
superconformal symmetry similar to the total spin of the rotation group.} 
$j$; it is proportional to the harmonic number 
\[\label{eq:Intro.Harmonic}
\algD_{12}\sim h(j)=\sum_{k=1}^j \frac{1}{k}\,.
\]\medskip

Inspired by the strongly constraining nature of the
superconformal algebra at one-loop, it is natural to expect it to
be very powerful at higher-loops as well. 
This is a very exciting prospect, since direct higher-loop 
computations are exceedingly labourious and not much is
known beyond the one-loop level.
Although one might think that one-loop accuracy is sufficient
for many purposes, one should keep in mind that 
it is only the first non-trivial order. 
Easily one can imagine some unexpected behaviour at 
next-to-leading order and, indeed, we shall encounter 
an example of a mismatch starting only at three-loops.
Furthermore, taken that the one-loop dilatation operator is 
completely constrained, there is hardly any freedom 
for the quantum theory to decide in either direction. 
Therefore, a one-loop computation does not provide much information about 
the \emph{quantum} theory itself.

The trick of writing down the most general structure 
for the dilatation operator with a number of undetermined
coefficients can be used at higher-loops as well.
We will, however, not try to generalise the complete dilatation operator 
to higher-loops. The derivation of the one-loop computation 
depends heavily on a particular feature of perturbation theory
which allows us to restrict to \emph{classical} 
superconformal invariance. 
Unfortunately, it does not apply at higher-loops and we would be left
with a very large number of independent coefficients to be fixed.
To obtain some higher-loop results with as little work as possible,
we may restrict to a subsector. 
The $\alSO(6)$ subsector of scalar fields, however, is not suitable,
there will be mixing with states involving fermions and other fields;
only at one-loop it happens to be closed.
To proceed to higher-loops, one could therefore 
restrict to an even smaller subsector. 
This so-called $\alSU(2)$ subsector consists of only two
charged scalar fields (which we shall denote by $\fldZ$ and $\phi$)
and charge conservation protects the states from mixing 
with more general states.
Here we can derive the two-loop dilatation 
operator by employing some known results without performing 
a full-fledged two-loop field theory computation
\cite{Beisert:2003tq}.

We cannot go much further at the moment because
there are no known results besides a few basic facts 
from representation theory. Symmetry is not very constraining in the
$\alSU(2)$ sector because the dilatation operator is
abelian and not part of a bigger algebra.
A better choice is the $\alSU(2|3)$ subsector: 
It consists of only five fields and 
the symmetry algebra includes the dilatation generator.
These properties make it both, convenient to handle and sufficiently constraining.
Furthermore, not only the dilatation generator, 
but also the other generators of the algebra
receive radiative corrections, a generic feature 
of the higher-loop algebra.
In \cite{Beisert:2003ys} this subsector was investigated 
in the planar limit and up to three-loops 
with an astonishing result:
Although there are \emph{hundreds} of 
independent coefficients at three-loops, closure of the symmetry
algebra 
\[\label{eq:Intro.Alg}
\bigscomm{\algJ_M(g)}{\algJ_N(g)}=\algstr_{MN}^P\, \algJ_P(g)
\]
constrains nearly all of them in such a way that only a \emph{handful}
remain. Moreover, all of them can be related to 
symmetries of the defining equations. Again, 
symmetry in combination with basic field theory 
provides a \emph{unique} answer.
\bigskip

Spectral studies of all the above radiative corrections 
to the dilatation operator reveal a property 
with tremendous importance: 
One finds a huge amount of 
\emph{pairs} of states $\Op_{\pm}$
whose scaling dimensions are exactly degenerate in 
the planar limit
\[\label{eq:Intro.Pairs}
D_+=D_-.
\]
This would not be remarkable if there was an obvious symmetry 
to relate those states. This symmetry, however, cannot be 
superconformal symmetry (or any subalgebra) for two reasons.
Firstly, the degeneracy is actually broken by non-planar corrections
while superconformal symmetry is exact.
Secondly, the degenerate states have a different \emph{parity}
which is preserved by superconformal transformations.
Here, as in the remainder of this \docphysrept[thesis]{work}, parity refers
to complex conjugation of the $\grSU(N)$ gauge group.
To explain the degeneracy we need some generator 
$\charge$ which inverts parity and commutes
with the dilatation generator. 

This curiosity of the spectrum is merely the tip of an iceberg;
it will turn out that the conjectured generator $\charge$ 
is part of an infinite set of commuting charges due to \emph{integrability}.
Integrability of a planar gauge theory
will be the other major topic of this dissertation.
The statement of integrability is equivalent to 
the existence of an unlimited number
of commuting scalar charges $\charge_r$
\[\label{eq:Intro.Commute}
\comm{\charge_r}{\charge_s}=\comm{\algJ}{\charge_r}=0.
\]
The planar dilatation operator $\algdD=g^2\charge_2$ 
is related to the second charge $\charge_2$.
It turns out that the odd charges are parity odd, therefore
the existence of the charge $\charge=\charge_3$ explains
the pairing of states. 
Only a few states have no partner and are unpaired.

Integrable structures play a crucial role in 
two dimensional field theories.  
One of the many intriguing features of two-dimensional CFT's is
that they are intimately connected to integrable 2+0 dimensional
lattice models in statistical mechanics or, equivalently, 
to 1+1 dimensional quantum spin chains. 
The infinite set of charges is directly related to the
infinite-dimensional conformal (Virasoro) algebra in $D=2$.
Given the huge success in understanding CFT's in \emph{two}
dimensions, one might hope that at least some of the aspects
allowing their treatment might fruitfully reappear in \emph{four}
dimensions.
One might wonder about standard no-go theorems that seem to 
suggest that integrability can never exist above $D=2$. These may be
potentially bypassed by the fact that there appears to be a hidden
`two-dimensionality' in $\grU(N)$ gauge theory when we look at it at large $N$
where Feynman diagrams can be classified in terms of 
two-dimensional surfaces. 
\medskip

The first signs of integrability in $\superN=4$ gauge theory 
were discovered by Minahan and Zarembo 
\cite{Minahan:2002ve}. 
They found that the planar one-loop dilatation operator
in the $\alSO(6)$ sector is isomorphic to 
the Hamiltonian of a $\alSO(6)$ integrable quantum spin chain.
The analogy between planar gauge theory and spin
chains is as follows: In the strict large~$N$ limit,
the structure of traces within local operators
cannot be changed and therefore we may consider each 
trace individually or, for simplicity, only
single-trace states. We then interpret the trace
as a cyclic spin chain and the fields within the trace
are the spin sites. For example, the $\alSU(2)$ sector
with two fields $\fldZ,\phi$ maps directly to
the Heisenberg spin chain, in which the spin at each
site can either point up ($\fldZ$) or down ($\phi$). 
For the $\alSO(6)$ sector one considers a more
general spin chain for which the spin can point
in six distinct abstract directions.
The spin chain Hamiltonian alias the planar one-loop 
dilatation generator acts on the spin chain and returns
a linear combination of states. 
The action is of a nearest-neighbour type, 
it can only modify two adjacent spins at a time.
Likewise the higher charges $\charge_r$ act on $r$ 
adjacent spins and are therefore local (along the spin chain).

Integrable spin chains had appeared before in four-dimensional 
gauge theories through
the pioneering work of Lipatov on high energy scattering in planar QCD
\cite{Lipatov:1994yb}. The model was subsequently identified as
a Heisenberg $\alSL(2)$ spin chain of non-compact spin zero
\cite{Faddeev:1995zg}. More recently, and physically closely
related to the present study, further integrable structures were
discovered in the computation of planar one-loop anomalous
dimensions of various types of operators in QCD  
\cite{Braun:1998id,Belitsky:1999qh,Braun:1999te,Belitsky:1999ru,Belitsky:1999bf,Derkachov:1999ze,Belitsky:2003ys,Ferretti:2004ba,Kirch:2004mk}
(see also the review \cite{Belitsky:2004cz}).%
\footnote{While QCD is surely not a conformally invariant quantum field
theory \cite{Callan:1970yg,Symanzik:1970rt}, 
it still behaves like one as far as one-loop anomalous
dimensions are concerned.} 

The full symmetry algebra of SYM is neither 
$\alSO(6)$ nor $\alSL(2)$, but the full superconformal algebra
$\alPSU(2,2|4)$.
If the discovered integrable structures 
are not accidental, we should expect that the $\alSO(6)$ results of
\cite{Minahan:2002ve} and the $\alSL(2)$ results suggested from one-loop QCD 
\cite{Braun:1998id,Belitsky:1999qh,Braun:1999te,Belitsky:1999ru,Belitsky:1999bf,Derkachov:1999ze,Belitsky:2003ys,Ferretti:2004ba,Kirch:2004mk,Belitsky:2004cz}
(see also \cite{Kotikov:2000pm,Kotikov:2001sc,Kotikov:2002ab,Kotikov:2003fb})
can be combined and `lifted' to a full $\alPSU(2,2|4)$ super spin chain.
Indirect evidence can be obtained by the investigation of the spectrum 
of anomalous dimensions. As we have mentioned above, 
the occurrence of pairs of states hints at the existence of 
at least one conserved charge. 
Indeed, the spectrum
of the complete one-loop planar dilatation operator displays 
many such pairs. 
Obviously, they are found in the $\alSO(6)$ and
$\alSL(2)$ subsectors
where integrability is manifest, 
but also generic states do pair up. 
Subsequently, it was shown in \cite{Beisert:2003yb} that 
the complete one-loop planar dilatation operator is 
isomorphic to a $\alPSU(2,2|4)$ supersymmetric spin chain.
\medskip

Integrability is not merely an academic issue, for 
it opens the gates for very precise tests 
of the AdS/CFT correspondence.
It is no longer necessary to compute and diagonalise 
the matrix of anomalous dimensions.
Instead, one may use the Bethe ansatz 
(c.f.~\cite{Faddeev:1996iy} for a pedagogical introduction)
to obtain the one-loop anomalous dimensions directly \cite{Minahan:2002ve,Beisert:2003yb}.
In the thermodynamic limit of very long spin chains, 
which is practically inaccessible by conventional methods,
the algebraic Bethe equations turn into integral equations. 
With the Bethe ansatz at hand, it became possible to compute anomalous dimensions
of operators with large spin quantum numbers
\cite{Beisert:2003xu}.

Via the AdS/CFT correspondence, these states correspond to 
highly spinning string configurations. 
Even though quantisation of string theory on $AdS_5\times S^5$ 
is an open problem,
these spinning strings can be treated in a classical fashion,
c.f.~\cite{Gubser:2002tv,Frolov:2002av},
when interested in the leading large spin behaviour.
It was shown by Frolov and Tseytlin \cite{Frolov:2003tu,Frolov:2003xy} 
that quantum ($1/\sqrt{\lambda}$) corrections in the string theory sigma model
are suppressed by powers of $1/J$, where $J$ is a large spin on the five-sphere $S^5$.
In direct analogy to the plane-wave limit, one obtains an effective
coupling constant 
\[\label{eq:Intro.Spinning}
\lambda'=\frac{\lambda}{J^2}\,.
\]
What makes the low-energy spinning string configurations very appealing 
is that their energies permit an expansion in integer powers of $\lambda'$ around 
$\lambda'=0$ \cite{Frolov:2003qc}.
Just as in the case of the plane-wave/BMN limit one can now 
compare to perturbative gauge theory in a quantitative fashion.
It was found that indeed string energies
and gauge theory scaling dimensions agree at first order in $\lambda'$
\cite{Beisert:2003ea,Arutyunov:2003uj,Khan:2003sm,Engquist:2003rn,Arutyunov:2003za,Larsen:2003tb,Kristjansen:2004ei,Ryang:2004tq,Lubcke:2004dg,Smedback:1998yn,Freyhult:2004iq,Kristjansen:2004za}.
Moreover, the comparison is not based on a single number, but
on a function of the ratio of two spins. Except in a
few special cases, this function is very non-trivial;
it involves solving equations of elliptic or even
hyperelliptic integrals.
The agreement can also be extended to the commuting charges
in \eqref{eq:Intro.Commute}, 
c.f.~\cite{Arutyunov:2003rg,Engquist:2004bx,Arutyunov:2004xy}.
These are merely tests of the spinning string correspondence
and there have been two recent proposals to prove the equivalence
of classical string theory and perturbative gauge theory 
in the thermodynamic limit.
The proposal of Kruczenski is based on 
comparing the string Hamiltonian 
to the dilatation operator 
\cite{Kruczenski:2003gt,Kruczenski:2004kw,Dimov:2004qv,Hernandez:2004uw,Stefanskijr.:2004cw,Kruczenski:2004cn,Tseytlin:2004cj}
(see also the related work \cite{Mateos:2003de,Mikhailov:2003gq,Mikhailov:2004qf,Mikhailov:2004xw})
while Kazakov, Marshakov, Minahan and Zarembo 
find a representation of string theory in terms of
integral equations and compare them to 
the Bethe ansatz \cite{Kazakov:2004qf}.
For a review of the topic of spinning strings 
please refer to \cite{Tseytlin:2003ii}.
\medskip

We have argued that integrability of the
planar gauge theory is, on the one hand, an interesting
theoretical aspect of $\superN=4$ SYM 
and, on the other hand, it allows for precision
tests of the AdS/CFT correspondence. 
So far, however, integrability is only a firm 
result at the one-loop level. 
At higher-loops, it may seem to be inhibited
for the following simple reason: 
The Hamiltonian of an integrable spin chain is usually of 
\emph{nearest-neighbour} type (as for one-loop gauge theories) or, 
at least, involves only two, non-neighbouring spins at a time
(as for the Haldane-Shastry and Inozemtsev integrable spin chains
\cite{Haldane:1988gg,SriramShastry:1988gh,Inozemtsev:1989yq,Inozemtsev:2002vb}).
This structure may appear to be required by the elastic scattering properties 
in integrable models. 
In contrast, higher-loop corrections to the dilatation generator
require interactions of \emph{more than two fields}.
Moreover, the \emph{number of fields is not even conserved} in general
(as in the $\alSU(2|3)$ subsector).
Nevertheless, there are two major reasons to believe in 
higher-loop integrability:
Firstly, the observation of pairing of states in the spectrum
of anomalous dimensions has been shown to extend to  
at least three-loops in the $\alSU(2|3)$ subsector \cite{Beisert:2003ys} 
(see \cite{Klose:2003qc} for the related issue
of integrability in the BMN matrix model)
\[\label{eq:Intro.HigherPairs}
D_+(g)=D_-(g).
\]
At one-loop this degeneracy is explained by integrability, but 
there is no obvious reason why it should extend to
higher-loops unless integrability does.%
\footnote{Pairing may appear to be a weaker statement,
but there are some indications that it is
sufficient to ensure integrability, 
see e.g.~\cite{Grabowski:1995rb,Beisert:2003tq}.}
Moreover, it is possible 
to construct a four-loop correction 
in the $\alSU(2)$ sector with this property
\cite{Beisert:2003tq,Beisert:2003jb}.
Secondly, one might interpret the AdS/CFT correspondence
as one important indication of the validity of integrability:
The classical world sheet theory,
highly non-trivial due to the curved $AdS_5 \times S^5$
background, is integrable  
\cite{Mandal:2002fs,Bena:2003wd,Vallilo:2003nx} 
(for the simpler but related case of plane-wave backgrounds see also
\cite{Maldacena:2002fy,Russo:2002qj,Bakas:2002kt,Alday:2003zb}).
\medskip

It seems that spin chains with interactions of many 
spins or \emph{dynamic} spin chains with a fluctuating number
of spin sites have not been considered so far.%
\footnote{The higher charges of an integrable spin chain
are indeed of non-nearest neighbour type. Nevertheless, they cannot yield 
higher-loop corrections because they commute among themselves,
whereas the higher-loop corrections in general do not.}
Yet, their apparent existence 
\cite{Beisert:2003tq,Klose:2003qc,Beisert:2003jb,Beisert:2003ys}
is fascinating. 
The novelty of such a model, however, comes along with a lack
of technology to investigate it. For instance,
we neither know how to construct higher commuting charges or
even prove integrability, nor is there an equivalent of
the Bethe ansatz to push the comparison with spinning strings 
to higher loops.

A first step to overcome those difficulties 
has been taken by Serban and Staudacher
who found a way to match the
Inozemtsev integrable spin chain \cite{Inozemtsev:1989yq,Inozemtsev:2002vb}
to the three-loop results in gauge theory \cite{Serban:2004jf}. 
The Bethe ansatz for the Inozemtsev spin
can thus be used to obtain exact planar three-loop anomalous
dimensions in gauge theory.
They have furthermore pushed 
the successful comparison of \cite{Beisert:2003ea} 
to higher-loops and
found that the agreement persists at two-loops.
The agreement was subsequently generalised to a
matching of integral equations or Hamiltonians in
\cite{Kazakov:2004qf,Kruczenski:2004kw}.

However, at three-loops the \emph{string theory prediction
turned out not to agree with gauge theory}. 
This parallels a discrepancy starting at three-loops which has been 
observed earlier in the near 
plane-wave/BMN correspondence \cite{Callan:2003xr,Callan:2004uv}.
These puzzles have not been resolved at the time this work was
written and we shall comment on some possible explanations,
such as \emph{an order of limits problem} and 
\emph{wrapping interactions}, in the main text. 
Here we mention only one, even if unlikely:
The AdS/CFT correspondence might not be exact after all. 
Irrespective of the final word on this issue,
we have learned that it is not always sufficient
to restrict to the leading, one-loop order, but there are
interesting and relevant effects to be found at higher-loops.

To deepen our understanding of the string/gauge correspondence,
whether or not exact, it would be useful to 
know the quantitative difference. 
Unfortunately, starting at four-loops, 
the Inozemtsev spin chain has a scaling behaviour in
the thermodynamic limit which does not agree with the one of string theory;
consequently it makes no sense to compare beyond three-loops. 
However, there is a proposal for an integrable spin chain
with the correct scaling behaviour even at four-loops \cite{Beisert:2003jb}.
In \cite{Beisert:2004hm} a Bethe ansatz is presented
which accurately reproduces the spectrum of the four-loop 
(and even five-loop) spin chain. 
What is more, the Bethe ansatz has a natural generalisation 
to all-loops, which incidentally reproduces the BMN energy formula 
\eqref{eq:Intro.BMNEnergy}.
In principle, this allows to compute scaling dimensions as
a true function of the coupling constant%
\footnote{The ansatz cannot deal with short 
states correctly, it should only be trusted when the 
number of constituent fields is larger than the loop order.}
and thus overcome some of the handicaps of perturbation theory.
One may hope that the ansatz gives some insight into
gauge theory away from the weak coupling regime.
\bigskip
\bigskip

\begin{docnophysrept}

\emph{Note added:}
This work is based on the author's PhD thesis,
which was submitted to Humboldt University, Berlin.

\end{docnophysrept}

\finishchapter 

\chapternonum{Overview}


This \docphysrept[thesis]{work} is organised as follows: 
The main text is divided into six chapters,
in the first two we investigate generic aspects of 
the dilatation operator and in the remaining
four we will explicitly construct one-loop
and higher-loop corrections and investigate
their integrability.
\begin{bulletlist}
\item[\ref{ch:N4}.]
We start by presenting the 
$\superN=4$ supersymmetric field theory
and review some useful results concerning the 
representation theory of
the superconformal algebra $\alPSU(2,2|4)$ on which we
will base the investigations of the following chapters.

\item[\ref{ch:Dila}.]
We will then investigate some
scaling dimensions and introduce the
dilatation operator as a means to measure them.
Most of the chapter is devoted to the 
discussion of various aspects
of the dilatation operator and its structure.
These include the behaviour in perturbation
theory and how one can consistently 
restrict to certain subsectors of states
in order to reduce complexity.
From an explicit and a conceptual 
computation of two-point functions in 
a subsector we shall learn about the 
structure of quantum corrections to the
dilatation generator. 
Finally, we will investigate the planar limit 
and introduce some notation.

\item[\ref{ch:One}.]
Having laid the foundations, 
we will now turn towards explicit algebraic constructions.
In this chapter we will derive the complete 
one-loop dilatation operator of $\superN=4$ SYM.
The derivation is similar to the
one presented in the article \cite{Beisert:2003jj}, but
here we improve it by replacing the field 
theory calculations by algebraic constraints.

\item[\ref{ch:Int}.]
Next, we introduce the notion of integrability
and a framework to investigate integrable quantum spin chains.
We will then prove the integrability of the 
just derived dilatation generator in the
planar limit. We extend the results of the
article \cite{Beisert:2003yb} by a proof of
a Yang-Baxter equation.
This allows us to write down
the Bethe ansatz for the corresponding 
supersymmetric quantum spin chain.

\item[\ref{ch:Higher}.]
At this point, the investigations of
one-loop scaling dimensions is complete and we proceed to higher-loops. 
For simplicity we will restrict to a subsector with finitely many
fields and the planar limit. 
We demand the closure of the pertinent symmetry algebra,
determine its most general three-loop deformations
\cite{Beisert:2003ys} and find an essentially unique result.
An interesting aspect of the deformations is that
they do not conserve the number of component fields within a state.

\item[\ref{ch:HighInt}.]
In the final chapter we consider
integrability at higher-loops
and argue why it should apply 
to planar $\superN=4$ SYM.
We will then construct deformations
to the Heisenberg spin chain to model
higher-loop interactions; they turn out
to be unique even at five-loops. 
Finally, we present an all-loop Bethe ansatz which 
reproduces the energies of this model.

\end{bulletlist}
The developed techniques are illustrated by several
sample calculations at various places in the text. 
In particular, we will present two important computations 
of scaling dimensions in the context of the AdS/CFT correspondence.
In \secref{sec:One.BMN} we shall compute the
genus-one energy shift of 
two-excitation BMN operators to be compared to 
strings on plane waves. 
The agreement represents the first dynamical test 
including string interactions.
In \secref{sec:Int.Spinning} we consider 
classical spinning strings on $AdS_5\times S^5$ 
and compare them to states with a large spin of $\alSO(6)$
to find an intricate functional agreement.

We then conclude and present 
a list of interesting open questions.
To expand on the main text we present some miscellaneous aspects
in the appendices: 
\begin{bulletlist}
\item[\ref{app:Example}.]
An example to illustrate the application of
the dilatation operator, at finite $N$ or 
in the planar limit.

\item[\ref{app:Spinors}.]
Spinor identities in four, six and ten dimensions.

\item[\ref{app:Ten}.]
A short review of the ten-dimensional supersymmetric gauge theory, 
either in superspace or in components.

\item[\ref{app:U224}.]
The algebra $\alU(2,2|4)$, its commutation relations
and the oscillator representation.

\item[\ref{app:SU2Tools}.]
Some \texttt{Mathematica} functions to deal with
planar interactions in the $\alSU(2)$ subsector
which can be used in the application and construction
of the dilatation operator.

\item[\ref{app:Harm}.]
The harmonic action to compute one-loop scaling dimensions in
a more convenient fashion than by using the abstract formula 
\eqref{eq:Intro.Harmonic}.

\end{bulletlist}

\finishchapter 

\chapter{Field Theory and Symmetry}
\label{ch:N4}

In this chapter we will discuss various,
loosely interrelated aspects 
of $\superN=4$ super Yang-Mills theory,
the superconformal algebra and its representation theory. 
We lay the foundations for the investigations of
the following chapters and introduce our notation, conventions
as well as important ideas.

We will start with a review of
classical $\superN=4$ SYM in \secref{sec:N4.D4}
and its path-integral quantisation in \secref{sec:N4.Quantum}.
In the following two sections we consider 
the gauge group (a generic group in \secref{sec:N4.Gauge}
or a group of large rank in \secref{sec:N4.LargeN})
in a quantum field theory.
In \secref{sec:N4.Alg} we introduce
the superconformal algebra, a central object of this \docphysrept[thesis]{work}.
The remainder of this chapter deals with representation theory.
Firstly, we present our notion of fields and local operators
and relate it to the algebra in \secref{sec:N4.States}.
In \secref{sec:N4.Modules,sec:N4.Split} we consider generic 
highest-weight modules and special properties of multiplets
close the unitarity bounds. 
The multiplet of fields and the current multiplet 
is investigated in \secref{sec:N4.Fund,sec:N4.Currents}.
Finally, in \secref{sec:N4.Corr} we review correlation
functions in a conformal field theory.

\section{$\superN=4$ Super Yang-Mills Theory}
\label{sec:N4.D4}

We start by defining the field theory on which we will focus 
in this work, ${\superN=4}$ maximally supersymmetric 
gauge theory in four dimensions \cite{Gliozzi:1977qd,Brink:1977bc}.%
\footnote{It is convenient to derive the four-dimensional theory with 
$\superN=4$ supersymmetry 
from a ten-dimensional theory with $\superN=1$ supersymmetry. 
In \appref{app:Ten} we shall present 
this ancestor theory.}
It consists of a covariant derivative $\cder$
constructed from the gauge field $\fldA$,
four spinors $\Psi$ as well as 
six scalars $\Phi$ to match the number of bosonic and fermionic 
on-shell degrees of freedom. 
We will collectively 
refer to the fields 
by the symbol $\fldW$ 
\footnote{Of course, the covariant derivative $\cder$ is not a field.
Instead of the gauge field $\fldA$, we shall place it here so that
all `fields' $\fldW$ have uniform gauge transformation properties.}
\[\label{eq:N4.D4.Fields}
\fldWf{A}=(\cder_\mu,\Psi_{\alpha a},\dot\Psi_{\dot\alpha}^a,\Phi_{m}).
\]
Our index conventions are as follows:
Greek letters refer to spacetime 
$\alSO(4)=\alSU(2)\times\alSU(2)$ 
symmetry.%
\footnote{As we are dealing with algebras only, global issues
such as the difference between a group and its double
covering need not concern us.}
Spacetime vector indices $\mu,\nu,\ldots$ take four values, 
spinor indices $\alpha,\beta,\ldots$ of one $\alSU(2)$
and spinor indices $\dot\alpha,\dot\beta,\ldots$ of 
the other $\alSU(2)$ take values $1,2$. 
Latin indices belong to the internal $\alSO(6)=\alSU(4)$ symmetry;
internal vector indices $m,n,\ldots$ take six values whereas 
spinor indices $a,b,\ldots$ take values $1,2,3,4$.
Calligraphic indices $\fldindn{A,B},\ldots$ 
label the fundamental fields in $\fldW$.

\begin{table}\centering
$
\begin{array}{|r|cccc|}\hline
\mbox{signature}             & \eta^{\mu\nu} & \eta^{mn} & \mbox{spacetime sym.} & \mbox{internal sym.} \\\hline
\mbox{physical}              & (3,1)         & (6,0)     & \alSL(2,\Comp)        & \alSU(4) \\
\mbox{Euclidian}             & (4,0)         & (5,1)     & \alSp(1)\times\alSp(1)& \alSL(2,\Quat) \\
\mbox{Minkowski, non-compact}& (3,1)         & (4,2)     & \alSL(2,\Comp)        & \alSU(2,2) \\
\mbox{maximally non-comapct} & (2,2)         & (3,3)     & \alSL(2,\Real)\times\alSL(2,\Real)&\alSL(4,\Real) \\\hline
\mbox{complex}               &   4           & 6         & \alSL(2,\Comp)\times\alSL(2,\Comp)&\alSL(4,\Comp) \\\hline
\end{array}
$
\caption{Possible signatures of spacetime, internal space
and symmetry algebras.}
\label{tab:N4.D4.Signatures}
\end{table}

Let us comment on the signature of the field theory and the algebras.
In order to write down a real-valued Lagrangian, 
the signatures of spacetime and internal space must be correlated,
we have listed the possible choices in \tabref{tab:N4.D4.Signatures}.
The physical choice has Minkowski signature 
and a positive-definite norm for internal space. 
The other choices require an 
internal metric of indefinite signature 
and possibly a spacetime with two time-like directions.
As far as perturbation theory and Feynman diagrams are concerned, 
the signature is irrelevant because we can perform Wick rotations 
at any point of the investigation.
It may therefore be convenient to work with the maximally non-compact signature 
which leads to a completely real theory 
and where conjugation does not play a role.
Alternatively, we can use a complexified spacetime and algebra.
In the following we will not pay much attention to signatures
and assume either the maximally non-compact or 
complex version of the algebra.

We define the covariant derivative 
\[\label{eq:N4.D4.Covariant}
\cder_\mu=\partial_\mu-ig\fldA_\mu,
\qquad
\cder_\mu \fldW:=\comm{\cder_\mu}{\fldW}=
\partial_\mu \fldW-ig \fldA_\mu \fldW+ig\fldW\fldA_\mu,
\]
where we have introduced a dimensionless coupling constant $g$.
Later on, in the quantum theory, $g$ will be an important 
parameter; however, on a classical level, we can 
absorb it completely by rescaling the fields, 
this corresponds to $g=1$.
Throughout this work we will assume the gauge group 
to be $\grSU(N)$ or $\grU(N)$
and represent all adjoint fields $\fldW$ by (traceless) 
hermitian $N\times N$ matrices.
Under a gauge transformation $U(x)\in\grU(N)$ the fields transform
canonically according to 
\[\label{eq:N4.D4.Gauge}
\fldW\mapsto U\fldW U^{-1},\qquad
\fldA_\mu\mapsto U\fldA_\mu U^{-1}-ig^{-1}\, \partial_\mu U \,U^{-1}.
\] 
The gauge field $\fldA$ transforms differently from the other fields
to compensate for the non-covariant 
transformation of the partial derivative within $\cder$. 
The covariant derivative $\cder$ is not truly a field, 
it must always act on some other field.
Nevertheless we can construct a field 
from the gauge connection alone, the field strength $\fldF$. 
Together with the associated Bianchi identity, it is given by
\[\label{eq:N4.D4.FieldStrength}
\fldF_{\mu\nu}=ig^{-1}\comm{\cder_\mu}{\cder_\nu}=
\partial_\mu \fldA_\nu-\partial_\nu \fldA_\mu-ig\comm{\fldA_\mu}{\fldA_\nu},\qquad
\cder_{[\rho}\fldF_{\mu\nu]}=0.
\]

After these preparations we can write down the 
Lagrangian of $\superN=4$ supersymmetric Yang-Mills theory. It is
\<\label{eq:N4.D4.Lagr}
\Lagr\indups{YM}[\fldW]\eq
\sfrac{1}{4}\Tr\fldF^{\mu\nu}\fldF_{\mu\nu}
+\sfrac{1}{2}\Tr \cder^\mu\Phi^n\cder_\mu\Phi_n
-\sfrac{1}{4}g^2\Tr \comm{\Phi^m}{\Phi^n}\comm{\Phi_m}{\Phi_n}
\nlnum\nonumber
+\Tr \dot\Psi^a_{\dot\alpha}\sigma_\mu^{\dot\alpha\beta}\cder^\mu\Psi_{\beta a}
-\sfrac{1}{2}ig\Tr\Psi_{\alpha a}\sigma_m^{ab}\varepsilon^{\alpha\beta}
\comm{\Phi^m}{\Psi_{\beta b}}
-\sfrac{1}{2}ig\Tr\dot\Psi^a_{\dot\alpha}
\sigma^m_{ab}\varepsilon^{\dot\alpha\dot\beta}
\comm{\Phi_m}{\dot\Psi^b_{\dot\beta}}.
\>
In addition to the standard kinetic terms for the gauge field,
spinors and scalars, there is a quartic coupling of the
scalars and a cubic coupling of a scalar and two 
spinors.
The symbols $\varepsilon$ are the totally antisymmetric 
tensors of $\alSU(2)$ and $\alSU(4)$. 
The matrices $\sigma^\mu$ and $\sigma^m$ are the chiral projections of the 
gamma matrices in four or six dimensions, respectively. 
They have the symmetry properties
$\sigma^\mu_{\dot\alpha\beta}=\sigma^\mu_{\beta\dot\alpha}$,
$\sigma^m_{ab}=-\sigma^m_{ba}$
and satisfy the relations%
\footnote{The brackets $\{\ldots\}$ at index level indicate a symmetric projection
of enclosed indices. Likewise $[\ldots]$ and $(\ldots)$ correspond to a 
antisymmetric and symmetric-traceless projection with respect to 
the metric $\eta$.}
\[\label{eq:N4.D4.Sigma}
\sigma^{\{\mu}\sigma^{\nu\}}=\eta^{\mu\nu},
\quad
\sigma^{\{m}\sigma^{n\}}=\eta^{mn},
\]
when considered as matrices which are 
summed over a pair of alike upper and lower intermediate indices.
Please refer to \appref{app:Spinors} for a number of useful identities and conventions. 
The equations of motion which follow from this action are
\<\label{eq:N4.D4.EOM}
\cder_\nu\fldF^{\mu\nu}\eq 
ig\comm{\Phi_n}{\cder^\mu \Phi^n}
-ig \sigma_\mu^{\dot\alpha\beta}
\acomm{\dot\Psi^a_{\dot\alpha}}{\Psi_{\beta a}},
\nln
\cder_\nu\cder^\nu\Phi^m\eq 
-g^2\comm{\Phi_n}{\comm{\Phi^n}{\Phi^m}}
+\sfrac{1}{2}ig\sigma^{m,ab}\varepsilon^{\alpha\beta}
\acomm{\Psi_{\alpha a}}{\Psi_{\beta b}}
+\sfrac{1}{2}ig\sigma^m_{ab}\varepsilon^{\dot\alpha\dot\beta}
\acomm{\dot\Psi^a_{\dot\alpha}}{\dot\Psi^b_{\dot\beta}},
\nln
\sigma_\mu^{\dot\alpha\beta}
\cder^\mu\Psi_{\beta a}\eq 
ig\varepsilon^{\dot\alpha\dot\beta}\sigma^m_{ab}
\comm{\Phi_m}{\dot\Psi_{\dot\beta}^b},
\nln
\sigma_\mu^{\alpha\dot\beta}
\cder^\mu\dot\Psi_{\dot\beta}^a\eq 
ig\varepsilon^{\alpha\beta}\sigma_m^{ab}
\comm{\Phi^m}{\Psi_{\beta b}}.
\>

It can be shown that the action and the equations of motion are
invariant under the $\superN=4$ super Poincar\'e algebra.
It consists of the manifest Lorentz and internal rotation symmetries
$\algL,\algLd,\algR$ of $\alSU(2)\times\alSU(2)\times\alSU(4)$
as well as the (super)translations $\algQ,\algQd,\algP$.
The (super)translation variations are parameterised by 
the fermionic and bosonic shifts 
$\epsilon^{\alpha}_{a},\dot\epsilon^{\dot\alpha a}$ and $e^\mu$
\[\label{eq:N4.D4.Vary}
\delta_{\epsilon,\dot\epsilon,e}=
\epsilon^{\alpha}_{a}\algQ^a{}_\alpha
+\dot\epsilon^{\dot\alpha a}\algQd_{\dot \alpha a}
+e^{\mu}\algP_{\mu}.
\]
The action of the variation on the fundamental fields 
$\delta_{\epsilon,\dot\epsilon,e}\fldW:=
\comm{\delta_{\epsilon,\dot\epsilon,e}}{\fldW}$
is given by
\<\label{eq:N4.D4.Trans}
\delta_{\epsilon,\dot\epsilon,e} \cder_\mu \eq 
 ig \epsilon^\alpha_{a}\varepsilon_{\alpha\beta}\sigma_\mu^{\beta\dot\gamma}\dot\Psi^a_{\dot\gamma}
+ig \dot\epsilon^{a \dot\alpha}\varepsilon_{\dot\alpha\dot\beta}\sigma_\mu^{\dot\beta\gamma}\Psi_{\gamma a}
+ig e^\nu\fldF_{\mu\nu},
\nln
\delta_{\epsilon,\dot\epsilon,e} \Phi_m \eq 
\epsilon^\alpha_{a}\sigma_m^{ab}\Psi_{\alpha b}
+\dot\epsilon^{a \dot\alpha}\sigma_{m,ab}\dot\Psi^b_{\dot\alpha}
+e^\mu \cder_\mu\Phi_m,
\nln
\delta_{\epsilon,\dot\epsilon,e} \Psi_{\alpha a}\eq
-\half \sigma^\mu_{\alpha\dot\beta}\varepsilon^{\dot\beta\dot\gamma}\sigma^\nu_{\dot\gamma\delta}
\epsilon^\delta_{a}\fldF_{\mu\nu}
+\half ig \sigma^m_{ab}\sigma_n^{bc}\varepsilon_{\alpha\beta}\epsilon^\beta_{c} \comm{\Phi_m}{\Phi^n}
\nl
+\sigma^n_{ab}\sigma^\mu_{\alpha\dot\beta}
\dot\epsilon^{b \dot\beta}\cder_\mu\Phi_n
+e^\mu \cder_\mu \Psi_{\alpha a}
,
\nln
\delta_{\epsilon,\dot\epsilon,e} \dot\Psi^a_{\dot\alpha}\eq
-\half \sigma^\mu_{\dot\alpha\beta} \varepsilon^{\beta\gamma} 
\sigma^\nu_{\gamma\dot\delta} 
\dot\epsilon^{a \dot\delta} \fldF_{\mu\nu}
+\half ig\sigma_m^{ab}\sigma^n_{bc}
\varepsilon_{\dot\alpha\dot\beta}  \dot\epsilon^{c \dot\beta} \comm{\Phi^m}{\Phi_n}
\nl
+\sigma_n^{ab}\sigma^\mu_{\dot\alpha\beta}
\epsilon^\beta_{b}\cder_\mu\Phi^n
+e^\mu \cder_\mu \dot\Psi^a_{\dot\alpha}.
\>
The algebra of supertranslations resulting from these variations is given by
\[\label{eq:N4.D4.Alg}
\arraycolsep0pt\begin{array}[b]{rclcrcl}
\acomm{\algQ^a{}_{\alpha}}{\algQ^b_{\beta}}\eq 
-2ig \epsilon_{\alpha\beta}\sigma_m^{ab}\Phi^m,
&\qquad&
\comm{\algP_\mu}{\algQ^a{}_\alpha}\eq 
-ig \varepsilon_{\alpha\beta}\sigma^{\beta\dot\gamma}_\mu \dot\Psi^a_{\dot\gamma},
\\[5pt]
\acomm{\algQd_{\dot\alpha a}}{\algQd_{\dot\beta b}}\eq 
-2ig \epsilon_{\dot\alpha\dot\beta}\sigma^m_{ab}\Phi_m,
&&
\comm{\algP_\mu}{\algQd_{\dot\alpha a}}\eq 
-ig \varepsilon_{\dot\alpha\dot\beta}\sigma^{\dot\beta\gamma}_\mu \Psi_{\gamma a},
\\[5pt]
\acomm{\algQ^a{}_{\alpha}}{\algQd_{b\dot\beta}}\eq 
2 \delta^a_b\sigma^\mu_{\alpha\dot\beta}\algP_\mu,
&&
\comm{\algP_\mu}{\algP_\nu}\eq -ig\fldF_{\mu\nu},
\end{array}
\]
up to terms proportional to the equations of motion.%
\footnote{It is a common feature of supersymmetric theories that
the algebra closes only on-shell. 
Here, it is related to the fact that the 
equations of motion \eqref{eq:Ten.Super.EOM}
are implied by the constraint \eqref{eq:Ten.Super.Constraint}
which is used in the reduction of superspace fields to 
their top level components.
For theories with less supersymmetry
one can introduce auxiliary fields or work in
superspace to achieve off-shell supersymmetry.}
Note that the action of the generators $\algJ$ 
on a combination of fields $X$ should be read as
$\comm{\algJ}{X}$. 
When $X$ is a covariant combination of fields,
the above commutators therefore yield $\comm{\fldW}{X}$,
where $\fldW$ is the field which appears on the 
right-hand side of \eqref{eq:N4.D4.Alg}.
For a gauge invariant combination $X$ all
fields drop out and only the momentum generator
$\algP$ acts non-trivially $\comm{\algP_\mu}{X}=\partial_\mu X$.

\begin{figure}\centering
\includegraphics{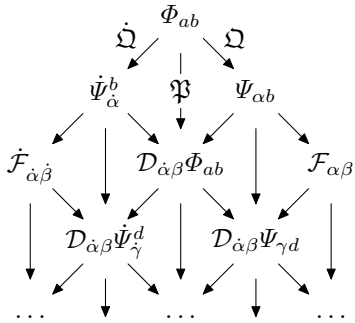}
\caption{Classical supertranslation variations of the fields in 
spinor notation. Left, vertical and right arrows correspond 
to generators $\algQd$, $\algP$ and $\algQ$, respectively.
We have dropped all commutators of fields
which are suppressed for a vanishing coupling constant $g$.}
\label{fig:N4.D4.Multiplet}
\end{figure}

As a more unified notation, it is possible to replace
all vector indices $\mu,\nu,\ldots,m,n,\ldots$ by a 
pair of spinor indices
by contracting with the $\sigma$ symbols
\<\label{eq:N4.D4.Spinors}
\cder_\mu\earel{\sim}\sigma_{\mu}^{\dot \alpha\beta}\cder_{\dot\alpha\beta},
\nln
\fldF_{\mu\nu}\earel{\sim}
\sigma_\mu^{\alpha\dot\gamma}\varepsilon_{\dot\gamma\dot\delta}
\sigma_\nu^{\dot\delta\beta} 
\fldF_{\alpha\beta}+
\sigma_\mu^{\dot\alpha\gamma}\varepsilon_{\gamma\delta}
\sigma_\nu^{\delta\dot\beta} 
\dot\fldF_{\dot\alpha\dot\beta},
\nln
\Phi_m\earel{\sim}\sigma_{m}^{ba}\Phi_{ab}.
\>
In this notation $\Phi_{ab}$ is antisymmetric while
$\fldF_{\alpha\beta}$ and $\dot\fldF_{\dot\alpha\dot\beta}$
are both symmetric.
Using identities in \appref{app:Spinors}, 
one can remove all explicit $\sigma$'s from 
the action and equations of motion and replace them by 
totally antisymmetric $\varepsilon$ tensors
of $\alSU(2),\alSU(2),\alSU(4)$. 
We will not do this explicitly here, but note that the 
set of fields (together with the covariant derivative) is given by 
\[\label{eq:N4.D4.FieldSpin}
\fldW=(\cder_{\dot\alpha\beta},
\Phi_{ab},
\Psi_{\alpha b},
\dot\Psi^b_{\dot\alpha},
\fldF_{\alpha\beta},
\dot\fldF_{\dot\alpha\dot\beta}),
\]
all of which are bi-spinors. For Minkowski signature the dotted fields 
would be related to the undotted ones by complex conjugation. 
Here we will consider them to be independent and real as
for a spacetime of signature $(2,2)$.
The structure of supersymmetry transformations of these fields,
depicted in \figref{fig:N4.D4.Multiplet}, 
is of elegant simplicity.
Generators $\algQ$ simply change a
$\alSU(4)$ index into an undotted $\alSU(2)$ index,
whereas generators $\algQd$ add both a $\alSU(4)$ index and 
a dotted $\alSU(2)$ index.
The momentum generator adds both an 
undotted $\alSU(2)$ index and a dotted $\alSU(2)$ index.
We will come back to this in \secref{sec:N4.Fund},
where we will represent the fields and generator in terms
of a set of harmonic oscillators.

The $\superN=4$ gauge theory is pure in the sense that
it consists only of the superspace gauge field,
c.f.~\appref{app:Ten.Super}.
As such it must be a massless theory and enjoys
an enhancement of Poincar\'e symmetry to
\emph{conformal} symmetry. 
Even more, conformal symmetry
and super(translation)symmetry join to form \emph{superconformal} symmetry.
We will discuss this symmetry in detail in \secref{sec:N4.Alg},
here we only note that it yields 
additional special conformal generators or boosts. 
The (super)boosts $\algS,\algSd,\algK$ are essentially the conjugate transformations
of (super)translations $\algQ,\algQd,\algP$.

\section{The Quantum Theory}
\label{sec:N4.Quantum}

There are various ways to quantise a field theory, we will 
consider only the path integral approach. 
The path integral measures the expectation value of
some operator functional $\Op[\fldW]$ by summing
over all field configuration weighted by the exponential 
of the action%
\footnote{We assume the signature of spacetime to be Euclidean.
For Minkowski signature the weight would be $\exp iS$.}
\[\label{eq:N4.Quantum.Path}
\bigvev{\Op[\fldW]}:=\int D\fldW\, \Op[\fldW]\, \exp \bigbrk{-S[\fldW]}.
\]
We assume the 
path integral to be normalised, $\vev{1}=1$.
The Yang-Mills action $S$ is the spacetime integral of the gauge theory 
Lagrangian \eqref{eq:N4.D4.Lagr}
\[\label{eq:N4.Quantum.Action}
S[\fldW]=\frac{2}{\gym^2}\int d^4 x \, \Lagr\indups{YM}[\fldW,g=1],
\]
where we have used the common definition of the Yang-Mills 
coupling constant $\gym$. 
For a reason to be explained in 
\secref{sec:N4.LargeN}, it will be more convenient to work with a 
different coupling constant 
\[\label{eq:N4.Quantum.Couple}
g^2:=\frac{\gym^2N}{8\pi^2}\,,
\]
where $N$ is the rank of the gauge group $\grU(N)$. 
We can easily recast the action 
in the following form
\[\label{eq:N4.Quantum.Action2}
S[\fldW]=
\frac{N}{4\pi^2}\int d^4 x \, \Lagr\indups{YM}[\fldW/g,g=\sqrt{\gym^2N/8\pi^2}\,].
\]
This form yields a convenient normalisation 
for spacetime correlators when 
the fields $\fldW$ are rescaled by $g$.
The rescaling can be absorbed into the
normalisation of the path integral and
we obtain the action to be used in this work
\[\label{eq:N4.Quantum.ActionNorm}
S[\fldW]=N\int \frac{d^4 x}{4\pi^2}\,\Lagr\indups{YM}[\fldW].
\]

There are various expectation values 
which one might wish to compute, let us state a few:
A frequent application is scattering of particles.
Particles are represented by fields
with well-prepared momenta $p_i$ and spins $\epsilon_i$.
One inserts these into the path integral
\[\label{eq:N4.Quantum.Scatter}
F(p_i,\epsilon_i)=\bigvev{\epsilon_1\cdott\Psi(p_1)\,\Phi(p_2)\,\ldots}
\]
and obtains the scattering function $F$ which 
describes the scattering process of the involved particles. 
Another possibility is to insert Wilson loops $\Op[\gamma]$
\[\label{eq:N4.Quantum.Wilson}
F[\gamma]=\bigvev{\Op[\gamma]}.
\]
Wilson loops are operators which are supported on a curve $x=\gamma(\tau)$ in 
spacetime. The function $F[\gamma]$ can, for example, 
be used to describe the potential between two heavy charged objects.
In this work we shall consider local operators $\Op(x)$, 
objects supported at a single point $x$ in spacetime,
and their correlators
\[\label{eq:N4.Quantum.nPoint}
F(x_i)=\bigvev{\Op(x_1)\,\Op(x_2)\,\ldots}.
\]
In particular we will focus on two-point functions
\[\label{eq:N4.Quantum.TwoPoint}
F(x_1,x_2)=\bigvev{\Op(x_1)\,\Op(x_2)},
\]
which are used to measure some generic properties 
of the local operators in question. 
They describe how a particle which is created/annihilated 
by that operator propagates through spacetime.
Local operators will be discussed in detail in \secref{sec:N4.States}.

The symmetries of the theory will be reflected 
by the correlation functions $F$. 
For example, due to translation invariance,
the Wilson-loop expectation value will not depend 
on a global shift of the contour,
$F[\gamma+c]=F[\gamma]$. For the same reason two-point functions 
can only depend on the distance of the two points
$F(x_1,x_2)=F(x_1-x_2)$. There are further constraints on 
two-point functions due to superconformal symmetry
which will be discussed in \secref{sec:N4.Corr}.

However, there is a possible catch about symmetries:
Classical symmetries of the action might not survive 
in the quantum theory. 
In the path integral formalism such \emph{anomalies} arise 
when it is impossible to consistently 
define a measure $D\fldW$ which obeys the symmetry.
In particular, conformal symmetry usually is anomalous.
When quantising a field theory, it is necessary to regularise it
first in order to remove divergencies;
this inevitably requires the introduction of a mass scale $\mu$. 
In the regularised theory $\mu$ breaks conformal symmetry for
which scale invariance is indispensable. 
When, after quantisation, the regulator is removed, 
the correlation functions $F$ usually still depend on the scale $\mu$.
Of course, a physically meaningful result must not depend on the
arbitrary scale. This apparent puzzle is resolved by assuming that the
parameters of the quantum theory also depend on the scale $\mu$
in such a way that the explicit and implicit dependence cancel out. 
In the case at hand, the only parameter is the coupling constant $g$
and its dependence on the scale is described by the 
beta function
\[\label{eq:N4.Quantum.BetaDef}
\beta=\mu\,\frac{\partial g}{\partial \mu}\,.
\]
The appearance of the beta function is related to the breakdown of
scale invariance and conformal symmetry in a massless gauge theory.
For $\superN=4$ SYM, however, the beta function is believed 
to vanish to all orders in perturbation theory as
well as non-perturbatively 
\cite{Sohnius:1981sn,Mandelstam:1983cb,Howe:1984sr,Brink:1983pd}
\[\label{eq:N4.Quantum.BetaZero}
\beta=0.
\]
In other words, (super)conformal symmetry is preserved even at the 
quantum level!
This does not imply, however, that there are no divergencies in $\superN=4$ SYM;
it merely means that, once the operators are properly renormalised, 
all divergencies and scale dependencies drop out in physically meaningful 
quantities.

Let us evaluate the expectation value 
$\vev{\Op[\fldW]}$ in perturbation theory.
Using standard path integral methods we find the generator of 
Feynman diagrams
%
\[\label{eq:N4.Quantum.Feyn}
\bigvev{\Op[\fldW]}=
\bigbrk{
\exp(W_0[\partial/\partial\fldW])
\exp(-S\indup{int}[g,\fldW])\,
\Op[\fldW]
}_{\fldW=0},\qquad
\]
where we have split up the action $S(g)=S_0+S\indup{int}(g)$
into the free part, quadratic in the fields,
and the interacting part,
which is (at least) cubic.%
\footnote{Apart from the trivial vacuum, 
in which all fields are identically zero,
other classical solutions to the equations of motion exist. 
For example, there are instantonic vacua
with non-trivial topological charge and 
non-conformal vacua in which some of the scalar fields
have a constant value. 
One can also expand around \emph{these} configurations
which leads to qualitatively very different results. 
For simplicity we shall only consider the trivial vacuum 
in this work.}
The free connected generating functional is given by 
\[\label{eq:N4.Quantum.FreeConnected}
W_0[\fldJ]=\frac{1}{N}\int dx\,dy\, \half\Tr\fldJ(x) \Delta(x,y) \fldJ(y).
\]
Here, $\Delta(x,y)$ is the free propagator which is the inverse
of the kinetic term in the free action $S_0$.
The source fields $\fldJ$ will usually be replaced by 
variations $\partial/\partial\fldW$.
The expression \eqref{eq:N4.Quantum.Feyn} can be read as follows,
see also \figref{fig:N4.Quantum.Feynman}:
\begin{figure}\centering
\includegraphics{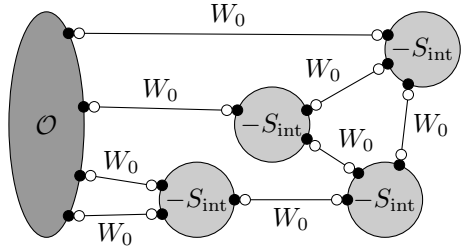}
\caption{A contribution to the 
quantum expectation value of the operator $\Op$
(Feynman graph).}
\label{fig:N4.Quantum.Feynman}
\end{figure}
There are arbitrarily many propagators $W_0$ and 
arbitrarily many vertices $S\indup{int}$.
Each propagator connects two fields
$\fldW$ within the vertices or the operator $\Op$.
In the end, all fields must be saturated.

For a perturbative treatment of a quantum gauge theory 
one must modify the action slightly. Firstly,
the divergencies which appear in a QFT need to be regularised. 
A convenient scheme which preserves most of the symmetries
is dimensional regularisation. 
In this scheme the number of spacetime dimensions is not fixed to four, 
but rather assumed to be $4-2\epsilon$ 
with a regularisation parameter $\epsilon$.
Correlators are thus analytic functions of $\epsilon$
and divergencies become manifest as poles at $\epsilon=0$.
The other issue is gauge fixing: Gauge invariance
leads to non-propagating modes of the gauge field
and a naive gauge field propagator is ill-defined.
We need to fix a gauge and a consistent treatment
may require the introduction of \emph{ghosts}.
The ghosts are auxiliary fermionic fields 
which interact with the gauge fields at a cubic vertex.
They are an artefact of the quantisation procedure 
and can appear only in the bulk of Feynman graphs;
they are forbidden in external states (operators).
These two issues are important for a consistent quantisation;
they will however hardly affect our investigations which
are algebraic in nature. 
We will merely have to assume that the perturbative contributions 
can be obtained consistently.

Let us comment on the counting of quantum loops. For simplicity, we will assume only 
cubic interactions. In gauge theories there are also quartic interactions,
but these may be represented by two cubic interactions connected by
an auxiliary field. This fits well with the fact that 
cubic interactions are suppressed by one power of
the coupling constant and quartic ones by two.
A Feynman graph can then be characterised by the number
of vertices $V$, propagators $I$, fields within the operator $E$ 
and connected components $C$. 
As the number of fields $\fldW$ and 
variations $\partial/\partial\fldW$ must match exactly,
we have $3V+E=2I$. 
Counting of momentum integrals $L$ (loops) 
furthermore implies $L=I-V-E+C$:
Each propagator introduces one new momentum variable, while
each vertex and external field introduces a constraint.
Due to momentum conservation the external momenta within each component 
must add up to zero, reducing the number of constraints
by one for each component.
In total we can write
\[\label{eq:N4.Quantum.CountVertex}
V=2L+2(E/2-C).
\]
In the free theory, there are neither vertices nor loops. 
Therefore we have $C_0=E/2$ independent pairwise 
contractions of fields. In the interacting theory 
$E/2-C=C_0-C$ gives the number of components that are now connected
due to interactions. The above formula states that it takes 
two vertices to construct a loop or to connect two components.
The number of vertices is important because it gives
the order in perturbation theory $g^V$.
We will consider a graph of order $g^{2\ell}$ 
in perturbation theory to be an `$\ell$-loop' graph
\[\label{eq:N4.Quantum.lLoop}
\mbox{`\emph{$\ell$-loop}'}:\quad \order{g^{2\ell}}.
\]
Note that these `loops' 
are not the momentum-loops counted by $L$.
The motivation for this terminology is that, 
when working in position space,
connecting two components of a graph may produce the same kind
of divergency as adding a loop.
This is quite different in momentum space, where divergencies can
only arise from true loops in the graph. 
At any rate, the counting scheme
is different there, as one usually considers only connected graphs 
with external propagators removed.
For Wilson loops the counting is again different, because each external 
leg also contributes one power of $g$.

\section{The Gauge Group}
\label{sec:N4.Gauge}

\begin{figure}\centering
\includegraphics{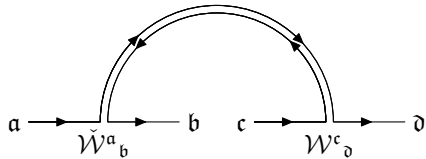}
\caption{The contraction of a matrix-valued
variation $\check\fldW$ and field $\fldW$.}
\label{fig:N4.Gauge.Contract}
\end{figure}

In the following we will present some useful notation to deal
with the matrix-valued fields $\fldWf{A}$. 
For a start, let us introduce explicit matrix indices for the fields
$(\fldWf{A})^{\gaugeind{a}}{}_{\gaugeind{b}}$.
For variations with respect to these fields
we introduce the notation $\fldWv{A}$, see also \figref{fig:N4.Gauge.Contract},%
\footnote{In a canonical quantisation scheme,
$\fldWf{}$ and $\fldWv{}$ correspond to
creation and annihilation operators.}
\[\label{eq:N4.Gauge.VaryCompUN}
(\fldWv{A})^{\gaugeind{a}}{}_{\gaugeind{b}}
:=\frac{\delta}{\delta (\fldWf{A})^{\gaugeind{b}}{}_{\gaugeind{a}}}\,,
\qquad
(\fldWv{A})^{\gaugeind{a}}{}_{\gaugeind{b}}
(\fldWf{B})^{\gaugeind{c}}{}_{\gaugeind{d}}
=
\delta^{\fldind{A}}_{\fldind{B}}
\delta^{\gaugeind{a}}_{\gaugeind{d}}
\delta^{\gaugeind{c}}_{\gaugeind{b}}.
\]
When, for the gauge group $\grSU(N)$, the matrices are
traceless $\fldW^{\gaugeind{a}}{}_{\gaugeind{a}}=0$,
the trace of the variation must vanish as well and we define
the variation by
\[\label{eq:N4.Gauge.VaryCompSUN}
(\fldWv{A})^{\gaugeind{a}}{}_{\gaugeind{b}}
(\fldWf{B})^{\gaugeind{c}}{}_{\gaugeind{d}}
=\delta^{\fldind{A}}_{\fldind{B}}
\delta^{\gaugeind{a}}_{\gaugeind{d}}
\delta^{\gaugeind{c}}_{\gaugeind{b}}-
N^{-1}\delta^{\fldind{A}}_{\fldind{B}}
\delta^{\gaugeind{a}}_{\gaugeind{b}}
\delta^{\gaugeind{c}}_{\gaugeind{d}}.
\]
We furthermore introduce normal ordering $\normord{\ldots}$
which suppresses all possible contractions
between fields and variations by moving all variations to the right, 
for example
\[\label{eq:N4.Gauge.NormalOrderComp}
\normord{
\ldots
(\fldWv{A})^{\gaugeind{a}}{}_{\gaugeind{b}} 
\ldots
(\fldWf{B})^{\gaugeind{c}}{}_{\gaugeind{d}}
\ldots
}
:=
\ldots
(\fldWf{B})^{\gaugeind{c}}{}_{\gaugeind{d}}
\ldots
(\fldWv{A})^{\gaugeind{a}}{}_{\gaugeind{b}}.
\]

For all practical purposes we need not write out the matrix indices
writing simply
\[\label{eq:N4.Gauge.Vary}
\fldWv{A}:=\frac{\delta}{\delta \fldWf{A}}\,.
\]
It is useful to write down the action of a variation
on a field \eqref{eq:N4.Gauge.VaryCompUN}
when both are inserted within traces. 
There are two cases to be considered:
The variation and field might be within 
different traces or within the same; these
are the fusion and fission rules, respectively
\< \label{eq:N4.Gauge.VaryUN}
\Tr X \fldWv{A} \Tr Y \fldWf{B}\eq\delta^{\fldind{A}}_{\fldind{B}} \Tr XY,
\nln
\Tr X \fldWv{A} Y \fldWf{B}\eq\delta^{\fldind{A}}_{\fldind{B}} \Tr X\Tr Y.
\>
Clearly, $\fldWv{}$ also acts on further fields $\fldW$ within $Y$ in the
same way. 
For the case of a gauge group $\grSU(N)$, 
the fusion and fission rules following from 
\eqref{eq:N4.Gauge.VaryCompSUN} are
\<\label{eq:N4.Gauge.VarySUN}
\Tr X \fldWv{A} \Tr Y \fldWf{B}\eq\delta^{\fldind{A}}_{\fldind{B}}
(\Tr XY-N^{-1}\Tr X\Tr Y),
\nln
\Tr X \fldWv{A} Y \fldWf{B}\eq\delta^{\fldind{A}}_{\fldind{B}}
(\Tr X \Tr Y-N^{-1}\Tr XY).
\>
Commonly, variations will appear within commutators only.
The appropriate rules are
\<\label{eq:N4.Gauge.VaryComm}
\Tr X\comm{Z}{\fldWv{A}} \Tr Y \fldWf{B}\eq\delta^{\fldind{A}}_{\fldind{B}}\Tr X\comm{Z}{Y},
\nln
\Tr X\comm{Z}{\fldWv{A}} Y \fldWf{B}\eq\delta^{\fldind{A}}_{\fldind{B}}(\Tr XZ \Tr Y-\Tr X\Tr ZY),
\>
which are valid for both, $\grU(N)$ and $\grSU(N)$ 
(the abelian trace does not contribute in commutators).
Note that when normal ordering expressions, 
it is sometimes impossible to simply move all variations 
to the right in this notation.
Instead, the possible contractions have to be removed
by hand, for example
\[\label{eq:N4.Gauge.NormalOrder}
\normord{\Tr\fldWf{A}\fldWv{B}\fldWf{C}\fldWv{D}}
=
\Tr\fldWf{A}\fldWv{B}\fldWf{C}\fldWv{D}-
\delta^{\fldind{B}}_{\fldind{C}}N\Tr\fldWf{A}\fldWv{D}.
\]
This notation is convenient to express, for example,
gauge transformations $\fldW\mapsto U\fldW U^{-1}$,
which are generated infinitesimally by 
\[\label{eq:N4.Gauge.Gauge}
\delta_\epsilon\fldW =i\comm{\epsilon}{\fldW}.
\]
Using our notation for matrix-valued variations this becomes
\[\label{eq:N4.Gauge.GaugeGen}
\delta_\epsilon=\Tr \epsilon\gaugerot,\quad\mbox{where}\quad
\gaugerot=i\normord{\comm{\fldWf{A}}{\fldWv{A}}}.
\]

We can also consider a more general gauge group. 
We will start with the gauge theory Lagrangian as defined in 
\eqref{eq:N4.D4.Lagr} for $\grSU(N)$.
Let us parameterise the fields using $\grSU(N)$ generators
$\gaugegen{m}$
\[\label{eq:N4.Gauge.FieldGen}
\fldWf{A}=\fldWf{A}^{\gaugeind{m}}\gaugegen{m}.
\]
We assume the generators and structure constants 
$\gaugestr{mn}^{\gaugeind{p}}$
to be normalised in a way such that 
\[\label{eq:N4.Gauge.GenNormal}
\Tr \gaugegen{m}\gaugegen{n}=\gaugemet{}_{mn},
\qquad
\comm{\gaugegen{m}}{\gaugegen{n}}=i\gaugestr{mn}^{\gaugeind{p}}\gaugegen{p}.
\]
The more general variations will be defined as 
\[\label{eq:N4.Gauge.GenVary}
\fldWv{A}:=\gaugegen{m}\gaugemet{mn}
\frac{\delta}{\delta \fldWf{A}^{\gaugeind{n}}},
\qquad \frac{\delta}{\delta \fldWf{A}^{\gaugeind{m}}}\, \fldWf{B}^{\gaugeind{n}}=
\delta^{\fldind{A}}_{\fldind{B}}
\delta_{\gaugeind{m}}^{\gaugeind{n}}.
\]
%
%
This allows us to rewrite the gauge theory
and all our results purely in terms of the 
metric $\gaugemet{mn}$ and the structure 
constants $\gaugestr{mn}^{\gaugeind{p}}$. 
In that form the results generalise to arbitrary gauge groups.
Nevertheless the matrix notation is most convenient and we will
stick to it in this work.
On rare occasions we shall use generators
$\gaugegen{m}$ to write down expressions valid for generic groups;
for example, it is better to write instead of \eqref{eq:N4.Gauge.NormalOrder}
\[\label{eq:N4.Gauge.NormalOrderGen}
\normord{\Tr\fldWf{A}\fldWv{B}\fldWf{C}\fldWv{D}}
=
\Tr\fldWf{A}\fldWv{B}\fldWf{C}\fldWv{D}-
\delta^{\fldind{B}}_{\fldind{C}}\gaugemet{mn}\Tr\fldWf{A}\gaugegen{m}\gaugegen{n}\fldWv{D}.
\]
%


For a unitary group we can define a \emph{parity operation}. 
It replaces a matrix by its negative transpose
\[\label{eq:N4.Gauge.Parity}
\mbox{`\emph{parity operation}'}:\quad \gaugepar \fldW\mapsto -\fldW^{\trans}.
\]
For hermitian matrices the conjugate equals the transpose,
therefore this parity is equivalent to charge conjugation.
Its eigenvalues $\pm 1$ will be denoted by the letter $P$.
It is easily seen that the Lagrangian \eqref{eq:N4.D4.Lagr} 
is invariant under this operation. 
Therefore parity is an exact symmetry of $\grU(N)$ or $\grSU(N)$ gauge theory.
Note that this parity is a unique feature of the 
unitary groups, it does not generalise to 
the orthogonal or symplectic groups.

\section{The 't Hooft Limit}
\label{sec:N4.LargeN}

A field theory with $\grU(N)$ gauge symmetry has remarkable
properties when $N$ is interpreted as an additional
coupling constant:
In the article \cite{'tHooft:1974jz}
't~Hooft realised that, in the large-$N$ limit, 
for any Feynman graph there is an associated two-dimensional surface.
The $N$-dependence of a graph is given by the Euler characteristic 
(genus) of the corresponding surface.
This makes the large-$N$ field theory 
very similar to a string field theory whose
coupling constant also counts the genus of the world sheet.

\begin{figure}\centering
\includegraphics{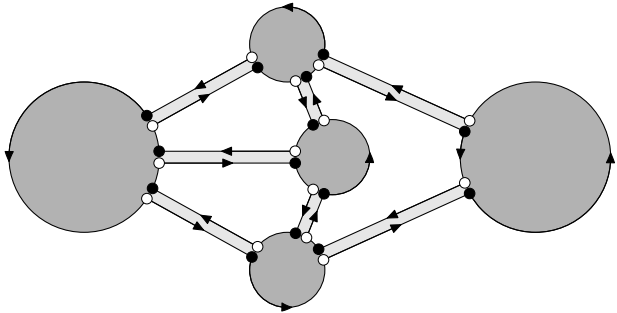}
\caption{A fat graph. The circles are $\grU(N)$ traces 
of the vertices and operators and 
the fat lines are propagators. For each face of the
graph there is a closed empty trace $\Tr 1=N$.}
\label{fig:N4.LargeN.Fat}
\end{figure}

Let us consider a gauge invariant Feynman graph and investigate its structure
in terms of the gauge group. 
The structure may be represented graphically using `double lines' or 
a `fat graph'. 
For that purpose we represent every upper (lower) $\grU(N)$ vector 
index of a field within the operators or vertices by a black
(white) dot. Consequently, every (adjoint) field provides one black and 
one white dot. For all external contractions between two vector indices, 
i.e.~those in the operator or in the vertex, 
draw an arrow from the black to the white dot. 
For a gauge invariant graph all indices must be contracted, 
so there are no unconnected dots.
Now we perform the contractions generated by the path integral,
i.e.~due to the propagators \eqref{eq:N4.Quantum.FreeConnected}
\[\label{eq:N4.LargeN.Propagator}
W_0[\fldJ]=\frac{1}{N}\int dx\,dy\, \half\fldJ^{\gaugeind{a}}{}_{\gaugeind{b}}(x) 
\Delta(x,y) \fldJ^{\gaugeind{b}}{}_{\gaugeind{a}}(y),
\]
where we made the gauge indices visible.
The propagator connects two fields, we should now
connect the corresponding two pairs of dots by 
antiparallel arrows along the propagator. Here, we will
draw the arrows from white dot to black dot.
In this way all propagators of the Feynman graph are represented by 
two parallel lines or, alternatively, a fat line,
c.f.~\figref{fig:N4.LargeN.Fat}.
In this representation it is easy to perform all $\grU(N)$ contractions:
All dots have exactly one incoming and one outgoing arrow. The arrows
thus form closed circles, which are known as index loops.
Each index loop provides $\delta^{\gaugeind{a}}_{\gaugeind{a}}=\Tr 1=N$ and 
furthermore there are explicit factors of $N$ in 
the action and propagators 
\eqref{eq:N4.Quantum.ActionNorm,eq:N4.LargeN.Propagator}.
The result is given by $N^{F+V-I}$ where $F$
is the number of closed (index) loops,
$V$ the number of vertices and $I$ the number of (double) lines.
Assume now that the graph has $C$ components and that each component will
be drawn on a surface of minimal genus without crossing of any lines.
Let $G$ be the total genus of all component surfaces 
and $T$ the number of traces within the local operators. 
Then Euler's theorem relates these
numbers as follows $T+V+F=I+2C-2G$. 
In total the $N$-dependence is given by%
\footnote{A contribution $N^V$ which commonly appears 
at $V/2$ quantum loops has already been 
absorbed into the definition of the
coupling constant \eqref{eq:N4.Quantum.Couple}.}
\[\label{eq:N4.LargeN.CountN}
\order{N^{2C-2G-T}}=\order{N^{\chi}}.
\]
Note that the coupling constant $g^2$ as defined in 
\eqref{eq:N4.Quantum.Couple} is proportional to the 
't~Hooft coupling $\lambda=\gym^2 N$
\[\label{eq:N4.LargeN.Couple}
g^2=\frac{\gym^2 N}{8\pi^2}=\frac{\lambda}{8\pi^2}\,.
\]
Note also that \eqref{eq:N4.LargeN.CountN} gives precisely the 
Euler characteristic $\chi=2C-2G-T$ of the set of surfaces.
This led 't~Hooft to his conjecture of the emergence 
of a string field theory in the large $N$ limit of a gauge theory:
In string field theory, an amplitude on a world sheet 
with Euler characteristic $\chi$ is proportional to
$g\indup{s}^{-\chi}$, where $g\indup{s}$ is the string coupling constant.
This matches with the $N$-dependence in gauge theory when we identify
\[g\indup{s}\sim \frac{1}{N}\,.\]

With some additional work, the large $N$ limit can also be taken 
for gauge groups $\grSU(N),\grSO(N),\grSp(N)$ as well as 
fields in the fundamental representation. 
Then, also unoriented surfaces as well as surfaces with 
boundaries%
\footnote{Here we mean boundaries which are not 
associated to an operator insertion.}
may appear.

\section{The Superconformal Algebra}
\label{sec:N4.Alg}

The Lagrangian \eqref{eq:N4.D4.Lagr} and action 
\eqref{eq:N4.Quantum.Action} of $\superN=4$ SYM 
in four spacetime dimensions do not involve any 
dimensionful coupling constants. Therefore the action is 
invariant under the scale transformation
\[\label{sec:N4.Alg.Scaling}
x\mapsto c^{-1}x,\quad
\fldA\mapsto c \fldA,\quad 
\Psi\mapsto c^{3/2}\Psi.
\]
For a gauge theory this implies also conformal invariance
and, in the case of a supersymmetric theory, also 
superconformal invariance. 
This symmetry is especially important for
$\superN=4$ SYM, because it is believed to be an 
exact symmetry even in the quantum theory,
where the beta function \eqref{eq:N4.Quantum.BetaZero}
is apparently exactly zero.

The super Poincar\'e symmetry algebra 
consisting of Lorentz rotations $\algL,\algLd$, internal rotations $\algR$ 
and (super)translations $\algQ,\algQd,\algP$
is enlarged by (super)conformal boosts $\algS,\algSd,\algK$
and the scaling operator $\algD$ which is also known as
\[\label{eq:N4.Alg.DilOp}
\mbox{`\emph{The Dilatation Operator}' }\algD.
\]
The boosts are essentially the conjugate 
transformations of the translations. 
The action of the translations and boosts
on the fields in the free theory is depicted in \figref{fig:N4.D4.Multiplet}.
The arrows correspond to momenta whereas 
boosts act in the inverse direction of the arrows.

The conformal symmetry algebra in four spacetime 
dimensions is $\alSO(4,2)=\alSU(2,2)$, 
the superconformal algebra is $\alSU(2,2|\superN)$. 
In the case of maximal $\superN=4$ supersymmetry, 
the algebra $\alSU(2,2|4)$ is reducible and 
the superconformal algebra is considered to be only
the irreducible part $\alPSU(2,2|4)$.
Let us, for the moment, consider the 
supermatrix algebra $\alU(2,2|4)$ and later restrict to $\alPSU(2,2|4)$.
It consists of the generators
\[\label{eq:N4.Alg.Gen}
\algJ\in\set{\algL,\algLd,\algR,\algP,\algK,\algD,\algB,\algC|
\algQ,\algQd,\algS,\algSd}.
\]
These are the $\alSU(2),\alSU(2),\alSU(4)$ rotations $\algL,\algLd,\algR$,
the (super)translations $\algQ,\algQd,\algP$, 
the (super)boosts $\algS,\algSd,\algK$
as well as the dilatation generator $\algD$, 
hypercharge $\algB$ and central charge $\algC$.
Please refer to \appref{app:U224} for details of this superalgebra and
its commutation relations. 
The signature of spacetime 
will not be important here; for algebraic purposes we can safely 
assume to work with a complexified algebra.
In the irreducible superconformal algebra $\alPSU(2,2|4)$,
the generators $\algB,\algC$ are absent:
The $\alU(1)$ hypercharge $\algB$ 
of $\alPU(2,2|4)=\alU(1)\ltimes\alPSU(2,2|4)$
is an external automorphism
which consistently assigns a charge to all the 
generators of $\alPSU(2,2|4)$.
The $\alU(1)$ central charge $\algC$ of $\alSU(2,2|4)=\alPSU(2,2|4)\ltimes\alU(1)$ 
must vanish to be able to reduce to $\alPSU(2,2|4)$.
The Lorentz algebra $\alSO(3,1)=\alSU(2)\times\alSU(2)$ is 
formed by $\algL,\algLd$. 
Together with $\algP,\algK,\algD$ one gets the conformal algebra 
$\alSO(4,2)=\alSU(2,2)$.

Note that only the Lorentz and internal 
symmetries, $\alSU(2)\times\alSU(2)$ and $\alSU(4)$,
are manifestly realised
in the quantum theory;
the other generators receive radiative corrections,
i.e.~they depend on the coupling constant $g$,
see \eqref{eq:N4.D4.Trans}.
In particular the dilatation generator $\algD$ receives
loop corrections. 
As we shall see, it makes sense to 
define an operator which measures the \emph{classical dimension}, the
\[\label{eq:N4.Alg.Classical}
\mbox{`\emph{Classical Dilatation Operator}'}\quad \algD_0,
\]
even in the quantum theory. The shift of scaling dimensions
by quantum effects, the \emph{anomalous dimension}, is
measured by the 
\[\label{eq:N4.Alg.Anomalous}
\mbox{`\emph{Anomalous Dilatation Operator}'}\quad \algdD=\algD-\algD_0.
\]
This is a $\alU(1)$ abelian generator. 
It is also reasonable to identify $\algdD$ with 
a `\emph{Hamiltonian}' $\ham(g)$ in the following way,
c.f.~\secref{sec:Dila.Pert.Anomalous}
\[\label{eq:N4.Alg.Hamiltonian}
\mbox{`\emph{The Hamiltonian}'}\quad \ham=g^{-2}\, \algdD.
\]
Its eigenvalues, the \emph{energies}, will be denoted by the letter $E$.

\begin{figure}\centering
\includegraphics{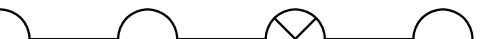}
\caption{The Dynkin diagram of $\alPSU(2,2|4)$ convenient for
$\superN=4$ SYM.}
\label{fig:N4.Alg.Dynkin}
\end{figure}

For a bosonic, semi-simple Lie algebra the Dynkin diagram is unique.
In the case of superalgebras, however, there is some freedom to distribute
the simple fermionic roots. 
Different choices of fermionic roots correspond to different 
assignments of positive and negative roots.
In the context of $\superN=4$ SYM 
one particular choice of Dynkin diagram 
turns out to be very useful \cite{Dobrev:1987qz},
see \figref{fig:N4.Alg.Dynkin}.%
\footnote{One might be tempted to denote the
superconformal algebra by $\alPSU(2|4|2)$.}
For this particular Dynkin diagram the generators associated to 
positive and negative roots and elements 
of the Cartan subalgebra $\algJ^+,\algJ^-,\algJ^0$
are given by 
\<\label{eq:N4.Alg.PosNeg}
\algJ^+\earel{\in}\set{\algK^{\alpha\dot \beta},\algS^\alpha{}_b,
\algSd^{a\dot\beta},\algL^\alpha{}_\beta\,(\alpha<\beta),
\algLd^{\dot\alpha}{}_{\beta}\,(\dot\alpha<\dot\beta),\algR^a{}_b\,(a<b)},
\nln
\algJ^0\earel{\in}\set{\algL^\alpha{}_\beta\, (\alpha=\beta),
\algLd^{\dot\alpha}{}_{\dot\beta}\, (\dot\alpha=\dot\beta),
\algR^a{}_b\,(a=b),\algD,\algB,\algC},
\nln
\algJ^-\earel{\in}\set{\algP_{\dot\alpha\beta},\algQ^a{}_\beta,\algQd_{\dot\alpha b},
\algL^\alpha{}_\beta\,(\alpha>\beta),
\algLd^{\dot\alpha}{}_{\beta}\,(\dot\alpha>\dot\beta),\algR^a{}_b\,(a>b)}.
\>
All the elements of the Cartan subalgebra, 
spanned by $\set{\algJ^0}$, commute among each other. 
One can therefore find simultaneous eigenstates with
respect to all its elements, the eigenvalues are the charges 
or `\emph{labels}' of that state.
There are many useful bases for the Cartan subalgebra
which give rise to different labellings of states,
we will use two of them.

Let us first note the Dynkin labels corresponding to 
the diagram in \figref{fig:N4.Alg.Dynkin} 
of $\alSU(2,2|4)$, see also \cite{Heslop:2003xu}%
\footnote{There is no obvious choice for the sign of the odd labels $r_1,r_2$.
Our choice implies, e.g., for the product of two
fundamental representations 
$[1;0;\ldots]\times [1;0;\ldots]=[2;0;\ldots]+[0;-1;\ldots]$.}
\[\label{eq:N4.Alg.DynkinLabel}
w=[s_1;r_1;q_1,p,q_2;r_2;s_2];\]
these are defined as the following linear combinations 
of the eigenvalues 
$L^\alpha{}_\beta,\dot L^{\dot \alpha}{}_{\dot\beta},R^a{}_b$ 
of Cartan generators
$\algL^\alpha{}_\beta,\algLd^{\dot \alpha}{}_{\dot\beta},\algR^a{}_b$
($\alpha=\beta, \dot\alpha=\dot\beta, a=b$)
\[\label{eq:N4.Alg.DynkinCartan}
\arraycolsep0pt\begin{array}{rclcrcl}
s_1 \eq L^2{}_2-L^1{}_1, 
&\qquad&
s_2 \eq \dot L^2{}_2-\dot L^1{}_1,
\\[5pt]
r_1 \eq \half D-\half C-L^1{}_1+R^1{}_1,
&&
r_2 \eq \half D+\half C-\dot L^1{}_1-R^4{}_4,
\\[5pt]
q_1 \eq R^2{}_2-R^1{}_1,
&&
q_2 \eq R^4{}_4-R^3{}_3,
\\[10pt]
p \eq R^3{}_3-R^2{}_2,
&&
r \eq -D+L^1{}_1+\dot L^1{}_1.
\end{array}
\]
The charges $[q_1,p,q_2]$ are the Dynkin labels of the $\alSU(4)$ subalgebra.
Equivalently $[s_1,s_2]$ are the Dynkin labels of the Lorentz algebra
$\alSO(3,1)=\alSU(2)\times \alSU(2)$. Together with $r$ they 
give the Dynkin labels $[s_1,r,s_2]$ of the conformal algebra $\alSU(2,2)$.
Note that we shall always use integer valued Dynkin labels $s_1,s_2$ instead
of the more common half-integer valued spin labels for $\alSU(2)$,
i.e.~$s_1,s_2$ equal \emph{twice} the spin.
The labels $q_1,q_2$ will also be integers, only for the labels 
$r_1,r_2,r$ we have to allow irrational numbers corresponding 
to anomalous dimensions. 
For the $\alSU(2,2|4)$ labels \eqref{eq:N4.Alg.DynkinLabel}
we do not need the label $r$ which is given by
$r=-r_1-q_1-p-q_2-r_2$.

Often, we will consider states
of the classical theory at $g=0$.%
\footnote{To use this set of labels makes sense even in the interacting theory:
There, the labels should be defined as the labels 
in the limit $g\to 0$.}
To label them we will use a notation 
which is based on more physical quantities
\[\label{eq:N4.Alg.Weight}
w=\weight{D_0;s_1,s_2;q_1,p,q_2;B,L}.\]
Here, $[q_1,p,q_2]$ and $[s_1,s_2]$ are as above.
The label $D_0$ is the classical scaling dimension as measured by $\algD_0$.
These are the six labels for a weight of 
the algebra $\alPSU(2,2|4)$ of rank six.
All of them are non-negative integers except $D_0$, which can take 
half-integer values.

We further introduce labels $B,C,L$ which do not belong to $\alPSU(2,2|4)$;
they are important to describe a state of the classical theory.
The hypercharge $B$, measured by $\algB$
of $\alPU(2,2|4)$, is half-integer valued and 
is defined via \tabref{tab:N4.States.Fields}.
The `\emph{length}' $L$, measured by the operator $\len$, 
counts the number of constituent fields of a state and
is therefore a positive integer. 
The central charge $C$, measured by $\algC$ of $\alSU(2,2|4)$, 
will always be zero.
The anomalous dimension $\delta D=D-D_0$ is not part of 
this set of labels, it will be given elsewhere.

\section{Fields and States}
\label{sec:N4.States}

In this work we will be interested in the properties of 
local, gauge invariant operators, which we will 
also refer to as `\emph{states}' and which will usually 
be denoted by the symbol `$\Op$'.
Local operators are constructed as linear combinations of products 
of the fundamental fields and their derivatives%
\footnote{The $\ast$'s refer to some unspecified indices}
\[\label{eq:N4.States.Local}
\mbox{`\emph{states}':}\quad
\Op(x)=\Phi_\ast(x)\cdott\Phi_\ast(x)\cdott\cder_\ast\cder_\ast\Psi_\ast(x)
\cdott\cder_\ast\fldF_\ast(x)\ldots+\ldots\,.
\]
Note that the coordinate $x$ is the same for all the fields and hence we can 
drop it altogether.%
\footnote{Moreover, we consider fields and states as abstract
objects which are not positioned at some point in space. They are 
merely elements of the space of fields or states, respectively.}
We demand states to be gauge invariant. Due to the inhomogeneous gauge 
transformation ($X\mapsto UXU^{-1}+\ldots$)
of the gauge field $\fldA$ and partial derivative $\partial$,
these cannot be used in the construction of states. Instead, we can
use field strengths $\fldF$ and covariant derivatives $\cder$, which 
transform homogeneously ($\fldW\mapsto U\fldW U^{-1}$). In the case of gauge groups 
$\grSU(N),\grSO(N),\grSp(N)$ and adjoint fields, 
a basis for the space of states is given by the multi-trace
operators
\[\label{eq:N4.States.Basis}
\mbox{`\emph{state basis}':}\quad
\bigset{\Tr \fldW_\ast\cdots \fldW_\ast\,\, 
\Tr \fldW_\ast\cdots \fldW_\ast \,\,\,\, \ldots },
\]
where each $\fldW$ is one of the \emph{fields}
\[\label{eq:N4.States.Fields}
\mbox{`\emph{field basis}':}\quad
\fldWf{A}\in\bigset{\cder^k\Phi,\cder^k\Psi,\cder^k\dot\Psi,\cder^k\fldF}.
\]
Here we slightly change the definition of the symbol $\fldWf{A}$:
By including arbitrarily many derivatives, in the sense of
a Taylor expansion $\fldW(x)=\fldW_0+x\fldW_1+\half x^2\fldW_2+\ldots$\,\,, 
we trade in spacetime-dependence for infinitely many components.%
\footnote{This is analogous to moving from a
superspace to ordinary spacetime when one trades in the dependence on
fermionic coordinates $\theta$ for component fields.}
Roughly speaking, the index $\fldindn{A}$ 
now comprises also the coordinate $x$, but in a fashion more suitable 
for local operators.
The basis \eqref{eq:N4.States.Basis} is in general over-complete. 
One reason is the Bianchi identity \eqref{eq:N4.D4.FieldStrength}
which defines the field strength
\[\label{eq:N4.States.FieldStrength}
\comm{\cder}{\cder}X\sim g\comm{\fldF}{X},
\]
which tells us that the left hand side is reducible, 
i.e.~it can be written as a product of fields.
As products of fields appear naturally within 
the basis states \eqref{eq:N4.States.Basis}, 
there is no reason to use reducible fields $\fldW$.
By avoiding them, 
one eliminates obscure identities between basis states \eqref{eq:N4.States.Basis}
and reduces the ambiguity.
Consequently, for an irreducible field, 
all derivatives in \eqref{eq:N4.States.Fields} should be totally symmetrised.
Furthermore, the Bianchi identity \eqref{eq:N4.D4.FieldStrength}
\[\label{eq:N4.States.FJacobi}
\cder\wedge\fldF=0\]
implies that none of the derivatives can be 
antisymmetrised with the field strength. Finally, the equations
of motion \eqref{eq:N4.D4.EOM}
\[\label{eq:N4.States.EOMs}
\cder\cdott\cder\Phi,
\cder\cdott\Psi,
\cder\cdott\dot\Psi,
\cder\cdott\fldF=\ldots,
\]
lead to further reducible terms $\Phi^3,\Phi\cder\Phi,\Psi^2,\ldots$\,\,.%
\footnote{In a quantum theory the equations of motion might be 
modified, but the modifications are again reducible.}
Therefore, contractions between
indices are not allowed for irreducible fields.
In total these constraints lead to the set of \emph{irreducible fields}
as presented in \tabref{tab:N4.States.Fields}. In the table
we have split up the field strengths into their
chiral and antichiral parts according to \eqref{eq:N4.D4.Spinors}. 
\begin{table}\centering
$\begin{array}{|c|ccccc|}\hline
\mathrm{field}    & D_0 & \alSU(2)\times\alSU(2)&\alSU(4)&B&L\\\hline
\cder^k \fldF     & k+2 & [k+2,k\phantom{\mathord{}+0}] & [0,0,0] &+1&1\\
\cder^k \Psi      & k+\sfrac{3}{2} & [k+1,k\phantom{\mathord{}+0}] & [0,0,1] &+\sfrac{1}{2}&1\\
\cder^k \Phi      & k+1 & [k\phantom{\mathord{}+0},k\phantom{\mathord{}+0}] & [0,1,0] &\pm 0&1\\
\cder^k \dot \Psi & k+\sfrac{3}{2} & [k\phantom{\mathord{}+0},k+1] & [1,0,0] &-\sfrac{1}{2}&1\\
\cder^k \dot \fldF& k+2 & [k\phantom{\mathord{}+0},k+2] & [0,0,0] &-1&1\\\hline
\end{array}$
\caption{Basis fields $\fldWf{A}$ of the $\superN=4$ SYM field strength multiplet $\mdlF$.}
\label{tab:N4.States.Fields}
\end{table}
We will use index letters $\fldindn{A,B,\ldots}$ to 
label precisely the elements of this set of irreducible fields.
For the rest of this work we will consider only 
irreducible fields and speak of `\emph{fields}' 
for short.

Matrix identities are 
another source of linear dependencies between 
the basis states \eqref{eq:N4.States.Basis} at finite $N$. 
These involve traces of $L>N$ fields and 
become irrelevant when $N$ is sufficiently large, 
e.g., in the large $N$ limit. 
Note also that traces are cyclic and states related by cyclic 
permutations within the traces are to be identified.

An alternative way of representing local operators is to use 
the state-operator map for $\superN=4$ SYM on $\Real\times S^3$,
which is conformally equivalent to flat $\Real^4$. 
When decomposing the fundamental fields into spherical harmonics on $S^3$
one gets precisely the same spectrum of fields $\fldW$
as in \tabref{tab:N4.States.Fields}.

\section{Highest-Weight Modules and Representations}
\label{sec:N4.Modules}

There are various types of representations of $\alU(2,2|4)$;
for example the defining one $\rep{4|4}$ or 
the adjoint $\rep{30|32}+\rep{1}+\rep{1}$ have finitely many components.
In the context of field theory we have to deal mainly with a different kind, 
namely non-compact or infinite-dimensional highest-weight representations.
A (Verma) module, i.e.~the vector space on which the representation acts,
is characterised by its \emph{highest-weight} or \emph{primary} state $\vac$.
In field theory this corresponds to a field or local operator,%
\footnote{A local operator can be viewed as an abstract object, 
i.e.~not based at some point in spacetime.}
for example the primary field $\fldZ$ or the 
primary Konishi state $\OpK$
\[\label{eq:N4.Modules.HighestEx}
\state{\fldZ}=\Phi_{34}\quad \mbox{or}\quad \state{\OpK}
=\varepsilon^{abcd}\Tr \Phi_{ab}\Phi_{cd}\sim\eta^{mn}\Tr\Phi_m\Phi_n.
\]
A highest-weight state $\vac$ is defined as being annihilated by 
all raising operators $\algJ^+$ in \eqref{eq:N4.Alg.PosNeg}
\[\label{eq:N4.Modules.Highest}
\algJ^+\vac=0,
\qquad \algJ^0\vac = w \vac;
\]
the action of the Cartan subalgebra $\algJ^0$ in \eqref{eq:N4.Alg.PosNeg} 
measures the charge vector $w$ of the highest weight,
see \eqref{eq:N4.Alg.DynkinLabel,eq:N4.Alg.Weight}.
Application of the lowering operators yields new states 
\[\label{eq:N4.Modules.States}
\vac,\quad \algJ^-\vac,\quad \algJ^- \algJ^-\vac,\quad\ldots\,,
\]
which belong to the highest-weight module. 
These are called \emph{descendants}.
For example
\[\label{eq:N4.Modules.StatesEx}
\algQ^{3}{}_2\state{\fldZ}=\Psi_{24}\quad\mbox{or}\quad
\algP_{22}\algP_{22}\state{\OpK}=
2\varepsilon^{abcd}
\bigbrk{\Tr \cder_{22}\Phi_{ab}\,\cder_{22}\Phi_{cd}+
\Tr \Phi_{ab}\,\cder_{22}\cder_{22}\Phi_{cd}}
\]
belong to the modules with highest weights $\state{\fldZ}$ or $\state{\OpK}$,
respectively. See \figref{fig:N4.D4.Multiplet} 
for an illustration of the module with highest weight $\state{\fldZ}$.

In general one can apply any number of lowering operators to the
highest weight and obtain an infinite multiplet of states.
The space spanned by the states is a module of $\alU(2,2|4)$
because applying any of the
generators yields another element: For lowering operators this is obvious
while raising and Cartan generators have to be commuted all the way to the
vacuum $\vac$ first. In the most general case, the obtained module is
irreducible. However, for very special highest weights, one will 
find further states which are annihilated by all the raising operators.
In that case the multiplet is reducible and the irreducible module
which contains $\vac$ is called short, see \secref{sec:N4.Split}
for the cases relevant to $\superN=4$ SYM.
Finite-dimensional representations are just extremely short: 
By repeatedly applying lowering operators to the highest weight, 
one will inevitably leave the irreducible module that belongs to the 
highest weight $\vac$.
In a conformal field theory, the modules will commonly
be very short, only a few of the shortening conditions 
are not satisfied. This means that, when one considers 
a fairly large subalgebra, here 
$\alSU(2)\times\alSU(2)\times\alSU(4)$, the modules will split 
into (infinitely many) finite-dimensional 
modules of the subalgebra.

\begin{figure}\centering
\includegraphics{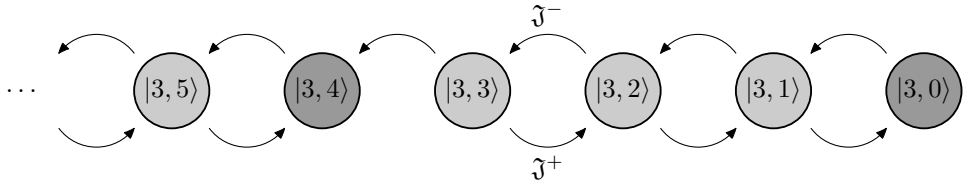}
\caption{A reducible highest-weight module.
All states can be obtained from the highest-weight state $\protect\state{3,0}$ 
by the action of $\algJ^-$, but there is no way back 
from $\protect\state{3,4}$ to $\protect\state{3,3}$ using $\algJ^+$.
Consequently $\protect\state{3,4}$ is a highest-weight state.}
\label{fig:N4.Module.Splitting}
\end{figure}

Let us demonstrate this feature in terms of a simple example
using the algebra of $\alSL(2)$ spanned by 
$\algJ^{\pm},\algJ^0$. The algebra of generators is
\[\label{eq:N4.Module.SU2}
\comm{\algJ^0}{\algJ^\pm}=\pm 2 \algJ^\pm,\qquad
\comm{\algJ^+}{\algJ^-}=J^0.
\]
We specify a highest-weight state $\state{s}$ by 
\[\label{eq:N4.Module.SU2Highest}
\algJ^+\state{s}=0,\qquad \algJ^0\state{s}=s\state{s}.
\]
A module is spanned by the states
\[\label{eq:N4.Module.SU2Module}
\state{s,k}=(\algJ^{-})^k\state{s}.
\]
Let us act with $\algJ^+$ on some state $\state{s,k}$, using the
algebra relations we find
\[\label{eq:N4.Module.SU2Raising}
\algJ^+\state{s,k}=k(s+1-k)\state{s,k-1}.
\]
The state $\state{s,0}=\state{s}$ with $k=0$ is a 
highest-weight state by construction. 
However, if $s$ is a non-negative integer, the state $\state{s,s+1}$ 
is another highest weight, see \figref{fig:N4.Module.Splitting}.
It has $\algJ^0$ charge $s'=-2-s$, it is therefore
equivalent to $\state{s'}$
\[\label{eq:N4.Module.SU2AnotherHighest}
\state{s,s+1}\hateq\state{-2-s}.
\]
The charge $s'$ of this highest weight is negative and the 
`lower' module therefore irreducible. By defining $\state{s,s+1}=0$ 
we can also make the `upper' highest weight module irreducible. 
As one can see, this is compatible with the 
algebra \eqref{eq:N4.Module.SU2}.

\section{Unitarity and Multiplet Shortenings}
\label{sec:N4.Split}

The real algebra $\alPSU(2,2|4)$ of
indefinite signature does not have any 
non-trivial finite-dimensional unitary representations. 
Unitary representations, which are relevant to 
quantum physics, are necessarily infinite-dimensional.
These have been classified \cite{Dobrev:1985qv} 
and the following two bounds have been found%
\footnote{Using the fermionic labels $r_1,r_2$
\eqref{eq:U224.Labels.FermiLabel} these conditions
simplify to $r_i-s_i\geq 1$ or $r_i=s_i=0$.}
\<\label{eq:N4.Split.Unitarity}
&\quad& D\geq 2+s_1+p+\sfrac{3}{2}q_1+\sfrac{1}{2}q_2
\quad\mbox{or}\quad D=p+\sfrac{3}{2}q_1+\sfrac{1}{2}q_2,\quad s_1=0,\nln
\mbox{and}&\qquad& D\geq 2+s_2+p+\sfrac{1}{2}q_1+\sfrac{3}{2}q_2
\quad\mbox{or}\quad D=p+\sfrac{1}{2}q_1+\sfrac{3}{2}q_2,\quad s_2=0.
\>

Unitary multiplets fall into different series,
the first one is the `half-BPS' series
with highest weights given by%
\footnote{Alternatively $w=[0;0;0,p,0;0;0]$.}
\[\label{eq:N4.Split.Half}
w=\weight{p;0,0;0,p,0;0,p}.
\]
These are the shortest physical multiplets.
Multiplets which are of this type in the classical 
theory are protected, the scaling dimension
of the classical theory
is preserved exactly in the quantum theory, $D=p$.
In field theory the highest weight state is composed only 
from fields $\fldZ$ and all states of this form are half-BPS.

Furthermore, there are two `eighth-BPS' conditions%
\footnote{Alternatively $r_i=s_i=0$.}
\[\label{eq:N4.Split.BPS}
\mbox{`\emph{eighth-BPS}':}\qquad
\begin{array}{rcl}
\mathrm{I}:&& D=p+\sfrac{3}{2}q_1+\sfrac{1}{2}q_2,\quad s_1=0,\\[3pt]
\mathrm{II}:&& D=p+\sfrac{1}{2}q_1+\sfrac{3}{2}q_2,\quad s_2=0,
\end{array}
\]
and two shortening%
\footnote{These are usually called semi-shortening conditions.
Here we shall distinguish between \emph{short} and \emph{BPS} multiplets.}
conditions%
\footnote{Alternatively $r_i-s_i=1$.}
\[\label{eq:N4.Split.Short}
\mbox{`\emph{short}':}\qquad
\begin{array}{rcl}
\mathrm{i}:&& D=2+s_1+p+\sfrac{3}{2}q_1+\sfrac{1}{2}q_2,\\[3pt]
\mathrm{ii}:&& D=2+s_2+p+\sfrac{1}{2}q_1+\sfrac{3}{2}q_2.
\end{array}
\]
These condition can be combined into a quarter-BPS condition (I+II),
two short-eighth-BPS conditions (i+II,I+ii) and 
a doubly-short condition (i+ii).

\begin{figure}\centering
\includegraphics{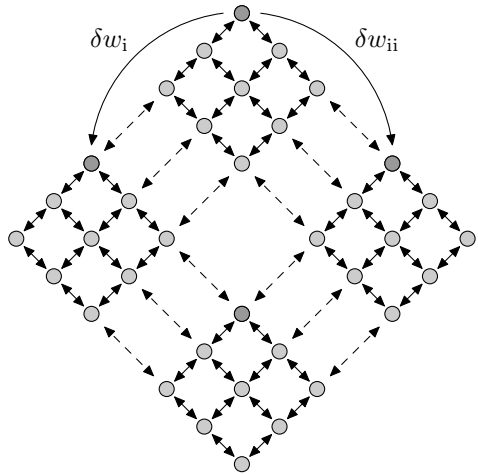}
\caption{A long multiplet can split in up to four short submultiplets
at the unitarity bounds. Short arrows correspond to 
$\order{1}$ generators whereas dashed arrows
correspond to $\order{\sqrt{D-D_0}}$ generators.}
\label{fig:N4.Split.Splitting}
\end{figure}

In perturbation theory multiplets close 
to the unitarity bound have some special features,
see for example \cite{Dobrev:1985qv,Andrianopoli:1999vr,Dolan:2002zh}.
Consider a multiplet whose classical dimension $D_0$ 
saturates one of the bounds in \eqref{eq:N4.Split.Unitarity}
and whose anomalous dimension $\delta D$ is non-zero.
When we send the coupling constant to zero, the anomalous 
dimension $\delta D$ vanishes and the highest-weight 
multiplet becomes short \eqref{eq:N4.Split.Short}. 
Nevertheless, the remaining states of the interacting long multiplet
cannot disappear, instead they form an additional short
highest-weight multiplet,
see \figref{fig:N4.Split.Splitting}.
For $s_{1,2}>0$ the highest weight of the submultiplet is offset
from the highest weight of the long multiplet by 
%
\<\label{eq:N4.Split.Offset}
\delta w\indup{i}\eq
\weight{+0.5;-1,0;+1,0,0;-0.5,+1},
\nln
\delta w\indup{ii}\eq
\weight{+0.5;0,-1;0,0,+1;+0.5,+1}.
\>
The new highest-weight multiplet is also of short type.
For $s_{1,2}=0$ the above shift would lead
to a negative spin; then the new highest weight is shifted further
by in total
\<\label{eq:N4.Split.OffsetBPS}
\delta w\indup{I}\eq
\weight{+1.0;0,0;+2,0,0;0.0,+1},
\nln
\delta w\indup{II}\eq
\weight{+1.0;0,0;0,0,+2;0.0,+1};
\>
this multiplet is of eighth-BPS type.

Naively, if one of the eighth-BPS conditions
is satisfied in the classical theory,
one might think that the dimension 
is protected because an anomalous dimension 
would violate unitarity \eqref{eq:N4.Split.Unitarity}.
However, the eighth-BPS multiplet may join, 
in similarity with the Higgs mechanism, with short multiplets
and form a generic, long multiplet. Note 
that the shift $\delta w\indup{I,II}$ 
yields multiplets with $q_{1,2}\geq 2$,
therefore eighth-BPS multiplets with $q_{1,2}=0,1$ are indeed protected.

\section{The Field-Strength Multiplet}
\label{sec:N4.Fund}

Let us now reconsider the fields $\fldWf{A}$ and understand 
their transformation properties.
For that purpose we have another look at \tabref{tab:N4.States.Fields}.
All representations of $\alSU(2)\times\alSU(2)$
are symmetric tensor products of the fundamental representation, while 
the representations of $\alSU(4)$ are antisymmetric. 
Using two bosonic oscillators 
$(\osca^\alpha,\osca^\dagger_\alpha)$, 
$(\oscb^{\dot \alpha},\oscb^\dagger_{\dot\alpha})$ 
with $\alpha,\dot\alpha=1,2$
and one fermionic oscillator $(\oscc^a,\oscc^\dagger_a)$
with $a=1,2,3,4$
we can thus write \cite{Gunaydin:1985fk}%
\footnote{In a complex algebra we can assume the
oscillators $(\osca^\alpha,\oscb^{\dot\alpha},\oscc^a)$ and
$(\osca^\dagger_\alpha,\oscb^\dagger_{\dot\alpha},\oscc^\dagger_a)$
to be independent.}
\<\label{eq:N4.Fund.Fields}
\cder^k \fldF \mathrel{}&&\hateq
  (\osca^\dagger)^{k+2} 
  (\oscb^\dagger)^{k\phantom{\mathord{}+0}}
  (\oscc^\dagger)^0 
  \vac, \nln
\cder^k \Psi \mathrel{}&&\hateq
  (\osca^\dagger)^{k+1} 
  (\oscb^\dagger)^{k\phantom{\mathord{}+0}}
  (\oscc^\dagger)^1 
  \vac, \nln
\cder^k \Phi \mathrel{}&&\hateq
  (\osca^\dagger)^{k\phantom{\mathord{}+0}} 
  (\oscb^\dagger)^{k\phantom{\mathord{}+0}} 
  (\oscc^\dagger)^2 
  \vac, \nln
\cder^k \dot\Psi \mathrel{}&&\hateq
  (\osca^\dagger)^{k\phantom{\mathord{}+0}} 
  (\oscb^\dagger)^{k+1} 
  (\oscc^\dagger)^3 
  \vac, \nln
\cder^k \dot \fldF \mathrel{}&&\hateq
  (\osca^\dagger)^{k\phantom{\mathord{}+0}}
  (\oscb^\dagger)^{k+2} 
  (\oscc^\dagger)^4 
  \vac .
\>
Each of the oscillators $\osca^\dagger_\alpha,\oscb^\dagger_{\dot\alpha},\oscc^\dagger_a$
carries one of the $\alSU(2),\alSU(2),\alSU(4)$ spinor indices of the fields,
for example
\[\label{eq:N4.Fund.Example}
\cder_{\dot\alpha\beta}\,\dot\Psi^a_{\dot\gamma}\sim
\varepsilon^{abcd}
\osca^\dagger_\beta
\oscb^\dagger_{\dot\alpha}\oscb^\dagger_{\dot\gamma}
\oscc^\dagger_b\oscc^\dagger_c\oscc^\dagger_d\vac.
\]
The statistics of the oscillators automatically symmetrises the
indices in the desired way 
as explained in \secref{sec:N4.States}.
We will further assume the commutation relations
\[\label{eq:N4.Fund.Comm}
\comm{\osca^{\alpha}}{\osca^\dagger_{\beta}}=\delta^\alpha_\beta,
\qquad
\comm{\oscb^{\dot\alpha}}{\oscb^\dagger_{\dot\beta}}=\delta^{\dot\alpha}_{\dot\beta},
\qquad
\acomm{\oscc^{a}}{\oscc^\dagger_{b}}=\delta^a_b.
\]
Finally, the oscillators $\osca^\alpha,\oscb^{\dot\alpha},\oscc^a$ are taken to
annihilate the state $\vac$.

Using oscillators we can construct 
a representation of the unitary superalgebra. 
We assemble the oscillators 
$(\osca,\oscb^\dagger|\oscc)$ into 
a $\rep{4|4}$-dimensional superoscillator $\oscA$,
whereas $\oscA^\dagger$ consists of 
$(\osca^\dagger,-\oscb|\oscc^\dagger)$.
The generators of $\alU(2,2|4)$ are then given by%
\footnote{$\scomm{A}{B}$ is the graded commutator:
It equals $\comm{A}{B}$ if $A$ or $B$ is bosonic and
$\acomm{A}{B}$ if both $A$ and $B$ are fermionic.}
\[\label{eq:N4.Fund.Generator}
\algJ_A{}^B=\oscA^\dagger_A\oscA^B,\qquad 
\bigscomm{\oscA^B}{\oscA^\dagger_A}=\delta_A^B.
\]
It is straightforward to verify that the generators satisfy the commutation
relations of a unitary superalgebra. 
In \appref{app:U224.Osc} we split up the generators into 
$\alSU(2)\times\alSU(2)\times\alSU(4)$ notation.

Using the expressions in \appref{app:U224.Osc} one can see that 
the set of fields \eqref{eq:N4.Fund.Fields} is closed under the action
of the generators, the fields thus span a module of $\alU(2,2|4)$.
This module will be denoted by $\mdlF$ and is spanned by all the
fields $\fldWf{A}$
\[\label{eq:N4.Fund.Module}
\mbox{`\emph{field-strength module}':}\quad
\mdlF=[\fldWf{A}]=[\cder^k \Phi,\cder^k \Psi,\cder^k \dot \Psi,
\cder^k \fldF,\cder^k \dot\fldF].
\]
This module is also called the \emph{singleton}.
It is furthermore a module of $\alPSU(2,2|4)$ because the central charge 
vanishes for all fields
\[\label{eq:N4.Fund.Central}
C=1-\half n_{\osca}+\half n_{\oscb}-\half n_{\oscc}=0.
\]
Let us now identify the highest weight. 
The supercharge $\algS$ transforms an $\osca^\dagger$ into a $\oscc^\dagger$.
Annihilation of a state by all $\algS$'s requires there to be no 
excitation of type $\osca^\dagger$ or the maximum number 
of $4$ excitations of type $\oscc^\dagger$.
Conversely, annihilation by $\algSd$ requires there to be
no excitation of type $\oscb^\dagger$ or no excitation of type $\oscc^\dagger$.
Among the fields \eqref{eq:N4.Fund.Fields}, 
these conditions are satisfied only by the scalar fields
$\Phi\hateq(\oscc^\dagger)^2\vac$. Annihilation by 
$\algR^a{}_b$ with $a<b$ picks out 
\[\label{eq:N4.Fund.Highest}
\Phi_{34}\hateq\oscc^\dagger_3\oscc^\dagger_4\vac
=\state{\fldZ}
\]
as the highest weight state or superconformal primary field.
The field $\state{\fldZ}$ is a scalar $[s_1,s_2]=[0,0]$, 
an $\alSO(6)$ vector $[q_1,p,q_2]=[0,1,0]$ and has dimension $D=1$
as well as central charge $C=0$.
The highest weight is therefore
\[\label{eq:N4.Fund.Label}
w\indups{F}=\weight{1;0,0;0,1,0;0,1}=
[0;0;0,1,0;0;0],\]
where we have defined the hypercharge $B=0$ and length $L=1$.

The vacuum state $\vac$ is invariant under $\alSU(4)$, but it is
not physical. Conversely, the highest weight state $\state{\fldZ}$ is
physical, but superficially breaks $\alSU(4)$ to $\alSU(2)\times\alSU(2)$.
When dealing with physical states it is convenient to employ a
notation suited for $\alSU(2)\times\alSU(2)$ invariance.
We define the oscillator $\oscd^{\dot a}$
with index $\dot a=1,2$
\[\label{eq:N4.Fund.Physical}
\oscd^\dagger_1=\oscc^4,\quad \oscd^\dagger_2=\oscc^3,\qquad
\oscd^1=\oscc^\dagger_4,\quad \oscd^2=\oscc^\dagger_3.
\]
The benefit of this notation is that now the highest weight 
state $\state{\fldZ}=\oscc^\dagger_3\oscc^\dagger_4\vac$,
see \eqref{eq:N4.Fund.Highest},
is annihilated by $\osca_{1,2},\oscb_{1,2},\oscc_{1,2},\oscd_{1,2}$.
The drawback is that the notation breaks the $\alSU(4)$ invariant
notation and the expressions for the $\alU(2,2|2+2)$ generators 
thus complicate. 
Let us also state the central charge constraint
\[\label{eq:N4.Fund.CentralPhys}
C=\half(n_{\oscb}+n_{\oscd})-\half(n_{\osca}-n_{\oscc})=0,
\]
i.e.~the number of $\osca^\dagger,\oscc^\dagger$'s
must equal the number of $\oscb^\dagger,\oscd^\dagger$'s.

In this context it is useful to know how
to represent a state 
in terms of excitations of the oscillators.
We introduce a multi-particle vacuum operator $\state{\fldZ,L}$ which 
is the tensor product of $L$ vacua $\state{\fldZ}$.
The oscillators 
$\osca^\dagger_{p,\alpha},\oscb^\dagger_{p,\dot\alpha},\oscc^\dagger_{p,a},\oscd^\dagger_{p,\dot a}$
now act on a site specified by $p$ and
commutators of two oscillators 
vanish unless they act on the same site.
Equivalently, we define the unphysical multi-particle vacuum state
$\state{0,L}$. A generic state is written as
\[\label{eq:N4.Fund.States}
(\osca^\dagger)^{n_{\osca}}(\oscb^\dagger)^{n_{\oscb}}
(\oscc^\dagger)^{n_{\oscc}}(\oscd^\dagger)^{n_{\oscd}}\state{\fldZ,L}
\qquad\mbox{or}\qquad
(\osca^\dagger)^{n_{\osca}}(\oscb^\dagger)^{n_{\oscb}}
(\oscc^\dagger)^{n_{\oscc}}\state{0,L}.
\]
The individual oscillator excitation numbers $n$ for a state with
given weight can be 
found in \appref{app:U224.Osc}.

\section{Correlation Functions}
\label{sec:N4.Corr}

In a conformal field theory, 
correlation functions of local operators obey certain relations due
to conformal symmetry. These are especially tight for 
two-point and three-point functions:
For example, two-point functions are allowed only between 
multiplets of equal labels and involve only one free parameter.
A similar result holds for superconformal symmetry, but we will
not consider it explicitly because it would require working in superspace.
For example, let us consider 
scalar primary or highest-weight operators $\Op_{1,2,3}$ 
with scaling dimensions $D_{1,2,3}$ at points $x_{1,2,3}$.
For two-point functions the dimensions must agree exactly $D=D_1=D_2$,
the correlator is
\[\label{eq:N4.Corr.TwoPoint}
\bigvev{\Op_1(x_1)\,\Op_2(x_2)}=\frac{M_{12}}{|x_{12}|^{2D}}\,,
\]
where $x_{ij}$ is the distance $x_i-x_j$.
Three-point functions are constrained to
\[\label{eq:N4.Corr.ThreePoint}
\bigvev{\Op_1(x_1)\,\Op_2(x_2)\,\Op_3(x_3)}=
\frac{C_{123}}
{|x_{12}|^{D_1+D_2-D_3}\,
 |x_{23}|^{D_2+D_3-D_1}\,
 |x_{31}|^{D_3+D_1-D_2}}\,.
\]
The structural uniqueness of those correlators can be understood by the fact
that all configurations of two or three (non coinciding) points
can be transformed to a standard set, say $\set{0,1}$ and $\set{-1,0,1}$,
by means of conformal transformations. 
The value of the correlator for this configuration 
determines the value for all configurations
when the points are shifted back to $\set{x_1,x_2}$ or $\set{x_1,x_2,x_3}$.

For non-scalar primary operators the story is similar, 
but we have to take care of spacetime indices.
Although we consider flat $\Real^4$, from the point
of view of conformal symmetry, spacetime is not flat;
it is rather the coset space of the
conformal group by the Poincar\'e
group and dilatations. 
As such we cannot simply compare the tangent spaces
at two different points, but we must introduce a connection.
For spinor indices the connection is 
\[\label{eq:N4.Corr.SpinConn}
J^{\dot\alpha\beta}_{12}=\frac{x_{12}^\mu \sigma_\mu^{\dot\alpha\beta}}{|x_{12}|}\,.
\]
A vector may be represented as a bi-spinor and the vector connection is
\[\label{eq:N4.Corr.VectorConn}
J^{\mu\nu}_{12}=-\half \sigma^\mu_{\dot\alpha\beta} J^{\dot\alpha\delta}_{12}J^{\beta\dot\gamma}_{12}
\sigma^\nu_{\dot\gamma\delta}
=\eta^{\mu\nu}-2\,\frac{x_{12}^\mu x^\nu_{12}}{|x_{12}|^2}\,.
\]
The two-point function for primary vector operators is thus
\[\label{eq:N4.Corr.TwoVector}
\bigvev{\Op^\mu_1(x_1)\,\Op^\nu_2(x_2)}=\frac{M_{12}\,J^{\mu\nu}_{12}}{|x_{12}|^{2D}}\,.
\]

In addition to primary operators there are also descendant operators
\[\label{eq:N4.Corr.Desc}
\Op'_{\mu\nu\ldots}=\algP_\mu \algP_\nu\ldots \Op.
\]
Although correlators of descendants follow immediately from the
corresponding correlators of primaries by differentiation 
\[\label{eq:N4.Corr.DescCorr}
\bigvev{\Op'_{\mu\nu\ldots}(x)\ldots}=
\partial_\mu\partial_\nu\ldots \bigvev{\Op(x)\ldots},
\]
it is sometime hard to distinguish between primaries 
and descendants when mixing occurs. 
Therefore it is useful to know the difference in 
correlation functions explicitly.
For example, the two point function of
descendants of a scalar operator 
of dimension $D-1$ is 
\[\label{eq:N4.Corr.TwoDesc}
\bigvev{\Op^{\prime\mu}_1(x_1)\,\Op^{\prime\nu}_2(x_2)}
=\partial^\mu_1\partial^\nu_2\frac{M_{12}}{|x_{12}|^{2D-2}}
=\frac{2(D-1) M_{12} \bigbrk{J^{\mu\nu}_{12}-2(D-1)\, x^\mu_{12}x^\nu_{12}/x_{12}^2}}
{|x_{12}|^{2D}}\,.
\]
Up to normalisation this is similar 
to \eqref{eq:N4.Corr.TwoVector} but for the extra piece 
in the numerator. If mixing between primaries and descendants has 
not been resolved, one will see traces of the extra piece 
in all correlators.

Starting with four-point functions, conformal invariants appear,
e.g.
\[\label{eq:N4.Corr.Invariants}
s=\frac{x_{13}^2x_{24}^2}{x_{12}^2x_{34}^2}\,,
\quad
t=\frac{x_{14}^2x_{23}^2}{x_{12}^2x_{34}^2}\,.
\]
Naturally, four-point functions may depend on $s,t$ and their form is
not fully restricted. 
However, in a conformal field theory one may expect to have an
operator product expansion (OPE), which enables one to write
products of two local operators at sufficiently 
close point $x$, $x+\delta x$ as a sum of local operators
at point $x$
\[\label{eq:N4.Corr.OPE}
\Op_1(x)\, \Op_2(x+\delta x)=\sum\nolimits_k F_{12}^k\, e_{12,k}(\delta x)\,\Op_k(x),
\]
where $e_{12,k}(\delta x)$ is
the conformal partial wave corresponding to the involved operators.
The structure constants $F_{12}^k$ can be determined by inserting this
expression in the three-point function
and comparing to the two-point function;
roughly speaking one obtains
\[\label{eq:N4.Corr.StrucConst}
F_{kl}^n M_{nm}\sim C_{kln}.\]
Equivalently, one obtains for a four-point
function where two pairs of points are close
\[\label{eq:N4.Corr.FourPoint}
\bigvev{\Op_1\,\Op_2\,\Op_3\,\Op_4}\sim
\sum\nolimits_{kl}
F_{12}^k F_{34}^l\,
e_{12,k}(\delta x_{12})
e_{34,l}(\delta x_{34})\,
\bigvev{\Op_k\,\Op_l}.
\]
%

\section{The Current Multiplet}
\label{sec:N4.Currents}

Superconformal symmetry is an exact global symmetry. 
As such it should give rise to one conserved current $\OpQ_\mu$
for each of its generators $\algJ$,
\[\label{eq:N4.Currents.Conserved}
\partial^\mu \OpQ_\mu =0.
\]
In the Hamiltonian picture, a conserved charge $Q$ is obtained as the 
integral of the time component $\OpQ_t$ over a time slice at $t_0$
\[\label{eq:N4.Currents.Charge}
Q=\int d^3 x\, \OpQ_t(t_0,x).
\]
The charge is indeed independent of the time slice $t_0$
due to the conservation of $\OpQ_\mu$. 
It acts as a symmetry generator $\algJ$ when inserted within 
Poisson brackets $\algJ=\{Q,\ldots\}$. 
Furthermore, the currents satisfy an 
algebra $\{\OpQ,\OpQ\}\sim\OpQ$, the \emph{current algebra}.
This gives rise to the symmetry algebra at the level of charges $\{Q,Q\}=\algstr\, Q$.
All this naturally translates 
into a quantum field theory in canonical quantisation.

The transformation properties of the current $\OpQ_{\mu\ldots}$ 
translate into the transformation properties of the generator $\algJ$.
For example, the conserved current associated to the momentum 
generator $\algP_\nu$ is the stress-energy tensor $\OpQ_{\mu\nu}$.
It has canonical dimension $4$, one for $\algP_{\nu}$ and three for $d^3 x$
in \eqref{eq:N4.Currents.Charge}. 
The stress-energy tensor also gives rise to the currents corresponding to 
the other generators of the conformal algebra%
\footnote{Note that the stress energy tensor 
is part of the reducible module with highest weight $[1,0,1]$,
the adjoint of $\alSU(4)$, as described in \secref{sec:N4.Modules}}
\[\label{eq:N4.Currents.Conformal}
\OpQ^{\algP}_{\mu\nu}\sim \OpQ_{\mu\nu},\quad
\OpQ^{\algL}_\mu{}^\rho{}_\sigma\sim\eta^{\rho\nu} \OpQ_{\mu[\nu}x_{\sigma]},\quad
\OpQ^{\algD}_\mu{}\sim\OpQ_{\mu\nu}x^\nu,\quad
\OpQ^{\algK}_{\mu\nu}\sim x^2\OpQ_{\mu\nu}-2\OpQ_{\mu\rho}x^\rho x_\nu.
\]
One easily verifies that all currents $\OpQ_\mu$ are indeed conserved
despite the appearance of $x_\nu$. 
However, conservation of
$\OpQ^{\algD}_\mu$ and $\OpQ^{\algK}_{\mu\nu}$ requires 
the trace of the stress-energy tensor to vanish
\[\label{eq:N4.Currents.Trace}
\eta^{\mu\nu}\OpQ_{\mu\nu}=0.
\]
In a quantum theory it is often impossible to construct 
a stress-energy tensor $\OpQ_{\mu\nu}$ which obeys
\eqref{eq:N4.Currents.Conserved} and \eqref{eq:N4.Currents.Trace}
at the same time. This indicates the breakdown of conformal 
symmetry; only Poincar\'e symmetry remains because 
conservation of $\OpQ^{\algL}_\mu{}^\rho{}_\sigma$ and 
$\OpQ^{\algP}_{\mu\nu}$ is independent of the tracelessness of $\OpQ_{\mu\nu}$.
In fact, the trace of the stress-energy tensor is related to
the beta function \eqref{eq:N4.Quantum.BetaDef}.

Before we proceed to superconformal symmetry we note that 
it is useful to write the conserved currents in spinor notation. 
The stress-energy tensor becomes 
$\OpQ_{\dot\alpha\dot\gamma\beta\delta}$
which is symmetric in both pairs of indices.
Now it is straightforward to construct the currents
by contracting the indices of $\OpQ_{\dot\alpha\dot\gamma\beta\delta}$
by $x^\mu \sigma_\mu$, for example
\[\label{eq:N4.Currents.Spinors}
\OpQ^{\algD}_{\dot\alpha\beta}\sim
x^\mu \sigma^{\dot\gamma\delta}_\mu \OpQ_{\dot\alpha\dot\gamma\beta\delta},
\qquad
\OpQ^{\algK}_{\dot\alpha\beta}{}^{\dot\gamma\delta}\sim
x^\mu x^\nu \sigma^{\dot\gamma\lambda}_\mu \sigma^{\delta\dot\kappa}_\nu
\OpQ_{\dot\alpha\dot\kappa\beta\lambda}.
\]
The currents are conserved 
due to the symmetry of the indices. 

For superconformal symmetry there are four conserved currents
\[\label{eq:N4.Currents.SuperCurrents}
\OpQ_{\dot\alpha\beta}{}^c{}_d,\quad
\OpQ^c_{\dot\alpha\beta\delta},\quad
\OpQ_{\dot\alpha\dot\gamma\beta d},\quad
\OpQ_{\dot\alpha\dot\gamma\beta\delta}.
\]
The first three correspond to rotations $\algR^c{}_d$
and supertranslations $\algQ^c{}_{\delta}$ and $\algQd_{\dot\gamma d}$.
Furthermore, upon contraction with one $x_\mu$ the 
second and third ones correspond to $\algSd^{c\dot\delta}$ and
$\algS^\gamma{}_d$.
All of these currents are part of the supercurrent multiplet with 
highest weight
\[\label{eq:N4.Currents.Multiplet}
w\indup{curr}=\weight{2;0,0;0,2,0;0,2}=[0;0;0,2,0;0;0].
\]
This multiplet, decomposed in terms of representations 
of $\alSU(2,2)\times\alSU(4)$, is presented in \tabref{tab:N4.Currents.Multi}.
The labels $[s_1,r,s_2]$ of $\alSU(2,2)$ are the spins $s_1,s_2$ as well
as $r=-D-\half s_1-\half s_2$. 
It is worth noting that the two scalars at $D=4$ are the on-shell
Lagrangian $\Lagr\indups{YM}$ and the topological charge density $\Tr \fldF\wedge\fldF$.
\begin{table}\centering
$\begin{array}{|l|l|}\hline
D   & \alSU(2,2)\times\alSU(4) \\\hline
2   & [0,-2,0]\times[0,2,0]  \\
2.5 & [1,-3,0]\times[0,1,1] + [0,-3,1]\times[1,1,0] \\
3   & [2,-4,0]\times[0,1,0] + [1,-4,1]\times[1,0,1] + [0,-4,2]\times[0,1,0] + \mathord{}\\
    & [0,-3,0]\times[0,0,2] + [0,-3,0]\times[2,0,0] \\
3.5 & [2,-5,1]\times[1,0,0] + [1,-5,2]\times[0,0,1] + \mathord{}\\
    & [1,-4,0]\times[0,0,1] + [0,-4,1]\times[1,0,0] \\
4   & [2,-6,2]\times[0,0,0] + [0,-4,0]\times[0,0,0] + [0,-4,0]\times[0,0,0]  \\\hline
\end{array}$

\caption{The supercurrent multiplet decomposed in $\alSU(2,2)\times\alSU(4)$.}
\label{tab:N4.Currents.Multi}
\end{table}

\finishchapter 

\chapter{The Dilatation Operator}
\label{ch:Dila}

The dilatation generator is a means to investigate scaling
dimensions in a conformal field theory. 
We will start in \secref{sec:Dila.Dim} by comparing different 
methods of obtaining scaling dimensions. 
We will then go on to study aspects of the dilatation operator 
which will be useful in the following chapters. 
In \secref{sec:Dila.Pert} we will consider 
the symmetry algebra and states in perturbation theory. 
\secref{sec:Dila.Sect} contains an investigation
of closed subsectors \cite{Beisert:2003jj} and 
in \secref{sec:Dila.SU2} we will compute 
the one-loop dilatation operator within 
a subsector \cite{Beisert:2003tq}.
General perturbative contributions in field theory
are investigated in \secref{sec:Dila.Theory}.
Finally, in \secref{sec:Dila.Planar} we will consider 
the planar limit and introduce the notation to be used 
in most parts of this \docphysrept[dissertation]{work}. 

\section{Scaling Dimensions}
\label{sec:Dila.Dim}

There are many ways to calculate scaling dimensions for
local operators in a conformal field theory. We will explain a few, 
paying special attention to two-point functions because 
their structure will be guiding us in the construction 
of the dilatation operator.

\subsection{Two-Point Functions}
\label{sec:Dila.Dim.TwoPoint}

In \secref{sec:N4.Corr} we have described how scaling dimensions 
affect correlation functions. It seems that they appear 
in the most direct way within two-point functions,
see \eqref{eq:N4.Corr.TwoPoint}.
Let us make the dependence on the coupling constant in
the two-point function explicit
\[\label{eq:Dila.Dim.TwoPoint}
\bigvev{\Op(x_1)\,\Op(x_2)}
=\frac{M(g)}{|x_{12}|^{2D(g)}}\,.
\]
We are aiming for a perturbative 
investigation and we can only expect to reproduce the 
form predicted by conformal symmetry in a series
expansion in $g$.
Here we pause and reconsider the above equation
noting that $|x_{12}|$ is a dimensionful quantity 
and its exponent $-2D(g)$ depends on $g$. 
A perturbative expansion in $g$ will 
lead to a formally meaningless 
expression involving $\log |x_{12}|$.
This is related to the fact that the mass dimension of 
the operator changes with $g$. 
The only fully consistent way to treat this issue in 
a series expansion is to introduce 
an arbitrary scale $\mu$ and rescale $\Op$ by 
$\mu^{-\delta D(g)}$ to a
fixed mass dimension $D_0=D(0)$. 
We can now expand and obtain
\[\label{eq:Dila.Dim.Expand}
\mu^{-2\delta D(g)}\bigvev{\Op(x_1)\,\Op(x_2)}
=\frac{M_0}{|x_{12}|^{2D_0}}
+g^2\frac{M_2+M_0 D_2\log |\mu x_{12}|^{-2}}{|x_{12}|^{2D_0}}
+\ldots\,.
\]
The very same problem occurs in perturbative quantum field theories 
and requires for the introduction of an auxiliary scale. 
Let us now go ahead and calculate the scaling dimension of the 
operator 
\[\label{eq:Dila.Dim.Op}
\Op_{mn}=\Tr \Phi_m\Phi_n.
\]
Using the free generating functional \eqref{eq:N4.Quantum.FreeConnected}
with scalar propagator 
\begin{figure}
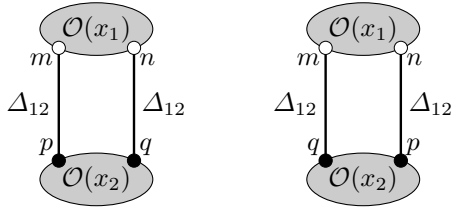
\centering
\parbox{3cm}{\centering\includegraphics{sec02.corr.0a.eps}}
\quad
\parbox{3cm}{\centering\includegraphics{sec02.corr.0b.eps}}
\caption{Tree-level contributions to the two-point function
of $\Op_{mn}=\Tr \Phi_m\Phi_n$.}
\label{fig:Dila.Dim.Tree}
\end{figure}
\[\label{eq:Dila.Dim.Prop}
\Delta(x,y)=\frac{1}{(x-y)^2}\,,
\]
the tree-level two point function is readily evaluated 
using $\grSU(N)$ as gauge group and the diagrams in 
\figref{fig:Dila.Dim.Tree}%
\footnote{We have neglected contractions between 
fields at the same point.
Their (divergent) contribution will have to be absorbed 
into the definition of $\Op$. 
This is always possible and we will assume that this step
has already been performed. In other words,
the operator $\Op$ is considered to be `normal ordered'.}
\<\label{eq:Dila.Dim.Tree}
\bigvev{\Op_{mn}(x_1)\,\Op_{pq}(x_2)}
\eq\frac{\eta_{mp}\eta_{nq}\,\gaugemet{\gaugeind{mp}}\gaugemet{\gaugeind{nq}}
\Tr \gaugegen{m}\gaugegen{n}\Tr \gaugegen{p}\gaugegen{q}}{N^2|x_{12}|^{4}}
+\frac{\eta_{mq}\eta_{np}\,\gaugemet{\gaugeind{mq}}\gaugemet{\gaugeind{np}}
\Tr \gaugegen{m}\gaugegen{n}\Tr \gaugegen{p}\gaugegen{q}}{N^2|x_{12}|^{4}}
+\order{g}
\nln
\eq\frac{2(1-N^{-2})\, \eta_{m\{p}\eta_{q\}n}}{|x_{12}|^{4}}
+\order{g}.
\>
We can read off the classical dimension $D_0=2$ from this expression.

\begin{figure}
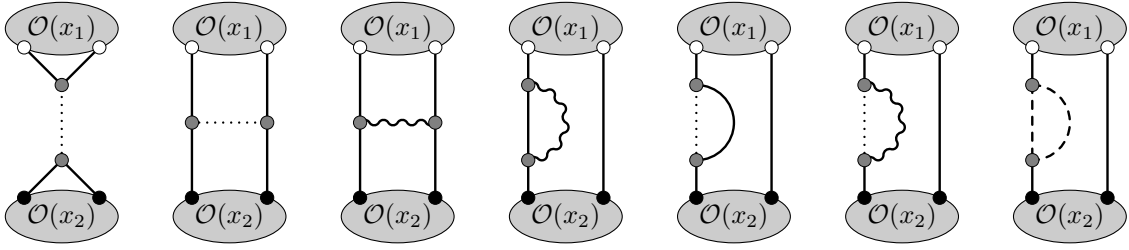
\centering
\parbox{1.6cm}{\centering\includegraphics{sec02.corr.a.eps}}
\hfil
\parbox{1.6cm}{\centering\includegraphics{sec02.corr.b.eps}}
\hfil
\parbox{1.6cm}{\centering\includegraphics{sec02.corr.c.eps}}
\hfil
\parbox{1.6cm}{\centering\includegraphics{sec02.corr.d1.eps}}
\hfil
\parbox{1.6cm}{\centering\includegraphics{sec02.corr.d2.eps}}
\hfil
\parbox{1.6cm}{\centering\includegraphics{sec02.corr.d3.eps}}
\hfil
\parbox{1.6cm}{\centering\includegraphics{sec02.corr.d4.eps}}
\caption{One-loop contributions to the two-point function.
The solid, wiggly, dashed lines represent scalars, gluons, fermions, respectively.
The dotted lines correspond to a non-propagating auxiliary field that 
represents a quartic interaction.}
\label{fig:Dila.Dim.OneLoop}
\end{figure}

Trying to compute the one-loop correction we will inevitably fail
and get a divergent result unless we first regularise the theory.
We will chose the dimensional regularisation/reduction scheme
in which we assume to have a $4-2\epsilon$ dimensional spacetime.
The difference between the two schemes is that in 
dimensional regularisation we work with $6$ internal directions, 
i.e.~$6$ flavours of scalars, whereas in dimensional reduction 
this number is assumed to be $6+2\epsilon$. The
dimensional reduction scheme \cite{Siegel:1979wq,Gates:1983nr}
is convenient for regularising 
extended supersymmetric theories.%
\footnote{Apparently, the dimensional reduction scheme
proposed in \cite{Siegel:1979wq} 
leads to problems at higher loops \cite{Siegel:1980qs}.
Certainly, at one-loop it is fine.}
We will work with the action
(see \appref{app:Ten.Comp} for the ten-dimensional Lagrangian) 
\[\label{eq:Dila.Dim.DimReg}
S\indups{DR}[\fldW]=N\int \frac{d^{4-2\epsilon} x}{(2\pi)^{2-\epsilon}} \, 
\Lagr\indups{YM}[\fldW,g\mu^\epsilon].
\]
This action is dimensionless if the fields $\Phi,\fldA$ have canonical 
dimensions $1-\epsilon$ and $\Psi$ has dimension
$\sfrac{3}{2}-\epsilon$.
The dimensionally regularised propagator is 
\[\label{eq:Dila.Dim.PropReg}
\Delta(x,y)=\frac{2^{-\epsilon}\,\Gammafn(1-\epsilon)}{|x-y|^{2-2\epsilon}}\,.
\]
We need to evaluate a couple of diagrams, see \figref{fig:Dila.Dim.OneLoop},
and find for the one-loop correlator
\<\label{eq:Dila.Dim.OneLoop}
\bigvev{\Op_{mn}(x_1)\,\Op_{pq}(x_2)}
\eq 
2(1-N^{-2})\, \eta_{m\{p}\eta_{q\}n}\,\Delta_{12}^2
\nlnum\nonumber
+(1-N^{-2})g^2\,
\bigbrk{\half \eta_{m\{p}\eta_{q\}n}\, \tilde H_{12,12}
        -\sfrac{1}{4} \eta_{mn}\eta_{pq}\, X_{1122}  }
+\order{g^3}.
\>
The following functions and integrals appear at the one-loop level
\<\label{eq:Dila.Dim.YXH}
I_{x_1x_2}\eq \half \Delta(x_1-x_2),
\nln
Y_{x_1x_2x_3}\eq\mu^{2\epsilon}\int\frac{d^{4-2\epsilon}z}{(2\pi)^{2-\epsilon}} I_{x_1z}I_{x_2z}I_{x_3z},
\nln
X_{x_1x_2x_3x_4}\eq\mu^{2\epsilon}\int\frac{d^{4-2\epsilon}z}{(2\pi)^{2-\epsilon}} I_{x_1z}I_{x_2z}I_{x_3z}I_{x_4z},
\\\nn
\tilde H_{x_1x_2,x_3x_4}\eq\half\mu^{2\epsilon}\lrbrk{\frac{\partial}{\partial x_1}+\frac{\partial}{\partial x_3}}^2
\int\frac{d^{4-2\epsilon}z_1\,d^{4-2\epsilon}z_2}{(2\pi)^{4-2\epsilon}} \,
I_{x_1z_1}I_{x_2z_1}I_{z_1z_2}I_{z_2x_3}I_{z_2x_4}.
\>
where the shape of the letter represents the connections in terms
of scalar propagators.
In two-point functions they evaluate to
 \cite{Kazakov:1984ns}
\[\label{eq:Dila.Dim.YXH2pt}
Y_{112}=
\frac{\xi\, I_{12}}{\epsilon(1-2\epsilon)}\,,\quad
X_{1122}=
\frac{2(1-3\epsilon)\kappa\,\xi \,I_{12}^2}{\epsilon(1-2\epsilon)^2}\,,\quad
\tilde H_{12,12}=
-\frac{2(1-3\epsilon)(\kappa-1)\,\xi\, I_{12}^2}{\epsilon^2(1-2\epsilon)}\,.
\]
These involve two convenient combinations $\xi,\kappa$ 
\[\label{eq:Dila.Dim.XiGamma}
\xi=\frac{\Gammafn(1-\epsilon)}{\bigabs{\half\mu^2 x_{12}^2}^{-\epsilon}}\,,
\qquad
\kappa=\frac{\Gammafn(1-\epsilon)\,\Gammafn(1+\epsilon)^2\,\Gammafn(1-3\epsilon)}
{\Gammafn(1-2\epsilon)^2\,\Gammafn(1+2\epsilon)}=1+6\zeta(3)\,\epsilon^3+\order{\epsilon^4}.
\]

At this point, it is useful to split up the operator into 
irreducible representations of $\alSO(6)$. There are two,
the symmetric-traceless $[0,2,0]$ and the singlet $[0,0,0]$
\[\label{eq:Dila.Dim.UnMix}
\OpQ_{mn}=\Op_{(mn)}=\Op_{mn}-\sfrac{1}{6+2\epsilon}\,\eta_{mn}\eta^{pq}\,\Op_{pq},
\qquad
\OpK=\eta^{mn}\,\Op_{mn}.
\]
These have classical weights
\[\label{eq:Dila.Dim.Weights}
w_{\OpQ}=\weight{2;0,0;0,2,0;0,2},\qquad
w_{\OpK}=\weight{2;0,0;0,0,0;0,2},
\]
which are indeed highest weights,
essentially because there are no states of lower dimension.
For the symmetric-traceless operator the correlator reduces to
\[\label{eq:Dila.Dim.BPS}
\bigvev{\OpQ_{mn}(x_1)\,\OpQ_{pq}(x_2)}
=
2(1-N^{-2})\, \eta_{m(p}\eta_{q)n}\,
\bigbrk{\Delta_{12}^2+\quarter g^2\tilde H_{12,12}}
+\order{g^3}.
\]
Here we can take the limit $\epsilon\to 0$, it turns out that the one-loop
correction vanishes identically and we obtain precisely the tree-level result
\cite{D'Hoker:1998tz}.
This remarkable cancellation is intimately related to the vanishing of the
beta function. 
The operator $\OpQ$ is part of the half-BPS multiplet
with weight \eqref{eq:Dila.Dim.Weights}, see \secref{sec:N4.Split}. 
In fact it is part of the 
current multiplet of superconformal symmetry, 
see also \secref{sec:N4.Currents}, were it not protected,
superconformal symmetry would be broken.

For the Konishi operator $\OpK$ the result is very different
\<\label{eq:Dila.Dim.Konishi}
\bigvev{\OpK(x_1)\,\OpK(x_2)}
\eq
4(1-N^{-2})(3+\epsilon)\,
\bigbrk{\Delta_{12}^2+\quarter g^2\tilde H_{12,12}-\quarter g^2(3+\epsilon)X_{1122}}
+\order{g^3}
\nln\eq
4(1-N^{-2})(3+\epsilon)\Delta_{12}^2 (1-g^2\gamma\,\xi/\epsilon)+\order{g^3}.
\>
We see that the coefficient 
\[\label{eq:Dila.Dim.KonishiDimEpsilon}
\gamma=
\frac{2(1-3\epsilon)(\kappa-1)}{\epsilon(1-2\epsilon)}
+\frac{2(3+\epsilon)(1-3\epsilon)\kappa}{(1-2\epsilon)^2}\to 6
\]
is finite in the limit $\epsilon\to 0$ and
the correlator is thus ill-defined.
We should first renormalise the operator
in order to remove the $1/\epsilon$ pole
\[\label{eq:Dila.Dim.Renorm}
Z\OpK=(1+\half g^2\gamma\, \xi_0/\epsilon)\,\OpK,
\quad
\mbox{where}
\quad
\xi_0=2^{-\epsilon}\,\Gammafn(1-\epsilon).
\]
In a correlator of $Z\OpK$'s, this replaces $\xi$ 
in \eqref{eq:Dila.Dim.Konishi} by $\xi-\xi_0$.
We can now evaluate the one-loop term
which is regular at $\epsilon\to 0$
\<
-g^2 \gamma \,\lim_{\epsilon\to 0}\frac{\xi-\xi_0}{\epsilon}
\eq -g^2 \gamma \,
\lim_{\epsilon\to 0}
\frac{\bigbrk{|\mu x_{12}|^{-2}}^{-\epsilon}-1}{\epsilon}\,\xi_0
\nln\eq
-g^2 \gamma\, 
\lim_{\epsilon\to 0}
\frac{\partial \bigbrk{|\mu x_{12}|^{-2}}^{-\epsilon}}{\partial \epsilon}
=g^2 \gamma\, \log |\mu x_{12}|^{-2}
\>
and take the limit. We find 
\[\label{eq:Dila.Dim.KonishiUnReg}
\bigvev{Z\OpK(x_1)\,Z\OpK(x_2)}=
\frac{12(1-N^{-2})}{|x_{12}|^{4}}\lrbrk{1+6 g^2 \log |\mu x_{12}|^{-2}}
+\order{g^3}.
\]
By comparing to \eqref{eq:Dila.Dim.Expand}
we obtain the one-loop
anomalous dimension $D_2=6$ or altogether,
after inserting the 
definition $g^2=\gym^2N/8\pi^2$ \eqref{eq:N4.Quantum.Action},
in agreement with \cite{Anselmi:1997mq}
\[\label{eq:Dila.Dim.KonishiDim}
D=2+6g^2+\order{g^3}=2+\frac{3\gym^2 N}{4\pi^2}+\order{g^3}.
\]

The calculation presented above
resulted in the simplest non-trivial scaling dimension. 
In a generic computation
one has to deal with more involved operator
mixing and many more diagrams. 
We have seen only a glimpse of that here, 
luckily representation theory alone was sufficient 
to resolve the mixing. 

\subsection{Higher-Point Functions}
\label{sec:Dila.Dim.FourPoint}

There are other ways in which to obtain scaling dimensions. 
One could, for example, calculate three-point functions.
They contain information not only about the scaling dimension
of all three involved operators, but at the same time 
also about the coefficients $C_{123}$. These are interesting 
because they are related to the structure constants $F_{12}^3$ of 
the operator product expansion.
The price one has to pay is the added difficulty due
to the additional spacetime point in the Feynman diagrams. 
In practice three-point functions are rarely considered.
More interesting are four-point functions, 
although they might appear even more difficult at first sight. 
The simplification comes about when one considers protected
operators at all four points \cite{Bianchi:1999ge}. 
Using superspace techniques
these correlators turn out to be manifestly finite
without the need to regularise or renormalise
\cite{Gonzalez-Rey:1998tk,Eden:1998hh,Eden:1999kh,Bianchi:1999ge}.
Furthermore, there are some constraints from 
superconformal field theory which can be used to 
reduce the complexity of the calculation \cite{Arutyunov:2003ae,Arutyunov:2003ad}.
Despite their simplicity, these four-point functions
are interesting due to the OPE (c.f.~\secref{sec:N4.Corr})
which allows for unprotected operators in the intermediate channel. 
A single four-point function 
can be shown to encode the information about scaling dimensions
and also structure constants of infinitely many local operators
\cite{Dolan:2001tt}.
A number of scaling dimensions, even at two-loops, 
have been obtained in this way
\cite{Bianchi:1999ge,Bianchi:2000hn,Dolan:2001tt,Bianchi:2002rw,Arutyunov:2002rs}.

\subsection{Violation of Current Conservation}
\label{sec:Dila.Dim.Anselmi}

A completely different method to evaluate scaling dimensions
due to Anselmi led to a few early results \cite{Anselmi:1998ms}. 
It is rather algebraic in nature
and does not require quantum field theoretic computations
as those presented above. 
It makes use of multiplet splitting at the unitarity bounds,
see \secref{sec:N4.Split}. 
Multiplet splitting occurs when the classical 
dimension $D_0$ is on one of the unitarity bounds.
When $\delta D$ is precisely zero, 
the multiplet splits up into several short multiplets.
A superconformal generator which would usually translate 
between states of different submultiplets, must annihilate the state. 
Therefore, in the interacting theory the 
action of this generator is 
proportional to $\sqrt{\delta D}\sim g$
when $g$ approaches zero.%
\footnote{This square root explains why multiplet splitting takes place at the unitarity
bound $\delta D\geq 0$: A negative $\delta D$ would yield an imaginary action
and thus violate unitarity.}
When the states are properly normalised, the
anomalous dimension $\delta D$ can be read off 
from the action of the generator.
In practice, to compute a one-loop anomalous dimension, 
this method requires to normalise the operators, 
i.e.~their two-point functions, at tree level.
For the generator one may use the semi-classical expression 
\eqref{eq:N4.D4.Trans} which does involve the coupling constant.
In principle, this trick allows also to 
obtain higher-loop anomalous dimensions from a 
field theory calculation at one loop below.
However, one has to take into account modifications of the
generators due to the Konishi anomaly \cite{Eden:2003sj}.

\subsection{The Dilatation Generator}
\label{sec:Dila.Dim.Dila}

The dilatation generator offers a different perspective on 
scaling dimensions. 
As described in \secref{sec:N4.Alg}, it measures the scaling
dimension of states transforming under 
the superconformal algebra. 
In \secref{sec:N4.States} we have emphasised that local operators
can be viewed as such states in an abstract space. 
Therefore the dilatation generator $\algD$ should yield
the scaling dimension $D$ when acting on an eigenstate $\Op$.
In particular, we have learned in \secref{sec:Dila.Dim.TwoPoint} 
that 
\[\label{eq:Dila.Dim.DilaQK}
\algD\,\OpQ_{mn}=2\, \OpQ_{mn},\qquad
\algD\,\OpK= (2+6g^2)\,\OpK+\order{g^3}.
\]
Clearly, the dilatation operator can act on any state, not just 
eigenstates. The action of $\algD$ on the mixed operator
$\Op_{mn}$ is%
\[\label{eq:Dila.Dim.DilaO}
\algD\,\Op_{mn}=(2\delta_m^p\delta_n^q+\eta_{mn}\eta^{pq}\, g^2)\,\Op_{pq}
+\order{g^3}.
\]
We obtain \eqref{eq:Dila.Dim.DilaQK} when
we project the indices to irreducible representations 
of $\alSO(6)$.

So far not much is gained by considering the dilatation generator;
we have merely rephrased the physical results of \secref{sec:Dila.Dim.TwoPoint}
into a single equation \eqref{eq:Dila.Dim.DilaO}.
Notice, however, that \eqref{eq:Dila.Dim.DilaO} describes the
eigenoperators along with their scaling dimensions. 
In contrast, a two-point function also contains the normalisation
coefficients. In practice, this fact is rather disadvantageous
because the normalisation coefficients obscure the scaling dimension
and their proper calculation usually involves a large amount of work. 
The dilatation generator clearly distinguishes between scaling dimensions 
and normalisation coefficients and thus avoids this complication.

To make true progress we need to find a way to obtain the action 
of the dilatation generator on the set of states in a more
direct fashion. 
There are several ways in which this could be done. 
To compute classical scaling dimensions is a rather trivial task,
we will describe how to implement this at the level of the 
classical dilatation operator $\algD_0$ in \secref{sec:Dila.Pert}.
Quantum corrections $\algdD$ to the dilatation generator are much 
harder to obtain. 
In \secref{sec:Dila.SU2} we will show how to extract some
information from the calculation of 
a two-point function of abstract operators.

\subsection{Canonical Quantisation}
\label{sec:Dila.Dim.Canonical}

From the path integral point of view there seems to be no obvious
way in which to represent the dilatation operator, but
in the Hamiltonian formalism and its canonical quantisation there is.
In that picture, the generators of the symmetry group correspond to conserved 
currents as explained in \secref{sec:N4.Currents}. In particular, the
dilatation operator is given by
\[\label{eq:Dila.Dim.ChargeDila}
\algD=\int d^3 x\, \OpQ_{t\mu}x^\mu,
\]
where $\OpQ_{\mu\nu}$ is the stress energy tensor of
$\superN=4$ SYM.
This we can apply to a local operator state 
$\state{\Op(x)}=\Op^\dagger(x)\vac$
\[\label{eq:Dila.Dim.ChargeApply}
\algD\,\state{\Op(x)}
\]
and thus obtain its scaling dimension.
However, it is questionable whether in practice this leads to
a reasonable simplification as compared to \secref{sec:Dila.Dim.TwoPoint}. 
As the eigenvalues of the dilatation operator are finite,
naively one might think that regularisation would be unnecessary.
Unfortunately, the bare $\algD$ can only 
act on renormalised states $Z\state{\Op}$. 
When the dilatation operator is intended to act on bare states, 
we need to renormalise it instead
\[\label{eq:Dila.Dim.ChargeRenorm}
\algD\indup{ren}= Z^{-1}\, \algD\indup{bare}\, Z.
\]
The renormalised $\algD\indup{ren}$ is finite and $\algD\indup{bare}$ and $Z$ do not commute, 
therefore $\algD\indup{bare}$ is expected to diverge.

\subsection{Matrix Quantum Mechanics}
\label{sec:Dila.Dim.QM}

A nice representation for the dilatation operator
is offered in gauge theory on the curved manifold $\Real\times S^3$,
which is conformally equivalent to flat $\Real^4$.
The map from $\Real^4$ to $\Real\times S^3$ is best described in 
radial coordinates on $\Real^4$.
The spherical coordinates map directly to $S^3$ whereas
the radial coordinate $r$ is mapped logarithmically to
the coordinate $t$ along $\Real$ of $\Real\times S^3$
\[\label{eq:Dila.Dim.RadialMap}
(r,\theta,\phi,\psi)\mapsto (t,\theta,\phi,\psi)
\quad\mbox{with}\quad
t=\log r.
\]
Therefore the dilatation generator,
which generates scale transformations 
$r\mapsto cr$, maps straightforwardly to the Hamiltonian,
i.e.~the generator of time translations, on $\Real\times S^3$. 
Spacetime rotations $\alSO(4)$ naturally map to rotations
of the sphere, whereas translations and boosts act on both,
$\Real$ and $S^3$. 
In this picture, it is natural to Kaluza-Klein decompose fields
on a time-slice, $S^3$, in terms of spherical harmonics. 
For $\superN=4$ SYM this yields precisely 
the spectrum of fields as given in \tabref{tab:N4.States.Fields}.
The decomposition turns the quantum field theory into a quantum
mechanical system of infinitely many matrices. 
This \emph{matrix quantum mechanics} is equivalent to 
$\superN=4$ SYM and one could attempt to derive the
dilatation operator in this system. 
Unfortunately, the Hamiltonian, which is derived as the 
Legendre transform of the Lagrangian, is not 
of the desired form, see \secref{sec:Dila.Pert.PreDiag}.
To perform the proposed diagonalisation might turn out to 
be very labourious in practice due to the infinite number
of matrices. 

A simpler model which appears to have a lot in common 
with $\superN=4$ SYM is the BMN matrix model \cite{Berenstein:2002jq}. 
It can be obtained from $\superN=4$ on $\Real\times S^3$
by removing all non-singlet fields under one of the 
$\alSU(2)$ symmetry algebras \cite{Kim:2003rz}. 
From the infinite set of
fields in \tabref{tab:N4.States.Fields}, only 
finitely many remain: $\cder^0\Phi,\cder^0\Psi,\cder^0\fldF$
\cite{Kim:2003rz}. 
Explicit calculations up to a relatively high order in perturbation 
theory are feasible in this model and they show 
qualitative agreement with $\superN=4$ 
\cite{Kim:2003rz,Klose:2003qc}, even if 
the results cannot agree in all cases due to the 
different multiplet structure.

\section{Perturbation Theory}
\label{sec:Dila.Pert}

In this section we will investigate 
the corrections to the generators 
of the symmetry algebra in perturbation theory.
Attention is payed to the dilatation operator
which will take a special role. 

\subsection{Quantum Representations}
\label{eq:Dila.Pert.Reps}

The superconformal symmetry algebra $\alPSU(2,2|4)$
is spanned by the generators $\algJ$. They satisfy the
algebra relations%
\footnote{Although $\alPSU(2,2|4)$ is a superalgebra,
for the sake of presentation, we shall 
assume that all operators and fields are bosonic. 
Everything generalises to fermions in a straightforward fashion,
but at the cost of obscure signs at various places.}
\[\label{eq:Dila.Pert.Alg}
\comm{\algJ_{\algind{A}}}{\algJ_{\algind{B}}}=
\algstr_{{\algind{AB}}}^{\algind{C}}\, \algJ_{\algind{C}},
\]
where $\algstr_{{\algind{AB}}}^{\algind{C}}$ are the 
structure constants of $\alPSU(2,2|4)$.
The generators can act on the set of states, 
or, more precisely, there is a representation 
which we shall also denote by $\algJ$. 
When quantum corrections are turned on, 
the transformation properties of states change.
In other words, the representation $\algJ(g)$ depends 
on the coupling constant $g$. 
For all values of $g$ the generators must satisfy 
the $\alPSU(2,2|4)$ algebra
\[\label{eq:Dila.Pert.AlgRep}
\comm{\algJ_{\algind{A}}(g)}{\algJ_{\algind{B}}(g)}=
\algstr_{{\algind{AB}}}^{\algind{C}}\, \algJ_{\algind{C}}(g).
\]
The structure constants are, in particular, 
independent of the coupling constant.
We will consider a perturbative quantum theory, therefore
we shall expand the (representation of) generators 
in powers of the coupling constant 
\[\label{eq:Dila.Pert.Expand}
\algJ(g)=\sum_{k=0}^\infty g^k \,\algJ_{k}.
\]
In perturbation theory the algebra relations can be written as
\[\label{eq:Dila.Pert.AlgPert}
\sum_{k=0}^l\comm{\algJ_{\algind{A},k}}{\algJ_{\algind{B},l-k}}=
\algstr_{{\algind{AB}}}^{\algind{C}}\, \algJ_{\algind{C},l}.
\]
Not all generators receive quantum corrections.
The Lorentz and internal rotations $\alSU(2)\times\alSU(2),\alSU(4)$
are manifest symmetries and thus independent of $g$. We do not intend to
modify them
\[\label{eq:Dila.Pert.Rotations}
\algL(g)^\alpha{}_\beta=\algL^\alpha{}_\beta,\quad
\algLd(g)^{\dot\alpha}{}_{\dot\beta}=\algLd^{\dot \alpha}{}_{\dot\beta},\quad
\algR(g)^a{}_b=\algR^a{}_b.
\]
%

\subsection{Tree-Level Algebra}
\label{eq:Dila.Pert.TreeLevel}

Let us start by investigating the classical algebra 
spanned by $\algJ_0=\algJ(0)$.
In the classical theory the fields transform among themselves
\[\label{eq:Dila.Pert.ActField}
\algJ_0\, \fldWf{A}=(\algJ_0)_{\fldind{A}}{}^{\fldind{B}} \fldWf{B}.
\]
When interactions are turned off, none of the fields can feel 
the presence of the others in the state. 
Therefore it is natural for a state to transform in the
tensor product representation of its composite fields. 
A generator $\algJ_0$ of $\alPSU(2,2|4)$ at tree-level 
can thus be written in terms of its
action on a single field $\fldWf{A}$ as
\[\label{eq:Dila.Pert.ActState}
\algJ_0 \, \fldWf{A} \cdots \fldWf{B}  = (\algJ_0)_{\fldind{A}}{}^{\fldind{C}} 
\fldWf{C} \cdots \fldWf{B} + \ldots 
+ (\algJ_0)_{\fldind{B}}{}^{\fldind{C}} \fldWf{A} \cdots \fldWf{C}.
\]
Using the notation of variations with respect to fields
introduced in \secref{sec:N4.Gauge} we shall write this as
\[\label{eq:Dila.Pert.Classical}
\algJ_0=(\algJ_0)_{\fldind{A}}{}^{\fldind{B}}  \Tr \fldWf{B} \fldWv{A}.
\]
The variation will pick any of the fields within the state
and replace it by the transformed field.
In particular, the tree-level dilatation generator is
\[\label{eq:Dila.Pert.DilClass}
\algD_0=\tsum_{\fldind{A}} \dim(\fldWf{A})\, \Tr \fldWf{A} \fldWv{A}.
\]
This isolates any of the fields and returns the same state multiplied
by the dimension of the field. When summed over all constituent fields,
the dilatation operator returns the same state multiplied by the 
total dimension being the sum of constituent dimensions
\[\label{eq:Dila.Pert.DilState}
\dim (\fldWf{A} \cdots \fldWf{B})=\dim(\fldWf{A})+\ldots+\dim(\fldWf{B}).
\]
Similarly, we can determine the classical dimension
of any operator $X$ acting on the set of states
\[\label{eq:Dila.Pert.DilOp}
X=\fldWf{A} \cdots \fldWf{B}\fldWv{C} \cdots \fldWv{D},\qquad
\comm{\algD_0}{X}=\dim (X)\, X,
\]
where the dimension is given by 
\[\label{eq:Dila.Pert.DimOp}
\dim(X)=\dim(\fldWf{A})+\ldots+\dim(\fldWf{B})
-\dim(\fldWf{C})-\ldots-\dim(\fldWf{D}).
\]
%

\subsection{Pre-Diagonalisation}
\label{sec:Dila.Pert.PreDiag}

Our aim is to diagonalise the full dilatation operator $\algD(g)$. 
We cannot expect this to be possible at the level of generators. 
However, as a first step, we can obtain a dilatation 
generator $\algD(g)$ which commutes with the classical dimension
\[\label{eq:Dila.Pert.Diag}
\comm{\algD_0}{\algD(g)}=0.
\]
This serves two purposes: On the practical side we will have 
to diagonalise $\algD(g)$ only on 
the subspace of states with equal classical dimension,
which is most easily determined through \eqref{eq:Dila.Pert.DilState}.
On the theoretical side, this removes the possibility of
states decaying into the vacuum or being created from it. 
This would be an obstacle for the definition of a planar limit. 
For the rest of this work we will assume \eqref{eq:Dila.Pert.Diag} to hold. 
This has an interesting side-effect, it specialises the dilatation 
generator $\algD$ with respect to the other generators $\algJ$ of the
superconformal algebra, see \secref{sec:Dila.Pert.Anomalous}.

In dimensional regularisation we can take \eqref{eq:Dila.Pert.Diag}
for granted. If \eqref{eq:Dila.Pert.Diag} does not hold
from the beginning,%
\footnote{This is the case for the canonically 
quantised matrix quantum mechanics of $\superN=4$ SYM 
on $\Real\times S^3$.}
we can diagonalise $\algD(g)$ perturbatively 
with respect to $\algD_0$ by means of a similarity transformation 
\[\label{eq:Dila.Pert.DiagSim}
\algJ(g)\mapsto T(g)\,\algJ(g)\,T^{-1}(g).
\]
This is possible on the operatorial level, 
i.e.~without acting on explicit states,
because all elementary interactions 
for the construction of $\algD_k$ have a 
definite dimension as given by \eqref{eq:Dila.Pert.DilOp,eq:Dila.Pert.DimOp}.
Let us state the resulting dilatation operator 
up to second order.
Assume $\delta \algD(g)=\algD(g)-\algD_0$ decomposes as 
\[\label{eq:Dila.Pert.DDim}
\algdD=\sum\nolimits_d \delta\algD_d\quad\mbox{with}\quad
\dim(\delta\algD_d)=d.
\]
Then the transformation 
\[\label{eq:Dila.Pert.DiagTrans}
T(g)=1+\sum\nolimits_{d\neq 0}\frac{1}{d}\, \delta \algD_d+\ldots
\]
yields the diagonalised dilatation operator
\[\label{eq:Dila.Pert.DPert}
\algdD\mapsto 
\algdD_0+\sum\nolimits_{d\neq 0}\, \delta\algD_d\,\frac{1}{d}\, \algdD_{-d}
+\ldots\,.
\]
Note that this is merely standard perturbation theory:
The first terms is the first order energy shift and
the second term is the second order energy shift
of two interactions connected by a propagator.
For a given order in $g$ the series terminates, because
$\algdD$ is at least of first order.

\subsection{The Hamiltonian}
\label{sec:Dila.Pert.Anomalous}

Conservation of classical dimensions by $\algD(g)$ also
implies that the other interacting generators 
have a definite classical dimension
\[\label{eq:Dila.Pert.JDiag}
\comm{\algD_0}{\algJ(g)}=\dim(\algJ) \,\algJ(g),
\]
which can be shown as follows:
Let $\Pi_{d}$ project to the states of classical dimension $d$.
Then $\Pi_{d}$ commutes with $\algD_m$ for arbitrary $d,m$
due to \eqref{eq:Dila.Pert.Diag}.
Now we project the algebra relation 
$\comm{\algD(g)}{\algJ(g)}=\dim(\algJ)\, \algJ(g)$ 
to subspaces of dimension $d,d'$ from the left and right, respectively, 
and expand in the coupling constant $g$.
The contribution at $\order{g^l}$ reads 
\[\label{eq:Dila.Pert.JClassInd}
\sum_{k=1}^{l} \Pi_{d} \comm{\algD_{k}}{\algJ_{l-k}} \Pi_{d'}
=\bigbrk{\dim(\algJ)-(d-d')}\Pi_{d}\, \algJ_{l}\, \Pi_{d'}.
\]
where we have moved the term with $k=0$ from the left to the right hand side
making use of $\Pi_d \algD_0=\algD_0\Pi_d=d\Pi_d$.
We assume that $\comm{\algD_0}{\algJ_k}=\dim(\algJ)\,\algJ_k$ for all $k<l$.
This is equivalent to the statement $\Pi_{d}\,\algJ_k \,\Pi_{d'}=0$ 
for all $d-d'\neq\dim (\algJ)$.
Choosing $d-d'\neq\dim (\algJ)$ in \eqref{eq:Dila.Pert.JClassInd} we find that 
$\Pi_{d}\,\algJ_l\, \Pi_{d'}$ must also vanish. The claim is proved by induction.

We can now combine \eqref{eq:Dila.Pert.JDiag} with the algebra relation
\eqref{eq:U224.Comm.Charge}
\[\label{eq:Dila.Pert.AlgDim}
\comm{\algD(g)}{\algJ(g)}=\dim(\algJ) \,\algJ(g)
\]
and infer that the anomalous dimension is 
conserved by the interacting algebra
\[\label{eq:Dila.Pert.AnoComm}
\comm{\algJ(g)}{\algdD(g)}=0.
\]
Thus we have constructed a $\alU(1)$ charge $\algdD$ 
in addition to the superconformal algebra $\alPSU(2,2|4)$. 
A very important consequence of \eqref{eq:Dila.Pert.AnoComm}
is that, 
\emph{at leading order, the anomalous dilatation operator $\algdD$ 
must commute with the classical algebra $\algJ_0$,}
\[\label{eq:Dila.Pert.AnoCommLeading}
\comm{\algD_0}{\algD_l}=0.
\]
We will see in \chref{ch:One} that the leading order 
is one-loop or $g^2$, i.e.~$l=2$.
For some purposes, it will therefore turn out that $\algD_{k+l}$ should be 
treated on equal footing with $\algJ_{k}$. To make this manifest
we introduce the notion of the `Hamiltonian' which is just the
anomalous dilatation order shifted by $l=2$ powers of $g$
\[\label{eq:Dila.Pert.Hamiltonian}
\mbox{`\emph{The Hamiltonian}':}\quad
\ham(g)=g^{-2} \algdD(g),\qquad
\comm{\algJ(g)}{\ham(g)}=0.
\]
The Hamiltonian is an invariant operator under superconformal symmetry. 
Note that its \emph{leading order} is $\ham_0$ and corresponds to \emph{one-loop},
$\ham_{0}=\algD_{2}$. 
The eigenvalues of the Hamiltonian are called `energies', $E$, and
are related to the scaling dimension by
\[\label{eq:Dila.Pert.Energy}
D(g)=D_0+g^2 E(g).\]

\subsection{Eigenstates}
\label{sec:Dila.Pert.Eigen}

Let us investigate the eigenstates of the dilatation operator.
For this purpose, we will introduce some basis of states $\OpE_{\fldind{M}}$.
We have seen that the classical dilatation operator $\algD_0$ 
commutes with $\algD(g)$. To find eigenstates of $\algD(g)$ we 
need to consider only a basis with fixed classical dimension $D_0$.%
\footnote{The generators $\algL,\algLd,\algR$ do not depend on $g$ 
and commute with $\algD(g)$. Therefore one can also restrict
to definite representations of $\alSU(2)\times\alSU(2)\times\alSU(4)$.}
As there are only a finite number of fields with a dimension
bounded from above, see \tabref{tab:N4.States.Fields},
also the basis $\OpE_{\fldind{M}}$ is finite.
When we expand the operator in the basis as 
$\Op=\Op^{\fldind{M}}\OpE_{\fldind{M}}$, 
we can write the eigenstate equation 
in a finite matrix form.
The matrix of scaling dimensions $D^{\fldind{M}}{}_{\fldind{N}}$
is obtained by acting with the dilatation operator on the basis
\[\label{eq:Dila.Pert.DilMatrix}
\algD\indup{op}(g)\, \OpE_{\fldind{M}}
=\OpE_{\fldind{N}}\, D^{\fldind{N}}{}_{\fldind{M}}(g).
\]
We will often find such a basis and write down the action of the
dilatation operator in matrix form.
The eigenstate equation is turned into an eigenvector equation
\[\label{eq:Dila.Pert.Eigen}
D^{\fldind{M}}{}_{\fldind{N}}(g)\, \Op^{\fldind{N}}(g)
=D\indup{ev}(g)\,\Op^{\fldind{M}}(g).
\]
In general the matrix depends on $g$ and so should an
eigenvector $\Op^{\fldind{M}}(g)$.
We can expand the eigenstate equation in powers of the coupling constant,
at $l$-th order we find
\[\label{eq:Dila.Pert.EigenPert}
\sum_{k=0}^l D^{\fldind{M}}_{k}{}^{}_{\fldind{N}}\,\Op_{l-k}^{\fldind{N}}
=\sum_{k=0}^l D\indup{ev,\mathnormal{k}}\, \Op_{l-k}^{\fldind{M}}\, .
\]
Note that we chose a basis of fixed classical dimension $D_0$,
therefore $D^{\fldind{M}}_{0}{}_{\fldind{N}}=\delta_{\fldind{N}}^{\fldind{M}}\,D_0$.
The equation for $l=0$ naturally requires $D\indup{0,ev}=D_0$. The 
equation at leading non-trivial order ($l=2$) simplifies to
\[\label{eq:Dila.Pert.EigenLeading}
 D^{\fldind{M}}_{l}{}_{\fldind{N}}\,\Op_{0}^{\fldind{N}}
= D\indup{ev,\mathnormal{l}}\, \Op_{0}^{\fldind{M}}\, .
\]
In general the eigenvalue problem is an algebraic equation which 
can only be solved numerically.
Once that is done and the spectrum of 
$D^{\fldind{M}}_{l}{}^{}_{\fldind{N}}$ happens to be non-degenerate, 
solving \eqref{eq:Dila.Pert.EigenPert} for any 
value of $l$ involves only linear algebra. 
If the leading order spectrum is degenerate,
the diagonalisation of $D^{\fldind{M}}_{l+1}{}^{}_{\fldind{N}}$
in the degenerate subspace is again an eigenvalue problem.
This continues as long as there are eigenvalues which are degenerate 
up to some order in perturbation theory.

The expansion of scaling dimensions is expected to
be in even powers of $g$, 
\[\label{eq:Dila.Pert.EigenEven}
D\indup{ev}(g)=D\indup{ev,0}+g^2 D\indup{ev,2}+g^4 D\indup{ev,4}+\ldots\,,
\]
even though $\algD\indup{op}$ involves also odd powers
\[\label{eq:Dila.Pert.OpAll}
\algD(g)=\algD_0+g^2 \algD_2
+g^3 \algD_3+g^4 \algD_4+\ldots\,.
\]
The odd powers play a special role as we will see in \chref{ch:Higher}.
This pattern may in principle
be broken when there are degenerate eigenvalues at leading order.
Similar problems occur in a double series expansion
in $g$ and $1/N$. There are cases in which the leading order
degeneracy is lifted by both, higher-loop and higher genus effects.
The eigenstates for expansion in $g$ and $1/N$ are not expected to agree,
consequently the double expansion will turn out to be inconsistent \cite{Beisert:2003tq}.

\section{Subsectors}
\label{sec:Dila.Sect}

In principle it would be desirable to derive the
dilatation operator which is 
valid for all fields of $\superN=4$ SYM. In most practical cases, however, 
this will turn out to be too involved. 
Therefore it is useful to know how to consistently restrict 
to subsectors of fields in such a way that
$\algD(g)$ closes on the subsector.
Within a subsector the number of fields as well as the symmetry 
algebra is reduced.
This reduction of complexity leads to a simplification
of the dilatation generator within the subsector. 
Thus, restricting to subsectors one can efficiently
compute anomalous dimensions.

\subsection{Construction of Subsectors}
\label{sec:Dila.Sect.Const}

To construct subsectors, we note that the number of excitations
in the oscillator picture, see 
\secref{sec:N4.Fund,app:U224.Osc} and \tabref{tab:U224.Osc.Numbers},
naturally puts constraints on the weights of operators. 
Certainly, there cannot be negative excitations.
Furthermore, the oscillators $\oscc^\dagger$ are fermionic, therefore
there can only be one excitation on each site. In total we find 
twelve bounds
\[\label{eq:Dila.Sect.Bounds}
n_{\osca}\geq 0,\quad
n_{\oscb}\geq 0,\quad
n_{\oscc}\geq 0,\quad
n_{\oscd}=L-n_{\oscc}\geq 0.
\]
All these excitation numbers will turn out to be conserved
by $\algD(g)$ at the one-loop level (c.f.~\secref{sec:One.Form.Symmetry}),
i.e.~they commute with $\algD_2$.
This means that the action of the one-loop dilatation operator closes on
operators with fixed excitation numbers.
Therefore, we can construct `one-loop subsectors' by 
considering operators for which several of the bounds are
met and thus some of the oscillators are not excited.
In some cases the subsectors remain closed even at higher loops.
We will refer to these as `closed subsectors'.

Let us investigate all closed subsectors.
Using \tabref{tab:U224.Osc.Numbers}
we can express the oscillator excitation numbers in terms of the
charges $D_0,s_1,s_2,p,q_1,q_2,B,L$.
We know that $\algD(g)$ commutes with the Cartan generators $s_1,s_2,p,q_1,q_2$ 
which are independent of the coupling constant.
Also the classical dimension $D_0$ is preserved by construction,
see \secref{sec:Dila.Pert.PreDiag}.
Only the charges $B$ and $L$ which are not part of $\alPSU(2,2|4)$ do 
not commute with $D(g)$ in general.
To construct a closed subsector we 
therefore need to find a positive 
linear combination of the bounds that
is independent of $B$ and $L$.
Put differently, it must be independent of 
$L-B$ and $L+B$. 
The number of excitations $n_{\osca}$
involves the combination $B-L$. This can only be
cancelled by $L-B$ in $n_{\oscc}$.
Therefore, we can remove oscillators of type $\osca$
if and only if we also remove oscillators of type $\oscc$.
Equivalently, we can remove oscillators of type $\oscb$
if and only if we also remove oscillators of type $\oscd$.

In the following we will construct all possible closed subsectors
and determine the set of fields as well
as the residual symmetry that transforms states within the subsector.
Note that for local operators we can enhance the 
superconformal algebra by the 
anomalous dimension operator $\algdD(g)=\algD(g)-\algD_0$
and consider 
\[\label{eq:Dila.Sect.FullSym}
\alPSU(2,2|4)\times \alU(1)
\]
as the full algebra.

\subsection{The Half-BPS Subsector}
\label{sec:Dila.Sect.HalfBPS}

Let us demonstrate how to obtain a rather trivial subsector.
We will consider the subsector of operators with no oscillator excitations
\[\label{eq:Dila.Sect.HalfBPS}
n_{\osca_{1}}=n_{\osca_{2}}=n_{\oscb_{1}}=n_{\oscb_{2}}=
n_{\oscc_{1}}=n_{\oscc_{2}}=n_{\oscd_{1}}=n_{\oscd_{2}}
=0.
\]
Using \tabref{tab:U224.Osc.Numbers}, 
the constraints \eqref{eq:Dila.Sect.HalfBPS}
force the weight to be
\[\label{eq:Dila.Sect.HalfBPSWeight}
w=\weight{L;0,0;0,L,0;0,L}.
\]
Here we have removed oscillators of all types, therefore the 
subsector is closed not only at one-loop but to all orders
in perturbation theory.
We can express the length in terms of a conserved charge, $L=p$, 
which implies that the length is protected even at higher loops. 
Equivalently, the hypercharge $B$ is exactly zero.

In conventional language the operators within this subsector consist only
of the highest weight of the field-strength multiplet
\[\label{eq:Dila.Sect.HalfBPSLetter}
\fldZ=\state{\fldZ}=\oscc^\dagger_3\oscc^\dagger_4\vac=\Phi_{34}.
\]
These are the half-BPS operators $\Tr \fldZ^L$ and 
its multi-trace cousins, the subsector will therefore be 
called `half-BPS' subsector.
The anomalous dilatation operator within this subsector vanishes identically,
as required by protectedness of half-BPS operators.
Note that almost all elements of a half-BPS multiplet
are outside this subsector. 
The important point is that \emph{every} half-BPS multiplet
has one component, its highest weight, within this subsector.
Due to superconformal invariance this is enough
to obtain information about the complete supermultiplet.

The subalgebra of $\alPSU(2,2|4)\times\alU(1)$ 
which closes on this subsector is 
$\alPSU(2|2)^2\times\alU(1)^3$. 
Effectively, however, the symmetry is only $\alU(1)$
which measures $p=D_0=D=L$, the other factors act trivially. 
Therefore we will only consider 
\[\label{eq:Dila.Sect.PoorU1}
\alU(1)\] 
as the residual symmetry.

\subsection{Short Subsectors}
\label{sec:Dila.Sect.Short}

\begin{figure}\centering
\includegraphics{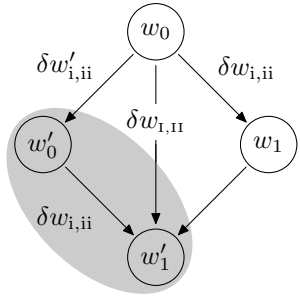}
\caption{Structure of highest weights in short subsectors. 
The superconformal highest weight is $w_0$. It is
at a unitary bound and $w_1$ is the highest weight of the
splitting submultiplet. The short subsector is marked as shaded.
The highest weight within is $w'_0$ and $w'_1$ 
is the highest weight of the submultiplet.
For an eighth-BPS state at $w_0$, the
multiplet at $w'_0$ is absent.}
\label{fig:Dila.Sect.Split}
\end{figure}

Suppose we require either i or ii in 
\<\label{eq:Dila.Sect.ShortCond}
\mathrm{i}:&\quad& n_{\osca_1}=n_{\oscc_1}=0,
\nln
\mathrm{ii}:&\quad& n_{\oscb_1}=n_{\oscd_1}=0,
\>
which is equivalent to 
\<\label{eq:Dila.Sect.ShortLabels}
\mathrm{i}:&\quad& 
D_0=s_1+\sfrac{1}{2}q_2+p+\sfrac{3}{2}q_1,\qquad L-B=D_0-s_1,
\nln
\mathrm{ii}:&\quad& 
D_0=s_2+\sfrac{1}{2}q_1+p+\sfrac{3}{2}q_2,\qquad L+B=D_0-s_2.
\>
In perturbation theory ($D_0\approx D$)
the weight is beyond the unitarity bound 
(c.f.~\secref{sec:N4.Split}) 
and cannot be the highest weight state of a unitary multiplet of $\alPSU(2,2|4)$.%
\footnote{The only exception is $s_{1,2}=0$ and $\delta D=0$ which is the highest
weight of an eighth-BPS multiplet.}
However, there is exactly one supersymmetry generator
that decreases the combination $D-s_1-\sfrac{1}{2}q_2-p-\sfrac{3}{2}q_1$
and one that decreases $D-s_2-\sfrac{1}{2}q_1-p-\sfrac{3}{2}q_2$.
These are $\algQ^1{}_2$ and $\algQd_{24}$ and they shift a weight by
\<\label{eq:Dila.Sect.HighestOffset}
\delta w'\indup{i}\eq
\weight{+0.5;+1,0;+1,0,0;+0.5,0},\qquad
\delta (D-s_1-\sfrac{1}{2}q_2-p-\sfrac{3}{2}q_1)=-2,
\nln
\delta w'\indup{ii}\eq
\weight{+0.5;0,+1;0,0,+1;-0.5,0},\qquad
\delta (D-s_2-\sfrac{1}{2}q_1-p-\sfrac{3}{2}q_2)=-2.
\>
Due to the fermionic nature of the generators, 
the shift can only be applied once and 
the highest weight must be 
close to the unitarity bound.
In the classical theory the dimensions are exactly at the
unitarity bound and the multiplets become short.
The subsectors i, ii will be called short subsectors, because all 
short multiplets of $\alPSU(2,2|4)$ are represented
by their highest weight shifted by the above $\delta w'\indup{i,ii}$.
Shortening also implies that the multiplet splits up,
the weight of the additional submultiplet is
reached from the highest weight by adding \eqref{eq:N4.Split.Offset}%
\footnote{In the case of $s_{1,2}=0$ for the highest weight,
the shift would lead to a negative $s_{1,2}$.
In this particular case, $\delta w\indup{i,ii}+\delta w'\indup{i,ii}$
leads to the highest weight of the other submultiplet which is eighth-BPS.}
\<\label{eq:Dila.Sect.SplitOffset}
\delta w\indup{i}\eq
\weight{+0.5;-1,0;+1,0,0;-0.5,+1},\qquad
\delta (D_0-s_1-\sfrac{1}{2}q_2-p-\sfrac{3}{2}q_1)=0,
\nln
\delta w\indup{ii}\eq
\weight{+0.5;0,-1;0,0,+1;+0.5,+1},\qquad
\delta (D_0-s_2-\sfrac{1}{2}q_1-p-\sfrac{3}{2}q_2)=0,
\>
which correspond to $\algQ^1{}_1$ and $\algQd_{14}$.%
\footnote{Note that the shifts have anomalous values $\delta B,\delta L$
and manifestly break the associated symmetries.}
An interesting aspect is that also 
the additional submultiplet has a descendant in the subsector.
The descendants of the submultiplets in the subsector are thus related by 
$\algQ^1{}_1$ and $\algQd_{14}$. 
In the classical theory these generators cannot act at all 
because the corresponding oscillators are disabled, see
\eqref{eq:Dila.Sect.ShortCond}.
In the quantum theory this changes and the submultiplets join.
The relationship between the above highest weights 
is illustrated in \figref{fig:Dila.Sect.Split}.

The residual symmetry within this subsector is
\[\label{eq:Dila.Sect.ShortSym}
\alU(1)\ltimes\alPSU(1,2|3)\times\alPSU(1|1)\ltimes\alU(1). 
\]
The $\alPSU(1,2|3)$ subgroup classically transforms all oscillators
except $\osca_1,\oscc_1$ or $\oscb_1,\oscd_1$. 
The $\alPSU(1|1)$ is associated to the supercharges
which shift by $\pm\delta w\indup{i}$ and
relate the two submultiplets. 
The $\alU(1)$ charge $L\mp B$
and the $\alU(1)$ anomalous dimension $\delta D$ 
are external automorphisms and 
central charges, respectively, for both $\alPSU$ factors.

In addition to 
$n_{\osca_{1}}=n_{\oscc_{1}}=0$
we can further demand 
(similarly for the other subsector)
\[\label{eq:Dila.Sect.ShorterCond}
n_{\oscc_{2}}=0,\quad
n_{\oscc_{2}}=n_{\oscc_{3}}=0\quad\mbox{or}\quad
n_{\oscc_{2}}=n_{\oscc_{3}}=n_{\oscc_{4}}=0.\]
This restricts to states which have charges
\[\label{eq:Dila.Sect.ShorterLabels}
q_1=0,\quad q_1=p=0\quad\mbox{or}\quad q_1=p=q_2=0
\]
and leads to even shorter subsectors
with residual symmetries
\[\label{eq:Dila.Sect.ShorterSym}
\alSU(1,2|2)\times\alU(1),\quad
\alSU(1,2|1)\times\alU(1)\quad\mbox{or}\quad
\alSU(1,2)\times\alU(1).\]

\subsection{BPS Subsectors}
\label{sec:Dila.Sect.BPS}

Assume we now remove both oscillators 
of either type $\osca$ or $\oscb$
\<\label{eq:Dila.Sect.BPSCond}
\mathrm{I}:&\quad& n_{\osca_1}=n_{\osca_2}=n_{\oscc_1}=0,
\nln
\mathrm{II}:&\quad& n_{\oscb_1}=n_{\oscb_2}=n_{\oscd_1}=0.
\>
Using \tabref{tab:U224.Osc.Numbers},
these conditions are equivalent to 
\<\label{eq:Dila.Sect.BPSLabels}
\mathrm{I}:&\quad& 
D_0=\sfrac{1}{2}q_2+p+\sfrac{3}{2}q_1,\quad s_1=0,\qquad L-B=D_0,
\nln
\mathrm{II}:&\quad& 
D_0=\sfrac{1}{2}q_1+p+\sfrac{3}{2}q_2,\quad s_2=0,\qquad L+B=D_0.
\>
For $D_0=D$ these are precisely the eighth-BPS conditions,
see \secref{sec:N4.Split},
and therefore every eighth-BPS multiplet has
components in these subsectors. Consequently we call
them eighth-BPS sectors.
In perturbation theory when $\delta D\approx 0$
the states are beyond the unitarity bound.
As discussed below \eqref{eq:Dila.Sect.SplitOffset}
we need to apply two supersymmetry generators 
$\varepsilon^{\alpha\beta}\algQ^1{}_{\alpha} \algQ^1{}_{\beta}$ or
$\varepsilon^{\dot\alpha\dot\beta}\algQd_{\dot\alpha 4}\algQd_{\dot\beta 4}$ 
to reach any state within the subsector from the highest weight.
The highest weight is shifted by 
\<\label{eq:Dila.Sect.BPSOffset}
\delta w\indup{I}\eq
\delta w\indup{i}+\delta w'\indup{i}=
\weight{1;0,0;2,0,0;0,1},
\nln
\delta w\indup{II}\eq
\delta w\indup{ii}+\delta w'\indup{ii}=
\weight{1;0,0;0,0,2;0,1}.
\>

The residual symmetry within this sector is
\[\label{eq:Dila.Sect.BPSSym}
\alSU(2|3)\times\alU(1),\]
where $\alU(1)$ corresponds to the anomalous dimension $\delta D$.
Note that, as there are no oscillators of either type $\osca$ or
type $\oscb$, we can only have a finite number of oscillator excitations 
for an elementary field. Therefore there are only finitely many fields
within this subsector, for type II they are
\[\label{eq:Dila.Sect.BPSFields}
\Phi_{a4}=\oscc_a^\dagger\oscc_4^\dagger\vac,\qquad 
\Psi_{\alpha 4}=\osca_\alpha^\dagger\oscc_4^\dagger\vac,
\]
with $a=1,2,3$, $\alpha=1,2$.
These transform in the fundamental representation of
$\alSU(2|3)$.
This sector will be discussed in detail in \chref{ch:Higher}.

As opposed to \secref{sec:Dila.Sect.Short}
we cannot disable more of the $\oscc$'s, here.
Requiring $n_{\osca_{1}}=n_{\osca_{2}}=n_{\oscc_{1}}=n_{\oscc_{2}}=0$ leads,
via the central charge constraint, c.f.~\secref{sec:N4.Fund}, to
$n_{\oscb_{1}}=n_{\oscb_{2}}=n_{\oscd_{1}}=n_{\oscd_{2}}=0$, 
i.e.~the half-BPS sector of \secref{sec:Dila.Sect.HalfBPS}.

\subsection{Combined Subsectors}
\label{sec:Dila.Sect.Combine}

We can also combine one of the restrictions on $\osca,\oscc$
with a restriction on $\oscb,\oscd$.
Let us denote the restrictions of \secref{sec:Dila.Sect.Short}
by $1,2,3,4$ depending on how many of the $\oscc$'s or $\oscd$'s
are removed. The BPS restriction of \secref{sec:Dila.Sect.BPS}
will be denoted by $1^+$. No restriction is denoted by $0$. 
The possible subsectors are given by a pair of symbols $(m,n)$.
Not all combinations are possible, we cannot remove and fully 
excite one oscillator of the type $\oscc$ at the same time
(fully exciting is equivalent to removing the 
corresponding oscillator of type $\oscd$).
This yields the bound $m+n\leq 4$.

We find the following cases:
\begin{bulletlist}
\item 
The only subsector which does not fit this scheme is
the half-BPS subsector $(2^+,2^+)$ discussed in \secref{sec:Dila.Sect.HalfBPS}.

\item
We have already discussed all subsectors of type 
$(0,n)$
in \secref{sec:Dila.Sect.Short} and
\secref{sec:Dila.Sect.BPS}.

\item
We can combine two eighth-BPS conditions ($1^+,1^+)$ to 
the quarter-BPS subsector.
We will discuss this one 
in \secref{sec:Dila.SU2}.

\item
We can combine a short condition with an eighth-BPS condition 
in $(n,1^+)$ for $n=1,2,3$.
The fields and residual symmetries are
\[\label{eq:Dila.Sect.ShortBPS}
\begin{array}[b]{ccrcl}
(1,1^+)&&
\{\oscc^\dagger_2,\oscc^\dagger_3,\osca^\dagger_2\}\oscc^\dagger_4\vac,&&
\alSU(1|2)\times\alU(1|1),
\\[5pt]
(2,1^+)&&
\{\oscc^\dagger_3,\osca^\dagger_2\}\oscc^\dagger_4\vac,&&
\alU(1|1)\times \alU(1),
\\[5pt]
(3,1^+)&&
\osca^\dagger_2\oscc^\dagger_4\vac,&&
\alU(1)\times\alU(1).
\end{array}\]
In particular the sector $(1,1^+)$ appears to be very interesting due
to its high amount of supersymmetry in combination with only
three fundamental fields. This might allow for higher loop 
calculations with a minimum amount of work, c.f.~the 
treatment of the $(0,1^+)$ 
sector in \chref{ch:Higher} 
of which this a subsector.
The sector $(2,1^+)$ has been investigated in 
\cite{Callan:2004dt} and found to be
equivalent to free fermions in the
one-loop approximation.

\item
There are four doubly-short sectors
$(1,1)$, $(2,1)$, $(2,2)$ and $(3,1)$.
We find the following fields and symmetries
\[\label{eq:Dila.Sect.ShortShort}
\begin{array}[b]{crcl}
(1,1)&
\{1,\oscc_2^\dagger\oscd_2^\dagger,\osca_2^\dagger\oscd_2^\dagger,\oscc_2^\dagger\oscb_2^\dagger\}
(\osca^\dagger_2\oscb^\dagger_2)^n\state{\fldZ},
&&
\alU(1)^2\ltimes\alPSU(1,1|2)\times\alPSU(1|1)^2\ltimes\alU(1),
\\[5pt]
(2,1)&
\{1,\osca^\dagger_2\oscd^\dagger_2\}(\osca^\dagger_2\oscb^\dagger_2)^n\state{\fldZ},&&
\alSU(1,1|1)\times \alU(1|1),
\\[5pt]
(3,1)&
\osca^\dagger_2\oscd^\dagger_2(\osca^\dagger_2\oscb^\dagger_2)^n\state{\fldZ},&&
\alSU(1,1)\times \alU(1|1),
\\[5pt]
(2,2)&
(\osca^\dagger_2\oscb^\dagger_2)^n\state{\fldZ},&&
\alSU(1,1)\times\alU(1)\times\alU(1).
\end{array}\]
Of particular interest is the sector $(3,1)$ which 
allows for a determination of the one-loop dilatation operator 
by purely algebraic means. It will be discussed in \secref{sec:One.Magic}. 
The sector $(2,2)$ is quite similar to $(3,1)$ and also very useful,
we will discuss it in \secref{sec:One.Baby}.
The sector $(2,1)$ combines the two.

\end{bulletlist}

\subsection{Excitation Subsector}
\label{sec:Dila.Sect.Excite}

Instead of removing oscillators of certain kinds, 
we can also fix the number of oscillator excitations to some value. 
Here we will consider only the total number
of oscillator excitations above the physical vacuum $\state{\fldZ,L}$;
this is an even number because oscillators can only be excited in 
pairs due to the central charge constraint. 
A state with $2M$ oscillator excitations will be
said to have \emph{$M$ excitations} 
\[\label{eq:Dila.Sect.nEx}
\mbox{`\emph{$M$ excitations}':}\quad 
(\oscA^\dagger)^{M}(\dot\oscA^\dagger)^{M}\state{\fldZ,L},
\]
where $\oscA=(\osca,\oscc)$ and $\dot \oscA=(\oscb,\oscd)$.
According to \tabref{tab:U224.Osc.Numbers}
the excitation number is related to the charges by
\[\label{eq:Dila.Sect.ExTotal}
M=\half (n_\osca+n_\oscb+n_\oscc+n_\oscd)=D_0-p-\half q_1-\half q_2,
\]
it is thus exactly conserved by the dilatation operator.
In other words, the sector of states with $M$ excitations is closed.
This type of sector is different from the above subsectors in 
that no type of oscillator is excluded. 
Instead, there is an upper bound on 
the number of excitations on a single field;
this also leads to a simplification of the representation of
generators.

The excitation subsectors are somewhat similar 
to the half-BPS subsector discussed in \secref{sec:Dila.Sect.HalfBPS}, 
which is in fact the sector with zero excitations. 
The residual symmetry in this type of subsector is
\[\label{eq:Dila.Sect.ExAlg}
\bigbrk{\alU(1)\ltimes \alPSU(2|2)\times \alPSU(2|2)\ltimes \alU(1)}\times \alU(1).
\]
The generators of $\alPSU(2|2)$ are given by a pair of
$\alSU(2)$ generators $\algL^{\alpha}{}_{\beta},\algR^{a}{}_{b}$
and a pair of supercharges $\algQ^a{}_\alpha,\algS^\alpha{}_a$.
Classically, they transform between oscillators $\osca$ and $\oscc$.
Equivalently, the other $\alPSU(2|2)$ is given by 
$\algLd^{\dot \alpha}{}_{\dot \beta},\algRd^{\dot a}{}_{\dot b},
\algQd^{\dot a}{}_{\dot\alpha},\algSd^{\dot\alpha}{}_{\dot a}$.
A $\alU(1)$ external automorphism for both $\alPSU(2|2)$'s is given by $\algD_0$.
The $\alU(1)$ central charge for both $\alPSU(2|2)$'s is given by $M+\algdD$.
Another central charge is $\algdD$.
The four sets of $\alSU(2)$ generators transform indices canonically. 
The non-vanishing anticommutators of supergenerators are given by
\<\label{eq:Dila.Sect.ExAlgRel}
\acomm{\algS^\alpha{}_a}{\algQ^b{}_\beta}\eq
  \delta^b_a \algL^\alpha{}_\beta
  +\delta_\beta^\alpha \algR^b{}_a
  +\half \delta_a^b \delta_\beta^\alpha (M+\algdD),
\nln
\acomm{\algSd^{\dot\alpha}{}_{\dot a}}{\algQd^{\dot b}{}_{\dot \beta}}\eq
  \delta^{\dot b}_{\dot a} \algLd^{\dot\alpha}{}_{\dot\beta}
  +\delta_{\dot \beta}^{\dot\alpha} \algRd^{\dot b}{}_{\dot a}
  +\half \delta_{\dot a}^{\dot b} \delta_{\dot\beta}^{\dot\alpha} (M+\algdD).
\>
The Dynkin labels of a weight of one of the $\alPSU(2|2)$'s are given by ($i=1,2$)
\[\label{eq:Dila.Sect.ExDynkin}
[s_i;r_i;q_i],\qquad r_i=\half M+\half \delta D+\half s_i-\half q_i.
\]
For a unitary representation the highest weights should obey
$r_i\geq s_i+1$ or $s_i=r_i=0$. 
A multiplet is short for $r_i=s_i+1$ and BPS for $s_i=r_i=0$.
At the unitarity bound $r_i=s_i+1$, 
a long multiplet $[s_i;r_i;q_i]$ splits off
a short multiplet $[s_i-1;r_i-1;q_i+1]$
or, when $s_i=0$, a BPS multiplet $[0;r_i-1;q_i+2]$.

A subsector of this kind is suited perfectly to investigate 
plane-wave physics and BMN operators \cite{Berenstein:2002jq}.
The number of excitations $M$ equals the 
classical BMN energy $D_0-J$ or \emph{impurity} number.
The residual symmetry in this sector maps directly to parts of 
the symmetries of the dual plane-wave string theory.

\section{The $\alSU(2)$ Quarter-BPS Sector}
\label{sec:Dila.SU2}

In this section we will demonstrate how to extract the 
dilatation generator from a perturbative calculation of 
the two-point function. We will restrict to the one-loop 
level and to the quarter-BPS subsector.


\subsection{The $\alSU(2)$ Subsector}
\label{sec:Dila.SU2.SU2}
First of all let us describe the subsector.
The quarter-BPS subsector is obtained by combining both
eighth-BPS conditions
described in \secref{sec:Dila.Sect.BPS}.
\[
\label{eq:Dila.SU2.Condition}
n_{\osca_{1}}=
n_{\osca_{2}}=
n_{\oscb_{1}}=
n_{\oscb_{2}}=
n_{\oscc_{1}}=
n_{\oscd_{1}}=0.
\]
There are only two charged scalar fields in this subsector
\<\label{eq:Dila.SU2.Fields}
\fldZ:=\varphi_1:=\Phi_{34}\eq\oscc^\dagger_3\oscc^\dagger_4\vac=\state{\fldZ},
\nln
\phi:=\varphi_2:=\Phi_{24}\eq\oscc^\dagger_2\oscc^\dagger_4\vac=\oscc^\dagger_2\oscd^\dagger_2\state{\fldZ},
\>
therefore it is the smallest non-trivial subsector
and we will often make use of it.
The possible weights are 
\[\label{eq:Dila.SU2.Weight}
w=\weight{L;0,0;K,L-2K,K;0,L},
\]
where $K$ counts the number of $\phi$'s
and $L$ is the total number of fields.
The residual symmetry is 
\[\label{eq:Dila.SU2.Sym}
\alSU(2)\times\alU(1)\times \alU(1).
\]
The $\alSU(2)$ factor
transforms $\varphi_1=\fldZ$ and $\varphi_2=\phi$ 
in the fundamental representation,
whereas the $\alU(1)$'s measure the 
classical dimension $D_0=L$ and the 
anomalous dimension $\delta D$.
With respect to $\alSU(2)\times \alU(1)\times\alU(1)$
a state is thus described by the charges 
\[\label{eq:Dila.SU2.Labels}
[L-2K],\quad L,\quad \delta D,
\]
where $[L-2K]$ is the Dynkin label%
\footnote{The Dynkin label for $\alSU(2)$ equals twice the spin.}
 of $\alSU(2)$ corresponding
to a third component of spin $L/2-K$.
In terms of the superconformal algebra, 
a state with $\delta D=0$ is (at least) quarter-BPS,
a generic state, however, will not be protected. 
In that case the weight $w$ is beyond 
the unitarity bounds and cannot be primary.
The highest weight within the subsector is obtained from the
highest weight of the $\alPSU(2,2|4)$ multiplet by a shift of 
\[\label{eq:Dila.SU2.Shift}
\delta w\indup{I+II}=\weight{2;0,0;2,0,2;0,2}.
\]
The $\alPSU(2,2|4)$ highest weight is on both unitarity bounds
and has no spin.

\subsection{Tree-Level}
\label{sec:Dila.SU2.Tree}

We will now compute the two-point function of 
states within the $\alSU(2)$ sector.
From the formal expression we will then extract the dilatation operator.
Let us start at tree-level. 
The state $\Op_\varphi$ at point $x_1$ is constructed from fields
$\varphi_i$ of the $\alSU(2)$ subsector. Conversely, the other operator
$\dot\Op_{\dot\varphi}$ at point $x_2$ is constructed from fields
$\dot\varphi^i$ of a conjugate $\alSU(2)$ subsector.
Note the charge conjugation requires us to 
use two different $\alSU(2)$ subsectors.
The operators are constructed as 
(not necessarily equal) products of traces of fields 
\<\label{eq:Dila.SU2.Ops}
\Op_\varphi[\fldW]\eq
\Tr \varphi_\ast\cdots\varphi_\ast \,
\Tr \varphi_\ast\cdots\varphi_\ast\,\ldots\,,
\qquad \varphi_i=\varphi_i(x_1),
\nln
\dot\Op_{\dot\varphi}[\fldW]\eq
\Tr \dot\varphi^\ast\cdots\dot\varphi^\ast \,
\Tr\dot\varphi^\ast\cdots\dot\varphi^\ast\,\ldots\,,
\qquad \dot\varphi^i=\dot\varphi^i(x_2).
\>
Written in this way, the operators become abstract objects 
in the tensor product space of fields and are not necessarily based 
at some point in spacetime.

According to the path integral (c.f.~\secref{sec:N4.Quantum}) 
the two-point function at 
tree-level is given by
\[\label{eq:Dila.SU2.FreeField}
\bigvev{\Op_\varphi\, \dot\Op_{\dot\varphi}}
=
\bigeval{
\exp(W_0[\partial/\partial \fldW])\,
\Op_\varphi[\fldW]\, \dot\Op_{\dot\varphi}[\fldW]}_{\fldW=0}
+\order{g}.
\]
In fact, we do not need to work with generic $x$-dependent fields $\fldW$,
but only the values of the scalar fields $\varphi,\dot\varphi$ 
at points $x_{1,2}$ are relevant. The correlator now becomes
\<\label{eq:Dila.SU2.FreeSU2}
\bigvev{\Op_\varphi\, \dot\Op_{\dot\varphi}}
\eq
\bigeval{
\exp\bigbrk{W_0(x_{12},\partial/\partial \varphi,\partial/\partial \dot\varphi)}\,
\Op_\varphi(\varphi)\, \dot\Op_{\dot\varphi}(\dot\varphi)}_{\varphi=\dot\varphi=0}
+\order{g}
\nln
\eq
\bigeval{
\exp\bigbrk{W_0(x_{12},\check\varphi,\check{\dot\varphi})}\,
\Op_\varphi\, \dot\Op_{\dot\varphi}}_{\varphi=\dot\varphi=0}
+\order{g},
\>
where $W_0$ is the free generating functional of connected Graphs
\[\label{eq:Dila.SU2.FreeGraph}
W_0(x_{12},\check\varphi,\check{\dot\varphi})=
N^{-1}\Delta_{12}
\Tr\check\varphi_i\check{\dot\varphi}^i.
\]
The scalar propagator $\Delta_{12}=\Delta(x_1,x_2)$ is 
defined in \eqref{eq:Dila.Dim.PropReg}.
Note that the second line in \eqref{eq:Dila.SU2.FreeSU2} 
merely involves performing ordinary derivatives $\check \varphi,\check{\dot\varphi}$
with respect to the matrices $\varphi,\dot\varphi$.
In order for the result to be non-vanishing, 
all the fields $\varphi$ in $\Op_\varphi$
need to be contracted to fields $\dot\varphi$ in $\dot\Op_{\dot\varphi}$ with 
propagators $\Delta_{12}$.
In particular, the numbers of fields of the two
states must be equal.

\subsection{One-Loop}
\label{sec:Dila.SU2.OneLoop}

To compute higher-loop correlators, we insert the interactions 
$S\indup{int}[g,\fldW]$ into the path integral
\[\label{eq:Dila.SU2.IntField}
\bigvev{\Op_\varphi\, \dot\Op_{\dot\varphi}}
=
\bigeval{
\exp(W_0[\partial/\partial \fldW])\,
\exp(-S\indup{int}[g,\fldW])\,
\Op_\varphi[\fldW]\, \dot\Op_{\dot\varphi}[\fldW]}_{\fldW=0}
.
\]
All the fields $\fldW$ in $S\indup{int}$ need to be contracted
to propagators before setting $\fldW=0$, 
therefore we can combine the first 
two exponentials into one and write 
\<\label{eq:Dila.SU2.IntConn}
\bigvev{\Op_\varphi\, \dot\Op_{\dot\varphi}}
\eq
\bigeval{
\exp(W[g,\partial/\partial \fldW])\,
\Op_\varphi[\fldW]\, \dot\Op_{\dot\varphi}[\fldW]}_{\fldW=0}
\nln\eq
\bigeval{
\exp\bigbrk{W(g,x_{12},\check\varphi,\check{\dot\varphi})}\,
\Op_\varphi\, \Op_{\dot\varphi}}_{\varphi=\dot\varphi=0}.
\>
Again, it will be sufficient to evaluate 
the full generating functional of connected graphs $W[g,\fldJ]$
only for fields $\varphi,\dot\varphi$ at points $x_{1,2}$. 

\begin{figure}\centering
\parbox{1.5cm}{\centering\includegraphics{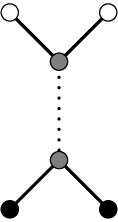}\par a}
\quad
\parbox{1.5cm}{\centering\includegraphics{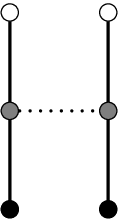}\par b}
\quad
\parbox{1.5cm}{\centering\includegraphics{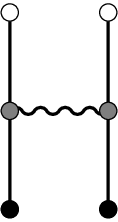}\par c}
\quad
\parbox{1.5cm}{\centering\includegraphics{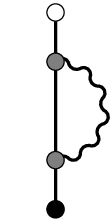}\par d1}
\quad
\parbox{1.5cm}{\centering\includegraphics{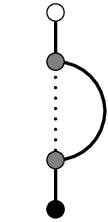}\par d2}
\quad
\parbox{1.5cm}{\centering\includegraphics{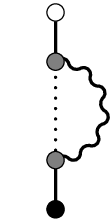}\par d3}
\quad
\parbox{1.5cm}{\centering\includegraphics{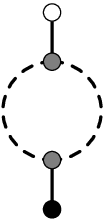}\par d4}
\caption{Connected graphs at one-loop. 
The solid, wiggly and dashed lines represent 
scalars, gluons and fermions, respectively.
The dotted lines correspond to a non-propagating auxiliary field that 
represents a quartic interaction.
The diagrams display the combinatorial structure with respect to the 
gauge group rather than their space-time configuration: 
The white and black dots are at the spacetime points $x_1$ and $x_2$, respectively.}
\label{fig:Dila.SU2.OneLoop}
\end{figure}

Let us now consider the connected graphs at one-loop.
There is no diagram at $\order{g}$ which conserves the charges.
The $\order{g^2}$ connected Green functions are 
depicted in \figref{fig:Dila.SU2.OneLoop}.
To evaluate them we make use of 
the regularised $\superN=4$ SYM action in \eqref{eq:Dila.Dim.DimReg,eq:N4.D4.Lagr}.
The one-loop Green functions evaluate to 
\<\label{eq:Dila.SU2.W2abcd}
W_{2,a}\eq 
\sfrac{1}{32}N^{-3}\, X_{1122} 
\Tr\comm{\check{\dot\varphi}_i}{\check{\dot\varphi}_j}
   \comm{\check\varphi^i}{\check\varphi^j},
\nln
W_{2,b}\eq
\sfrac{1}{32}N^{-3}\, X_{1122} 
\Tr\comm{\check{\dot\varphi}_i}{\check\varphi^j}
   \comm{\check\varphi^i}{\check{\dot\varphi}_j},
\nln
W_{2,c}\eq 
\sfrac{1}{32}N^{-3}\,\bigbrk{-2\tilde H_{12,12}-4Y_{112}I_{12}+X_{1122}}
\Tr\comm{\check{\dot\varphi}_i}{\check\varphi^i}
   \comm{\check{\dot\varphi}_j}{\check\varphi^j},
\nln
W_{2,d}\eq 
-\sfrac{1}{4}N^{-2} \,Y_{112}\, \gaugemet{mn}
\Tr\comm{\check{\dot\varphi}_i}{\gaugegen{m}}
   \comm{\gaugegen{n}}{\check\varphi^i}
\>
with the integrals $X,Y,\tilde H$ defined in \eqref{eq:Dila.Dim.YXH}.
We use a Jacobi identity to transform the second structure
in $W_{2,b}$
\[\label{eq:Dila.SU2.Jacobi}
\Tr\comm{\check{\dot\varphi}_i}{\check\varphi^j}
   \comm{\check\varphi^i}{\check{\dot\varphi}_j}
=
 \Tr\comm{\check{\dot\varphi}_i}{\check{\dot\varphi}_j}
    \comm{\check\varphi^i}{\check\varphi^j}
-\Tr\comm{\check\varphi^i}{\check{\dot\varphi}_i}
    \comm{\check\varphi^j}{\check{\dot\varphi}_j}
\]
and order the terms according to their spacetime integrals
\<\label{eq:Dila.SU2.Conn2XHIY}
W_{2,X}\eq 
\sfrac{1}{16}N^{-3}\, X_{1122} 
\Tr\comm{\check{\dot\varphi}_i}{\check{\dot\varphi}_j}
   \comm{\check\varphi^i}{\check\varphi^j},
\nln
W_{2,H}\eq 
-\sfrac{1}{16}N^{-3}\,\tilde H_{12,12}
\Tr\comm{\check{\dot\varphi}_i}{\check\varphi^i}
   \comm{\check{\dot\varphi}_j}{\check\varphi^j},
\nln
W_{2,IY}\eq 
-\sfrac{1}{8}N^{-3}\,I_{12}Y_{112}
\bigbrk{
\Tr\comm{\check{\dot\varphi}_i}{\check\varphi^i}
   \comm{\check{\dot\varphi}_j}{\check\varphi^j}
+N\Delta^{-1}_{12} \gaugemet{mn}
\Tr\comm{\check{\dot\varphi}_i}{\gaugegen{m}}
   \comm{\gaugegen{n}}{\check\varphi^i}}.
\>
We refrain from evaluating these functions until later 
and insert them as they stand into the expression for 
the one-loop correlator
\[\label{eq:Dila.SU2.OneLoopConn}
\bigvev{\Op_\varphi\, \dot\Op_{\dot\varphi}}=
\bigeval{\exp\bigbrk{W_0(x_{12},\check\varphi,\check{\dot\varphi})}
\bigbrk{1+g^2 W_2(x,\check\varphi,\check{\dot\varphi})}\,
\Op_\varphi\,\dot\Op_{\dot\varphi}}_{\varphi=\dot\varphi=0}
+\order{g^3}.
\]
We now change the argument $\check{\dot\varphi}$ of $W_2$ 
to $N\Delta^{-1}_{12}\varphi$.
This can be done because the result vanishes unless
every $\varphi$ is removed by some $\check\varphi$ 
before the fields $\varphi$ are set to zero.
Here, the only possibility is to contract with $W_0$ 
which effectively changes 
$N\Delta^{-1}_{12}\varphi$ back to $\check{\dot\varphi}$. 
In doing so we need to make sure
that no new contractions appear between the 
arguments $\varphi$ and $\check\varphi$ of $W_2$. 
Formally, this is achieved by `normal ordering'.
The correlator becomes
\[\label{eq:Dila.SU2.OneLoopVertex}
\bigvev{\Op_\varphi\, \dot\Op_{\dot\varphi}}=
\bigeval{\exp\bigbrk{W_0(x,\check\varphi,\check{\dot\varphi})}
\bigbrk{1+g^2 V_{2,\varphi}(x_{12})}\,
\Op_\varphi\, \dot\Op_{\dot\varphi}}_{\varphi=\dot\varphi=0}
+\order{g^3}
\]
with the one-loop effective vertex
\[\label{eq:Dila.SU2.Vertex}
V_{2,\varphi}(x_{12})=\normord{W_2(x_{12},\check\varphi,N\Delta_{12}^{-1}\varphi)}.
\]
We transform the explicit expressions for the connected graphs 
\eqref{eq:Dila.SU2.Conn2XHIY} and obtain 
\<\label{eq:Dila.SU2.VertexXHIY}
V_{2,X}\eq 
\sfrac{1}{4}N^{-1}
\, X_{1122} I_{12}^{-2}
\,\normord{\Tr\comm{\varphi_i}{\varphi_j}\comm{\check\varphi^i}{\check\varphi^j}},
\nln
V_{2,H}\eq 
-\sfrac{1}{4}N^{-1}
\,\tilde H_{12,12}I_{12}^{-2}
\,\normord{\Tr\comm{\varphi_i}{\check\varphi^i}\comm{\varphi_j}{\check\varphi^j}}
,
\nln
V_{2,IY}\eq 
-\sfrac{1}{2}N^{-1}Y_{112}I_{12}^{-1}
\bigbrk{
\normord{\Tr\comm{\varphi_i}{\check\varphi^i}\comm{\varphi_j}{\check\varphi^j}}
+
\gaugemet{mn}\normord{\Tr\comm{\varphi_i}{\gaugegen{m}}\comm{\gaugegen{n}}{\check\varphi^i}}}.
\>
We can change the normal ordering in the first term of $V_{2,IY}$ 
in order to absorb the second, see \eqref{eq:N4.Gauge.NormalOrderGen}
\[\label{eq:Dila.SU2.VertexGauge}
V_{2,IY}=
-\sfrac{1}{2}N^{-1}Y_{112}I_{12}^{-1}\,
\Tr\normord{\comm{\varphi_i}{\check\varphi^i}}
   \normord{\comm{\varphi_j}{\check\varphi^j}}
=
\sfrac{1}{2}N^{-1}Y_{112}I_{12}^{-1}\,\Tr \gaugerot\gaugerot.
\]
We can thus write $V_{2,IY}$ in terms of the generator of 
gauge rotations $\gaugerot=i\normord{\comm{\varphi_i}{\check\varphi^i}}$
within the $\alSU(2)$ subsector, 
see \eqref{eq:N4.Gauge.GaugeGen}.
Therefore $V_{2,IY}$ does not act on gauge invariant objects
such as the states $\Op$ and we can drop it altogether,
$V_{2,IY}\hateq 0$.

Instead of replacing $\check{\dot\varphi}$ we could also have 
replaced $\check \varphi$ resulting in the effective vertex
\[\label{eq:Dila.SU2.VertexConj}
\dot V_{2,\dot\varphi}(x_{12})=
\normord{W_2(x_{12},N\Delta_{12}^{-1}\dot\varphi,\check{\dot\varphi})}.
\]
This shows that in \eqref{eq:Dila.SU2.OneLoopVertex}
$V_{2,\varphi}$ is equivalent to $\dot V_{2,\dot\varphi}$ 
\[\label{eq:Dila.SU2.VertexSelfAdj}
V_{2,\varphi}\hateq\dot V_{2,\dot\varphi}.
\]
The form of this $\dot V_{2,\dot\varphi}$ is 
the same as in \eqref{eq:Dila.SU2.VertexXHIY}
upon conjugation of $\alSU(2)$ indices.
In other words, $V_2$ is self-adjoint 
with respect to the tree-level scalar product.

In a renormalised theory we should compute 
the correlator of renormalised states $Z\Op$.
At this point it is possible to guess the operator $Z$
for the renormalisation of states
\[\label{eq:Dila.SU2.ZFactor}
Z=1-\half g^2 V_2(1/\mu)+\order{g^3}.
\]
We insert this into \eqref{eq:Dila.SU2.OneLoopVertex} and
use the equivalence of $V_{2,\varphi}$ and $\dot V_{2,\dot\varphi}$ to find
\[\label{eq:Dila.SU2.OneLoopRenorm}
\bigvev{Z\Op_\varphi\, \dot Z\dot\Op_{\dot\varphi}}=
\bigeval{\exp\bigbrk{W_0(x_{12},\check\varphi,\check{\dot\varphi})}
\bigbrk{1+g^2 V_{2,\varphi}(x_{12})-
g^2 V_{2,\varphi}(1/\mu)}\,
\Op_\varphi\, \dot\Op_{\dot\varphi}}_{\varphi=\dot\varphi=0}
+\order{g^3}.
\]
A closer look at $V_2(x_{12})$ reveals that the $x_{12}$-dependence 
is only through $\xi$ as defined in \eqref{eq:Dila.Dim.XiGamma}.
This is a manifest property of a renormalisable field theory 
in dimensional regularisation.
We can thus write
\[\label{eq:Dila.SU2.VertexX}
V_2(x_{12})=\xi V_2=\frac{\Gammafn(1-\epsilon)}{\bigabs{\half\mu^2 x_{12}^2}^{-\epsilon}}\,
V_2.
\]
We send the regulator to zero and find
\[\label{eq:Dila.SU2.VertexLimit}
\lim_{\epsilon\to 0}\bigbrk{V_2(x_{12})-V_2(1/\mu)}
=
\log |\mu x_{12}|^{-2}\,D_2
\]
with 
\[\label{eq:Dila.SU2.DilLimit}
\algD_2=-\lim_{\epsilon\to 0} \epsilon V_2.\]
In the case at hand, we obtain from \eqref{eq:Dila.SU2.VertexXHIY}
\[\label{eq:Dila.SU2.DilaFirst}
\algD_{2}=-\sfrac{1}{2}N^{-1}
\normord{\Tr\comm{\phi_i}{\phi_j}\comm{\check\phi^i}{\check\phi^j}},
\]
where we have used the following expansion in $\epsilon$,
see \eqref{eq:Dila.Dim.YXH2pt},
for the functions appearing in $V_2$
\<\label{eq:Dila.SU2.XHexpand}
X_{00xx}I_{0x}^{-2} \xi^{-1}\eq
2\epsilon^{-1}+2+\order{\epsilon^2},\nln
\tilde H_{0x,0x}I_{0x}^{-2}\xi^{-1}\eq
-48\zeta(3)\,\epsilon+\order{\epsilon^2}.
\>
The final answer for the renormalised correlator 
at $\epsilon=0$ is
\[\label{eq:Dila.SU2.OneLoopFinal}
\bigvev{Z\Op_\varphi\, \dot Z\dot\Op_{\dot\varphi}}=
\bigeval{\exp(W_0)
\exp\bigbrk{\log |\mu x_{12}|^{-2} g^2 \algD_{2,\varphi}}\,
\Op_\varphi\,\dot\Op_{\dot\varphi}}_{\varphi=\dot\varphi=0}
+\order{g^3},
\]
in agreement with the form predicted by conformal field theory.%
\footnote{In fact, the mass dimension of the operators has
not changed from its classical value, hence the 
residual $\mu$-dependence. The fully renormalised operator to be 
inserted into the path integral would be $\mu^{\delta D(g)}Z\Op$,
but formally this cannot be expanded into a series
as emphasised above.}
The operator $\algD_2$ is the one-loop correction
to the dilatation generator.
Furthermore, the coefficient of the correlator
is given by its tree-level value.
Notice that although we are interested in correlators of renormalised operators
$Z \Op$ as on the left-hand side of \eqref{eq:Dila.SU2.OneLoopFinal},
we can work with bare operators $\Op$ as on the right hand side 
of \eqref{eq:Dila.SU2.OneLoopFinal}. 
In other words, we choose to renormalise the dilatation operator
instead of the states.

\subsection{Application}
\label{sec:Dila.SU2.Apply}

In the last section we have derived the one-loop dilatation generator
\eqref{eq:Dila.SU2.DilaFirst}
for the $\alSU(2)$ subsector. When we write it in components 
$(\fldZ,\phi)=(\varphi_1,\varphi_2)$ it reads
\[\label{eq:Dila.SU2.Dila}
\algD_{2}=-N^{-1}\normord{\Tr\comm{\fldZ}{\phi}\comm{\check \fldZ}{\check\phi}}.
\]
Using the rules in \secref{sec:N4.Alg} we can determine its
action on 
any operator of the form $\Tr \fldZ\fldZ\phi \fldZ\phi \Tr \phi\ldots$\,.

Let us now apply $\algD_2$ to rederive the results
of \secref{sec:Dila.Dim.TwoPoint}.
The first observation is that $D_2$ acts on $\fldZ$ and $\phi$ 
simultaneously. If either of them is absent in the state,
$\algD_2$ will annihilate it, therefore
\[\label{eq:Dila.SU2.HalfBPS}
\algD_2\,\Tr \fldZ\cdots \fldZ\,\Tr \fldZ\cdots \fldZ \ldots=0.
\]
As emphasised in \secref{sec:Dila.Sect.HalfBPS}, these states
are half-BPS and thus protected from quantum corrections.
In particular, the operator $\OpQ_{mn}$ 
has one component, $\Tr \fldZ\fldZ$, of this type.
The other state discussed in \secref{eq:Dila.Dim.TwoPoint}, 
the Konishi operator $\OpK$, is not part of the $\alSU(2)$ sector. 
However, it is on both unitarity bounds and has spin zero. 
Therefore, it has a descendant within the subsector whose weight
is given by \eqref{eq:Dila.Dim.Weights,eq:Dila.SU2.Shift}
\[\label{eq:Dila.SU2.WeightKon}
w_{\OpK}'=
w_{\OpK}+\delta w\indup{I+II}=
\weight{4;0,0;2,0,2;0,4}.\]
This is a state of length $L=4$ with $K=2$ fields
of type $\phi$, see also \cite{Bianchi:1999ge}. 
Let us write down a basis for all such states 
in $\grSU(N)$ gauge theory
(the line separates single from double-trace states)
\[\label{eq:Dila.SU2.Basis}
\OpE^\trans=\matr{c}{\Tr \fldZ\fldZ\phi\phi\\\Tr \fldZ\phi \fldZ\phi\\\hline
\Tr \fldZ\fldZ\,\Tr\phi\phi\\\Tr \fldZ\phi\,\Tr \fldZ\phi}.
\]
We apply the one-loop dilatation operator to the basis,
$\algD_2 \OpE=\OpE D_2$,
and obtain the matrix of anomalous dimensions 
\[\label{eq:Dila.SU2.DilaMat}
D_2=\matr{cc|cc}
{+2&-4&+\frac{\vphantom{\dot 8}8}{N}&-\frac{4}{N}\\
-2&+4&-\frac{8}{N}&+\frac{4}{N}\\\hline
0&0&0&0\\
0&0&0&0}.
\]
Its eigenvectors are 
\[\label{eq:Dila.SU2.DilaEv}
(-2,2,0,0)^\trans,\quad(2,1,0,0)^\trans,\quad(0,0,1,2)^\trans,\quad
(-\sfrac{2}{N},\sfrac{2}{N},1,-1)^\trans.\]
The first one corresponds to the Konishi descendant
\[\label{eq:Dila.SU2.EigenKonishi}
\OpK'=-2\Tr \fldZ\phi \fldZ\phi+2\Tr \fldZ\phi\phi \fldZ
=\Tr \comm{\fldZ}{\phi}\comm{\fldZ}{\phi},\qquad
D(g)=2+6g^2+\order{g^3}
\]
with eigenvalue $D_2=6$. The other three states have vanishing 
anomalous dimension. The first two,
$2\Tr \fldZ\fldZ\phi\phi+\Tr \fldZ\phi \fldZ\phi$ and 
$\Tr \fldZ\fldZ\Tr\phi\phi+2\Tr \fldZ\phi\Tr \fldZ\phi$ 
are related to $\Tr \fldZ\fldZ\fldZ\fldZ$ and 
$\Tr \fldZ\fldZ \Tr \fldZ\fldZ$ by $\alSU(2)$ rotations;
this explains $D_2=0$. The last operator
\[\label{eq:Dila.SU2.EigenQuarterBPS}
\Op=\Tr \fldZ\fldZ\Tr \phi\phi-\Tr \fldZ\phi \Tr \fldZ \phi
+N^{-1}\Tr \comm{\fldZ}{\phi}\comm{\fldZ}{\phi},\qquad
D_2=0
\]
is indeed a highest weight state of $\alPSU(2,2|4)$,
as such it is, unlike $\OpK'$, quarter-BPS
and protected \cite{Bianchi:1999ge,Bianchi:2000hn,Ryzhov:2001bp}.

\section{Field Theoretic Considerations}
\label{sec:Dila.Theory}

In this section we will investigate the structure of 
the dilatation operator at higher orders in perturbation theory 
without actually computing it. 
This will yield important structural constraints for 
the algebraic construction pursued in the following chapters.
%

\subsection{Two-Point Functions at Higher-Loops}
\label{sec:Dila.Theory.Higher}

Here, we would like to continue the 
investigation of the last section at higher loops and see 
how the dilatation operator can be extracted. 
We will show how to resolve some 
complications which appear starting at four loops and which are due to the fact that
the various loop contributions to the dilatation operator
do not commute with each other, e.g.~$\comm{\algD_2}{\algD_4}\neq 0$.
We will not compute higher-loop amplitudes explicitly.

To obtain the arbitrary loop correlator we insert
all $\ell$-loop connected Green functions $W_{2\ell}$ 
in the correlator
\[\label{eq:Dila.Theory.AllLoopW}
\bigvev{\Op_\varphi\, \dot\Op_{\dot\varphi}}=
\bigeval{\exp(W_0)
\exp\lrbrk{\tsum_{\ell=1}^\infty g^{2\ell} 
  W_{2\ell}(x_{12},\check\varphi,\check{\dot\varphi})}
\,\Op_\varphi\,\dot\Op_{\dot\varphi}}_{\varphi=\dot\varphi=0}.
\]
In analogy to \eqref{eq:Dila.SU2.OneLoopVertex} we change 
the argument $\check{\dot\varphi}$ of 
$W_{2\ell}(x,\check\varphi,\check{\dot\varphi})$
to $N\Delta_{12}^{-1}\varphi$
\[\label{eq:Dila.Theory.AllLoopWSub}
\bigvev{\Op_\varphi\, \dot\Op_{\dot\varphi}}=
\bigeval{\exp(W_0)\,
\normord{\exp\lrbrk{\tsum_{\ell=1}^\infty g^{2\ell} 
W_{2\ell}(x_{12},\check\varphi,N\Delta_{12}^{-1}\varphi)}}
\,\Op_\varphi\,\dot\Op_{\dot\varphi}}_{\varphi=\dot\varphi=0}.
\]
Alternatively, we could change the argument
$\check \varphi$ to $N\Delta^{-1}_{12}\dot\varphi$.
We would then like to rewrite
\eqref{eq:Dila.Theory.AllLoopWSub} in a convenient form for
the conformal structure of the correlator:
\<\label{eq:Dila.Theory.nLoopV}
\bigvev{\Op_\varphi\,\dot\Op_{\dot\varphi}}\eq
\bigeval{\exp(W_0)
\exp\bigbrk{V_\varphi(x_{12})}\,
\Op_\varphi\,\dot\Op_{\dot\varphi}}_{\varphi=\dot\varphi=0}
\nln\eq
\bigeval{\exp(W_0)
\exp\bigbrk{\dot V_{\dot\varphi}(x_{12})}\,
\Op_\varphi\,\dot\Op_{\dot\varphi}}_{\varphi=\dot\varphi=0},
\>
where%
\footnote{The commutator term was included for 
convenience, it could have been 
included in $V_8$, we will explain this issue below.}
\[\label{eq:Dila.Theory.VnDef}
V(x_{12})=\sum_{\ell=1}^\infty g^{2\ell}V_{2\ell}(x_{12}) 
-\sfrac{1}{48}g^8\bigcomm{V_2(x_{12})}{\comm{V_2(x_{12})}{V_4(x_{12})}}+\ldots.
\]
The terms $V_{2\ell}$ are defined by the equality of 
\eqref{eq:Dila.Theory.AllLoopWSub} and \eqref{eq:Dila.Theory.nLoopV}
\[\label{eq:Dila.Theory.Vn}
\exp\bigbrk{V(x_{12})}=
\normord{\exp\lrbrk{\tsum_{\ell=1}^\infty g^{2\ell} 
W_{2\ell}(x,\check\varphi,N\Delta^{-1}_{12}\varphi)}},
\]
which will have to be solved perturbatively. All
the terms that arise due to normal ordering of the exponential
and the commutator terms in \eqref{eq:Dila.Theory.VnDef} need to
be absorbed into the definition of higher order vertices.
For example, the two-loop effective vertex is
\[\label{eq:Dila.Theory.V4}
V_4(x)=\normord{W_4(x_{12},\check\varphi,N\Delta_{12}^{-1}\varphi)}
-\half \bigbrk{V_2(x_{12})V_2(x_{12})-\normord{V_2(x_{12})V_2(x_{12})}}.
\]

Let us introduce a transpose operation 
on a generator $X$ by the definition 
\[\label{eq:Dila.Theory.Transpose}
\exp(W_0)\,X(\varphi,\check\varphi)=
\exp(W_0)\,X^{\trans}(\dot\varphi,\check{\dot\varphi}).
\]
In other words, letting $X$ act on $\varphi$
is equivalent to letting $X^\trans$ act on $\dot\varphi$.
The alternative forms of \eqref{eq:Dila.Theory.nLoopV} lead to 
\[\label{eq:Dila.Theory.VSymTranspose}
V^\trans_{2\ell}(x_{12})=\dot V_{2\ell}(x_{12}).
\]
In a real field theory $W_{2\ell}(x_{12},\check\varphi,\check{\dot\varphi})$ 
must be hermitian in the arguments $\check\varphi$ and $\check{\dot\varphi}$. 
Therefore $\dot V_{2\ell}$ is indeed the complex
conjugate of $V_{2\ell}$ and \eqref{eq:Dila.Theory.VSymTranspose}
shows that $V_{2\ell}$ is self-adjoint.

We renormalise the operators according to 
\<\label{eq:Dila.Theory.Zn}
Z\eq\exp\Bigbrk{
-\half\tsum_{\ell=1}^\infty g^{2\ell} V_{2\ell}(1/\mu)
+\sfrac{1}{24}g^6 \bigcomm{V_2(1/\mu)}{V_4(1/\mu)}+\ldots},
\nln
\dot Z\eq\exp\Bigbrk{
-\half\tsum_{\ell=1}^\infty g^{2\ell} \dot V_{2\ell}(1/\mu)
+\sfrac{1}{24}g^6 \bigcomm{\dot V_2(1/\mu)}{\dot V_4(1/\mu)}+\ldots}.
\>
This gives
\[\label{eq:Dila.Theory.nLoopVRenorm}
\bigvev{Z\Op_\varphi\,\dot Z\dot\Op_{\dot\varphi}}=
\bigeval{\exp(W_0)
\exp\bigbrk{V_\varphi(x)} Z_\varphi \Op_\varphi \,
\dot Z_{\dot\varphi} \dot\Op_{\dot\varphi} }_{\phi=\varphi=0}.
\]
We can commute objects that depend only on $\varphi$ with 
objects that depend only on $\dot\varphi$ freely.
Then we use the transpose operation \eqref{eq:Dila.Theory.Transpose}
to make $\dot Z_{\dot\varphi}$ act on $\varphi$ instead.
We get
\[\label{eq:Dila.Theory.nLoopVRenorm2}
\bigvev{Z\Op_\phi\,Z\Op_\varphi}=
\bigeval{\exp(W_0)\,
\dot Z^\trans_{\varphi}\exp\bigbrk{V_\varphi(x_{12})}Z_\varphi \,
\Op_\varphi\,\dot\Op_{\dot\varphi}}_{\varphi=\dot\varphi=0}.
\]
The vertices $V_{2\ell}(1/\mu)$ in $Z$ are hermitian, 
\eqref{eq:Dila.Theory.VSymTranspose}, 
only the commutator in \eqref{eq:Dila.Theory.Zn} requires special 
care, because $V_2$ and $V_4$ need to be transformed
consecutively. This effectively inverts their order and
flips the sign of the commutator:
\[\label{eq:Dila.Theory.ZtnDef}
\dot Z^\trans=
\exp\lrbrk{
-\half \tsum_{\ell=1}^\infty g^{2\ell} V_{2\ell}(1/\mu)
-\sfrac{1}{24}g^6 \bigcomm{V_2(1/\mu)}{V_4(1/\mu)}+\ldots}.
\]
In a renormalisable field theory the dependence of $V_{2\ell}$ on $x_{12}$ is
determined, we write
\[\label{eq:Dila.Theory.VnX}
V_{2\ell}(x_{12})=\xi^\ell V_{2\ell}.
\]
We combine the exponentials in \eqref{eq:Dila.Theory.nLoopVRenorm2}
with the surrounding $Z$'s into a single exponent
\[\label{eq:Dila.Theory.nLoopVRenormExp}
\sum_{\ell=1}^{\infty} (\xi^\ell-\xi_0^\ell) g^{2\ell}V_{2\ell,\varphi} 
-\sfrac{1}{48}g^8(\xi-\xi_0)^4
\bigcomm{V_{2,\varphi}}{\comm{V_{2,\varphi}}{V_{4,\varphi}}}
+\ldots\,.
\]
The $\ell$-loop Green function $W_{2\ell}$ is expected to have multiple poles
at $\epsilon=0$.
In a conformal field theory, however, these poles
must have cancelled in the combination $V_{2\ell}$
as given by \eqref{eq:Dila.Theory.VnDef,eq:Dila.Theory.Vn}.
If so, we can finally send the regulator to zero and find 
\[\label{eq:Dila.Theory.nLoopDRenorm}
\bigvev{Z\Op_\varphi\,\dot Z\dot\Op_{\dot\varphi}}=
\bigeval{\exp(W_0)
\exp\bigbrk{\log |\mu x_{12}|^{-2}
\tsum_{\ell=1}^\infty g^{2\ell} \algD_{2\ell,\varphi}}
\,\Op_\varphi\,\dot\Op_{\dot\varphi}}_{\varphi=\dot\varphi=0}
\]
with 
\[\label{eq:Dila.Theory.Dn}
\algD_{2\ell}=-\ell\lim_{\epsilon\to 0} \epsilon V_{2\ell}.
\]
Note that the commutator term in 
\eqref{eq:Dila.Theory.nLoopVRenormExp} vanishes due to four powers
of $\epsilon$ from $(\xi-\xi_0)^4$ as opposed to only three powers
of $1/\epsilon$ from the $V_{2\ell}$.
For this cancellation to happen the commutator terms in 
\eqref{eq:Dila.Theory.VnDef} and \eqref{eq:Dila.Theory.Zn} are necessary:
We have investigated all possible terms that can arise in a four-loop
computation. 
We find that precisely the commutator structure in \eqref{eq:Dila.Theory.VnDef} 
is required to obtain a finite, conformally covariant
correlator.

Some comments about the renormalisation programme are
in order. Firstly, the programme ensures that the coefficient
of the two-point function is given by 
free contractions of the unrenormalised operators.
Secondly, the effective vertices $V_{2\ell}$ are 
self-adjoint with respect to the scalar product induced
by free contractions, see \eqref{eq:Dila.Theory.VSymTranspose}.
The same holds for the dilatation generator
which consequently has real eigenvalues.
Notice that in some case there may appear to be complex eigenvalues.
However, a more careful analysis will show that the 
corresponding eigenstate is zero. 
This may happen if the rank of the group is small
compared to the size of the operators and
group identities lead to 
non-trivial linear dependencies in the basis of operators.

\subsection{Two-Point Functions of Non-Scalar Operators}%
\label{sec:Dila.Theory.NonScalar}

Correlation functions of non-scalar operators are not as 
easy to handle as their scalar counterparts. This is due
to their spacetime indices which can not only be
contracted among themselves but also with $x_{12}$.
Furthermore, there are qualitative differences
between primaries and descendants, 
see \secref{sec:N4.Corr}. Therefore the form predicted by 
conformal symmetry is not as simple as \eqref{eq:Dila.Theory.nLoopDRenorm}.
It certainly involves the symmetry generators $\algK$ and $\algP$ to be
able to distinguish between primaries and descendants. 
These generators also receive quantum corrections, which would
have to be found at the same time. 

However, in some cases the dilatation generator 
may be obtained anyway without taking these complications into account.
A crucial observation is that, 
although \eqref{eq:N4.Corr.TwoVector} and \eqref{eq:N4.Corr.TwoDesc} are
different, the difference is only in the part that multiplies
$x_{12}^\mu x_{12}^\nu$. The `direct' contraction via $\eta_{\mu\nu}$ is 
the same for both. 
If all contractions between the operator indices and $x_{12}$ are dropped, 
the operators behave as though they were a set of scalars.
In \cite{Beisert:2003jj} this simplification made a computation of the
one-loop dilatation operator possible within the non-scalar subsector $(2,2)$.

Note that the covariant derivatives acting on a field are 
just ordinary partial derivatives at leading order. The appearance of the
gauge connection should be treated as an interaction that takes
place at the point of the field (boundary) 
and everywhere in spacetime (bulk). 
Algebraically, the structure of boundary interactions is
the same as in the bulk, the gauge field couples
to the field via the gauge group structure constants and one 
power of the coupling constant.

\subsection{Feynman Diagrams}
\label{sec:Dila.Theory.Feyn}

In \secref{sec:Dila.Theory.Higher}
we have seen how the corrections to the dilatation operator arise 
from divergent Feynman diagrams. Here we would like to investigate
the structure of interacting contributions to the dilatation operator
and other generators of the symmetry algebra.
This will be an important constraint for the constructions 
in the following chapters. 
The `\emph{interactions}', i.e.~the contributions to the
group generators are constructed from 
fields $\fldWf{}$, variations $\fldWv{}$
and structure constants $\gaugestr{}$ of the gauge group.
Notice that due to the form of the Lagrangian \eqref{eq:N4.D4.Lagr} there is 
exactly one power of the coupling constant $g$ for each
$\gaugestr{}$ (before gauge group identities are used).

Our first claim is that the generators $\algJ(g)$ are \emph{connected}.
Here, connectedness refers to the gauge algebra. 
It means that all gauge group indices are contracted so that 
the symbols form a connected graph.
The connectedness can be inferred from \secref{sec:Dila.Theory.Higher}:
The effective vertices $V_{2\ell}$ are connected diagrams.
They are generated from the Green functions $W_{2\ell}$ 
by removing the normal ordering of an exponential
\eqref{eq:Dila.Theory.Vn}
and adding commutators \eqref{eq:Dila.Theory.VnDef}.
One can easily convince oneself that these operations
produce only connected diagrams.
The same is true also for the dilatation generator $\algD$.
Connectedness can also be seen in $\superN=4$ SYM on $\Real\times S^3$.
There the bare Hamiltonian is clearly connected, but it does
not obey $\comm{\ham(0)}{\ham(g)}=0$ \eqref{eq:Dila.Pert.Diag}. 
In this work we will require this identity and therefore need
to diagonalise the Hamiltonian first. This diagonalisation procedure 
described in \secref{sec:Dila.Pert.PreDiag} produces only commutator terms
and thus connected diagrams.
This can be seen in \eqref{eq:Dila.Pert.DPert} by rewriting it as
\[\label{eq:Dila.Theory.DiagComm}
\algdD\mapsto 
\algdD_0+
\sum\nolimits_{d>0}\frac{1}{d}\,\comm{\algdD_d}{\algdD_{-d}}
+\ldots\,.
\]

\begin{figure}\centering
\includegraphics{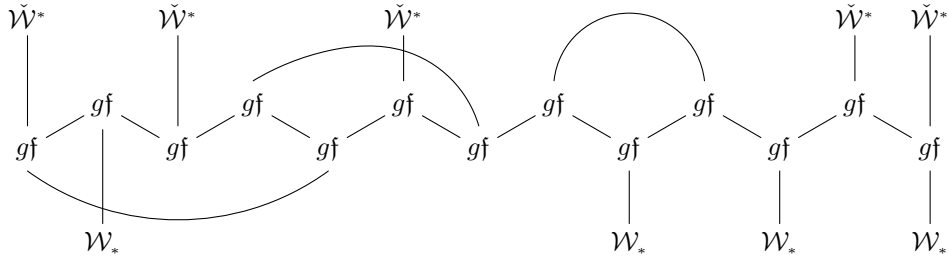}
\caption{A generic interaction with 
$E\indup{i}=5$ variations, $E\indup{o}=4$ fields, $L'=3$ index loops
and $V=13$ structure constants and powers of the 
coupling constant.}
\label{fig:N4.Theory.Interact}
\end{figure}

Secondly, we can count the number of external legs $E$, 
i.e.~the number of fields $E\indup{o}$ plus 
the number of variations $E\indup{i}$. 
According to \eqref{eq:N4.Quantum.CountVertex}
this equals 
\[\label{eq:Dila.Theory.Legs}
E=V+2-2L',
\]
where $V$ is the number of structure constants and $L'$ is the number
of adjoint index loops. For each structure constant there is 
precisely one power of $g$. A contribution of $\order{g^V}$ 
therefore has no more than $V+2$ legs.%
\footnote{The definition of the Hamiltonian $\ham$ involves two powers of $g$,
therefore $\ham$ generically has two legs more at a given order of $g$.}
A useful basis for interactions which can be achieved by making use of 
Jacobi identities is
\[\label{eq:Dila.Theory.Interact}
\mbox{`\emph{interactions}':}\quad \algJ(g)\sim
g^{E\indups{i}+E\indups{o}+2L'-2}\,
\gaugestr{m\ldots m}^{E\indups{i}+E\indups{o}+2L'-2}
(\gaugemet{mm})^{L'}
(\fldWf{\ast}{}^{\gaugeind{m}})^{E\indups{o}} 
(\fldWv{\ast}{}^{\gaugeind{m}})^{E\indups{i}} ,
\]
where the linear contraction of $V$ structure constants 
$\gaugestr{m\ldots m}^{V}$
is given by, see \figref{fig:N4.Theory.Interact},
\[\label{eq:Dila.Theory.LinearF}
\gaugestr{m\ldots m}^V
=
\gaugemet{}{}^{}_{\gaugeind{m}_1 \gaugeind{n}_1}
\gaugestr{}{}_{\gaugeind{m}_2 \gaugeind{n}_2}^{\gaugeind{n}_1}
\gaugestr{}{}_{\gaugeind{m}_3 \gaugeind{n}_3}^{\gaugeind{n}_2}
\cdots
\gaugestr{}{}_{\gaugeind{m}_{V+1} \gaugeind{m}_{V+2}}^{\gaugeind{n}_{V}}
=i^V
\Tr \gaugegen{}{}_{\gaugeind{m}_1} 
   [\gaugegen{}{}_{\gaugeind{m}_2},
   [\gaugegen{}{}_{\gaugeind{m}_3},[\ldots,
   [\gaugegen{}{}_{\gaugeind{m}_{V+1}},
    \gaugegen{}{}_{\gaugeind{m}_{V+2}}]\ldots]]].
\]
These interactions preserve the parity operation defined
in \secref{sec:N4.Gauge} as one can confirm easily.
This is because they are only composed of structure constants 
which have positive parity.

Finally, there is a peculiar feature of maximal scalar diagrams
which will be important to select the right terms later on.
These are diagrams without index loops $L'=0$ which therefore have
the maximum number of external scalar legs $V+2$ at order $V$ 
in perturbation theory.
We are interested in the flow of $\alSO(6)$ vector indices 
across the diagram.
As this is a tree diagram, the internal lines can 
only be scalars or gauge fields. 
Gauge fields are singlets of $\alSO(6)$ therefore 
only scalar fields can support the flow.
Only at quartic interactions of the scalars two
lines of flow can cross.
This shows that at order $V$ there cannot be more than $V/2$
crossings of $\alSO(6)$ vector flow lines.

\section{The Planar Limit and Spin Chains}
\label{sec:Dila.Planar}

Generic interactions have a very complicated structure
due to a large number of possible contractions between
the indices in \eqref{eq:Dila.Theory.Interact}.
Most of the time it is therefore useful to restrict 
to the planar limit, see \secref{sec:N4.LargeN}.

\subsection{States}
\label{sec:Dila.Planar.States}

In the large $N$ limit, field theory diagrams are suppressed 
unless $2C-2G-T=0$, see \eqref{eq:N4.LargeN.CountN}. 
As each component requires at least two traces, one incoming
and one outgoing (there are no vacuum diagrams), 
we need $T=2C$ and $G=0$. In other words, the diagrams
may connect only two single trace operators
and cannot have handles.
Therefore it makes sense to consider only single trace states 
\[\label{eq:Dila.Planar.State}
\mbox{`\emph{single-trace state}':}\quad 
\state{\fldindn{A}_1\ldots \fldindn{A}_L}
:=
\Tr \fldWf{}{}_{{\fldind{A}_{1}}}\cdots\fldWf{}{}_{\fldind{A}_{L}}.
\]
The cyclicity of the trace gives rise to cyclic identifications
\[\label{eq:Dila.Planar.StateCyclic}
\state{\fldindn{A}_1\ldots \fldindn{A}_p \fldindn{A}_{p+1}\ldots\fldindn{A}_L}
=(-1)^{(\fldindn{A}_1\ldots \fldindn{A}_{p})(\fldindn{A}_{p+1}\ldots\fldindn{A}_L)}
\state{\fldindn{A}_{p+1}\ldots\fldindn{A}_L\fldindn{A}_1\ldots \fldindn{A}_{p}}.
\]
The sign is due to statistics of the fields:
$(-1)^{XY}$ equals $-1$ if both, $X$ and $Y$, are fermionic
and $+1$ otherwise.
In particular, some states are incompatible with this symmetry
\[\label{eq:Dila.Planar.NoState}
\state{\fldindn{A}_1\ldots \fldindn{A}_{L/2}
\fldindn{A}_1\ldots \fldindn{A}_{L/2}}=0,
\qquad
\mbox{if $\fldindn{A}_1\ldots \fldindn{A}_{L/2}$ is fermionic}.
\]
A generic state is a linear superposition of the above basis states
\[\label{eq:Dila.Planar.LinearComb}
\Op=
c\,\state{\fldindn{A}_1\ldots \fldindn{A}_L}+
c'\,\state{\fldindn{A'}_1\ldots \fldindn{A'}_{L'}}
+\ldots\,,
\]
where mixing of states with different length is explicitly allowed.

\subsection{Interactions}
\label{sec:Dila.Planar.Interact}

\begin{figure}\centering
\includegraphics{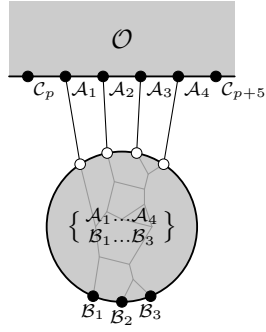}
\caption{Insertion of a planar interaction.
The black dots correspond to fields, the white dots to variations.
Inside the blob there is some unspecified planar diagram
that connects the dots.}
\label{fig:N4.Planar.Interact}
\end{figure}
For planar interactions the precise structure of internal connections 
does not play a role as long as it is planar, c.f.~\figref{fig:N4.Planar.Interact}.
The only relevant structure 
is the order of external legs. 
In the planar limit it is therefore
sufficient to consider interactions of the type
\[\label{eq:Dila.Planar.Interact}
\mbox{`\emph{planar interactions}':}\quad
\ITerm{\fldindn{A}_{1}\ldots \fldindn{A}_{E\indup{i}}}
{\fldindn{B}_{1}\ldots \fldindn{B}_{E\indup{o}}}
:=
N^{1-E\indup{i}}\,
\Tr \fldWf{}{}_{\fldind{B}_1}\ldots \fldWf{}{}_{{\fldind{B}_{E\indup{o}}}}
\fldWv{}{}^{\fldind{A}_{E\indup{i}}}\ldots \fldWv{}{}^{{\fldind{A}_{1}}},
\]
which searches for the sequence of fields 
$\fldWf{}{}_{\fldind{A}_1}\ldots\fldWf{}{}_{\fldind{A}_{E\indup{i}}}$
within a state and replaces it by the sequence
$\fldWf{}{}_{\fldind{B}_1}\ldots \fldWf{}{}_{\fldind{B}_{E\indup{o}}}$.
More explicitly, the action on a state $\state{\fldindn{C}_1\ldots \fldindn{C}_L}$ is
\[\label{eq:Dila.Planar.Action}
\sum_{p=1}^{L} 
(-1)^{(\fldindn{C}_1\ldots \fldindn{C}_{p-1})(\fldindn{B}_{1}\ldots \fldindn{B}_{E\indup{o}}\fldindn{A}_{1}\ldots \fldindn{A}_{E\indup{i}})}
\delta^{\fldindn{A}_1}_{\fldindn{C}_{p}}\ldots \delta^{\fldindn{A}_{E\indup{i}}}_{\fldindn{C}_{p+E\indup{i}-1}}
\state{\fldindn{C}_1 \ldots \fldindn{C}_{p-1} \fldindn{B}_1 \ldots \fldindn{B}_{E\indup{o}} 
       \fldindn{C}_{p+E\indup{i}}\ldots \fldindn{C}_L}.
\]
A sample action is
\[\label{eq:Dila.Planar.ActionEx}
\ITerm{\fldindn{A}\fldindn{B}}{\fldindn{B}\fldindn{A}}\state{12345}
=\state{21345}\pm\state{13245}\pm\state{12435}\pm\state{12354}\pm\state{52341}.
\]

The order of an interaction in perturbation theory 
is given by $V=E\indup{i}+E\indup{o}+2L'-2$, therefore
$\algJ(g)\sim g^{E\indups{i}+E\indups{o}+2L'-2}
\ITerm{\fldindn{A}_{1}\ldots \fldindn{A}_{E\indups{i}}}
{\fldindn{B}_{1}\ldots \fldindn{B}_{E\indups{o}}}$.
We see that for planar interactions, 
adding an index loop simply increases the loop order by one. 
At a fixed loop order this leads to diagrams with fewer external legs.
To reduce the complexity, we can install a pair of legs by means
of a gauge transformation. For that purpose we insert the generator of gauge
rotations $\gaugerot$ into some interaction
\<\label{eq:Dila.Planar.GaugeAdd}
0\earel{\widehat{=}}
-i\Tr \gaugerot\,\,\fldWf{}{}_{\fldind{B}_1}\ldots \fldWf{}{}_{{\fldind{B}_{E\indups{o}}}} 
\fldWv{}{}^{\fldind{A}_{E\indups{i}}}\ldots \fldWv{}{}^{{\fldind{A}_{1}}}
\nln\eq
\Tr \normord{ \fldWf{C}\fldWv{C}   }
\fldWf{}{}_{\fldind{B}_1}\ldots \fldWf{}{}_{{\fldind{B}_{E\indups{o}}}} 
\fldWv{}{}^{\fldind{A}_{E\indups{i}}}\ldots \fldWv{}{}^{{\fldind{A}_{1}}}
-\Tr \normord{ \fldWv{C} \fldWf{C}  }
\fldWf{}{}_{\fldind{B}_1}\ldots \fldWf{}{}_{{\fldind{B}_{E\indups{o}}}} 
\fldWv{}{}^{\fldind{A}_{E\indups{i}}}\ldots \fldWv{}{}^{{\fldind{A}_{1}}}
\nln\earel{\widehat{=}}
N\Tr \fldWf{}{}_{\fldind{B}_1}\ldots \fldWf{}{}_{{\fldind{B}_{E\indups{o}}}} 
\fldWv{}{}^{\fldind{A}_{E\indups{i}}}\ldots \fldWv{}{}^{{\fldind{A}_{1}}}
\mp \Tr  \fldWf{C} 
\fldWf{}{}_{\fldind{B}_1}\ldots \fldWf{}{}_{{\fldind{B}_{E\indups{o}}}} 
\fldWv{}{}^{\fldind{A}_{E\indups{i}}}\ldots \fldWv{}{}^{{\fldind{A}_{1}}}
\fldWv{C}.
\>
The equivalence in the last line is for planar insertions only. 
This means that adding a pair of legs to the left of the interaction 
has no effect. Equivalently, we can add a pair of legs to the
right of the interaction
\[\label{eq:Dila.Planar.Spectator}
\ITerm{\fldindn{A}_{1}\ldots \fldindn{A}_{E\indups{i}}}
{\fldindn{B}_{1}\ldots \fldindn{B}_{E\indups{o}}}
\hateq
\ITerm{\fldindn{A}_{1}\ldots \fldindn{A}_{E\indups{i}}\fldindn{C}}
{\fldindn{B}_{1}\ldots \fldindn{B}_{E\indups{o}}\fldindn{C}}
\hateq
(-1)^{\fldindn{C}(\fldindn{A}_{1}\ldots \fldindn{A}_{E\indups{i}}\fldindn{B}_{1}\ldots \fldindn{B}_{E\indups{o}})}
\ITerm{\fldindn{C}\fldindn{A}_{1}\ldots \fldindn{A}_{E\indups{i}}}
{\fldindn{C}\fldindn{B}_{1}\ldots \fldindn{B}_{E\indups{o}}}.
\]
This is obvious because the additional pair
of legs does not change the field at that position, 
it is only a \emph{spectator}. 
We can now add $L'$ pairs of spectator legs to an interaction
and thus drop the index loop parameter%
\footnote{Note that for $\ham_k$
as defined in \secref{sec:Dila.Pert.Anomalous} 
there should be $E\indup{i}+E\indup{o}=k+4$ legs.}
\[\label{eq:Dila.Planar.LegCount}
\algJ_k\sim
\ITerm{\fldindn{A}_{1}\ldots \fldindn{A}_{E\indups{i}}}
{\fldindn{B}_{1}\ldots \fldindn{B}_{E\indups{o}}},
\quad \mbox{with}\quad 
E\indup{i}+E\indup{o}=k+2.
\]
%

\subsection{Wrapping Interactions}
\label{sec:Dila.Planar.Wrapping}

\begin{figure}\centering
\includegraphics{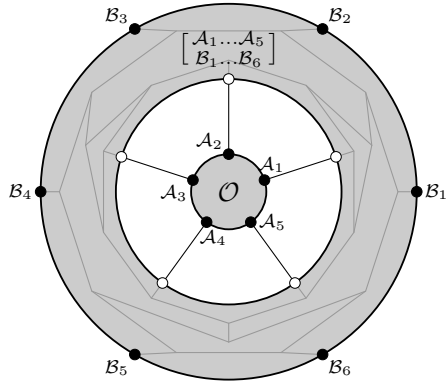}
\caption{The action of a wrapping interaction.
For a planar insertion, 
the wrapping interaction must surround the state
and the number of fields $L$ must match the number
of variations $E\indup{i}$.}
\label{fig:N4.Planar.Wrapping}
\end{figure}
This can, however, not be completely true: If the number of 
variations, $E\indup{i}$, equals the length of the state, $L$,
we cannot add a spectator pair of legs. In fact, there is a
subtlety in the second equivalence in \eqref{eq:Dila.Planar.GaugeAdd}: 
When in the second term the variation 
hits the field $\fldWf{}{}_{{\fldind{B}_{E\indups{o}}}}$ we get
a `\emph{wrapping diagram}'. It can be represented by 
the following symbol and trace structure
\[\label{eq:Dila.Planar.Wrapping}
\mbox{`\emph{wrapping interactions}':}\quad
\WTerm{\fldindn{A}_{1}\ldots \fldindn{A}_{E\indup{i}}}
{\fldindn{B}_{1}\ldots \fldindn{B}_{E\indup{o}}}
:=
N^{-E\indup{i}}\,
\Tr \fldWf{}{}_{\fldind{B}_1}\ldots \fldWf{}{}_{{\fldind{B}_{E\indup{o}}}}\,
\Tr \fldWv{}{}^{\fldind{A}_{E\indup{i}}}\ldots \fldWv{}{}^{{\fldind{A}_{1}}}.
\]
Wrapping interactions remove the state as a whole and replace
it by a new one. They are best understood graphically, see 
\figref{fig:N4.Planar.Wrapping}. 
Wrapping diagrams are generically non-planar,
but when applied to a state of the minimally required length,
the action becomes planar. This is because the diagram can be wrapped
fully around the trace. If, however, some uncontracted fields remain
within the trace, they are disconnected from the fields
of the interaction and the action is non-planar.
An improved version of \eqref{eq:Dila.Planar.Spectator} 
which takes states of finite length into account is
\[\label{eq:Dila.Planar.SpectatorWrap}
\ITerm{\fldindn{A}_{1}\ldots \fldindn{A}_{E\indups{i}}}
{\fldindn{B}_{1}\ldots \fldindn{B}_{E\indups{o}}}
=
\ITerm{\fldindn{A}_{1}\ldots \fldindn{A}_{E\indups{i}}\fldindn{C}}
{\fldindn{B}_{1}\ldots \fldindn{B}_{E\indups{o}}\fldindn{C}}
+\WTerm{\fldindn{A}_{1}\ldots \fldindn{A}_{E\indups{i}}}
{\fldindn{B}_{1}\ldots \fldindn{B}_{E\indups{o}}}
=
(-1)^{\fldindn{C}(\fldindn{A}_{1}\ldots \fldindn{A}_{E\indups{i}}\fldindn{B}_{1}\ldots \fldindn{B}_{E\indups{o}})}
\ITerm{\fldindn{C}\fldindn{A}_{1}\ldots \fldindn{A}_{E\indups{i}}}
{\fldindn{C}\fldindn{B}_{1}\ldots \fldindn{B}_{E\indups{o}}}
+\WTerm{\fldindn{A}_{1}\ldots \fldindn{A}_{E\indups{i}}}
{\fldindn{B}_{1}\ldots \fldindn{B}_{E\indups{o}}}
\]

For wrapping diagrams the order in perturbation theory is given by 
\[\label{eq:Dila.Planar.WrapLoop}
\algJ_k\sim
\WTerm{\fldindn{A}_{1}\ldots \fldindn{A}_{E\indups{i}}}
{\fldindn{B}_{1}\ldots \fldindn{B}_{E\indups{o}}}
\quad\mbox{with}\quad
E\indups{i}+E\indups{o}=k+2-2L',\quad
L'\geq 1.
\]
They act only on states with length $L=E\indup{i}$ and 
need at least one index loop $L'$.
Therefore they appear only at rather high loop orders,
especially for long states.
Unfortunately, there are no obvious structural constraints 
on wrapping interactions as the one described 
at the end of \secref{sec:Dila.Theory.Feyn}.
This makes them rather hard to handle and 
we will not make quantitative statements in this work.

\subsection{Parity}
\label{sec:Dila.Planar.Parity}

In \secref{sec:N4.Gauge} we have defined a parity operation
for a unitary gauge group. 
It replaces all fields by their negative transpose.
Transposing all matrices within a trace simply 
reverses their order. We find that parity acts on a state as
\[\label{eq:Dila.Planar.ParityState}
\gaugepar\,\state{\fldindn{A}_1\ldots \fldindn{A}_L}=
(-1)^{L+f(f-1)/2}\state{\fldindn{A}_L\ldots \fldindn{A}_1},
\]
where $f$ is the number of fermionic fields in the trace.
Effectively we can use this definition of parity 
also for gauge groups $\grSO(N)$ and $\grSp(N)$. 
There, however, parity must act trivially 
and only states of positive parity are allowed.

The parity operation for interactions is
($f\indup{i}$ and $f\indup{o}$ are the numbers of fermions in $A_1\ldots A_{E\indups{i}}$ and
$B_1\ldots B_{E\indups{o}}$, respectively)
\[\label{eq:Dila.Planar.ParityInteract}
\gaugepar\,
\ITerm{\fldindn{A}_1\ldots \fldindn{A}_{E\indups{i}}}
      {\fldindn{B}_1\ldots \fldindn{B}_{E\indups{o}}}\,
\gaugepar^{-1}=
(-1)^{{E\indups{i}}+{E\indups{o}}+f\indup{i}(f\indup{i}-1)/2+f\indup{o}(f\indup{o}-1)/2} 
  \ITerm{\fldindn{A}_{E\indups{i}}\ldots \fldindn{A}_1}
        {\fldindn{B}_{E\indups{o}}\ldots \fldindn{B}_i}.
\]
For the interactions within algebra generators $\algJ(g)$ 
parity must be positive. Nevertheless, we will also
make contact with generators of negative parity later on.

\subsection{Scalar Product}
\label{sec:Dila.Planar.Adjoint}

Our investigations in this work are independent of the
definition of a scalar product.
Nevertheless, it is useful to know 
how to construct a meaningful norm
because the dilatation operator
will be self-adjoint with respect to this norm
and thus have real eigenvalues.
We will sketch how the norm should look like.

The construction in \secref{sec:Dila.Theory.Higher} shows that
states can be renormalised in such a way as to preserve the 
classical scalar product. At tree-level
the scalar product is given by pairwise contractions 
$\langle\fldindn{A}|\fldindn{B}\rangle$ between
fields of both states. In the planar limit all contractions 
must be parallel. 
Therefore the planar scalar product of two states 
%
\[\label{eq:Dila.Planar.ScalarProduct}
\langle\fldindn{A}_1\ldots \fldindn{A}_{L}|
\fldindn{B}_1\ldots \fldindn{B}_{L'}\rangle = 
\delta_{L=L'}\sum_{p'=1}^L 
(\pm 1)
\prod_{p=1}^L
\langle\fldindn{A}_{p}|\fldindn{B}_{p'-p}\rangle
\]
vanishes unless $L=L'$ and both states are related 
by a cyclic permutation.
For generic overlapping states the elementary scalar products 
in \eqref{eq:Dila.Planar.ScalarProduct} are non-zero
for all $p$ only for one very specific value of $p'$. 
However, for a state which can be written as
$\state{(\fldindn{A}_1\ldots\fldindn{A}_{L/n})^n}$
with $n$ as large as possible, 
there are $n$ possible values for $p'$. 
The square norm for a state is thus proportional to $\pm n$
\[\label{eq:Dila.Planar.Norm}
\state{(\fldindn{A}_1\ldots \fldindn{A}_{L/n})^{n}} \sim \sqrt{n}\,.
\]
An adjoint operation for interactions compatible 
with the above scalar product for states
should interchange the two rows in the interaction symbol%
\footnote{The reverse ordering of the adjoint is
related to the scalar product. One could combine the
adjoint with parity to define a different
adjoint operation which only interchanges both rows.}
\[\label{eq:Dila.Planar.AdjointInt}
\ITerm{\fldindn{A}_1\ldots \fldindn{A}_{E\indups{i}}}
      {\fldindn{B}_1\ldots \fldindn{B}_{E\indups{o}}}^\dagger \sim 
\ITerm{\fldindn{B}_{E\indups{o}}\ldots \fldindn{B}_1}
      {\fldindn{A}_{E\indups{i}}\ldots \fldindn{A}_1}.
\]
Note, however, that the action of a self-adjoint interaction
on a set of states is only equivalent 
to a hermitian matrix if all 
states are normalised to one with respect to \eqref{eq:Dila.Planar.Norm}.
Otherwise the matrix is only self-adjoint with respect to
the norm on the set of states.
For example, this is the case for the 
asymmetric matrix \eqref{eq:Dila.SU2.DilaMat}.

\subsection{Spin Chains}
\label{sec:Dila.Planar.SpinChain}

Single-trace local operators can be viewed as states of a 
dynamic, cyclic, quantum \emph{spin chain} \cite{Minahan:2002ve}. 
A cyclic spin chain is a set of $L$ \emph{spin sites}
with a cyclic adjacency property.
In a quantum spin chain, the \emph{spin} at each site is a module 
of the symmetry algebra of the system and
the Hilbert space is the tensor product of
$L$ spin modules.
For a \emph{dynamic} spin chain the number of sites $L$ is not fixed
\cite{Beisert:2003ys};
the full Hilbert space is the tensor product of all Hilbert spaces 
of a fixed length.
The basic quantum spin chain is the Heisenberg chain. 
Its symmetry group is $\alSU(2)$ and 
all spins transform in the fundamental representation. 
A basis for the Hilbert space is given by those states 
for which the spin at each site points either `up' or `down'. 
The Hilbert space is thus $\Comp^{2^L}$.
In a more general spin chain, the spin can point in 
more than just two directions, in most cases even infinitely many.
Note that the cyclic identification 
\eqref{eq:Dila.Planar.StateCyclic} of
field theory states is an additional 
constraint on cyclic spin chains.
For example, the field theory Hilbert space 
corresponding to the Heisenberg chain is $\Comp^{2^L}/\Integers_L$.
Physical states are identified by a trivial shift operator,
states with non-zero momentum are unphysical.

\begin{figure}\centering
\parbox{5cm}{\centering\includegraphics{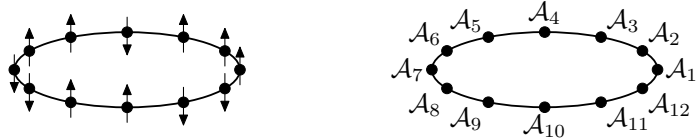}}
\quad
\parbox{5cm}{\centering\includegraphics{sec02.chain.su224.eps}}
\caption{A single-trace operator as a spin chain.
For the simplest spin chain, the spin can take two
alignments, for $\superN=4$ the `spin' can take infinitely many.}
\label{fig:Dila.Planar.Chains}
\end{figure}

In the spin chain picture each field is identified with 
one site of the chain. 
The alignment of the spin at that site corresponds to the component
of the multiplet of fields, c.f.~\figref{fig:Dila.Planar.Chains}.
For $\superN=4$ SYM, the spin chain is a $\alPSU(2,2|4)$ 
cyclic super spin chain with spins transforming in the 
representation $[0;0;0,1,0;0;0]$, see \secref{sec:N4.Fund},
\cite{Beisert:2003yb}.
When working in the planar limit, we will commonly make use of
spin chain terminology. 
In particular, the quantum correction to the dilatation generator 
will be called the `\emph{Hamiltonian}' $\ham=g^{-2}\algdD$ 
and  anomalous dimensions are synonymous for `\emph{energies}' $E=g^{-2}\delta D$,
see \tabref{tab:Dila.Planar.SpinChain} for a small dictionary.
For the other generators of the superconformal group 
we use the same symbols as in the non-planar case.
\begin{table}\centering
\begin{tabular}{|l|l|}\hline
planar $\superN=4$ SYM & $\alPSU(2,2|4)$ spin chain \\\hline
single trace operator & cyclic spin chain \\
field & spin site \\
anomalous dilatation operator $g^{-2}\delta\algD$ & Hamiltonian $\ham$ \\
anomalous dimension $g^{-2}\delta D$ & energy $E$ \\
cyclicity constraint & zero-momentum condition $U=1$ \\\hline
\end{tabular}
\caption{Dictionary for $\superN=4$ SYM and the spin chain picture.}
\label{tab:Dila.Planar.SpinChain}
\end{table}

\finishchapter 

\chapter{One-Loop}
\label{ch:One}

In this chapter we will derive the complete one-loop dilatation 
operator of $\superN=4$ Super Yang-Mills Theory.
The text is based on the article \cite{Beisert:2003jj},
but we present a new derivation of the coefficients $C_j$.
In \cite{Beisert:2003jj} the coefficients have been obtained
in a quantum field theory calculation, 
here we will merely employ the superconformal algebra.
The spectral and plane-wave investigations have been compiled from the articles
\cite{Beisert:2002ff,Beisert:2002tn,Beisert:2003tq,Beisert:2003jj,Beisert:2004di}.

\section{The Form of the Dilatation Generator}
\label{sec:One.Form}

We start by investigating the general form of the one-loop
dilatation generator. We will see that
representation theory of the symmetry group as
well as Feynman diagrammatics put tight constraints on the form.
What remains is a sequence of undetermined coefficients $C_j$,
one for each value of `total spin'. 

\subsection{One-Loop as Leading Order}
\label{sec:One.Form.Leading}

In \secref{sec:Dila.Pert.Anomalous} we have learned that 
the leading order anomalous dilatation operator $D_l$ is 
invariant under classical superconformal transformations $\algJ_0$. 
It is impossible to construct an invariant operator $\algD_1$
at first order of the coupling constant $g$, therefore
the leading order is one-loop, $l=2$.
We will come back to this point after having reviewed some
representation theory at the end of \secref{sec:One.Form.Symmetry}.
In what follows we will consider only 
the classical $\alPSU(2,2|4)$ algebra of generators $\algJ_0$;
the one-loop anomalous dilatation generator $\algD_2$ will
be considered an independent object;
we will refer to it as the Hamiltonian $\ham$,
\[\label{eq:One.Form.JHDef}
\algJ(g)=\algJ+\order{g},\qquad 
\algD(g)=\algD+g^2\, \ham+\order{g^3},\qquad 
\comm{\algJ}{\ham}=0.
\]

\subsection{Generic Form}
\label{sec:One.Form.Generic}

The Hamiltonian 
has the following generic form%
\footnote{These expressions are valid for bosonic fields $\fldWf{A}$ only.
They do generalise to fermions, 
but only at the cost of obscure signs at various places.}
\begin{figure}\centering
\parbox{1.5cm}{\centering\includegraphics{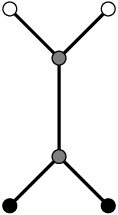}\par a}
\quad
\parbox{1.5cm}{\centering\includegraphics{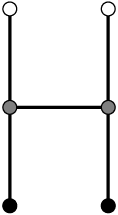}\par b}
\quad
\parbox{1.5cm}{\centering\includegraphics{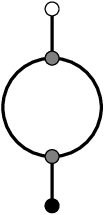}\par c}
\caption{Algebraic structure of the one-loop diagrams contributing 
to the anomalous dimension.
The lines correspond to any of the fundamental fields of the theory.}
\label{fig:One.Form.Diagrams}
\end{figure}%
\<\label{eq:One.Form.HDiagram}
\ham\eq
-N^{-1}(C\indup{a})^{\fldind{AB}}_{\fldind{CD}} \,\,
\normord{\Tr\bigcomm{\fldWf{A}}{\fldWv{C}}\bigcomm{\fldWf{B}}{\fldWv{D}}}
\nl
-N^{-1}(C\indup{b})^{\fldind{AB}}_{\fldind{CD}} \,\,
\normord{\Tr\bigcomm{\fldWf{A}}{\fldWf{B}}\bigcomm{\fldWv{C}}{\fldWv{D}}}
\nl
+N^{-1}(C\indup{c})^{\fldind{A}}_{\fldind{B}} \,\,
\gaugemet{mn}
\normord{\Tr\bigcomm{\fldWf{A}}{\gaugegen{m}}\bigcomm{\gaugegen{n}}{\fldWv{B}}}.
\>
These terms correspond to the three basic types of 
divergent Feynman diagrams which arise at the one-loop level, 
see \figref{fig:One.Form.Diagrams}.
As before, in \secref{sec:Dila.SU2.OneLoop}, we can transform 
the term of type c by means of gauge invariance.
The generator of gauge transformations (see \secref{sec:N4.Gauge})
is $\gaugerot=i\normord{\comm{\fldWf{C}}{\fldWv{C}}}$ 
and it annihilates gauge invariant operators.
Therefore we can write 
(note the change of normal orderings)
\[\label{eq:One.Form.Gauge}
0\mathrel{\widehat{=}}-i\Tr \gaugerot\, \normord{\bigcomm{\fldWf{A}}{\fldWv{B}}}  =
\normord{\Tr \bigcomm{\fldWf{C}}{\fldWv{C}} \bigcomm{\fldWf{A}}{\fldWv{B}} }
+\gaugemet{mn}\normord{\Tr \bigcomm{\fldWf{A}}{\gaugegen{m}} \bigcomm{\gaugegen{n}}{\fldWv{B}}},
\]
which allows us to write the term of type c as a term of type a.
Furthermore the term of type b can be transformed by 
means of a Jacobi-identity
\[\label{eq:One.Form.Jacobi}
\normord{\Tr\bigcomm{\fldWf{A}}{\fldWf{B}}\bigcomm{\fldWv{C}}{\fldWv{D}}}
=\normord{\Tr\bigcomm{\fldWf{A}}{\fldWv{C}}\bigcomm{\fldWf{B}}{\fldWv{D}}}
-\normord{\Tr\bigcomm{\fldWf{A}}{\fldWv{D}}\bigcomm{\fldWf{B}}{\fldWv{C}}}.
\]
We combine all coefficients 
into a single one of type a
\[\label{eq:One.Form.CoeffCombine}
C^{\fldind{AB}}_{\fldind{CD}}=
-\bigbrk{(C\indup{a})^{\fldind{AB}}_{\fldind{CD}}
+(C\indup{b})^{\fldind{AB}}_{\fldind{CD}}
-(C\indup{b})^{\fldind{AB}}_{\fldind{DC}}
+\half\delta^{\fldind{A}}_{\fldind{C}}(C\indup{c})^{\fldind{B}}_{\fldind{D}}
+\half(C\indup{c})^{\fldind{A}}_{\fldind{C}}\delta^{\fldind{B}}_{\fldind{D}}}.
\]
The total Hamiltonian is
\[\label{eq:One.Form.HForm}
\ham=-N^{-1}C^{\fldind{AB}}_{\fldind{CD}} \,\,
\normord{\Tr\bigcomm{\fldWf{A}}{\fldWv{C}}\bigcomm{\fldWf{B}}{\fldWv{D}}}
\]
with some yet undetermined coefficient $C^{\fldind{AB}}_{\fldind{CD}}$.

\subsection{Symmetry}
\label{sec:One.Form.Symmetry}

The combined coefficient $C^{\fldind{AB}}_{\fldind{CD}}$ must be invariant
under the classical superconformal algebra, 
it describes an \emph{intertwining map} $\mdlF\times\mdlF\mapsto \mdlF\times \mdlF$.
This requirement puts tight constraints on the coefficients, the independent 
components can be obtained by investigating the irreducible modules 
in the tensor product $\mdlF\times\mdlF$. 
The tensor product of two $\mdlF$ is given by
(see e.g.~\cite{Beisert:2004di})
\[\label{eq:One.Form.Tensor}
\mdlF\times \mdlF=
\sum_{j=0}^\infty \mdl_{j},
\]
where $\mdl_j$ are the modules with primary weights
\<\label{eq:One.Form.Irreps}
w_0\eq \weight{2;0,0;0,2,0;0,2},\nln
w_1\eq \weight{2;0,0;1,0,1;0,2},\nln
w_j\eq \weight{j;j-2,j-2;0,0,0;0,2}.
\>
The module $\mdl_0$ is the half-BPS current multiplet,
$\mdl_1$ is quarter-BPS and the other ones are doubly-short.
As an aside, it is interesting to see that the quadratic Casimir
(c.f.~\appref{app:U224.Casimir}) for these modules is given by
\[\label{eq:One.Form.Casimir}
\algJ^2_{12}\,\mdl_{j}=j(j+1)\,\mdl_j,
\]
just as if we were dealing with $\alSL(2)$ alone.
Due to invariance, $C^{\fldind{AB}}_{\fldind{CD}}$ must be 
of block-diagonal form: All states of a module $\mdl_j$ must be
mapped to the same type of module $\mdl_j$ with equal coefficients.
In our case, each module $\mdl_j$ appears only once in the tensor product,
therefore we can assign only one coefficient $C_j$ for each $\mdl_j$.
Let $(\fldproj_j)^{\fldind{AB}}_{\fldind{CD}}$ project 
two fields $\fldWf{A},\fldWf{B}$ to the module $\mdl_j$.
Then the most general invariant coefficients
can be written as
\[\label{eq:One.Form.IrredCoeff}
C^{\fldind{AB}}_{\fldind{CD}}=
\sum_{j=0}^\infty C_j\, (\fldproj_j)^{\fldind{AB}}_{\fldind{CD}}.
\]
Note that the decomposition \eqref{eq:One.Form.Tensor} is also 
valid for the group $\alPU(2,2|4)$. 
Therefore, the hypercharge $\algB$ is preserved by $\ham$. 
Obviously, also the length, measured by the operator $\len$, is conserved
\[\label{eq:One.Form.ConserveBL}
\comm{\ham}{\algB}=\comm{\ham}{\len}=0.
\]
This will clearly not be the case for higher-loop
corrections to the dilatation generator, 
which act on more than two fields at the same time.
At higher loops, 
the Konishi anomaly~\cite{Konishi:1984hf} 
mixes operators of different hypercharges. 
The same points also hold for the length $L$ of a state. 
Nevertheless, it makes perfect sense to speak of the 
leading order hypercharge and length to describe a state. 
Mixing with states of different hypercharges
or lengths is sub-leading, because the one-loop dilatation generator 
conserves these.

At this point we can also exclude 
the possibility of a `half-loop' 
contribution $\algD_1$ or a length non-preserving contribution to $\algD_2$ 
on algebraic grounds:
There is no overlap between the irreducible modules in the
in and out channels
\[\label{eq:One.Form.Overlap}
\mdlF^0\not\in\mdlF^3,\quad \mdlF\not\in \mdlF^2,
\qquad
\mdlF^0\not\in\mdlF^4,\quad \mdlF\not\in \mdlF^3,
\]
which can be seen by comparing the
scaling dimensions.
The only possible contributions up to second order in $g$ are
$\mdlF^2\mapsto \mdlF^2$ or $\mdlF\mapsto \mdlF$ as
assumed in \eqref{eq:One.Form.HDiagram}.

\subsection{Planar Limit}
\label{sec:One.Form.Planar}

We can now take the planar limit of \eqref{eq:One.Form.HForm}
\[\label{eq:One.Form.PlanarSym}
\ham=
\sum_{j=0}^\infty 2\,C_j\, (\fldproj_j)_{\fldind{CD}}^{\fldind{AB}}\,
\ITerm{\fldindn{C}\fldindn{D}}{\fldindn{A}\fldindn{B}}
\]
in the notation introduced in \secref{sec:Dila.Planar.Interact}.
In this chapter we will use a slightly different notation
which assumes that $\ham$ acts on a spin chain of length $L$
and transforms two adjacent fields%
\footnote{We assume cyclic site indices, i.e.~$\ham_{L,L+1}=\ham_{L,1}$.}
\[\label{eq:One.Form.HFormPlanar}
\ham=\sum_{p=1}^L \ham_{p,p+1},\qquad
\ham_{p,p+1}=\sum_{j=0}^\infty 2\,C_j \fldproj_{p,p+1,j}.\]
The symbol $\fldproj_{p,p+1,j}$ projects
the fields at positions $p,p+1$ to the module $\mdl_j$.
We see that all coefficients $C_j$ can be read off
from this Hamiltonian.
Therefore, the \emph{Hamiltonian density} $\ham_{12}$ generalises uniquely
to the non-planar Hamiltonian $\ham$ 
in \eqref{eq:One.Form.HForm}.
In what follows we can safely restrict 
ourselves to the investigation of $\ham_{12}$ alone.

To simplify some expressions, 
we introduce the $\alPSU(2,2|4)$ invariant 
total spin operator $\fldspin_{12}$ by the implicit definition
\[\label{eq:One.Form.SpinFunctionAct}
\fldspin_{12}\, \mdl_j= j\, \mdl_j.
\]
We can now define a function $f(\fldspin_{12})$ of this operator by 
\[\label{eq:One.Form.SpinFunction}
f(\fldspin_{12})=\sum_{j=0}^\infty
f(j) \,\fldproj_{12,j}.
\]
%
Using the short notation the Hamiltonian density becomes simply
\[\label{eq:One.Form.Short}
\ham_{12}=2\,C(\fldspin_{12}).
\]
%

\section{The Fermionic $\alSU(1,1)\times\alU(1|1)$ Subsector}
\label{sec:One.Magic}

It remains to determine the coefficients $C_j$. 
To accomplish this task we will consider the closed subsector $(1,3)$ 
of $\superN=4$ SYM (c.f.~\secref{sec:Dila.Sect.Combine})
and show how to derive the Hamiltonian 
from the algebraic constraints.

\subsection{Fields and States}
\label{sec:One.Magic.States}

The fields in this subsector consist only of 
the fermion $\Psi=\Psi_{42}$ with 
$K$ derivatives $\cder=\cder_{22}$ acting on it,
see \secref{sec:Dila.Sect.Combine}.
In the oscillator notation of \secref{sec:N4.Fund}
they can be written as
\[\label{eq:One.Magic.Fields}
\state{k}:=\frac{1}{(k+1)!}\,(\osca^\dagger_2\oscb^\dagger_2)^k\state{\Psi}
         =\frac{1}{(k+1)!}\,(\osca^\dagger_2\oscb^\dagger_2)^k\osca^\dagger_2 \oscd^\dagger_2\state{\fldZ}.
\]
States are constructed as tensor products of the fields
\[\label{eq:One.Magic.State}
\state{k_1,\ldots,k_L}
\]
with the cyclic identifications 
(the sign is due to statistics)
\[\label{eq:One.Magic.StateCycle}
\state{k_1,\ldots k_p,k_{p+1},\ldots,k_L}
=(-1)^{p(L-p)}\state{k_{p+1},\ldots,k_L,k_1,\ldots k_p}.
\]
The identifications exclude states of the form
\[\label{eq:One.Magic.NoState}
\state{k_1,\ldots, k_{L/2},k_1,\ldots ,k_{L/2}}=0.
\]

The weight of a state with a total number 
of $K$ excitations is given by
\[\label{eq:One.Magic.Weight}
w=\weight{3L/2+K;K+L,K;0,0,L;L/2,L}.
\]
This weight is beyond a unitarity bound of $\alPSU(2,2|4)$
and cannot be primary. 
The generic shift from the highest superconformal weight to
the highest weight within the subsector is given by
\[\label{eq:One.Magic.Shift}
\delta w=\delta w'\indup{i}+\delta w'\indup{ii}+
\weight{1;-2, 0;-1,0,1;1,0}
=
\weight{2;-1,+1; 0,0,2;1,0}.
\]
The shifts $\delta w'\indup{i,ii}$ take the weight beyond 
the unitarity bound and should be omitted for 
quarter-BPS multiplets.
The additional shift is related to 
the two additional conditions $n_{\oscc_2}=n_{\oscc_3}=0$ in the 
definition of the subsector.

\subsection{Symmetry}
\label{sec:One.Magic.Sym}

The subsector is invariant under an $\alSU(1,1)\times\alU(1|1)$ subalgebra
of the superconformal algebra. 
In the fully interacting theory, 
the $\alSU(1,1)$ algebra consists of the generators%
\footnote{The precise form of $\algJ_3$ can be obtained from the 
commutator of $\algP_{22}$ and $\algK^{22}$ in \appref{app:U224.Comm} noting that
we can set $\algL^2{}_2=\half \algD_0-\sfrac{1}{4}\len$ and
$\algLd^2{}_2=\half \algD-\sfrac{3}{4}\len$ in this sector.}
\[\label{eq:One.Magic.SU11Gen}
\algJ'_+(g)=\algK^{22}(g),\quad
\algJ'_-(g)=\algP_{22}(g),\quad
\algJ'_0(g)=\len-2\algD_0-\algdD(g).
\]
Note that the dilatation generator $\algD$ is part of the algebra.
At higher loops, one should keep in mind that only half of 
the anomalous piece appears. The $\alSU(1,1)$ algebra is
\[\label{eq:One.Magic.SU11Alg}
\comm{\algJ'_+(g)}{\algJ'_-(g)}=-\algJ'_0(g),\qquad 
\comm{\algJ'_0(g)}{\algJ'_\pm(g)}=\pm 2\algJ'_\pm(g).
\]
%

The $\alU(1|1)$ algebra is generated by 
\[\label{eq:One.Magic.U11Gen}
\len,\quad
\algQ'_-(g)=\algQd^1{}_4(g),\quad 
\algQ'_+(g)=\algSd^4{}_1(g),\quad 
\algdD(g)
\]
and the non-zero commutators are%
\footnote{Note that $\algL^1{}_1=-\half \algD_0+\sfrac{1}{4}\len$ and 
$\algR^1{}_1=-\sfrac{1}{4}\len$ in the sector.}
\[\label{eq:One.Magic.U11Alg}
\comm{\len}{\algQ'_{\pm}(g)}=\mp \algQ'_{\pm}(g),\qquad
\acomm{\algQ'_+(g)}{\algQ'_-(g)}=\half \algdD(g).
\]
The generators of $\alSU(1,1)$ and $\alU(1|1)$ commute with each other
\[\label{eq:One.Magic.SU11U11Alg}
\comm{\algJ'(g)}{\len}=\comm{\algJ'(g)}{\algQ'_\pm(g)}=\comm{\algJ'(g)}{\algdD(g)}=0.
\]

In the classical limit, the algebra $\alU(1|1)$ is trivial,
$\algQ'_{\pm}(0)=0$, it transforms between oscillators
of type $\oscb^\dagger_1$ and $\oscd^\dagger_1$,
both of which are absent in this subsector. 
In the interacting theory, however, the
generators $\algQ'_{\pm}(g)$ must receive 
non-trivial corrections 
(see also \figref{fig:N4.Split.Splitting}) because they close on $\algdD(g)$. 
In particular, they must produce $\algD_2$ which is 
possible only if $\algQ'_{\pm,1}\neq 0$.

Let us now restrict to the leading orders of all
generators as in \secref{sec:One.Form.Leading}
\[\label{eq:One.Magic.LeadingGen}
\algJ'_{\pm}:=\algJ'_{\pm ,0},\quad
\algJ'_{0}:=\algJ'_{0,0},\quad
\algQ'_{\pm}:=\algQ'_{\pm ,1},\quad
\ham':=\algD_2.
\]
The resulting non-trivial commutators of $\alU(1|1)$ are
\[\label{eq:One.Magic.LeadingAlg}
\comm{\len}{\algQ'_\pm}=\mp \algQ'_{\pm},\qquad
\acomm{\algQ'_+}{\algQ'_-}=\half \ham'.
\]
%

\subsection{Representations}
\label{sec:One.Magic.Reps}

The fields transform under $\alSU(1,1)$ as
\[\label{eq:One.Magic.FieldGen}
\algJ'_- \state{k}=(k+2)\state{k+1},
\quad
\algJ'_+ \state{k}=k\state{k-1},
\quad
\algJ'_0 \state{k}=-2(k+1)\state{k},
\]
as can be inferred from the oscillator representation.
All of the fields can be transformed into each other, they therefore
span an irreducible module $\mdlF^\prime$ of 
$\alSU(1,1)$. The Dynkin label of the highest weight $\state{0}$, 
measured by $\algJ'_0$, is 
\[\label{eq:One.Magic.FieldWeight}
w'\indups{F}=[-2],\]
in other words, the fields transform in the 
spin $-1$ irreducible representation.

The Hamiltonian density $\ham'_{12}$ is 
$\alSU(1,1)$ invariant and acts 
on two fields at a time.
Of particular interest is therefore the tensor 
product of two $\mdlF^\prime$'s, 
by standard $\alSL(2)$ rules it splits into
modules of spin $-1-j$, $j\geq 1$
\[\label{eq:One.Magic.Tensor}
\mdlF^\prime\times \mdlF^\prime=\sum_{j=1}^\infty \mdl'_j,
\qquad\mbox{with}\quad w'_j=[-2-2j].
\]
All irreducible modules
have multiplicity one and we can write the invariant Hamiltonian
as 
\[\label{eq:One.Magic.SpinFunction}
\ham'_{12}=C'(\fldspin'_{12}),
\]
where the total spin operator $\fldspin'_{12}$ is defined implicitly by
\[\label{eq:One.Magic.TotalSpin}
\fldspin'_{12} \,\mdl'_j = j\, \mdl'_j.
\]
%

\subsection{Supercharges}
\label{sec:One.Magic.Super}

In order to find the Hamiltonian $\ham'$, it suffices to 
find the supercharges $\algQ'_{\pm}$; via the supersymmetry
relation \eqref{eq:One.Magic.LeadingAlg} we can 
generate $\ham'$ later. The supercharges $\algQ'_{\pm}$ are of order $g$ 
in perturbation theory, therefore they should have three legs 
(c.f.~\secref{sec:Dila.Theory.Feyn}).
We already know from \eqref{eq:One.Magic.LeadingAlg} 
that $\algQ'_-$ increases the length
by one and $\algQ'_+$ decreases it. 
Consequently, we make the ansatz%
\footnote{Here, $\state{n}$ is considered to be a field 
within a state. A single field should be annihilated by $\algQ'_-$.}
\[\label{eq:One.Magic.QForm}
\algQ'_- \state{m} = \sum_{k=0}^{m-1} c^-_{m,k} \state{k,m-1-k}.
\]
The supercharge should commute with all generators $\algJ'$  
because they belong to distinct algebras. 
Therefore $\algQ'_-$ conserves the $\algJ'_0$ charge and 
$\algQ'_- \state{m}$ may only yield  
states of the form
$\state{k,m-1-k}$.
The commutator of $\algQ'_-$ with $\algJ'_-$ is
\[\label{eq:One.Magic.QComm}
\comm{\algQ'_-}{\algJ'_-}\state{m}=
\sum_{k=0}^m \bigbrk{c^-_{m+1,k}(n+2)
       -\delta_{k\neq 0} c^-_{m,k-1} (k+1)
       -\delta_{k\neq m}c^-_{m,k} (m-k+1) }\state{k,m-k}.
\]
The coefficients can be computed recursively
and one easily confirms that 
the only possibility to make $\comm{\algQ'_-}{\algJ'_-}$
vanish identically is $c^-_{m,k}=0$. For that purpose, start with 
$m=k=0$ and find that $c^-_{1,0}=0$; then continue with
$m=1$, $k=0,1$ and so on. In terms of representation theory this is understood
because $\mdlF'$ and $\mdlF'\times\mdlF'$ have no irreducible modules in 
common. This might seem to be disastrous for it leads to $\algQ'_-=\ham'=0$. 
However, we do not need to require that $\comm{\algQ'_-}{\algJ'_-}$
vanishes identically, but only that its action annihilates all states.
In particular, this allows $\comm{\algQ'_-}{\algJ'_-}$ to generate 
a gauge transformation which annihilates gauge invariant states.
The only suitable gauge transformation to match 
$\state{m}\mapsto \state{k,m-k}$ is 
$\cder^m\Psi\mapsto \acomm{\Psi}{\cder^m\Psi}$.
Therefore we should merely require
\[\label{eq:One.Magic.QCommGauge}
\comm{\algQ'_-}{\algJ'_-}\state{m}=c_- \state{m,0}+ c_-\state{0,m}.
\]
It is not difficult to see that there is a unique solution to
this equation, namely $c^-_{m,k}=c_-$.

By making a similar ansatz for $\algQ'_+$ we find in total
\<\label{eq:One.Magic.QRes}
\algQ'_- \state{m} \earel{\sim} \frac{1}{2}\sum_{k=0}^{m-1} \state{k,m-1-k},
\nln
\algQ'_+ \state{k,m-k} \earel{\sim} 
\lrbrk{\frac{1}{k+1}+\frac{1}{m-k+1}}\state{m+1}.
\>
Again, $\algQ'_+$ commutes with $\algJ'_+$ only up to a gauge transformation
\[\label{eq:One.Magic.QCommGauge2}
\comm{\algQ'_+}{\algJ'_+}\state{k,m-k}\sim (\delta_{k=0}+\delta_{k=m})\state{m}.
\]
The other commutators $\comm{\algQ'_+}{\algJ'_-}$ and $\comm{\algQ'_-}{\algJ'_+}$ 
turn out to vanish identically.

\subsection{The Hamiltonian}
\label{sec:One.Magic.Ham}

\begin{figure}
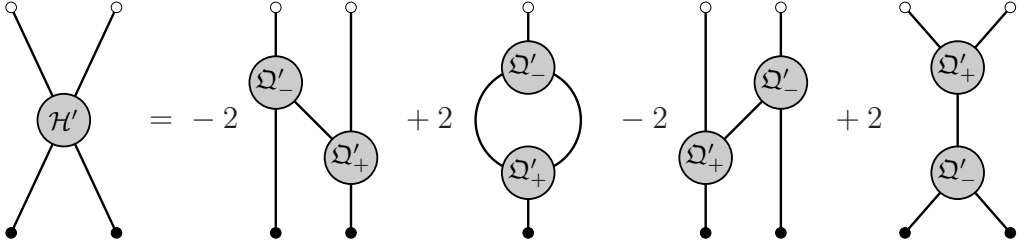
\centering
\parbox{2.0cm}{\centering\includegraphics{sec03.comm.e.eps}}
$=$
$\mathord{}-2$\parbox{2.0cm}{\centering\includegraphics{sec03.comm.a.eps}}
$\mathord{}+2$\parbox{2.0cm}{\centering\includegraphics{sec03.comm.c.eps}}
$\mathord{}-2$\parbox{2.0cm}{\centering\includegraphics{sec03.comm.b.eps}}
$\mathord{}+2$\parbox{2.0cm}{\centering\includegraphics{sec03.comm.d.eps}}
\caption{Diagrammatic representation of the commutator 
$\ham'=2\protect\acomm{\algQ'_+}{\algQ'_-}$. The diagrams are to be inserted 
to a state at the white dots.}
\label{fig:One.Magic.Diag}
\end{figure}

We are now ready to compute the Hamiltonian $\ham'=2\acomm{\algQ'_+}{\algQ'_-}$.
For definiteness, we will assume unit proportionality constants 
in \eqref{eq:One.Magic.QRes}.
In total there are four types of diagrams to represent the
anticommutator, see \figref{fig:One.Magic.Diag}.
To compute $\algQ'_+ \algQ'_-$ there are three different ways in which $\algQ'_+$
could act. Let us therefore add labels to indicate the fields
on which each generator acts.
Firstly, $\algQ'_+$ could act on one of the fields generated by $\algQ'_-$
and one adjacent field 
(the sign is due to statistics).
\<\label{eq:One.Magic.QQDiag1}
\algQ'_{+,23}\algQ'_{-,1}\state{k,m-k}
\eq-\frac{1}{2}\sum_{k'=0}^{k-1} \lrbrk{\frac{1}{k-k'}+\frac{1}{m-k+1}}\state{k',m-k'},
\nln
\algQ'_{+,12}\algQ'_{-,2}\state{k,m-k}
\eq-\frac{1}{2}\sum_{k'=k+1}^m \lrbrk{\frac{1}{k+1}+\frac{1}{k'-k}}\state{k',m-k'}.
\>
Secondly, it could act on both fields that result from 
the action of $\algQ'_-$ 
\[\label{eq:One.Magic.QQDiag2a}
\algQ'_{+,12} \algQ'_{-,1} \state{k,m-k}=\frac{1}{2}\sum_{k'=0}^{k-1}
\lrbrk{\frac{1}{k'+1}+\frac{1}{k-k'}}\state{k,m-k}
=h(k)\state{k,m-k},
\]
where the \emph{harmonic numbers} $h(m)$ are defined as
\[\label{eq:One.Magic.Harm}
h(m):=\sum_{k=1}^m\frac{1}{k}\,.
\]
We want all interactions to act on two adjacent sites,
therefore we should evenly distribute this contribution between both fields
\[\label{eq:One.Magic.QQDiag2}
\bigbrk{\half \algQ'_{+,12} \algQ'_{-,1}+\half \algQ'_{+,23} \algQ'_{-,2}} \state{k,m-k}
=\half \bigbrk{h(k)+h(m-k)}\state{k,m-k}.
\]
Finally, $Q'_- Q'_+$ is easily computed 
\[\label{eq:One.Magic.QQDiag3}
\algQ'_{1,-} \algQ'_{+,12} \state{k,m-k}=
\frac{1}{2}\lrbrk{\frac{1}{k+1}+\frac{1}{m-k+1}}
\sum_{k'=0}^{m}
\state{k',m-k'}.
\]
In total the action of the Hamiltonian density is 
\[\label{eq:One.Magic.HFromQQ}
\ham'_{12}=
2\algQ'_{+,23} \algQ'_{-,1}
+2\algQ'_{+,12}\algQ'_{-,2}
+2\algQ'_{1,-} \algQ'_{+,12}
+\algQ'_{+,12} \algQ'_{-,1}
+\algQ'_{+,23} \algQ'_{-,2};
\]
summing the terms in 
\eqref{eq:One.Magic.QQDiag1,eq:One.Magic.QQDiag2,eq:One.Magic.QQDiag3} we get 
\<\label{eq:One.Magic.Ham}
\ham'_{12}\state{k,m-k}\earel{\sim}
\sum_{k'=0}^{k-1} \lrbrk{\frac{1}{k+1}-\frac{1}{k-k'}}\state{k',m-k'}
\nl
+\bigbrk{h(k+1)+h(m-k+1)}\state{k,m-k}
\nl
+\sum_{k'=k+1}^{m} \lrbrk{\frac{1}{m-k+1}-\frac{1}{k'-k}}\state{k',m-k'}.
\>
The $\alSU(1,1)$ invariance of $\ham'_{12}$ is
inherited from $\algQ'_{\pm}$.

\subsection{Eigenvalues of the Hamiltonian}
\label{sec:One.Magic.Eigen}

In order to transform this expression 
into the concise form \eqref{eq:One.Form.Short}
we need to find the eigenvalue of the Hamiltonian on module $\mdl'_j$.
The highest weight state of $\mdl'_j$,
which is annihilated by $\algJ'_{1,+}+\algJ'_{2,+}$, is
\[\label{eq:One.Magic.HighestSpin}
\state{j}=\sum_{k=0}^{j-1}\frac{(-1)^k (j-1)!}{k!(j-k-1)!}\,\state{k,j-k-1}.\]
We know that $\state{j}$ is an eigenstate of $\ham'_{12}$ because 
$\ham'_{12}$ is invariant under $\alSU(1,1)$. Therefore we only need
to compute the coefficient of $\state{0,j-1}$ in 
$\ham'_{12}\state{j}$; we obtain
\[\label{eq:One.Magic.HOnSpin}
\ham'_{12}\state{j}\sim
\lrbrk{
1+h(j)+
\sum_{k=1}^{j-1}\frac{(-1)^{k} (j-1)!}{k!(j-k-1)!}
\lrbrk{\frac{1}{k+1}-\frac{1}{k}}}\state{0,j-1}
+\ldots\,.
\]
The first part of the sum is easily performed by extending the range and
thus completing the binomial $(1-1)^j$
\[\label{eq:One.Magic.Sum1}
\sum_{k=1}^{j-1}\frac{(-1)^{k} (j-1)!}{(k+1)!(j-k-1)!}=
\sum_{k=-1}^{j-1}\frac{(-1)^{k} (j-1)!}{(k+1)!(j-k-1)!}
-1+\frac{1}{j}
=-1+\frac{1}{j}\,.
\]
For the second part we replace $1/k$ by 
$1/(j-1)+(j-k-1)/k(j-1)$ and get 
\[\label{eq:One.Magic.Sum2}
-\sum_{k=1}^{j-1}\frac{(-1)^{k} (j-1)!}{k!(j-k-1)!k}
=
-\sum_{k=1}^{j-1}\frac{(-1)^{k} (j-2)!}{k!(j-k-1)!}
-\sum_{k=1}^{j-1}\frac{(-1)^{k} (j-2)!}{k!(j-k-2)!k}\,.
\]
The first of the resulting sums is done by completing $(1-1)^{j-1}$ and 
evaluates to $1/(j-1)$.
The second sum is the same as above with $j$ replaced by $j-1$. 
By induction we thus get
\[\label{eq:One.Magic.Sum}
-\sum_{k=1}^{j-1}\frac{(-1)^{k} (j-1)!}{k!(j-k-1)!k}
=\frac{1}{j-1}+\frac{1}{j-2}+\ldots+\frac{1}{1}=h(j-1).
\]
Summing up we obtain
\[\label{eq:One.Magic.Eigen}
\ham'_{12}\state{j}\sim 
\bigbrk{1+h(j)-1+1/j+h(j-1)}\state{0,j-1}+\ldots
=2h(j)\,\state{j}.
\]
where we have reconstructed all other terms of $\state{j}$ by means of invariance. 
We have therefore determined the Hamiltonian in this subsector
up to an overall constant
\[\label{eq:One.Magic.Short}
\ham'_{12}\sim 2h(\fldspin'_{12}).
\]
%

\section{The Lift to $\alPSU(2,2|4)$}
\label{sec:One.Lift}

Let us now compare the results of the preceding sections.
This allows us to derive the complete one-loop dilatation operator
of $\superN=4$ supersymmetric gauge theory.
The state $\state{j}$ has length $L=2$ and $K=j-1$ excitations.
According to \eqref{eq:One.Magic.Weight,eq:One.Magic.Shift}
the highest weight of the superconformal multiplet 
that includes $\state{j}$ is 
\[\label{eq:One.Lift.TwistTwoSub}
w'_j=\weight{j;j-2,j-2;0,0,0;0,2}.
\]
For $j=1$ the shift is exceptional,
it excludes $\delta w'\indup{i}+\delta w'\indup{ii}$
in \eqref{eq:One.Magic.Shift} because the multiplet is quarter-BPS.
The corresponding highest superconformal weight is
\[\label{eq:One.Lift.TwistTwo}
w'_1=\weight{2;0,0;1,0,1;0,2}.
\]
These are precisely the highest weights of the 
superconformal modules $\mdl_j$, $j\geq 1$ in \eqref{eq:One.Form.Tensor}.
There is a one-to-one correspondence between
the modules $\mdl_j$, $j\geq 1$ \eqref{eq:One.Form.Irreps} and $\mdl'_j$
\[\label{eq:One.Lift.SubMod}
\mdl'_j\subset \mdl_j, \quad j\geq 1.
\]
Using the fact that the two Hamiltonians must agree within 
the subsector we find
\[\label{eq:One.Lift.CoeffMatch}
2\,C_j \,\state{j}=\ham_{12}\, \state{j}=\ham'_{12}\,\state{j}
\sim 2h(j)\, \state{j}, \quad j\geq 1.
\]
This leaves one overall constant and the coefficient $C_0$ 
to be determined. 
The multiplet $\mdl_0$ is half-BPS and thus protected, it cannot
acquire an anomalous dimension; we must set
\[\label{eq:One.Lift.CoeffBPS}
C_0=h(0)=0.
\]
The overall constant cannot possibly be fixed by algebraic considerations,
we need to match it to a field theory calculation. 
For example, we can use the anomalous dimension of the Konishi operator, 
$D_2=6$ \eqref{eq:Dila.Dim.KonishiDim}, as input. 
In the free theory, it is part of the multiplet $\mdl_2$. 
The Hamiltonian acting on a state of length $L=2$
is $\ham=\ham_{12}+\ham_{21}$. For the Konishi state we 
therefore get $D_2=4C_2$
and set $C_2=\frac{3}{2}=h(2)$. 
The resulting `Hamiltonian density' $\ham_{12}$ for $\superN=4$ SYM is thus
\[\label{eq:One.Lift.Ham}
\ham_{12}=\sum_{j=0}^\infty
2h(j) \,\fldproj_{12,j}
=2h(\fldspin_{12}).
\]

To conclude, the one-loop dilatation generator of $\superN=4$ 
can be written as
\[\label{eq:One.Lift.Dil}
\algD(g)=\algD_0-\frac{\gym^2}{8\pi^2}\sum_{j=0}^\infty
h(j) \,(\fldproj_j)_{\fldind{CD}}^{\fldind{AB}} \,\,
\normord{\Tr \bigcomm{\fldWf{A}}{\fldWv{C}}\bigcomm{\fldWf{B}}{\fldWv{D}}}
+\order{g^3},
\]
where we have inserted the conventional Yang-Mills coupling constant 
$\gym^2=8\pi^2g^2/N$.
This is the principal result of this chapter.
The coefficients are the \emph{harmonic numbers} $h(j)$,
elements of the harmonic series
\[\label{eq:One.Lift.Harm}
\mbox{`\emph{harmonic numbers}'}:\quad
h(m):=\sum_{k=1}^m\frac{1}{k}=\Psifn(m+1)-\Psifn(1),
\]
which can also be expressed in terms of 
the digamma function $\Psifn(x)=\Gammafn'(x)/\Gammafn(x)$.
In \appref{app:Harm} we will present the 
\emph{harmonic action}, a practical implementation of
the action of $\ham_{12}=2h(\fldspin_{12})$.

\section{The Bosonic $\alSU(1,1)$ Subsector}
\label{sec:One.Baby}

In this section we will consider the closed subsector $(2,2)$;
it is a nice sector, quite similar to the fermionic subsector $(3,1)$, see
\secref{sec:One.Magic}, and we will make use of it later.  
The fields in this subsector consist only of 
the field $\fldZ=\Phi_{34}$ with 
$K$ derivatives $\cder=\cder_{22}$ acting on it.
They can be written as
\[\label{eq:One.Baby.Fields}
\state{k}:=\frac{1}{k!}\,(\osca^\dagger_2\oscb^\dagger_2)^k\state{\fldZ}
=\frac{1}{k!}\,\cder^k \fldZ.
\]
States are constructed as tensor products of the fields
modulo cyclicity of the trace
\[\label{eq:One.Baby.State}
\state{k_1,\ldots,k_L}=\state{k_{p+1},\ldots,k_L,k_1,\ldots k_p}.
\]

The weight of a state with 
a total number of $K$ excitations is given by 
\[\label{eq:One.Baby.Weight}
w=\weight{L+K;K,K;0,L,0;0,L}.
\]
The generic shift from the highest superconformal weight to
the highest weight within the subsector is given by
\[\label{eq:One.Baby.Shift}
\delta w=\delta w'\indup{i}+\delta w'\indup{ii}+
\weight{1;-1,-1;-1,+2,-1;0,0}
=
\weight{2;-2,-2; 0,-2,0;0,0}.
\]
The shifts $\delta w'\indup{i,ii}$ take the weight beyond 
the unitarity bound and should be omitted for 
quarter-BPS multiplets with $K=1$.
The additional shift is related to 
the two additional conditions $n_{\oscc_2}=n_{\oscd_2}=0$ in the 
definition of the subsector; 
it should be omitted for half-BPS multiplets with $K=0$.

The subsector is invariant under 
an $\alSU(1,1)\times\alU(1)\times\alU(1)$ subalgebra
of the superconformal algebra. 
The $\alSU(1,1)$ algebra of generators $\algJ''$ is exactly the same 
as in \eqref{eq:One.Magic.SU11Gen,eq:One.Magic.SU11Alg}.
The two $\alU(1)$ charges are the length $L$ 
and anomalous dimension $\delta D(g)$. 

The fields transform under $\alSU(1,1)$ as
\[\label{eq:One.Baby.FieldGen}
\algJ''_- \state{k}=(k+1)\state{k+1},
\quad
\algJ''_+ \state{k}=k\state{k-1},
\quad
\algJ''_0 \state{k}=-(2k+1)\state{k}.
\]
All fields can be transformed into each other, they therefore
span an irreducible module $\mdlF''$ of $\alSU(1,1)$. 
The Dynkin label of the highest weight $\state{0}$ is 
\[\label{eq:One.Baby.FieldWeight}
w''\indups{F}=[-1],\]
in other words, the fields transform in the 
spin $-\frac{1}{2}$ irreducible representation.

The tensor product of two $\mdlF''$ is as in 
\eqref{eq:One.Magic.Tensor}, but here also the module
$\mdl''_0$ with $j=0$ appears.
There is a one-to-one correspondence between the 
modules $\mdl''_j$ and the 
irreducible modules of the superconformal algebra
\[\label{eq:One.Baby.OneToOne}
\mdl''_j\subset \mdl_j.
\]

In \cite{Beisert:2003jj} the Hamiltonian density
was obtained from a direct field theory computation
\[\label{eq:One.Baby.Ham}
\ham''_{12}\, \state{m,m-k}=
\sum_{k'=0}^m
\lrbrk{\delta_{k=k'}\bigbrk{h(k)+h(m-k)}-\frac{\delta_{k\neq k'}}{|k-k'|}}
\state{k',m-k'},
\]
It is straightforward to verify
that $\ham''_{12}$ is invariant under the generators $\algJ''_{12}$.
As in \secref{sec:One.Magic.Eigen}, one can 
show that \eqref{eq:One.Baby.Ham} is equivalent to 
\[\label{eq:One.Baby.SpinHam}
\ham''_{12}=2h(\fldspin''_{12}).
\]
This result can be lifted to $\superN=4$ SYM as well to obtain the complete 
one-loop dilatation operator \cite{Beisert:2003jj}.

\section{Planar Spectrum}
\label{sec:One.Spec}

In this section we will apply the planar, one-loop 
dilatation generator (Hamiltonian)
to find some anomalous dimensions (energies).

\subsection{Lowest-Lying States}
\label{sec:One.Spec.Low}

\begin{table}\centering
$\begin{array}{|c|cccc|c|l|}\hline
D_0&\alSU(2)^2&\alSU(4)&B&L&\alPSU(2,2|4)&E^{P}\\\hline
2&[0,0]&[0,2,0]&0&2&[0;0;0,2,0;0;0]&0^+ \\
 &[0,0]&[0,0,0]&0&2&[0;1;0,0,0;1;0]&6^+ \\
\hline
3&[0,0]&[0,3,0]&0&3&[0;0;0,3,0;0;0]&0^- \\
 &[0,0]&[0,1,0]&0&3&[0;1;0,1,0;1;0]&4^- \\
\hline
4&[0,0]&[0,4,0]&0&4&[0;0;0,4,0;0;0]&0^+ \\
 &[0,0]&[0,2,0]&0&4&[0;1;0,2,0;1;0]&(10E-20)^+ \\
 &[0,0]&[1,0,1]&0&4&[0;1;1,0,1;1;0]&6^- \\
 &[0,0]&[0,0,0]&0&4&[0;2;0,0,0;2;0]&(13E-32)^+ \\
 &[2,0]&[0,0,0]&1&3&[2;3;0,0,0;2;0]&9^- \hfill+\mathord{\mbox{conj.}}\\
 &[1,1]&[0,1,0]&0&3&[1;2;0,1,0;2;1]&\frac{15}{2}^{\pm}\\
 &[2,2]&[0,0,0]&0&2&[2;3;0,0,0;3;2]&\frac{25}{3}^{+}\\
\hline
5&[0,0]&[0,5,0]&0&5&[0;0;0,5,0;0;0]&0^- \\
 &[0,0]&[0,3,0]&0&5&[0;1;0,3,0;1;0]&2^-,6^- \\
 &[0,0]&[1,1,1]&0&5&[0;1;1,1,1;1;0]&5^\pm \\
 &[0,0]&[0,0,2]&0&5&[0;2;0,0,2;1;0]&(14E-36)^+  \hfill+\mathord{\mbox{conj.}} \\
 &[2,0]&[0,0,2]&1&4&[2;3;0,0,2;1;0]&10^- \hfill+\mathord{\mbox{conj.}}\\
 &[0,0]&[0,1,0]&0&5&[0;2;0,1,0;2;0]&10^-,10^-,(10E-20)^- \\
 &[2,0]&[0,1,0]&1&4&[2;3;0,1,0;2;0]&(16E-62)^+  \hfill+\mathord{\mbox{conj.}} \\
 &[1,1]&[0,2,0]&0&4&[1;2;0,2,0;2;1]&6^\pm \\
 &[1,1]&[1,0,1]&0&4&[1;2;1,0,1;2;1]&5^\pm,10^\pm \\
 &[1,1]&[0,0,0]&0&4&[1;3;0,0,0;3;1]&9^\pm \\
 &[2,2]&[0,1,0]&0&3&[2;3;0,1,0;3;2]&6^-\\
\hline
5.5&[1,0]&[0,2,1]&\half&5&[1;2;0,2,1;1;0]&8^\pm  \hfill+\mathord{\mbox{conj.}}\\
   &[1,0]&[1,1,0]&\half&5&[1;2;1,1,0;2;0]&(16E-62)^\pm  \hfill+\mathord{\mbox{conj.}}\\  
   &[1,0]&[0,0,1]&\half&5&[1;3;0,0,1;2;0]&(\frac{35}{2}E-\frac{305}{4})^\pm  \hfill+\mathord{\mbox{conj.}}\\
   &[2,1]&[0,1,1]&\half&4&[2;3;0,1,1;2;1]&9^\pm  \hfill+\mathord{\mbox{conj.}}\\
   &[2,1]&[1,0,0]&\half&4&[2;3;1,0,0;3;1]&(\frac{37}{2}E-\frac{333}{4})^\pm  \hfill+\mathord{\mbox{conj.}}\\
   &[3,2]&[0,0,1]&\half&3&[3;4;0,0,1;3;2]&10^\pm  \hfill+\mathord{\mbox{conj.}}\\
\hline
\end{array}$
\caption{All one-loop planar anomalous dimensions of primary operators with $D_0\leq 5.5$.
The label $P$ refers to parity, $P=\pm$ indicates
a degenerate pair of states.
The label `+conj.' represents conjugate states with
$\alSU(2)^2,\alSU(4),\alPSU(2,2|4)$ labels reversed and opposite hypercharge $B$.}
\label{tab:One.Spec.Main}
\end{table}
In \tabref{tab:One.Spec.Main} we present the spectrum 
of lowest-lying states in $\superN=4$ SYM.
For a given highest weight of the classical algebra 
we write the anomalous dimensions 
along with the parity $P$.
The parity is defined such that 
for a $\grSO(N)$ or $\grSp(N)$ gauge group 
the states with negative parity are projected out.
Parity $P=\pm$ indicates a pair of states with opposite parity
and degenerate energies.
Furthermore, we have indicated states with conjugate representations
for which the order of $\alSU(2)^2$, $\alSU(4)$ and $\alPSU(2,2|4)$ labels
as well as the hypercharge $B$ are inverted.
Generically, the one-loop energy shifts are not fractional numbers
but solutions to some algebraic equations.
We refrain from solving
these numerically, but instead give the equations. 
In the table such states are represented by polynomials 
$X(E)$ of degree $k-1$.
The true energies $E$ are obtained as solutions to the equation
\[\label{eq:One.Spec.Poly}
E^k=X(E).
\]
For example, the quadratic polynomial for the state 
with weight $\weight{4;0,0;0,2,0;0,4}$ is $X(E)=10E-20$.
It translates to the energy (see also \cite{Bianchi:2002rw})
\[\label{eq:One.Spec.PolyEx}
E^2=10E-20,\qquad E=5\pm\sqrt{5}.\]

The table was computed as follows:
A \texttt{C++} programme was used to determine all
highest weight states up to and including 
classical dimension $5.5$ as well as their descendants. 
In analogy to the sieve of Eratostene
the algorithm \cite{Bianchi:2003wx} subsequently removes 
descendants from the set of all states. 
What remains, are the primary states.
Please refer to \cite{Beisert:2003te} for details of the implementation 
of the sieve algorithm.
For each multiplet we pick one state and compute the
total excitation numbers using 
\tabref{tab:U224.Osc.Numbers}. 
Here it is crucial to choose a descendant for 
which the mixing problem is minimised.
This reduces the size of the energy matrix to 
be computed and diagonalised. 
For these purposes, a good descendant usually has 
as few different types of oscillators as possible.

In a \texttt{Mathematica} programme all states with 
a given set of oscillator excitations were collected:
We spread the oscillators on the sites of the spin chain 
in all possible ways taking the central charge constraint at each site
into account. Identical states with respect to 
cyclicity of the trace are dropped.
In a second step, the harmonic action, c.f.~\appref{app:Harm}, 
was applied to all the states to
determine the matrix of anomalous dimensions.
For all the descendants which were 
removed in the sieve algorithm, we remove the corresponding energy eigenvalues.
The remaining eigenvalues are the one-loop planar anomalous dimensions 
of highest weight states.

To go to higher canonical dimensions involves obtaining
and diagonalising bigger and bigger matrices. 
One can reduce the complexity by going to certain subsectors. 
The smallest subsector is the $\alSU(2)$ subsector,
see \secref{sec:Dila.SU2}.
There are only two fields, $\fldZ,\phi$, which we might 
indicate in a planar notation as
\[\label{eq:One.Spec.SU2Fields}
\state{0}=\fldZ,\quad \state{1}=\phi.
\]
The Hamiltonian density acts on two adjacent fields,
within this sector it is \eqref{eq:Dila.SU2.DilaFirst}
\[\label{eq:One.Spec.SU2Ham}
\ham_{12}'''\, \state{k_1,k_2} = \state{k_1,k_2} - \state{k_2,k_1}.
\]
Here there are far less states and it is much easier to 
compute the energy matrix. In \tabref{tab:One.Spec.SU2} we show
a complete table of states and energies up to classical dimension
${D_0\leq 9}$. 
We have omitted the vacuum states with $K=0$;
there is one for each length $L$ and its energy vanishes.
The states and their energies can be 
obtained conveniently using a computer algebra system.
In \appref{app:SU2Tools} we present a couple of 
\texttt{Mathematica} functions to deal with 
the $\alSU(2)$ subsector.
Similarly, we can obtain the spectrum for the bosonic 
$\alSU(1,1)$ subsector.
The expression \eqref{eq:One.Baby.Ham}
can be used to calculate any
one-loop anomalous dimension within this subsector. 
We display our results in \tabref{tab:One.Spec.Baby}.

There are two points to be observed in the spectra in 
\tabref{tab:One.Spec.Main,tab:One.Spec.SU2,tab:One.Spec.Baby}.
Firstly, we note the appearance of
pairs of states with degenerate energy and opposite
parities $P=\pm$
\[\label{eq:One.Spec.Pair}
\mbox{`\emph{paired state}':}\quad E_+=E_-.
\]
This will be an important issue for integrability discussed in \chref{ch:Int}.
Secondly, we find some overlapping primaries in
\tabref{tab:One.Spec.SU2,tab:One.Spec.Baby},
clearly their energies do agree. 
What is more, we find that a couple
of energies repeatedly occur. 
These are for example, $6,10,5,9$,
but also $10E-20$ and $13E-32$.
As these states are primaries transforming in different representations,
they cannot be related by $\alPSU(2,2|4)$. 
Of course, these degeneracies could merely be a coincidence of small numbers.
Nevertheless the reappearance of e.g.~$13E-32$ is somewhat striking.
This could hint at yet another symmetry enhancement
of the planar one-loop Hamiltonian.
It might also turn out to be a consequence of integrability.
Furthermore, one might speculate that it is some
remnant of the broken higher spin symmetry of the free theory, 
see e.g.~\cite{Konshtein:1989yg,Sundborg:2000wp,Sezgin:2001zs,Mikhailov:2002bp}
and references in \cite{Bianchi:2003wx}.

\begin{table}\centering
$\begin{array}[t]{|cc|l|}\hline
L&K& E^{P} \\\hline
4&2&6^+ \\\hline
5&2&4^- \\\hline
6&2&(10E-20)^+ \\
 &3&6^- \\\hline
7&2&2^-, 6^- \\
 &3&5^\pm \\\hline
\end{array}
\qquad
\begin{array}[t]{|cc|l|}\hline
L&K& E^{P} \\\hline
8&2&(14E^2-56E+56)^+\\
 &3&4^\pm, 6^- \\
 &4&(20E^2-116E+200)^+ \\\hline
9&2&(8E-8)^-,4^-\\
 &3&(17E^2-90E+147)^\pm \\
 &4&5^\pm,(12E-24)^- \\\hline
\end{array}$
\caption{
The lowest-lying states within the $\alSU(2)$ subsector \protect\cite{Beisert:2003tq}.
The weights of the corresponding primaries are 
$\protect\weight{L-2;0,0;K-2,L-2K,K-2;0,L-2}$.}
\label{tab:One.Spec.SU2}
\end{table}
\begin{table}\centering
$\begin{array}[t]{|c|cc|l|}\hline
D_0&L&K& E^{P}\\\hline
4&2&2&6^+ \\\hline
5&3&2&4^- \\\hline
6&4&2&(10E-20)^+ \\
 &3&3&\frac{15}{2}^\pm \\
 &2&4&\frac{25}{3}^+ \\\hline
7&5&2&2^-, 6^- \\
 &4&3&6^\pm \\
 &3&4&6^- \\\hline
\end{array}
\qquad
\begin{array}[t]{|c|cc|l|}\hline
D_0&L&K& E^{P}\\\hline
8&6&2&(14E^2-56E+56)^+\\
 &5&3&(\frac{25}{2}E-\frac{147}{4})^\pm  \\
 &4&4&\frac{23}{3}^\pm, (\frac{73}{3}E^2-\frac{553}{3}E+\frac{1274}{3})^+ \\
 &3&5&\frac{35}{4}^\pm \\
 &2&6&\frac{49}{5}^+ \\\hline
9&7&2&(8E-8)^-,4^-\\
 &6&3&(19E^2-\frac{459}{4}E+216)^\pm \\
 &5&4&(13E-32)^-, (\frac{97}{6}E-\frac{2291}{36})^\pm \\
 &4&5&(\frac{35}{2}E-\frac{665}{9})^\pm \\
 &3&6&\frac{22}{3}^-,\frac{227}{20}^\pm \\\hline
\end{array}$
\caption{
The first few states within the bosonic $\alSU(1,1)$ subsector
\protect\cite{Beisert:2003jj}.
The weights of the corresponding primaries are 
$\protect\weight{L+K-2;K-2,K-2;0,L-2,0;0,L}$.}
\label{tab:One.Spec.Baby}
\end{table}

\subsection{Two Partons}
\label{sec:One.Spec.TwistTwo}

A straightforward exercise is to determine the spectrum of
states of length $L=2$. These so-called twist-two states 
can conveniently be written as
\[\label{eq:One.Spec.TwistTwoOp}
\Op_{j,\fldind{AB}}=(\fldproj_j)^{\fldind{CD}}_{\fldind{AB}} 
\Tr \fldWf{C}\fldWf{D}.
\]
Note that $j$ must be even due to cyclicity of the trace.
Using \eqref{eq:One.Lift.Ham} we find%
\footnote{Note that $\ham=\ham_{12}+\ham_{21}=2\ham_{12}=4h(\fldspin_{12})$.}
\[\label{eq:One.Spec.TwistTwoEng}
E=4h(j),\qquad \delta D=\frac{\gym^2 N}{2\pi^2}\,h(j)+\order{g^3}
\]
in agreement with the results of \cite{Kotikov:2000pm,Dolan:2001tt}.
Twist-two states have positive parity.

\subsection{Three Partons}
\label{sec:One.Spec.LengthThree}

For states of length $L=3$ the following multiplets are found
within a trace \cite{Beisert:2004di}
\[\label{eq:One.Spec.TensorThree}
\Tr \mdlF\times\mdlF\times\mdlF =
\sum_{m=-\infty}^\infty\sum_{n=0}^\infty
\bigbrk{\mdl_{2m,2n}^- +\mdl_{2m+1,2n}^+  + c_n \mdl_{m,n+3}^\pm } ,
\]
where $c_{0,1,2,3,4,5}=(1,0,1,1,1,1)$ and $c_{n+6}=c_n+1$.
The modules $\mdl_{m,n}$ have highest weights%
\<\label{eq:One.Spec.WeightThree}
w_{0,0}\eq \weight{3;0,0;0,3,0;0,3},
\nln
w_{0,n}\eq \weight{n+1;n-2,n-2;0,1,0;0,3},
\nln
w_{1,0}\eq \weight{3;0,0;0,0,0;0,3},
\nln
w_{1,n}\eq \weight{n+5/2;n,n-1;0,0,1;1/2,3},
\nln
w_{m,n}\eq \weight{n+2m;n+2m-2,n+m-2;0,0,0;1,3}
\>
and the conjugate $w_{-m,n}$ has
reversed $\alSU(2)^2,\alSU(4)$ labels and opposite hypercharge.
The multiplets $\mdl_{0,n}$ have components in the
subsector $(2,2)$, see \eqref{sec:One.Baby},
the multiplets $\mdl_{1,n}$ have components
in the fermionic subsector $(3,1)$, see \eqref{sec:One.Magic}, and
all the other $\mdl_{m,n}$ are represented in the sector $(4,0)$.

By inspecting the spectrum of lowest-lying states 
and their energies, we find that almost all
of them form pairs with degenerate energies. 
We list the pairs in \tabref{tab:One.Spec.Length3Pair}.%
\footnote{The energies are all rational numbers because 
there is always just a single pair up to $n\leq 8$ \eqref{eq:One.Spec.TensorThree}. 
Starting from $n=9$ there is more than one pair and the energies
become irrational.}
Concerning the unpaired states, 
there is one for each even $n$, it has parity $(-1)^{m+1}$.
For the unpaired states one can observe a pattern
in the table of energies, \tabref{tab:One.Spec.Length3Single}.
We find that all energies agree with the formula
\[\label{tab:One.Spec.Length3SingleDim}
E=2 h(\half m-\half)
+2 h(m+\half n)
+2 h(\half m+\half n)
-2 h(-\half).
\]
In particular, for $m=1$ the energies are
\[\label{tab:One.Spec.Length3TwistTwo}
E=2h(1+\half n)
+2h(\half+\half n)
-2h(-\half)
=4h(n+2),\]
which agrees precisely with the energy \eqref{eq:One.Spec.TwistTwoEng} of 
the short twist-two multiplet $\mdl_{2n+2}$.
Superconformal invariance requires this degeneracy so that
the short multiplets can join to form a long multiplet.
The cases $m=0$ and $n=0$ also seem interesting, we find $E=4 h(\half n)$
and $E=6 h(m)$.
\begin{table}\centering
$\begin{array}[t]{|c|ccccccc|}\hline
n\backslash m &0&1&2&3&4&5&6
\\\hline
3\vphantom{\frac{\hat1}{1}}&\frac{15}{2} & 10 & \frac{47}{4} & \frac{131}{10} & \frac{71}{5} & \frac{1059}{70} & \frac{4461}{280} \\
5\vphantom{\frac{\hat1}{1}}&\frac{35}{4} & \frac{133}{12} & \frac{761}{60} & \frac{487}{35} & \frac{12533}{840} & \frac{39749}{2520} & \frac{13873}{840} \\
6\vphantom{\frac{\hat1}{1}}&\frac{227}{20} & \frac{761}{60} & \frac{967}{70} & \frac{2069}{140} & \frac{39349}{2520} & \frac{2747}{168} & \frac{3929}{231} \\ 
7\vphantom{\frac{\hat1}{1}}&\frac{581}{60} & \frac{179}{15} & \frac{3763}{280} & \frac{18383}{1260} & \frac{39133}{2520} & \frac{7543}{462} & \frac{94373}{5544} \\
8\vphantom{\frac{\hat1}{g}}&\frac{5087}{420} & \frac{1403}{105} & \frac{18187}{1260} & \frac{38677}{2520} & \frac{49711}{3080} & \frac{2593}{154} & \frac{629227}{36036} \\
\hline
\end{array}$
\caption{First few paired anomalous dimensions for $\mdl_{m,n}$ \protect\cite{Beisert:2004di}.}
\label{tab:One.Spec.Length3Pair}
\end{table}
\begin{table}\centering
$\begin{array}[t]{|c|ccccccc|}\hline 
n\backslash m &0&1&2&3&4&5&6
\\\hline
0\vphantom{\frac{\hat1}{1}}&            0 & 6 & 9 & 11 & \frac{25}{2} & \frac{137}{10} & \frac{147}{10} \\
2\vphantom{\frac{\hat1}{1}}&            4 & \frac{25}{3} & \frac{32}{3} & \frac{123}{10} & \frac{407}{30} & \frac{3067}{210} & \frac{542}{35} \\
4\vphantom{\frac{\hat1}{1}}&            6 & \frac{49}{5} & \frac{71}{6} & \frac{929}{70} & \frac{72}{5} & \frac{9661}{630} & \frac{2259}{140} \\ 
6\vphantom{\frac{\hat1}{1}}& \frac{22}{3} & \frac{761}{70} & \frac{191}{15} & \frac{8851}{630} & \frac{528}{35} & \frac{221047}{13860} & \frac{21031}{1260} \\
8\vphantom{\frac{\hat1}{1}}& \frac{25}{3} & \frac{7381}{630} & \frac{202}{15} & \frac{101861}{6930} & \frac{6581}{420} & \frac{329899}{20020} & \frac{21643}{1260} \\
10\vphantom{\frac{\hat1}{g}}&\frac{137}{15} & \frac{86021}{6930} & \frac{493}{35} & \frac{2748871}{180180} & \frac{20383}{1260} & \frac{203545}{12012} & \frac{122029}{6930} \\
\hline
\end{array}$
\caption{First few unpaired anomalous dimensions for $\mdl_{m,n}$ \protect\cite{Beisert:2004di}. 
The parity is $P=(-1)^{m+1}$.}
\label{tab:One.Spec.Length3Single}
\end{table}
%

\subsection{Two Excitations}
\label{sec:One.Spec.TwoEx}

Instead of considering a fixed number of fields,
one can also consider the $L$-particle vacuum state 
$\state{\fldZ,L}$ and add a small number of excitations,
see \secref{sec:Dila.Sect.Excite}.
A state without excitations is just the half-BPS vacuum and 
a state with one excitation is related to the vacuum by one 
of the lowering operators. 
The first interesting case is two excitations \cite{Berenstein:2002jq}.
Assume we consider four oscillator excitations 
of type $\osca_2^\dagger,\osca_2^\dagger,\oscb_2^\dagger,\oscb_2^\dagger$. 
This corresponds to a state of the bosonic $\alSU(1,1)$ sector with 
a total of two derivatives $\cder$ acting on $L$ fields $\fldZ$. 
A useful basis of states is thus 
\[\label{eq:One.Spec.TwoBasis}
\OpE^L_1= \Tr \cder \cder \fldZ\,\fldZ^{L-1},\qquad
\OpE^L_p= \Tr \cder \fldZ\,\fldZ^{p-2}\,\cder \fldZ\, \fldZ^{L-p}.
\]
Note that we should identify $\OpE^L_p$ and $\OpE^L_{L+2-p}$ 
due to cyclicity of the trace and 
consider a matrix with half the number of rows and columns.
Equivalently, we may choose to restrict to vectors which are
symmetric under $p\leftrightarrow L+2-p$.
Using the Hamiltonian \eqref{eq:One.Baby.Ham},
we find the matrix of anomalous dimensions
in this basis
\[\label{eq:One.Spec.TwoHam}
H=\left(\begin{array}{c|ccccc}
+2&-1&  &  &  &-1\\\hline
-2&+4&-2&  &  &  \\
  &-2&\ddots&\ddots& &  \\
  &  &\ddots&\ddots&-2&\\
  &  &  &-2&+4&-2\\
-2&  &  &  &-2&+4 \\
\end{array}\right).
\]

The bulk of the matrix has precisely the form of a second 
lattice derivative. The appropriate ansatz to diagonalise it,
is a vector with elements $\cos(ap+b)$. 
The boundary contributions together with the symmetry determine
the constants $a$ and $b$.
The matrix \eqref{eq:One.Spec.TwoHam}
has the following exact eigenvectors  
\[\label{eq:One.Spec.TwoEigen}
\Op^{L}_n=\frac{1}{L}\sum_{p=1}^{L}
\cos\lrbrk{\frac{\pi n(2p-1)}{L+1}} \OpE^{L}_p.
\]
Note that $\Op^{L}_{n}=\Op^{L}_{-n}=-\Op^{L}_{L+1-n}$. 
Thus the set of independent modes 
is given by the mode numbers $0\leq n< (L+1)/2$.
The corresponding exact planar one-loop anomalous dimension is 
\cite{Beisert:2002tn}
\[\label{eq:One.Spec.TwoEnergy}
E^{L}_n=8\sin^2\frac{\pi n}{L+1},\qquad
\delta D=\frac{\gym^2 N}{\pi^2}\,\sin^2\frac{\pi n}{L+1}+\order{g^3}.
\]

This is just one component of a multiplet 
of the residual symmetry $\alPSU(2|2)\times \alPSU(2|2)$
within the sector. The oscillators $\oscA^\dagger=(\osca^\dagger,\oscc^\dagger)$ 
(c.f.~\secref{sec:Dila.Sect.Excite}) transform in
the fundamental representation $[0;0;1]$ of one of
the $\alPSU(2|2)$'s. 
For two excitations we should consider the tensor
product of two fundamental modules which is
$[0;0;1]\times [0;0;1]=[0;0;2]_++[0;1;0]_-$. 
These two correspond to the symmetric and antisymmetric 
combination of two indices $A,B$. The same applies to
the oscillators $\dot\oscA^\dagger=(\oscb^\dagger,\oscd^\dagger)$.
In total we find four multiplets corresponding to 
the combined symmetrisations $++,+-,-+,--$:
\<\label{eq:One.Spec.TwoMulti}
\Op^L_{n,\{AB]\{\dot C\dot D]}\eq
\sum_{p=1}^L \cos\biggbrk{\frac{\pi n(2p-1)}{L+1}}\, \Tr 
\oscA_{1,\{A}^\dagger\oscA_{p,B]}^\dagger\, 
\dot\oscA_{1,\{\dot C}^\dagger\dot\oscA_{p,\dot D]}^\dagger\,
\state{\fldZ,L},
\nln
\Op^{L+1}_{n,\{AB][\dot C\dot D\}}\eq
\sum_{p=2}^{L+1} \sin\biggbrk{\frac{\pi n(2p-2)}{L+1}}\, \Tr 
\oscA_{1,\{A}^\dagger\oscA_{p,B]}^\dagger\, 
\dot\oscA_{1,[\dot C}^\dagger\dot\oscA_{p,\dot D\}}^\dagger\,
\state{\fldZ,L+1},
\nln
\Op^{L+1}_{n,[AB\}[\dot C\dot D]}\eq
\sum_{p=2}^{L+1} \sin\biggbrk{\frac{\pi n(2p-2)}{L+1}}\, \Tr 
\oscA_{1,[A}^\dagger\oscA_{p,B\}}^\dagger\, 
\dot\oscA_{1,\{\dot C}^\dagger\dot\oscA_{p,\dot D]}^\dagger\,
\state{\fldZ,L+1},
\nln
\Op^{L+2}_{n,[AB\}[\dot C\dot D\}}\eq
\sum_{p=2}^{L+2} \cos\biggbrk{\frac{\pi n(2p-3)}{L+1}}\, \Tr 
\oscA_{1,[A}^\dagger\oscA_{p,B\}}^\dagger\, 
\dot\oscA_{1,[\dot C}^\dagger\dot\oscA_{p,\dot D\}}^\dagger\,
\state{\fldZ,L+2}.
\>
All of these have the same energy $E_n^L$.
This is related to the fact that all short multiplets
join to form a long multiplet in the interacting theory 
\cite{Beisert:2002tn} (unless $n=0$).
For $\alPSU(2|2)$ the interacting multiplet
${[0;1+\half \delta D;0]}$ is at the unitarity bound $r_1\approx s_1+1$.
When $\delta D$ approaches zero, the interacting multiplet
splits up into a short and a BPS multiplet
${[0;1+\half \delta D;0]}\to {[0;1;0]}+{[0;0;2]}$.
In total, the highest weight of the long multiplet is
therefore
\[\label{eq:One.Spec.TwoLabels}
[0;1+\half \delta D;0]\times [0;1+\half \delta D;0].
\]
In $\alPSU(2,2|4)$ the classical highest weight for
states $n\neq 0$ is 
$\weight{L;0,0;0,L-2,0;0,0}={[0;1;0,L-2,0;1;0]}$.
The protected states for $n=0$ are part of the half-BPS multiplet
$\weight{L;0,0;0,L,0;0,0}$.
The states have parity $(-1)^L$.

\subsection{Three Excitations}
\label{sec:One.Spec.ThreeEx}

Let us investigate the states with three excitations. 
We find that such states almost always form pairs
with degenerate planar energies. The only exceptions from this rule
are states with weight
\[\label{eq:One.Spec.ThreeWeight}
\weight{2m+4;0,0;1,2m,1;0,2m+4}
=[0;1;1,2m,1;1;0],
\quad m\geq0.
\]
They have a descendant in the $\alSU(2)$ subsector which is given by 
\[\label{eq:One.Spec.ThreeOp}
\Op=\sum_{k=1}^{2m+2}(-1)^k\Tr \phi\phi \fldZ^k\phi \fldZ^{2m+3-k},\qquad E=6,
\qquad P=-1.\]
Interestingly, two of the excitations are always adjacent
in this leading order approximation.
Further states of this kind with more excitations can be found.

\section{Plane Wave Physics}
\label{sec:One.BMN}

In this section we would like to demonstrate the use of
the dilatation operator to find 
\emph{non-planar corrections}, 
i.e.~corrections in $1/N$, to the scaling dimensions.
Here the dilatation operator brings about a major simplification \cite{Beisert:2002ff}
as opposed to the computation of correlation functions
\cite{Kristjansen:2002bb,Constable:2002hw,Chu:2002pd,Beisert:2002bb,Constable:2002vq,Gross:2002mh}
because it allows to derive scaling dimensions
independently of two-point normalisation constants.
In particular we will derive the genus-one correction
to the scaling dimension of two-excitation BMN operators
in the BMN limit.

\subsection{The BMN Limit}
\label{sec:One.BMN.BMNLimit}

Berenstein, Maldacena and Nastase (BMN) \cite{Berenstein:2002jq} 
suggested to investigate operators of a large dimension $D_0$
and a nearly equally large charge $J$ of $\alSU(4)$
\[\label{eq:One.BMN.JCharge}
J=p-\half q_1-\half q_2.\]
Then the relevant states constitute long strings of 
$\fldZ$-fields with $D_0-J$ excitations or \emph{impurities} scattered in%
\footnote{The excitation subsectors in \secref{sec:Dila.Sect.Excite} 
were constructed to describe states of this kind.}
\[\label{eq:One.BMN.Op}
\Tr \fldZ\ldots \fldZ\, \phi \,
    \fldZ\ldots \fldZ\,\cder \fldZ\, 
    \fldZ\ldots \fldZ\, \psi\, 
    \fldZ\ldots \fldZ,
\]
which became known as BMN operators.
As particular examples, BMN investigated operators 
with zero, one and two excitations of scalar type.
The operators with less than two excitations belong to
half-BPS multiplets and are thus protected. 
Starting with two excitations there are states 
whose scaling dimension changes in the quantum theory.
For large $J$ one finds that the smallest 
one-loop planar anomalous dimensions 
scale as $1/J^2$, 
\[\label{eq:One.BMN.EnergyScale}
E=\order{1/J^2}.
\]
as confirmed by the exact values in the case of two excitations
\eqref{eq:One.Spec.TwoEnergy}.
BMN proposed to absorb the dependence on $J$ into an 
effective coupling constant $\lambda'$ 
\[\label{eq:One.BMN.LambdaPrime}
\lambda':=\frac{\lambda}{J^2}=\frac{\gym^2 N}{J^2}\,,
\]
for our purposes it seems convenient to use the combination $\hat g$
\[\label{eq:One.BMN.g1}
\hat g:=\frac{g}{J}\,,\qquad \lambda'=8\pi^2\hat g^2.
\]
BMN conjectured that this would lead to 
finite planar eigenvalues for the 
BMN energy operator $\algD-J$ (as a function of $\hat g$)
in the large $J$ limit, even beyond one-loop.

Moreover, it was found \cite{Kristjansen:2002bb,Constable:2002hw} 
that also the genus counting parameter $1/N$ can be renormalised
in such a way as to obtain finite results 
for non-planar correlators
\[\label{eq:One.BMN.g2}
\hat g\indup{s}=g_2:=\frac{J^2}{N}\,.
\]
The non-planar BMN limit can be defined as the \emph{double-scaling limit}
\[\label{eq:One.BMN.Limit}
\mbox{`\emph{BMN limit}':}\qquad
J,N,g,\lambda\rightarrow \infty\quad\mbox{with}\quad \hat g,\hat g\indup{s}\quad\mbox{fixed}.
\]

The physical significance of the above lies in 
the \emph{BMN correspondence}, which is a limit
of the celebrated AdS/CFT correspondence.
The statement of the correspondence is that BMN operators 
are dual to states of string theory on the \emph{plane-wave} background.
The scaling dimensions of BMN operators minus their charge $J$ 
should match the light-cone energies of the corresponding string states.
In an operatorial form, the correspondence can be written as
\[\label{eq:One.BMN.Corres}
\mbox{`\emph{BMN Correspondence}':}\qquad
H\indups{LC}=\algD-J+\order{1/J}.
\]
The planar limit corresponds to a non-interacting string theory
and it is fairly easy to derive the light-cone energy eigenvalues
\cite{Berenstein:2002jq}
\[\label{eq:One.BMN.Energy}
E\indups{LC}=\sum_{k=1}^M \sqrt{1+\lambda'\,n_k^2}
=\sum_{k=1}^M \sqrt{1+2\hat g^2(2\pi n_k)^2}\,.
\]
The numbers $n_k$ are the mode numbers (positive, negative or zero) 
of $M$ string oscillator excitations
and are subject to the level matching constraint $\sum_{k=1}^M n_k=0$.
There are some indications 
that this all-loop prediction for gauge theory might indeed be true
\cite{Gross:2002su,Santambrogio:2002sb}.

There exists an exceedingly large literature on 
the BMN correspondence, see 
\cite{Pankiewicz:2003pg,Plefka:2003nb,Kristjansen:2003uy,Sadri:2003pr,Russo:2004kr}
for reviews.

\subsection{Basis of States}
\label{sec:One.BMN.Basis}

Starting from here, we will only consider operators with two excitations. 
As shown in \secref{sec:One.Spec.TwoEx}, it makes perfect sense
to consider these operators also for arbitrary finite values of $J$. 
All two-excitation states form a single multiplet of
superconformal symmetry, so we are free to choose a particular descendant
to be used in our investigation.
In particular there is one descendant in the $\alSU(2)$ sector
that can be written in terms of $J$ fields $\fldZ$ and two 
excitations of type $\phi$.
Generic multi-trace operators with two excitations can have
both excitations in one trace
\[\label{eq:One.BMN.Multi}
\OpE_p^{J_0;J_1,\ldots,J_K}= \Tr \phi \fldZ^p \phi \fldZ^{J_0-p}\,
\prod_{k=1}^K \Tr \fldZ^{J_k},
\]
or the two excitations separated in different traces
\[\label{eq:One.BMN.MultiBPS}
\OpQ^{J_0,J_1;J_2,\ldots,J_K}=
\Tr\phi \fldZ^{J_0}\,\Tr \phi \fldZ^{J_1}
\prod_{k=2}^K \Tr \fldZ^{J_k},
\]
with $\sum_{k=0}^K J_k=J$.
Both series of operators are symmetric under the interchange of 
sizes $J_k$ of traces $\Tr \fldZ^{J_k}$, $\OpE$ is symmetric under
$p\to J_0-p$ and $\OpQ$ is symmetric under $J_0\leftrightarrow J_1$.

\subsection{The Action of the Dilatation Generator}
\label{sec:One.BMN.Dila}

The non-planar dilatation generator \eqref{eq:Dila.SU2.Dila} 
\[\label{eq:One.BMN.Dila}
\ham=\algD_{2}=-N^{-1}\normord{\Tr\comm{\fldZ}{\phi}\comm{\check \fldZ}{\check\phi}}
\]
can be seen to act as 
\[\label{eq:One.BMN.DilaAct}
\ham \matr{cc}{\OpE_p&\OpQ}=
\matr{cc}{\OpE_p&\OpQ}
\matr{cc}{\ast&\ast\\0&0},
\]
i.e.~operators of type $\OpQ$ are never produced. 
This follows from the fact that all produced 
objects will contain a commutator $\comm{\fldZ}{\phi}$ 
in some trace and this trace
will vanish unless it contains another $\phi$.
It immediately follows that for every $\OpQ$ there is 
one protected quarter-BPS operator.
Its leading part is given by $\OpQ$ itself, plus a $1/N$ 
correction from the operators $\OpE_p$ 
\cite{Ryzhov:2001bp,D'Hoker:2003vf,Beisert:2002bb,Constable:2002vq}.%
\footnote{We note that all quarter-BPS states 
in \cite{Ryzhov:2001bp,D'Hoker:2003vf} are annihilated by 
the operator $\comm{\check \fldZ}{\check\phi}$ which
is part of $D_2$. It is also part of a
superboost which relates would-be quarter-BPS states
with their partners in a long multiplet. 
For true quarter-BPS states, this must not happen and 
$\comm{\check \fldZ}{\check\phi}$ annihilates them.}
On the other hand, the operators $\OpE_p$ are in general not protected
and we will investigate their spectrum of anomalous dimensions in what follows. 
From the form of the dilatation matrix we 
infer that operators of type $\OpE_p$ 
do not receive corrections from operators of type $\OpQ$; 
the latter therefore completely decouple
as far as the consideration of the $\OpE_p$'s is concerned.

\begin{figure}
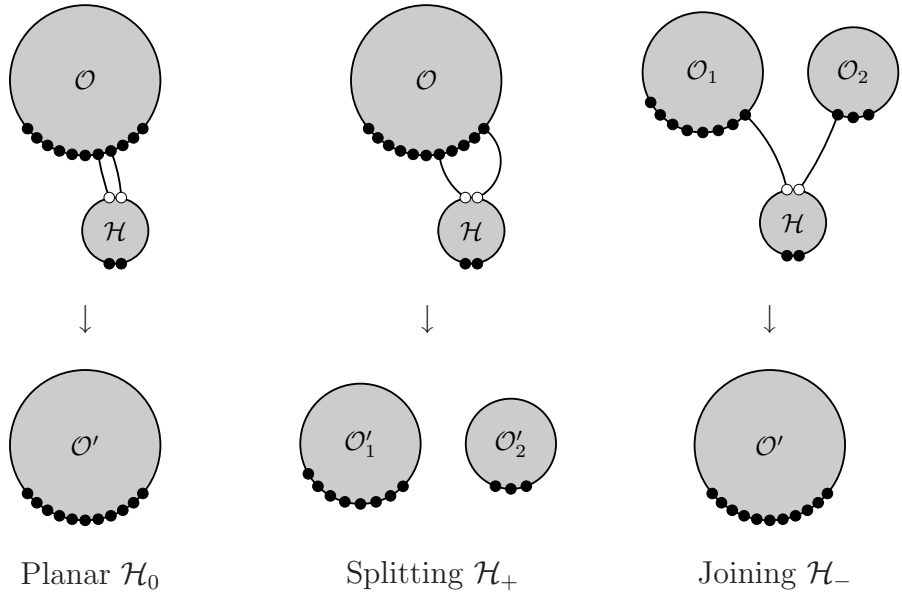
\centering
\parbox{4cm}{\centering\includegraphics{sec03.glue.in.planar.eps}}
\quad
\parbox{4cm}{\centering\includegraphics{sec03.glue.in.split.eps}}
\quad
\parbox{4cm}{\centering\includegraphics{sec03.glue.in.join.eps}}
\bigskip\par
\parbox{4cm}{\centering$\downarrow$}
\quad
\parbox{4cm}{\centering$\downarrow$}
\quad
\parbox{4cm}{\centering$\downarrow$}
\bigskip\par
\parbox{4cm}{\centering\includegraphics{sec03.glue.out.planar.eps}}
\quad
\parbox{4cm}{\centering\includegraphics{sec03.glue.out.split.eps}}
\quad
\parbox{4cm}{\centering\includegraphics{sec03.glue.out.join.eps}}
\bigskip\par
\parbox{4cm}{\centering Planar $\ham_0$}
\quad
\parbox{4cm}{\centering Splitting  $\ham_+$}
\quad
\parbox{4cm}{\centering Joining  $\ham_-$}
\caption{Topological structures of the action of the one-loop dilatation operator.}
\label{fig:One.BMN.Diagrams}
\end{figure}

It is easy to write down the exact expression for $\ham\OpE_p$. 
Let us define
\[\label{eq:One.BMN.NonPlanar}
\ham=\ham_0+N^{-1}\ham_+ + N^{-1}\ham_-,
\]
where $\ham_0$ is trace conserving and $\ham_{+}$ and $\ham_{-}$
respectively increases and decreases the number of traces by one. 
These three different contributions arise from three different 
contractions of the variations in the dilatation generator
with the fields in the states, see \figref{fig:One.BMN.Diagrams}.
Contractions to adjacent fields
within a trace lead to planar contributions. 
Contractions to non-adjacent fields will split up the trace. 
Contractions to different traces will join them. 
We find
\<\label{eq:One.BMN.Ham}
\ham_0\, \OpE_p^{J_0;J_1,\ldots,J_K}\eq
-2\bigbrk{
  \delta_{p\neq J_0}                      \OpE^{J_0;J_1,\ldots,J_K}_{p+1} 
-(\delta_{p\neq J_0}+\delta_{p\neq 0}) \OpE^{J_0;J_1,\ldots,J_K}_p 
+\delta_{p\neq 0}                        \OpE^{J_0;J_1,\ldots,J_K}_{p-1}} \,
 ,
\nln
\ham_+\, \OpE_p^{J_0;J_1,\ldots,J_K}\eq
\mathbin{\phantom{+}}
  \sum_{J_{K+1}=1}^{p-1}\, 
2\bigbrk{\OpE^{J_0-J_{K+1};J_1,\ldots,J_{K+1}}_{p-J_{K+1}} 
       -\OpE^{J_0-J_{K+1};J_1,\ldots,J_{K+1}}_{p-1-J_{K+1}} }
\nl
-\sum_{J_{K+1}=1}^{J_0-p-1}\,
2\bigbrk{\OpE^{J_0-J_{K+1};J_1,\ldots,J_{K+1}}_{p+1}
       -\OpE^{J_0-J_{K+1};J_1,\ldots,J_{K+1}}_p },
\nln
\ham_-\, \OpE_p^{J_0;J_1,\ldots,J_K}\eq
\mathbin{\phantom{+}}
 \sum_{k=1}^K
2J_k\, 
\bigbrk{\OpE^{J_0+J_k;J_1,\ldots,\makebox[0pt]{\,\,\,\,$\times$}J_{k},\ldots,J_K}_{J_k+p} 
       -\OpE^{J_0+J_k;J_1,\ldots,\makebox[0pt]{\,\,\,\,$\times$}J_{k},\ldots,J_K}_{J_k+p-1}}
\nl
-\sum_{k=1}^K
2J_k\, 
\bigbrk{\OpE^{J_0+J_k;J_1,\ldots,\makebox[0pt]{\,\,\,\,$\times$}J_{k},\ldots,J_K}_{p+1}
       -\OpE^{J_0+J_k;J_1,\ldots,\makebox[0pt]{\,\,\,\,$\times$}J_{k},\ldots,J_K}_p}. 
\>
In view of the AdS/CFT and BMN correspondence this is very suggestive.
The one-loop dilatation operator can either not change the structure of traces,
split one trace into two, or join two into one. 
This is in qualitative agreement with string field theory when
traces are interpreted as strings. Also the parameter $1/N$ 
appears in the right places to be interpreted as the string coupling
constant.

\subsection{The BMN Limit of Two Excitation Operators}
\label{sec:One.BMN.Limit}

With $J$ being very large
we can view $\hat p=p/J$ and $\hat J_k=J_k/J$ as continuum
variables and replace the discrete set of states in equation \eqref{eq:One.BMN.Multi}
by a set of continuum states
\[\label{eq:One.BMN.States}
\OpE^{J_0;J_1,\ldots,J_K}_p \rightarrow 
\state{\hat p;\hat J_1, \ldots, \hat J_K }
=\state{\hat J_0-\hat p;\hat J_1, \ldots, \hat J_K },
\]
where
\[\label{eq:One.BMN.Ranges}
\hat p \in [0,\hat J_0],\qquad
\hat J_0,\hat J_k\in [0,1] \qquad \mbox{and}\qquad 
\hat J_0=1-(\hat J_1+ \ldots +\hat J_K).
\]
It is understood that 
$\state{\hat p;\hat J_1, \ldots, \hat J_K}=\state{\hat p;\hat J_{\pi(1)}, \ldots, \hat J_{\pi(K)} }$
with $\pi$ an arbitrary permutation of $K$ elements.

Absorbing the $J$-dependence into the definition of the Hamiltonian
\[\label{eq:One.BMN.BMNDila}
\hat \ham=J^2 \ham\quad\mbox{and}\quad
\hat \ham=\hat \ham_{0}+\hat g\indup{s} \hat \ham_{+}+\hat g\indup{s} \hat \ham_{-},
\]
we impose the BMN limit \eqref{eq:One.BMN.Limit} and get
a continuum version of \eqref{eq:One.BMN.Ham}
\<\label{eq:One.BMN.BMNHam}
\hat \ham_0\, \state{\hat p;\hat J_1, \ldots, \hat J_K}
\eq -2\partial_{\hat p}^2\,\state{\hat p;\hat J_1, \ldots,\hat J_K},
\\\nn
\hat \ham_+\, \state{\hat p;\hat J_1, \ldots, \hat J_K} 
\eq\mathord{}\mathbin{\phantom{+}}
\int\limits_0^{\hat p} d\hat J_{K+1}\, 
2\partial_{\hat p} \state{\hat p-\hat J_{K+1};\hat J_1, \ldots, \hat J_{K+1} }
\nl
-\int\limits_0^{\hat J_0-\hat p} d\hat J_{K+1}\, 
2\partial_{\hat p} \state{\hat J;\hat J_1, \ldots, \hat J_{K+1} } ,
\nln
\hat \ham_{-} \state{\hat p;\hat J_1, \ldots, \hat J_K} 
\eq\phantom{+}
\sum_{k=1}^K 2\hat J_k\,\partial_{\hat p}
\state{\hat p+\hat J_k;\hat J_1, \ldots, \makebox[0pt]{\,\,\,\,\,$\times$}\hat J_{k},\ldots,\hat J_K }
\nl
-\sum_{k=1}^K 2\hat J_k\,\partial_{\hat p}\,
\state{\hat p;\hat J_1, \ldots, \makebox[0pt]{\,\,\,\,\,$\times$}\hat J_{k},\ldots,\hat J_K }.
\nn
\>
%

\subsection{The Mode Decomposition}
\label{sec:One.BMN.Modes}

The $(K+1)$-trace eigenstates at $\hat g\indup{s}=0$ are
\[
\label{eq:One.BMN.Fourier}
\state{n;\hat J_1, \ldots, \hat J_K } = \frac{1}{\hat J_0}\,
\int_0^{\hat J_0} d\hat p\, \cos \bigbrk{\sfrac{2\pi n}{\hat J_0} \hat p}\, 
\state{\hat p;\hat J_1, \ldots, \hat J_K },\qquad n=0,1,2,\ldots\,.
\]
This is of course in accordance with the nature of the exact eigenstates
at finite $J$, c.f.~\secref{sec:One.Spec.TwoEx}.
The inverse transformation of \eqref{eq:One.BMN.Fourier}
reads
\[\label{eq:One.BMN.FourierInv}
\state{\hat p;\hat J_1, \ldots, \hat J_K} =
\state{0;\hat J_1, \ldots, \hat J_K}+
2\sum_{n=1}^{\infty}\cos\bigbrk{\sfrac{2\pi n}{\hat J_0} \hat p}
\state{n;\hat J_1, \ldots, \hat J_K}.
\]
In the basis \eqref{eq:One.BMN.Fourier}, the action of the operator $\hat \ham$ reads
\<\label{eq:One.BMN.HamFourier}
\hat \ham_0\, \state{n;\hat J_1,\ldots, \hat J_K} \eq
2\bigbrk{\sfrac{2\pi n}{\hat J_0}}^2\, \state{n;\hat J_1,\ldots, \hat J_K},
\\
\hat \ham_+\, \state{n;\hat J_1, \ldots, \hat J_K } \eq
\frac{16}{\hat J_0}\int_0^{\hat J_0} d\hat J_{K+1}\, \sum_{n'=1}^{\infty}
\frac{\bigbrk{\frac{2\pi n'}{\hat J_0-\hat J_{K+1}}}^2\sin^2\bigbrk{\pi n\sfrac{\hat J_{K+1}}{\hat J_0}}}
{\bigbrk{\frac{2\pi n'}{\hat J_0-\hat J_{K+1}}}^2-\bigbrk{\frac{2\pi n}{\hat J_0}}^2}\,
\state{n';\hat J_1, \ldots, \hat J_{K+1} } ,
\nln\nn
\hat \ham_-  \state{n;\hat J_1, \ldots, \hat J_K } \eq
16\sum_{k=1}^K \frac{\hat J_k}{\hat J_0} \sum_{n'=1}^{\infty}
\frac{\bigbrk{\frac{2\pi n'}{\hat J_0+\hat J_k}}^2 \sin^2\bigbrk{\pi n'\sfrac{\hat J_k}{\hat J_0+\hat J_k}}}
{\bigbrk{\frac{2\pi n'}{\hat J_0+\hat J_k}}^2-\bigbrk{\frac{2\pi n}{\hat J_0}}^2}\,
\state{n';\hat J_1, \ldots, \makebox[0pt]{\,\,\,\,\,$\times$}\hat J_{k},\ldots,\hat J_K } .
\>

In interacting plane-waves string theory similar expressions have been derived 
\cite{Spradlin:2002ar,Spradlin:2002rv,Chu:2002eu,Pankiewicz:2002gs,Pankiewicz:2002tg,Chu:2002wj,He:2002zu,Roiban:2002xr,Pankiewicz:2003kj,DiVecchia:2003yp,Pankiewicz:2003ap,Lucietti:2004wy}.
The Hamiltonians of both theories should however not be 
compared directly, but only modulo a similarity transformation.
A proposal for the change of basis was given in 
\cite{Pearson:2002zs,Gomis:2002wi,Gomis:2003kj} and applied in 
\cite{Spradlin:2003bw} to show the equivalence of the Hamiltonians in
the one-loop approximation. 
Up to some assumptions \cite{Roiban:2002xr} 
(which appear to be inconsistent \cite{Gutjahr:2004dv})
regarding excitation number non-preserving amplitudes in string theory,
it proves the BMN correspondence \eqref{eq:One.BMN.Corres}
at first order in $\hat g^2$ (one-loop)
and all orders in $\hat g\indup{s}$ (all-genus)
for single trace states with two excitations.
A similar statement for three excitation states 
was investigated in \cite{Gutjahr:2004qj}, but
a generalisation to arbitrarily many excitations of arbitrary type
has not been attempted yet.

\subsection{The Genus-One Energy Shift}
\label{sec:One.BMN.Torus}

Now the scene is set for determining the spectrum of the full one-loop
Hamiltonian order by order in $\hat g\indup{s}$ by standard quantum mechanical
perturbation theory.
The leading non-planar correction to the energy 
$\hat E_{n,0}=2(2\pi n)^2$ of a single trace state $\state{n}$ is obtained by
second-order perturbation theory
\[\label{eq:One.BMN.TorusPert}
\hat E_{n,2}\state{n} = \pi_n \,\hat\ham_-
     \frac{1}{\hat E_{n,0}-\hat \ham_0}\,\hat \ham_+\,\state{n},
\]
where $\pi_n$ projects to $\state{n}$. We now insert \eqref{eq:One.BMN.HamFourier}
and get the genus-one (torus) correction to the energy
\[\label{eq:One.BMN.Torus}
\hat E_{n,2}=\int_0^{1} d\hat J_1\, \sum_{n'=1}^{\infty}
\frac{128\,\hat J_1\brk{\frac{2\pi n'}{1-\hat J_1}}^2 (2\pi n)^2 \sin^4 (\pi n \hat J_1)}
{(1-\hat J_1)\bigbrk{\brk{\frac{2\pi n'}{1-\hat J_1}}^2-(2\pi n)^2}^3}
=
\frac{1}{6}+\frac{35}{4(2\pi n)^2}\,.
\]
The total scaling dimension is thus
\<\label{eq:One.BMN.GenusOne}
D\eq J+2+2(2\pi n)^2 \hat g^2
+\lrbrk{\frac{1}{6}+\frac{35}{4(2\pi n)^2}} \hat g^2 \hat g\indup{s}^2+\ldots
\nln\eq
J+2+\lambda'\,n^2
+\lambda' \hat g\indup{s}^2\lrbrk{\frac{1}{48\pi^2}+\frac{35}{128\pi^4 n^2}} 
+\ldots\,.
\>
This genus-one result was first derived by computing 
gauge theory correlation functions 
\cite{Beisert:2002bb,Constable:2002vq}
and confirmed in string theory \cite{Roiban:2002xr}, 
see also \cite{Pearson:2002zs}.
Here it was assumed that one can restrict to
excitation number preserving amplitudes in string theory,
however, there are doubts that this 
assumption is consistent \cite{Gutjahr:2004dv}.
Subsequently, the formula \eqref{eq:One.BMN.Torus} 
was rederived by Janik by considering matrix elements 
of the dilatation generator \cite{Janik:2002bd}.
This lead to a great simplification of the calculation.

\finishchapter 

\chapter{Integrability}
\label{ch:Int}

In \secref{sec:Dila.Planar} we have demonstrated how, in the planar limit, 
local operators can be interpreted as quantum spin chains.
In that picture, the planar dilatation operator is represented by 
the spin chain Hamiltonian.
Minahan and Zarembo realised that the one-loop
dilatation operator of $\superN=4$ SYM for states composed 
from only scalar fields (the one-loop $\alSO(6)$ sector) is precisely 
the Hamiltonian of an \emph{integrable} spin chain \cite{Minahan:2002ve}.
This parallels earlier discoveries of integrable spin chains
in generic, non-supersymmetric gauge theories
at one-loop and in the large $N$ limit 
when dealing with states composed mostly from covariant derivatives
\cite{Lipatov:1994yb,Faddeev:1995zg,Braun:1998id,Belitsky:1999qh,Braun:1999te,Belitsky:1999ru,Belitsky:1999bf,Derkachov:1999ze,Belitsky:2003ys,Ferretti:2004ba,Kirch:2004mk}
(see also the review \cite{Belitsky:2004cz}).

In this chapter we will show how these two lines of 
development can be combined into 
a $\alPSU(2,2|4)$ \emph{supersymmetric spin chain} \cite{Beisert:2003yb}. 
We will start by introducing the notion of integrable spin chains
and later present the algebraic Bethe ansatz technique.
As an application, we shall derive the one-loop anomalous
dimension of a state dual to a macroscopic spinning string 
in $AdS_5\times S^5$ and find a remarkable agreement
\cite{Beisert:2003xu,Beisert:2003ea}.

\section{Integrable Spin Chains}
\label{sec:Int.Chains}

A quantum integrable system is a quantum mechanical
system with an infinite number%
\footnote{The precise counting is somewhat unclear in a quantum 
system. In a classical system one needs exactly half the number 
of phase-space dimensions.
Here, the spin chains can be arbitrarily long which 
gives rise to an arbitrarily large number of conserved charges.
This is what is meant by infinitely many.}
of mutually commuting scalar charges $\charge_r$
\[\label{eq:Int.Chains.Def}
\comm{\charge_r}{\charge_s}=\comm{\algJ}{\charge_r}=0.
\]
In other words, the naive symmetry algebra is enlarged by
infinitely many abelian generators constituting the
algebra $\alU(1)^\infty$. 
The Hamiltonian $\ham$, a $\alU(1)$ generator invariant 
under the symmetry algebra, 
will turn out to be one of the charges, $\ham=\charge_2$,
and is absorbed into $\alU(1)^\infty$. The symmetry
enhancement might thus be stated as 
\[\label{eq:Int.Chains.IntAlg}
\alU(1)\longrightarrow \alU(1)^\infty.
\]
In this section we will discuss the integrable structures found
at the one-loop level.

\subsection{The R-Matrix}
\label{sec:Int.Chains.RMatrix}

\begin{figure}\centering
\includegraphics{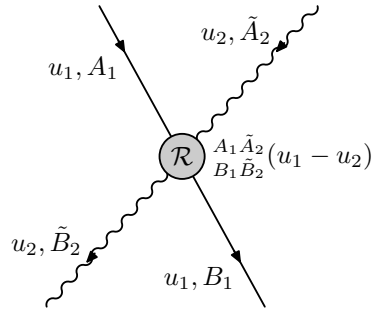}
\caption{A scattering process of two particles is described by the R-matrix.}
\label{fig:Int.Chains.RMatrix}
\end{figure}

\begin{figure}
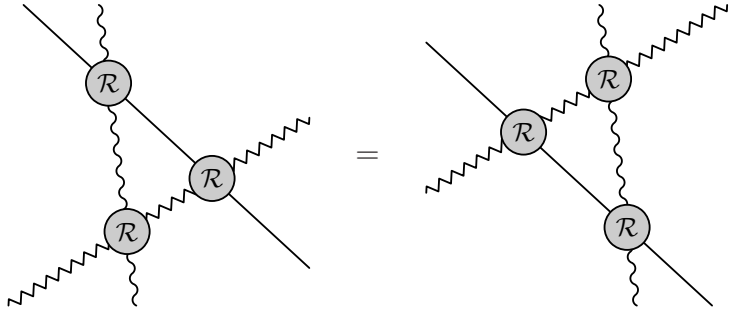
\centering
$
\parbox{5cm}{\centering\includegraphics{sec04.yangbaxter.a.eps}}
=
\parbox{5cm}{\centering\includegraphics{sec04.yangbaxter.b.eps}}
$
\caption{The Yang-Baxter equation.}
\label{fig:Int.Chains.YBE}
\end{figure}

A spin chain is composed from $L$ modules 
transforming in some representation of a symmetry algebra.
We will assume the symmetry algebra to be
of unitary type, $\alSU(M)$.
To understand the integrable model, it makes
sense to consider the individual spins as `particles'. 
A particle $X_{A}(u)$
is thus defined as an element of a module of the symmetry group
together with a spectral parameter $u$.
The central object of the integrable model 
is the R-matrix, it describes the `scattering' of particles. 
The R-matrix rotates two modules depending
on their representations and difference of spectral parameters,
c.f.~\figref{fig:Int.Chains.RMatrix}
\[\label{eq:Int.Chains.Scatter}
\tilde X'_{2,\tilde B_2}(u_2)\,X'_{1,B_1}(u_1) = 
\Rmatrix^{A_1 \tilde A_2}_{B_1 \tilde B_2}(u_1-u_2) \, 
X_{1,A_1}(u_1)\, \tilde X_{2,\tilde A_2}(u_2).
\]
The scattering is elastic in the sense that 
neither the representation nor the spectral parameters are changed;
the only effect is a generalised phase shift described by the R-matrix.
In an integrable system, the order in which particles scatter
does not matter. For the scattering of three particles this fact
is described by the Yang-Baxter equation
\<\label{eq:Int.Chains.YangBaxter}
\earel{}
\Rmatrix^{A_1 \tilde A_2}_{B_1 \tilde B_2}(u_1-u_2)\,
\Rmatrix^{B_1 \hat A_3}_{C_1 \hat B_3}(u_1-u_3)\,
\Rmatrix^{\tilde B_2 \hat B_3}_{\tilde C_2 \hat C_3}(u_2-u_3)
\nln
\eq
\Rmatrix^{\tilde A_2 \hat A_3}_{\tilde B_2 \hat B_3}(u_2-u_3)\,
\Rmatrix^{A_1 \hat B_3}_{B_1 \hat C_3}(u_1-u_3)\,
\Rmatrix^{B_1 \tilde B_2}_{C_1 \tilde C_2}(u_1-u_2)
\>
or $\Rmatrix_{12}\Rmatrix_{13}\Rmatrix_{23}=
\Rmatrix_{23}\Rmatrix_{13}\Rmatrix_{12}$ for short.
The Yang-Baxter equation is most intuitively represented 
in a diagrammatic fashion, see \figref{fig:Int.Chains.YBE};
it implies that the particle lines can be moved around freely,
even past other interactions.
From this it follows that, also for a larger 
number of particles, the order of scatterings does not matter.

\begin{figure}\centering
$
\parbox{5cm}{\centering\includegraphics{sec04.inverse.a.eps}}
=
\parbox{5cm}{\centering\includegraphics{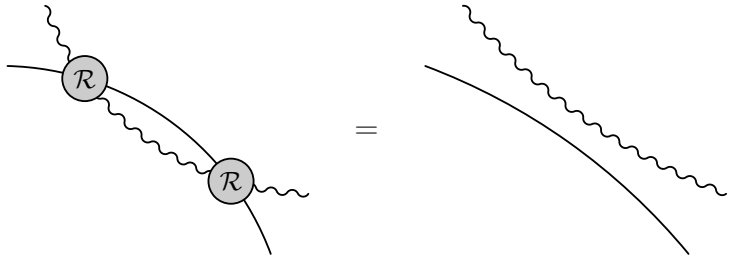}}
$
\caption{The R-matrix and its inverse.}
\label{fig:Int.Chains.RInverse}
\end{figure}%
In addition to the Yang-Baxter equation, 
there is the relation,
see \figref{fig:Int.Chains.RInverse}
\[\label{eq:Int.Chains.RInv}
\Rmatrix^{A_1\tilde{A}_2}_{B_1\tilde{B}_2}(u_1-u_2)\,
\Rmatrix^{\tilde B_2 B_1}_{\tilde C_2 C_1}(u_2-u_1)
=\delta^{A_1}_{C_1}\delta^{\tilde A_2}_{\tilde C_2},
\]
which defines the inverse of the R-matrix.

The case of all particles transforming in the fundamental 
representation is particularly easy to solve.
The solution of the bi-fundamental Yang-Baxter equation is
\[\label{eq:Int.Chains.RFundA}
\Rmatrix^{ab}_{cd}(u)=\frac{u}{u+i}\,\delta^a_c\delta^b_d
+\frac{i}{u+i}\,\delta^a_d\delta^b_c
\qquad\mbox{or}\qquad
\Rmatrix_{12}(u)=\frac{u}{u+i}\,\fldident_{12}+\frac{i}{u+i}\,\fldperm_{12},
\]
where $\fldident_{12}$ is the identity acting on particles $1,2$ and
$\fldperm_{12}$ is the permutation.
It is useful to write this in a mixed notation where we keep
one index manifest and suppress the other in a matrix notation
\[\label{eq:Int.Chains.RFund}
\Rmatrix^{a}_{b}(u)
=\frac{u+i/M}{u+i}\,\delta^a_b+\frac{i}{u+i}\,\algJ^a{}_b,
\]
where we make use the symmetry generator 
$(\algJ^a{}_b{})^{c}{}_{d}=\delta^a_d\delta^c_b-\delta^a_b\delta^c_d/M$
in the fundamental representation.
The fundamental R-matrix, 
where one particle transforms in the fundamental representation
and the other in an arbitrary one, is given by a
similar expression as \eqref{eq:Int.Chains.RFund}
using the symmetry generators.

\subsection{Transfer Matrices}
\label{sec:Int.Chains.Transfer}

\begin{figure}
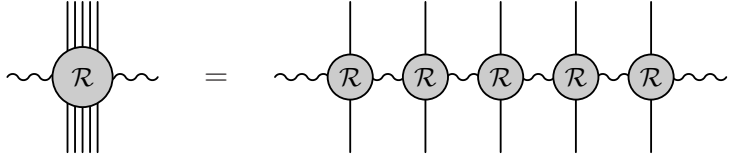
\centering
$
\parbox{3cm}{\centering\includegraphics{sec04.monodromy.eps}}
=
\parbox{7cm}{\centering\includegraphics{sec04.monodromy.r.eps}}
$
\caption{The monodromy matrix for a composite particle.}
\label{fig:Int.Chains.Monodromy}
\end{figure}

Several particles can be grouped into a composite particle
$X_{A_1\ldots A_L}(v_1,\ldots,v_L)$.%
\footnote{Commonly, all spectral parameters will be aligned $v_p=v$
and constitute a homogeneous chain.
In \secref{sec:HighInt.Ansatz.Ansatz} we will however 
encounter an inhomogeneous chain with different $v_p$'s.}
The spin chain is just such a composite particle.
For a composite particle one can define a composite R-matrix 
(monodromy matrix) by, see \figref{fig:Int.Chains.Monodromy} 
\[\label{eq:Int.Chains.Monodromy}
\tilde \Rmatrix^{\tilde A,A_1\ldots A_L}_{\tilde B,B_1\ldots B_L}(u)
=
\Rmatrix^{\tilde A \,\,A_1}_{\tilde C_2 B_1}(u-v_1)\,
\Rmatrix^{\tilde C_2 A_2}_{\tilde C_3 B_2}(u-v_2)
\ldots
\Rmatrix^{\tilde C_{L} A_L}_{\tilde B\,\,\,\, B_L}(u-v_L).
\]
This R-matrix naturally satisfies the Yang-Baxter equation.
Usually one suppresses most indices and spectral parameters
$\Rmatrix^{A}_{B}=
(\Rmatrix_1)^{A}_{C_2}\,
(\Rmatrix_2)^{C_2}_{C_3}\,
\ldots
(\Rmatrix_L)^{C_{L}}_{B}$.
Let us write the monodromy matrix 
for the fundamental representation 
using \eqref{eq:Int.Chains.RFund}
with all spectral parameters aligned $v_p=0$
\[\label{eq:Int.Chains.Wilson}
\Rmatrix^{a}_{b}(u)=
\lrbrk{\frac{u+i/M}{u+i}\,\delta^a_{c_2} +\frac{i}{u+i}\,\algJ_1{}^{a}{}_{c_2}}
\cdots
\lrbrk{\frac{u+i/M}{u+i}\,\delta^{c_{L}}_{b} +\frac{i}{u+u_L}\,\algJ_L{}^{c_{L}}{}_{b}}\,.
\]
This expression reveals an interesting interpretation
of the monodromy matrix for a spin chain: 
The generator of rotations, $\algJ$, may be considered
as the component of some gauge field in the direction of the spin chain. 
Then, the monodromy matrix has a great similarity to a
Wilson line along the spin chain. 
In that picture, an elementary R-matrix is just the
monodromy of the gauge field across one spin chain site. 
Moreover, there is not only a single 
gauge field, but a family of gauge fields,
parameterised by the spectral parameter $u$.
When viewed in this way, the integrable structure
is very similar to the one found in string theory
on $AdS_5\times S^5$, see \cite{Mandal:2002fs,Bena:2003wd,Alday:2003zb}.
This similarity allowed the authors of \cite{Dolan:2003uh,Dolan:2004ps} to 
promote the Yangian structure from string theory to gauge theory.

For a cyclic spin chain it is natural to close the
Wilson line to a loop and take the trace. 
One obtains the \emph{transfer matrix}
\[\label{eq:Int.Chains.Loop}
\tilde \transfer(u)=
\tilde \Rmatrix^{\tilde A}_{\tilde A}(u)
=
(\Rmatrix_1)^{\tilde C_1}_{\tilde C_2}\,
(\Rmatrix_2)^{\tilde C_2}_{\tilde C_3}\,
\ldots
(\Rmatrix_L)^{\tilde C_L}_{\tilde C_1}.
\]
The transfer matrix can be taken for any representation circulating
around the Wilson loop and for any spectral parameter.
In this work, however, we shall restrict to the
equal representations of the spins and the Wilson loop.
The interesting aspect of transfer matrices is that
all of them commute 
\[\label{eq:Int.Chains.Commute}
\comm{\tilde \transfer(u)}{\hat \transfer(v)}=0.
\]
This statement can be shown easily by inserting a R-matrix and its inverse
into the traces, see \figref{fig:Int.Chains.RInverse}.
Using the Yang-Baxter equation, the R-matrix is commuted around the traces
interchanging the order of monodromy matrices 
\<
\tilde\transfer(u)\,\hat\transfer(v)\eq
\tilde\Rmatrix^{\tilde A}_{\tilde A}(u)\,
\hat\Rmatrix^{\hat A}_{\hat A}(v)
=
\Rmatrix^{\hat A\tilde A}_{\hat B\tilde B}(v-u)\,
\Rmatrix^{\tilde B\hat B}_{\tilde C\hat C}(u-v)\,
\tilde\Rmatrix^{\tilde C}_{\tilde A}(u)\,
\hat\Rmatrix^{\hat C}_{\hat A}(v)
\nln\eq
\Rmatrix^{\hat A\tilde A}_{\hat B\tilde B}(v-u)\,
\hat\Rmatrix^{\hat B}_{\hat C}(v)\,
\tilde\Rmatrix^{\tilde B}_{\tilde C}(u)\,
\Rmatrix^{\tilde C\hat C}_{\tilde A\hat A}(u-v)\,
\nln\eq
\hat\Rmatrix^{\hat B}_{\hat B}(v)\,
\tilde\Rmatrix^{\tilde B}_{\tilde B}(u)\,
=
\hat\transfer(v)\,\tilde\transfer(u).
\>
Afterwards the R-matrix and its inverse cancel out and 
the transfer matrices are interchanged.

\subsection{The Local Charges}
\label{sec:Int.Chains.Charges}

There are many uses for monodromy and transfer matrices. 
A particular one is the Yangian, an associative Hopf algebra 
which enlarges the symmetry algebra, see e.g.~\cite{Bernard:1993ya}. 
The Yangian is an important object for integrable systems. 
The elements of the Yangian are given 
by the monodromy matrix in the fundamental representation
\eqref{eq:Int.Chains.Monodromy}. 
Commonly, the Yangian is expanded around $u=\infty$. 
In the leading two orders one finds the identity and the generators
of the symmetry algebra, $\algJ$, acting on the full spin chain. 
At the next order, the first non-trivial elements of the Yangian appear. 
They are bi-local along the spin-chain and 
can be used to generate all higher elements.
In $\superN=4$ SYM we deal with cyclic spin chains and 
the open Wilson line of the Yangian breaks cyclic symmetry. 
At the moment it is not clear how to make direct use of the Yangian 
for the study of scaling dimension and 
we will not consider it further. 
See \cite{Dolan:2003uh,Dolan:2004ps} for a
treatment of the Yangian in $\superN=4$ SYM.

Here we would like to investigate the transfer matrices. 
These are closed Wilson loops and they preserve cyclic symmetry. 
The transfer matrix can be used as a generating function for 
the charges $\shift,\charge_n$ when expanded in the spectral parameter
\[\label{eq:Int.Chains.TransCharge}
\transfer(u)=\shift\,\exp \sum_{r=2}^\infty i u^{r-1} \charge_r.
\]
All of these charges commute with each other due
to commuting of the transfer matrices at different values of 
the spectral parameters \eqref{eq:Int.Chains.Commute}.

\begin{figure}
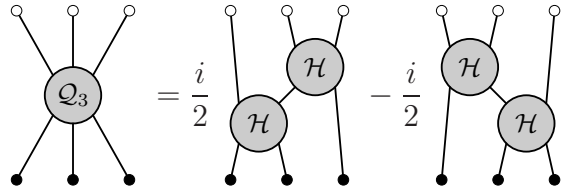
\centering
$\displaystyle
\parbox{2.cm}{\centering\includegraphics{sec04.charge3.eps}}
=\frac{i}{2}
\parbox{2.cm}{\centering\includegraphics{sec04.charge3.a.eps}}
-\frac{i}{2}
\parbox{2.cm}{\centering\includegraphics{sec04.charge3.b.eps}}
$
\caption{The third charge density is
composed from two copies of the Hamiltonian density.}
\label{fig:Int.Chains.Charge3}
\end{figure}

For spin chains with equal spin representations at 
each site, it is useful to pick the
same representation to circle around the Wilson loop as well.
We will furthermore assume that all R-matrices for the construction of
the transfer matrix are the same and have a specific value at $u=0$%
\[\label{eq:Int.Chains.RatZero}
\Rmatrix^{A_1A_2}_{B_1B_2}(0)=\delta^{A_1}_{B_2}\delta^{A_2}_{B_1},
\]
i.e.~they permute the modules.%
\footnote{One might have to redefine the R-matrix slightly
exploiting the symmetries 
of the Yang-Baxter equation \eqref{eq:Int.Chains.YangBaxter}: 
Firstly, we can rescale the R-matrix by a function of 
the spectral parameter. Secondly, we can shift and rescale the
spectral parameters by a constant. 
The bi-fundamental $R$-matrix in 
\eqref{eq:Int.Chains.RFund} is 
already in this from.}
Let us now expand the transfer matrix in $u$. 
At $u=0$ we find the cyclic shift operator
\[\label{eq:Int.Chains.Shift}
\shift{}^{A_1\ldots A_L}_{B_1\ldots B_L}
=\transfer{}^{A_1\ldots A_L}_{B_1\ldots B_L}(0)
=\delta^{A_1}_{B_2}\delta^{A_2}_{B_3}\ldots\delta^{A_{L-1}}_{B_L}\delta^{A_L}_{B_1}.
\]
Expanding to first order in $u$ we find that we have
to insert a derivative $\Rmatrix'=\partial\Rmatrix/\partial u$ 
of the R-matrix into the shift operator
and sum over all insertion points,
\[\label{eq:Int.Chains.ExpandOne}
\transfer{}^{A_1\ldots A_L}_{B_1\ldots B_L}(u)
=\shift{}^{A_1\ldots A_L}_{B_1\ldots B_L}
+u\sum_{p=1}^L\delta^{A_1}_{B_2}\ldots 
\Rmatrix'\,{}^{A_p A_{p+1}}_{B_{p+2} B_{p+1}}(0)\ldots\delta^{A_L}_{B_1}
+\order{u^2}.
\]
Let us define the charge density
\[\label{eq:Int.Chains.Q2Dens}
\charge^{}_{2,}{}^{A_1A_2}_{B_1B_2}
=-i \Rmatrix'\,{}^{A_1A_2}_{B_2B_1}(0)\quad\mbox{or}\quad
\charge_{2,12}=-i\,\fldperm_{12}\, \Rmatrix'_{12}(0),
\]
where the second form is short for the first. 
The permutation $\fldperm_{12}$ interchanges the spins at two sites.
According to the definition \eqref{eq:Int.Chains.TransCharge} 
we should absorb the 
cyclic shift in \eqref{eq:Int.Chains.ExpandOne} 
into $\shift$ and the second charge $\charge_2$ is simply 
\[\label{eq:Int.Chains.Q2}
\charge^{}_{2,}{}^{A_1\ldots A_L}_{B_1\ldots B_L}=
\sum_{p=1}^L \delta^{A_1}_{B_1}\ldots 
\charge^{}_{2,}{}^{A_p A_{p+1}}_{B_{p} B_{p+1}}
\ldots\delta^{A_L}_{B_L}
\quad\mbox{or}\quad
\charge_2=\sum_{p=1}^L \charge_{2,p,p+1}.
\]
It is very suggestive to interpret this charge as the Hamiltonian,
it has a nearest-neighbour type interaction as desired for a spin chain
\[\label{eq:Int.Chains.Ham}
\ham=\charge_2=\sum_{p=1}^L \ham_{p,p+1},\qquad
\ham_{12}=\charge_{2,12}=-i\, \fldperm_{12}\,\Rmatrix'_{12}(0).
\]
Expanding $\transfer(u)$ to quadratic order in $u$ we find many 
terms. There are some disconnected terms which should be absorbed 
into $-\half u^2 \shift\charge_2^2$
from the expansion of the exponential
in \eqref{eq:Int.Chains.TransCharge}. 
To complete the square $\charge_2^2$ 
we need the identity $\Rmatrix''_{12}(0)=\Rmatrix'_{12}(0)\fldperm_{12}\Rmatrix'_{12}(0)$ 
due to the Yang-Baxter equation. The remaining terms give rise
to the third charge, see \figref{fig:Int.Chains.Charge3}
\[\label{eq:Int.Chains.Q3}
\charge_{3}=\sum_{p=1}^L \charge_{3,p,p+1,p+2},\qquad
\charge_{3,123}=\sfrac{i}{2}(\ham_{12}\ham_{23}-\ham_{23}\ham_{12}).
\]
One can go on constructing the higher charges $\charge_r$ in this way 
and finds that they can all be written in terms 
of the Hamiltonian density $\ham_{12}$.%
\footnote{A more efficient way is to use the
boost operator $\chainop{B}=\sum_{p=1}^L ip\, \ham_{p,p+1}$. 
It generates the higher charges 
recursively via $\comm{\chainop{B}}{\charge_r}=r\charge_{r+1}$.
This can be deduced by assigning different 
spectral parameters to the individual spins, $u_p=u+p\epsilon$.
Note however, that the boost leaves some
undesired boundary terms which are, in particular, 
incompatible with the cyclic nature of the spin chain. 
These should be dropped.}

\subsection{Parity and Pairs}
\label{sec:Int.Chains.Pairs}

Above we have constructed two charges of the spin chain, 
$\ham=\charge_2$ and $\charge_3$. 
From \eqref{eq:Int.Chains.Def,eq:Int.Chains.Commute}
we know that they commute
\[\label{eq:Int.Chains.HQ3}
\comm{\ham}{\charge_3}=0,
\]
even though this statement is labourious to verify explicitly.
Let us find out what happens when we invert the order of spins
within the spin chain. This is equivalent to the
parity operation $\gaugepar$ defined in 
\secref{sec:N4.Gauge,sec:Dila.Planar.Parity}.
up to a factor of $(-1)^L$. 
The Hamiltonian density will be assumed to have positive
parity 
\[\label{eq:Int.Chains.HParity}
\gaugepar \,\ham\, \gaugepar^{-1}=\ham,\qquad 
\comm{\gaugepar}{\ham}=0.
\]
From this it immediately follows that the third charge
has negative parity
(it has negative mirror symmetry with respect to the 
vertical axis, see \figref{fig:Int.Chains.Charge3})
\[\label{eq:Int.Chains.Q3Parity}
\gaugepar\, \charge_3\, \gaugepar^{-1}=-\charge_3,\qquad
\acomm{\gaugepar}{\charge_3}=0.
\]
Similarly one finds for the higher charges 
\[\label{eq:Int.Chains.QnParity}
\gaugepar\, \charge_{r}\,\gaugepar^{-1} =(-1)^r \charge_r.
\]

A consequence of 
\eqref{eq:Int.Chains.HParity,eq:Int.Chains.Q3Parity,eq:Int.Chains.QnParity}
is that the spectrum of $\ham$ will display a 
degeneracy of states $\state{\pm}$ with opposite parities 
\cite{Doikou:1998jh}
\[\label{eq:Int.Chains.Paired}
\mbox{`\emph{paired state}':}\qquad \bigset{\state{+},\state{-}}\quad\mbox{with}\quad E_+=E_-.
\]
This is a very non-trivial statement because $\ham$ conserves parity
and thus cannot relate states with opposite parities in any way.
Assume we find a state of positive parity $\state{+}$ and energy $E_+$. 
Then the state $\state{-}=\charge_3\state{+}$ has negative parity
and energy $E_+$
\[\label{eq:Int.Chains.PairedEnergy}
\ham\,\state{-}
=\ham\,\charge_3\,\state{+}
=\charge_3\,\ham\,\state{+}
=E_+\,\charge_3\,\state{+}
=E_+\,\state{-}.
\]
Of course, we cannot exclude that $\charge_3$ annihilates $\state{+}$
and $\state{-}=0$.
In this case the state of definite parity $P=+$ or $P=-$ is unpaired
\[\label{eq:Int.Chains.Unpaired}
\emph{`unpaired state':}\qquad \state{P}\quad 
\mbox{with}\quad \charge_3 \state{P}=0.
\]
Such states exist when,
for example, the numbers of positive and negative parity states 
do not agree.

We have seen that the third charge of the integrable spin chain
has important consequences. It is thus natural to 
investigate the higher charges $\charge_r$. 
In contrast to $\charge_3$ we find that $\charge_4$ does not pair up
operators, it simply assigns a number (charge) to each operator.
This is in fact what might be expected. 
The reason why $\charge_3$ was interesting is that it 
anticommutes with $\gaugepar$, while $\ham$ commutes, thus giving rise to pairs. 
The next charge, $\charge_5$, does again anticommute with parity. 
This generator will relate the same pairs, 
only with different coefficients (charges).

\section{One-Loop Integrability}
\label{sec:Int.OneLoop}

In this section we derive the R-matrix 
for the integrable spin chain 
considered in this chapter. 
For this purpose we make use of a special subsector 
of the spin chain
with residual $\alSU(1,1)$ symmetry
and show how to 
lift the universal $\alSL(2)=\alSU(1,1)$ R-matrix 
to an $\alPSU(2,2|4)$ invariant R-matrix.
The derived Hamiltonian is shown
to agree with the complete one-loop planar dilatation generator 
of $\superN=4$ SYM, thus proving the integrability of the latter.

\subsection{Planar Parity Pairs}
\label{sec:Int.OneLoop.Pairs}

Let us have a look at the tables of 
one-loop planar spectra in \secref{sec:One.Spec}.
One observes a large number of degenerate pairs
of states with opposite parity which are indicated by $\gaugepar=\pm$.
In fact, in no representation of the symmetry group 
unpaired states of both parities can be found. 
In other words, it appears that the only possibility 
for an unpaired state to exist, 
is the absence of a suitable partner. This picture is not expected 
to continue strictly at higher dimensions, but it shows that
the pairing of states is very systematic and not merely a coincidence.
A simple explanation for the pairing of states would be the
existence of a conserved charge that anticommutes with parity, 
just like $\charge_3$, as explained in \secref{sec:Int.Chains.Pairs}.
Indeed, pairing of states is a useful criterion for
integrability:
The planar one-loop spectrum of $\superN=4$ displays pairing
and is thus a candidate integrable system. 
Moreover, there is phenomenological evidence that 
paired spectra can only arise in an integrable system,
see also \cite{Grabowski:1995rb}.
We will discuss this point in \secref{sec:HighInt.SU2.HigherLoops}. 
In this section we will use the methods of integrable spin chains
introduced in \secref{sec:Int.Chains} to find that 
planar one-loop $\superN=4$ is indeed integrable.

One may wonder whether the degeneracy also holds at the non-planar 
level. In an example we show that this is not the case.
There are three unprotected multiplets 
with highest weights $w=\weight{5;0,0;1,1,1;0,5}$, two single-trace
and one double-trace state. 
They are at both unitarity bounds and have zero spin. 
As such they have descendants in the quarter-BPS $\alSU(2)$ sector.
The states have length $L=7$ and excitation number $K=3$, 
i.e.~they are of the form $\fldZ^4\phi^3$.
Two have negative parity
and one has positive parity, the line separates between them
\[\label{eq:Int.OneLoop.NonPlanarOp}
\OpE^\trans=\left(\begin{array}{c}
2\Tr \fldZ^4\phi^3
+2\Tr \fldZ^2\phi \fldZ^2\phi^2
+2\Tr \fldZ^2\phi \fldZ\phi \fldZ\phi
-3 \Tr \fldZ^3\phi\acomm{\phi}{\fldZ}\phi\\
\Tr \fldZ\phi \Tr \fldZ^2\comm{\phi}{\fldZ}\phi
-\Tr \fldZ^2 \Tr \fldZ\comm{\phi}{\fldZ}\phi^2\\\hline
\Tr \comm{\phi}{\fldZ}\comm{\phi}{\fldZ}\comm{\phi}{\fldZ}\fldZ,
\end{array}\right).
\]
The dilatation operator \eqref{eq:Dila.SU2.Dila} acts on these 
as (note $\ham\OpE=\OpE H$)
\[\label{eq:Int.OneLoop.NonPlanarHam}
H=\left(\begin{array}{cc|c}
5&\sfrac{10}{N}&0\\
\sfrac{4}{N}&4&0\\\hline
0&0&5
\end{array}\right).
\]
This corresponds to the scaling dimensions 
exact for all values of $N$ \cite{Ryzhov:2001bp}
\[\label{eq:Int.OneLoop.NonPlanarEnergy}
E_+=5,\qquad
E_-=\sfrac{9}{2}\pm\sqrt{\sfrac{1}{4}+\sfrac{40}{N^2}}\,.
\]
We find that the scaling dimensions of the two 
single-trace operators are degenerate at $N=\infty$.
For finite $N$ or in an expansion in powers of $1/N$ we
find that the degeneracy is broken.
Therefore integrability, as defined above, 
can only hold in the planar limit and breaks down
when topological interactions take place.
This is in agreement with the picture of a Wilson loop
as a generating function for the charges. 
The Wilson loop of a flat connection
can be moved around freely on the `world-sheet' of the spin chain. 
It cannot, however, be moved past points of topological changes.
This would require to cut open the loop and glue 
the ends in a different order, thus modifying the 
Wilson loop.
Still, one may hope for some aspects of integrability 
to survive even when non-planar corrections are taken into account:
The family of gauge connections (alias the R-matrix) 
underlying the Wilson loop is a
local object and does not depend on the global structure
of the world-sheet.

\subsection{The Bosonic $\alSU(1,1)$ Subsector}
\label{sec:Int.OneLoop.SL2}

We will use a similar trick as in 
\secref{sec:One.Lift} to derive the R-matrix
of the complete $\alPSU(2,2|4)$ spin chain. 
Here, we shall use the Hamiltonian within the 
bosonic $\alSU(1,1)$ subsector introduced 
in \secref{sec:One.Baby} to obtain 
an expression for the R-matrix which is
subsequently lifted to the full theory.

The Hamiltonian density \eqref{eq:One.Baby.SpinHam} 
\[\label{eq:Int.OneLoop.SL2Ham}
\ham''_{12}=2h(\fldspin''_{12}):=
\sum_{j=0}^\infty
2 h(j)\, \fldproj''_{12,j}
\]
equals the one of the so-called Heisenberg XXX$_{-1/2}$ spin chain.%
\footnote{The integrable $\alSL(2)$ spin chain 
with fundamental spin representation $[s]$ (spin $s/2$) 
is called the `Heisenberg XXX$_{s/2}$ spin chain'}
Recall that the spins belong to $\mdlF''$ which is the 
highest-weight module $[-1]$ (spin $-1/2$)
and the tensor product of two $\mdlF''$ decomposes into 
$\mdl''_j$ with highest weight $[-2-2j]$ (spin $-1-j$).
The operator $\fldproj''_{12,j}$ projects a two-spin state
to the module $\mdl''_{j}$ and $\fldspin''_{12}$ measures
the label $j$ of $\mdl''_{j}$. The function $h(j)$ gives
the harmonic numbers.

Let us show that the above Hamiltonian is integrable.
To accomplish this, we make use of the universal 
R-matrix of $\alSL(2)$ spin chains. This $\alSL(2)$ invariant 
operator can be decomposed into its irreducible components 
corresponding to the modules $\mdl''_j$
\[\label{eq:Int.OneLoop.SL2R}
\Rmatrix''_{12}(u)=\sum_{j=0}^\infty \Rmatrix''_j(u)\, \fldproj''_{12,j}.
\]
The eigenvalues $\Rmatrix''_j(u)$ of the $\alSL(2)$ universal R-matrix 
were determined in \cite{Kulish:1981gi}. 
In a spin $-1-j$ representation the eigenvalue is 
\[\label{eq:Int.OneLoop.SL2REigen}
\Rmatrix''_{j}(u)=(-1)^{j+1}\,\frac{\Gammafn(-j-cu)}{\Gammafn(-j+cu)}\,\frac{f(+cu)}{f(-cu)}\,.
\]
The arbitrary function $f(u)$ and normalisation constant $c$ 
reflect trivial symmetries of the 
Yang-Baxter equation. We choose the function and constant to be%
\footnote{The normalization for \eqref{eq:Int.Chains.RFundA} 
uses $c=+i$. For non-compact representations it is however more
convenient to use a different sign $c=-i$.}
\[\label{eq:Int.OneLoop.SL2RFix}
f(cu)=\Gammafn(1+cu)\,,\quad c=-i.
\]
This enables us to find rational expressions 
for $\Rmatrix''_j$ and its derivative when $j$ is integer
\[\label{eq:Int.OneLoop.SL2RExpand}
\Rmatrix''_{j}(u)=
\prod_{k=1}^j \frac{u-ik}{u+ik}\,,
\qquad
\frac{\partial\Rmatrix''_{j}}{\partial u}(u)=
\Rmatrix''_{j}(u)\sum_{k=1}^j \frac{2ik}{k^2+u^2}\,.
\]
We note that for even (odd) $j$ the composite module
$\mdl''_j$ is a (anti)symmetric combination 
of two $\mdlF''$, consequently the permutation acts as
\[\label{eq:Int.OneLoop.SL2Permute}
\fldperm_{12}\,\mdl''_j=(-1)^j\,\mdl''_j.
\]
In other words, the R-matrix at $u=0$, whose elements
equal $(-1)^j$, is a permutation
\[\label{eq:Int.OneLoop.SL2RZero}
\Rmatrix''_{12}(0)=\fldperm_{12}.
\]
We now obtain the induced Hamiltonian density using
\eqref{eq:Int.Chains.Ham,eq:Int.OneLoop.SL2RExpand}
\[\label{eq:Int.OneLoop.SL2RHam}
\ham''_{12}=-i\,\fldperm_{12}\,\frac{\partial\Rmatrix''_{j}}{\partial u}(0)
=2h(\fldspin''_{12})\, \fldperm_{12}\,\Rmatrix''_{12}(0) 
=2h(\fldspin''_{12}).
\]
This proves the integrability of the Hamiltonian density
$\ham''_{12}$.

\subsection{The Complete R-matrix}
\label{sec:Int.OneLoop.Complete}

To derive an R-matrix for the full $\alPSU(2,2|4)$ spin chain 
we will assume that for given representations of the
symmetry algebra there exists a unique R-matrix 
which satisfies the Yang-Baxter equation
(modulo the symmetries of the YBE).
This claim \cite{Kulish:1981gi}
is supported by the existence and uniqueness 
of the algebraic Bethe ansatz procedure in \secref{sec:Int.Bethe}.
Let $\Rmatrix_{12}$ be this R-matrix for the
$\alPSU(2,2|4)$ integrable spin chain.
The R-matrix is an invariant operator,
thus it can be reduced to its irreducible components 
corresponding to the modules $\mdl_j$
\[\label{eq:Int.OneLoop.R}
\Rmatrix_{12}(u)=\sum_{j=0}^\infty \Rmatrix_{j}(u)\, \fldproj_{12,j}.
\]
The restriction $\Rmatrix''$ of the R-matrix to the bosonic $\alSU(1,1)$ sector 
must also satisfy the Yang-Baxter equation.
The unique solution for 
the eigenvalues of $\Rmatrix''$ is \eqref{eq:Int.OneLoop.SL2REigen}.
Due to the one-to-one correspondence of modules $\mdl_j$ and $\mdl''_j$, 
c.f.~\secref{sec:One.Baby}, 
the eigenvalues of the unique $\alPSU(2,2|4)$ 
R-matrix must be
\[\label{eq:Int.OneLoop.REigen}
\Rmatrix_{j}(u)=
\Rmatrix''_{j}(u)=
(-1)^{j+1}\, \frac{\Gammafn(-j-cu)}{\Gammafn(-j+cu)}\, \frac{f(+cu)}{f(-cu)}\,.\]
For the choice \eqref{eq:Int.OneLoop.SL2RFix} of $f$ and $c$,
this R-matrix yields \eqref{eq:Int.OneLoop.SL2RHam}
\[\label{eq:Int.OneLoop.RHam}
\ham_{12}=2h(\fldspin_{12}).
\]
This is just the one-loop Hamiltonian density of $\superN=4$ SYM,
c.f.~\secref{sec:One.Lift},
which in turn shows that the planar one-loop
dilatation generator of $\superN=4$ is integrable.
Note, however, that this proof is based on the assumption
of the existence of a unique R-matrix.

Let us verify that the R-matrix satisfies the Yang-Baxter equation
involving two multiplets $\mdlF$ and one fundamental module.
We shall use the fundamental R-matrix
\[\label{eq:Int.OneLoop.RFundFull}
\Rmatrix_{p}(u_{p})=\frac{u_p}{u_p-i}-\frac{i}{u_p-i}\,\algJ_p,
\]
which obeys the Yang-Baxter equation with two 
fundamental particles 
and is similar to the bi-fundamental R-matrix \eqref{eq:Int.Chains.RFund}.%
\footnote{For non-compact representations 
it is convenient to flip the sign of $u$.}
The generator $\algJ_p$ is a matrix of operators,
the operators act on $\mdlF$ at site $p$ and the matrix 
is bi-fundamental.
We now substitute this into the Yang-Baxter equation
and expand (we suppress all indices)
\<\label{eq:Int.OneLoop.YBE}
0\eq
\Rmatrix_{12}(u_1-u_2)\,
\Rmatrix_{1}(u_1)\,
\Rmatrix_{2}(u_2)
-
\Rmatrix_{2}(u_2)\,
\Rmatrix_{1}(u_1)\,
\Rmatrix_{12}(u_1-u_2)
\nln
\eq
-\frac{i(u_1+u_2)}{2(u_1-i)(u_2-i)}\,\comm{\Rmatrix_{12}}{\algJ_{12}}
-\frac{1}{2(u_1-i)(u_2-i)}\,\comm{\Rmatrix_{12}}{\chainop{C}_{12}}
\nl
+\frac{i(u_1-u_2)}{2(u_1-i)(u_2-i)}\,\comm{\Rmatrix_{12}}{\chainop{A}_{12}}
-\frac{1}{2(u_1-i)(u_2-i)}\,\acomm{\Rmatrix_{12}}{\chainop{B}_{12}},
\>
where we have defined the bi-fundamental matrices of operators%
\footnote{The commutator in $Q_{12}$ does not vanish, because
the operators $\algJ_1$ and $\algJ_2$ are matrices.}
\[\label{eq:Int.OneLoop.Maps}
\algJ_{12}=\algJ_1+\algJ_2,\quad
\chainop{A}_{12}=\algJ_1-\algJ_2,\quad
\chainop{B}_{12}=\comm{\algJ_1}{\algJ_2},\quad
\chainop{C}_{12}=\acomm{\algJ_1}{\algJ_2}.
\]
The action of these operators on the modules $\mdl_j$ was investigated
in \cite{Dolan:2003uh,Dolan:2004ps}. To understand how these operators act, 
it is useful to know their parity. 
It is straightforward to see that $\algJ_{12}$ and $\chainop{C}_{12}$ have
positive parity, while $\chainop{A}_{12}$ and $\chainop{B}_{12}$ have negative parity.
Therefore $\algJ_{12},\chainop{C}_{12}$ map between modules of the
same parity and $\chainop{A}_{12},\chainop{B}_{12}$ invert the parity
\[\label{eq:Int.OneLoop.MapParity}
\algJ_{12},\chainop{C}_{12}:\mdl_j\to \mdl_{j+2n},\qquad
\chainop{A}_{12},\chainop{B}_{12}:\mdl_j\to \mdl_{j+2n+1}.
\]
Furthermore, all operators are invariant under $\alPSU(2,2|4)$ if
one simultaneously rotates the modules $\mdlF$ and the 
bi-fundamental matrix. The bi-fundamental 
representation is just the adjoint, which 
can shift the highest weight of the module by not more than one step
\[\label{eq:Int.OneLoop.MapAdjoint}
\algJ_{12},\chainop{A}_{12},\chainop{C}_{12},\chainop{B}_{12}:\mdl_j\to \mdl_{j-1},\mdl_{j},\mdl_{j+1}.
\]
Together this teaches us that $\algJ_{12},\chainop{C}_{12}$ do not change
the spin $j$ while $\chainop{A}_{12},\chainop{B}_{12}$ change the spin $j$ by one.
We can immediately see that the first two commutators 
in \eqref{eq:Int.OneLoop.YBE} vanish%
\footnote{The commutator $\comm{\Rmatrix_{12}}{\algJ_{12}}$ is trivially zero
by invariance of the R-matrix.}
because $\Rmatrix_{12}$ depends only on the total spin $j$. 

We will now choose some state $\state{j}$ from the module 
$\mdl_j$. Then $\chainop{A}_{12}$ must change the spin by one
$\chainop{A}_{12}\state{j}=\state{j+1}+\state{j-1}$
with some states $\state{j+1},\state{j-1}$ 
from the modules $\mdl_{j+1},\mdl_{j-1}$.
We note a useful identity \cite{Dolan:2003uh,Dolan:2004ps} 
to express $\chainop{B}_{12}$ in terms of the 
quadratic Casimir $\algJ_{12}^2$, c.f.~\appref{app:U224.Casimir}, 
\[\label{eq:Int.OneLoop.QIdent}
\chainop{B}_{12}=-\half\comm{\chainop{A}_{12}}{\algJ_{12}^2}.
\]
Now we can compute $\chainop{B}_{12}$ acting on $\state{j}$
making use of $\algJ_{12}^2\state{j}=j(j+1)\state{j}$,
see \eqref{eq:One.Form.Casimir},
\[\label{eq:Int.OneLoop.QAction}
\chainop{B}_{12}\state{j}=-\half\comm{\chainop{A}_{12}}{\algJ_{12}^2}\state{j}
=(j+1)\state{j+1}-j\state{j-1}.
\]
Let us now determine the remaining two terms
in \eqref{eq:Int.OneLoop.YBE} with $u=u_1-u_2$
\<\label{eq:Int.OneLoop.YBESolve}
0\eq
\bigbrk{u\comm{\Rmatrix_{12}}{\chainop{A}_{12}}+i\acomm{\Rmatrix_{12}}{\chainop{B}_{12}}}\state{j}
\nln\eq
+\bigbrk{(u+i(j+1))\Rmatrix_{j+1}-(u-i(j+1))\Rmatrix_{j}}\state{j+1}
\nl
-\bigbrk{(u+ij)\Rmatrix_{j}-(u-ij)\Rmatrix_{j-1}}\state{j-1}.
\>
Due to \eqref{eq:Int.OneLoop.SL2RExpand}, the R-matrix satisfies 
the recursion relation
\[\label{eq:Int.OneLoop.RRecurse}
\Rmatrix_{j+1}(u)=\frac{u-i(j+1)}{u+i(j+1)}\,\Rmatrix_{j}(u),
\]
which completes the proof of the Yang-Baxter equation.

\section{The Algebraic Bethe Ansatz}
\label{sec:Int.Bethe}

The Bethe ansatz determines the 
energy eigenvalues of a quantum integrable spin chain. 
It is very different from the direct diagonalisation of the
Hamiltonian in that it does not involve finding a 
matrix representation for the Hamiltonian on some basis of states.
Instead, it gives a set of algebraic equations whose solution
directly leads to the energies as well as the eigenvalues of the higher charges. 

\subsection{The Heisenberg Chain}
\label{sec:Int.Bethe.SL2}

Let us explain the Bethe ansatz in the simplest case of 
an $\alSL(2)$ chain, the so-called XXX$_{s/2}$ Heisenberg chain.  
(For a very pedagogical introduction, see \cite{Faddeev:1996iy}).
The results apply directly to the $\alSU(2)$ subsector
of \secref{sec:Dila.SU2} when $s=1$ (spin $1/2$) and
the $\alSU(1,1)$ subsectors of 
\secref{sec:One.Magic,sec:One.Baby} when 
$s=-2$ or $s=-1$.
Each eigenstate of the Hamiltonian is 
uniquely characterised by a set 
of complex Bethe roots $u_k$, $k=1,\ldots,K$,
\[\label{eq:Int.Bethe.SU2Roots}
\mbox{`\emph{Bethe roots}':}\quad \set{u_1,\ldots,u_K},
\qquad u_k\in \Comp.\]
These determine the energy $E$ and eigenvalue $U$ 
of the shift operator $\shift$ of the state by%
\footnote{The absolute value for $s$ is used for convenience; 
it makes the energy positive, 
but requires a redefinition of $u_k$ when changing the sign of $s$.}
\[\label{eq:Int.Bethe.SU2Energy}
E=\sum_{k=1}^{K} \frac{|s|}{u_k^2+\quarter s^2}\,,\qquad
U=\prod_{k=1}^{K} \frac{u_k+\frac{i}{2}|s|}{u_k-\frac{i}{2}|s|}\,.
\]
More generally, the matrix elements of the 
transfer matrix in a spin $t/2$ representation 
for a given set of roots are determined by
\<\label{eq:Int.Bethe.SU2GenTransfer}
T_t(u)\eq
\sum_{m=0}^{t}
\lrbrk{\frac{\Gammafn(iu\sign s-\half t+\half s+m)}{\Gammafn(iu\sign s-\half t-\half s+m)}\,
\frac{\Gammafn(iu\sign s-\half t-\half s)}{\Gammafn(iu\sign s-\half t+\half s)}}^L
\nlnum\nn\qquad
\times\prod_{k=1}^K
\lrbrk{\frac{u-u_k+\frac{i}{2}(-t)\sign s}{u-u_k+\frac{i}{2}(t-2m)\sign s}\,
\frac{u-u_k+\frac{i}{2}(t+2)\sign s}{u-u_k+\frac{i}{2}(t-2m+2)\sign s}}.
\>
Here the upper limit of the sum should be extended to infinity 
whenever $t$ is not a positive integer.
From the transfer matrix $T(u)=T_s(u)$ in the spin representation, $t=s$, 
we can read off the higher charges $Q_r$ via \eqref{eq:Int.Chains.TransCharge}
\[\label{eq:Int.Bethe.SU2Charges}
Q_r=\frac{i}{r-1}\sum_{k=1}^K
\lrbrk{\frac{1}{(u_k+\sfrac{i}{2}|s|)^{r-1}}
      -\frac{1}{(u_k-\sfrac{i}{2}|s|)^{r-1}}}\,,
\quad
T(u)=\prod_{k=1}^K
\frac{u-u_k-\sfrac{i}{2}|s|}{u-u_k+\sfrac{i}{2}|s|}
+\ldots\,.
\]
The charges are only valid for $r\leq L$ due to the neglected 
terms in $T(u)$ with $m\neq 0$. 

The Bethe roots are found by solving the Bethe equations
for $k=1,\ldots,K$
\[\label{eq:Int.Bethe.SU2Bethe}
\mbox{`\emph{Bethe equations}':}\qquad
\lrbrk{\frac{u_k-\sfrac{i}{2}s}{u_k+\sfrac{i}{2}s}}^L=
\prod_{\textstyle\atopfrac{l=1}{l\neq k}}^K\frac{u_k-u_l-i}{u_k-u_l+i}\,.
\]
These equations should be solved subject to the constraint that 
no two roots coincide. Furthermore, roots at infinity correspond
to descendants;
for highest-weight states there are no roots at infinity.
Note that the above Bethe equations 
follow from \eqref{eq:Int.Bethe.SU2GenTransfer} 
by cancellation of poles in $T_t(u)$ at
$u=u_k-\frac{i}{2}(t-2m+2)\sign s$.

Note that the Bethe ansatz conceptually agrees with the
particle picture presented in \secref{sec:Int.Chains.RMatrix}:
Each Bethe root can be considered as a particle. 
The right-hand side of the Bethe equations
\eqref{eq:Int.Bethe.SU2Bethe} corresponds to
scattering of two particles, while the 
left-hand side corresponds to the propagation 
of the particle across $L$ spin chain sites.
There is no interaction of more than two particles.
The phase-shifts due to these interactions
must agree for an eigenstate. 
The total energy \eqref{eq:Int.Bethe.SU2Energy}
is just the sum of the energies of the particles within the system.

We will start with the simplest example: The $\alSU(2)$ sector with 
spin representation $s=1$ (spin $1/2$) and states of the form
\[\label{eq:Int.Bethe.SU2StateSU2}
\Tr \fldZ^{L-K}\phi^K+\ldots\,.
\]
In this particular model, the spin at each site can either 
point up ($\fldZ$) or down ($\phi$).
The vacuum state with no excitations, $K=0$,
is the half-BPS state
\[\label{eq:Int.Bethe.SU2Vac}
\state{\fldZ,L}=\fldZ^L,
\]
with all spins aligned.
This is the ferromagnetic ground state of the chain. 
The excitation number $K$, giving the total number of roots, 
counts the number of $\phi$'s or
down-spins along the chain.
Assuming excitations are generated by some creation operator $\chainop{B}(u)$,%
\footnote{The operator $\chainop{B}$ is an element of the
monodromy matrix in the fundamental representation.}
a generic state has the form
\[\label{eq:Int.Bethe.SU2State}
\set{u_1,\ldots,u_K}\quad\leftrightarrow\quad
\chainop{B}(u_1)\ldots \chainop{B}(u_K)\state{\fldZ,L}\quad\leftrightarrow\quad
\fldZ^{L-K}\phi^K+\ldots\,.
\]
There is an additional constraint on the Bethe roots:
\[\label{eq:Int.Bethe.SU2Momentum}
1=U=\prod_{k=1}^K\frac{u_k+\sfrac{i}{2}}{u_k-\sfrac{i}{2}}\,.
\]
For the spin chain, it means that we have periodic boundary conditions
and we are only looking for zero-momentum states. In the
gauge theory interpretation it expresses the
cyclicity of the trace
\[\label{eq:Int.Bethe.SU2Trace}
\set{u_1,\ldots,u_K}\mbox{ with \eqref{eq:Int.Bethe.SU2Momentum}}
\quad\leftrightarrow\quad
\Tr \chainop{B}(u_1)\ldots \chainop{B}(u_K)\state{\fldZ,L}\quad\leftrightarrow\quad
\Tr\fldZ^{L-K}\phi^K+\ldots\,.
\]
For the $s=1$ Bethe ansatz, the exact eigenvalue of the 
transfer matrix \eqref{eq:Int.Bethe.SU2GenTransfer} is
\[\label{eq:Int.Bethe.SU2Transfer}
T(u)=
\prod_{k=0}^K
\frac{u-u_k-\frac{i}{2}}{u-u_k+\frac{i}{2}}
+
\lrbrk{\frac{u}{u+i}}^L
\prod_{k=0}^K
\frac{u-u_k+\frac{3i}{2}}{u-u_k+\frac{i}{2}}\,.
\]
Note that we can derive the Bethe equations from 
this expression by demanding that 
$(u+i)^L T(u)$ has no singularities.

The Dynkin label of a solution with $K$ excitations
is $[L-2K]$. Therefore, one should consider only 
solutions with $K\leq L/2$, there are clearly 
no highest-weight states with more excited spin sites.
However, the Bethe equations do have solutions
also for $K>L/2$. It is interesting to see
that for a solution with $K\leq L/2$, there exists also
a mirror solution with $K'=L+1-K> L/2$.%
\footnote{This solution has norm zero, thus it 
is not realised as a spin chain state.}
In terms of Dynkin labels, the solutions are related by $s'=-s-2$.
There is an explanation for this behaviour in terms of multiplet shortening,
see \secref{sec:N4.Modules}.
We are considering a spin chain with 
a finite (short) $s=1$ multiplet at each site,
consequently also the eigenstates form finite multiplets.
In the Bethe ansatz, shortening is not taken into account and all multiplets
are assumed to be infinite (long). 
The relevant solutions are therefore 
highest weights of reducible multiplets 
which split into two irreducible components. 
Interestingly, the Bethe ansatz finds the highest
weight states of both submultiplets and naturally
the energies and changes must agree. In some cases this 
peculiarity can be made use of by solving for the mirror 
states.

The second simplest example concerns the bosonic $\alSU(1,1)$ subsector
with spin representation $s=-1$ (spin $-1/2$) and states of the form
\[\label{eq:Int.Bethe.SU2StateBaby}
\Tr (\cder^{n_1}\fldZ) \cdots
(\cder^{n_L}\fldZ) .	
\]
Here the spins at each lattice site $p$ may take any value
$n_p=0,1,2,\ldots$, as we have an infinite $[-1]$
representation of $\alSL(2)$. 
Furthermore, the total excitation number $K=\sum n_k$ is not bounded
as in the above example.
The vacuum is still $\fldZ^L$. 
Again, the energies of the states \eqref{eq:Int.Bethe.SU2StateBaby} 
with momentum $\shift=1$ are given via 
\eqref{eq:Int.Bethe.SU2Energy,eq:Int.Bethe.SU2Bethe}.

The third example is the
fermionic $\alSU(1,1)$ subsector
with spin representation $s=-2$ (spin $-1$) and states of the form
\[\label{eq:Int.Bethe.SU2StateMagic}
\Tr  (\cder^{n_1}\Psi) \cdots
(\cder^{n_L}\Psi) .
\]
There are two chief differences as compared to the 
other subsectors:
Firstly, the fermionic nature of the fields
requires a modified cyclicity condition
\[\label{eq:Int.Bethe.SU2MomentumMagic}
(-1)^{L+1}=U=\prod_{k=1}^K\frac{u_k+i}{u_k-i}\,.
\]
For example, the ground state exists only for odd $L$,
as for even $L$ we have $\Tr \Psi^L=0$.
Secondly, the ground state does not have zero energy, but $E=2L$
\[\label{eq:Int.Bethe.SU2EnergyMagic}
E=2L+\sum_{k=1}^{K} \frac{2}{u_k^2+1}\,.
\]
In particular the ground state $\Tr\Psi^3$ is a Konishi descendant 
with $E=6$.

\subsection{Generic Algebras}
\label{sec:Int.Bethe.Generic}

In the above example the algebra was $\alSU(2)$ and thus of rank one.
There is a beautiful extension of the Bethe equations to 
an arbitrary symmetry algebra
and arbitrary representation due to Reshetikhin and
Ogievetsky, Wiegmann
\cite{Reshetikhin:1983vw,Reshetikhin:1985vd,Ogievetsky:1986hu}.
The general form also extends to the case of super algebras, 
see \cite{Saleur:1999cx} and references therein,
and is precisely what we need for $\superN=4$ SYM at one-loop.
There, we should expect Bethe equations for the 
superalgebra $\alPSU(2,2|4)$ to generate the correct spectrum.
The general equation is based on knowing the Dynkin diagram of the algebra.
The Dynkin diagram of $\alPSU(2,2|4)$ contains seven dots corresponding
to a choice of seven simple roots.
Consider a total of $K$ excitations.
For each of the corresponding Bethe roots $u_k$, $k=1,\ldots,K$,
we specify by $j_k=1,\ldots,7$
which of the seven simple roots is excited.
The Bethe equations for $k=1,\ldots,K$
can then be written in the compact form 
\[\label{eq:Int.Bethe.Bethe}
\lrbrk{\frac{u_k-\sfrac{i}{2}V_{j_k}}{u_k+\sfrac{i}{2}V_{j_k}}}^L=
\prod_{\textstyle\atopfrac{l=1}{l\neq k}}^K
\frac{u_k-u_l-\sfrac{i}{2}M_{j_k,j_l}}{u_k-u_l+\sfrac{i}{2}M_{j_k,j_l}}\,.
\]
Here, $M$ is the Cartan matrix of the algebra and $V$ 
are the Dynkin labels of the spin representation.
Furthermore, we still consider a cyclic spin chain with zero
total momentum. This gives the additional constraint%
\footnote{For a fermionic vacuum there is an additional sign
as in \eqref{eq:Int.Bethe.SU2MomentumMagic}.}
\[\label{eq:Int.Bethe.Momentum}
1=U=\prod_{k=1}^K
\frac{u_k+\sfrac{i}{2}V_{j_k}}{u_k-\sfrac{i}{2}V_{j_k}}\,.
\]
The energy of a configuration of roots that
satisfies the Bethe equations is now given by%
\footnote{In fact, the Bethe equations determine the energy 
only up to scale $c$ and a shift $e L$ 
as in \eqref{eq:Int.Bethe.SU2EnergyMagic}.}
\[\label{eq:Int.Bethe.Energy}
E=\sum_{k=1}^K
\frac{V_{j_k}}{u_k^2+\sfrac{1}{4}V_{j_k}^2}\,.
\]
Apparently, also the higher charges ($r\leq L$) 
and transfer matrix
can be obtained \cite{Reshetikhin:1987bz} 
\[\label{eq:Int.Bethe.Charges}
Q_r=\frac{i}{r-1}\sum_{k=1}^K
\lrbrk{\frac{1}{(u_k+\sfrac{i}{2}V_{j_k})^{r-1}}
      -\frac{1}{(u_k-\sfrac{i}{2}V_{j_k})^{r-1}}}\,,
\quad
T(u)=\prod_{k=1}^K
\frac{u-u_k-\sfrac{i}{2}V_{j_k}}{u-u_k+\sfrac{i}{2}V_{j_k}}
+\ldots\,.
\]

It is easily seen that restricting these equations to the Dynkin
diagram of the algebra $\alSO(6)$ reproduces the Bethe equations of
\cite{Minahan:2002ve}. It will turn out, see below, that these 
general equations,
which are well known in the literature on integrable spin chains,
indeed solve the entire problem of computing planar anomalous
dimensions in $\superN=4$ SYM, once we (\emph{i}) identify the 
correct representations of the fundamental fields on the lattice
sites, and (\emph{ii}) after resolving certain subtleties concerning
Dynkin diagrams for superalgebras. 

\subsection{The Complete Bethe Ansatz}
\label{sec:Int.Bethe.Beauty}

In \secref{sec:Int.OneLoop} we have established that
the planar one-loop dilatation operator 
of $\superN=4$ SYM is integrable. We therefore expect the general
Bethe ansatz equations \eqref{eq:Int.Bethe.Bethe} to hold. 
However, for them to be useful, we still need to specify the Dynkin 
labels, the Cartan matrix and precise form of the energy
\eqref{eq:Int.Bethe.Energy}.
Furthermore, we will perform a check of the validity of this 
$\alPSU(2,2|4)$ Bethe ansatz
which goes beyond the $\alSO(6)$ spin chain 
investigated in \cite{Minahan:2002ve}.

\begin{figure}\centering
\includegraphics{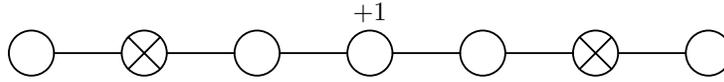}
\caption{Dynkin diagram and spin representation vector 
for the $\alPSU(2,2|4)$ Bethe ansatz.}
\label{fig:Int.Bethe.BeautyDynkin}
\end{figure}

First, we need to specify the Cartan matrix,
determined by the Dynkin diagram, and the Dynkin labels of the
spin representation corresponding to the module $\mdlF$.
For a classical semi-simple Lie algebra the Dynkin diagram is unique.
In the case of superalgebras, however, there is some freedom to distribute
the simple fermionic roots. 
For $\superN=4$ SYM the Dynkin diagram 
\figref{fig:N4.Alg.Dynkin,fig:Int.Bethe.BeautyDynkin} turns
out to be very convenient. 
On top of the Dynkin diagram \figref{fig:Int.Bethe.BeautyDynkin}
we have indicated the
Dynkin labels of the spin representation.
We write the Cartan matrix corresponding to this choice of Dynkin diagram
and the representation vector as%
\footnote{In fact, the Cartan matrix is obtained 
from this by inverting some lines.
The Bethe equations are invariant under the
inversion and it is slightly more
convenient to work with a symmetric matrix $M$.}
\[\label{eq:Int.Bethe.BeautyMatrix}
M=\left(\begin{array}{c|c|ccc|c|c}
-2&+1&  &  &  &  &   \\\hline   
+1&  &-1&  &  &  &   \\\hline
  &-1&+2&-1&  &  &   \\
  &  &-1&+2&-1&  &   \\
  &  &  &-1&+2&-1&   \\\hline
  &  &  &  &-1&  &+1 \\\hline
  &  &  &  &  &+1&-2
\end{array}\right),\qquad
V=\left(\begin{array}{r}
0\\\hline0\\\hline0\\1\\0\\\hline0\\\hline0
\end{array}\right).
\]

\begin{figure}\centering
\includegraphics{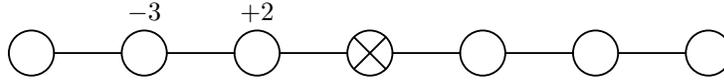}
\caption{A different Dynkin diagram and spin representation vector 
for the $\alPSU(2,2|4)$ Bethe ansatz.}
\label{fig:Int.Bethe.BeastDynkin}
\end{figure}

There exist other choices of Dynkin diagrams. E.g.~the `distinguished' one is
depicted in \figref{fig:Int.Bethe.BeastDynkin}.
We have indicated the Dynkin labels of $\mdlF$ on top.
The energy is given by \eqref{eq:Int.Bethe.Energy}, 
except for a vacuum energy shift of $3L$.
The ansatz is rather odd and
appears hardly helpful in terms of physics.
Nevertheless, it was investigated in \cite{Beisert:2003yb} and 
shown to yield the same spectrum by means of example,
a good confirmation of the validity of the Bethe ansatz methods.

\subsection{Excitation Numbers}
\label{sec:Int.Bethe.Excitations}

\begin{figure}\centering
\includegraphics{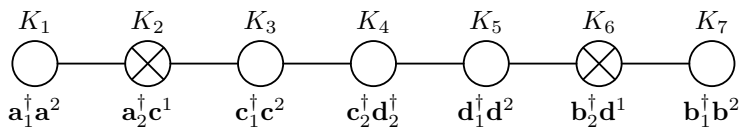}
\caption{Excitation numbers for the Bethe roots 
and associated oscillator representation.}
\label{fig:Int.Bethe.BeautyRoots}
\end{figure}

Finally, we need to obtain the number of excitations $K_j$,
$j=1,\ldots,7$, of the individual simple roots 
for a state with a given weight
\[\label{eq:Int.Bethe.BeautyWeight}
w=\weight{D_0;s_1,s_2;q_1,p,q_2;B,L}.
\]
This is most easily seen in the oscillator 
picture in \secref{sec:N4.Fund}
using the physical vacuum $\state{\fldZ,L}$.
We present the action 
of the generators
corresponding to the simple roots
in terms of creation and annihilation operators
in \figref{fig:Int.Bethe.BeautyRoots}.
It is now clear that 
$K_1=n_{\osca_1}$,
$K_2=n_{\osca_1}+n_{\osca_2}$
and so on. Using the formulas in \tabref{tab:U224.Osc.Numbers}
we write down the corresponding excitation
numbers of the simple roots
\[\label{eq:Int.Bethe.BeautyExcite}
K_j=\left(\begin{array}{l}
\half D_0-\sfrac{1}{2}(L-B)-\half s_1\\
\phantom{\half}D_0-\phantom{\half}(L-B)\\
\phantom{\half}D_0-\half(L-B)-\half p-\sfrac{3}{4}q_1-\sfrac{1}{4}q_2\\
\phantom{\half}D_0\phantom{\mathord{}-\sfrac{0}{2}(L-B)}-\phantom{\half}p-\half q_1-\half q_2\\
\phantom{\half}D_0-\half(L+B)-\half p-\sfrac{1}{4}q_1-\sfrac{3}{4}q_2\\
\phantom{\half}D_0-\phantom{\half}(L+B)\\
\half D_0-\half(L+B)-\half s_2
\end{array}\right).\]
Not all excitations of the simple roots correspond to
physical states. Obviously, the excitation numbers of the
oscillators must be non-negative, this gives the bounds
\[\label{eq:Int.Bethe.BeautyBounds}
0\leq K_1\leq K_2\leq K_3\leq 
K_4\geq K_5\geq K_6\geq K_7\geq 0.
\]
Furthermore, each fermionic oscillator cannot be excited more than
once, this gives the bounds%
\footnote{Superconformal primaries reside in the 
fundamental Weyl chamber defined by the bounds
$-2K_1+K_2>-1$, 
$K_2-2K_3+K_4>-1$, 
$K_3-2K_4+K_5+L>-1$, 
$K_4-2K_5+K_6>-1$, 
$K_6-2K_7>-1$.
Together with \eqref{eq:Int.Bethe.BeautyBounds} this implies, among other relations, 
\eqref{eq:Int.Bethe.BeautyBounds2}.
Solutions of the Bethe equations outside the fundamental 
domain apparently correspond to mirror images of primary states 
due to reflections at the chamber boundaries.}
\[\label{eq:Int.Bethe.BeautyBounds2}
K_2+2L\geq K_3+L\geq K_4\leq K_5+L\leq K_6+2L.\]

Certainly, we should obtain the $\alSO(6)$ subsector 
studied by Minahan and Zarembo \cite{Minahan:2002ve}
when we remove the outer four simple roots from the
Dynkin diagram in \figref{fig:Int.Bethe.BeautyDynkin}.
When we restrict to the states of this 
subsector the number of excitations \eqref{eq:Int.Bethe.BeautyExcite} of
the outer four roots is trivially zero. They become
irrelevant for the Bethe ansatz and can be discarded.
Thus all solutions to the $\alSO(6)$ Bethe equations
are also solutions to the $\alPSU(2,2|4)$ Bethe equations.
What is more, we can apply this Bethe ansatz to a 
wider range of operators, in fact, to \emph{all}
single-trace operators of $\superN=4$ SYM.

\subsection{Multiplet Splitting}
\label{sec:Int.Bethe.Splitting}

Now we can write down and try to solve the Bethe equations 
for any state in $\superN=4$ SYM. 
Note, however, that the Bethe equations need to be solved 
only for highest weight states. 
All descendants of a highest weight state are obtained by
adding Bethe roots at infinity, $u_k=\infty$. 
In other words, the solutions to the Bethe equations
corresponding to highest weight states
are distinguished in that they have no roots $u_k$ at infinity.
Nevertheless, there is one subtlety related to this point
which can be used to our advantage.
Namely this is multiplet splitting at the unitarity bounds
as discussed in \secref{sec:N4.Split}.
We assume that the spin chain of $L$ sites transforms in the
tensor product of $L$ spin representations. 
The corrections $\delta D$ to the scaling dimension
induced by the Hamiltonian $\ham$ are not included in this picture.
Thus, in terms of the spin chain, 
only the classical $\alPSU(2,2|4)$ algebra applies
where the scaling dimension is exactly $D_0$.
The shortening conditions given in \secref{sec:N4.Split}
can also be expressed in terms
of excitations of simple roots, we find 
\<\label{eq:Int.Bethe.BeautyShorten}
\mathrm{i}:&\quad& K_1+K_3=K_2+1,\nln
\mathrm{ii}:&\quad& K_7+K_5=K_6+1.
\>
The corresponding offsets translate into
\<\label{eq:Int.Bethe.BeautyOffsets}
\delta K\indup{i}\eq
(\phantom{+}0,-1,-1,\phantom{+}0,\phantom{+}0,\phantom{+}0,\phantom{+}0),\quad \delta L=1,\qquad\mbox{for }K_2>K_1,
\nln
\delta K\indup{I}\eq
(\phantom{+}0,\phantom{+}0,-1,\phantom{+}0,\phantom{+}0,\phantom{+}0,\phantom{+}0),\quad \delta L=1,\qquad\mbox{for }K_2=K_1,
\nln
\delta K\indup{ii}\eq
(\phantom{+}0,\phantom{+}0,\phantom{+}0,\phantom{+}0,-1,-1,\phantom{+}0),\quad \delta L=1,\qquad\mbox{for }K_6>K_7,
\nln
\delta K\indup{II}\eq
(\phantom{+}0,\phantom{+}0,\phantom{+}0,\phantom{+}0,-1,\phantom{+}0,\phantom{+}0),\quad \delta L=1,\qquad\mbox{for }K_6=K_7.
\>
We thus see that in the case of multiplet shortening,
the primaries of higher submultiplets have less excitations. 
In an explicit calculation this may reduce the complexity 
of the Bethe equations somewhat as we shall see in an example below.

Multiplet splitting is an extremely interesting
issue from the point of view of integrability.
Let us consider some operator acting on a
spin chain. Assume the operator
is invariant under the classical 
algebra $\alPSU(2,2|4)$. In the most 
general case, this operator can assign 
a different value to all irreducible multiplets
of states.
In particular this is so for the submultiplets 
of a long multiplet at the unitarity bound,
see \secref{sec:N4.Split}.
Now, if we impose integrability on the operator,
we obtain the one-loop planar correction to the 
dilatation operator of $\superN=4$ SYM.
In $\superN=4$ SYM the
submultiplets rejoin into a long multiplet
and for consistency they must be degenerate
(note that the momentum constraint $U=1$ is crucial
for this observation).
A priori, from the point of view of the spin chain, 
this seems like a miracle,
especially in view of the fact that the submultiplets
have a different number of spin sites $L$!
Why should integrability imply \emph{this} degeneracy?
For a simple manifestation of this fact, one may consider
the fermionic $\alSU(1,1)$ subsector discussed in 
\secref{sec:One.Magic}. This subsector has an additional
$\alU(1|1)$ supersymmetry, which relates states of
different length and which was used to 
construct the complete one-loop dilatation generator. 
Obviously the solutions to the Bethe ansatz
for this system, c.f.~\secref{sec:Int.Bethe.SL2}, must
display this symmetry (note the momentum constraint).
What is the origin of this symmetry (putting $\superN=4$ SYM aside)?
It almost seems as if integrability selects the one invariant operator
which is suitable as a consistent deformation
of the dilatation generator!
Then, clearly the miracle would turn into
the condition for integrability.

\subsection{Degenerate Pairs}
\label{sec:Int.Bethe.Pairs}

The Bethe equations are invariant under 
the map 
\[\label{eq:Int.Bethe.Parity}
\set{u_k}\mapsto \set{-u_k}.
\]
Also the energy and all even charges are invariant,
the odd charges change sign,
\[\label{eq:Int.Bethe.ChargeParity}
Q_r\mapsto (-1)^r Q_r.
\]
This operation is
most naturally identified with parity $\gaugepar$.
Therefore, for every solution $\set{u_k}$ there is another
solution $\set{-u_k}$ with degenerate energy and even charges, but
negative odd charges
\[\label{eq:Int.Bethe.Paired}
\mbox{`\emph{paired states}':}\quad \set{u_k}\neq \set{-u_k}.
\]
Unpaired states are such states for which
\[\label{eq:Int.Bethe.Unpaired}
\mbox{`\emph{unpaired state}':}\quad \set{u_k}=\set{-u_k}.
\]
This is the manifestation of the findings of
\secref{sec:Int.Chains.Pairs} within the Bethe ansatz.%
\footnote{The parity eigenvalue $P$ seems to be determined
by the number of Bethe roots at zero and the length.}

\section{Spectrum}
\label{sec:Int.Spec}

In order to illustrate the application of the Bethe ansatz, 
we shall repeat the investigation of the spectrum in
\secref{sec:One.Spec} with the Bethe ansatz.
We will see that, except in a few examples,
it is rather tedious to find exact solutions to the Bethe 
equations. 
In the following section, however, we will
investigate a class of states for which the 
Bethe ansatz is of tremendous importance.

\subsection{Example}
\label{sec:Int.Spec.Example}

In the following, we will apply the 
complete Bethe ansatz to the two-parton state with 
highest weight (c.f.~\secref{sec:One.Spec.TwistTwo})
\[\label{eq:Int.Spec.Twist}
w=\weight{4;2,2;0,0,0;0,2}.
\]
Using \eqref{eq:Int.Bethe.BeautyExcite}
we find the excitation numbers and length
\[\label{eq:Int.Spec.TwistExcite}
K_{0,j}=(0,2,3,4,3,2,0),\quad L_0=2.
\]
This weight is 
on both unitarity bounds, c.f.~\eqref{eq:Int.Bethe.BeautyShorten}, 
the excitation numbers 
of the highest submultiplet, c.f.~\eqref{eq:Int.Bethe.BeautyOffsets}, 
are
\[\label{eq:Int.Spec.TwistExcite2}
K_j=K_{0,j}+\delta K\indup{I}{}_{,j}+\delta K\indup{II}{}_{,j}
=(0,1,2,4,2,1,0),\quad L=L_0+2\delta L=4.
\]
We therefore configure the simple roots as follows
\[\label{eq:Int.Spec.TwistConfig}
j_k=(2,3,3,4,4,4,4,5,5,6).
\]
Now we note that twist-two states are unpaired states.
Therefore the configurations of Bethe roots
must be invariant under the symmetry $\set{u_k}\mapsto \set{-u_k}$. 
This tells us
\[\label{eq:Int.Spec.TwistSym}
u_1=u_{10}=0,\quad
u_2=-u_3,\quad
u_4=-u_5,\quad
u_6=-u_7,\quad
u_8=-u_9
\]
and the momentum constraint \eqref{eq:Int.Bethe.Momentum}
is automatically satisfied.
Furthermore the excitations \eqref{eq:Int.Spec.TwistExcite2}
are invariant under flipping the Dynkin diagram,
$K_j\mapsto K_{8-j}$. This suggests the ansatz
\[\label{eq:Int.Spec.TwistSym2}
u_2=u_8.
\]
The Bethe equations \eqref{eq:Int.Bethe.Bethe} are then solved exactly by 
\[\label{eq:Int.Spec.TwistRoots}
u_2=\sqrt{\frac{5}{7}}\,,
\quad
u_{4,6}=\sqrt{\frac{65\pm 4\sqrt{205}}{140}}\,,
\]
which yields the energy \eqref{eq:Int.Bethe.Energy}
\[\label{eq:Int.Spec.TwistEnergy}
E=\frac{25}{3}\,.
\]
This is indeed the energy of the twist-two state at dimension four,
c.f.~\secref{sec:One.Spec.TwistTwo} and \cite{Anselmi:1998ms}.

Note that in some cases it may be more convenient to use
a different Dynkin diagram from the one in
\figref{fig:Int.Bethe.BeautyDynkin}. In this example,
the distinguished Dynkin diagram 
\figref{fig:Int.Bethe.BeastDynkin} would require only 
two Bethe roots \cite{Beisert:2003yb}. 
Alternatively, one might consider one of the 
$\alSU(1,1)$ subsectors of
\secref{sec:One.Magic,sec:One.Baby} to simplify the investigation.

\subsection{Two Excitations}
\label{sec:Int.Spec.TwoEx}

States with two excitations, see \secref{sec:One.Spec.TwoEx}, 
are the simplest solutions to the Bethe equations \cite{Minahan:2002ve}.
Let us consider the two-excitation state 
of the $\alSU(2)$ subsector first. For this purpose, we can restrict
to the Bethe ansatz for the Heisenberg XXX$_{1/2}$ spin chain.
We should solve the Bethe equations for two roots $u_{1,2}$.
Let us start with the momentum constraint
\[\label{eq:Int.Spec.TwoMom}
1=U=
\frac{u_1+\sfrac{i}{2}}{u_1-\sfrac{i}{2}}\,
\frac{u_2+\sfrac{i}{2}}{u_2-\sfrac{i}{2}}\,,
\]
this requires $u_2=-u_1$. Now the Bethe equations for $u_1$ and $u_2$ 
collapse to the single equation
\[\label{eq:Int.Spec.TwoBethe}
\lrbrk{\frac{u_1-\sfrac{i}{2}}{u_1+\sfrac{i}{2}}}^L=
\frac{2u_1-i}{2u_1+i}\qquad
\mbox{or}\qquad
\lrbrk{\frac{u_1-\sfrac{i}{2}}{u_1+\sfrac{i}{2}}}^{L-1}=1.
\]
This equation has the solutions  
\[\label{eq:Int.Spec.TwoSol}
u_{1,2}=\pm \frac{1}{2} \cot \frac{\pi n}{L-1}\,
\]
for $0\leq n< (L-1)/2$.
The energy of this solution is
\[\label{eq:Int.Spec.TwoEng}
E=8\sin^2\frac{\pi n}{L-1}\,,
\]
in agreement with \secref{sec:One.Spec.TwoEx}.
In addition to the energy, we can also compute the values of
the higher charges
\[\label{eq:Int.Spec.TwoCharge}
Q_{r}=\frac{\bigbrk{1+(-1)^r}2^{r}}{r-1}\,
\sin\frac{\pi (r-1)n}{L-1}\,\sin^{r-1}\frac{\pi n}{L-1}\,.
\]

The two-excitation multiplets are actually at both unitarity
bounds and split up in the classical theory. As far as
the one-loop Bethe ansatz is concerned, the classical 
symmetry algebra applies and we should be able to find 
further solutions corresponding to the three other 
submultiplets, \secref{sec:One.Spec.TwoEx}.
The highest weight of the top multiplet is 
$\weight{L;0,0;0,L-2,0;0,L}$
According to \eqref{eq:Int.Bethe.BeautyExcite}, it all requires
two excitations of type $4$ and one excitation of types $3,5$ each.
We will configure the Bethe roots as $j_k=(4,4,3,5)$. 
The solution to the Bethe equations is found straightforwardly
\[\label{eq:Int.Spec.TwoTop}
u_{1,2}=\pm \frac{1}{2} \cot \frac{\pi n}{L+1}\,,\quad
u_3=u_4=0,
\qquad
E=8\sin^2\frac{\pi n}{L+1}\,.
\]
Finally, the highest weights of the middle submultiplets are
$\weight{L;0,0;2,L-3,0;0,L}$ and its conjugate. The solutions
to the root configuration $j_k=(4,4,3)$ or $j_k=(4,4,5)$ are
\[\label{eq:Int.Spec.TwoMiddle}
u_{1,2}=\pm \frac{1}{2} \cot \frac{\pi n}{L}\,,\quad
u_3=0,
\qquad
E=8\sin^2\frac{\pi n}{L}\,.
\]
Their energies agree precisely with the
results of \secref{sec:One.Spec.TwoEx}.

\subsection{Three Excitations}
\label{sec:Int.Spec.ThreeEx}

In \secref{sec:One.Spec.ThreeEx} we have investigated a peculiar set
of states with three impurities in the $\alSU(2)$ sector 
and found their exact
planar one-loop energies and eigenstates. 
The energy of all states turned out to be the same for all
states, $E=6$. 
This is best understood in terms of the Bethe equations
of which they are very special solutions. 
The states are unpaired and therefore we should
expect the Bethe roots to be invariant under
$\set{u_k}\mapsto \set{-u_k}$. This requires that $u_3=0$ is 
one of the roots and $u_1=-u_2$. Unfortunately, this seems
to imply $U=-1$ and violate the trace condition.
However, the singular points $u_{1,2}=\pm \sfrac{i}{2}$
can invert the momentum once again.%
\footnote{The singular roots lead to states with
sticky excitations which are always on adjacent
spin chain sites, c.f.~\eqref{eq:One.Spec.ThreeOp}.}
Therefore the roots must be 
\[\label{eq:Int.Spec.ThreeRoots}
u_{1,2}=\pm \sfrac{i}{2}\,,\qquad u_3=0.
\]
The singularity needs to be regularised,
e.g.~the Bethe equations and the energy formula are naively divergent.
It is best to consider the transfer matrix 
\<\label{eq:Int.Spec.ThreeTrans}
T(u)\eq
\frac{u}{(u+i)}\,
\frac{(u-i)}{u}\,
\frac{(u-\frac{i}{2})}{(u+\frac{i}{2})}
+
\lrbrk{\frac{u}{u+i}}^L
\frac{(u+2i)}{(u+i)}\,
\frac{(u+i)}{u}\,
\frac{(u+\frac{3i}{2})}{(u+\frac{i}{2})}
\nln\eq
\frac{(u-i)}{(u+i)}\,
\frac{(u-\frac{i}{2})}{(u+\frac{i}{2})}+
\lrbrk{\frac{u}{u+i}}^{L-1}
\frac{(u+2i)}{(u+i)}\,
\frac{(u+\frac{3i}{2})}{(u+\frac{i}{2})}\,.
\>
It is easy to confirm that $(u+i)^L T(u)$ has no poles
for even $L$ and thus $\set{u_{1,2,3}}$ is indeed a solution, even if 
the Bethe equations and energy formula appear divergent.
From this expression it is also straightforward to
derive the energy $E=6$ and higher charges $Q_r$ 
\[\label{eq:Int.Spec.ThreeCharges}
Q_r=\frac{\bigbrk{(+i)^{r-2}+(-i)^{r-2}}\bigbrk{2^{r-1}+1}}{(r-1)}\,,\qquad
r\leq L-2,
\]
which can clearly seen
to be independent of $L$ for small $r$.

\section{The Thermodynamic Limit}
\label{sec:Int.Thermo}

The BMN limit (c.f.~\secref{sec:One.BMN}) is very interesting for 
the AdS/CFT correspondence
because it allows 
to make contact to (plane-wave) string theory
on a quantitative level. 
In the BMN limit, the length of the spin chain 
approaches infinity, $J\sim L\to\infty$, while 
the number of excitations is fixed at a finite value.
This requires some rescaling of energies.

The thermodynamic limit is a generalisation of the BMN limit 
in that the spin chain grows very long while focusing 
on the low energy spectrum. The difference to the BMN limit is that 
the number of excitations is 
proportional to $L$ and also approaches infinity.
In this case, the Bethe equations 
turn into integral equations, similar to the
ones found in matrix models.
As in the BMN limit, one can make contact to string theory
as will be seen in the following section. Here, we will 
lay the foundation for this comparison on a general level. 
For a beautiful review of the thermodynamic limit of the Bethe equations
and the arising Riemann surfaces, see \cite{Kazakov:2004qf}.

\subsection{The Heisenberg Chain}
\label{sec:Int.Thermo.SU2}

Here we will outline the thermodynamic limit of the 
Bethe ansatz system of equations
\eqref{eq:Int.Bethe.SU2Energy,eq:Int.Bethe.SU2Bethe} 
for the case of the XXX$_{s/2}$
Heisenberg spin chain
with length $L$ and $K$ excitations
(c.f.~\cite{Sutherland:1995aa})
\[\label{eq:Int.Thermo.BetheExact}
E=\sum_{k=1}^{K} \frac{|s|}{u_k^2+\quarter s^2}\,,
\qquad
1=\prod_{k=1}^K
\frac{u_k+\sfrac{i}{2}|s|}{u_k-\sfrac{i}{2}|s|}
\,,\qquad
\lrbrk{\frac{u_k-\sfrac{i}{2}s}{u_k+\sfrac{i}{2}s}}^L=
\prod_{\textstyle\atopfrac{l=1}{l\neq k}}^K\frac{u_k-u_l-i}{u_k-u_l+i}\,.
\]
For a large length $L$
and solutions of a sufficiently low energy,
we expect that the positions of the roots $u_k$
scale as $L$, see \eqref{eq:Int.Spec.TwoSol} \cite{Minahan:2002ve}. 
Let us therefore define $u_k=L \tilde u_k$.%
\footnote{Interestingly, we might include $s$ in the rescaling 
and remove it completely from the equations.}
We then take the logarithm of the equations \eqref{eq:Int.Thermo.BetheExact}
and obtain for large $L$ 
\[\label{eq:Int.Thermo.BetheThermo}
\tilde E=LE=\frac{1}{L}\sum_{k=1}^{K}
\frac{|s|}{\tilde u_k^2}\,,
\qquad
2\pi n=\frac{1}{L}\sum_{k=1}^{K}
\frac{|s|}{\tilde u_k}\,,\qquad
2\pi n_k-\frac{s}{\tilde u_k}=\frac{1}{L}
\sum_{\textstyle\atopfrac{l=1}{l\neq k}}^{K} 
\frac{2}{\tilde u_l-\tilde u_k}\,.
\]
The integer mode numbers $n_k,n$ 
enumerate the possible branches of the
logarithm. The rescaled energy $\tilde E=LE$ was defined such that
there is one power of $1/L$ in front of the sum. This will cancel
against the $\order{L}$ terms of the sum.%
\footnote{This is the chief difference
to the BMN limit, where there are only finitely many excitations.
Consequently, in the BMN limit one would define $\hat E=L^2E$.}
The total rescaled dimension is thus given by 
\[\label{eq:Int.Thermo.DimensionThermo}
\tilde D=D/L=\tilde D_0+\tilde g^2 \tilde E+\order{\tilde g^4}\,
\]
where we have introduced the effective 
coupling constant $\tilde g^2$ similar
to the BMN coupling $\hat g^2\sim\lambda'$, see \secref{sec:One.BMN}%
\footnote{The length $L$ in this section corresponds 
to the combination $J+M$ from discussion of the BMN limit. 
Hence, we distinguish between the coupling constant $\tilde g=g/L$ 
for the thermodynamic limit
and $\hat g=g/J$ from \secref{sec:One.BMN}.}
\[\label{eq:Int.Thermo.Coupling}
\tilde g^2=\frac{g^2}{L^2}=
\frac{\lambda}{8\pi^2 L^2}\,.\]
Likewise, the charges and transfer matrix 
in the thermodynamic limit are given by \cite{Arutyunov:2003rg}%
\footnote{For the BMN limit one would define
$\hat Q_r=L^{r} Q_r$ and 
$\hat T(\hat u)=T(\hat u L)^L$
to account for the different scaling of the number of excitations.}
\[\label{eq:Int.Thermo.ChargesThermo}
\tilde Q_r=L^{r-1} Q_r=
\frac{1}{L}\sum_{k=1}^K\frac{|s|}{\tilde u_k^{r}}\,,
\quad
-i\log \tilde T(\tilde u)
=-i\log T(\tilde u L)
=\frac{1}{L}\sum_{k=1}^K \frac{|s|}{\tilde u_k-\tilde u}+
\ldots\,.
\]

We shall start by assuming that
in the large $L$ limit the Bethe roots accumulate on $A$ smooth
contours $\contour_a$, the so-called `Bethe-strings'
\[\label{eq:Int.Thermo.Contours}
\mbox{`\emph{Bethe-strings}':}\quad
\contour_1, \ldots ,\contour_A.
\]
It is reasonable, therefore, to replace the discrete
root positions $\tilde u_k$ by a smooth continuum variable $\tilde u$
and introduce a density $\rho(\tilde u)$ describing the distribution
of the roots in the complex $\tilde u$-plane:
\[\label{eq:Int.Thermo.Density}
\frac{1}{L} \sum_{k=1}^K \longrightarrow
\int_{\contour} d\tilde u\,\rho(\tilde u),
\]
where $\contour$ is the support of the density, i.e.~the 
union of contours $\contour_a$ along which the roots are distributed.
The density is normalised to the filling fraction
$\tilde K=K/L$, 
\[\label{eq:Int.Thermo.Filling}
\int_{\contour}d\tilde u\, \rho(\tilde u)=\tilde K.\]
Moreover, we may specify solutions by 
the contour filling fractions, 
i.e. the numbers of roots $L\tilde K_a$ residing
on each contour $\contour_a$, by
\[\label{eq:Int.Thermo.ContourFilling}
\int_{\contour_a}d\tilde u\, \rho(\tilde u)=\tilde K_a.\]

The Bethe equations \eqref{eq:Int.Thermo.BetheThermo} 
in the `thermodynamic limit'
then conveniently turn into singular integral equations:
\[\label{eq:Int.Thermo.SU2Bethe}
\tilde E=|s|\int_{\mathcal{C}}\frac{d\tilde u\,\rho(\tilde u)}{\tilde u^2}\,,
\qquad
2\pi n=|s|\int_{\mathcal{C}}\, \frac{d\tilde u\, \rho(\tilde u)}{\tilde u}\,,
\qquad
2\pi n_{\tilde u}-\frac{s}{\tilde u}=
2\pint_{\mathcal{C}}\frac{d\tilde v\, \rho(\tilde v)}{\tilde v-\tilde u}\,,
\]
where $n_{\tilde u}$ is the mode number $n_k$ at point $\tilde u=\tilde u_k$.
It is expected to be constant, 
$n_{\tilde u}=n_a$, along each contour $\contour_a$ and
contours are distinguished by their mode number.
Here and in the following, the slash through the integral sign
implies a principal part prescription.
In addition, we have 
a consistency condition derived from
the right of \eqref{eq:Int.Thermo.SU2Bethe} by integrating both sides 
over $\contour$ and using 
\eqref{eq:Int.Thermo.ContourFilling}:
$n=\sign s\sum_{a=1}^A \tilde K_a n_a$.
Finally, we can compute the eigenvalues of the higher charges
\eqref{eq:Int.Bethe.Charges},
they read \cite{Arutyunov:2003rg}
\[
\label{eq:Int.Thermo.SU2Charges}
\tilde Q_r=
|s|\int_{\contour}\frac{\,d\tilde u\,\rho(\tilde u)}{\tilde u^{r}}\,,\qquad
G(\tilde u)=
|s|\int_{\mathcal{C}}\frac{d\tilde v\,\rho(\tilde v)}{\tilde v-\tilde u}\,.
\]
The resolvent $G(\tilde u)$ is a central object
of a solution. 
It is defined by
\[
\label{eq:Int.Thermo.SU2Resolv}
G(\tilde u)=\sum_{r=1}^{\infty}\tilde u^{r-1}\tilde Q_r\qquad
\mbox{with}\quad Q_1=-i\log U=2\pi n,
\]
so naively one might think $G(\tilde u)=-i\log \tilde T(\tilde u)$.
This is not quite true due to the omitted terms in the eigenvalue
of the transfer matrix \eqref{eq:Int.Bethe.SU2Charges}.
The additional term is of $\order{u^L}$, so in the large $L$ limit one might
be tempted to drop it. However, let us see what happens for $s=1$ for which we
know the exact transfer matrix \eqref{eq:Int.Bethe.SU2Transfer}.
The second term is multiplied by $u^L/(u+i)^L$. In the thermodynamic limit
this becomes
\[
\label{eq:Int.Thermo.SU2NonPert}
\lrbrk{\frac{u}{u+i}}^L=
\lrbrk{\frac{\tilde u L}{\tilde u L+i}}^L=
\lrbrk{1+\frac{i}{L\tilde u}}^{-L}\longrightarrow
\exp\lrbrk{-\frac{i}{\tilde u}},
\]
which is indeed non-zero despite the suppression by $u^L$.
In total we obtain for the eigenvalue of the transfer matrix
\[
\label{eq:Int.Thermo.SU2TransferExact}
\tilde T(\tilde u)=\exp\bigbrk{iG(\tilde u)}
+\exp\bigbrk{-iG(\tilde u)-i/\tilde u}
=\exp\bigbrk{-i/2\tilde u}\,
2\cos \bigbrk{G\indup{sing}(\tilde u)}.
\]
The exponential prefactor may now be absorbed into the definition 
of $\tilde T$ and we merely have $2\cos G\indup{sing}(\tilde u)$
with the singular resolvent
\[
\label{eq:Int.Thermo.SU2ResolvSing}
G\indup{sing}(\tilde u)=G(\tilde u)+\frac{1}{2\tilde u}\,.
\]
Gladly, the additional terms only change the form of 
the transfer matrix, all physically relevant information 
is encoded into the non-singular resolvent $G$.
The resolvent $G$ may therefore be obtained 
even in ignorance of the additional 
terms in $\tilde T$.

Note that the Bethe equation
\eqref{eq:Int.Thermo.BetheThermo}
can alternatively be obtained as a consistency
condition on the transfer matrix $\tilde T(\tilde u)$.
The resolvent has many sheets, but $2\cos G'\indup{sing}(\tilde u)$ 
must be single-valued on the complex $\tilde u$ plane. 
This requires
\[\label{eq:Int.Thermo.BetheResolv}
G\indup{sing}(\tilde u+i\epsilon)+G'\indup{sing}(\tilde u-i\epsilon)=2\pi n_{\tilde u}
\]
across a branch cut of $G$ at $\tilde u$,
which is an equivalent formulation of the
Bethe equation \eqref{eq:Int.Thermo.BetheThermo}.

\subsection{Generic Algebras}
\label{sec:Int.Thermo.Generic}

Let us briefly state the generalisation of the thermodynamic limit for 
arbitrary groups with Cartan matrix $M$ and representation labels $V$. 
In addition to the mode numbers $n_a$, here we have to
specify for each contour $\contour_a$ to 
which simple root $j_a$ of the algebra it belongs.
The energy, momentum constraint and Bethe equations are 
\[\label{eq:Int.Thermo.Bethe}
\tilde E=\int\frac{d\tilde u\,\rho(\tilde u)\,V_{\tilde u}}{\tilde u^2}\,,
\qquad
2 \pi n=\int\frac{d\tilde u\,\rho(\tilde u)\,V_{\tilde u}}{\tilde u}\,,
\qquad
2\pi n_{\tilde u}-
\frac{V_{\tilde u}}{\tilde u}=
\pint\frac{d\tilde v\, \rho(\tilde v)\, M_{\tilde u,\tilde v}}{\tilde v-\tilde u}\,.
\]
Here we have used the short notation 
$M_{\tilde u,\tilde v}=M_{j_a,j_{a'}}$ 
or $V_{\tilde u}=V_{j_a}$ 
for the element of the Cartan matrix or representation vector
corresponding to the simple roots of the contours 
$\tilde u\in \contour_a, \tilde v\in \contour_{a'}$.
The higher charges \eqref{eq:Int.Bethe.Charges} 
and resolvent \eqref{eq:Int.Thermo.SU2Resolv} as
their generating function, 
$G(\tilde u)\approx -i\log \tilde T(\tilde u)$,
are given by 
\[\label{eq:Int.Thermo.Charges}
\tilde Q_r=\int\frac{d\tilde u\,\rho(\tilde u)\,V_{\tilde u}}{\tilde u^r}\,,
\qquad
G(\tilde u)=
\int\frac{d\tilde v\,\rho(\tilde v)\,V_{\tilde v}}{\tilde v-\tilde u}\,.
\]
%

\section{Stringing Spins}
\label{sec:Int.Spinning}

Following the work \cite{Gubser:2002tv},
Frolov and Tseytlin proposed
a novel possibility for a quantitative comparison 
of string theory and gauge theory \cite{Frolov:2003qc,Frolov:2003tu},
see \cite{Tseytlin:2003ii} for a nice review of the subject.
They suggested to investigate states with 
large charges (angular momenta) of both, the conformal symmetry $\alSU(2,2)$ and 
the internal symmetry $\alSU(4)$. 
In the case of string theory it was understood 
\cite{Gubser:2002tv,Frolov:2002av} 
(see also \cite{Russo:2002sr,Minahan:2002rc,Tseytlin:2002ny,Alishahiha:2002fi,Aleksandrova:2003nn,Alishahiha:2004vi,Dimov:2004xi,Bigazzi:2004yt}) 
that the string sigma model
can be efficiently treated by semi-classical methods. 
On the gauge theory side, it was realised that such states 
can be treated with the Bethe ansatz in the thermodynamic limit
\cite{Beisert:2003xu}. 
This lead to a remarkable agreement at the one-loop
level \cite{Beisert:2003xu,Beisert:2003ea}. 
We shall use this example to illustrate the
use of the Bethe ansatz. 
First, we will shortly review the string theory 
computations, then derive the gauge theory result and compare. 

\subsection{String Theory Details}

We will investigate a folded string
(a closed string, which is folded to a line)
which stretches along a spatial direction of $AdS_5$. 
It rotates with angular momentum (spin) $S$ around its centre of mass
and moves in the time direction with energy $D$
as well as on a great circle of $S^5$ with angular momentum (charge) $J$.
For this string configuration, 
we would like to find the dependence of the energy $D$
on the charges $S,J$
\[\label{eq:Int.Spinning.Aim}
D=D(S,J).
\]

We will make the ansatz that 
the embedding coordinates of the string world sheet, 
parameterised by $\tau,\sigma$, are given by
\[\label{eq:Int.Spinning.Coords}
t=\kappa \tau,\quad 
\phi_1=\phi=\omega \tau,\quad
\varphi_3=\varphi=w \tau,\quad
\rho=\rho(\sigma)=\rho(\sigma+2\pi),
\]
all the other coordinates are zero. In fact, the
string moves only in a subspace $AdS_3\times S^1$ of $AdS_5\times S^5$.
The relevant part of the $AdS_5\times S^5$ metric is
\[\label{eq:Int.Spinning.Metric}
ds^2 = d\rho^2-\cosh^2 \rho \,dt^2 + \sinh^2\rho \, d\phi^2 +d\varphi^2 .
\]

The string theory sigma model is given by the Polyakov action
\[\label{eq:Int.Spinning.SigmaModel}
S\indup{string}=\sqrt{\lambda}
\int d\tau\int\frac{d\sigma}{2\pi}\,\half\, G_{MN}(\dot X^M\dot X^N-X^{\prime\,M} X^{\prime\,N})
\]
together with the Virasoro constraints
\[\label{eq:Int.Spinning.Virasoro}
G_{MN}(\dot X^M\dot X^N+X^{\prime\,M} X^{\prime\,N})=G_{MN}\dot X^M X^{\prime\,N}=0.
\]
The equations of motion following from the action are
\[\label{eq:Int.Spinning.EOM}
\frac{\partial}{\partial \tau}(G_{MN}\dot X^N)-
\frac{\partial}{\partial \sigma}(G_{MN} X^{\prime N})=0.
\]
The conserved charges $D,S,J$ corresponding to 
$t,\phi,\varphi$ are determined using 
\eqref{eq:Int.Spinning.SigmaModel,eq:Int.Spinning.Metric}
\[\label{eq:Int.Spinning.Charges}
D=\kappa \sqrt{\lambda}\int \frac{d\sigma}{2\pi}\,\cosh^2 \rho,\quad
S=\omega \sqrt{\lambda}\int \frac{d\sigma}{2\pi}\,\sinh^2 \rho,\quad
J=w\sqrt{\lambda}\,.
\]
From the prefactor of the action we can now infer 
that quantum loops are counted by $1/\sqrt{\lambda}$ if $\lambda$ is
large. Furthermore, we see that, as $J=w\sqrt{\lambda}$, 
quantum loops are effectively counted by $1/J$ if we fix $w$.
Therefore, if we content ourselves with the leading order in 
an expansion with respect to $1/J$ and $w$ fixed, we can neglect
all quantum loops and consider the classical string theory.

In the classical model, the parameter $\lambda$ can be absorbed 
into the definition of the charges, we use
\[\label{eq:Int.Spinning.Curly}
\Dcurl=\frac{D}{\sqrt{\lambda}}=\kappa \int \frac{d\sigma}{2\pi}\,\cosh^2 \rho,\quad
\Scurl=\frac{S}{\sqrt{\lambda}}=\omega \int \frac{d\sigma}{2\pi}\,\sinh^2 \rho,\quad
\Jcurl=\frac{J}{\sqrt{\lambda}}=w.
\]
The only non-trivial equation of motion is
\[\label{eq:Int.Spinning.EOMAdS3}
\rho''-(\kappa^2-\omega^2)\sinh \rho\cosh \rho=0.
\]
and the non-trivial Virasoro constraint is
\[\label{eq:Int.Spinning.ViraAdS3}
\rho^{\prime\,2}-\kappa^2 \cosh^2\rho+\omega^2\sinh^2 \rho+w^2=0.
\]

To solve the system, we will assume that 
$\rho(\sigma)$ is a periodic function 
stretching between $\pm\rho_0$. By inverting the function 
$\rho(\sigma)$ to $\sigma(\rho)$ (for half of the period)
we can rewrite the worldsheet integrals as
\[\label{eq:Int.Spinning.IntTrans}
\int_{0}^{2\pi} \frac{d\sigma}{2\pi}=
\frac{2}{2\pi}\int_{-\rho_0}^{\rho_0} \frac{d\rho}{\rho'}\,.
\]
There are two points to be taken into account. 
Firstly, an integral without an integrand should yield $1$
and secondly, $\rho'$ must be zero at $\pm \rho_0$,
this leads to two new constraints.
We can now solve the Virasoro constraint for $\rho'$ 
and compute the integrals $\Dcurl$ and $\Scurl$.
At this point we have five equations in total:
the definition of the 
three charges $\Dcurl,\Scurl,\Jcurl$
and two constraints from the change of parameters.
It is now possible to solve three equations for $\kappa,\omega,w$,
see \cite{Beisert:2003ea} for details which we omit here.
The two remaining equations are
\[\label{eq:Int.Spinning.StringSol}
\lrbrk{\frac{\Jcurl}{\ellK(x)}}^2
-\lrbrk{\frac{\Dcurl}{\ellE(x)}}^2=\frac{4}{\pi^2}\,x,
\quad
\lrbrk{\frac{\Scurl}{\ellK(x)-\ellE(x)}}^2
-\lrbrk{\frac{\Jcurl}{\ellK(x)}}^2
=\frac{4}{\pi^2}\,(1-x),
\]
where $x=-\sinh^2 \rho_0$ is related to the end-points of the string.
The functions $\ellK(x)$ and $\ellE(x)$ are the elliptic integrals 
of the first and second kind, respectively
\[\label{eq:Int.Spinning.EllipticKE}
\ellK(x)=\int_0^1\frac{dy}{\sqrt{1-y^2}}\,\frac{1}{\sqrt{1-xy^2}}\,,\qquad
\ellE(x)=\int_0^1\frac{dy}{\sqrt{1-y^2}}\,\sqrt{1-xy^2}\,.
\]
The first equation in \eqref{eq:Int.Spinning.StringSol}
determines the energy $\Dcurl$ 
in terms of the charges $\Scurl,\Jcurl$ and the
parameter $x$. The parameter $x$ is fixed by the second equation.
In total we obtain the energy as a function of the charges as
\[\label{eq:Int.Spinning.ParSol}
\Dcurl(\Scurl,\Jcurl)=\Dcurl(\Scurl,\Jcurl,x(\Scurl,\Jcurl)).
\]
Frolov and Tseytlin noticed that $\Dcurl$ admits an expansion 
in powers of $1/\Jcurl$ when we fix 
the ratio of the charges $\alpha=\Scurl/\Jcurl$
\[\label{eq:Int.Spinning.ExpandCurly}
\Dcurl(\Scurl,\Jcurl)=\delta_0(\alpha)\,\Jcurl
+\frac{\delta_1(\alpha)}{8\pi^2\Jcurl}
+\frac{\delta_2(\alpha)}{64\pi^4\Jcurl^3}
+\frac{\delta_3(\alpha)}{512\pi^6\Jcurl^5}
+\ldots\,,\qquad
\alpha=\frac{\Scurl}{\Jcurl}\,.
\]
Using the original charges $D,S,J$ we can write
\[\label{eq:Int.Spinning.Expand}
D(S,J,g)=J\,\bigbrk{\delta_0(\alpha)
+\tilde g^2\,\delta_1(\alpha)
+\tilde g^4\,\delta_2(\alpha)
+\tilde g^6\,\delta_3(\alpha)
+\ldots}\,,\qquad
\alpha=\frac{S}{J}\,.
\]
where we have used the effective 
coupling constant $\tilde g^2$, see \eqref{eq:Int.Thermo.Coupling},
in the thermodynamic limit
(note that $L=J$ in this case)
\[\label{eq:Int.Spinning.Coupling}
\tilde g^2=\frac{g^2}{J^2}=
\frac{\lambda}{8\pi^2 J^2}=
\frac{1}{8\pi^2\Jcurl^2}\,.\]
The expression \eqref{eq:Int.Spinning.Expand}
suggests that, when $g\sim \sqrt{\lambda}$ is assumed to be small, 
we can compare to perturbative gauge theory!
Nevertheless, a word of caution is in order here: 
We have started out assuming that $\lambda$ is indeed large. 
One may hope that, due to analyticity, the function
$D(S,J,g)$ is valid even for small $g$, but there might
be some additional terms which can be neglected
for large $g$, but become relevant for small $g$ 
\cite{Serban:2004jf,Kazakov:2004qf}.
We will comment on this issue in 
\secref{sec:HighInt.Stringing}.

Extracting the leading-order or `one-loop'
term $\delta_1$ from the relations 
\eqref{eq:Int.Spinning.StringSol} is straightforward. 
For large $\Jcurl$ one sets
$x=x_0+x_1/\Jcurl^2+\ldots$
and solves the resulting transcendental equation for $x_0$. 
One then finds 
\[\label{eq:Int.Spinning.Tree}
\delta_0(\alpha)=1+\alpha
\]
and the parametric solution 
\[\label{eq:Int.Spinning.OneLoop}
\delta_1=-16\,\ellK(x_0)\bigbrk{\ellE(x_0)-(1-x_0)\ellK(x_0)},\qquad
\alpha=\frac{S}{J}=\frac{\ellE(x_0)}{\ellK(x_0)}-1.
\]
%

\subsection{Gauge Theory Details}

Let us see whether we can obtain an expression for $\delta_1$ 
from gauge theory. We shall be interested in a state with 
large charge $J$ of $\alSU(4)$,
large charge (spin) $S$ of $\alSU(2,2)$ as well as a large dimension $D$.
In the classical theory, the charges 
should obey the relations \eqref{eq:Int.Spinning.Tree}
\[\label{eq:Int.Spinning.Dim0}
D_0=J+S.
\]
The weight of such a state is 
\[\label{eq:Int.Spinning.Weight}
w=\weight{J+S;S,S;0,J,0;0,J}
\]
it belongs to the bosonic $\alSU(1,1)$ sector, 
c.f.~\secref{sec:One.Baby},
and the state has the form $\Tr \cder^S\fldZ^J$.

As the charges $S,J$ in string theory are very large
while the one-loop energy is small, 
we can use the thermodynamic limit of the Bethe equations
as explained in \secref{sec:Int.Thermo}.
We expect the roots for the ground state to lie on the real axis
(this may be verified by explicit solution of the exact
Bethe equations for small values of $J$). Furthermore, 
we assume the distribution of roots to be symmetric,  
$d\tilde u\,\rho(\tilde u)=d\tilde u'\,\rho(\tilde u')$ 
with $\tilde u'=-\tilde u$, which implies $n=0$. 
For the ground state we expect
the support of the root density to 
split into two disjoint intervals 
$\mathcal{C}=\mathcal{C}_- \cup \mathcal{C}_+$ with
$\mathcal{C}_- =[-b,-a]$ and $\mathcal{C}_+ =[a,b]$,
where $a<b$ are both real.
The mode numbers should be 
$n_\pm=\mp 1$ on $\mathcal{C}^{\pm}$
and the filling fractions should be $\tilde K_\pm=S/2J$.
The total filling filling fraction will be denoted by
\[\label{eq:Int.Spinning.AlphaKappa}
\alpha=\tilde K=\frac{S}{J}\,.
\]
For this distribution of roots, the Bethe equations
\eqref{eq:Int.Thermo.SU2Bethe} become
\[\label{eq:Int.Spinning.AirFoil}
\pint_a^b \frac{d\tilde v\,\rho(\tilde v)\,\tilde u^2}{\tilde v^2-\tilde u^2}
=\frac{1}{4}-\frac{\pi}{2}\,\tilde u\ , 
\qquad
\tilde E=2\int_a^b\frac{d\tilde u\,\rho(\tilde u)}{\tilde u^2} \,.
\]
The solution of the integral equation
(see, e.g.,~\cite{Kostov:1992pn,Muskhelishvili}),
yielding the density $\rho(\tilde u)$, may be obtained
explicitly; it
reads
\[\label{eq:Int.Spinning.DensInt}
\rho(\tilde u)=\frac{2 }{\pi \tilde u} \pint_a^b 
\frac{d\tilde v\, \tilde v^2}{\tilde v^2-\tilde u^2}
\sqrt{\frac{(b^2-\tilde u^2)(\tilde u^2-a^2)}{(b^2-\tilde v^2)(\tilde v^2-a^2)}}\,.
\] 
This density may be expressed explicitly through standard functions: 
\[\label{eq:Int.Spinning.Density}
\rho(\tilde u)=\frac{1}{2 \pi b \tilde u} 
\sqrt{\frac{\tilde u^2-a^2}{b^2-\tilde u^2}}
\lrbrk{\frac{b^2}{a}-4 \tilde u^2 \ellPi \Bigbrk{\frac{b^2-\tilde u^2}{b^2},q} },
\qquad
q=\frac{b^2-a^2}{b^2}\,,
\]
where we introduced the modulus $q$ which is related to the
end-points $a,b$ of the `strings' of Bethe roots; it
plays the role of an auxiliary parameter. The function
$\ellPi$ is the elliptic integral of the third kind
\[\label{eq:Int.Spinning.EllipticK}
\ellPi(m,q)= 
\int_0^{1}\frac{dy}{\sqrt{1-y^2}}\,
\frac{1}{(1-m y^2)\sqrt{1-qy^2}}\ .
\]
Furthermore, we may derive two conditions determining the
interval boundaries $a,b$ as a function of the filling fraction
$\alpha$:
\[\label{eq:Int.Spinning.EndPointsInt}
\int_a^b \frac{ d\tilde u\ \tilde u^2}{\sqrt{(b^2-\tilde u^2)(\tilde u^2-a^2)}}
=\frac{1+2 \alpha}{4}\qquad \mbox{and} \qquad
\int_a^b \frac{d\tilde u}{\sqrt{(b^2-\tilde u^2)(\tilde u^2-a^2)}}=\frac{1}{4 a b}\,.
\]
The first is derived from the normalisation condition 
\eqref{eq:Int.Thermo.Filling}, 
while the second is a consistency condition
assuring the positivity of the density.
These may be reexpressed through standard elliptic 
integrals of, respectively, the second and the first kind 
\eqref{eq:Int.Spinning.EllipticKE}.
It is straightforward to eliminate the interval boundaries $a,b$
from these equations via
\[\label{eq:Int.Spinning.EndPoints}
a=\frac{1}{4 \ellK(q)}\,,\qquad 
b=\frac{1}{4 \sqrt{1-q}\,\ellK(q)}\,.
\]
Furthermore, we can integrate the density and 
compute the energy $\tilde E$ from the right equation in
\eqref{eq:Int.Spinning.AirFoil}
\[\label{eq:Int.Spinning.GaugeRes}
\tilde E=-4 \,\ellK(q)\bigbrk{2\ellE(q)-(2-q)\ellK(q)} 
,\qquad 
\alpha=\frac{S}{J}=
\frac{1}{2\sqrt{1-q}}\,\frac{\ellE(q)}{\ellK(q)}-\frac{1}{2}\,. 
\]
In total the scaling dimension of our solution is
\[\label{eq:Int.Spinning.GaugeDim}
D(S,J,g)=S+J+\frac{g^2}{J}\,\tilde E(\alpha)+\ldots=
J\bigbrk{1+\alpha+\tilde g^2\,\tilde E(\alpha)+\ldots}+\ldots
\]
with the effective coupling $\tilde g$ defined in 
\eqref{eq:Int.Thermo.Coupling,eq:Int.Spinning.Coupling}.

Finally, we can compute the resolvent 
\eqref{eq:Int.Thermo.SU2Resolv}
\cite{Arutyunov:2003rg}%
\footnote{
Note that the first term in $G(\tilde u)$ is due to 
the shift in
$G\indup{sing}(\tilde u)=G(\tilde u)+s/2\tilde u$,
the generalisation of \eqref{eq:Int.Thermo.SU2ResolvSing}
to arbitrary spin $s$.
The ambiguities $\pm$ and $n$ drop out in 
$2\cos G\indup{sing}(\tilde u)$.
}
\[\label{eq:Int.Spinning.Resolvent}
-i\log \tilde T(\tilde u)\approx 
G(\tilde u)=
\frac{1}{2\tilde u}\mp\frac{2a^2}{b\tilde u}
\sqrt{-\frac{b^2-\tilde u^2}{\tilde u^2-a^2}}\,
\ellPi\lrbrk{\frac{q\,\tilde u^2}{\tilde u^2-a^2},q}
+ 2\pi n.
\]
The resolvent $G(\tilde u)$ is a central object in
the investigation of the solution.
It is multi-valued
on the complex plane and has branch cuts with the
discontinuity proportional to the density $\rho(\tilde u)$. 
Furthermore, it encodes the values of all rescaled charges $\tilde Q_r$ 
when expanded around $\tilde u=0$.

\subsection{Comparison}

Let us now compare the string theory
system \eqref{eq:Int.Spinning.OneLoop}
for the classical energy and the gauge theory system 
\eqref{eq:Int.Spinning.GaugeRes} for the one-loop anomalous dimension.
Both systems are parametric, i.e.~finding energy/dimension as a function of
spins involves elimination of auxiliary parameters. 
They look similar, but superficially they are not identical. 
However, if we relate the
auxiliary parameters $x_0$ and $q$ by
\[\label{eq:Int.Spinning.Moduli}
x_0=-\frac{(1-\sqrt{1-q})^2}{4\sqrt{1-q}}\,, 
\]
one can show, using the elliptic integral modular transformation relations 
\[\label{eq:Int.Spinning.Modular}
\ellK(x_0)=(1-q)^{1/4}\ellK(q),\quad
\ellE(x_0)=\half (1-q)^{-1/4} \ellE(q)+\half (1-q)^{1/4}\ellK(q)\,, 
\]
that the systems 
\eqref{eq:Int.Spinning.OneLoop} and \eqref{eq:Int.Spinning.GaugeRes}
are, in fact, exactly the same. 
As a result, their solutions $\delta_1(\alpha)=\tilde E(\alpha)$ do become identical!
We have thus demonstrated the equivalence 
between the string theory and gauge theory results 
for a particular two-spin part of the spectrum 
at the full \emph{functional} level. 
In \cite{Arutyunov:2003rg} it was furthermore shown, 
that not only the energy, 
but also the set of all higher charges agrees with string theory!

Recently, Kazakov, Marshakov, Minahan and Zarembo
have proposed a proof for the complete agreement between string theory 
and gauge theory at the one-loop (and two-loop) level 
in the dual case of strings
spinning on $S^5$ instead of $AdS_5$, 
i.e.~the $\alSU(2)$ subsector \cite{Kazakov:2004qf}.
We will comment on the comparison 
at higher-loops in \secref{sec:HighInt.Stringing}.
In another line of work initiated by Kruczenski
similar statements can be made \cite{Kruczenski:2003gt,Kruczenski:2004kw}.
These are based on a coherent state picture and 
independent of integrability.

\finishchapter 

\chapter{Higher-Loops}
\label{ch:Higher}

In \chref{ch:One} we have seen how to make use of the
algebra to find the complete one-loop dilatation operator
and in the previous chapter we have seen that its
integrability in the planar limit enables
a precise comparison to string theory 
within the AdS/CFT correspondence.
It is exciting to see whether these ideas
may be extended to higher loops. 
In this chapter we will aim at the construction
of higher-loop corrections to the
dilatation generator. Higher-loop 
integrability will be the subject of the next chapter. 

At one-loop the analysis was simplified due to
the preserved classical algebra; at higher-loops
this is not the case and a derivation of the
complete dilatation operator would require 
a large amount of work.
We will therefore restrict to 
the $\alSU(2|3)$ subsector of $\superN=4$ SYM with
a finite number of fundamental fields and a
smaller supersymmetry algebra, 
which includes the dilatation operator.

Here we will find and investigate 
deformations $\algJ(g)$ of the classical representation 
$\algJ_0$ of the symmetry algebra on the space of states.
These deformations are furnished in such a way that they are compatible
($i$) with the symmetry algebra and 
($ii$) with $\superN=4$ SYM field theory and its Feynman diagrams.
The text is based on the article \cite{Beisert:2003ys} and
contains excerpts from \cite{Beisert:2003tq,Beisert:2003jb}.

\section{The $\alSU(2|3)$ Eighth-BPS Sector}
\label{sec:Higher.SU23}

The model discussed in this chapter is the 
$(0,1^+)$ subsector with $\alSU(2|3)\times \alU(1)$
symmetry, c.f.~\secref{sec:Dila.Sect.BPS}, in the planar limit.  
In the large $N$ limit, the gauge theory turns 
into a quantum spin chain as 
described in \secref{sec:Dila.Planar.SpinChain} and we will 
use spin chain terminology.
Note that the model is a subsector, not only of $\superN=4$ SYM, but also
of the BMN matrix model,
which was briefly introduced in \secref{sec:Dila.Dim.QM}.
Therefore, all results obtained in this chapter apply equally
well to the BMN matrix model.

In the following, we shall describe the model in terms of the 
space of states, symmetry and how it is related to $\superN=4$ 
gauge theory.

\subsection{Fields, States and Interactions}
\label{sec:Higher.SU23.Notation}

The subsector consists of three complex scalars $\phi_a$ 
(Latin indices take the values $1,2,3$)
and two complex fermions $\psi_\alpha$ 
(Greek indices take the values $1,2$)
\[\label{eq:Higher.SU23.Fields}
\phi_a\quad (a=1,2,3),\qquad
\psi_\alpha\quad (\alpha=1,2).
\]
These can be combined into a supermultiplet $\fldW_A$ 
(capital indices range from $1$ to $5$)
of fields 
\[\label{eq:Higher.SU23.SuperField}
\fldW_{1,2,3}=\phi_{1,2,3},\quad
\fldW_{4,5}=\psi_{1,2}.\]
We shall use the notation introduced in 
\secref{sec:Dila.Planar} to describe (single-trace) states
and interactions 
\[\label{eq:Higher.SU23.Single}
\state{A_1\ldots A_L}=\Tr \fldW_{A_1}\ldots \fldW_{A_L},
\qquad
\ITerm{A_1\ldots A_{E\indup{i}}}
{B_1\ldots B_{E\indup{o}}}.
\]
To distinguish between bosons and fermions in the interaction 
symbols, we use Latin and Greek letters.
For example, the interaction $\ITerm{\alpha b c}{c \alpha b}$ 
searches for one fermion followed by two bosons within the trace.
Wherever they can be found these three fields are
permuted such that the last boson comes first,
next the fermion and the other boson last.
A sample action is
\[\label{eq:Higher.SU23.IntSample}
\ITerm{\alpha b c}{c \alpha b} \state{142334452}
=\state{134234452}+\state{242334415}.
\]

\subsection{The Algebra}
\label{sec:Higher.SU23.Alg}

The fields $\fldW_A$ transform canonically in a fundamental
$\rep{3|2}$ representation of $\alSU(2|3)$. 
Let us start by defining this algebra. 
The $\alSU(2|3)\times \alU(1)$ algebra consists of the generators 
\[\label{eq:Higher.SU23.Alg}
\algJ=\{\algL^\alpha{}_\beta,\algR^a{}_b,\algD_0,\ham|\algQ^a{}_\alpha,\algS^\alpha{}_a\}.
\]
The bar separates bosonic from fermionic operators.
The $\alSU(2)$ and $\alSU(3)$ generators $\algL^\alpha{}_\beta$ and $\algR^a{}_b$ 
are traceless, $\algL^\alpha{}_\alpha=\algR^a{}_a=0$.
The commutators are defined as follows:
Under the rotations $\algL,\algR$, the indices of any generator $\algJ$
transform canonically according to the rules
\[\label{eq:Higher.SU23.AlgRot}
\arraycolsep0pt
\begin{array}{rclcrcl}
\comm{\algL^\alpha{}_\beta}{\algJ_\gamma}\eq
\delta^\alpha_\gamma \algJ_\beta
-\half \delta^\alpha_\beta \algJ_\gamma,
&\quad&
\comm{\algL^\alpha{}_\beta}{\algJ^\gamma}\eq
-\delta^\gamma_\beta \algJ^\alpha
+\half \delta^\alpha_\beta \algJ^\gamma,
\\
\comm{\algR^a{}_b}{\algJ_c}\eq
\delta^a_c \algJ_b
-\sfrac{1}{3} \delta^a_b \algJ_c,
&&
\comm{\algR^a{}_b}{\algJ^c}\eq
-\delta_b^c \algJ^a
+\sfrac{1}{3} \delta^a_b \algJ^c.
\end{array}\]
The commutators of the dilatation operator $\algD_0$ and 
the Hamiltonian $\ham$ are given by
\[\label{eq:Higher.SU23.AlgCharge}
\comm{\algD_0}{\algJ}=\dim(\algJ)\, \algJ,\qquad \comm{\ham}{\algJ}=0.
\]
In other words, $\ham$ is the central $\alU(1)$ generator and 
the non-vanishing dimensions are
\[\label{eq:Higher.SU23.AlgDim}
\dim(\algQ)=-\dim(\algS)=\half.
\]
The supercharges anticommuting into rotations are given by
\[\label{eq:Higher.SU23.AlgMomRot}
\acomm{\algS^\alpha{}_a}{\algQ^b{}_\beta}=
  \delta^b_a \algL^\alpha{}_\beta
  +\delta_\beta^\alpha \algR^b{}_a
  +\delta_a^b \delta_\beta^\alpha (\sfrac{1}{3}\algD_0+\sfrac{1}{2}g^2\ham).
\]
This implies that the linear combination $\algD_0+\sfrac{3}{2}g^2 \ham$ 
belongs to the algebra $\alSU(2|3)$.
Furthermore, we demand a parity even algebra
\[\label{eq:Higher.SU23.Parity}
\gaugepar\, \algJ\, \gaugepar^{-1}=\algJ \quad\mbox{or}\quad 
\comm{\gaugepar}{\algJ}=0.
\]

It is straightforward to find the fundamental $\rep{3|2}$ 
representation acting on the fundamental module 
(we will do this explicitly in \secref{sec:Higher.Tree})
\[\label{eq:Higher.SU23.Module}
\mdlF=[\fldW_1,\fldW_2,\fldW_3,\fldW_4,\fldW_5].
\]
As states are constructed from the fundamental fields $\fldW_A$ there is
an induced representation on the space of states; 
this is simply a tensor product representation and we will denote it by $\algJ_0$.
The higher order corrections $\algJ_k$ will act on more than one field at a time.

\subsection{Representations}
\label{sec:Higher.SU23.Reps}

In terms of representation theory, a state is characterised 
by the charges 
\[\label{eq:Higher.SU23.Charges}
D_0,\quad s,\quad [q,p],\quad E,\]
where $D_0$ is the classical dimension, 
$E$ is the energy (i.e. the eigenvalue of the Hamiltonian $\ham$),
$s$ is \emph{twice} the $\alSU(2)$ spin
and
$[q,p]$ are the $\alSU(3)$ Dynkin labels.
These can be arranged into Dynkin labels%
\footnote{The sign of $r$ was chosen such that
$[1;0;0,0]\times[1;0;0,0]=[2;0;0,0]+[0;-1;0,0]$.}
of $\alSU(2|3)$ 
\[\label{eq:Higher.SU23.Dynkin}
w=[s;r;q,p],
\qquad r=\sfrac{1}{3} D_0+\half g^2 E+\sfrac{1}{2}s-\sfrac{1}{3}p-\sfrac{2}{3}q.\]
Although it is sufficient to give either the dimension or the label $r$, we
will usually state both for convenience.
The labels $s,q,p$ are integer-valued, whereas
the fermionic label $r$ can be any real number.%
\footnote{Nevertheless we will usually write 
its value at $g=0$ and state the irrational part $E(g)$
separately.}
Representations are characterised by their highest weight.
For instance, the highest weight of the fundamental module
$\mdlF$ is
\[\label{eq:Higher.SU23.FundWeight}
w\indups{F}=[0;0;0,1].\]

It is helpful to know how to construct a state with given 
charges $D_0,s,p,q$ and length $L$ from the 
fundamental fields $\phi_{1,2,3},\psi_{1,2}$. 
The numbers of constituents of each kind are given by
\[\label{eq:Higher.SU23.Numbers}
n_{\phi}=n_{1,2,3}=\left(\begin{array}{l}
L-\sfrac{2}{3}D_0+\sfrac{2}{3}p+\sfrac{1}{3}q\\
L-\sfrac{2}{3}D_0-\sfrac{1}{3}p+\sfrac{1}{3}q\\
L-\sfrac{2}{3}D_0-\sfrac{1}{3}p-\sfrac{2}{3}q
\end{array}\right),
\qquad
n_{\psi}=n_{4,5}=
\left(\begin{array}{l}
D_0-L+\half s\\
D_0-L-\half s
\end{array}\right).
\]

The following `unitarity' bound%
\footnote{We use the terminology of $\superN=4$ SYM even 
if some terms might be inappropriate.}
applies to multiplets of $\alSU(2|3)$
\[\label{eq:Higher.SU23.Bound}
D_0+\sfrac{3}{2}g^2 E\geq 3+\sfrac{3}{2}s+p+2q,\quad
r-s\geq 1,\qquad
\mbox{or}\qquad
D_0+\sfrac{3}{2}g^2 E=p+2q,\quad r=s=0.
\]
A (typical) multiplet of $\alSU(2|3)$ 
away from the bound consists of 
\[\label{eq:Higher.SU23.Dimension}
(32|32)\times (s+1)\times \half (p+1)(q+1)(p+q+2)\]
components.
However, under certain conditions on the dimension, 
the multiplet is shortened (atypical).
We find three conditions relevant to the spin chain.
The first one is the `half-BPS'%
\footnote{In fact, $4$ out of $6$ supercharges annihilate the state.}
condition%
\[\label{eq:Higher.SU23.HalfBPS}
D_0+\sfrac{3}{2}g^2 E=p,\quad
s=r=q=0,\qquad n_{2}=n_{3}=n_{4}=n_{5}=0,
\]
where we have also displayed the condition in terms of the number of fields
\eqref{eq:Higher.SU23.Numbers}.
Such a multiplet has $1+p(p+1)|p(p+1)$ components.
The second one is the `quarter-BPS'%
\footnote{In fact, $2$ out of $6$ supercharges annihilate the state.
Multiplets of this kind have states belonging to the $\alSU(2)$ subsector
of just two complex bosonic fields $\phi_{1,2}$.}
condition
\[\label{eq:Higher.SU23.QuarterBPS}
D_0+\sfrac{3}{2} g^2 E=p+2q,\quad s=r=0,
\qquad n_{3}=n_{4}=n_{5}=0.
\]
Although a quarter-BPS multiplet is beyond the
unitarity bound, it can acquire a non-zero energy 
if it joins with another multiplet to form a long one.
The last condition determines short 
(usually called semi-short) multiplets%
\[\label{eq:Higher.SU23.Short}
D_0+\sfrac{3}{2}g^2 E=3+\sfrac{3}{2}s+p+2q,\quad
r-s=1,\qquad
n_{3}+n_{5}=1.
\]

A long multiplet whose energy approaches the
unitarity bound \eqref{eq:Higher.SU23.Bound}
splits in two at \eqref{eq:Higher.SU23.Short}.
If $s>0$, the highest weight of the 
upper short submultiplet is shifted by 
\[\label{eq:Higher.SU23.SplitShort}
\delta D_0=+\half,\quad \delta w\indup{i}=[-1;-1;+1,0],\qquad \delta L=+1.
\]
For $s=0$ the upper submultiplet is
quarter-BPS and its highest weight is shifted by%
%
\[\label{eq:Higher.SU23.SplitBPS}
\delta D_0=+1,\quad \delta w\indup{I}=[0;-1;+2,0],\qquad \delta L=+1.
\]
Multiplet shortening will turn out to be important later on.
This is because the generators which relate both submultiplets
must act as $\order{g}$ so that the multiplet can indeed split at $g=0$.

\subsection{Fluctuations in Length}
\label{sec:Higher.SU23.Dynamics}

Note that all three bosons together have vanishing 
$\alSU(3)$ charges and dimension $3$. 
Similarly, both fermions have vanishing $\alSU(2)$ spin
and dimension $3$, i.e.~the same quantum numbers
\[\label{eq:Higher.SU23.EqualCharges}
\phi_{[1}\phi_2\phi_{3]}\sim \psi_{[1}\psi_{2]}.
\]
Therefore one can expect fluctuations between these two configurations. 
In field theory these are closely related to 
the Konishi anomaly \cite{Konishi:1984hf,Konishi:1985tu}.
A state composed from $n_1\geq n_2\geq n_3$ bosons and
$n_4\geq n_5$ fermions can mix
with states
\[\label{eq:Higher.SU23.DynamicRange}
(n_1-k,n_2-k,n_3-k;n_4+k,n_5+k),\qquad   -n_5\leq k\leq n_3.
\]
Note that the length $L=n_1+n_2+n_3+n_4+n_5$ decreases by $k$. 
We will refer to this aspect of the spin chain as \emph{dynamic}.
The length can fluctuate by as
much as $n_3+n_5=r-s$ units, a quantity directly related to
the distance to the unitarity bound.

Length fluctuations are especially interesting for multiplet shortenings.
The highest weight state of a half-BPS or quarter-BPS multiplet 
has fixed length due to $n_3=n_5=0$.
For short multiplets we have $n_3+n_5=1$. 
This means that the length fluctuates by one unit for the 
highest weight state. 
Two of the six supercharges transform a $\phi_3$ into a fermion $\psi_{1,2}$.
Naively, both cannot act at the same time 
because there is only one $\phi_3$
(we will always have $n_3=1$ for a long multiplet),
and the multiplet becomes short. 
However, we could simultaneously replace the resulting $\psi_{[1}\psi_{2]}$ 
by $\phi_{[1}\phi_2\phi_{3]}$ and thus fill up the $\phi_3$-hole.
A suitable transformation rule is%
\footnote{In fact, this is part of the `classical' supersymmetry
variation.}
\[\label{eq:Higher.SU23.SuperFluct}
\algQ^3{}_{2}\,\psi_1 \sim g\,\phi_{[1}\phi_{2]}.
\]
This is also the step between the two short submultiplets
\eqref{eq:Higher.SU23.SplitShort}. This property was used in 
\cite{Eden:2003sj} to determine two-loop 
scaling dimensions for operators at the unitarity bounds
from a one-loop field-theory calculation.
Note that $\algQ^3{}_{2}$ annihilates the highest weight 
when $n_4=0$; we need to apply $\algQ^3{}_{1}$ first to produce a $\psi_1$. 
In this case the upper submultiplet is
quarter-BPS \eqref{eq:Higher.SU23.SplitBPS}.
Furthermore note that when we apply $\algQ^3{}_1$ first, there are no 
more $\phi_3$'s and $\psi_2$'s left and length
fluctuations are ruled out. 
Therefore, in a BPS or short multiplet we can always find 
states with fixed length; fluctuations are 
frozen at the unitarity bound.
In contrast, all states in a multiplet away 
from the unitarity bound \eqref{eq:Higher.SU23.Bound} are mixtures 
of states of different lengths.

\subsection{From $\superN=4$ SYM to $\alSU(2|3)$}
\label{sec:Higher.SU23.N4}

A state of free $\superN=4$ SYM is characterised by the 
classical dimension $D_0$, the $\alSU(2)^2$ labels
$[s,s_2]$, the $\alSU(4)$ Dynkin labels $[q,p,q_2]$,
the $\alU(1)$ hypercharge $B$ as well as the length $L$.
The $\alSU(2|3)$ subsector is obtained by restricting to states with 
(c.f.~\secref{sec:Dila.Sect.BPS})
\[\label{eq:Higher.SU23.EighthBPS}
D_0=p+\sfrac{1}{2}q+\sfrac{3}{2}q_2,\qquad s_2=0.
\]
This also implies $D_0=B+L$.
We write these as relations of the corresponding generators
\[\label{eq:Higher.SU23.Sector}
(\algR_{\alSU(4)})^4{}_4=\half \algD_0,\quad 
\algLd^{\dot \alpha}{}_{\dot\beta}=0,\quad
\algD_0=\len+\algB.
\]
Furthermore, we express the $\alSU(4)$ generator 
$\algR_{\alSU(4)}$ in terms of an $\alSU(3)$ generator $\algR$
\[\label{eq:Higher.SU23.RotRel}
(\algR_{\alSU(4)})^a{}_b=\algR^a{}_b-\sfrac{1}{6}\delta^a_b \algD_0.
\]
Now we can reduce the $\alPSU(2,2|4)$ algebra 
as given in \appref{app:U224.Comm} to the 
$\alSU(2|3)$ subsector and find 
precisely the $\alSU(2|3)$ relations 
(c.f.~\secref{sec:Higher.SU23.Alg})
if the Hamiltonian $\ham$ is identified with 
the anomalous dilatation generator as follows
\[\label{eq:Higher.SU23.Ham}
\algdD=g^2 \ham.
\]
As we would like to compare directly to $\superN=4$ SYM,
we write one of the generators of $\alSU(2|3)$ 
as $\algD_0+\sfrac{3}{2}g^2 \ham$ instead of
assigning a new letter.

We note that the states in this subsector are (classically) eighth-BPS in
terms of ${\superN=4}$ SYM 
(in \secref{sec:Higher.Spec.EighthBPS} we will
present a true eighth-BPS state). 
Unprotected primary states of the subsector
can therefore not be primary states of $\alPSU(2,2|4)$. 
To shift from the corresponding superconformal primary 
to the highest weight in the subsector we have to shift by
(c.f.~\secref{sec:Dila.Sect.BPS})
\[\label{eq:Higher.SU23.N4Shift}
\delta w\indup{II}=\weight{+1;0,0;0,0,+2;0,+1}.
\]
Note that in terms of the Dynkin labels 
$[s;r;q,p,q_2;r_2;s_2]$ 
the last two are zero in this subsector $r_2=s_2=0$.
We then simply restrict to the first four labels $[s;r;q,p]$.

\section{Tree-Level}
\label{sec:Higher.Tree}

We would like to construct a representation $\algJ(g)$
of $\alSU(2|3)\times \alU(1)$ on the spin chain.
The generators must satisfy the
algebra relations 
\[\label{eq:Higher.Tree.Alg}
\bigscomm{\algJ_M(g)}{\algJ_N(g)}=\algstr_{MN}^P\, \algJ_P(g)
\]
with $\algstr_{MN}^P$ the structure constants of the symmetry algebra
as given in \secref{sec:Higher.SU23.Alg}.

Let us illustrate the procedure for the generators at tree-level.
At tree-level, composite states transform in tensor product representations
of the fundamental representation $\rep{3|2}$.
The generators therefore act on one field at a time. 
We write down the most general form of generators
that respects $\alSU(3)\times\alSU(2)$ symmetry
\<\label{eq:Higher.Tree.Struct}
\algR^{a}{}_{b}\eq c_1\ITerm{a}{b}+c_2\delta^a_b\ITerm{c}{c},
\nln
\algL^{\alpha}{}_{\beta}\eq c_3\ITerm{\alpha}{\beta}+c_4\delta^\alpha_\beta\ITerm{\gamma}{\gamma},
\nln
\algD_0\eq
 c_5\ITerm{a}{a}+c_6\ITerm{\alpha}{\alpha},
\nln
(\algQ_0)^{a}{}_{\alpha}\eq c_7\ITerm{a}{\alpha},
\nln
(\algS_0)^{\alpha}{}_{a}\eq c_8\ITerm{\alpha}{a}.
\>
The algebra relations have two solutions. One is the trivial 
solution $c_k=0$ corresponding to the trivial representation.
The other solution requires
\[\label{eq:Higher.Tree.Coeff}
c_1=c_3=c_5=1,\quad
c_2=-\sfrac{1}{3},\quad
c_4=-\sfrac{1}{2},\quad
c_6=\sfrac{3}{2},\quad
c_7=e^{i\beta_1},\quad
c_8=e^{-i\beta_1}.
\]
As expected, we find that the bosons and fermions have dimension
$1$ and $\sfrac{3}{2}$, respectively
\[\label{eq:Higher.Tree.Ham}
\algD_0=\ITerm{a}{a}+\sfrac{3}{2}\ITerm{\alpha}{\alpha}.
\]
The appearance of a free parameter $\beta_1$ is related to
a possible rescaling of the bosons and fermions.
This can be represented in terms of a similarity 
transformation on the algebra 
\[\label{eq:Higher.Tree.Alpha}
\algJ_0\mapsto \exp\bigbrk{2i\beta_1 \algD_0}\,\algJ_0\,
\exp\bigbrk{-2i\beta_1 \algD_0}.
\]
Obviously, the algebra relations in \secref{sec:Higher.SU23.Alg} 
are invariant under such a transformation.
The only other $\alSU(3)\times\alSU(2)$ invariant similarity transformation
besides \eqref{eq:Higher.Tree.Alpha} is
\[\label{eq:Higher.Tree.Eps}
\algJ_0\mapsto \exp\bigbrk{i\beta_2 \len}\,\algJ_0\,
\exp\bigbrk{-i\beta_2 \len},
\]
where $\len$ is the length operator
\[\label{eq:Higher.Tree.Length}
\len=\ITerm{a}{a}+\ITerm{\alpha}{\alpha}\qquad\mbox{or simply}\quad
\len=\ITerm{\cdot}{\cdot}.
\]
The transformation \eqref{eq:Higher.Tree.Eps} is trivial 
and does not give rise to a new parameter 
at tree-level
because the length is conserved there
\[\label{eq:Higher.Tree.LengthSym}
\comm{\len}{\algJ_0}=0.
\]
%

\section{One-Loop}
\label{sec:Higher.One}

In this section we construct deformations of the algebra generators
$\algJ(g)$ obeying the algebra relations in \secref{sec:Higher.SU23.Alg}. 
Here, we will proceed up to $\order{g}$
for the deformations of the Hamiltonian $\ham(g)$. 
This can still be done conveniently by hand 
without the help of computer algebra systems.
This section is meant to illustrate the methods of this chapter
in a simple context before we proceed to 
higher-loops in the sections to follow.

The most important one of the algebra relations is
the invariance of the interaction Hamiltonian
\[\label{eq:Higher.One.Comm}
\comm{\algJ_M(g)}{\ham(g)}=0.
\]
Moreover we will assume the $\alSU(2),\alSU(3)$ rotation generators 
$\algR^a{}_b$ and $\algL^\alpha{}_\beta$ to receive no corrections. 
This is natural, for the rotation symmetries are preserved by 
the quantisation procedure. 

\subsection{Pre-Leading Order}
\label{sec:Higher.One.First}

Let us restrict \eqref{eq:Higher.One.Comm} to its leading order
\[\label{eq:Higher.One.Leading}
\comm{\algJ_0}{\ham_l}=0,
\]
in other words, the leading order of the Hamiltonian
at some $\order{g^l}$ is conserved by the classical algebra.
The leading order for $\ham$ will be $l=0$ and we 
shall now exclude a correction to $\ham$ at order $l=-1$
\footnote{Note that $\ham$ is shifted by two
orders in perturbation theory due to $\algdD=g^2 \ham$.
It therefore makes sense to consider $\ham_{-1}=\algD_1$.}
by representation theory
in analogy to \secref{sec:One.Form.Symmetry}: 
At this order the interactions have three legs and 
the possible ways to distribute them among the
in and out channels are 
\[\label{eq:Higher.One.NoFirst}
\ITerm{\cdot}{ABC},\quad\ITerm{C}{AB},\quad\ITerm{BC}{A},\quad\ITerm{ABC}{\cdot}.
\]
The indices cannot be contracted fully, hence there is no 
invariant interaction at this order. In other words there
is no common irreducible representation of the free algebra
among the in and the out channel
\[\label{eq:Higher.One.NoIntertwiner}
\mdlF^0\not\in\mdlF^3,\quad
\mdlF^1\not\in\mdlF^2.
\]

\subsection{Leading Order}
\label{sec:Higher.One.Second}

\begin{figure}\centering
\parbox[c]{1.5cm}{\centering\includegraphics{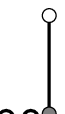}}
\parbox[c]{1.5cm}{\centering\includegraphics{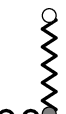}}
\parbox[c]{1.5cm}{\centering\includegraphics{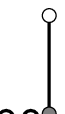}}
\parbox[c]{1.5cm}{\centering\includegraphics{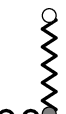}}
\parbox[c]{1.5cm}{\centering\includegraphics{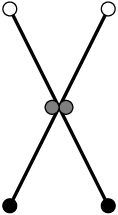}}
\parbox[c]{1.5cm}{\centering\includegraphics{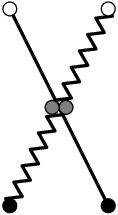}}
\parbox[c]{1.5cm}{\centering\includegraphics{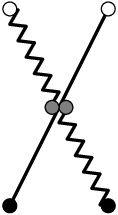}}
\parbox[c]{1.5cm}{\centering\includegraphics{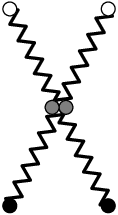}}
\caption{The structures for the construction of $\ham_0$. Straight and zigzag lines 
correspond to bosons and fermions, respectively.}
\label{fig:Higher.One.H2SU2SU3}
\end{figure}

\begin{figure}\centering
$\ham_0=c_1\parbox[c]{1.5cm}{\centering\includegraphics{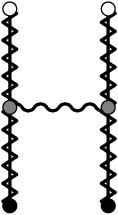}}+c_4
\parbox[c]{1.5cm}{\centering\includegraphics{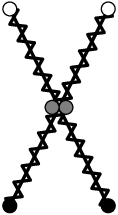}}$
\caption{The structures of $\ham_0$ which are compatible
with $\alSU(2|3)$ symmetry at leading order.
A straight+zigzag line correspond to a supermultiplet.}
\label{fig:Higher.One.H2SU23}
\end{figure}

A similar argument is used to show that at leading order 
we must evenly distribute the four fields among the in and out channel,
i.e.
\[\label{eq:Higher.One.Intertwiner}
\mdlF^0\not\in \mdlF^4,\quad
\mdlF^1\not\in \mdlF^3,\quad\mbox{but }
\mdlF^2=\mdlF^2.
\]
The most general form of $\ham_0$,
expressed as an action on bosons ($a,b$) and fermions ($\alpha,\beta$)
is therefore
\<\label{eq:Higher.One.H0}
\ham_0\eq
 c_1\ITerm{ab}{ab}
+c_2\ITerm{a\beta}{a\beta}
+c_2'\ITerm{\alpha b}{\alpha b}
+c_3\ITerm{\alpha \beta}{\alpha\beta}
\nl
+c_4\ITerm{ab}{ba}
+c_5\ITerm{a\beta}{\beta a}
+c_5'\ITerm{\alpha b}{b \alpha}
+c_6\ITerm{\alpha \beta}{\beta\alpha},
\>
see also \figref{fig:Higher.One.H2SU2SU3}.
First of all we demand 
that $\ham_0$ conserves parity \eqref{eq:Higher.SU23.Parity},
\[\label{eq:Higher.One.H0Parity}
\gaugepar\, \ham_0\, \gaugepar^{-1}=\ham_0.
\]
As can be seen easily, this requires
\[\label{eq:Higher.One.ParityCoeff}
c_2=c_2',\quad c_5=c_5'.\]
We now commute $\algQ_0$ with $\ham_0$ and find 
\<\label{eq:Higher.One.QH2}
\comm{(\algQ_0)^{a}{}_{\alpha}}{\ham_0}\eq
e^{i\beta_1}(c_1-c_2)\bigbrk{\ITerm{a b}{\alpha b}+\ITerm{ba}{b\alpha}}
+e^{i\beta_1}(c_4-c_5)\bigbrk{\ITerm{b a}{\alpha b}+\ITerm{ab}{b \alpha}}
\nlnum\nonumber
+e^{i\beta_1}(c_2-c_3)\bigbrk{\ITerm{a \beta}{\alpha \beta}-\ITerm{\beta a}{\beta\alpha}}
-e^{i\beta_1}(c_5+c_6)\bigbrk{\ITerm{a \beta}{\beta\alpha}-\ITerm{\beta a}{\alpha \beta}}.
\>
According to \eqref{eq:Higher.One.Leading} this must vanish, so we set
\[\label{eq:Higher.One.Coeff2}
c_1=c_2=c_3,\quad c_4=c_5=-c_6.
\]
The commutator $\comm{\algS_0}{\ham_0}$ leads to the same set of constraints.
The two independent constants correspond to the two
irreducible representations in the tensor product 
(see \figref{fig:Higher.One.H2SU23})
\[\label{eq:Higher.One.Tensor}
\mdlF\times \mdlF=
[0;0;0,2]_+ + [0;0;1,0]_-\,.
\]
More explicitly, $c_1+c_4$ corresponds to the symmetric product
$[0;0;0,2]$ which is half-BPS 
and $c_1-c_4$ to the antisymmetric one $[0;0;1,0]$ which is
quarter-BPS.

\subsection{First Order}
\label{sec:Higher.One.Third}

The virtue of a classically invariant interaction 
applies only to the leading order, for $\ham_1$ we should break it.
However, we do not wish to break classical $\alSU(2|3)$ in the most general way,
but assume that the classical $\alSU(3)\times \alSU(2)$ invariance is conserved.
In field theory these correspond to symmetries compatible with the 
regularisation scheme.

\begin{figure}\centering
$\ham_0\sim
\parbox[c]{1.5cm}{\centering\includegraphics{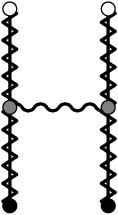}}-
\parbox[c]{1.5cm}{\centering\includegraphics{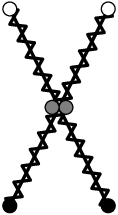}}$
\caption{Closure of the algebra at $\order{g^3}$ fixes 
the relative coefficients within $\ham_0$.}
\label{fig:Higher.One.H2SU23Final}
\end{figure}

The possible first order corrections involve the 
totally antisymmetric tensors of $\alSU(3)$ and $\alSU(2)$,
see \figref{fig:Higher.One.H3}:
\begin{figure}\centering
$\ham_1=c_7\parbox[c]{2.5cm}{\centering\includegraphics{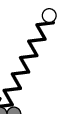}}
+c_8\parbox[c]{2.5cm}{\centering\includegraphics{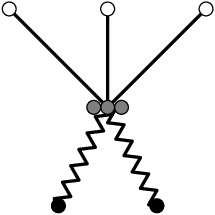}}$,\quad
$\algQ_1=c_9\parbox[c]{1.5cm}{\centering\includegraphics{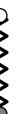}}$,\quad
$\algS_1=c_{10}\parbox[c]{1.5cm}{\centering\includegraphics{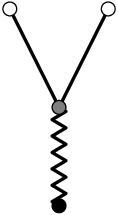}}$
\caption{The structures for the construction of $\ham_1,\algQ_1,\algS_1$.
The number of spin sites is not conserved here.}
\label{fig:Higher.One.H3}
\end{figure}
\<\label{eq:Higher.One.H3}
\ham_1\eq
c_7 \,\varepsilon_{\alpha\beta}\varepsilon^{abc}\ITerm{\alpha\beta}{abc}
+c_8 \,\varepsilon_{abc}\varepsilon^{\alpha\beta}\ITerm{abc}{\alpha\beta},
\nln
(\algQ_1)^{a}{}_{\alpha}\eq c_{9}\,\varepsilon_{\alpha\beta}\varepsilon^{abc}\ITerm{\beta}{bc},
\nln
(\algS_1)^{\alpha}{}_{a}\eq c_{10}\,\varepsilon_{abc}\varepsilon^{\alpha\beta}\ITerm{bc}{\beta}.
\>
With these expressions it is possible, yet tedious, to work out 
the commutators at first order by hand.
It is useful to note a version of the 
gauge invariance identity \eqref{eq:Dila.Planar.Spectator} adapted to 
this particular situation
\[\label{eq:Higher.One.Gauge}
\ITerm{\beta d}{bcd}+\ITerm{\beta\delta}{bc\delta}=
\ITerm{d \beta}{dbc}-\ITerm{\delta\beta}{\delta bc}=
\ITerm{\beta}{bc}.
\]
Furthermore we will employ some identities of the totally antisymmetric tensors 
$\varepsilon^{abc}$ and $\varepsilon_{\alpha\beta}$
and find for the commutator $\comm{\algQ}{\ham}$ at $\order{g}$
\<\label{eq:Higher.One.QH3}
\comm{(\algQ_0)^{a}{}_{\alpha}}{\ham_1}+\comm{(\algQ_1)^{a}{}_{\alpha}}{\ham_0}\eq
(c_4c_{9}-e^{i\beta_1}c_7)\,\varepsilon^{bcd}\varepsilon_{\alpha\beta}
\bigbrk{\ITerm{a\beta}{bcd}-\ITerm{\beta a}{bcd}}
\nl
+(c_4c_{9}-e^{i\beta_1}c_7)\,\varepsilon^{abc}\varepsilon_{\beta\gamma}
\bigbrk{-\ITerm{\beta \gamma}{\alpha bc}+\ITerm{\beta \gamma}{b \alpha c}-\ITerm{\beta \gamma}{bc \alpha}}
\nl
-(c_1+c_4)c_{9}\,\varepsilon^{abc}\varepsilon_{\alpha\beta}
\ITerm{\beta}{bc}.
\>
To satisfy \eqref{eq:Higher.One.Comm} this must vanish. 
The commutator $\comm{\algS}{\ham}$ gives similar constraints and 
closure of the algebra requires
\[\label{eq:Higher.One.Coeff3}
c_1=-c_4,\quad e^{i\beta_1}c_7=c_4c_{9},\quad e^{-i\beta_1}c_8=c_4c_{10}.
\]
Here, there are two types of constraints. 
The latter two fix the 
coefficients of $\algQ_1$ and $\algS_1$. The first one
is more interesting, it fixes a coefficient of $\ham_0$ from one order below,
see \figref{fig:Higher.One.H2SU23Final}.
This is related to the fact that 
$\ham_0$ was constructed to assign equal energies 
to all states of a multiplet of the free algebra.
In a superalgebra, several atypical multiplets of the free theory
can join to form one typical multiplet in the interacting theory,
see \secref{sec:Higher.SU23.Reps} and \figref{fig:N4.Split.Splitting}.
A consistency requirement for this to happen is that the 
energy shift of the submultiplets agree.
In this case it is achieved by $c_1=-c_4$.
To ensure agreement of energies in terms of commutators,
we need to consider one additional power of the coupling constant, 
which is required to move between the submultiplets.
Furthermore, we note that the constraint $c_1=-c_4$ 
assigns a zero eigenvalue to 
the representation ${[0;0;0,2]}$ in \eqref{eq:Higher.One.Tensor}. 
This is essential, because ${[0;0;0,2]}$ is in fact half-BPS and must 
have zero energy. 
It is good to see though, that the protectedness of
half-BPS states follows from the algebraic constraints; 
we will not have to impose it by hand.

\subsection{Conclusions}
\label{sec:Higher.One.Concl}

We now set the remaining independent constants $c_1,c_9,c_{10}$
to
\[\label{eq:Higher.One.Coeffs}
c_1=\alpha_1^2, \quad c_9=\sfrac{1}{\sqrt{2}}\, \alpha_1\, e^{i\beta_1+i\beta_2},\quad 
c_{10}=\sfrac{1}{\sqrt{2}}\, \alpha_1\, e^{-i\beta_1-i\beta'_2}.\]
In total we find the deformations at first order
\<\label{eq:Higher.One.Total}
\ham_0\eq
 \alpha_1^2\ITerm{ab}{ab}
+\alpha_1^2\bigbrk{\ITerm{a\beta}{a\beta}+\ITerm{\alpha b}{\alpha b}}
+\alpha_1^2\ITerm{\alpha \beta}{\alpha\beta}
\nl
-\alpha_1^2\ITerm{ab}{ba}
-\alpha_1^2\bigbrk{\ITerm{a\beta}{\beta a}+\ITerm{\alpha b}{b \alpha}}
+\alpha_1^2\ITerm{\alpha \beta}{\beta\alpha},
\nln
\ham_1\eq
-\sfrac{1}{\sqrt{2}}\,\alpha_1^3\, e^{i\beta_2}\,\varepsilon_{\alpha\beta}\varepsilon^{abc}\ITerm{\alpha\beta}{abc}
-\sfrac{1}{\sqrt{2}}\,\alpha_1^3\, e^{-i\beta'_2}\,\varepsilon_{abc}\varepsilon^{\alpha\beta}\ITerm{abc}{\alpha\beta},
\nln
(\algQ_1)^{a}{}_{\alpha}\eq \sfrac{1}{\sqrt{2}}\,\alpha_1\, e^{i\beta_1+i\beta_2}\,\varepsilon_{\alpha\beta}\varepsilon^{abc}\ITerm{\beta}{bc},
\nln
(\algS_1)^{\alpha}{}_{a}\eq \sfrac{1}{\sqrt{2}}\,\alpha_1\, e^{-i\beta_1-i\beta'_2}\,\varepsilon_{abc}\varepsilon^{\alpha\beta}\ITerm{bc}{\beta}.
\>
Let us discuss the free parameters. 
The parameter $\beta'_2$ will in fact be determined 
by a constraint from fourth order, see the following
section: $c_9c_{10}=\half c_1$ or
\[\label{eq:Higher.One.FromHigher}
\beta'_2=\beta_2.\]
As shown in \eqref{eq:Higher.Tree.Alpha,eq:Higher.Tree.Eps}, 
the coefficients $\beta_{1,2}$ correspond 
to a similarity transformation of the algebra
\[\label{eq:Higher.One.Sim}
\algJ_0\mapsto \exp\bigbrk{2i\beta_1 \algD_0+i\beta_2 \len}\,\algJ_0\,
\exp\bigbrk{-2i\beta_1 \algD_0-i\beta_2 \len}.
\]
The algebra relations \eqref{eq:Higher.Tree.Alg} are invariant under similarity 
transformations, so $\beta_{1,2}$ can take arbitrary values.
For convenience, we might fix a gauge and set $\beta_1=\beta_2=0$,
but we will refrain from doing that here. 
Last but not least, the parameter $\alpha_1$ corresponds to a rescaling 
of the coupling constant
\[\label{eq:Higher.One.Renorm}
g\mapsto \alpha_1\, g.
\]
The algebra relations \eqref{eq:Higher.Tree.Alg} are also 
invariant under this redefinition.

In a real form of the algebra we get a few additional constraints.
There, the algebra should be self-adjoint
which imposes some reality constraint on $\alpha_1,\beta_{1,2}$. 
For a real $\alSU(2|3)$ they have 
to be real and $\alpha_1^2$ needs to be positive. 
This ensures positive planar energies
as required by the unitarity bound.

In conclusion we have found that the deformations of the generators
are uniquely fixed at one-loop. 
Note that $\ham_0$ agrees with the complete one-loop dilatation 
operator found in \chref{ch:One}.
Here, it is understood that some parameters cannot
be fixed due to symmetries of the algebra relations.
In determining the coefficients we saw that $\comm{\ham(g)}{\algJ(g)}=0$ 
at order $\order{g^{2\ell-2}}$ makes the $\ell$-loop energy shift
agree within short multiplets, whereas $\order{g^{2\ell-1}}$ joins 
up short multiplets into long multiplets.
Note that the anticommutator of supercharges at
first order is trivially satisfied due to the 
flavours of incoming and outgoing fields
\[\label{eq:Higher.One.SQ1}
\acomm{(\algS_1)^{\alpha}{}_{a}}{(\algQ_0)^{b}{}_{\beta}}
+\acomm{(\algS_0)^{\alpha}{}_{a}}{(\algQ_1)^{b}{}_{\beta}}
=0=
\half \delta^b_a \delta^\alpha_\beta \ham_{-1}.\]
%

\section{Two-Loops}
\label{sec:Higher.Two}

In this section we will discuss the restrictions from 
the algebra at two-loops, i.e.~up to third order.
The steps are straightforward, but involve very lengthy expressions. 
We have relied on the algebra system \texttt{Mathematica} 
to perform the necessary computations.

\subsection{Structures}
\label{sec:Higher.Two.Struc}

At second order we need to determine $\ham_2,\algQ_2,\algS_2$. 
For $\ham_2$ the $\alSU(3)\times \alSU(2)$ invariant interactions
which preserve the dimension also preserve the number
of fields, i.e.~three fields are mapped into three fields.
Similarly, for $\algQ_2,\algS_2$ we need two fields going into two fields
\[\label{eq:Higher.Two.Struc}
\ham_2\sim \ITerm{A_1A_2A_3}{B_1B_2B_3},\quad 
\algQ_2,\algS_2\sim \ITerm{A_1A_2}{B_1B_2}.\]
It is an easy exercise to count the number of
structures in $\ham_2,\algQ_2,\algS_2$. For $\ham_2$ there $2^3=8$
ways to determine the statistics of $A_1A_2A_3$ 
and $3!=6$ ways to permute the fields 
(each $A$ must be contracted to one of the $B$'s).
In total there are $6\cdot 8=48$ structures for $\ham_2$ and $8$ 
for $\algQ_2,\algS_2$ each.
We now demand parity conservation. This restricts the number of 
independent structures to $28$ and $4$ for $\ham_2$ 
and $\algQ_2,\algS_2$, respectively.

At third order we need to determine $\ham_3,\algQ_3,\algS_3$. 
Like $\ham_1,\algQ_1,\algS_1$, 
all of these involve the totally antisymmetric tensors for
$\alSU(3),\alSU(2)$ and change the number of fields by one. 
Counting of independent structures is also straightforward,
we find $48$ for $\ham_3$ and $12$ for $\algQ_3,\algS_3$ each.
Parity conservation halves each of these numbers.

\subsection{Coefficients}
\label{sec:Higher.Two.Coeff}

Now we demand that 
energy shifts are conserved at
third order
\[\label{eq:Higher.Two.QH4}
\comm{\algQ^a{}_\alpha}{\ham}=\comm{\algS^\alpha{}_a}{\ham}=\order{g^4}.
\]
This fixes the remaining coefficient $\beta'_2$ at first order 
\eqref{eq:Higher.One.FromHigher} and 
many coefficients at second and third order.
The anticommutator of supercharges 
\[\label{eq:Higher.Two.SQ2}
\acomm{\algS^\alpha{}_a}{\algQ^b{}_\beta}=
  \delta^b_a \algL^\alpha{}_\beta
  +\delta_\beta^\alpha \algR^b{}_a
  +\delta_a^b \delta_\beta^\alpha (\sfrac{1}{3}\algD_0+\sfrac{1}{2}g^2 \ham)+\order{g^4}
\]
does not lead to additional constraints.
The resulting deformations of the generators up to second order are presented
in \tabref{tab:Higher.Two.Vertex}.
In the remainder of this subsection we shall discuss the 
undetermined coefficients $\alpha,\gamma,\delta$ 
and we shall find an explanation for each of them.
\begin{table}
\<
\algR^{a}{}_{b}\eq \ITerm{a}{b}-\sfrac{1}{3}\delta^a_b\ITerm{c}{c},
\nln
\algL^{\alpha}{}_{\beta}\eq \ITerm{\alpha}{\beta}-\sfrac{1}{2}\delta^\alpha_\beta\ITerm{\gamma}{\gamma},
\nln
\algD_0\eq
 \ITerm{a}{a}+\sfrac{3}{2} \ITerm{\alpha}{\alpha},
\nln
\ham_0\eq
 \alpha_1^2\ITerm{ab}{ab}
+\alpha_1^2\bigbrk{\ITerm{a\beta}{a\beta}+\ITerm{\alpha b}{\alpha b}}
+\alpha_1^2\ITerm{\alpha \beta}{\alpha\beta}
\nl
-\alpha_1^2\ITerm{ab}{ba}
-\alpha_1^2\bigbrk{\ITerm{a\beta}{\beta a}+\ITerm{\alpha b}{b \alpha}}
+\alpha_1^2\ITerm{\alpha \beta}{\beta\alpha},
\nln
\ham_1\eq
-\sfrac{1}{\sqrt{2}}\,\alpha_1^3\,e^{i\beta_2}\,\varepsilon_{\alpha\beta}\varepsilon^{abc}\ITerm{\alpha\beta}{abc}
-\sfrac{1}{\sqrt{2}}\,\alpha_1^3\,e^{-i\beta_2}\,\varepsilon_{abc}\varepsilon^{\alpha\beta}\ITerm{abc}{\alpha\beta},
\nln
\ham_2\eq
(-2\alpha_1^4+2\alpha_1\alpha_3)\ITerm{a b c}{a b c}
+(-\half \alpha_1^4+2\alpha_1\alpha_3+\delta_3)\ITerm{\alpha \beta \gamma}{\alpha \beta \gamma}
\nl
+(\sfrac{1}{2}\alpha_1^4+2\alpha_1\alpha_3+2\delta_2)\ITerm{a \beta c}{a \beta c}
+(-4\alpha_1^4 +2\alpha_1\alpha_3- 2\delta_2)\ITerm{\alpha b\gamma}{\alpha b\gamma}
\nl
+(-\sfrac{11}{4}\alpha_1^4+2\alpha_1\alpha_3-\delta_2)\bigbrk{\ITerm{a b \gamma}{a b \gamma}+\ITerm{\alpha b c}{\alpha b c}}
+(2\alpha_1\alpha_3+\delta_2)\bigbrk{\ITerm{a \beta \gamma}{a \beta \gamma}+\ITerm{\alpha \beta c}{\alpha \beta c}}
\nl
+(\sfrac{3}{2}\alpha_1^4-\alpha_1\alpha_3)\bigbrk{\ITerm{a b c}{b a c}+\ITerm{a b c}{a c b}}
+(\alpha_1^4-\alpha_1\alpha_3)\bigbrk{\ITerm{a b \gamma}{b a \gamma}+\ITerm{\alpha b c}{\alpha c b}}
\nl
+(\sfrac{5}{4}\alpha_1^4-\alpha_1\alpha_3+i\alpha_1^2\gamma_3+i\delta_1)\bigbrk{\ITerm{\alpha b c}{b \alpha c}+\ITerm{a b \gamma}{a \gamma b}}
\nl
+(\sfrac{5}{4}\alpha_1^4-\alpha_1\alpha_3-i\alpha_1^2\gamma_3-i\delta_1)\bigbrk{\ITerm{a \beta c}{\beta a c}+\ITerm{a \beta c}{a c \beta}}
\nl
+(\alpha_1^4-\alpha_1\alpha_3+i\delta_1)\bigbrk{\ITerm{\alpha b \gamma}{b \alpha \gamma}+\ITerm{\alpha b \gamma}{\alpha \gamma b}}
\nl
+(\alpha_1^4-\alpha_1\alpha_3-i\delta_1)\bigbrk{\ITerm{a \beta \gamma}{\beta a \gamma}+\ITerm{\alpha \beta c}{\alpha c \beta}}
\nl
+(-\sfrac{7}{4}\alpha_1^4+\alpha_1\alpha_3)\bigbrk{\ITerm{\alpha \beta c}{\beta \alpha c}+\ITerm{a \beta \gamma}{a \gamma \beta}}
+(-\sfrac{7}{4}\alpha_1^4+\alpha_1\alpha_3-\delta_3)\bigbrk{\ITerm{\alpha \beta \gamma}{\beta \alpha \gamma}+\ITerm{\alpha \beta \gamma}{\alpha \gamma \beta}}
\nl
-\half \alpha_1^4 \bigbrk{\ITerm{a b c}{c a b}+\ITerm{a b c}{b c a}}
+\delta_3\bigbrk{\ITerm{\alpha \beta\gamma}{\gamma \alpha \beta}+\ITerm{\alpha \beta\gamma}{\beta \gamma \alpha}}
\nl
+(-\sfrac{1}{4}\alpha_1^4+i\alpha_1^2\gamma_1)\bigbrk{\ITerm{\alpha b c}{c \alpha b}+\ITerm{a b\gamma}{b \gamma a}}
+(\sfrac{1}{4} \alpha_1^4+i\alpha_1^2\gamma_2)\bigbrk{\ITerm{a \beta \gamma}{\gamma a \beta}+\ITerm{\alpha \beta c}{\beta c \alpha}}
\nl
+(-\sfrac{1}{4}\alpha_1^4-i\alpha_1^2\gamma_1)\bigbrk{\ITerm{a \beta c}{c a \beta}+\ITerm{a \beta c}{\beta c a}}
+(\sfrac{1}{4} \alpha_1^4-i\alpha_1^2\gamma_2)\bigbrk{\ITerm{\alpha b\gamma}{\gamma \alpha b}+\ITerm{\alpha b\gamma}{b \gamma \alpha}}
\nl
-\half \alpha_1^4 \bigbrk{\ITerm{a b \gamma}{\gamma b a}+\ITerm{\alpha b c}{c b \alpha}}
+\half \alpha_1^4 \ITerm{\alpha b \gamma}{\gamma b \alpha}
-\delta_3 \ITerm{\alpha \beta \gamma}{\gamma \beta \alpha}
\nl
+\sfrac{3}{2}\alpha_1^4\bigbrk{\WTerm{\alpha\beta}{\alpha\beta}-\WTerm{\alpha\beta}{\beta\alpha}}
,
\nln
(\algQ_0)^{a}{}_{\alpha}\eq e^{i\beta_1}\ITerm{a}{\alpha},
\nln
(\algQ_1)^{a}{}_{\alpha}\eq \sfrac{1}{\sqrt{2}}\,\alpha_1\,e^{i\beta_1+i\beta_2}\varepsilon_{\alpha\beta}\varepsilon^{abc}\ITerm{\beta}{bc},
\nln
(\algQ_2)^{a}{}_{\alpha}\eq 
e^{i\beta_1}(\sfrac{1}{4}\alpha_1^2-i\half \gamma_3+i\half \gamma_4)\bigbrk{\ITerm{a b}{\alpha b}+\ITerm{b a}{b \alpha}}
+e^{i\beta_1}(\half i\gamma_3+\half i\gamma_4)\bigbrk{\ITerm{a\beta}{\alpha\beta}-\ITerm{\beta a}{\beta\alpha}}
\nl
+e^{i\beta_1}(-\sfrac{1}{4} \alpha_1^2-i\gamma_1)\bigbrk{\ITerm{a b}{b \alpha}+\ITerm{b a}{\alpha b}}
+e^{i\beta_1}(\sfrac{1}{4} \alpha_1^2+i\gamma_2)\bigbrk{\ITerm{a \beta}{\beta \alpha}-\ITerm{\beta a}{\alpha \beta}},
\nln
(\algS_0)^{\alpha}{}_{a}\eq e^{-i\beta_1}\ITerm{\alpha}{a},
\nln
(\algS_1)^{\alpha}{}_{a}\eq \sfrac{1}{\sqrt{2}}\,\alpha_1\, e^{-i\beta_1-i\beta_2} \varepsilon_{abc}\varepsilon^{\alpha\beta}\ITerm{bc}{\beta},
\nln
(\algS_2)^{\alpha}{}_{a}\eq 
e^{-i\beta_1}(\sfrac{1}{4} \alpha_1^2+\half i\gamma_3-\half i\gamma_4)\bigbrk{\ITerm{\alpha b}{a b}+\ITerm{b \alpha}{b a}}
+e^{-i\beta_1}(-\half i\gamma_3-\half i\gamma_4)\bigbrk{\ITerm{\alpha \beta}{a \beta}-\ITerm{\beta \alpha}{\beta a}}
\nl
+e^{-i\beta_1}(-\sfrac{1}{4}\alpha_1^2+i\gamma_1)\bigbrk{\ITerm{\alpha b}{b a}+\ITerm{b \alpha}{a b}}
+e^{-i\beta_1}(\sfrac{1}{4}\alpha_1^2-i\gamma_2)\bigbrk{-\ITerm{\alpha \beta}{\beta a}+\ITerm{\beta \alpha}{a \beta}}.
\nonumber
\>
\caption{Two-loop deformations of the generators}
\label{tab:Higher.Two.Vertex}
\end{table}
\begin{bulletlist}
\item
Firstly, the constants $\delta_{1,2}$ multiply a structure
which has a spectator leg on either side of the interaction
\[\label{eq:Higher.Two.Spectators}
\ITerm{A_{1}\ldots A_{E\indups{i}}C}
{B_{1}\ldots B_{E\indups{o}}C}
-
(-1)^{C(A_{1}\ldots A_{E\indups{i}}B_{1}\ldots B_{E\indups{o}})}
\ITerm{CA_{1}\ldots A_{E\indups{i}}}
{CB_{1}\ldots B_{E\indups{o}}},
\]
such that both interactions cancel out 
in a cyclic state.

\item
Secondly, the constant $\delta_3$ multiplies a structure
which is zero due to an $\alSU(2)$ identity.
We cannot antisymmetrise more than two
fundamental representations of $\alSU(2)$
\[\label{eq:Higher.Two.SU2Ident}
\ITerm{\alpha\beta\gamma}{[\alpha\beta\gamma]}=0.
\]

\item
Thirdly, we can use a similarity transformation to
modify the generators
\[\label{eq:Higher.Two.Sim}
\algJ(g)\mapsto T(g)\,\algJ(g)\,T(g)^{-1}.
\]
In \secref{sec:Higher.One.Concl}, we have used a transformation
which is independent of the coupling constant, 
here we consider a transformation $T(g)=1+g^2 T_2+\ldots$ 
proportional to $g^2$.
For consistency with the algebra, the transformation 
will have to be $\alSU(3)\times \alSU(2)$ invariant and preserve 
the dimension as well as parity. 
Also, according to \secref{sec:Dila.Planar.Interact}, 
it should involve four fields. 
These are exactly the requirements for the form of $\ham_0$,
the $6$ independent structures are given 
in \eqref{eq:Higher.One.H0,eq:Higher.One.H0Parity}.
Out of these six, there are two special combinations:
One of them is $\ham_0$ itself and the other one 
is equivalent to the length operator
\[\label{eq:Higher.Two.LenGauge}
\len=\ITerm{ab}{ab}+\ITerm{a\beta}{a\beta}
+\ITerm{\alpha b}{\alpha b}+\ITerm{\alpha\beta}{\alpha\beta}
\]
up to gauge transformations. 
The similarity transformation amounts to adding 
commutators with $\ham_0,\algQ_0,\algS_0$
\[\label{eq:Higher.Two.GenSim}
\ham_2\mapsto \ham_2+\comm{T_2}{\ham_0},\quad
\algJ_2\mapsto \algJ_2+\comm{T_2}{\algJ_0}.
\]
These commutators vanish for $\ham_0$ and $\len$
\[\label{eq:Higher.Two.SimSym}
\comm{\ham_0}{\algJ_0}=\comm{\len}{\algJ_0}=\comm{\ham_0}{\ham_0}=\comm{\len}{\ham_0}=0.
\]
In other words, conjugation with $g^2\ham_0$ and $g^2\len$ will have no effect on
$\ham_2,\algQ_2,\algS_2$.
The remaining four structures in \eqref{eq:Higher.One.H0,eq:Higher.One.H0Parity}
do not commute with $\ham_0,\algQ_0,\algS_0$
and amount to the constants $\gamma_{1,2,3,4}$. 
Note that $\gamma_4$ is related to the structure $\algD_0$ 
and does not appear in $\ham_2$ because of $\comm{\algD_0}{\ham_0}=0$.

\item
Finally, we are allowed to perform 
a transformation of the coupling constant 
\[\label{eq:Higher.Two.Renorm}
\algJ(g)\mapsto \algJ(f(g)).
\]
If we use the function $f(g)=\alpha_1 g+\alpha_3 g^3$
we find that 
\[\label{eq:Higher.Two.ActRenorm}
\ham_2\mapsto \alpha_1^4 \ham_2+2\alpha_1\alpha_3 \ham_0,
\]
which explains the degree of freedom $\alpha_3$.
\end{bulletlist}

\subsection{Short States and Wrapping Interactions}
\label{sec:Higher.Two.Short}

The second order interactions $\ham_2$ act on three fields.
We should also determine its action on the states of length two%
\footnote{Length-one states are $\grU(1)$ fields 
and do not interact at all.}
\[\label{eq:Higher.Two.ShortStates}
\Op_{(ab)}=\state{ab}=\Tr \phi_a\phi_b,\quad
\Op_{a\beta}=\state{a\beta}=\Tr \phi_a\psi_\beta,\quad
\Op_1=\varepsilon^{\alpha\beta}\state{\alpha\beta}=
\varepsilon^{\alpha\beta}\Tr \psi_\alpha\psi_\beta.
\]
Together, these form the protected half-BPS multiplet $[0;0;0,2]$.
It is therefore reassuring to see that $\Op_{(ab)}$ and $\Op_{a\beta}$
are annihilated by $\ham_0,\ham_1$;
just as well they should be annihilated by $\ham_2$.
For $\Op_1$ the situation is different: It is annihilated by 
$\ham_0$, but $\ham_1$ produces the operator
\[\label{eq:Higher.Two.H3ActShort}
\Op_2=\varepsilon^{abc}\state{abc}=\varepsilon^{abc}\Tr\phi_a\phi_b\phi_c.
\]
The action of $\ham(g)$ on these two operators up to second order is given by
\[\label{eq:Higher.Two.ShortAct}
\ham(g)\matr{c}{\Op_1\\\Op_2}=\matr{cc}{\epsilon g^2 & -2\sqrt{2}\,e^{i\beta_2}\alpha_1^3 g\\ 
-9\sqrt{2}\,e^{i\beta_2}e^{-i\beta_2} \alpha_1^3 g
&6\alpha_1^2-18 \alpha_1^2 g^2+12\alpha_1\alpha_3 g^2}
\matr{c}{\Op_1\\\Op_2},
\]
where we have assumed that $\ham_2 \Op_1=\epsilon\Op_1$.
The eigenvalues of this matrix at fourth order are given by 
\[\label{eq:Higher.Two.ShortEng}
E_1=\epsilon g^2-6\alpha_1^4 g^2,\quad
E_2=6\alpha_1^2-12\alpha_1^4g^2+12\alpha_1\alpha_3g^2.
\]
Due to its half-BPS nature, the energy of the diagonalised $\Op_1$ 
must be exactly zero, 
$E_1=0$, and we set
\[\label{eq:Higher.Two.EpsRes}
\epsilon=6\alpha_1^4.
\]
The second order Hamiltonian for states of length two should thus
annihilate the states $\Op_{(ab)},\Op_{a\beta}$ and yield 
$6\alpha_1^4\Op_1$ when acting on $\Op_1$. This is achieved by
a wrapping interaction, c.f.~\secref{sec:Dila.Planar.Wrapping},
\[\label{eq:Higher.Two.H4Short}
\ham_2=
\ldots+
\sfrac{3}{2}\alpha_1^4
\bigbrk{\WTerm{\alpha\beta}{\alpha\beta}-\WTerm{\alpha\beta}{\beta\alpha}}.
\]

\subsection{Conclusions}
\label{sec:Higher.Two.Concl}

We see that for all free parameters in \tabref{tab:Higher.Two.Vertex}
there is an associated symmetry of the algebra relations
and we can say that the two-loop contribution is uniquely fixed. 
The only parameter that influences energies is $\alpha_3$;
we cannot remove it by algebraic considerations.
The parameters $\gamma_{1,2,3,4}$ rotate only the eigenstates.
Finally, the parameters $\delta_{1,2,3}$ are there only because
we were not careful enough in finding \emph{independent} structures
(for $\ham_2$ there are only $25=28-3$ independent structures).
They have no effect at all.

\section{Three-Loops}
\label{sec:Higher.Three}

For the fourth order contributions $\ham_4,\algQ_4,\algS_4$ we find in total 
$208+56+56$ parity conserving structures; they all conserve the number of fields%
\footnote{At sixth order the number of fields can be changed by two
using four antisymmetric tensors.}.
Of these only $173+32+32$ are independent due to identities
as discussed above.
We impose the constraint \eqref{eq:Higher.One.Comm} at fourth order 
\[\label{eq:Higher.Three.QHSH6}
\comm{\algQ}{\ham}=\comm{\algS}{\ham}=\order{g^5}\]
and find that the algebra relations fix $202$ coefficients 
(plus one coefficient at third order). This leaves $35$ free coefficients.
The anticommutator of supercharges \eqref{eq:Higher.SU23.AlgMomRot}
at fourth order
\[\label{eq:Higher.Three.SQ4}
\acomm{\algS^\alpha{}_a}{\algQ^b{}_\beta}=
  \delta^b_a \algL^\alpha{}_\beta
  +\delta_\beta^\alpha \algR^b{}_a
  +\delta_a^b \delta_\beta^\alpha (\sfrac{1}{3}\algD_0+\sfrac{1}{2}g^2 \ham)+\order{g^5}
\]
is satisfied automatically.

As we have learned above, the commutators at fourth order are not
sufficient to ensure consistency for splitting multiplets at the
unitarity bound, we should also consider fifth order. 
To perform those commutators would be even harder. 
We therefore consider a set of probe multiplets at the unitarity bound. 
By requiring that the three-loop energy shifts coincide within 
submultiplets we are able to fix another $8$ coefficients. 

\begin{table}\centering
$\begin{array}{|c|rrrrr|}\hline
      k                           & 0 & 1 & 2 & 3 & 4   \\ \hline
\ham_{k}                          & 6 & 2 & 25& 18&173  \\ 
\algQ_{k}                         & 1 & 1 &  4&  6& 32  \\
\algS_{k}                         & 1 & 1 &  4&  6& 32  \\\hline
\mbox{total}                      & 8 & 4 & 33& 30&237  \\
\mbox{fixed at $\order{g^k}$}     & 5 & 2 & 25& 26&202  \\
\mbox{fixed at $\order{g^{k+1}}$} & 1 & 1 &  3&  1&  8  \\
\mbox{relevant}                   & 1 & 0 &  1&  0&  4  \\\hline
\mbox{irrelevant}                 & 1 & 1 &  4&  3& 23  \\
\mbox{symmetries}                 & 1 &-1 &  2& -1&  2  \\\hline
\ham_{k-2}                        & 2 & 0 &  6&  2& 25  \\
\hline
\end{array}$
\caption{Number of coefficients. $\ham_k,\algQ_k,\algS_k$ give the number 
of independent structures that can be used for the construction of
generators.
The algebra relations fix a certain number of coefficients.
Of the remaining coefficients, some are relevant for energies
and some correspond to similarity transformations generated by 
the structures in $\ham_{k-2}$.
Some of the similarity transformations are symmetries.}
\label{tab:Higher.Three.Count}
\end{table}%
Still this leaves $27$ coefficients to be fixed, however, 
almost all of them rotate the space of states.
Experimentally, we found that only $4$ coefficients affect the energies. 
The remaining $23$ coefficients can be attributed to similarity transformations.
As before, the number of similarity transformations equals the number
of structures for $\ham_2$, i.e.~$25$. This means that there must be
$2$ commuting generators which are readily found to be $g^4 \ham_0$ and $g^4 L$. 
We summarise our findings concerning the number of 
coefficients in \tabref{tab:Higher.Three.Count}. The symmetries indicated 
in the table refer to $\len$ (which is conserved at leading order but
broken at first order, hence the $-1$ at $k=1$), $g^2\len$
(broken at third order), $g^2 \ham$,
$g^4\len$ (will break at fifth order) and $g^4 \ham$.
This sequence of symmetries will continue at higher orders, but
there will be additional ones due to integrability,
see \secref{sec:HighInt.Higher.Revisit}.

Let us now discuss the relevant coefficients.
One coefficient is due to a redefinition of the coupling constant and
cannot be fixed algebraically. 
To constrain the other three we will need further input.
Unfortunately, the resulting generators are too lengthy to
be displayed here. Instead, let us have a look at the set 
of totally bosonic states.
In this subsector (which is closed only when further 
restricted to two flavours, i.e.~the $\alSU(2)$ subsector) 
$\ham_4$ is presented in \tabref{tab:Higher.Three.H6SU3}.

The coefficients in \tabref{tab:Higher.Three.H6SU3} can be understood as follows.
The coefficients $\sigma_{1,2,3,4}$ are relevant.
One of them, $\sigma_1$, multiplies the structure $\ham_0$ 
(up to spectator legs)
and therefore corresponds to a redefinition of the
coupling constant as in \eqref{eq:Higher.Two.Renorm}.
The coefficients $\zeta_{1,2,3}$ multiply structures which are actually zero: 
More explicitly, $\zeta_1$ multiplies $\ITerm{abcd}{[abcd]}$
and $\zeta_{2,3}$ can be gauged away by removing spectator legs.
Finally, the coefficients $\xi_{1,2,3}$ are related 
to similarity transformations and have no effect on the
energy shifts.
\begin{table}
\<
\ham_4\eq
\bigbrk{\sfrac{15}{2}\alpha_1^6-8\alpha_1^3\alpha_3+\sigma_1-\sfrac{1}{3}\sigma_2+12\sigma_3-2\sigma_4+\xi_2+\zeta_1}\ITerm{abcd}{abcd}\nl
+\bigbrk{-\sfrac{13}{4}\alpha_1^6+3\alpha_1^3\alpha_3-\sfrac{1}{4}\sigma_1+\sfrac{1}{6}\sigma_2-3\sigma_3+\sigma_4-\sfrac{1}{2}\xi_2-\zeta_1+\zeta_3}\bigbrk{\ITerm{abcd}{abdc}+\ITerm{abcd}{bacd}}\nl
+\bigbrk{-\sfrac{13}{2}\alpha_1^6+6\alpha_1^3\alpha_3-\sfrac{1}{2}\sigma_1+\sfrac{1}{3}\sigma_2-6\sigma_3+2\sigma_4-\xi_2-\zeta_1-2\zeta_3}\ITerm{abcd}{acbd}\nl
+\bigbrk{\sfrac{3}{2}\alpha_1^6-\alpha_1^3\alpha_3-\sigma_4+\sfrac{1}{2}\xi_2+\zeta_1-i\zeta_2}\bigbrk{\ITerm{abcd}{acdb}+\ITerm{abcd}{cabd}}\nl
+\bigbrk{\sfrac{3}{2}\alpha_1^6-\alpha_1^3\alpha_3-\sigma_4+\sfrac{1}{2}\xi_2+\zeta_1+i\zeta_2}\bigbrk{\ITerm{abcd}{adbc}+\ITerm{abcd}{bcad}}\nl
+\bigbrk{-\sfrac{1}{6}\sigma_2-\sfrac{1}{2}\xi_2-\zeta_1}\bigbrk{\ITerm{abcd}{adcb}+\ITerm{abcd}{cbad}}
+\bigbrk{\sfrac{1}{2}\alpha_1^6-2\sigma_3+\zeta_1}\ITerm{abcd}{badc}\nl
+\bigbrk{-\sfrac{1}{2}\alpha_1^6+\sfrac{1}{3}\sigma_2-4\sigma_3+2\sigma_4-\zeta_1}\bigbrk{\ITerm{abcd}{bcda}+\ITerm{abcd}{dabc}}\nl
+\bigbrk{-\sfrac{1}{3}\sigma_2+4\sigma_3+\sigma_4-i\xi_3-\zeta_1}\ITerm{abcd}{bdac}
+\bigbrk{-\sfrac{1}{3}\sigma_2+4\sigma_3+\sigma_4+i\xi_3-\zeta_1}\ITerm{abcd}{cadb}\nl
+\bigbrk{\sigma_3-\sigma_4+i\xi_1+\zeta_1}\bigbrk{\ITerm{abcd}{bdca}+\ITerm{abcd}{dbac}}
+\bigbrk{\sigma_3-\sigma_4-i\xi_1+\zeta_1}\bigbrk{\ITerm{abcd}{cbda}+\ITerm{abcd}{dacb}}\nl
+\bigbrk{-2\sigma_3-2\sigma_4+\zeta_1}\ITerm{abcd}{cdab}
+\bigbrk{-\zeta_1}\bigbrk{\ITerm{abcd}{cdba}+\ITerm{abcd}{dcab}}\nl
+\bigbrk{2\sigma_4-\zeta_1}\ITerm{abcd}{dbca}
+\bigbrk{\zeta_1}\ITerm{abcd}{dcba}.
\nonumber
\>\vspace{-0.8cm}
\caption{$\ham_4$ acting on bosonic states.}
\label{tab:Higher.Three.H6SU3}
\end{table}

The crucial point is that we want $\ham_4$ to be generated by Feynman diagrams. 
Here we can make use of a special property of the scalar sector, 
c.f.~\secref{sec:Dila.Theory.Feyn}. 
The Feynman diagrams with the maximum number of eight legs do not 
have internal index loops.
In the planar case, such diagrams must be iterated one-loop diagrams.
This implies that we can only have three permutations of \emph{adjacent}
fields. The structures
\[\label{eq:Higher.Three.NoNo}
\ITerm{abcd}{cdab},\ITerm{abcd}{bdca},\ITerm{abcd}{dbac},
\ITerm{abcd}{cbda},\ITerm{abcd}{dacb},\quad
\ITerm{abcd}{cdba},\ITerm{abcd}{dcab},\ITerm{abcd}{dbca},
\quad\ITerm{abcd}{dcba}
\]
consist of four, five or six crossings of adjacent fields and are therefore excluded. 
We must set their coefficients to zero
\[\label{eq:Higher.Three.NoNoCoeff}
\sigma_3=\sigma_4=0,\qquad \zeta_1=\xi_1=0.\]
The final relevant coefficient $\sigma_2$ multiplies a structure
$\charge_{4,0}$ which commutes with $\ham_0$. This issue is 
related to integrability, see \secref{sec:HighInt.Higher.Revisit}.
At this point we cannot determine $\sigma_2$, but believe that 
it will be fixed due 
to the anticommutator $\acomm{\algQ}{\algS}$
at $\order{g^6}$ (four-loops).

\section{Spectrum}
\label{sec:Higher.Spec}

In this section we fix the remaining degrees of freedom within
the Hamiltonian and apply it to a number of states to find their energies. 

\subsection{The Remaining Coefficients}
\label{sec:Higher.Spec.Coeff}

First of all, we would like to fix the remaining relevant
coefficients. 
We cannot do this algebraically because 
most of them correspond to symmetries of the 
commutation relations. One of them is
a redefinition of the coupling constant 
\[\label{eq:Higher.Spec.Renorm}
g\mapsto f(g).
\]
Unlike the other symmetries, this transformation has relevant consequences, 
it implies that energies are changed according to
\[\label{eq:Higher.Spec.RenormEng}
E(g)\mapsto E(f(g)).
\]
In order to match these degrees of freedom to 
$\superN=4$ SYM we should use some scaling dimension
that is known to all orders in perturbation theory.
In fact it is sufficient to use the scaling behaviour
in the BMN limit, c.f.~\secref{sec:One.BMN}, 
i.e.~for large $J$, all properly rescaled quantities should
depend only on 
\[\label{eq:Higher.Spec.LambdaPrime}
\hat g=\frac{g}{J}\qquad \mbox{or}\qquad 
\lambda'=\frac{\lambda}{J^2}=\frac{8\pi^2g^2}{J^2}=8\pi^2\hat g^2.
\]
Let us assume this to be the case.
If we redefine the coupling constant $g$
we obtain for the rescaled coupling constant 
\[\label{eq:Higher.Spec.LambdaPrimeRenorm}
\hat g\mapsto \frac{f(g)}{J}
=\frac{f_1 g+f_3 g^3+\ldots}{J}
=f_1 \hat g+f_3 \hat g^3 J^2+\ldots\,\,.
\]
The problem is that all the higher expansion 
coefficients of $f$ yield divergent contributions in the BMN limit
$J\to\infty$.
Thus all $\hat g$ dependent quantities will also become divergent.
The only degree of freedom compatible with BMN scaling behaviour
is to change $g$ by a constant factor $f_1$.

In our model we would like to define the coupling constant by
fixing $\alpha_3,\sigma_1$ in
such a way as to obtain a good scaling behaviour of energies in the BMN limit.
It is not possible to achieve proper scaling by adjusting 
$\alpha_3,\sigma_1$ alone; also $\sigma_2$ multiplies
a structure with a wrong scaling. 
This is fortunate, because it
allows us to determine $\sigma_2$ as well, in total we find
\[\label{eq:Higher.Spec.BMNCoeff}
\alpha_3=0,\quad \sigma_1=0,\quad \sigma_2=0.
\]
Afterwards we can only change $\alpha_1$. This final degree of freedom is 
eliminated by a single known scaling dimension, e.g.~the one of the 
Konishi multiplet \eqref{eq:Dila.Dim.KonishiDim} $E_0=6$,
or using the quantitative BMN energy formula \eqref{eq:One.BMN.Energy}.
It fixes $\alpha_1$ to unity
\[\label{eq:Higher.Spec.FinalCoeff}
\alpha_1=1.
\]

We conclude that \emph{the planar three-loop Hamiltonian is uniquely fixed} 
by the symmetry algebra, field theory and the BMN scaling behaviour.%
\footnote{Without BMN scaling
the constants $\alpha_1,\alpha_3,\sigma_1,\sigma_2$ remain unknown.
However, $\alpha_1,\alpha_3,\sigma_1$ are related to 
a redefinition of the coupling constant and, as we shall see
in \secref{sec:HighInt.Higher.Revisit}, there is a natural
explanation for $\sigma_2$ in terms of integrability. 
It therefore makes sense to say that the Hamiltonian is 
uniquely fixed (up to symmetries) even without making use of BMN scaling.}
Together with the fact that this model is a closed subsector of
$\superN=4$ SYM we have derived the planar dilatation generator 
in the $\alSU(2|3)$ subsector at three-loops.
Similarly, this model is a closed subsector of the BMN matrix model
and the two Hamiltonians must agree up to three loops
(after a redefinition of the coupling constant and provided that 
the BMN matrix model has a BMN limit).
This is indeed the case as shown in \cite{Klose:2003qc}.

\subsection{Lowest-Lying States}
\label{sec:Higher.Spec.Low}

We are now ready to compute numerical values for some energies. 
For this we should consider the charges 
$D_0,s,p,q,L$ of a state 
and compute the number of constituent fields
according to \eqref{eq:Higher.SU23.Numbers}.
These are arranged within a trace in all possible ways
\[\label{eq:Higher.Spec.AllOps}
\OpE_n=
(
\Tr
\phi_1^{n_1}
\phi_2^{n_2}
\phi_3^{n_3}
\psi_1^{n_4}
\psi_2^{n_5},
\ldots
).
\]
Note that the length $L$ is not a good quantum number at
$\order{g}$, so we must include states of all admissible lengths
in \eqref{eq:Higher.Spec.AllOps}.
In practice this means that we may replace a complete set 
of bosons $\phi_1\phi_2\phi_3$ by a complete set of fermions
$\psi_1\psi_2$, \eqref{eq:Higher.SU23.EqualCharges}. 
Due to conservation of charges, 
the Hamiltonian closes on this set of states
and we can evaluate its matrix elements%
\footnote{Although the 
Hilbert space is infinite-dimensional,
the Hamiltonian acts on a space of
fixed dimension $D_0$. 
Therefore the matrix $H^j{}_i(g)$ has a finite size.}
\[\label{eq:Higher.Spec.HMatrix}
\ham(g)\, \OpE_i= \OpE_j \,H^j{}_i(g).
\]
It is a straightforward task to find the eigenvalues and their perturbations
\[\label{eq:Higher.Spec.PerturbSplit}
\ham(g)=\ham_0+V(g),\qquad
\mbox{with }V=\order{g}.
\]
Diagonalising the leading order matrix $\ham_0$ is a non-linear problem. 
The resulting eigenvalues represent the one-loop energies $E_0$. 
Now we pick an eigenvalue $e=E_0$ of $\ham_0$ and consider the subspace of states
with energy $e$.
The higher-order energy shifts are given by
(in contrast to the formula in \cite{Klose:2003qc} 
$V'$ was constructed such that conjugation symmetry is preserved)
\<\label{eq:Higher.Spec.PerturbDiag}
V'\eq 
\smash{\sum\nolimits_{e}\Pi_e\Big[}
 V 
+ V\Delta_e V 
+ \bigbrk{V \Delta_e V\Delta_e V
-\half V \Delta_e^2 V\Pi_e V
-\half V \Pi_e V\Delta^2_e V}
\nl\qquad\qquad
+V\Delta_e V\Delta_e V\Delta_e V
-\half V\Delta_e^2 V\Pi_e V\Delta_e V
-\half V\Delta_e V\Pi_e V\Delta_e^2 V
\nl\qquad\qquad 
-\half V\Delta_e^2 V\Delta_e V\Pi_e V
-\half V\Pi_e V\Delta_e  V\Delta_e^2 V
\nlnum\qquad\qquad
-\half V\Delta_e V\Delta_e^2 V\Pi_e V
-\half V\Pi_e V\Delta_e^2 V\Delta_e V
\nl\qquad\qquad
+\sfrac{1}{3} V\Delta_e^3 V\Pi_e V\Pi_e  V
+\sfrac{1}{3} V\Pi_e V\Delta_e^3 V\Pi_e  V
+\sfrac{1}{3} V\Pi_e V\Pi_e V\Delta_e^3 V
+\ldots \smash{\Big]} \Pi_e.
\nonumber
\>
The propagator $\Delta_e$ is given by
\[\label{eq:Higher.Spec.PerturbProp}
\Delta_e= \frac{1-\Pi_e}{e-\ham_0}
\]
and $\Pi_e$ projects to the subspace
with leading correction $e$.
If there is only a single state with one-loop energy $e$,
\eqref{eq:Higher.Spec.PerturbDiag} gives its higher order corrections. 
For degenerate states at one-loop, 
\eqref{eq:Higher.Spec.PerturbSplit,eq:Higher.Spec.PerturbDiag} must be applied 
iteratively until the resulting matrix $V'$ becomes diagonal%
\footnote{In principle it could happen that states
with equal leading order energy $e$ have matrix
elements at $\order{g}$. 
In this case the energy would have an expansion in terms of $g\sim\sqrt{\lambda}$ 
instead of $g^2\sim \lambda$
similar to the peculiarities noticed in \cite{Beisert:2003tq}.
It would be interesting to see if this 
does indeed happen or, if not, why?}.

Next, it is important to know the multiplets
of states. In the interacting theory there
are two types of single-trace multiplets, half-BPS and
long ones.
The half-BPS multiplets are easily identified,
there is one multiplet with labels
\[\label{eq:Higher.Spec.HalfBPS}
D_0=L=p,\quad E=0,\quad [0;0;0,p],\quad P=(-1)^p
\]
for each $p$, they receive no corrections to their energy.
Long multiplets are not so easy to find. 
By means of a \texttt{C++} computer programme
we have constructed the spectrum of 
all states explicitly (up to some energy bound) 
and iteratively removed the multiplets
corresponding to the leftover highest weight state
(this `sieve' algorithm, also reminiscent of the
standard algorithm for division, is described in more
detail in \cite{Bianchi:2003wx,Beisert:2003te}).
For a set of states with given charges as in \eqref{eq:Higher.Spec.AllOps}
this also tells us how many representatives there are 
from each of the multiplets and allows us to
identify the energy we are interested in.

Finally, to obtain the energy shift of a given multiplet, 
a lot of work can be saved by choosing a suitable representative. 
Resolving the mixing problem for the highest weight state 
is usually more involved than for a descendant. 
For instance, highest weight states involve all
three flavours of bosons, $n_1,n_2,n_3\geq 1$.
This increases the number
of permutations in \eqref{eq:Higher.Spec.AllOps}
and also gives rise to mixing between states 
of different lengths. 
The matrix $H^m{}_n$ will be unnecessarily large.
If, instead, one applies three supergenerators
$\algQ^1{}_4 \algQ^2{}_4 \algQ^3{}_4$, i.e.
\[\label{eq:Higher.Spec.Optimise}
n_1\mapsto n_1-1,\quad
n_2\mapsto n_2-1,\quad
n_3\mapsto n_3-1,\quad
n_4\mapsto n_4+3,
\]
the state becomes more uniform.
This decreases the number of permutations and,
in the case of multiplets at the unitarity bound \eqref{eq:Higher.SU23.Bound},
mixing between states of different lengths is
prevented due to $n_3=n_5=0$.

\begin{table}\centering
$\begin{array}{|l|lc|l|}\hline
D_0&\multicolumn{1}{c}{\alSU(2|3)}&L&\bigbrk{E_0,E_2,E_4}^{P}\\\hline
2  &[0;0;0,2]^{\dagger\bullet}&2&(0,0,0)^+ \\
\hline
3  &[0;0;0,3]^{\dagger\bullet}&3&(0,0,0)^- \\
3  &[0;1;0,0]^{\ast\bullet}&3&\bigbrk{6,-12,42}^+ \\
\hline
4  &[0;0;0,4]^{\dagger\bullet}&4&(0,0,0)^+ \\
4  &[0;1;0,1]^{\ast\bullet}&4&\bigbrk{4,-6,17}^- \\
\hline
5  &[0;0;0,5]^{\dagger\bullet}&5&(0,0,0)^- \\
5  &[0;1;0,2]^{\ast\bullet}&5&\bigbrk{10E-20,-17E+60,\frac{117}{2}E-230}^+ \\
5  &[0;1;1,0]^{\ast\bullet}&5&\bigbrk{6,-9,\frac{63}{2}}^- \\
\hline
6  &[0;0;0,6]^{\dagger\bullet}&6&(0,0,0)^+ \\
6  &[0;1;0,3]^{\ast\bullet}&6&\bigbrk{2,-\frac{3}{2},\frac{37}{16}}^-,\bigbrk{6,-\frac{21}{2},\frac{555}{16}}^- \\
6  &[0;1;1,1]^{\ast\bullet}&6&\bigbrk{5,-\frac{15}{2},25}^\pm \\
6  &[0;2;0,0]&6&\bigbrk{14E-36,-24E+90,\frac{173}{2}E-315}^+\\
6  &[2;3;0,0]^\ast&5&\bigbrk{10,-20,\frac{145}{2}}^-\\
\hline
6.5&[1;2;0,2]^\ast&6&\bigbrk{8,-14,49}^\pm \\
\hline
7  &[0;0;0,7]^{\dagger\bullet}&7&\bigbrk{0,0,0}^- \\
7  &[0;1;0,4]^{\ast\bullet}&7&\bigbrk{\scriptstyle 14E^2-56E+56,\,-23E^2+172E-224,\,79E^2-695E+966}^+ \\
7  &[0;1;1,2]^{\ast\bullet}&7&\bigbrk{4,-5,14}^\pm,\bigbrk{6,-9,33}^- \\
7  &[0;1;2,0]^{\ast\bullet}&7&\bigbrk{\scriptstyle 20E^2-116E+200,\,-32E^2+340E-800,\,112E^2-1400E+3600}^+ \\
7  &[0;2;0,1]&7&\bigbrk{\scriptstyle22E^2-144E+248,\,-37E^2+460E-1016,\,125E^2-1893E+4438}^- \\
7  &[2;3;0,1]^\ast&6&\bigbrk{8,-14,46}^+ \\
\hline
7.5&[1;2;0,3]^\ast&7&\bigbrk{7,-12,\frac{83}{2}}^\pm \\
7.5&[1;2;1,1]^\ast&7&\bigbrk{6,-\frac{33}{4},\frac{1557}{64}}^\pm,\bigbrk{10,-\frac{75}{4},\frac{4315}{64}}^\pm \\
7.5&[1;3;0,0]&7&\bigbrk{9,-15,51}^\pm \\
\hline
8  &[0;0;0,8]^{\dagger\bullet}&8&\bigbrk{0,0,0}^+ \\
8  &[0;1;0,5]^{\ast\bullet}&8&\bigbrk{4,-5,\frac{49}{4}}^-,
                              \bigbrk{8E-8,-13E+18,\frac{179}{4}E-61}^- \\
8  &[0;1;1,3]^{\ast\bullet}&8&\bigbrk{\scriptstyle 17E^2-90E+147,\,-\frac{51}{2}E^2+\frac{525}{2}E-\frac{1239}{2},\,\frac{169}{2}E^2-\frac{2091}{2}E+\frac{5649}{2}}^\pm \\
8  &[0;1;2,1]^{\ast\bullet}&8&\bigbrk{5,-\frac{15}{2},\frac{55}{2}}^\pm,
                              \bigbrk{12E-24,-18E+54,57E-171}^- \\
8  &[0;2;0,2]&8&\bigbrk{7,-\frac{19}{2},\frac{59}{2}}^\pm,
                \bigbrk{\scriptstyle 44E^5-768E^4+6752E^3-31168E^2+70528E-60224,\,
                        A,\,B}^+ \\
8  &[0;2;1,0]&8&\bigbrk{9,-\frac{31}{2},\frac{103}{2}}^\pm,
                \bigbrk{\scriptstyle 24E^2-172E+344,\,-39E^2+524E-1372,\,138E^2-2209E+6198}^- \\
8  &[2;3;0,2]^\ast&7&\bigbrk{\scriptstyle28E^2-252E+728,\,-51E^2+906E-3864,\,179E^2-3965E+20090}^- \\
8  &[2;3;1,0]^\ast&7&\bigbrk{8,-\frac{25}{2},\frac{687}{16}}^+,\bigbrk{12,-\frac{45}{2},\frac{1281}{16}}^+ \\
\hline
8.5&[1;2;0,4]^\ast&8&\bigbrk{6,-\frac{19}{2},\frac{247}{8}}^\pm,\bigbrk{8,-\frac{29}{2},\frac{427}{8}}^\pm \\
8.5&[1;2;1,2]^\ast&8&\bigbrk{\scriptstyle 31E^3-350E^2+1704E-3016,\,
                             -50E^3+1111E^2-7971E+18452,\,
                             C}^\pm \\
8.5&[1;2;2,0]^\ast&8&\bigbrk{8,-13,\frac{343}{8}}^\pm,
                     \bigbrk{15E-48,-23E+135,\frac{595}{8}E-\frac{4023}{8}}^\pm \\
8.5&[1;3;0,1]&8&\bigbrk{8,-13,\frac{173}{4}}^\pm,
                \bigbrk{10,-\frac{67}{4},\frac{3725}{64}}^\pm,
                \bigbrk{\scriptstyle 19E-86,\,-\frac{133}{4}E+\frac{1169}{4},\,\frac{7395}{64}E-\frac{79503}{64}}^\pm \\
\hline
\end{array}$
{\scriptsize\<
A\eq -73E^5+2486E^4-31804E^3+188280E^2-506048E+487104\nln
B\eq 251E^5-10452E^4+156202E^3-1041992E^2+3055168E-3125328\nln
C\eq \textstyle\frac{337}{2}E^3-\frac{18363}{4}E^2+38740E-102390\nonumber
\>\vspace{-1cm}}%
\caption{Spectrum of highest weight states with $D_0\leq 8.5$ in the 
dynamic $\alSU(2|3)$ spin chain. 
Please refer to the end of \protect\secref{sec:Higher.Spec.Low} for explanations.}
\label{tab:Higher.Spec.Low}
\end{table}

We summarise our findings for states of 
dimension $D_0\leq 8.5$ in \tabref{tab:Higher.Spec.Low}.
We have labelled the states by their dimension $D_0$, 
classical $\alSU(2|3)$ Dynkin labels, and classical length $L$.
For each multiplet we have given its energy
$E=E_0+g^2 E_2+g^4 E_4+\order{g^6}$
up to three-loops and parity $P$. 
A pair of degenerate states with opposite parity
is labelled by $P=\pm$. 
For convenience we have indicated the shortening conditions
relevant for the $\alSU(2|3)$ representations:
Half-BPS multiplets and multiplets at the unitarity bound (which split at $g=0$) 
are labelled by $^\dagger$ and $^\ast$, respectively. 
For $s=0$ some of the components are in the $\alSU(2)$ subsector,
such multiplets are indicated
by $^\bullet$.

Generically, the one-loop energies are not fractional numbers,
but solutions to some algebraic equations. We refrain from solving
them (numerically), but instead give the equations. 
In the table such states are indicated as polynomials 
$X_{0,2,4}(E)$ of degree $k-1$. 
The energies are obtained as solutions to the equation
\[\label{eq:Higher.Spec.Poly}
E^k=X_0(E)+g^2 X_2(E)+g^4\, X_4(E)+\order{g^6},
\]
see also \secref{sec:One.Spec.Low}.
The scaling dimension of the corresponding gauge theory states
are given by 
$D(g)=D_0+g^2 E(g)$.

For example, the three-loop planar scaling dimension of
the Konishi operator $\OpK=\eta^{mn}\Tr \Phi_m\Phi_n$
introduced in \secref{sec:Dila.Dim.TwoPoint} is given by 
(see \figref{tab:Higher.Spec.Low}, line 3 corresponds to a descendant of $\OpK$)
\[\label{eq:Higher.Spec.Konishi}
D=2+6g^2-12g^4+42g^6+\order{g^8}
=2+\frac{3\gym^2N}{4\pi^2}
-\frac{3\gym^2N}{16\pi^2}
+\frac{21\gym^2N}{256\pi^2}+\ldots\,.
\]
The two-loop result was computed in \cite{Bianchi:2000hn}
and the three-loop coefficient was first conjectured 
in \cite{Beisert:2003tq}.
It was later derived in \cite{Beisert:2003ys} using the
methods described in the current chapter.
This result was recently confirmed by independent arguments
based on extracting the $\superN=4$ SYM anomalous dimensions
of twist-two operators (c.f.~\secref{sec:One.Spec.TwistTwo}) 
from the exact QCD result.
The three-loop QCD result became available after an impressive,
full-fledged and rigorous field theoretic computation
by Moch, Vermaseren and Vogt \cite{Moch:2004pa,Vogt:2004mw}.
To generalise to $\superN=4$ SYM, it was observed that 
in maximally supersymmetric gauge theory only 
terms of `highest transcendentality' seem to arise. 
Here, terms similar to $\zeta(k)$ have transcendentality $k$
and an $\ell$-loop anomalous dimension in $\superN=4$ SYM should have
transcendentality $2\ell-1$.
Even more remarkably, for purely gluonic amplitudes, 
the contributions of highest transcendentality appear to 
independent of the matter content. 
If true, one can truncate to highest transcendentality \cite{Kotikov:2004er}
to obtain the anomalous dimensions of twist-two operators from QCD.
The conjecture of \cite{Kotikov:2004er} for the lowest 
twist-two operator, which is part of the Konishi multiplet, 
\emph{agrees} with the result \eqref{eq:Higher.Spec.Konishi} 
in a spectacular fashion.

\subsection{Two Excitations}
\label{sec:Higher.Spec.TwoEx}

We can use our above results to find the energy 
of two-excitation states up to three-loops.
In this subsector they are represented by the highest weight modules 
with Dynkin labels ${[0;1;0,J-2]}$ and length $L=J+1$.
All `flavours' of two-excitation states 
are part of the same multiplet, c.f.~\secref{sec:One.Spec.TwoEx}, 
and it is convenient to use a descendant in the 
$\alSU(2)$ subsector as in \secref{sec:One.BMN}
\[\label{eq:Higher.Spec.TwoBasis}
\OpE^J_p= \Tr \phi\,\fldZ^{p}\,\phi\, \fldZ^{J-p}.
\]
The action of the one-loop Hamiltonian was found in 
\eqref{eq:One.BMN.Ham} 
\[\label{eq:Higher.Spec.TwoOneLoop}
\ham_0\, \OpE^J_p =
-2\delta_{p\neq J}\,\OpE^J_{p+1}+2(\delta_{p\neq J}+\delta_{p\neq 0})\,\OpE^J_{p}
-2\delta_{p\neq 0}\,\OpE^J_{p-1}
\]
and the exact eigenstates are given in 
\eqref{eq:One.Spec.TwoMulti}
\[\label{eq:Higher.Spec.TwoOneEigen}
\Op^J_{0,n}=\frac{1}{J+1}\sum_{p=0}^J \cos\frac{\pi n(2p+1)}{J+1}\,\OpE^J_p
\]
with the exact one-loop energy \eqref{eq:One.Spec.TwoEnergy}
\[\label{eq:Higher.Spec.TwoOneEnergy}
E^J_{0,n}=8\sin^2 \frac{\pi n}{J+1}\,.\]
Let us state the inverse transformation of the 
discrete cosine transform \eqref{eq:Higher.Spec.TwoOneEigen}
\[\label{eq:Higher.Spec.TwoOneInverse}
\OpE_p^J=\Op_{0,0}^J+2\sum_{n=1}^{[J/2]}
\cos\left(\frac{\pi n(2p+1)}{J+1}\right)\, \Op_{0,n}^J.
\]
We act with the two-loop Hamiltonian and find
that it mostly equals the square of the one-loop Hamiltonian
\[\label{eq:Higher.Spec.TwoTwoLoop}
\ham_{2}\, \OpE_p^{J}=-\sfrac{1}{4} \ham_0^2\, \OpE_p^{J}+V_{2}\,\OpE_p^{J},
\]
up to a contact-interaction of the two excitations
\[\label{eq:Higher.Spec.TwoContact}
V_{2}\,\OpE_p^{J}=
(\delta_{p,0}+\delta_{p,J}-\delta_{p,1}-\delta_{p,J-1})
\lrbrk{\OpE^{J}_{1}-\OpE^{J}_{0}}.
\]
We now face the problem that the
states $\Op^J_{0,n}$ are no longer eigenstates of $\ham_{2}$, since
$\ham_{0}$ and $V_{2}$ do not commute. 
We find
\[\label{eq:Higher.Spec.TwoContactModes}
V_{2} \Op_{0,n}^J=
-\frac{64}{J+1}
\sin^2\frac{\pi n}{J+1} \cos\frac{\pi n}{J+1}
\sum_{n'=1}^{[J/2]}
\sin^2\frac{\pi n'}{J+1} \cos\frac{\pi n'}{J+1}\,
\Op_{0,n'}^J.
\]
However, we can treat $\ham_{2}$ as a perturbation and thus find
that the two-loop part of the planar anomalous dimension 
is the diagonal ($m=m'$) piece of $\ham_{2}$.
We obtain the following two-loop energy shift
\[\label{eq:Higher.Spec.TwoTwoEnergy}
E^J_{2,n}=
64\,\sin^4\frac{\pi n}{J+1}
\lrbrk{-\frac{1}{4}-\frac{\cos^2\frac{\pi n}{J+1}}{J+1}}.
\]
Furthermore, using standard perturbation theory, we can also find the 
perturbative correction to the 
eigenstates:
They involve the coupling constant dependent redefinition
\[\label{eq:Higher.Spec.TwoReder}
\Op^J_{n}=\Op^J_{0,n}+g^2 \Op^J_{2,n}+\ldots
\]
with 
\[\label{eq:Higher.Spec.TwoDeform}
\Op_{2,n}^J = - \frac{64}{J+1}\,
\sum_{\textstyle\atopfrac{n'=1}{n'\neq n}}^{[J/2]}
\frac{\sin^2\frac{\pi n}{J+1} \cos\frac{\pi n}{J+1}
\sin^2\frac{\pi n'}{J+1} \cos\frac{\pi n'}{J+1}}
{\sin^2\frac{\pi n}{J+1} - \sin^2\frac{\pi n'}{J+1}}~
\Op_{0,n'}^J.
\]
This mixing of modes is a complicating feature that 
we can expect at each further quantum loop order; remarkably,
it is absent in the large $J$ (BMN) limit.

We move on to three-loops and find that 
the result agrees with a general formula
\[\label{eq:Higher.Spec.TwoAllLoop}
D^J_n=D_0+g^2 E=J+2+\sum_{\ell=1}^\infty 
\lrbrk{8g^2\sin^2 \frac{\pi n}{J+1}}^{\ell}
\lrbrk{c_\ell+\sum_{k,l=1}^{\ell-1}c_{\ell,k,l}
         \frac{\cos^{2l} \frac{\pi n}{J+1}}{(J+1)^k}}
\]
with the coefficients up to three-loops given by
\cite{Beisert:2003tq}
\[\label{eq:Higher.Spec.TwoCoeffs}
c_1=1,\qquad
c_2=-\sfrac{1}{4},\quad
c_{2,1,1}=-1,\qquad
c_3=\sfrac{1}{8},\quad
c_{3,k,l}=\matr{ll}{+\sfrac{3}{4}&+\sfrac{1}{2}\\[2pt]-\sfrac{3}{4}&+\sfrac{5}{2}}.
\]
As expected, the formula reproduces the energies of 
all states ${[0;1;0,J-2]}$ in 
\tabref{tab:Higher.Spec.Low}.

The coefficients $c_\ell$ agree with the prediction of string theory 
on plane-waves \eqref{eq:One.BMN.Energy,eq:One.BMN.Corres}%
\[\label{eq:Higher.Spec.TwoLeading}
D^J_n=J+2\sqrt{1+\lambda' n^2}+\order{J^{-1}},\qquad
\lambda'=8\pi^2 g^2/J^2.
\]
This is a non-trivial result: Although we have only made use of the  
qualitative BMN limit, the quantitative BMN energy formula seems
to be the outcome.

We can also compare our result to string theory on 
a near plane-wave background \cite{Parnachev:2002kk,Callan:2003xr,Callan:2004uv}.
This corresponds to an expansion of the results in 
powers of $1/J$. Let us expand our result 
\eqref{eq:Higher.Spec.TwoAllLoop,eq:Higher.Spec.TwoCoeffs}
to first order 
\<\label{eq:Higher.Spec.TwoNear}
D^J_n\eq J+2
+\sum_{\ell=1}^\infty 
\lrbrk{\lambda'n^2}^{\ell}
\lrbrk{c_\ell+J^{-1}\lrbrk{-2\ell c_\ell+\tsum_{l=1}^{\ell-1} c_{\ell,1,l}}+\order{J^{-2}}}
\\\nonumber\eq
J+2
+\lrbrk{\lambda'n^2}
\lrbrk{1-2J^{-1}}
+\lrbrk{\lambda'n^2}^{2}
\lrbrk{-\sfrac{1}{4}+0J^{-1}}
+\lrbrk{\lambda'n^2}^{3}
\lrbrk{\sfrac{1}{8}+\half J^{-1}}
+\ldots\,.
\>
This is to be compared to the near plane-wave string theory result
\cite{Callan:2003xr,Callan:2004uv} 
(the comparison takes place at level 4 of the multiplet)
\<\label{eq:Higher.Spec.TwoString}
D^J_n\eq J+2\sqrt{1+\lambda' n^2}
-2\lambda' n^2\, J^{-1}+\order{J^{-2}}
\\\nonumber\eq
J+2
+\lrbrk{\lambda'n^2}
\lrbrk{1-2J^{-1}}
+\lrbrk{\lambda'n^2}^{2}
\lrbrk{-\sfrac{1}{4}+0J^{-1}}
+\lrbrk{\lambda'n^2}^{3}
\lrbrk{\sfrac{1}{8}+0J^{-1}}
+\ldots\,.
\>
Structurally, both expression are equivalent and 
all coefficients agree except single one 
at $\order{\lambda^{\prime\, 3}J^{-1}}$. 
The same kind of disagreement was also
observed for three excitations
\cite{Callan:2004ev,Callan:2004dt}
and arbitrarily many of scalar type
\cite{Arutyunov:2004vx,McLoughlin:2004dh}.
We will see further evidence of a disagreement between string theory
and gauge theory starting at three loops in 
\secref{sec:HighInt.Stringing};
we will discuss this issue there.

\subsection{An Eighth-BPS state}
\label{sec:Higher.Spec.EighthBPS}

Let us take a peek at non-planar physics within this sector:
The lowest-dimensional eighth-BPS state
is expected to be a triple-trace state
with weight $w=\weight{6;0,0;0,0,4;0,6}$.
Using the non-planar, one-loop Hamiltonian 
we find this protected state
\<\label{eq:Higher.Spec.EighthBPS}
\Op\indup{1/8-BPS}\eq
\varepsilon^{abc}\varepsilon^{def}\big[
N(N^2-3)\Tr \phi_a \phi_d \Tr \phi_b \phi_e \Tr \phi_c \phi_f
\nl\qquad\qquad
+6(N^2-1)\Tr \phi_a \phi_d \Tr \phi_b \phi_c \phi_e \phi_f
-12N\Tr \phi_a \phi_b \phi_c \phi_d \phi_e \phi_f
\nl\qquad\qquad
+8N\Tr \phi_a \phi_d \phi_b \phi_e \phi_c \phi_f
+4\Tr \phi_a \phi_b \phi_c \Tr \phi_d \phi_e \phi_f\big].
\>
It is annihilated by the operators
\[\label{sec:Higher.Spec.EighthBPSAnni}
\varepsilon_{abc}\comm{\check\phi^b}{\check\phi^c},
\qquad
\varepsilon_{abc}\varepsilon^{\alpha\beta}\Tr \psi_\alpha\comm{\check\phi^a}{\comm{\check\phi^b}{\comm{\check\phi^c}{\psi_\beta}}},
\]
which are part of the non-planar 
generalisations of 
$\algS_1,\ham_0$ and $\ham_1$. This 
implies that the state is protected (at least at one-loop).

It would be interesting to generalise some of the results
of this chapter to include non-planar corrections. 
At two-loops this might indeed be feasible as there are
only few non-planar graphs.

\finishchapter 

\chapter{Higher-Loop Integrability}
\label{ch:HighInt}

In this final chapter we would like to put together
the results of the previous two chapters.
In \chref{ch:Higher} we have seen how to make use of the
interacting algebra to find higher-loop corrections.
In \chref{ch:Int} we have investigated the 
integrability of planar $\superN=4$ SYM at the one-loop level
and demonstrated its usefulness.
An obvious question is whether the integrable structures
persist even at higher-loops. This will be the subject of 
the current chapter. 

To start off, we shall introduce higher-loop integrability
and argue that ${\superN=4}$ gauge theory (or, more precisely, 
the subsector discussed in \chref{ch:Higher})
is indeed integrable at higher-loops.
The main part of the chapter is devoted to the investigation of 
an integrable model in the $\alSU(2)$ subsector.
By making some assumptions on the form of interactions,
we will find that this model is \emph{uniquely} 
determined at five-loops and, excitingly, 
reproduces the plane wave energy formula.
What is more, we find the corresponding Bethe ansatz to compute 
the spectrum at an arbitrary order in perturbation theory!

\section{Higher-Loop Spin Chains}
\label{sec:HighInt.Higher}

First of all, we would like to describe the notion of integrability
for spin chains at higher-loops. We then go on by explaining
why we believe that these structures should apply to higher-loop 
$\superN=4$ SYM. Finally we will investigate
the scaling behaviour of charges in the thermodynamic limit and
describe how they can be defined canonically.

\subsection{Aspects of Higher-Loop Integrability}
\label{sec:HighInt.Higher.Aspects}

To describe higher-loop corrections to scaling dimensions
we have promoted the Hamiltonian $\ham=\ham_0$ of \chref{ch:One} to
a function of the coupling constant $\ham(g)$ in \chref{ch:Higher}.
At zero coupling one recovers the one-loop Hamiltonian $\ham(0)=\ham_0$. 
For higher-loop integrability we do the same and promote
the charges $\charge_r=\charge_{r,0}$ to functions 
$\charge_r(g)$ with $\charge_r(0)=\charge_{r,0}$. 
A Hamiltonian $\ham(g)=\charge_2(g)$ is considered to be higher-loop integrable 
if there exist conserved charges $\charge_r(g)$ with
\[\label{eq:HighInt.Higher.Def}
\mbox{`\emph{higher-loop integrability}':}\quad
\comm{\charge_r(g)}{\charge_s(g)}=\comm{\algJ(g)}{\charge_r(g)}=0.
\]

In the case of $\superN=4$ SYM, the
symmetry algebra is $\alPSU(2,2|4)\times\alU(1)$. 
The $\alU(1)$ factor corresponds to the anomalous piece of the 
dilatation operator $\algdD(g)$ which is conserved 
in two-point functions. 
We will argue that $\superN=4$ SYM
in the planar limit might be integrable 
for arbitrary values of the coupling constant $g$.
The higher charges $\charge_r(g)$ form an 
abelian algebra which enlarges the 
symmetry in the planar limit 
\[\label{eq:HighInt.Higher.IntAlg}
\alPSU(2,2|4)\times\alU(1)\longrightarrow
\alPSU(2,2|4)\times\alU(1)^\infty.
\]
Here, the $\alU(1)$ anomalous dimension is absorbed
into the set of integrable charges by the
identification $\algdD(g)=g^2 \ham(g)=g^2\charge_2(g)$.

An obstacle to the investigation of higher-loop integrability is that 
it apparently cannot be described 
with the formalism introduced in \secref{sec:Int.Chains}.
The reason is that higher-loop interactions are between several nearby fields, 
whereas an ordinary integrable spin chain
involves nearest-neighbour interactions only.%
\footnote{
Although non-nearest neighbour interactions of several spins are included
in the tower of higher charges $\charge_r$, these cannot be related to the
higher orders of the Hamiltonian because the charges
commute among themselves, whereas the $\ham_k$'s in general do not.}
What is more, higher-loop interactions can change the length 
of the spin chain giving rise to completely new structures.
In order to construct a higher-loop integrable spin chain, 
the R-matrix of \secref{sec:Int.Chains} appears to be 
not suitable. For instance, it describes the scattering of 
two elementary spins and it is not yet understood 
how to generalise it to the interactions that occur at higher-loops. 
Consequently, we cannot even attempt to prove 
the associated Yang-Baxter equation, which would
make higher-loop integrability manifest. 
Finally, there is not yet a constructive means 
to obtain higher charges.

In \secref{sec:Int.Chains.Charges} we have shown how to
extract the charges $\charge_{r,0}$ from the 
transfer matrix $\transfer(u,0)$.  
Here, we can take the opposite direction and 
package all charges into a transfer matrix 
\[\label{eq:HighInt.Higher.Transfer}
\transfer(u,g)
=\exp i\sum_{r=2}^\infty u^{r-1}\,\charge_{r}(g),
\]
which should satisfy the equivalent of 
\eqref{eq:Int.Chains.Commute}
\[\label{eq:HighInt.Higher.Commute}
\comm{\algJ(g)}{\transfer(u,g)}=\comm{\transfer(u,g)}{\transfer(v,g)}=0.
\]
An interpretation of the transfer matrix might be as follows:
An ordinary spin chain can be considered to be a
composite object assembled from individual fields/spins/particles.
The transfer matrix is derived from scattering processes
of the individual particles. 
At higher-loops it is not known how to make sense of
parton scattering processes. Nevertheless, it might still
be useful to consider the transfer matrix as 
a Wilson loop around the composite object. 
The identification of particles with fields/spins, however, 
would be lost for two reasons: Interactions take place between more than two
fields and the number of fields is not even preserved.
Instead, the Bethe ansatz (c.f.~\secref{sec:Int.Bethe}) offers an alternative 
notion of particles: 
The composite object has some vacuum state 
and its excitations correspond to particles. 
The interactions of such excitation particles 
are pairwise (right-hand side of the Bethe equations) and 
can therefore be integrable.
This picture may be generalised to higher-loops without complications.

\subsection{The Local Charges}
\label{sec:HighInt.Higher.Charges}

We do not know how to obtain the higher charges $\charge_r$
explicitly and are therefore forced to construct them by hand
such that they satisfy \eqref{eq:HighInt.Higher.Def}.
We cannot expect this to be feasible for finite values of $g$
and restrict ourselves to a perturbative treatment. 
In fact, we know (parts of) $\ham(g)$ only up to order $g^4$
(three-loops)
and we may construct the charges $\charge_r(g)$ only up to
the same order. 
The algebra need not be satisfied exactly, 
but only up to terms of higher order in perturbation theory
\[\label{eq:HighInt.Higher.PertAlg}
\comm{\charge_{r}(g)}{\charge_{s}(g)}=\comm{\algJ(g)}{\charge_{r}(g)}
=\order{g^{2\ell}}.
\]
For all-loop integrability the remaining higher-order terms 
would have to be cancelled 
by higher-order corrections to the charges.
We now expand the charges in powers of the coupling 
constant $g$ 
\[\label{eq:HighInt.Higher.PertCharge}
\charge_r(g)=\sum_{k=0}^\infty g^k\,\charge_{r,k}.\]
The integrability condition \eqref{eq:HighInt.Higher.Def}
translates to the statement of \emph{perturbative integrability}
\[\label{eq:HighInt.Higher.PertInt}
\mbox{`\emph{perturbative integrability}':}\quad
\sum_{k=0}^l\comm{\charge_{r,k}}{\charge_{s,l-k}}
=
\sum_{k=0}^l\comm{\algJ_{k}}{\charge_{r,l-k}}
=0.
\]

One important consideration for the construction of charges 
is their representation as an interaction acting 
on the spin chain, see \secref{sec:Dila.Planar.Interact}.
In \secref{sec:Int.Chains.Charges} we have learned that 
the local charges $\charge_{r,0}$ act on $r$ adjacent spins.
In other words, the charge $\charge_{r,0}$ has $2r$ legs, 
$r$ incoming and $r$ outgoing ones.
Although the higher-loop form of interactions 
described in \secref{sec:Dila.Planar.Interact}
applies to quantities that appear in correlators,
it seems natural to generalise it to the charges. 
An order $g^k$ correction to a generator $\algJ$ involves
$k+2$ legs and we conclude that for each power of $g$ 
we should have one leg.
In total, a charge $\charge_{r,k}$ should have $2r+k$ legs
\[\label{eq:HighInt.Higher.Interact}
\charge_{r,k}\sim
\ITerm{\fldindn{A}_{1}\ldots \fldindn{A}_{E\indups{i}}}
{\fldindn{B}_{1}\ldots \fldindn{B}_{E\indups{o}}},
\quad \mbox{with}\quad 
E\indup{i}+E\indup{o}=2r+k.
\]
It is also natural to assume that the charges
have a definite parity, the same as at leading order
\[\label{eq:HighInt.Higher.Parity}
\gaugepar \,\charge_{r}(g) \,\gaugepar^{-1}=(-1)^r \charge_{r}(g).
\]
Finally, the charges preserve the classical dimension
\[\label{eq:HighInt.Higher.Dim}
\comm{\algD_0}{\charge_r(g)}=0
\]
when we identify the second charge with 
the anomalous dimension $\delta D(g)=g^2 \charge_2(g)$,
because $\algD(g)$ and $\charge_2(g)$ commute with all charges
and so does their difference.

Let us now comment on the role of the Hamiltonian $\ham(g)$. 
On the one hand, it belongs to the symmetry algebra $\alPSU(2,2|4)$
when combined with $\algD_0$ 
\[\label{eq:HighInt.Higher.HinPSU224}
\algD_0+g^2 \ham(g)=\algD_0+\algdD(g)=\algD(g)\in \alPSU(2,2|4).
\]
On the other hand, $\ham(g)$ is also one of the integrable charges. 
As such it is a generator of the abelian algebra $\alU(1)^\infty$
defined by \eqref{eq:HighInt.Higher.Def}
\[\label{eq:HighInt.Higher.HinAlg}
\ham(g)=\charge_2(g)\in \alU(1)^\infty.
\]
This is somewhat different from the situation at one-loop
where the symmetry algebra is taken strictly at $g=0$.
The one-loop anomalous dilatation generator is an independent object
and belongs only to the abelian algebra $\alU(1)^\infty$.

\subsection{Parity Pairs}
\label{sec:HighInt.Higher.Pairs}

As we have seen in \secref{sec:Int.Chains.Pairs},
the integrable structure gives rise to
pairs of states with degenerate energies and 
opposite parities.
We have proved integrability only at one-loop,
but a closer look at \tabref{tab:Higher.Spec.Low} 
reveals that the degeneracy of all one-loop pairs ($^\pm$) is
preserved even at three-loops!%
\footnote{Let us emphasise that, even if the table was computed assuming BMN scaling
to fix the values of $\alpha_1,\alpha_3,\sigma_1,\sigma_2$,
the pairing holds for generic values.}
This is so for the pairs of 
the $\alSU(2)$ sector ($^\bullet$) \cite{Beisert:2003tq},
for pairs at the unitarity bound ($^\ast$), but
also, and most importantly, for pairs
away from the unitarity bound (unmarked). 
As discussed at the end of \secref{sec:Higher.SU23.Dynamics}
all states of such a multiplet are superpositions of 
states of different lengths.
This is interesting because it shows
that also for truly \emph{dynamic spin chains}
with a fluctuating number of sites,
integrability is an option.

We do not know how to use the R-matrix formalism 
beyond one-loop, if this is possible at all. 
Therefore we cannot rigorously prove higher-loop integrability
by means of a Yang-Baxter equation. 
One might construct several of the higher charges
explicitly at higher-loops and thus make integrability 
very plausible, but this would not constitute a proof. 
Instead, we shall be satisfied by demonstrating the preservation of 
degenerate pairs at three-loops. 
This is certainly a necessary condition for
integrability, but at first sight it appears not to 
imply the existence of commuting charges.
Nevertheless, there are some indications that 
pairing is indeed sufficient to guarantee integrability.
First of all, a systematic pairing is most naturally 
explained by the following set of identities
\[\label{eq:HighInt.Higher.Pairs}
\comm{\ham(g)}{\charge(g)}=\comm{\gaugepar}{\ham(g)}=\acomm{\gaugepar}{\charge(g)}=0
\]
among the Hamiltonian $\ham(g)$, parity $\gaugepar$ and some
charge $\charge(g)$. 
The investigations for the model in  
\secref{sec:HighInt.SU2} have shown that indeed we can
construct a charge $\charge(g)=\charge_3(g)$ whenever 
the spectrum has sufficiently many pairs without imposing
further constraints on the Hamiltonian $\ham(g)$.
More remarkably, it always turned out to be possible
to construct conserved charges $\charge_r(g)$ as
soon as the Hamiltonian pairs up states. 
The charges do not only commute with the Hamiltonian,
but also among themselves, i.e.~they automatically
satisfy \eqref{eq:HighInt.Higher.Def}.
This parallels earlier observations \cite{Grabowski:1995rb} that it 
appears close to impossible to construct systems with 
$\comm{\charge_2}{\charge_3}=0$ which 
do not have arbitrarily many other commuting charges $\charge_r$,
i.e. which are not integrable.

It is therefore likely that 
planar $\superN=4$ SYM (at least) in the $\alSU(2|3)$ subsector
and (at least) at three-loops is integrable.
In agreement with the findings of 
\cite{Klose:2003qc} we conclude that 
integrability appears to be a consequence
of field theory combined with symmetry and does not
depend on the specific model very much.
It strongly supports the idea
of all-loop integrability in planar 
$\superN=4$ SYM. What is more, the dynamic aspects
of higher-loop spin chains appear to 
be no obstruction.
Let us emphasise, though, that 
a rigorous proof of higher-loop integrability 
remains a challenge.

\subsection{The $\alSU(2|3)$ Sector Revisited}
\label{sec:HighInt.Higher.Revisit}

At this point we can reinvestigate the undetermined coefficients
of the $\alSU(2|3)$ spin chain at three-loops,
c.f.~\secref{sec:Higher.Three}. 
By imposing
\[\label{eq:HighInt.Higher.ThreeLoop}
\comm{\algQ(g)}{\ham(g)}=\comm{\algS(g)}{\ham(g)}=\order{g^6}
\]
we found that $\ham_4$ depends on four relevant coefficients 
$\sigma_{1,2,3,4}$. The coefficients $\sigma_{3,4}$ multiply 
invalid structures, 
whereas $\sigma_1$ corresponds to a redefinition of the coupling constant. 
The remaining coefficient $\sigma_2$ multiplies 
$\charge_{4,0}$ which is structurally equivalent to $\ham_4$
and satisfies $\comm{\algJ(g)}{\charge_4(g)}=0$ as well.
In fact, by imposing $\comm{\algQ(g)}{X(g)}=\comm{\algS(g)}{X(g)}=0$ we 
do not only find $\ham(g)$, but 
also all the other even generators $\charge_{r}(g)$ of the abelian algebra
of integrable charges $\alU(1)^\infty$.
Thus $\sigma_2$ corresponds to the transformation
\[\label{eq:HighInt.Higher.Q4}
\ham(g)\mapsto \ham(g)+\sigma_2 \,g^4\, \charge_{4}(g),
\]
which has no influence on \eqref{eq:HighInt.Higher.ThreeLoop} due to 
$\comm{\algJ(g)}{\charge_r(g)}=0$. 
This degree of freedom may be fixed by considering the 
anticommutator of supercharges \eqref{eq:Higher.SU23.AlgMomRot} 
at order $g^6$
\[\label{eq:HighInt.Higher.MomRot7}
\sum_{k=0}^6
\bigacomm{(\algS_k)^\alpha{}_a}{(\algQ_{6-k})^b{}_\beta}=
  \sfrac{1}{2} \delta_a^b \delta_\beta^\alpha \ham_4.
\]
Unfortunately, this equation involves $\algQ_6,\algS_6$ which are part of a four-loop 
calculation and 
out of reach here. We believe that \eqref{eq:HighInt.Higher.MomRot7} will force the
coefficient $\sigma_2$ to vanish, and in conclusion
all corrections up to three loops would be determined
uniquely (up to a redefinition of the coupling constant).
At higher loops this picture is expected to continue:
While $\comm{\algQ(g)}{X(g)}=\comm{\algS(g)}{X(g)}=0$ determines
the even elements $\charge_{r}$ of $\alU(1)^\infty$, 
the anticommutator $\acomm{\algS(g)}{\algQ(g)}$ yields 
the one element $\charge_2(g)=\ham(g)$ 
which is also associated to $\alSU(2|3)$ 
as the generator $\algD_0+\sfrac{3}{2}g^2 \ham(g)$.
It is reasonable to assume that $\sigma_2$ 
may alternatively be fixed by the non-planar algebra
where conservation of the charge $\charge_4$ is lost.

Let us comment on the 
effect of integrability on the
degrees of freedom for similarity transformations. 
Similarity transformations are symmetries of the
algebra relations and thus give rise to 
undetermined coefficients in the construction 
of the most general deformation of generators. 
In \secref{sec:Higher.Three} 
we argued that the coefficients which arise
for the generators $\algJ_k$ are in one-to-one correspondence
with the structures that can be used for the construction
of $\ham_{k-2}$, see also \tabref{tab:Higher.Three.Count}.
However there are also some similarity transformations
which do not change the generators. These are generated by
invariant operators such as $\ham(g)$ and, 
to some extent, the length $\len$. In an integrable system there are
more invariant operators: The charges $\charge_r(g)$. 
Only the even charges $\charge_{r}$ are compatible
with the structure of $\ham_{k-2}$. 
For example, the fourth charge will appear as a symmetry in 
\tabref{tab:Higher.Three.Count} starting
at sixth order (four-loops).

\subsection{The Thermodynamic Limit}
\label{sec:HighInt.Higher.Scaling}

The thermodynamic limit is the limit 
in which the length of the spin chain $L$ 
as well as the number of excitations is taken 
to infinity while focusing on the the low-energy spectrum,
c.f.~\secref{sec:Int.Thermo}.
In this limit it was observed 
that the $r$-th charge $Q_{r,0}$ at one-loop
scales as $L^{1-r}$ \cite{Arutyunov:2003rg}.
Here, we would like to generalise the thermodynamic limit to higher-loops.
From the investigation of the closely related BMN limit 
(c.f.~\secref{sec:One.BMN}) as well as 
from classical spinning strings (c.f.~\secref{sec:Int.Spinning}), 
we infer that each power of the
coupling constant $g$ should be accompanied by one power of $1/L$.
It is common belief that this scaling behaviour holds for 
perturbative gauge theory, but it is clearly not 
a firm fact. 
We shall assume its validity for several reasons:
Firstly, it was not only confirmed at one-loop, but also
at two-loops \cite{Gross:2002su,Beisert:2003tq}. 
It is a nice structure and conceptually it would be somewhat 
disappointing if broken at some higher loop order.
Secondly, the AdS/CFT correspondence seems to suggest it.
Thirdly, it will allow us to define charges uniquely, 
see below, and
arrive at a \emph{unique} result in \secref{sec:HighInt.SU2}. 

In conclusion, the scaling of charges in the thermodynamic limit is given by%
\footnote{In the BMN we would account for the
finite number of excitations by the slightly modified definitions
$\hat \charge_{r,k}=L^{k+r}\charge_{r,k}$ and
$\hat \transfer(\tilde u,\tilde g)=\transfer(L\tilde u,L\tilde g)^L$}
\[\label{eq:HighInt.Higher.Thermo}
\tilde \charge_{r,k}=\lim_{L\to\infty} L^{k+r-1}\charge_{r,k},\qquad
\tilde \transfer(\tilde u,\tilde g)=\lim_{L\to\infty}\transfer(L\tilde u,L\tilde g).
\]
%

\subsection{Canonical Charges}
\label{sec:HighInt.Higher.Canonical}

As the charges form an abelian algebra, one can replace 
$\charge_r(g)$ by some polynomial $\charge'_r(g)$ in the charges 
without changing the algebra. 
We will now prove the uniqueness of the `canonical' set of charges with good
structural (c.f.~\secref{sec:HighInt.Higher.Charges}) 
and scaling properties 
(c.f.~\secref{sec:HighInt.Higher.Scaling}). 
We will start by assuming that the charges $\charge_r(g)$ 
are canonical and show that we cannot change them
without spoiling the one of the properties.

The charges can be written as a local interaction
$\charge_{r,k}=\sum_{p=1}^L \charge_{r,k,p\ldots}$ with
the local charge density $\charge_{r,k,p\ldots}$.
A generic polynomial transformation would therefore make $\charge'_r$ 
multi-local in general.
To preserve locality we are restricted to linear transformations 
which are generated by 
\[\label{eq:HighInt.Higher.Redef}
\charge'_r(g)=\charge_r(g)+\alpha_{r,s,l}\, g^{2l}\, \charge_{r+2s}(g).
\]

We now find two constraints on $l$ and $s$: 
On the one hand, there is the structural constraint
\eqref{eq:HighInt.Higher.Interact} which tells that $\charge'_{r,k}$
can only have $2r+k$ legs.
This requires 
\[\label{eq:HighInt.Higher.Bound1}
l\geq 2s.
\] 
On the other hand, a correct scaling 
in the thermodynamic limit \eqref{eq:HighInt.Higher.Thermo}
requires that $\charge'_{r,k}$ scales as $\order{L^{1-k-r}}$.
In order not to spoil scaling, we need 
\[\label{eq:HighInt.Higher.Bound2}
l\leq s.\]
Together, these two
constraints imply $l\leq 0$, but we do not
allow negative powers of $g$. In total we get $l=s=0$ or, 
in other words, $\charge_r$ can only be rescaled by a constant $\alpha_{r,0,0}$.
Finally, this constant can be fixed by using the canonical transfer 
matrix $\transfer(u,0)$ of the one-loop spin-chain, 
c.f.~\secref{sec:Int.Chains.Charges}.

In conclusion, the canonical definition for $\charge_r(g)$ is unique
(if it exists).
As $\ham(g)$ is subject to the same constraints as $\charge_2(g)$,
both of them must be equal $\ham(g)=\charge_2(g)$.

\section{The $\alSU(2)$ Sector at Higher-Loops}
\label{sec:HighInt.SU2}

In this section we will construct a model for
higher-loop anomalous dimensions in the 
$\alSU(2)$ subsector of $\superN=4$ SYM.
We will rely on three assumptions on the form of interactions:
($i$) Integrability, ($ii$) the thermodynamic limit
and ($iii$) some constraints inspired by Feynman diagrammatics.
Note that none of these assumptions should be taken as a firm fact.
Whether or not they are fully justified in
(perturbative) $\superN=4$ SYM 
will not be the subject of this chapter, 
but we believe that the model 
shares several features with higher-loop gauge theory 
and therefore deserves an investigation.
Intriguingly, it will turn out to be \emph{unique}
up to (at least) five-loops and 
agree with the excitation energy formula in the BMN limit!
At any rate, this makes it a very interesting model 
to consider in its own right.
With some luck, the assumptions ($i$-$iii$) will turn out
to be valid for $\superN=4$ SYM and we have constructed the 
planar, five-loop%
\footnote{We disregard wrapping interactions, see
\secref{sec:Dila.Planar.Wrapping}, 
i.e.~this applies only to states of
length $L\geq 6$.} 
dilatation generator in the $\alSU(2)$ subsector.

\subsection{Interactions}
\label{sec:HighInt.SU2.Interact}

In the $\alSU(2)$ sector, the number of field sites
is conserved. 
In particular, this implies that odd powers of $g$ 
are not allowed.
Furthermore, rotations
are manifestly realised, the $\alSU(2)$ generators
do not receive radiative corrections.
Therefore the interactions can only be
of the form (see~\secref{sec:Dila.Planar.Interact})
\[\label{eq:HighInt.SU2.Interact}
\ITerm{\textstyle a_{1}\ldots a_{E}}{\textstyle a_{\pi(1)}\ldots a_{\pi(E)}}
\]
with $\pi$ some permutation of $E$ elements.%
\footnote{Note that we will consider the states to be sufficiently long and 
drop wrapping interactions (c.f.~\secref{sec:Dila.Planar.Wrapping}).
We will comment on the role 
of wrappings in \secref{sec:HighInt.Spinning.Wrappings}.}
Any permutation can be represented in terms
of elementary permutations $\fldperm_{p,p+1}$ of 
adjacent fields. A generic term will be written as
\[\label{eq:HighInt.SU2.Perm}
\PTerm{p_1,p_2,\ldots}=
\sum_{p=1}^L 
\fldperm_{p+p_1,p+p_1+1}
\fldperm_{p+p_2,p+p_2+1}\ldots\,.
\]
For example, in this notation the one-loop dilatation generator
\eqref{eq:One.Spec.SU2Ham} is given by
\[\label{eq:HighInt.SU2.PermH0}
\ham_0=\charge_{2,0}=\bigbrk{\PTerm{}-\PTerm{1}}.
\]
This notation is useful due to the nature of maximal scalar diagrams
as discussed at the end of \secref{sec:Dila.Theory.Feyn}:
An interaction of scalars at $\ell$ loops 
with the maximal number of $2+2\ell$ legs can be composed from 
$\ell$ crossings of scalar lines. 
In the planar limit, the crossings correspond to elementary
permutations and at $\ell$-loops there should be no more
than $\ell$ permutations. 
In field theory this is a feature of maximal diagrams, 
but here we will assume the pattern to hold in general.
Furthermore, in \secref{sec:Int.Chains.Charges} we have learned that 
the $r$-th charge at leading (one-loop) order can be constructed
from $r-1$ copies of the Hamiltonian density 
which, in this case \eqref{eq:HighInt.SU2.PermH0}, 
is essentially an elementary permutation.
We will therefore assume the contributions to the charges 
to be of the form 
\[\label{eq:HighInt.SU2.ChargePerm}
\charge_{r,2\ell-2}\sim\PTerm{p_1,\ldots,p_m}\quad
\mbox{with }m\leq r+\ell-2\mbox{ and }1\leq p_i\leq r+\ell-2.
\]

Finally, the even (odd) charges should be 
parity even (odd) and (anti)symmetric.%
\footnote{In fact, the Hamiltonian $\ham(g)$ and charges
$\charge_r(g)$ should be hermitian. The coefficients
of the interaction structures should therefore be real (imaginary)
for even (odd) $r$. Reality of the Hamiltonian follows from the 
equivalence of the Hamiltonian for the $\alSU(2)$ sector and its
conjugate.}
Parity acts on the interactions as
\[\label{eq:HighInt.SU2.PermParity}
\gaugepar\,\PTerm{p_1,\ldots,p_m}\,\gaugepar^{-1}
=\PTerm{-p_1,\ldots,-p_m},
\]
whereas symmetry acts as 
\[\label{eq:HighInt.SU2.PermSymmetry}
\PTerm{p_1,\ldots,p_m}^{\trans}=\PTerm{p_m,\ldots,p_1}.
\]
Symmetry will ensure that the eigenvalues of the 
charges are real.

Note that the interaction symbols are subject to several identities 
which can be used to bring them into some normal form. 
One identity involves a repeated elementary 
permutation
\[\label{eq:HighInt.SU2.PermRep}
\PTerm{\ldots,p,p,\ldots}=\PTerm{\ldots,\ldots}.
\]
Another obvious identity
\[\label{eq:HighInt.SU2.PermPerm}
\PTerm{\ldots,p,p',\ldots}=\PTerm{\ldots,p',p\ldots}\quad
\mbox{if}\quad |p-p'|\geq 2
\]
allows to commute two non-overlapping elementary permutations.
A third identity is due to gauge invariance or 
cyclic invariance of interactions
\[\label{eq:HighInt.SU2.PermShift}
\PTerm{p_1+p',\ldots,p_m+p'}=\PTerm{p_1,\ldots,p_m}.
\]
Finally, the spin at each site can take two different values and 
we cannot antisymmetrise more than two spins.
This leads to the the identity 
\<\label{eq:HighInt.SU2.PermSU2}
\PTerm{\ldots,p,p\pm 1,p,\ldots}\eq\PTerm{\ldots,\ldots}
-\PTerm{\ldots,p,\ldots}-\PTerm{\ldots,p\pm 1,\ldots}
\nl
+\PTerm{\ldots,p,p\pm 1,\ldots}+\PTerm{\ldots,p\pm 1,p,\ldots}.
\>
%

\subsection{The Higher Charges}
\label{sec:HighInt.SU2.Charges}

We would now like to construct some of the higher charges 
for the model in \chref{ch:Higher} explicitly. 
Let us start by writing down the Hamiltonian in
\tabref{tab:Higher.Two.Vertex,tab:Higher.Three.H6SU3},
restricted to the $\alSU(2)$ subsector in the notation introduced above
\<\label{eq:HighInt.SU2.SU23}
\ham_0\eq
 \PTerm{}
-\PTerm{1},
\nln
\ham_2\eq
(-2+2\alpha_3)\PTerm{}
+(3-2\alpha_3)\PTerm{1}
-\half \bigbrk{\PTerm{1,2}+\PTerm{2,1}},
\nln
\ham_4\eq
\bigbrk{\sfrac{15}{2}-8\alpha_3+\sigma_1-\sfrac{2}{3}\sigma_2}\PTerm{}
+\bigbrk{-13+12\alpha_3-\sigma_1+\sfrac{4}{3}\sigma_2}\PTerm{1}
+\sfrac{1}{2}\PTerm{1,3}
\nl
+\bigbrk{3-2\alpha_3-\sfrac{1}{3}\sigma_2}\bigbrk{\PTerm{1,2}+\PTerm{2,1}}
+\bigbrk{-\sfrac{1}{2}+\sfrac{1}{3}\sigma_2}\bigbrk{\PTerm{1,2,3}+\PTerm{3,2,1}}
\nl
+\bigbrk{-\sfrac{1}{3}\sigma_2-i\xi_3}\PTerm{2,1,3}
+\bigbrk{-\sfrac{1}{3}\sigma_2+i\xi_3}\PTerm{1,3,2}.
\>
Here, we should set $\sigma_1=\sigma_2=\alpha_3=0$ to
obtain the correct scaling behaviour in the 
thermodynamic limit, see \secref{sec:Higher.Spec.Coeff}.
Furthermore, the coefficient $\xi_3$ is related to a similarity
transformation; it consequently does not affect scaling dimensions
and we set it to zero.%
\footnote{Furthermore, the coefficients of interaction structures
can be assumed to be real for the Hamiltonian.}
According to \secref{sec:Int.Chains.Charges}, 
the leading order third charge 
is given by 
\[\label{eq:HighInt.SU2.Q30}
\charge_{3,0}=\sfrac{i}{2}\bigbrk{\PTerm{1,2}-\PTerm{2,1}}.
\]
As discussed in \secref{sec:HighInt.Higher.Revisit}, the leading fourth 
charge can be read off from $\ham_{4}$ 
as the structure multiplied by $\sigma_2$
\<\label{eq:HighInt.SU2.Q40}
\charge_{4,0}\eq 
-\sfrac{2}{3}\PTerm{}
+\sfrac{4}{3}\PTerm{1}
-\sfrac{1}{3}\lrbrk{\PTerm{1,2}+\PTerm{2,1}}
\nl
-\sfrac{1}{3}\bigbrk{\PTerm{1,3,2}+\PTerm{2,1,3}}
+\sfrac{1}{3}\bigbrk{\PTerm{1,2,3}+\PTerm{3,2,1}}.
\>
Both of them commute with $\ham_{0}$ and among themselves.

Let us now go ahead and compute the first correction
to a higher charge. 
For $\charge_{3,2}$ the only suitable
structures are
$\brk{\PTerm{1,2}-\PTerm{2,1}}$ and 
$\brk{\PTerm{1,2,3}-\PTerm{3,2,1}}$.
We demand that $\charge_{3}(g)$ commutes with 
$\ham(g)$ in perturbation theory
\[\label{eq:HighInt.SU2.HQ32}
\comm{\ham_{0}}{\charge_{3,2}}
+\comm{\ham_{2}}{\charge_{3,0}}=0
\]
and find the coefficient of
$\brk{\PTerm{1,2,3}-\PTerm{3,2,1}}$
to be fixed to $i/2$. The coefficient of 
$\brk{\PTerm{1,2}-\PTerm{2,1}}$ cannot be determined
because this structure is proportional to
$\charge_{3,0}$ and commutes with 
$\ham_{0}$ by construction.
We can only fix it by demanding 
a correct scaling behaviour in the thermodynamic limit,
c.f.~\secref{sec:HighInt.Higher.Scaling,sec:HighInt.Higher.Canonical}
and obtain
\[\label{eq:HighInt.SU2.Q32}
\charge_{3,2}=-2i\bigbrk{\PTerm{1,2}-\PTerm{2,1}}
+\sfrac{i}{2}\bigbrk{\PTerm{1,2,3}-\PTerm{3,2,1}}.
\]
We proceed in the same way to determine the 
integrable charges $\charge_{3},\charge_{4}$ up 
to $\order{g^4}$ (three-loops). The unique solution
with correct scaling in the thermodynamic limit is
presented in \tabref{tab:HighInt.SU2.Charges}.

\begin{table}\centering
\<
\ham_{0}\eq 
+\PTerm{}-\PTerm{1},
\nln
\ham_{2}\eq 
-2\PTerm{}+3\PTerm{1}-\sfrac{1}{2}\bigbrk{\PTerm{1,2}+\PTerm{2,1}},
\nln
\ham_{4}\eq 
+\sfrac{15}{2}\PTerm{}
-13\PTerm{1}
+\sfrac{1}{2}\PTerm{1,3}
+3\lrbrk{\PTerm{1,2}+\PTerm{2,1}}
-\sfrac{1}{2}\bigbrk{\PTerm{1,2,3}+\PTerm{3,2,1}},
\nonumber
\\[12pt]
\charge_{3,0}\eq 
+\sfrac{i}{2}\bigbrk{\PTerm{1,2}-\PTerm{2,1}},
\nln
\charge_{3,2}\eq 
-2i\bigbrk{\PTerm{1,2}-\PTerm{2,1}}
+\sfrac{i}{2}\bigbrk{\PTerm{1,2,3}-\PTerm{3,2,1}},
\nln
\charge_{3,4}\eq
+\sfrac{73i}{8}\bigbrk{\PTerm{1,2}-\PTerm{2,1}}
-\sfrac{i}{4}\bigbrk{\PTerm{1,2,4}+\PTerm{1,3,4}-\PTerm{1,4,3}-\PTerm{2,1,4}}
\nl
-\sfrac{7i}{2}\bigbrk{\PTerm{1,2,3}-\PTerm{3,2,1}} 
-\sfrac{i}{8}\bigbrk{\PTerm{1,3,2,4}-\PTerm{2,1,4,3}}
\nl
-\sfrac{i}{8}\bigbrk{\PTerm{1,2,4,3}-\PTerm{1,4,3,2}+\PTerm{2,1,3,4}-\PTerm{3,2,1,4}}
\nl
+\sfrac{5i}{8}\bigbrk{\PTerm{1,2,3,4}-\PTerm{4,3,2,1}},
\nonumber
\\[12pt]
\charge_{4,0}\eq 
-\sfrac{2}{3}\PTerm{}
+\sfrac{4}{3}\PTerm{1}
-\sfrac{1}{3}\lrbrk{\PTerm{1,2}+\PTerm{2,1}}
\nl
-\sfrac{1}{3}\bigbrk{\PTerm{1,3,2}+\PTerm{2,1,3}}
+\sfrac{1}{3}\bigbrk{\PTerm{1,2,3}+\PTerm{3,2,1}}
,
\nln
\charge_{4,2}\eq
+5  \PTerm{}
-\sfrac{31}{3}\PTerm{1}
+\sfrac{2}{3} \PTerm{1,3}
+\sfrac{17}{6} \bigbrk{\PTerm{1,2}+\PTerm{2,1}} 
+\sfrac{11}{6}\bigbrk{\PTerm{1,3,2}+\PTerm{2,1,3}} 
\nl
-\sfrac{13}{6} \bigbrk{\PTerm{1,2,3}+\PTerm{3,2,1}}
-\sfrac{1}{3}\PTerm{2,1,3,2}
-\sfrac{1}{6}\bigbrk{\PTerm{1,3,2,4}+\PTerm{2,1,4,3}} 
\nl
-\sfrac{1}{6}\bigbrk{\PTerm{1,2,4,3}+\PTerm{1,4,3,2}+\PTerm{2,1,3,4}+\PTerm{3,2,1,4}}  
\nl
+\sfrac{1}{2}\bigbrk{\PTerm{1,2,3,4}+\PTerm{4,3,2,1}},
\nln
\charge_{4,4}\eq
-\sfrac{63}{2}\PTerm{}
+\sfrac{401}{6}\PTerm{1}
-\sfrac{20}{3}\PTerm{1,3}
-\sfrac{5}{6}\PTerm{1,4}
-\sfrac{77}{4}\bigbrk{\PTerm{1,2}+\PTerm{2,1}}
\nl
-\sfrac{61}{6}\bigbrk{\PTerm{1,3,2}+\PTerm{2,1,3}}
+\sfrac{1}{2}\bigbrk{\PTerm{1,2,4}+\PTerm{1,3,4}+\PTerm{1,4,3}+\PTerm{2,1,4}}
\nl
+\sfrac{83}{6}\bigbrk{\PTerm{1,2,3}+\PTerm{3,2,1}}
+\sfrac{8}{3}\PTerm{2,1,3,2}
-\sfrac{1}{6}\bigbrk{\PTerm{1,2,4,5}+\PTerm{2,1,5,4}}
\nl
+\sfrac{1}{6}\bigbrk{\PTerm{1,3,2,5}+\PTerm{1,3,5,4}+\PTerm{1,4,3,5}+\PTerm{2,1,3,5}}
\nl
+\sfrac{19}{12}\bigbrk{\PTerm{1,3,2,4}+\PTerm{2,1,4,3}}
+\sfrac{1}{6}\bigbrk{\PTerm{1,2,5,4}+\PTerm{2,1,4,5}}
\nl
+\sfrac{17}{12}\bigbrk{\PTerm{1,2,4,3}+\PTerm{1,4,3,2}+\PTerm{2,1,3,4}+\PTerm{3,2,1,4}}
\nl
-\sfrac{1}{6}\bigbrk{\PTerm{1,2,3,5}+\PTerm{1,3,4,5}+\PTerm{1,5,4,3}+\PTerm{3,2,1,5}}
\nl
-\sfrac{19}{4}\bigbrk{\PTerm{1,2,3,4}+\PTerm{4,3,2,1}}
+\sfrac{1}{12}\bigbrk{\PTerm{1,4,3,2,5}+\PTerm{2,1,3,5,4}}
\nl
+\sfrac{1}{12}\bigbrk{\PTerm{1,3,2,5,4}+\PTerm{2,1,4,3,5}}
+\sfrac{1}{12}\bigbrk{\PTerm{1,2,5,4,3}+\PTerm{3,2,1,4,5}}
\nl
-\sfrac{1}{6}\bigbrk{\PTerm{1,3,2,4,3}+\PTerm{2,1,3,2,4}+\PTerm{2,1,4,3,2}+\PTerm{3,2,1,4,3}}
\nl
-\sfrac{1}{4}\bigbrk{\PTerm{1,2,4,3,5}+\PTerm{1,3,2,4,5}+\PTerm{2,1,5,4,3}+\PTerm{3,2,1,5,4}}
\nl
-\sfrac{1}{4}\bigbrk{\PTerm{1,2,3,5,4}+\PTerm{1,5,4,3,2}+\PTerm{2,1,3,4,5}+\PTerm{4,3,2,1,5}}
\nl
+\sfrac{3}{4}\bigbrk{\PTerm{1,2,3,4,5}+\PTerm{5,4,3,2,1}}.
\nonumber
\>
\caption{The Hamiltonian $\ham=\charge_2$ and
the charges $\charge_3,\charge_4$ at three-loops.}
\label{tab:HighInt.SU2.Charges}
\end{table}

\subsection{Higher-Loop Construction}
\label{sec:HighInt.SU2.HigherLoops}

Here we would like to construct 
the most general ($i$) integrable higher-loop spin chain 
with ($ii$) the proposed scaling in the thermodynamic limit. 
For that purpose, 
we make the most general ansatz for the charges
in terms of ($iii$) permutation symbols $\PTerm{\ldots}$ 
multiplied by undetermined coefficients.
We then demand that the charges mutually commute
and have the right scaling behaviour in the 
thermodynamic limit. 
We will use the computer algebra system 
\texttt{Mathematica} to preform all
necessary commutators and solve the
arising sets of linear equations to determine
the coefficients. Some of the methods
used in the construction are given in 
\appref{app:SU2Tools}.
Let us describe the details of our construction 
of the integrable model:%
\footnote{We keep track of the number of free coefficients in 
\tabref{tab:HighInt.SU2.Coeffs}.
We shall also indicate in the text at what loop order $\ell$
the individual calculations have been performed.}%
\begin{table}\centering
$\begin{array}[t]{|l|ccccc|}\hline
\ell
  &1&2&3& 4& 5 \\\hline
\mbox{structures for }\ham_{2\ell-2}
  &2&3&6&12&27 \\
\mbox{integrability}
  &0&0&2& 5&17 \\\hline
\mbox{integrable }\ham_{2\ell-2}
  &2&3&4& 7&10 \\
\mbox{structures for }T_{2\ell-2}
  &0&0&0& 1& 3 \\\hline
\ham_{2\ell-2}\mbox{ relevant}
  &2&3&4& 6& 7 \\
\mbox{propagation} 
  &1&2&3& 4& 5 \\
\mbox{two-spin interaction}
  &0&0&1& 2& 2(+1) \\\hline
\mbox{remaining d.o.f.}
  &1&1&0& 0& 0 \\\hline
\end{array}\quad
\begin{array}[t]{|l|ccccc|}\hline
\ell
  &1&2&3& 4& 5\\\hline
\mbox{structures for }\charge_{3,2\ell-2}
  &1&2&6&15&46 \\
\mbox{integrability}
  &0&1&4&13&43 \\\hline
\charge_{3,2\ell-2}
  &1&1&2& 2& 3 \\\hline
\hline
\mbox{structures for }\charge_{4,2\ell-2}
  &6&12&27&63& \\
\mbox{integrability}
  &3&9&23&59 &\\\hline
\charge_{4,2\ell-2}
  &3&3&4& 4 & \\\hline
\end{array}$
\caption{Number of coefficients for 
the higher-loop integrable $\alSU(2)$ spin chain}
\label{tab:HighInt.SU2.Coeffs}
\end{table}%
\begin{bulletlist}
\item
We make the ansatz that the charges 
$\charge_{r,2\ell-2}$ with $r$ even (odd), 
have even (odd) parity and are (anti)symmetric.
They consist of no more than $r+\ell-2$ elementary
permutations ranging over $r+\ell-1$ adjacent sites,
see \secref{sec:HighInt.SU2.Interact}.

\item
We compute the commutator 
of $\ham$ and $\charge_3$ 
(up to five-loops).
By demanding that it should vanish, we obtain a set of linear 
equations among the coefficients. We solve it
for coefficients of $\charge_3$ as far as possible,
but some equations relate coefficients of $\charge_2$ only among themselves.

\item
Alternatively, we may 
ignore the charge $\charge_3$ and only 
demand that all pairs in the spectrum of 
$\ham$ remain degenerate at higher-loops.
This yields the same set of constraints for the
coefficients of $\ham$ 
(up to four-loops).

\item
We then compute the commutator of
$\ham$ and $\charge_r$
(for $r=3,4$ up to $\ell=4$ and for $r=5,6$ up to $\ell=2$).
This constrains most coefficients of $\charge_r$ and
remarkably yields no new constraints for $\ham$.

\item
All the obtained charges $\charge_r$ commute among themselves
without further constraints. We notice that
the remaining degrees of freedom correspond precisely
to linear redefinitions of charges,
c.f.~the right hand side of \tabref{tab:HighInt.SU2.Coeffs}:
We can rescale the charge $\charge_{3}$ by a function
of the coupling constant, this yields one degree of freedom
at each loop order.
We may also add $g^4 f(g^2)\charge_{5}(g)$ 
which is structurally equivalent to $\charge_3(g)$,
this yields one degree of freedom starting from $\ell=3$.
For $\charge_4$ the story is equivalent. 
Here, we can always add the length operator $\len$ as well as the
the Hamiltonian $\ham$ 
or rescale by a function. 
This gives three degrees of freedom for all loop orders.
Starting from $\ell=3$ there are further degrees of freedom
due to $\charge_6$.
All in all, this is just the expected number of coefficients. 
In the thermodynamic limit (should it exist at all), 
all coefficients would be fixed for the canonical charges, 
see \secref{sec:HighInt.Higher.Canonical}.

\end{bulletlist}
In this way we have established the most general integrable 
system for the assumed set of interactions.
We conclude that, indeed, the pairing of
states appears to be a \emph{sufficient} 
condition for integrability (in this model).
Next we would like to impose the thermodynamic limit. 
For the thermodynamic limit we make use of two processes: 
Propagation of a single excitation
and interaction of two excitations. 
These should yield the correct scaling 
behaviour at each loop order.
\begin{bulletlist}
\item
In \secref{sec:One.BMN} we have seen that the planar one-loop
dilatation operator acts on the position of a single excitation 
as a lattice Laplacian $\square$. 
The resulting eigenstates are Fourier modes. 
The lowest, non-zero eigenvalue of the Laplacian
is proportional to $1/L^2$, 
exactly the right behaviour for $\ham_{0}$.
Due to the form of the interaction,
$\ham_{2\ell-2}$ must act as a polynomial in $\square$ 
of degree $\ell$. For the correct scaling behaviour, 
$1/L^{2\ell}$, all terms $\square^{k}$ with $k<\ell$ 
should be absent. In general, this determines $\ell$ 
coefficients, see \tabref{tab:HighInt.SU2.Coeffs}.
We shall not fix the coefficient of $\square^{\ell}$
although the quantitative BMN excitation 
energy formula $\ham(g)=(\sqrt{1-2g^2\square}-1)/g^2$
predicts it 
\[\label{eq:HighInt.SU2.Propagate}
\ham_{2\ell-2}\sim -2^{-\ell-1}C_{\ell-1}\square^{\ell}
\]
with $C_k=(2k)!/k!(k+1)!$ the Catalan numbers governing 
the expansion of the square root.

\item
The interaction of two excitations is a more delicate issue,
it is obtained by acting with 
$\ham_{2\ell-2}$ on two excitations and subtracting the 
contribution from the propagation of 
the individual excitations. The remainder can only be non-zero 
if both excitations are close 
(the distance depends on the loop order), 
in other words the remainder is a contact interaction. 
This interaction must also be suppressed by sufficiently 
high powers of $1/L$.

Let us investigate the first order effect 
of the contact term for states with only two excitations,
see \secref{sec:Higher.Spec.TwoEx}.
The first order is determined by 
diagonal scattering, we therefore compute 
the matrix element of $\Op_{0,n}$ going into itself. 
First of all, this is suppressed by $1/L$ due to 
phase space considerations. 
Furthermore, there is a suppression of $n^2n^{\prime\,2}/L^4$ for
the process $\Op_{0,n}\to\Op_{0,n'}$. 
This is due to the zero mode state $\Op_{0,0}$ 
which must be annihilated and can never be produced.
In total there is a suppression of $1/L^5$ for the contact term,
this is sufficient for $\ell=2$. 
Starting from three-loops, the contact term may violate the 
scaling behaviour and there will be additional constraints.
At three-loops a single constraint is enough to remove terms of
order $1/L^5$.
At four-loops we need to remove terms of orders $1/L^5$ and $1/L^7$ yielding 
two constraints. 
At $\ell=5$ there are three constraints, but only two
independent coefficients which may influence the scaling
behaviour. Miraculously, the three constraints seem to
be degenerate such that the scaling in the 
thermodynamic limit appears to be fine at five-loops.

To impose the constraints is not a trivial task.%
\footnote{It would be interesting to find a general criterion 
which determines whether an interaction
of two or more excitations has the correct scaling or not.}
The problem is that also iterated contact terms 
of perturbation theory
may violate the scaling in the thermodynamic limit.
These must be cancelled by higher order contact terms.
We will therefore consider only states with exactly two
excitations for which closed expression can be found. 
We will assume that the 
conjectured energy formula \eqref{eq:Higher.Spec.TwoAllLoop} 
holds to all orders. We will then compute the energies
of several two-excitation states at higher-loops and match them 
with the formula. 
This can only be possible if 
a qualitative BMN limit exists and the coefficients
are adjusted such that 
Hamiltonian provides the correct scaling. 

\item 
At this point nearly all relevant coefficients are fixed.
However, starting at four-loops,
there are some free coefficients left
which have no influence on the scaling dimensions. 
These are due to the freedom to rotate the space 
of states with an orthogonal transformation generated 
by an antisymmetric operator $A$.
For the four-loop interactions
there is precisely one antisymmetric operator $A_6$,
it happens to be proportional to $\comm{\charge_{2,0}}{\charge_{2,2}}$.
It gives rise to the following similarity transformation 
\[\label{eq:HighInt.SU2.Rot}
\charge_r(g)\mapsto \exp(\alpha g^6A_6)\,
\charge_r(g)\,
\exp(-\alpha g^6A_6).\]
The leading term in $\ham=\charge_2$ due to the transformation is
\[\label{eq:HighInt.SU2.RotComm6}
\ham_{6}\mapsto
\ham_{6}+
\alpha\comm{A_6}{\ham_{0}}.
\]
Consequently, the eigenvalues of $\ham(g)$ are not changed 
and $\alpha$ only affects the eigenvectors.
Similarly, at five-loops there are 
three even antisymmetric operators for the construction of $A_8$.

\end{bulletlist}
There are some interesting points to be mentioned regarding this solution.
First of all, integrability and the thermodynamic limit
fix exactly the right number of 
coefficients for a unique solution. 
For this solution, the contribution $\delta \hat E$ of one excitation 
to the energy in the BMN limit
is given by (recall that $\hat g^2=g^2/J^2=\lambda'/8\pi^2$ and $D=D_0+\hat g^2\hat E$)
\[\label{eq:HighInt.SU2.BMN}
\delta\hat E_n=\frac{c_1}{\hat g^2}\lrbrk{\sqrt{1+c_2 \,8\pi^2 n^2 \hat g^2}-1}
+\order{g^{10}}
\]
The constants $c_1,c_2$ correspond to symmetries of the equations, they 
can therefore not be fixed by algebraic arguments. 
We will set them to their physical values, $c_1=c_2=1$. 
It is interesting to observe 
that the BMN quantitative square-root formula for the
energy of one excitation is predicted correctly; we have only demanded
that the thermodynamic (i.e. qualitative BMN limit) limit exists.
Finally, let us mention that the three-loop contribution agrees 
precisely with the calculation of \chref{ch:Higher}.
For the physical choice of $c_1,c_2$ we present the four-loop
and five-loop 
contribution to the Hamiltonian in \tabref{tab:HighInt.SU2.FourFive}.

\begin{table}
\<
\ham_{6}\eq
-35 \PTerm{}
+\bigbrk{67+4\alpha}\PTerm{1}
+\bigbrk{-\sfrac{21}{4}-2\alpha} \PTerm{1,3}
-\sfrac{1}{4} \PTerm{1, 4}
\nl
+\bigbrk{-\sfrac{151}{8}-4\alpha} \bigbrk{\PTerm{1,2}+\PTerm{2,1}} 
+2\alpha \bigbrk{\PTerm{1,3,2}+\PTerm{2,1,3}} 
\nl
+\sfrac{1}{4}\bigbrk{\PTerm{1,2,4}+\PTerm{1,3,4}+\PTerm{1,4,3}+\PTerm{2,1,4}}
+\bigbrk{6+2\alpha} \bigbrk{\PTerm{1,2,3}+\PTerm{3,2,1}}
\nl
+\bigbrk{-\sfrac{3}{4}-2\alpha} \PTerm{2, 1, 3, 2}
+\bigbrk{\sfrac{9}{8}+2\alpha} \bigbrk{\PTerm{1,3,2,4}+\PTerm{2,1,4,3}} 
\nl
+\bigbrk{-\sfrac{1}{2}-\alpha} \bigbrk{\PTerm{1,2,4,3}+\PTerm{1,4,3,2}+\PTerm{2,1,3,4}+\PTerm{3,2,1,4}}  
\nl
-\sfrac{5}{8}\bigbrk{\PTerm{1, 2, 3, 4} + \PTerm{4, 3, 2, 1}},
\nonumber
\\[12pt]
\ham_8\eq
+\sfrac{1479}{8}\PTerm{}
+\bigbrk{-\sfrac{1043}{4}-12\alpha+4\beta_1}\PTerm{1}
+\bigbrk{-19+8\alpha-2\beta_1-4\beta_2}\PTerm{1,3}
\nl
+\bigbrk{5+2\alpha+4\beta_2+4\beta_3}\PTerm{1,4}
+\sfrac{1}{8}\PTerm{1,5}
+\bigbrk{11\alpha-4\beta_1+2\beta_3}\bigbrk{\PTerm{1,2}+\PTerm{2,1}}
\nl
-\sfrac{1}{4}\PTerm{1,3,5}
+\bigbrk{\sfrac{251}{4}-5\alpha+2\beta_1-2\beta_3}\bigbrk{\PTerm{1,3,2}+\PTerm{2,1,3}}
\nl
+\bigbrk{-3-\alpha-2\beta_3}\bigbrk{\PTerm{1,2,4}+\PTerm{1,3,4}+\PTerm{1,4,3}+\PTerm{2,1,4}}
\nl
-\sfrac{1}{8}\bigbrk{\PTerm{1,2,5}+\PTerm{1,4,5}+\PTerm{1,5,4}+\PTerm{2,1,5}}
\nl
+\bigbrk{\sfrac{41}{4}-6\alpha+2\beta_1-4\beta_3}\bigbrk{\PTerm{1,2,3}+\PTerm{3,2,1}}
+\bigbrk{-\sfrac{107}{2}+4\alpha-2\beta_1}\PTerm{2,1,3,2}
\nl
+\bigbrk{\sfrac{1}{4}+\beta_2}\bigbrk{\PTerm{1,3,2,5}+\PTerm{1,3,5,4}+\PTerm{1,4,3,5}+\PTerm{2,1,3,5}}
\nl
+\bigbrk{\sfrac{183}{4}-6\alpha+2\beta_1-2\beta_2}\bigbrk{\PTerm{1,3,2,4}+\PTerm{2,1,4,3}}
\nl
+\bigbrk{-\sfrac{3}{4}-2\beta_2}\bigbrk{\PTerm{1,2,5,4}+\PTerm{2,1,4,5}}
+\bigbrk{1+2\beta_2}\bigbrk{\PTerm{1,2,4,5}+\PTerm{2,1,5,4}}
\nl
+\bigbrk{-\sfrac{51}{2}+\sfrac{5}{2}\alpha-\beta_1+\beta_2+3\beta_3}\bigbrk{\PTerm{1,2,4,3}+\PTerm{1,4,3,2}+\PTerm{2,1,3,4}+\PTerm{3,2,1,4}}
\nl
-\beta_2\bigbrk{\PTerm{1,2,3,5}+\PTerm{1,3,4,5}+\PTerm{1,5,4,3}+\PTerm{3,2,1,5}}
\nl
+\bigbrk{\sfrac{35}{4}+\alpha+2\beta_3}\bigbrk{\PTerm{1,2,3,4}+\PTerm{4,3,2,1}}
\nl
+\bigbrk{-\sfrac{7}{8}-\alpha+2\beta_3}\bigbrk{\PTerm{1,4,3,2,5}+\PTerm{2,1,3,5,4}}
\nl
+\bigbrk{\sfrac{1}{2}+\alpha}\bigbrk{\PTerm{1,3,2,5,4}+\PTerm{2,1,4,3,5}}
\nl
+\bigbrk{\sfrac{5}{8}+\half\alpha-\beta_3}\bigbrk{\PTerm{1,3,2,4,3}+\PTerm{2,1,3,2,4}+\PTerm{2,1,4,3,2}+\PTerm{3,2,1,4,3}}
\nl
+\bigbrk{\sfrac{1}{4}-2\beta_3}\bigbrk{\PTerm{1,2,5,4,3}+\PTerm{3,2,1,4,5}}
\nl
+\bigbrk{\sfrac{1}{4}+\half\alpha+\beta_3}\bigbrk{\PTerm{1,2,4,3,5}+\PTerm{1,3,2,4,5}+\PTerm{2,1,5,4,3}+\PTerm{3,2,1,5,4}}
\nl
+\bigbrk{-\half\alpha-\beta_3}\bigbrk{\PTerm{1,2,3,5,4}+\PTerm{1,5,4,3,2}+\PTerm{2,1,3,4,5}+\PTerm{4,3,2,1,5}}
\nl
-\sfrac{7}{8}\bigbrk{\PTerm{1,2,3,4,5}+\PTerm{5,4,3,2,1}}
\nonumber
\>

\caption{Four-loop and five-loop contributions to the Hamiltonian.}
\label{tab:HighInt.SU2.FourFive}
\end{table}

\section{Spectrum}
\label{sec:HighInt.Spec}

We can now apply the higher-loop integrable Hamiltonian to 
obtain some energies. In addition we can evaluate the
integrable charges on the eigenstates of the Hamiltonian.

\subsection{Lowest-Lying States}
\label{sec:HighInt.Spec.Low}

In preparation for the next section, it will be
helpful to know the spectrum of lowest-lying modes
for our spin chain.
To obtain a matrix representation for the operators,
we have used standard higher order quantum mechanical
perturbation theory.
We have applied the Hamiltonian $\ham$ and charges 
$\charge_{3},\charge_{4}$ up to four-loops 
to all states with a given length $L$ and number of excitations $K$.
The computations were performed using the routines
in \appref{app:SU2Tools}.
Then, the leading order energy matrix was diagonalised
to obtain the leading order energy eigenvalues.
Next, the off-diagonal terms at higher-loops were
removed iteratively by a sequence of similarity transformations,
see also \secref{sec:Higher.Spec.Low}.
Afterwards, the Hamiltonian is diagonal and we can read off the
energy eigenvalues.
The same similarity transformations which were used
to make $\ham$ diagonal also diagonalise
$\charge_{3},\charge_{4}$ and we may  
read off their eigenvalues.

We present our findings up to $L=8$ in 
\tabref{tab:HighInt.Spec.Low} 
(we omit the protected states with $K=0$)
which is read as follows:
For each state there is a polynomial 
and we write down the coefficients up to $\order{g^8}$ and
$\order{x^2}$.
For single states the polynomial $X(x,g)$ equals simply
\[\label{eq:HighInt.Spec.LowSingle}
X(x,g)=E(g)+x^2 Q_4(g).
\]
If there is more than one state transforming in the same representation,
the eigenvalues are solutions to algebraic equations. These could be solved
numerically, here we prefer to state the exact algebraic equation in 
terms of a polynomial $X(\omega,x,g)$
of degree $k-1$ in $\omega$ ($k$ is also the number of lines in one block, 
one for each coefficient of the polynomial). 
The energy and charge eigenvalues are determined through 
the formula
\[\label{eq:HighInt.Spec.LowMulti}
\omega=E(g)+x Q_3(g)+x^2 Q_4(g),
\qquad
\omega^k=X(\omega,x,g).
\]
At first sight the terms linear in $x$ may appear spurious and
the corresponding charge $Q_3(g)$ would have to be zero.
For unpaired states with non-degenerate
$Q_2(g)$ this is true, but not so for pairs of degenerate states.
Then the solution of the algebraic equation leads to terms
of the sort $\sqrt{0+x^2}=\pm x$, where the $0$ is meant
to represent the degeneracy.
Note that for some states the interaction is longer 
than the state. 
In such a case, indicated by $\ast$ in the table, 
we do not know the energy/charge eigenvalue,
see also \secref{sec:HighInt.Spinning.Wrappings}.

\begin{table}\centering
$\begin{array}{|ccc|llll|llll|}\hline
L&K&P    & g^0x^0     & g^2x^0         & g^4x^0                &  g^6x^0               & g^0x^2                &  g^2 x^2             & g^4 x^2               & g^6 x^2                 \vphantom{\hat A^A_{g_g}}\\\hline\hline
4&2& +   & +6         & -12            & +42                   & \ast                  & +0                    & \ast                 & \ast                  & \ast                    \vphantom{\hat A^A_{g_g}}\\\hline
5&2& -   & +4         & -6             & +17                   & -\frac{115}{2}        & +\frac{8}{3}          & -8                   & \ast                  & \ast                    \vphantom{\hat A^A_{g_g}}\\\hline
6&2& +   & +10\omega  & -17\omega      & +\frac{117}{2}\omega  & -\frac{1037}{4}\omega & -\frac{10}{3}\omega   & +30\omega            & -\frac{381}{2}\omega  & \ast                    \vphantom{\hat A^A_{g_g}}\\
 & &     & -20        & +60            & -230                  & +1025                 & +0                    & -\frac{140}{3}       & +420                  & \ast                    \vphantom{\hat A^A_{g_g}}\\\hline
6&3& -   & +6         & -9             & +\frac{63}{2}         & -\frac{621}{4}        & -6                    & +36                  & +0                    & \ast                    \vphantom{\hat A^A_{g_g}}\\\hline
7&2& -   & +2         & -\frac{3}{2}   & +\frac{37}{16}        & -\frac{283}{64}       & +\frac{4}{3}          & -\frac{5}{2}         & +\frac{81}{16}        & -\frac{707}{64}         \vphantom{\hat A^A_{g_g}}\\\hline
7&2& -   & +6         & -\frac{21}{2}  & +\frac{555}{16}       & -\frac{8997}{64}      & +0                    & +\frac{9}{2}         & -\frac{513}{16}       & +\frac{11907}{64}       \vphantom{\hat A^A_{g_g}}\\\hline
7&3& \pm & +10\omega  & -15\omega      & +50\omega             & -\frac{875}{4}\omega  & -\frac{10}{3}\omega   & +25\omega            & -\frac{285}{2}\omega  & +\frac{1615}{2}\omega   \vphantom{\hat A^A_{g_g}}\\
 & &     & -25        & +75            & -\frac{1225}{4}       & +\frac{5875}{4}       & +\frac{245}{12}       & -180                 & +\frac{28145}{24}     & -\frac{86875}{12}       \vphantom{\hat A^A_{g_g}}\\\hline
8&2& +   & +14\omega^2& -23\omega^2    & +79\omega^2           & -349\omega^2          & -\frac{14}{3}\omega^2 & +39\omega^2          & -250\omega^2          & +\frac{4691}{3}\omega^2 \vphantom{\hat A^A_{g_g}}\\
 & &     & -56\omega  & +172\omega     & -695\omega            & +3254\omega           & +\frac{56}{3}\omega   & -\frac{700}{3}\omega & +\frac{5258}{3}\omega & -11822\omega            \vphantom{\hat A^A_{g_g}}\\
 & &     & +56        & -224           & +966                  & -4585                 & -0                    & +168                 & -\frac{5054}{3}       & +\frac{38269}{3}        \vphantom{\hat A^A_{g_g}}\\\hline
8&3& -   & +6         & -9             & +33                   & -162                  & -6                    & +33                  & -192                  & +1191                   \vphantom{\hat A^A_{g_g}}\\\hline
8&3& \pm & +8\omega   & -10\omega      & +28\omega             & -102\omega            & +\frac{4}{3}\omega    & -2\omega             & +4\omega              & -\frac{26}{3}\omega     \vphantom{\hat A^A_{g_g}}\\
 & &     & -16        & +40            & -137                  & +548                  & -\frac{4}{3}          & -\frac{40}{3}        & +\frac{328}{3}        & -\frac{1948}{3}         \vphantom{\hat A^A_{g_g}}\\\hline
8&4& +   & +20\omega^2& -32\omega^2    & +112\omega^2          & -511\omega^2          & -\frac{32}{3}\omega^2 & +72\omega^2          & -442\omega^2          & +\frac{8264}{3}\omega^2 \vphantom{\hat A^A_{g_g}}\\
 & &     & -116\omega & +340\omega     & -1400\omega           & +6938\omega           & +\frac{392}{3}\omega  & -\frac{3100}{3}\omega& +\frac{20708}{3}\omega& -45348\omega            \vphantom{\hat A^A_{g_g}}\\
 & &     & +200       & -800           & +3600                 & -18400                & -320                  & +2800                & -\frac{58400}{3}      & +\frac{389680}{3}       \vphantom{\hat A^A_{g_g}}\\\hline
\end{array}$

\caption{Four-loop energies and charges $Q_{3,4}$.
See \protect\secref{sec:HighInt.Spec.Low} for an explanation.}
\label{tab:HighInt.Spec.Low}
\end{table}

\subsection{Two Excitations}
\label{sec:HighInt.Spec.TwoEx}

Now that the Hamiltonian is known up to five loops,
we may continue the analysis of 
the two-excitation states in \secref{sec:Higher.Spec.TwoEx}.
In principle, we should diagonalise the energy
in perturbation theory, however, this is very labourious. 
Instead we will assume the all-loop
formula \eqref{eq:Higher.Spec.TwoAllLoop} to be correct
and match the coefficients to sufficiently many 
two-excitation states.%
\footnote{In fact, the five-loop coefficients have been obtained in a more
convenient way, see \secref{sec:HighInt.Ansatz.Results}.}
When the coefficients have been determined, 
we may compare the formula to further
states and find agreement. We take this as compelling evidence
that the obtained formula and coefficients are indeed correct.
We present a summary of findings in \tabref{tab:HighInt.Spec.TwoEx}.

\begin{table}\centering
%
$\displaystyle
D^J_n=J+2
+\sum_{\ell=1}^\infty 
\lrbrk{\frac{\gym^2 N}{\pi^2}\sin^2 \frac{\pi n}{J+1}}^{\ell}
\lrbrk{c_\ell+\sum_{k,l=1}^{\ell-1}c_{\ell,k,l}\frac{\cos^{2l} \frac{\pi n}{J+1}}{(J+1)^k}},
$\bigskip\par
$\begin{array}{lll}
\displaystyle c_1=+1,&\quad&\\[3pt]
\displaystyle c_2=-\frac{1}{4},&\quad&
c_{2,1,1}=-1,\\[0.25cm]
\displaystyle c_3=+\frac{1}{8},&\quad&
c_{3,k,l}=\matr{ll}{+\sfrac{3}{4}&+\sfrac{1}{2}\\[1.5pt]-\sfrac{3}{4}&+\sfrac{5}{2}
},\\[0.4cm]
\displaystyle c_4=-\frac{5}{64},&\quad&
c_{4,k,l}=\matr{lll}{
-\sfrac{5}{8}&-\sfrac{5}{12}&-\sfrac{1}{3}\\[1.5pt]
+\sfrac{3}{4}&- \sfrac{7}{4}&-\sfrac{7}{2}\\[1.5pt]
-\sfrac{1}{2}&+\sfrac{59}{12}&-\sfrac{49}{6}
},\\[0.8cm]
\displaystyle c_5=+\frac{7}{128},&\quad&
c_{5,k,l}=\matr{llll}{
+\frac{35}{64}&+\frac{35}{96}&+\frac{7}{24}&+\frac{1}{4}\\[1.5pt]
-\frac{45}{64}&+\frac{185}{96}&+\frac{131}{48}&+\frac{33}{8}\\[1.5pt]
+\frac{5}{8}&-\frac{125}{24}&-\frac{13}{24}&+\frac{81}{4}\\[1.5pt]
-\frac{5}{16}&+\frac{305}{48}&-\frac{1319}{48}&+\frac{243}{8}
}.
\end{array}$
\caption{Planar scaling dimension of two-excitation states.}
\label{tab:HighInt.Spec.TwoEx}
\end{table}

An application of the exact energies of
two-excitation operators is the 
near BMN limit of $\order{1/J}$ corrections. 
Some inspired guessing yields an all-loop expression for the near BMN limit
which agrees with \tabref{tab:HighInt.Spec.TwoEx} at five-loops
\[\label{eq:HighInt.Spec.TwoNearBMN}
D^J_n=J+2\sqrt{1+\lambda'\, n^2}
-\frac{4\lambda'\,n^2}{J\sqrt{1+\lambda'\,n^2}}
+\frac{2\lambda'\,n^2}{J(1+\lambda'\,n^2)}
+\order{1/J^2}.
\]
The first $1/J$ term can be regarded as a renormalisation 
of the term $\lambda' n^2$ in the first square root. 
For instance, we might replace $J$ in the definition of $\lambda'$ by
$L=J+2$ to absorb the second term into the leading order energy.
Incidentally, this yields precisely the coupling constant 
$\tilde g=g/L$ for the thermodynamic limit (see \secref{sec:HighInt.Ansatz.Thermo}).
Unfortunately, as we have seen in \secref{sec:Higher.Spec.TwoEx},
this formula does not agree with the 
expression for the near plane-wave limit 
\eqref{eq:Higher.Spec.TwoString}
derived in \cite{Callan:2003xr,Callan:2004uv}
\[\label{eq:HighInt.Spec.TwoNearPlane}
D^J_n=J+
2\sqrt{1+\lambda'\, n^2}
-\frac{2\lambda'\,n^2}{J}
+\order{1/J^2}.
\]
A curious observation is that the coefficient 
$c_{\ell,1,1}$ equals $2\ell c_\ell$. At order $1/J$,
it cancels the effect of the expansion of the leading order sine. 
Only at one-loop there is no $c_{\ell,1,1}$
to cancel $2c_{1}$. 
We find exactly the string theory prediction 
when we set $c_{\ell,1,l}=0$ for $l>1$.

\subsection{Three Excitations}
\label{sec:HighInt.Spec.ThreeEx}

Let us continue the analysis of unpaired three-excitation
states at higher loops. We find for the scaling dimensions
\<\label{eq:HighInt.Spec.ThreeEng}
D\eq \phantom{0}2,\nln
D\eq \phantom{0}4+6g^2-12g^4+\sfrac{84}{2}g^6+\ldots\,,\nln
D\eq \phantom{0}6+6g^2-\phantom{1}9g^4+\sfrac{63}{2}g^6-\sfrac{621}{4}g^8+\sfrac{7047}{8}g^{10}+\ldots\,,\nln
D\eq \phantom{0}8+6g^2-\phantom{1}9g^4+\sfrac{66}{2}g^6-\sfrac{648}{4}g^8+\sfrac{7212}{8}g^{10}+\ldots\,,\nln
D\eq 10+6g^2-\phantom{1}9g^4+\sfrac{66}{2}g^6-\sfrac{645}{4}g^8+\sfrac{7179}{8}g^{10}+\ldots\,,\nln
D\eq 12+6g^2-\phantom{1}9g^4+\sfrac{66}{2}g^6-\sfrac{645}{4}g^8+\sfrac{7182}{8}g^{10}+\ldots\,,\nln
D\eq 14+6g^2-\phantom{1}9g^4+\sfrac{66}{2}g^6-\sfrac{645}{4}g^8+\sfrac{7182}{8}g^{10}+\ldots\,,\nln
&&\ldots\,,
\>
where we have added the dimension-two half-BPS state and 
a Konishi descendant which appear to be the natural first two elements 
of this sequence.
Note that the exact one-loop form of the eigenstates 
is corrected at higher-loops.
 
We observe that all corrections $D_k$ to the scaling dimensions
below the `diagonal' $k\leq L-2$ , are equal. Incidentally, the coefficients
agree with the formula
\[\label{eq:HighInt.Spec.ThreeAllLoop}
D(g)=L+\bigbrk{\sqrt{1+8g^2}-1}+\bigbrk{\sqrt{1+2g^2}-1}
+\bigbrk{\sqrt{1+2g^2}-1}.
\]
We may interpret the three terms in parenthesis
as the energies of the three excitations.
Then this form can be taken as a clear confirmation of an integrable
system with elastic scattering of excitations.

Only if the loop order is at least half the 
classical dimension at $\order{g^L}$ the pattern breaks down.
Interestingly, if the loop order is exactly half the classical
dimension, the coefficient is decreased by $3\cdot 2^{2-\ell}$.
It would be of great importance to understand the 
changes further away from the diagonal. This might provide
us with clues about wrapping interactions, which, in the above example,
obscure the scaling dimension of the Konishi state beyond three-loops.

For completeness, we state a similar all-loop conjecture for the
higher charges \cite{Beisert:2004hm} 
to generalise \eqref{eq:Int.Spec.ThreeCharges}
\[\label{eq:HighInt.Spec.ThreeCharges}
Q_r(g)=\frac{i}{r-1}
\lrbrk{\frac{1+(-1)^r}{\bigbrk{\sfrac{i}{4}+\sfrac{i}{4}\sqrt{1+8g^2}\,}^{r-1}} 
+\frac{1+(-1)^r}{\bigbrk{\sfrac{i}{2}+\sfrac{i}{2}\sqrt{1+2g^2}\,}^{r-1}}}.
\]
Note that $Q_{r}$ is accurate only 
up order $g^{L-2-r}$.
The corresponding transfer matrix, 
to be compared to the one-loop counterpart
\eqref{eq:Int.Spec.ThreeTrans}, is
\[\label{eq:HighInt.Spec.ThreeTransfer}
T(x,g)=
\frac{x-\bigbrk{\sfrac{i}{4}+\sfrac{i}{4}\sqrt{1+8g^2}\,}}
     {x+\bigbrk{\sfrac{i}{4}+\sfrac{i}{4}\sqrt{1+8g^2}\,}}\,\,
\frac{x-\bigbrk{\sfrac{i}{2}+\sfrac{i}{2}\sqrt{1+2g^2}\,}}
     {x+\bigbrk{\sfrac{i}{2}+\sfrac{i}{2}\sqrt{1+2g^2}\,}}
+\ldots\,.
\]
Here we have used the symbol $x$ instead of $u$ 
for the spectral parameter; the reason will become more
apparent in the next section.

\section{Long-Range Bethe Ansatz}
\label{sec:HighInt.Ansatz}

In \secref{sec:HighInt.SU2} we have investigated 
the $\alSU(2)$ subsector up to five-loops assuming that
higher-loop integrability holds and that the thermodynamic limits exists.
Remarkably, these requirements were sufficient 
to obtain a \emph{unique} system!
For an integrable system we might hope for a Bethe ansatz to describe
the energy eigenvalues. 
Serban and Staudacher have shown that the Inozemtsev long-range
spin chain and associated \emph{asymptotic} Bethe ansatz 
\cite{Inozemtsev:1989yq,Inozemtsev:2002vb}
can be used to reproduce
this model up to three-loops \cite{Serban:2004jf}.
At four-loops there is, however, a fundamental difference and the 
scaling in the thermodynamic limit breaks down in the Inozemtsev chain. 
In this context, asymptotic refers to the fact that 
the Bethe ansatz is only reliable up to 
$L$ loops, where $L$ is the length of the chain. 

\subsection{Ansatz}
\label{sec:HighInt.Ansatz.Ansatz}

Without further ado, let us write down an ansatz 
\cite{Beisert:2004hm} to reproduce 
the results of the previous section. 
The universal Bethe equations are
the same as for the Inozemtsev chain 
\cite{Inozemtsev:1989yq,Inozemtsev:2002vb}
proposed in \cite{Serban:2004jf}
\[\label{eq:HighInt.Ansatz.Ansatz}
\frac{P_L(u_k-\sfrac{i}{2})}{P_L(u_k+\sfrac{i}{2})}=
\prod_{\textstyle\atopfrac{l=1}{l\neq k}}^K
\frac{u_k-u_l-i}{u_k-u_l+i}\,.
\]
Here we use the Bethe roots $u_k$ instead of momenta $p_k$
as the fundamental variables, see \cite{Beisert:2004hm}
for a description of the transformation.
The precise model is specified by 
the function $P_L(u)$. 
For the model derived in \secref{sec:HighInt.SU2}
we suggest%
\footnote{The alternative model specified by $P'_L(u)=x(u)^L$ is equivalent 
to our model at the desired accuracy. 
Beyond that order, one of the two functions might be preferred, 
but probably the model changes substantially.
For the singular solutions in \secref{sec:Int.Spec.ThreeEx},
the simplified, non-polynomial function $P'_L(u)$ leads to problems due to overlapping
divergencies.}
\[\label{eq:HighInt.Ansatz.Phase}
P_L(u)=x(u)^L+\lrbrk{\frac{g^{2}}{2x(u)}}^L
\]
with the function $x(u)$ defined as%
\footnote{At $g=0$ we reproduce the one-loop Bethe ansatz.}
\[\label{eq:HighInt.Ansatz.XofU}
x(u)=\half u+\half u\sqrt{1-2g^2/u^2}\,.
\]
This relation is the main difference to the Inozemtsev chain.
Inspired by the findings of \secref{sec:HighInt.Spec.ThreeEx}
we propose the energy to be given by 
\[\label{eq:HighInt.Ansatz.Energy}
E=\sum_{k=1}^K
\lrbrk{
\frac{i}{x(u_k+\sfrac{i}{2})}
-
\frac{i}{x(u_k-\sfrac{i}{2})}
}+\order{g^{2L-2}},\qquad
D=L+g^2 E.
\]
The unknown terms of order $g^{2L-2}$ 
are related to the asymptotic
nature of our Bethe ansatz. 

Furthermore, the charges are apparently given by
\[\label{eq:HighInt.Ansatz.Charges}
Q_r=\sum_{k=1}^K
\frac{i}{r-1}
\lrbrk{
\frac{1}{x(u_k+\sfrac{i}{2})^{r-1}}
-
\frac{1}{x(u_k-\sfrac{i}{2})^{r-1}}
}+\order{g^{2L-2r+2}}.
\]
They can be summed up into a transfer matrix 
\[\label{eq:HighInt.Ansatz.Transfer}
T(x)=
U\exp\sum_{r=2}^\infty iu^{r-1}\charge_r
+\ldots
=
\prod_{k=1}^K
\frac{x-x(u_k+\frac{i}{2})}{x-x(u_k-\frac{i}{2})}
+\ldots\,,
\]
where the dots indicate further possible terms like $x^L$ or $g^{2L}$ 
which cannot be seen for the lower charges or at lower loop orders.
The transfer matrix at $u=0$ gives the shift eigenvalue
\[\label{eq:HighInt.Ansatz.Shift}
1=U=T(0)=\prod_{k=1}^K \frac{x(u_k+\sfrac{i}{2})}{x(u_k-\sfrac{i}{2})}
\]
which should equal $U=1$ for gauge theory states with cyclic symmetry.

The function $P_L(u)$ is indeed a polynomial of degree $L$ in $u$,
which can be derived from the inverse of \eqref{eq:HighInt.Ansatz.XofU} 
\[\label{eq:HighInt.Ansatz.UofX}
u(x)=x+\frac{g^2}{2x}\,.
\]
Therefore, the equation \eqref{eq:HighInt.Ansatz.Ansatz} is the 
Bethe equation of an inhomogeneous spin chain, 
see \secref{sec:Int.Chains.Transfer} and 
\cite{Faddeev:1996iy}.
The polynomial can be factorised and we obtain for the 
inhomogeneities $v_p$ 
\[\label{eq:HighInt.Ansatz.Inhomogeneities}
P_L(u)
=\prod_{p=1}^{L}(u-v_p)\quad
\mbox{with}\quad
v_p=\sqrt{2}\,g\cos\frac{\pi(2p-1)}{2L}\,.
\]
Now it can be noticed that the physical transfer matrix $T(x)$
is not the natural transfer matrix $T'(u)$ associated to the
inhomogeneous spin chain
\[\label{eq:HighInt.Ansatz.BarTransfer}
T'(u)=
\prod_{k=1}^K
\frac{u-u_k-\frac{i}{2}}{u-u_k+\frac{i}{2}}
+
\frac{P_L(u)}{P_L(u+i)}
\prod_{k=1}^K
\frac{u-u_k+\frac{3i}{2}}{u-u_k+\frac{i}{2}}\,.
\]
The Bethe equations follow from this transfer matrix
by cancellation of poles at $u_k-\frac{i}{2}$.
The charges $Q'_r$ derived from $T'(u)$ 
are given as in \eqref{eq:Int.Bethe.SU2Charges}.
In perturbation theory we can relate
these charges to the physical charges $Q_r$ by
\[\label{eq:HighInt.Ansatz.RelateCharges}
Q_r=Q'_r
+\sfrac{1}{2}(r+1)g^2 Q'_{r+2}
+\sfrac{1}{8}(r+2)(r+3)g^4 Q'_{r+4}
+\ldots\,.
\]

Let us first of all comment on the inhomogeneity.
Our spin chain is homogeneous, how can the Bethe ansatz of
an inhomogeneous spin chain describe our model?
First of all, the equation \eqref{eq:HighInt.Ansatz.RelateCharges}
is merely an eigenvalue equation, it does not directly relate 
the homogeneous and inhomogeneous charges, $\charge_r$ and $\charge'_s$;
it merely tells us that there is a similarity transformation 
between the two. 
Similar maps are encountered within the inhomogeneous spin chain itself:
On the one hand, the order of the
inhomogeneities $v_p$ does not matter for the Bethe ansatz and thus
for the eigenvalues $Q'_r$ of the charges.
On the other hand, it should certainly influence 
the charge operators $\charge'_r$.
Consequently, the eigenstates should be related by a similarity
transformation.%
\footnote{The inhomogeneities $v_p$ and $v_{p+1}$ can
be interchanged by conjugation with $\Rmatrix_{p,p+1}(v_p-v_{p+1})$.}
To understand our model better, it would be essential
to investigate this point further and find the map
between our homogeneous spin chain model and the
common inhomogeneous spin chain.

In our equations, the map between $x$ and $u$ 
\eqref{eq:HighInt.Ansatz.XofU,eq:HighInt.Ansatz.UofX}
plays a major role.
It is a double covering map,
for every value of $u$ there are two corresponding values of $x$,
namely
\[\label{eq:HighInt.Ansatz.DoubleCover}
u\longleftrightarrow
\lrset{x,\frac{g^2}{2x}}.
\]
For small values of $g$, where the asymptotic Bethe ansatz
describes the long-range spin chain, 
we will always assume that $x\approx u$.
When $g$ is taken to be large
(if this makes sense at all is a different question), however,
special care would be needed in selecting the appropriate branch.
The double covering map for $x$ and $u$ has an analog
for the transfer matrices $T(x)$ and $T'(u)$.
We find the relation
\[\label{eq:HighInt.Ansatz.TransferDoubleCover}
\frac{T(x)\,T(g^2/2x)}{T(0,g)}
\approx T'(u).
\]
which holds if the second term in \eqref{eq:HighInt.Ansatz.BarTransfer} is
dropped. It can be proved by using the double covering relation
\[\label{eq:HighInt.Ansatz.ProductDouble}
(x-x')\lrbrk{1-\frac{g^2}{2xx'}}
=
\lrbrk{x+\frac{g^2}{2x}}-
\lrbrk{x'+\frac{g^2}{2x'}}
=
u-u'.
\]
We believe it is important to further study the implications of the double
covering maps. This might lead to insight into the definition of our
model, possibly even beyond wrapping order.

Now we have totally self-consistent Bethe equations 
with associated transfer matrix elements $T'(u)$.
Unfortunately, $T'(u)$ does not directly describe physical quantities,
such as the energy $E$.
These are encoded in the physical transfer matrix elements $T(x)$
which involve the function $x(u)$ and are
ambiguous due to the two branches of the square root.
This is not a problem in perturbation theory, however,
even there inconsistencies are observed at higher order in $g$
\cite{Beisert:2004hm}.
Remarkably, these appear precisely at the order where
wrapping interactions start to contribute
and our asymptotic Bethe ansatz is fully consistent
to the desired accuracy.
Conversely, there are signs of the missing of wrapping terms.
We hope that finding a cure for the
problems beyond wrapping order
might help to find a generalization of the
Bethe equations which include wrapping interactions.
Presumably these equations will have a substantially different form,
see \secref{sec:HighInt.Ansatz.Others}.

\subsection{Results}
\label{sec:HighInt.Ansatz.Results}

Here we summarise the results of a comparison
of the above Bethe ansatz with the spectrum of the 
spin chain model. 
For the details of the comparison 
we refer the reader to the article \cite{Beisert:2004hm}.

\begin{bulletlist}
\item
The energy of states with two excitations agrees
with the formula given in \tabref{tab:HighInt.Spec.TwoEx}.
In fact, the five-loop result was obtained using
the Bethe ansatz, where this is a straightforward task.
It was shown to agree with the model in a number of cases.
Furthermore, it is possible to derive all-loop
results in the near BMN limit. 
The energy as an analytic function in $\lambda'$ 
is indeed given by the conjectured formula
\eqref{eq:HighInt.Spec.TwoNearBMN}.

\item
The general BMN energy formula 
\eqref{eq:One.BMN.Energy,eq:One.BMN.Corres}
is easily confirmed. 

\item
The unpaired three-excitation states are singular.
We can treat these solutions by 
demanding cancellation of singularities 
in the transfer matrix $T'(u)$. 
Remarkably, the results agree with their respective mirror solutions, 
see \secref{sec:Int.Bethe.SL2}.
These have $L-2$ excitations instead of $3$ and are regular.
Up to $L=10$ their energies do agree with \eqref{eq:HighInt.Spec.ThreeEng}.

\item
All states with $L\leq 8$ and 
all unpaired ones with $L\leq 10$ 
have been computed in the Bethe ansatz. Their energies agree with 
\tabref{tab:HighInt.Spec.Low}.

\item
We have also compared some higher charges of the Bethe ansatz with 
the corresponding explicit computations.
They agree.

\end{bulletlist}

In conclusion, we can say that for all considered examples, the Bethe ansatz
yields precisely the same spectrum as 
the integrable spin chain model constructed in \secref{sec:HighInt.SU2}.
It shows that an integrable spin chain 
with a well-defined thermodynamic limit (see \secref{sec:HighInt.Higher.Scaling}) 
is very likely to exist in contrast to the doubts raised in 
\cite{Serban:2004jf}.
In terms of the long-range Bethe ansatz there may seem to be many such models.
These would be obtained by replacing the phase relation 
\eqref{eq:HighInt.Ansatz.Phase}
and energy formula \eqref{eq:HighInt.Ansatz.Energy} 
by some other function that is well-behaved in the limit.
If we however demand that the model is related to
an inhomogeneous spin chain, 
we find a unique model with thermodynamic scaling behaviour, 
see \cite{Beisert:2004hm} for details.

The upshot for the integrable spin chain model is similar: 
In its construction we have assumed a very specific form of interactions
and the obtained Hamiltonian has turned out to be unique (at five loops).
In other words, the very relations
\eqref{eq:HighInt.Ansatz.Phase,eq:HighInt.Ansatz.Energy}
are special and correspond to the assumed form 
of interactions ($iii$).%
\footnote{This picture is rather similar to the Inozemtsev spin chain
where the requirement of pairwise interactions of spins
at a distance was shown to lead to the phase relation of
the Inozemtsev-Bethe ansatz.}
At any rate, 
the relations \eqref{eq:HighInt.Ansatz.Phase,eq:HighInt.Ansatz.Energy}
are very suggestive in view of a correspondence 
to string theory on plane waves, see \secref{sec:One.BMN}.
It is therefore not inconceivable, that the Bethe ansatz indeed 
describes planar $\superN=4$ gauge theory in the $\alSU(2)$ subsector
at higher-loops.

\subsection{The Thermodynamic Limit}
\label{sec:HighInt.Ansatz.Thermo}

Here we shall only present the thermodynamic limit of 
the equations in 
\eqref{sec:HighInt.Ansatz.Ansatz}.
The overall structure is the same as
described in \secref{sec:Int.Thermo} only
that we now introduce dependence on the
effective coupling constant
\[\label{sec:HighInt.Ansatz.ThermoCoup}
\tilde g=g/L.
\]
An application and sample calculation is found in
\cite{Beisert:2004hm}.

As before, we shall assume that the roots $\tilde u_k=u_k/L$ condense on a
disconnected contour $\contour$ in the complex plane
with density function $\rho(\tilde u)$.
The density is again normalised by the total filling fraction $\tilde K=K/L$
\[\label{sec:HighInt.Ansatz.ThermoNorm}
\int_{\contour}d \tilde u\,\rho(\tilde u)=\tilde K.
\]
We find for the energy and the momentum constraint 
\[\label{sec:HighInt.Ansatz.ThermoEnergyMom}
\tilde E=\int_{\contour}\frac{d\tilde u\,\rho(\tilde u)}{\tilde u\sqrt{1-2 \tilde g^2/\tilde u^2}}
\,\frac{1}{\tilde x(\tilde u)}\,,
\qquad
2\pi n=\int_{\contour}\frac{d\tilde u\,\rho(\tilde u)}{\tilde u\sqrt{1-2 \tilde g^2/\tilde u^2}}
\,,
\]
with the map between $\tilde x$ and $\tilde u$ is given by%
\footnote{Note the useful identity $\tilde x-g^2/2\tilde x=\tilde u\sqrt{1-2\tilde g^2/\tilde u^2}$\,.}
\[\label{sec:HighInt.Ansatz.ThermoMap}
\tilde x(\tilde u)=\half \tilde u+\half \tilde u\sqrt{1-2\tilde g^2/\tilde u^2}
\,,
\qquad
\tilde u(\tilde x)=\tilde x+\frac{g^2}{2\tilde x}\,,
\]
whereas the higher-loop generalisation of the Bethe equation reads
\[\label{sec:HighInt.Ansatz.ThermoBethe}
2\pi n_{\tilde u}
-\frac{1}{\tilde u\sqrt{1-2 \tilde g^2/\tilde u^2}}
=
2\pint_{\contour} \frac{d\tilde v\,\rho(\tilde v)}{\tilde v-\tilde u}\,.
\]
In the thermodynamic limit, the physical charges and resolvent are given by 
\[\label{sec:HighInt.Ansatz.ThermoCharges}
\tilde Q_r=\int_{\contour}\frac{d \tilde u\,\rho(\tilde u)}{\tilde u\sqrt{1-2 \tilde g^2/\tilde u^2}}
\,
\frac{1}{\tilde x(\tilde u)^{r-1}}\,,
\qquad
G(\tilde x)=\int_{\contour}\frac{d \tilde u\,\rho(\tilde u)}{\tilde u\sqrt{1-2 \tilde g^2/\tilde u^2}}
\,\frac{\tilde x(\tilde u)}{\tilde x(\tilde u)-\tilde x}\,.
\]

As in the one-loop case, 
the Bethe equation \eqref{sec:HighInt.Ansatz.ThermoBethe}
can alternatively be written as a consistency
condition on the singular transfer matrix 
$\tilde T'(\tilde u)\sim 2\cos G'\indup{sing}(\tilde u)$ with
\[\label{sec:HighInt.Ansatz.ThermoResolvInhomo}
G'\indup{sing}(\tilde u)
=G'(\tilde u)+\frac{1}{2\tilde u\sqrt{1-2\tilde g^2/\tilde u^2}}
\qquad
\mbox{and}
\qquad
G'(\tilde u)=\int_{\contour}\frac{d\tilde v\,\rho(\tilde v)}{\tilde v-\tilde u}\,.
\]
The function $2\cos G'\indup{sing}(\tilde u)$ is single-valued
if it obeys \eqref{eq:Int.Thermo.SU2ResolvSing}
\[\label{sec:HighInt.Ansatz.ThermoBetheResolv}
G'\indup{sing}(\tilde u+i\epsilon)+G'\indup{sing}(\tilde u-i\epsilon)=2\pi n_{\tilde u}
\]
across a cut of $G'$ at $\tilde u$.
At this point,
it is however not clear how the physical transfer matrix
$\tilde T(x)$ is related to the physical
resolvent $G(x)$ and if there is also a
consistency requirement which leads to the
Bethe equations.
This is largely related to mirror cuts
in $\tilde T(\tilde g^2/2\tilde x)$ which
are due to the double covering map
\eqref{eq:HighInt.Ansatz.ProductDouble}.

\subsection{Bethe Ans\"atze for Bigger Subsectors}
\label{sec:HighInt.Ansatz.Others}

The Bethe ansatz has proved to be a very powerful tool 
in obtaining the spectrum at high loop orders. It would therefore be 
extremely interesting and important to generalise
it to bigger subsectors than the $\alSU(2)$ subsector,
preferably to the complete $\alPSU(2,2|4)$ spin chain.
Despite some attempts we have not succeeded in finding suitable
equations beyond the $\alSU(2)$ sector.%
\footnote{In the thermodynamic limit there are simplifications 
which allow to guess the Bethe equations \cite{Minahan:2004ds}.}
We would thus like to present a number of considerations for
the construction of a complete all-loop Bethe ansatz.
\begin{bulletlist}
\item
The most important issue seems to be multiplet shortening, 
see also \secref{sec:N4.Split,sec:Int.Bethe.Splitting}.
The spectrum contains a number of multiplets which are 
short in the free theory. A short multiplet cannot acquire 
an anomalous dimension (energy) unless it
combines with other short multiplets to form a long one
(in analogy to the Higgs mechanism).

The one-loop Bethe ansatz was not constructed to respect 
multiplet joining, nevertheless it does display this feature.
For all short, non-protected multiplets compatible short multiplets
can be found. All of these have not only equal anomalous dimension
but also equal higher charges
so that they can join in the interacting theory.
The one-loop Bethe ansatz has a solution for all highest-weight states
of the multiplets. In particular, there are multiple solutions for
splitting multiplets. At higher-loops these multiplets join, 
consequently the complete Bethe ansatz should only find the 
highest weight of the long multiplet. The highest weights of the
submultiplets should not generalise or display some other
kind of inconsistency beyond one-loop.

\item
There is another issue related to multiplet shortening.
The Bethe ansatz not only yields solutions corresponding 
to gauge theory states, but also solutions with non-zero
momentum, $U\neq 1$. A naive generalisation to higher-loops
would also produce these states. This, however, would be inconsistent
for a simple reason: Short multiplets in the non-zero momentum sector
generically do not have suitable partners to join up.
Let us consider the state
\[\label{eq:HighInt.Ansatz.MomentumState}
\state{X}=\state{\fldZ\phi}-\state{\phi\fldZ}.
\]
This state does not obey the momentum constraint,
i.e.~$\Tr\state{X}=0$ or $\shift \state{X}=-\state{X}$.
It is not physical, but nevertheless reproduced by the Bethe ansatz.
Its Dynkin labels are $[0;0;1,0,1;0;0]$, 
it is the primary of a BPS multiplet. 
In principle this multiplet could join with another short multiplet of lower
dimension to form a long one. 
Here this is not possible, there are no potential partners. 
Consequently, this multiplet would have to be protected, but
a direct computation using the one-loop Hamiltonian or Bethe ansatz yields $E=4$
which is inconsistent.
A prospective Bethe ansatz should take this into account and 
exclusively yield solutions with zero momentum. 

\item
There is a complication which applies to non-compact and
supersymmetric representations: The Dynkin labels $r,r_1,r_2$
contain the anomalous dimension $\delta D$ and thus
change with $g$. 
For the Bethe ansatz, the labels also specify the number
of excitations for any given state (c.f.~\secref{sec:Int.Bethe.Excitations}) 
which certainly must be positive integers. 
It is not clear if and how these two points can be combined.
If possible, it is reasonable to believe that the Bethe equations 
will be a self-consistency equation on the energy
(this is somewhat similar to the integral equations 
which appear in string theory \cite{Kazakov:2004qf}).
In perturbation theory at each order, the corrected energy would have to be 
used as input for the next order.

\item
The length $L$ and hypercharge $B$ are not good quantum numbers
in dynamic spin chains.
However, they are also used as input for the Bethe equations,
see \secref{sec:Int.Bethe.Excitations}. 
It is not clear how to identify states when $L$ and $B$ cannot be fixed;
ideally the Bethe ansatz for a dynamic spin chain should not 
distinguish between states with different $L,B$.

\end{bulletlist}
All in all this suggests that the prospective 
all-loop $\alPSU(2,2|4)$ Bethe ansatz, 
if it exists,%
\footnote{Given that $\superN=4$ SYM is likely to be an integrable model,
even beyond one-loop, we can hope for some associated Bethe ansatz.}
has a rather different structure than 
the one-loop approximation. Even more, the one-loop
ansatz might turn out not to be a smooth limit of it.
This could indeed be a blessing in disguise because a substantially different 
Bethe ansatz might allow for the inclusion of wrapping interactions,
see \secref{sec:HighInt.Spinning.Wrappings}
or even incorporate them naturally.

The arguments presented above do not apply to the $\alSU(2)$ subsector 
because there is no multiplet shortening
and the $\alSU(2)$ labels are not affected by the anomalous dimension.
This explains why it was relatively easy to find our 
all-loop generalisation.

For a generalisation one might take a more pragmatic point of view and 
merely consider the classical algebra as the symmetry algebra. 
All classical representation labels would 
be perfectly well-defined and integer.
The Bethe ansatz should yield some energies 
which we interpret as the corrections to the scaling dimensions.
It would then be essential that all submultiplets have precisely the
same energy and charges. Solutions with non-zero momentum would have 
to be ignored. 
The only shortcoming of such an ansatz would be that it does
not explain the truly interacting structure of the algebra.

\section{Stringing Spins at Higher Loops}
\label{sec:HighInt.Stringing}

Now that we have a Bethe ansatz for higher-loop scaling dimensions
we may continue the comparison of spinning strings and gauge theory 
started in \secref{sec:Int.Spinning}.

\subsection{Spinning out of Control}
\label{sec:HighInt.Spinning.Trouble}

The first steps in this direction have been performed in \cite{Serban:2004jf}
using the Inozemtsev spin chain, which is consistent 
with the results of the previous and current chapter up to three-loops.
Further progress was made in \cite{Kazakov:2004qf,Arutyunov:2004xy,Beisert:2004hm}.
Here we shall only summarise the results. 

In the thermodynamic limit the perturbative Bethe equations 
reduce to expressions similar to the ones given in \secref{sec:Int.Thermo},
but with a few additional $g$-dependent terms, 
c.f.~\secref{sec:HighInt.Ansatz.Thermo}.
These terms modify the solution in two ways. On the one hand,
the contour will experience a perturbative deformation
and, on the other hand, the energy formula receives radiative corrections.
Together, these determine the higher-loop contributions to the
energy, either implicitly or explicitly.
It was then found that the two-loop correction indeed coincides with
the prediction from string theory \cite{Serban:2004jf}. 
Moreover the higher charges do agree \cite{Arutyunov:2004xy}.
This result was subsequently generalized to all solutions
within the $\alSU(2)$ subsector 
by comparing their Bethe ans\"atze \cite{Kazakov:2004qf,Beisert:2004hm}.

Interestingly, 
\emph{the agreement does not persist at the three-loop level} \cite{Serban:2004jf}. 
One might argue that this due to a flawed gauge theory 
Hamiltonian. Although this is a possibility, it would
not explain the discrepancy: 
The authors of \cite{Serban:2004jf}
investigated whether agreement can be achieved by modifying 
the phase relation and expression for the energy in the most general way 
compatible with the scaling behaviour of string theory: 
The outcome was negative. 
Therefore it may seem impossible to construct
a weak coupling integrable spin chain to reproduce 
string theory at `three-loops'.
Giving up on integrability is not an option either,
because (classical) string theory on $AdS_5\times S^5$ is integrable
\cite{Bena:2003wd} and the spectra could not possibly agree. 

The problem parallels the earlier three-loop disagreement with
near plane-wave string theory \cite{Callan:2003xr,Callan:2004uv}
discussed in \secref{sec:Higher.Spec.TwoEx}.
In fact, it appears that the mismatch in these two examples is 
related: Both of them constitute a deviation from the 
BMN limit, either by considering many excitations or a 
state where the length is not strictly infinite.%
\footnote{One could extract the expansion
of the function $\tilde E(\alpha)$ for small $\alpha$,
(c.f.~\secref{sec:Int.Spinning}) from a 
large $L$ expansion. For that one would 
consider an arbitrary number of excitations, $K$. 
For the $1/L^n$ correction to the energy one should find 
no more than $n$ powers of $K$. 
The term $K^n/L^n$ is to be interpreted as $\alpha^n$, whereas 
all lower powers of $K$ would have to be dropped.}

\subsection{Order of Limits}
\label{sec:HighInt.Spinning.Limits}

\begin{figure}\centering
\includegraphics[scale=1.2]{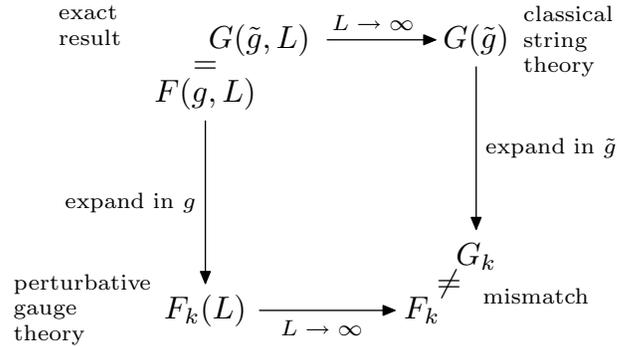}
\caption{A possible explanation for both the near BMN and the FT spinning strings
disagreement. $F_\ell$ excludes gauge theory wrapping effects,
while $G_\ell$ is expected to include them.}
\label{fig:HighInt.Spinning.NonCommute}
\end{figure}

The above problems suggest that either 
the correspondence between string theory and gauge theory 
breaks down at three-loops 
or some subtlety has not been taken into account properly
\cite{Serban:2004jf,Kazakov:2004qf,Beisert:2004hm}.
Indeed, there may be a fundamental problem in the comparison: 

The comparison takes place in the thermodynamic limit
$L\to\infty$ and in an expansion around $\tilde g=g/L=0$.
However starting with an exact function $F(g,L)$,
we must decide which limit is taken first.
It turns out that for classical string theory,
the thermodynamic limit $L\to\infty$ is a basic assumption.
The resulting energy may then be expanded in powers of $\tilde g$.
In contrast, gauge theory takes the other path.
The computations are based on perturbation theory around $g=0$.
This expansion coincides with the
expansion in $\tilde g$ because the coefficients turn out to be
suppressed by sufficiently many powers of $1/L$.
Now the order of limits does potentially matter.
This is best illustrated 
in the non-commutative diagram \figref{fig:HighInt.Spinning.NonCommute}
and the example in \secref{sec:HighInt.Spinning.Example}.
Semi-classical string theory corresponds to the upper right corner
of the diagram, i.e.~it requires the large spin limit.
Conversely, perturbative gauge theory is situated at
the lower left corner, where the length $L$ is finite, but
only the first few orders in $g$ are known.%
\footnote{We recall that
the number of known terms grows with $L$, if our spin
chain ansatz is correct.}

The BMN and FT proposals are both based on the assumption that the
diagram in \figref{fig:HighInt.Spinning.NonCommute} does commute. In other words one
should be able to compare, order by order, the gauge theory loop
expansion with the string theory expansion in $\tilde g$. 
That this might in fact not be true was first hinted at in 
another context in \cite{Klebanov:2002mp}. 
Another, more closely related, instance where the different limiting
procedures lead to different results can be found in \cite{Serban:2004jf}. 
For the hyperbolic Inozemtsev spin chain it
was shown that the order of limits does matter.
In the `gauge theory{}' order, this spin chain appears to have no proper
thermodynamic limit. For the `string theory{}' order, i.e.~when
the thermodynamic limit is taken right from the start, it is
meaningful!

In order to make contact with string theory we propose to sum up 
the perturbation series in $\lambda$ before taking the thermodynamic limit. 
In this case, the comparison would take place at the
upper right corner of \figref{fig:HighInt.Spinning.NonCommute}.%
\footnote{If we wish to compare at the lower left corner
we should sum up all quantum corrections in string theory
before we compare to perturbative gauge theory.
There the $1/L$ alias $1/g$ suppression 
of quantum effects was derived assuming a large $g$. 
For small $g$ this simplification is not justified and
additional effects may contribute.}
With the all-loop spin chain at hand this may indeed be feasible. 
In contrast to the Inozemtsev chain, there appears to be no
difference between the two orders of limits because
the thermodynamic limit is well-behaved in perturbation theory.
However one has to take into account wrapping interactions
which could violate thermodynamic scaling behaviour. 
These arise at higher loop orders $\ell$ when the interaction stretches
all around the state, i.e.~when $\ell\geq L$. We will discuss them
in \secref{sec:HighInt.Spinning.Wrappings} after an example, 
which illustrates the potential importance of these interactions.

\subsection{Example}
\label{sec:HighInt.Spinning.Example}

Here we present an example where one can see the importance
of the order of limits.
We choose a function
\[\label{eq:HighInt.Spinning.Example}
F(g,L)=\frac{g^{2L}}{(c+g^2)^L}
=\lrbrk{1+\frac{c}{\tilde g^2 L^2}}^{-L}=G(\tilde g,L).
\]
In perturbation theory around $g=0$ we find that
the function vanishes at $L$ leading loop orders
\[\label{eq:HighInt.Spinning.ExampleExpand}
F(g,L)=\sum_{k=0}^\infty F_k\, g^k=
\frac{g^{2L}}{c^L}
-\frac{g^{2L+2}}{c^{L+1}}
+\frac{g^{2L+4}}{c^{L+2}}
+\ldots,
\qquad
\mbox{i.e.}\quad
F_k(L)=0\quad
\mbox{for}\quad k<2L.
\]
The leading factor $g^{2L}$ mimics the effect
of wrapping interactions in gauge theory as
explained below.
When we now go to the thermodynamic limit, $L\to\infty$,
we see that all coefficients $F_k$ are zero.

Now let us take the thermodynamic limit first. The large $L$ limit
of $G(\tilde g,L)=\sum_k G_k\,\tilde g^k$
yields $G(\tilde g)=1$ in a straightforward fashion. 
This result depends crucially on the function $F(g,L)$.
Currently, we do not know how to incorporate wrapping
interactions, but $g^{2L}$ alone would not have a sensible
thermodynamic limit. To compensate this, we have introduced some
function ${1/(c+g^2)^L}$. Clearly we cannot currently prove
that gauge theory produces a function like this, but it appears to be
a definite possibility. In our toy example, the expansion in
$\tilde g$ gives $G_0=1$ and $G_k=0$ otherwise.

In conclusion we find $G_0=1$
while $F_0=0$ which demonstrates the non-commutativity
of the diagram in \figref{fig:HighInt.Spinning.NonCommute} 
in an example potentially relevant to our context.
It is not hard to construct a function $F(g,L)$ which yields
arbitrary coefficients $G_k$ while all $F_k$ remain zero.

Note however that there is a sign of the non-commutativity
in \eqref{eq:HighInt.Spinning.ExampleExpand}:
A correct scaling behaviour would require the coefficient
$F_k$ to scale as $L^{-k}$. In particular for
$k=2L$, the coefficient should scale as $L^{-2L}$ instead of
$c^{-L}$. Therefore one can say that the function $F$ violates
the scaling law even at weak coupling, but in a mild way that is easily
overlooked.
This parallels the above observations for the Inozemtsev spin chain
that a proper scaling behaviour might be obscured
in perturbation theory.

\subsection{Wrapping Interactions}
\label{sec:HighInt.Spinning.Wrappings}

One very important aspect of the spin chain are wrapping 
interactions (c.f.~\secref{sec:Dila.Planar.Wrapping}). 
These interactions appear when the 
expected length of the interactions, $\ell+1$, exceeds 
the length of the state $L$.
Let us note that the \emph{asymptotic} Bethe ansatz 
for the Inozemtsev spin chain
apparently does not incorporate wrapping interactions correctly. 
As our ansatz is very similar, we expect the same to apply here.

For a fixed length $L$, the wrapping interactions
are irrelevant at lower loop orders, but 
at higher-loops they are the only contribution.
Therefore, at finite values of the coupling constant,
wrapping interactions dominate and 
the dependence on the coupling constant may change completely.
This is very appealing from the point of view of
the AdS/CFT correspondence, which predicts the
scaling dimension of a generic, unprotected 
state to grow like $\sqrt{g}\sim \lambda^{1/4}$ 
for large $g$ \cite{Gubser:1998bc}.%
Instead, the energy formula 
\eqref{eq:HighInt.Ansatz.Energy}
seems to suggest a linear growth in $g$,
but it is valid only for sufficiently low
loop orders.
Therefore we suspect that wrapping interactions
may be responsible for the conjectured $\sqrt{g}$ behaviour at large $g$. 
Note that a generalisation of the 
Bethe equations for classical string theory \cite{Kazakov:2004qf}
to towards the quantum regime has been conjectured
in \cite{Arutyunov:2004vx}. These equations 
reproduce the $\lambda^{1/4}$ behaviour 
as well as the near plane wave results of 
\cite{Callan:2003xr,Callan:2004uv,Callan:2004ev,Callan:2004dt,McLoughlin:2004dh}.

In the same spirit, wrappings 
may be important for the comparison 
between gauge theory 
and string theory in the case of spinning
strings and the near BMN limit \cite{Beisert:2004hm}
where they become an order-of-limits issue:
When we assume the length of the state to be sufficiently large,
wrapping interactions are suppressed.
Then we find that $g$ and $1/L$ combine 
and the energy is a function of $\tilde g$.
However, for a comparison to string theory, we might wish to take 
$g$ finite for a fixed length. 
Again we should find that $g$ and $1/L$ combine 
(possibly for a completely different reason) 
and obtain a function of $\tilde g$.
Here we may expect a qualitative difference
because wrappings dominate the complete tail of
the perturbative expansion for any fixed length.
Summing up the perturbative series we might get a totally
different function, as demonstrated by the example
in \secref{sec:HighInt.Spinning.Example}.

Unfortunately, we currently have no handle on wrapping interactions.
In the algebraic analysis of \chref{ch:Higher}, there seem to 
be few constraints on their form. 
In fact, virtually any higher-loop anomalous dimensions
can be assigned to multiplets with a small length by means
of wrapping interactions. For the Bethe ansatz the inclusion 
of wrapping interactions will probably require a substantially
different form. For instance, an exact, \emph{non-asymptotic} 
Bethe ansatz for the Inozemtsev chain 
is not known \cite{Inozemtsev:1989yq,Inozemtsev:2002vb}.

All in all, a better understanding of wrapping interactions
might be of great importance to the 
AdS/CFT correspondence and our understanding
of higher-loop conformal gauge theory.

\finishchapter 

\chapternonum{Conclusions}

In this \docphysrept[dissertation]{work} we have developed 
means to efficiently compute and investigate
scaling dimensions of local operators 
in a perturbative four-dimensional conformal field theory.
The central object is the \emph{dilatation operator};
it is one of the generators of the (super)conformal algebra and
it measures scaling dimensions.
In the example of $\superN=4$ supersymmetric gauge theory,
we have made use of the symmetry algebra and structural properties of
Feynman graphs to derive this generator up to a few quantum loops;
the analysis is purely algebraic, 
no actual (higher-loop) computations were required.
The obtained results have allowed us to prove that 
the planar dilatation operator is 
completely \emph{integrable}, not only at one-loop, 
but possibly even at higher-loops and 
for finite values of the coupling constant.

Apart from generic interest in the structure of field theories
at higher-loops, our investigations are motivated by the 
AdS/CFT correspondence. One prediction of this conjecture
is the agreement of the spectrum of scaling dimensions of local 
operators in $\superN=4$ SYM with the spectrum of energies 
of IIB string theory on $AdS_5\times S^5$.
The dilatation operator has become 
a versatile tool for testing and even proving parts of 
the AdS/CFT correspondence.

There are various ways to deduce scaling dimensions of local operators 
and their quantum corrections
(i.e.~the anomalous dimensions) from field theory correlators. 
They are convenient when interested in particular and
rather simple local operators at low loop orders.
Unfortunately, the AdS/CFT correspondence relates this regime of 
operators to an essentially inaccessible strong-coupling regime 
in string theory. Conversely, the perturbative regime of
string theory is usually mapped to an inaccessible regime in gauge theory.
In recent years, it has emerged that this incompatibility can be 
overcome when focusing on states with a large spin quantum number 
on $S^5$, or equivalently of $\alSO(6)$.
In gauge theory this requires operators with a large number of constituents
in which case the direct approach requires too much work. 
It is therefore desirable to have some technology to deal with such states
in an efficient way.

The dilatation generator is perfectly 
suited to investigate scaling dimensions.
As compared to conventional methods, it offers several advantages:
Once it is obtained from field theory,
the computation of scaling dimensions is turned into
a combinatorial exercise. Furthermore, the dilatation
operator is an algebraic object and one can save
a significant amount of work by simplifying it,
before it is applied to specific states.
Moreover, one can try to obtain the dilatation operator
without performing labourious field theory computations; 
this constitutes one of the key results of the current work.
Finally, whereas scaling dimensions are just a set of data,
the dilatation generator comprises the \emph{whole spectrum};
it allows to compare on an abstract level and thus 
prove the equivalence of certain spectra instead of performing
tests for an (inevitably small) set of states.
For instance, when non-planar corrections are taken into account,
the dilatation generator can split and join traces (alias strings) 
making it very reminiscent of a stringy Hamiltonian.
What is more, in $\superN=4$ SYM in the BMN limit, it 
was shown to coincide also \emph{quantitatively} with 
the plane-wave string field theory Hamiltonian.
This proves the agreement of arbitrary-genus contributions to
scaling dimensions/energy and constitutes a strong test
of the AdS/CFT or plane-wave/BMN correspondence.

Here, we have chosen $\superN=4$ supersymmetric gauge theory
as a model quantum conformal field theory in four 
spacetime dimensions. For this model we have 
first investigated the dilatation operator in the one-loop
approximation. We have started by making the most general ansatz
compatible with field theory, which involves 
infinitely many undetermined coefficients. 
Subsequently, we have used conformal invariance to reduce the independent 
coefficients, first to one infinite sequence, then to just a
single one. Being related to a rescaling of the coupling
constant, it is not possible to assign a value to the leftover 
coefficient except by actually computing it in field theory.
Therefore, the complete one-loop
dilatation operator of $\superN=4$ SYM is entirely fixed
by symmetry (up to obvious transformations).

Similar conclusions hold also at higher-loops:
A general treatment would have required very much work, therefore
we have first investigated subsectors on which the dilatation
operator closes. In an interesting one, the $\alSU(2|3)$ subsector, 
we were able to obtain planar three-loop corrections
by algebraic means. 
Again, the result has turned out to be unique up to symmetries
of the defining relations. 
This enables us to perform a very exact test of AdS/CFT correspondence and
the near plane-wave/BMN limit thereof.
Remarkably, this test has revealed a discrepancy starting 
only at three-loops. 
We have presented a possible explanation 
in terms of an order of limits problem,
but more importantly, it demonstrates that we
can find relevant and unexpected physics even in a 
higher-loop computation!

The dilatation operator is not only useful to obtain
scaling dimensions, but it is interesting in itself.
One exciting feature of the planar dilatation operator,
with very important consequences, is its apparent integrability.
As far as the spectrum is concerned, integrability merely leads to 
a curiosity:
For generic multiplets of local operators there is
a partner with exactly degenerate anomalous dimension. 
Below the surface, however, the existence of
arbitrarily many conserved commuting charges strongly constrains
the dilatation operator. This leads to a major simplification in computing 
scaling dimensions due to the algebraic Bethe ansatz.

The Bethe ansatz is especially powerful in the thermodynamic limit, 
i.e.~for local operators with a large number of constituent fields.
This limit is practically inaccessible by conventional methods,
however, here the Bethe equations turn into integral equations
which can still be solved in practice. 
The AdS/CFT correspondence relates the thermodynamic limit 
of $\superN=4$ SYM to classical spinning string 
configurations on $AdS_5\times S^5$.
The energy is usually given by intricate 
algebraic, elliptic or hyperelliptic functions of 
the ratios of the involved spins. 
In a number of cases, agreement between
gauge theory and string theory could be shown,
which confirms the correspondence with unprecedented accuracy.
Even more, the Bethe equations in the thermodynamic limit 
could be shown to coincide with integral equations derived from
string theory!

Integrability of the planar dilatation operator 
was first found at one-loop and for certain subsectors of states.
More accurately, it was shown that the dilatation operator 
is isomorphic to the Hamiltonian of an integrable quantum spin chain
with $\alSO(6)$ or $\alSL(2)$ symmetry.
In this work we have extended these one-loop results to
the complete spectrum of states
and full $\alPSU(2,2|4)$ superconformal symmetry.
Even more exciting is a generalisation of integrability to higher-loops,
an entirely new topic. 
We have found several indications for integrability beyond one-loop
although a framework to investigate, prove and 
exploit it, is yet to be established. 
Most importantly,
the three-loop corrections to the dilatation operator within the
$\alSU(2|3)$ subsector preserve the above-mentioned degeneracy of pairs. 
Furthermore, the integrable spin chain due to Inozemtsev reproduces 
three-loop planar scaling dimensions correctly. 
Finally, the sigma model of string theory on $AdS_5\times S^5$ is 
integrable and, via the AdS/CFT correspondence, one expects the same
feature for the corresponding gauge theory.

In order to investigate higher-loop integrability, 
we have constructed a deformation of 
the Heisenberg integrable spin chain model.
The assumed form of interactions is inspired by field theory
and conjectures about $\superN=4$ SYM. 
If all our conjectures are justified,
the model will describe planar anomalous dimensions. 
Independently of that question, the model displays some remarkable features:
Our assumptions have turned out to be sufficiently
constraining such that we obtain a unique result up to at least five-loops.
Intriguingly, it reproduces the BMN energy formula correctly.
Moreover, we have found a long-range Bethe ansatz, very similar to
the one describing the Inozemtsev spin chain, which 
reproduces the spectrum of the spin chain exactly.
But this is not all: It is valid for \emph{all loop orders}, at least
if the spin chain is sufficiently long! 
Have we hereby opened up a window for finite coupling constant?

In conclusion, we have presented a versatile technology 
to investigate scaling dimensions in a conformal field theory.
We have completed the one-loop calculation and 
even had a glimpse of higher-loop physics. 
Hopefully, making full use of integrability
will finally allow us to leave the weak coupling regime.

\finishchapter 

\chapternonum{Outlook}

There is a wide range of open questions and problems which can be addressed
with the ideas and methods presented in this work;
we will group them according to the topics presented in 
the individual chapters.
Let us start with \emph{the dilatation operator in general}
which has been considered in \chref{ch:Dila}:
\begin{bulletlist}
\item 
It would be very helpful to develop techniques, similar to the
ones presented here, for the efficient evaluation of structure
constants of the operator product expansion,
see \secref{sec:N4.Corr}.
Apart from the scaling dimensions, 
they are the other central quantity in a conformal field theory 
of local operators. 
The structure constants may be obtained from
three-point functions, but a direct computation is
`contaminated' by useless finite and divergent 
contributions from the perturbative expansion of the scaling dimensions. 
One therefore wonders whether one may
generalise our methodology and develop purely algebraic techniques 
for directly finding the structure constants.
See e.g.~\cite{D'Hoker:2001bq,Okuyama:2004bd,Roiban:2004va} 
for some work in that direction. Especially within the BMN
three-point functions are important for the comparison 
to the plane-wave string theory
\cite{Chu:2003qd,Yoneya:2003mu,Dobashi:2004nm}, 
c.f.~\cite{Chu:2002pd,Beisert:2002bb,Constable:2002vq,Georgiou:2003aa,Chu:2003ji,Georgiou:2004ty}
for some results.

\item
It might be interesting to extend the current analysis to
non-perturbative effects like instantons. 
Possibly the symmetry algebra also puts constraints 
on these and a direct computation as in \cite{Kovacs:2003rt}
might be simplified or even bypassed. 

\end{bulletlist}
%
In \chref{ch:One} we have investigated the 
\emph{dilatation operator at one-loop}:
\begin{bulletlist}
\item
We have focussed on $\superN=4$ SYM in this work,
but there are a few further four-dimensional 
conformal quantum field theories
with $\superN=2$ supersymmetry. 
For these, the determination 
of the dilatation generator might shed some light on
holographic dualities
away from the well-studied case of $AdS_5\times S^5$.
Even if the superconformal algebra is smaller, we 
expect that it is similarly constraining and our
results should generalise in a rather straightforward fashion.
Some advances in this direction have been made in 
e.g.~\cite{Gomis:2003kb,Wang:2003cu}.

\item
Even in a QFT with broken conformal invariance
\cite{Callan:1970yg,Symanzik:1970rt}, the techniques 
developed in this work
can be used to investigate logarithmic corrections 
to two-point functions and scattering amplitudes 
in a systematic way.
In particular in QCD at large $N\indup{c}$ 
and deep inelastic scattering, 
similar techniques are at use 
(see e.g.~\cite{Lipatov:1994yb,Faddeev:1995zg,Braun:1998id,Belitsky:1999qh,Braun:1999te,Belitsky:1999ru,Belitsky:1999bf,Derkachov:1999ze,Belitsky:2003ys,Ferretti:2004ba,Kirch:2004mk,Belitsky:2004cz}).
\end{bulletlist}
%
Questions related to the \emph{dilatation operator at higher-loops} 
in \chref{ch:Higher}:
\begin{bulletlist}
\item
The complete one-loop contribution to the dilatation operator has turned out 
to be totally fixed by superconformal symmetry;
the same might apply at two or even higher-loops.
This conjecture is not unreasonable, as the action is unique
and entirely determined by $\alPSU(2,2|4)$. 
Whether or not the conjecture is true, it would be great 
to derive the complete two-loop deformation.

\item
In \chref{ch:One,ch:Higher}
we have considered the algebras $\alPSU(2,2|4)$ and $\alSU(2|3)$
with the fundamental fields transforming in certain representations.
An interesting mathematical question is, which algebras and 
representations are suited for consistent deformations 
of the algebra generators? Are they all related to 
(conformal) field theories?

\item
A very important issue is the form of 
wrapping interactions.
For those, our methods appear to be 
not very constraining at higher loop orders.
A non-planar treatment might solve the problem,
but the complexity will increase drastically due to
the large amount of such digrams.
For operators of a finite length,
wrappings dominate the
tail of the perturbation series. They are therefore
of tremendous importance for the finite coupling regime. 

\item
It would be nice to confirm some of the higher-loop 
results of this work explicitly in field theory. 
Although we believe our computations are rigorous, 
we had to rely on some basic assumptions
(e.g.~the BMN-limit) which are not firm facts of gauge theory yet.

\item
A curious observation is that all the
anomalous dimensions we have found are solutions to 
algebraic equations. In contrast, higher-loop
amplitudes usually involve transcendental numbers
such as $\zeta(3),\zeta(5),\ldots$\,\,. 
Are these merely a renormalisation artefact or
do they appear at some higher loop order?

A related question is whether the
coupling constant is renormalised by a finite
amount. For instance, this happens in the BMN matrix
model (c.f.~\secref{sec:Dila.Dim.QM}), where 
a redefinition of the coupling constant
is required to achieve a proper scaling in
the BMN limit \cite{Klose:2003qc} 
(it is a non-trivial result that 
this is possible at all).

\end{bulletlist}
%
\emph{One-loop integrability} was the subject of \chref{ch:Int}:
\begin{bulletlist}
\item
Clearly, the deep question is, \emph{why} does integrability 
emerge from the planar $\superN=4$ gauge theory?
Of course, via the the AdS/CFT correspondence,
one could take integrability of the classical string sigma model 
\cite{Mandal:2002fs,Bena:2003wd,Vallilo:2003nx} 
as compelling evidence, nevertheless we believe there should
also be an intrinsically field theoretical reason.

\item
We know that the spectrum 
of $\charge_2$ is related to the spectrum of anomalous dimensions. 
A natural question to ask is whether
the spectra of the higher charges $\charge_r$ 
have a physical meaning in the gauge theory. 

\item
Integrability implies the appearance of degenerate pairs 
in the planar spectrum due to an interplay of
two charges and parity. 
However, there appears to
be no deeper reason for the pairs in terms
of representation theory. 
It would be very desirable to understand the degeneracy of pairs
better, in terms of $\superN=4$ SYM as well as in terms
of the AdS/CFT correspondence. 
A possible explanation would be that symmetry extends 
by some compact $\grSO(2)$ group whose representations are either
uncharged singlets or charged doublets.
The integrable charges $\charge_r$ are not 
suitable $\grSO(2)$ generators because
their spectrum is non-integer.
The $\grSO(2)$ symmetry naturally combines with
the parity $\Integers_2$ to $\grO(2)$
and $1/N$ corrections break it to $\Integers_2$. 

Could the conjectured $\grSO(2)$ symmetry
be related to the conjectured modular invariance
of $\superN=4$ SYM? 
In the unbroken form, $\grSL(2,\Real)$, the
$\grSO(2)$ subgroup of modular invariance
would pair up states. When broken to
$\grSL(2,\Integers)$ by higher-genus corrections,
there is no $\grSO(2)$ group to protect the
pairing and the degeneracy is lifted. 

\item
Integrability seems to apply to a wider range of 
field theories. 
The obvious candidates are conformal field theories,
see e.g.~\cite{Chen:2004yf}.
However, in theories where conformal invariance is broken 
by quantum effects, such as massless QCD, 
one may still investigate the one-loop dilatation operator,
for which conformal symmetry applies, 
see e.g.~\cite{Wang:2003cu,Belitsky:2004yg,Berenstein:2004ys,DiVecchia:2004jw}.
In QCD, following pioneering work of Lipatov \cite{Lipatov:1994yb}, 
methods of integrability have also had much impact,
see e.g.~\cite{Faddeev:1995zg,Braun:1998id,Belitsky:1999qh,Braun:1999te,Belitsky:1999ru,Belitsky:1999bf,Derkachov:1999ze,Belitsky:2003ys,Ferretti:2004ba,Kirch:2004mk,Belitsky:2004cz}.

\item
On the one hand, we have demonstrated in 
\secref{sec:One.Lift} that the superconformal 
algebra leads to a unique one-loop anomalous dilatation operator.
On the other hand, there is a unique standard
spin chain Hamiltonian with $\alPSU(2,2|4)$ symmetry. 
In fact, both operators turn out to be the same, which shows that
conformal symmetry and integrability go hand in hand.
This is remarkable because we consider a four-dimensional field
theory here. For a two-dimensional theory this relationship 
is well-understood.
For more details, see the end of \secref{sec:Int.Bethe.Splitting}.

\end{bulletlist}
%
Questions related to \emph{higher-loop integrability} in \chref{ch:HighInt}:
\begin{bulletlist}
\item
An improved notion of higher-loop integrability and,
even more urgently,
a better understanding of the long-range spin chain in \chref{ch:HighInt}
or the dynamic spin chain in \chref{ch:Higher}
is required.

\item
Can we find a Bethe ansatz for
the dynamic spin chain in \chref{ch:Higher}?
If so, can we generalise it to the complete
spin chain for $\superN=4$ SYM?
This presumably would be a non-compact, dynamic $\alPSU(2,2|4)$ 
super spin chain with long-range interactions.
See also \secref{sec:HighInt.Ansatz.Others} for further comments.

\item
The Bethe ans\"atze for the Inozemtsev spin chain 
and our long-range spin chain investigated in \secref{sec:HighInt.Ansatz}
apparently do not incorporate wrapping interactions
(see \secref{sec:Higher.Two.Short,sec:HighInt.Spinning.Wrappings}). 
A key to unravel planar $\superN=4$ gauge theory at all loops
would be to modify the equations to account for wrappings.
Unfortunately, it is hard to find the correct 
wrapping interactions in field theory.
Here, the investigation of the unknown terms 
in the physical transfer matrix \eqref{eq:HighInt.Ansatz.Transfer}
(from which the Bethe equations should follow as a consistency condition)
or a better understanding of the double covering map 
\eqref{eq:HighInt.Ansatz.DoubleCover,eq:HighInt.Ansatz.TransferDoubleCover}
might help.

\item
Is integrability related to the closure of the interacting algebra
or to some property of field theory renormalisation?
Can we \emph{prove} higher-loop integrability in some way?

\item
An interesting model is the BMN matrix model.
It behaves quite similarly to $\superN=4$ SYM, but
consists of only a finite number of fields.
Explicit higher-loop computations are feasible
\cite{Kim:2003rz,Klose:2003qc}, and show partial 
agreement with $\superN=4$ SYM.
Here we might learn about wrapping-interactions and
dynamic aspects explicitly.

\end{bulletlist}
%
Questions regarding \emph{the AdS/CFT correspondence} in
the context of the plane-wave/BMN correspondence and
spinning strings:
\begin{bulletlist}
\item
The plane-wave/BMN correspondence and 
the topic of spinning strings can be investigated
for theories with less supersymmetry. 
They are especially interesting because they involve also
open strings/traces as opposed to the maximally supersymmetric
case with only closed strings.
Many investigations are devoted to this topic, 
e.g.~\cite{Imamura:2002wz,Skenderis:2002wx,Niarchos:2002fc,Skenderis:2002ps,Stefanski:2003zc,Gomis:2003kb,Schvellinger:2003vz,Lucietti:2003ki,Stefanski:2003qr,DeWolfe:2004zt,Chen:2004yf}, 
but so far there are no higher-loop results for gauge theory.

\item
The equivalence of the dilatation operator in the BMN limit and
the plane-wave string field theory Hamiltonian has been shown 
at one-loop in the case of two 
\cite{Spradlin:2003bw}
and three \cite{Gutjahr:2004qj} excitations (impurities).
One could try to prove the equivalence for an arbitrary number of 
excitations. 
A generalisation to higher-loops would be interesting:
At two-loops one might have to consider a
$g$-dependent change of basis to avoid the
mismatch reported in \cite{Spradlin:2003bw}. 
At three-loops it would be exciting to see if problems 
of the kind encountered for the near plane-wave background
\cite{Callan:2003xr,Callan:2004uv} or for spinning strings
\cite{Serban:2004jf} also exist for non-planar corrections.

\item
In \secref{sec:HighInt.Stringing} we have offered a 
possible explanation for 
the apparent disagreement
of gauge theory and string theory in the case of 
near plane-waves \cite{Callan:2003xr,Callan:2004uv} or 
spinning strings \cite{Serban:2004jf}. 
The discrepancy, however, turns out to starts merely at three loops;
why do we find agreement at one loop and two loops?
At lower loop orders, the structure might allow only for 
a unique answer on either side.
Together with the structural equivalence of both models,
one being a lattice discretisation of the other, 
the agreement may be explained.

\end{bulletlist}

\finishchapter 

\docphysrept{\chapternonum{Acknowledgements}

Here, I would like to express my gratitude to
my advisor, Matthias Staudacher, for all the support,
encouragement, advice, time,
stimulating discussions, disputes, chocolates 
and fantastic collaboration 
throughout the last two years.
Without his keen sense of interesting physical problems,
this work would not be the same.
\begin{docnophysrept}%
Also, I thank him for being the sociable person he is,
this made my time in Potsdam as well as on joint travels 
both exciting and pleasant.
\end{docnophysrept}%

I have also benefitted very much from everyday discussions
with Gleb Arutyunov, Charlotte Kristjansen and Jan Plefka
and thank them for sharing their knowledge on various aspects of physics.
\begin{docnophysrept}%
Especially, I will never forget the afternoons spent 
in Matthias's office drinking tea and discussing the latest 
discoveries with each other.
\end{docnophysrept}%

I am grateful to
Massimo Bianchi,
Virginia Dippel,
Sergey Frolov,
Charlotte Kristjansen, 
Joe Minahan, 
Francisco Morales, 
Jan Plefka, 
Henning Samtleben,
Gordon Semenoff,
Matthias Staudacher,
Arkady Tseytlin and
Kostya Zarembo
for the fruitful collaboration 
on the projects on which 
this \docphysrept[thesis]{work} is based in part.

I thank Virginia Dippel, Markus P\"ossel, Matthias Staudacher and
Arkady Tseytlin for proof-reading the manuscript and 
suggesting valuable improvements 
to make your reading of this text more enjoyable. 
\begin{docdophysrept}%
I also thank Curtis Callan, Hermann Nicolai and Jan Plefka
for spending a substantial amount of time on carefully reading 
this manuscript from the beginning to the end.
\end{docdophysrept}%
\begin{docnophysrept}%
I also thank the referees of the dissertation, 
Curtis Callan, Hermann Nicolai and Jan Plefka, for their time.
\end{docnophysrept}%
Again, thanks are due to Matthias
for all the help while preparing earlier articles.

For interesting discussions of physics and otherwise, 
it is a pleasure to thank
Vladimir Bazhanov, 
Andrei Belitsky,
David Berenstein, 
Vladimir Braun, 
Chong-Sun Chu,
Neil Constable, 
Luise Dolan, 
Thomas Fischbacher,
Dan Freedman,
Jaume Gomis,
Alexander Gorsky, 
Christine Gottschalkson,
Petra Gutjahr,
Paul Howe,
Hikaru Kawai,
Volodya Kazakov, 
Nakwoo Kim,
Axel Kleinschmidt,
Thomas Klose,
Gregory Korchemsky,
David Kosower,
Stefano Kovacs,
Juan Maldacena,
Shiraz Minwalla,
Chiara Nappi,
Ari Pan\-kie\-wicz,
Christina Pappa,
Kasper Peeters,
Markus P\"ossel, 
Thomas Quella, 
Rodolfo Russo,
Anton Ryzhov,
Kazuhiro Sakai,
Dierk Schleicher,
John Schwarz, 
Didina Serban, 
Corneliu Sochichiu,
Emery Sokatchev, 
Marcus Spradlin,
Bogdan Stefa\'nski,
Ian Swanson,
Diana Vaman, 
Herman Verlinde,
Edward Witten, 
Marija Zamaklar and
fellow physicists I have met at seminars, schools and conferences
as well as everybody I have unfortunately missed.

I would like to thank my office-mates,
colleagues and members of staff at the Institute 
for the pleasant, inspiring atmosphere 
and for making my work so enjoyable. 
In particular, I acknowledge the prompt
help by the computer support 
when my laptop went to the happy hunting grounds 
after four and a half years of loyal service around the world. 

Ich m\"ochte mich ganz herzlich bei Anna, Thomas, Maissi, Miriam
und meinen Freunden
f\"ur ihre fort\-w\"ahrende Unterst\"utzung bedanken, ohne
die diese Arbeit in der jetzigen Form nicht m\"oglich gewesen w\"are.

Schlie{\ss}lich bedanke ich mich bei
der \emph{Studien\-stif\-tung des deut\-schen Vol\-kes} 
f\"ur die Unter\-st\"utzung durch ein 
Promotions\-f\"orderungs\-stipendium, sowie auch bei
mei\-nem Vertrauensdozenten Prof.~Dr.~Harald Uhlig 
und Mit-Stiftis f\"ur ihr Engagement. 
\begin{docnophysrept}%
Die Sommerakademie 2003 in St.~Johann hat sehr zur Erweiterung meiner 
Horizonte beigetragen.
Ebenso hoffe ich auf dem nun folgenden Sprachkurs 
in Amboise nicht nur meine Fran\-z\"o\-sisch\-kenntnisse verbessern zu 
k\"onnen, sondern mich auch von den Strapazen 
des Anfertigens dieser Arbeit zu erholen.
\end{docnophysrept}%
\bigskip

\hfill\emph{\large Thank You!}

\finishchapter 
}

\appendix
\chapter{An Example}
\label{app:Example}

Here we would like to demonstrate how to apply the dilatation operator
step-by-step in order to introduce our notation. 
In the following two sections we will present some essential matrix 
model and spin chain technology.

\section{Non-Planar Application}
\label{app:Example.NonPlanar}

Consider two $N\times N$ matrices $\fldZ$ and $\phi$.
Their elements are given by the variables $\fldZ^{\gaugeind{a}}{}_{\gaugeind{b}}$ and 
$\phi^{\gaugeind{a}}{}_{\gaugeind{b}}$ with indices 
$\gaugeind{a},\gaugeind{b},\ldots$ ranging from $1$ to $N$. 
We would now like to write down a polynomial $\Op=\Op(\fldZ,\phi)$ in the elements of 
$\fldZ$ and $\phi$ that is invariant under similarity transformations
$\fldZ\mapsto T\fldZ T^{-1}, \phi\mapsto T\phi T^{-1}$.
This is achieved conveniently by taking traces of matrices such as 
\[\label{eq:Example.NonPlanar.Op}
\Op(\fldZ,\phi) = \Tr \fldZ\phi\fldZ\phi-\Tr \fldZ\fldZ\phi \phi
= \sum_{\gaugeind{a},\gaugeind{b},\gaugeind{c},\gaugeind{d}=1}^N
\bigbrk{
\fldZ^{\gaugeind{a}}{}_{\gaugeind{b}}\,
\phi^{\gaugeind{b}}{}_{\gaugeind{c}}\,
\fldZ^{\gaugeind{c}}{}_{\gaugeind{d}}\,
\phi^{\gaugeind{d}}{}_{\gaugeind{a}}
-\fldZ^{\gaugeind{a}}{}_{\gaugeind{b}}\,
\fldZ^{\gaugeind{b}}{}_{\gaugeind{c}}\,
\phi^{\gaugeind{c}}{}_{\gaugeind{d}}\,
\phi^{\gaugeind{d}}{}_{\gaugeind{a}}
}.
\]
We now introduce a differential operator $\ham$ 
on polynomials of the matrix elements
\[\label{eq:Example.NonPlanar.Ham}
\ham=-N^{-1}\Tr \comm{\fldZ}{\phi}\comm{\check\fldZ}{\check\phi},
\]
where the derivatives $\check \fldZ$ and $\check \phi$ are defined as follows%
\footnote{In the language of canonical 
quantisation, the fields $\fldZ,\phi$ and the 
variations $\check\fldZ,\check\phi$ correspond 
to creation and annihilation operators, respectively.}
\[\label{eq:Example.NonPlanar.Der}
\check \fldZ^{\gaugeind{a}}{}_{\gaugeind{b}}=
\frac{\partial}{\partial \fldZ^{\gaugeind{b}}{}_{\gaugeind{a}}}\,,
\qquad
\check \phi^{\gaugeind{a}}{}_{\gaugeind{b}}=
\frac{\partial}{\partial \phi^{\gaugeind{b}}{}_{\gaugeind{a}}}\,.
\]
Let us act with $\ham$ on $\Op$. In this elementary form it is quite
tedious, so let us restrict to the first terms
\<\label{eq:Example.NonPlanar.ActLong}
\Tr \fldZ\phi\check\fldZ\check\phi\, \Tr\fldZ\phi\fldZ \phi\eq
\smash{\sum_{\gaugeind{a},\gaugeind{b},\gaugeind{c},\gaugeind{d},
\gaugeind{e},\gaugeind{f},\gaugeind{g},\gaugeind{h}=1}^N}
\smash{\Big[}\quad\,\,\,
\fldZ^{\gaugeind{a}}{}_{\gaugeind{b}}\,
\phi^{\gaugeind{b}}{}_{\gaugeind{c}}\,
\delta^{\gaugeind{c}}{}_{\gaugeind{f}}\,
\delta^{\gaugeind{d}}{}_{\gaugeind{g}}\quad
\delta^{\gaugeind{e}}{}_{\gaugeind{d}}\,
\delta^{\gaugeind{f}}{}_{\gaugeind{a}}\,
\fldZ^{\gaugeind{g}}{}_{\gaugeind{h}}\,
\phi^{\gaugeind{h}}{}_{\gaugeind{e}}
\nl\qquad\qquad\qquad
+\fldZ^{\gaugeind{a}}{}_{\gaugeind{b}}\,
\phi^{\gaugeind{b}}{}_{\gaugeind{c}}\,
\delta^{\gaugeind{c}}{}_{\gaugeind{f}}\,
\delta^{\gaugeind{d}}{}_{\gaugeind{e}}\quad
\delta^{\gaugeind{e}}{}_{\gaugeind{d}}\,
\phi^{\gaugeind{f}}{}_{\gaugeind{g}}\,
\fldZ^{\gaugeind{g}}{}_{\gaugeind{h}}\,
\delta^{\gaugeind{h}}{}_{\gaugeind{a}}
\nl\qquad\qquad\qquad
+\fldZ^{\gaugeind{a}}{}_{\gaugeind{b}}\,
\phi^{\gaugeind{b}}{}_{\gaugeind{c}}\,
\delta^{\gaugeind{c}}{}_{\gaugeind{h}}\,
\delta^{\gaugeind{d}}{}_{\gaugeind{g}}\quad
\fldZ^{\gaugeind{e}}{}_{\gaugeind{f}}\,
\delta^{\gaugeind{f}}{}_{\gaugeind{a}}\,
\delta^{\gaugeind{g}}{}_{\gaugeind{d}}\,
\phi^{\gaugeind{h}}{}_{\gaugeind{e}}
\nl\qquad\qquad\qquad
+\fldZ^{\gaugeind{a}}{}_{\gaugeind{b}}\,
\phi^{\gaugeind{b}}{}_{\gaugeind{c}}\,
\delta^{\gaugeind{c}}{}_{\gaugeind{h}}\,
\delta^{\gaugeind{d}}{}_{\gaugeind{e}}\quad
\fldZ^{\gaugeind{e}}{}_{\gaugeind{f}}\,
\phi^{\gaugeind{f}}{}_{\gaugeind{g}}\,
\delta^{\gaugeind{g}}{}_{\gaugeind{d}}\,
\delta^{\gaugeind{h}}{}_{\gaugeind{a}}
\quad\smash{\Big]}
\nln\eq
\sum_{\gaugeind{a},\gaugeind{b},
\gaugeind{e},\gaugeind{f}=1}^N
\Bigbrk{
2\fldZ^{\gaugeind{a}}{}_{\gaugeind{b}}\,
\phi^{\gaugeind{b}}{}_{\gaugeind{a}}\,
\fldZ^{\gaugeind{e}}{}_{\gaugeind{f}}\,
\phi^{\gaugeind{f}}{}_{\gaugeind{e}}
+2N\fldZ^{\gaugeind{a}}{}_{\gaugeind{b}}\,
\phi^{\gaugeind{b}}{}_{\gaugeind{f}}\,
\phi^{\gaugeind{f}}{}_{\gaugeind{e}}\,
\fldZ^{\gaugeind{e}}{}_{\gaugeind{a}}
}
\nln\eq
2\Tr \fldZ\phi\,\,\Tr\fldZ\phi
+2N\Tr \fldZ\phi\phi\fldZ.
\>
This calculation can be significantly abbreviated by parameterising the matrices by 
$\grU(N)$ generators $\gaugegen{m}$ and using the fusion and fission rules
\[\label{eq:Example.NonPlanar.FuFi}
\gaugemet{mn}
\Tr X\gaugegen{m}\Tr Y\gaugegen{n}
=
\Tr XY,\qquad
\gaugemet{mn}
\Tr X\gaugegen{m} Y\gaugegen{n}
=
\Tr X\Tr Y.
\]
The action \eqref{eq:Example.NonPlanar.ActLong} is now
\<\label{eq:Example.ActShort}
\Tr \fldZ\phi\check\fldZ\check\phi\, \Tr\fldZ\phi\fldZ \phi\eq
2\gaugemet{mp}\gaugemet{nq}\Tr \fldZ\phi\,\gaugegen{m}\gaugegen{n}\,\Tr \gaugegen{p}\gaugegen{q}\fldZ \phi
+2\gaugemet{mp}\gaugemet{nq}\Tr \fldZ\phi\,\gaugegen{m}\gaugegen{n}\,\Tr \gaugegen{p}\phi\fldZ\, \gaugegen{q}
\nln\eq
2\gaugemet{nq}\Tr \fldZ\phi\,\gaugegen{q}\fldZ \phi\,\gaugegen{n}
+2\gaugemet{nq}\Tr \fldZ\phi\phi\fldZ\, \gaugegen{q}\gaugegen{n}
\nln\eq
2\Tr \fldZ\phi\Tr\fldZ \phi
+2\Tr \fldZ\phi\phi\fldZ \Tr 1.
\>
Summing up all contributions in $\ham\,\Op$ we get
\[\label{eq:Example.NonPlanar.ActEigen}
\ham\,\Op=
6\Tr \fldZ\phi\fldZ\phi-6\Tr \fldZ\fldZ\phi \phi=E\, \Op,\qquad E=6.
\]

Now it is time to interpret our calculations in terms of physics.
The polynomial $\Op$ is a gauge invariant local operator (state)
and $\ham=\algD_2$ is the one-loop dilatation operator. We have thus found
that $\Op$ is an eigenstate of $\ham$ with energy is $6$; its 
anomalous dimension therefore $6g^2$. 
We note our definition of coupling constant $g$ in terms
of the ordinary Yang-Mills coupling constant $\gym$ and rank $N$ of 
the $\grU(N)$ gauge group
\[\label{eq:Example.NonPlanar.Coupling}
g^2=\frac{\gym^2 N}{8\pi^2}=\frac{\lambda}{8\pi^2}\,.
\]
The classical dimension of $\Op$ is computed using the 
operator $\algD_0$
\[\label{eq:Example.NonPlanar.Classical}
\algD_0=\Tr \fldZ\check \fldZ+\Tr \phi\check \phi,\qquad
\algD_0\,\Op =4\, \Op,
\]
here it just counts the number of constituent fields.
In conclusion, the scaling dimension of $\Op$ up to one-loop is given by
\[\label{eq:Example.NonPlanar.Dim}
D=4+6g^2+\ldots=4+\frac{3\gym^2 N}{4\pi^2}\,.
\]
The state $\Op$ is a descendant of the Konishi operator
$\Tr \Phi_m\Phi_m$, see e.g.~\secref{sec:Dila.SU2.Apply}.

\section{Planar Application}
\label{app:Example.Planar}

Let us repeat the example in the planar limit.
Consider basis states of the type $\state{0010110\ldots10}$.  
The labels are identified cyclically, e.g.~$\state{00010110\ldots1}$ 
represents the same state.
A generic state $\Op$ is a linear combination of these, for instance
\[\label{eq:Example.Planar.State}
\Op=\state{0101}-\state{0011}.
\]
Now consider a linear operator $\ham$ on the space of states
\[\label{eq:Example.Planar.Ham}
\ham=\sum_{p=1}^L \ham_{p,p+1},\qquad \ham_{12}=1-\fldperm_{12}.
\]
The operator acts on all pairs of adjacent labels (enumerated by $p$) 
within a state.
For each pair, it returns the same state, $1$, minus the state with both
labels interchanged, $\fldperm_{p,p+1}$. Note that $\ham_{L,L+1}$ is to be interpreted as
$\ham_{L,1}$ due to the cyclic nature of states. 
Furthermore, it suffices to give $\ham_{12}$ acting on the first two labels.
The action of $\ham_{p,p+1}$ on the other labels is equivalent.
For example
\[\label{eq:Example.Planar.Act}
\ham_{12}\state{0101}=\state{0101}-\state{1001}.
\]
In total we get for $\ham\,\Op$
\<\label{eq:Example.Planar.Eigen}
\ham \Op\eq
\bigbrk{\ham_{12}+\ham_{23}+\ham_{34}+\ham_{41}}
\state{0101}
-\bigbrk{\ham_{12}+\ham_{23}+\ham_{34}+\ham_{41}}
\state{0011}
\nln\eq
+\,\state{0101}
-\state{1001}
+\state{0101}
-\state{0011}
+\state{0101}
-\state{0110}
+\state{0101}
-\state{1100}
\nl
-\state{0011}
+\state{0011}
-\state{0011}
+\state{0101}
-\state{0011}
+\state{0011}
-\state{0011}
+\state{1010}
\nln\eq
+6\state{0101}
-6\state{0011}
=6\,\Op.
\>
The physical interpretation is as in the previous section.
The major difference is that 
double-trace terms which arise in a non-planar computation, 
c.f.~\eqref{eq:Example.NonPlanar.ActLong},
are discarded right away in the planar limit,
c.f.~\eqref{eq:Example.Planar.Act}.
In this particular example, the non-planar terms cancel in the end
and therefore the planar approximation happens to be exact.

\finishchapter 

\chapter{Spinors in Various Dimensions}
\label{app:Spinors}

In this appendix we present a selection of useful identities 
when dealing with chiral spinors in four, six and ten dimensions.

\section{Four Dimensions}
\label{app:Spinors.Four}

There are two types of spinor indices, 
$\alpha=1,2$ and $\dot\alpha=1,2$ belonging
to the two $\alSU(2)$ factors of $\alSO(4)$.
There are two types of invariant objects, 
$\varepsilon$ and $\sigma$. 
There are four types of totally antisymmetric tensors 
$\varepsilon_{\alpha\beta},\varepsilon^{\alpha\beta},
\varepsilon_{\dot\alpha\dot\beta},
\varepsilon^{\dot\alpha\dot\beta}$ and it is convenient 
to use four types of sigma symbols (Pauli matrices)
$\sigma^\mu_{\dot\alpha\beta},\sigma^\mu_{\beta\dot\alpha},
\sigma_\mu^{\dot\alpha\beta},\sigma_\mu^{\beta\dot\alpha}$. 
We can now suppress spinor indices and use a matrix notation,
in all cases it should be clear which symbol to use.
The $\sigma$'s are defined by the relation
\[\label{eq:Spinors.Four.Cliff}
\sigma_{\{\mu}\sigma_{\nu\}}=\eta_{\mu\nu}.
\]
The different ordering of spinor indices was
introduced artificially, we remove it by the identification
\[\label{eq:Spinors.Four.Sym}
\sigma^\mu_{\dot\alpha\beta}=
\sigma^\mu_{\beta\dot\alpha},
\qquad
\sigma_\mu^{\dot\alpha\beta}=
\sigma_\mu^{\beta\dot\alpha}.
\]
Here are some identities involving $\varepsilon$'s in matrix notation
\[\label{eq:Spinors.Four.Eps}
\varepsilon^\trans=-\varepsilon,\qquad
\varepsilon\varepsilon=-1,\qquad
\varepsilon\sigma_{\mu}=\sigma_{\mu}\varepsilon.
\]
The Fierz identities for the $\sigma$'s read 
\[\label{eq:Spinors.Four.Fierz}
\sigma^{\dot\alpha\beta}_{\mu}\sigma_{\dot\gamma\delta}^{\mu}
=2\delta^{\dot\alpha}_{\dot\gamma}\delta^{\beta}_{\delta},
\qquad
\sigma^{\dot\alpha\beta}_{\mu}\sigma^{\mu,\dot\gamma\delta}
=2\varepsilon^{\dot\alpha\dot\gamma}\varepsilon^{\beta\delta},
\qquad
\sigma_{\mu,\dot\alpha\beta}\sigma^\mu_{\dot\gamma\delta}
=2\varepsilon_{\dot\alpha\dot\gamma}\varepsilon_{\beta\delta}
\]
and the completeness relation for antisymmetric tensors is
\[\label{eq:Spinors.Four.EpsIdent}
\varepsilon_{\alpha\beta}
\varepsilon^{\gamma\delta}
=\delta_\alpha^\gamma\delta_\beta^\delta
-\delta_\alpha^\delta\delta_\beta^\gamma
=2\delta_{[\alpha}^{\gamma}\delta_{\beta]}^{\delta}.
\]
%

\section{Six Dimensions}
\label{app:Spinors.Six}

In six dimensions there is one type of spinor index $a=1,2,3,4$, 
two totally antisymmetric tensors
$\varepsilon_{abcd},\varepsilon^{abcd}$ and
two sigma symbols $\sigma^m_{ab}$ and $\sigma_m^{ab}$.
Again we can suppress spinor indices in a matrix notation.
The sigma symbols are antisymmetric
\[\label{eq:Spinors.Six.Sym}
\sigma_m^\trans=-\sigma_m
\]
and its indices can be raised or lowered by the
totally antisymmetric tensor
\[\label{eq:Spinors.Six.Conj}
\sigma^{m,ab}=
\half\varepsilon^{abcd}\sigma^m_{cd},\qquad
\sigma_{m,ab}=
\half\varepsilon_{abcd}\sigma_m^{cd}.
\]
They satisfy the Clifford algebra
\[\label{eq:Spinors.Six.Cliff}
\sigma_{\{m}\sigma_{n\}}=\eta_{mn}.
\]
Finally, we note the Fierz identities for the $\sigma's$
\[\label{eq:Spinors.Six.Fierz}
\sigma_m^{ab}\sigma^m_{cd}
=2\delta^a_d\delta^b_c-2\delta^a_c\delta^b_d,
\qquad
\sigma_m^{ab}\sigma^{m,cd}=-2\varepsilon^{abcd},
\qquad
\sigma_{m,ab}\sigma^m_{cd}=-2\varepsilon_{abcd}.
\]
%

\section{Ten Dimensions}
\label{app:Spinors.Ten}

We will denote spinor indices in ten dimensions by $A,B,\ldots=1,\ldots, 16$.
There are two sigma symbols $\Sigma^M_{AB}$ and $\Sigma_M^{AB}$
and we can suppress spinor indices.
The sigma symbols are symmetric
\[\label{eq:Spinors.Ten.Sym}
\Sigma_M^\trans=\Sigma_M
\]
and satisfy
\[\label{eq:Spinors.Ten.Cliff}
\Sigma_{\{M}\Sigma_{N\}}=\eta_{MN}.
\]
For the construction of supersymmetric gauge theory,
there is one particularly useful identity
\[\label{eq:Spinors.Ten.Ident}
\Sigma_{M,AB}\Sigma^M_{CD}
+\Sigma_{M,AC}\Sigma^M_{DB}
+\Sigma_{M,AD}\Sigma^M_{BC}=0.
\]

In order to obtain $\superN=4$ SYM from the ten-dimensional
supersymmetric gauge theory we reduce the ten-dimensional spacetime
to four spacetime and six internal dimensions.
We will assume that a spinor $\Psi^A$ in ten dimensions
decomposes into $\Psi_{\alpha a}+\Psi_{\dot\alpha}^a$ 
in $4+6$ dimensions. Accordingly, the
sigma symbols in ten dimensions split as follows
\<\label{eq:Spinors.Ten.Reduce}
\Sigma^{AB}_\mu\eq 
\sigma_{\mu,\alpha\dot\beta}\delta_a^b
+\sigma_{\mu,\dot\alpha\beta}\delta^b_a,
\nln
\Sigma_{\mu,AB}\eq
\sigma_\mu^{\alpha\dot\beta}\delta_b^a
+\sigma_\mu^{\dot\alpha\beta}\delta_a^b,
\nln
\Sigma_{m}^{AB}\eq
-\sigma_{m,ab}\varepsilon_{\alpha\beta}
-\sigma_m^{ab}\varepsilon_{\dot\alpha\dot\beta},
\nln
\Sigma_{m,AB}\eq
\sigma_m^{ab}\varepsilon^{\alpha\beta}
+\sigma_{m,ab}\varepsilon^{\dot\alpha\dot\beta}.
\>

\finishchapter 

\chapter{SYM in Ten Dimensions}
\label{app:Ten}

Four-dimensional maximally supersymmetric
Yang-Mills theory is most conveniently derived from 
its ten-dimensional supersymmetric counterpart. 
We will therefore present the ten-dimensional theory,
in superspace or in components, in the following two
sections. In this work we have not made use of these theories,
except maybe for the dimensional reduction scheme, which relies on the
component theory. 

\section{Ten-Dimensional Gauge Theory in Superspace}
\label{app:Ten.Super}

Let us first consider $\superN=1$ gauge theory in ten-dimensional superspace
\cite{Gates:1987is,Sohnius:1978wk,Gates:1980jv,Gates:1980wg}.
Superspace is parameterised by 
ten bosonic coordinates $X^M$ and 
sixteen fermionic coordinates $\Theta^A$. 
Here, indices $M,N,\ldots$ refer to ten-component vectors
and indices $A,B,\ldots$ to sixteen-component spinors of $\alSO(10)$.
The covariant derivatives on this space are defined as
\[\label{eq:Ten.Super.Derivative}
D_M=\partial_M,\qquad
D_A=\partial_A+\Sigma^M_{AB}\Theta^B \partial_M.
\]
The fermionic derivatives satisfy the anticommutation relation
\[\label{eq:Ten.Super.Algebra}
\acomm{D_A}{D_B}=2\Sigma^M_{AB}D_M,
\]
while commutators with a bosonic derivative $D_M$ vanish.%
\footnote{This space may also be 
considered as the quotient space $G/H$ 
of the super Poincar\'e group $G$ as defined by 
\eqref{eq:Ten.Super.Derivative} and the Lorentz group $H$.}
In other words, superspace is a space with torsion given by $\Sigma^M_{AB}$. 
The matrices $\Sigma^M$ are the chiral projections of the
gamma matrices of $\alSO(10)$.
They are symmetric $\Sigma^M_{AB}=\Sigma^M_{BA}$ and obey 
\[\label{eq:Ten.Super.Sigma}
\Sigma^{\{M}_{AB}\Sigma^{N\},BC}_{\vphantom{B}}=\delta_A^C \delta^{MN}.
\]
In \appref{app:Spinors.Ten} we present our notation 
and some useful identities.

To have real fermionic coordinates $\Theta^A$, the signature 
of spacetime must be either $(9,1)$ or $(5,5)$. 
Whereas Minkowski space $(9,1)$ is certainly the correct choice 
in terms of physics, it may be more useful to work in 
Euclidean space $(10,0)$ when computing Feynman diagrams.
In fact, the actual signature of spacetime does not matter 
for all the algebraic considerations in this work as we can
do Wick rotations at any point. 
We will therefore not distinguish between different signatures
of spacetime and algebras.
Alternatively, one may consider a complexified space/algebra where 
the signature is irrelevant.

On this space we define a gauge theory with the supercovariant
derivatives
\[\label{eq:Ten.Super.Covariant}
\cder_M=D_M-ig\fldA_M,\qquad\cder_A=D_A-ig\fldA_A.
\]
We will assume the gauge group to be $\grSU(N)$ or $\grU(N)$
and all adjoint fields $\fldA$ to be (traceless) Hermitian $N\times N$ matrices.
Under a gauge transformation $U(X,\Theta)\in\grU(N)$ the gauge fields transform
canonically according to 
\[\label{eq:Ten.Super.GaugeTrans}
\fldA_M\mapsto U\fldA_M U^{-1}-ig^{-1}\, D_M U \,U^{-1},\qquad
\fldA_A\mapsto U\fldA_A U^{-1}-ig^{-1}\, D_A U \,U^{-1}.
\] 
The covariant field strengths of the gauge field are
\<\label{eq:Ten.Super.FieldStrength}
\acomm{\cder_A}{\cder_B}\eq 2\Sigma^M_{AB}\cder_M-ig\fldF_{AB},\nln
\comm{\cder_A}{\cder_M}\eq -ig\fldF_{AM},\nln
\comm{\cder_M}{\cder_N}\eq -ig\fldF_{MN}.
\>
%
%
We can now impose a constraint on the gauge field, namely that the 
field strength $\fldF_{AB}$ vanishes
\<\label{eq:Ten.Super.Constraint}
\fldF_{AB}=0.
\>
This field strength can be decomposed 
into two $\alSO(10)$ irreducible modules, $\rep{10}$ and $\rep{126}$.
The vanishing of the $\rep{10}$ part determines the bosonic gauge field
$\fldA_M$ in terms of the fermionic one. The $\rep{126}$ part
has much more drastic consequences as it 
not only reduces the number of independent components, but also
implies equations of motion for the gauge field.
Before stating these, we 
present two important consequences of the constraint
and the Jacobi identities
\<\label{eq:Ten.Super.SuperSymmetry}
\comm{\cder_A}{\cder_M}\eq -ig\fldF_{AM}=ig\Sigma_{M,AB}\Psi^B,
\nln
\acomm{\cder_A}{\Psi^B}\eq
\sfrac{i}{2}g^{-1}\Sigma^{M,BC}\Sigma^N_{CA}[\cder_M,\cder_N]
=\sfrac{1}{2}\Sigma^{M,BC}\Sigma^N_{CA}\fldF_{MN}.
\>
The first shows that the $\rep{144}$ part of the 
field strength $\fldF_{AM}$ is zero, it 
can be proved by using the Jacobi identity and 
inserting the constraint.
The second one can be proved by projecting on
the $\rep{1}$, $\rep{45}$, $\rep{210}$ parts and using 
the Jacobi identity and constraint. 
The equations of motion which follow from 
\eqref{eq:Ten.Super.Constraint}
in much the same way as 
\eqref{eq:Ten.Super.SuperSymmetry}
are
\<\label{eq:Ten.Super.EOM}
\comm{\cder_{N}}{\fldF^{MN}}
\eq -\sfrac{i}{2}g\Sigma^M_{AB}\acomm{\Psi^A}{\Psi^B},
\nln
\Sigma^M_{AB}\comm{\cder_M}{\Psi^B}\eq 0.
\>
%

\section{Ten-Dimensional Gauge Theory in Components}
\label{app:Ten.Comp}

Let us now leave superspace and represent superfields
by a collection of ordinary fields. 
It has been shown that the only independent components of the
gauge field are the $\Theta=0$ components of 
$\fldA_M$ and $\Psi^{B}$ once we impose the constraint 
\eqref{eq:Ten.Super.Constraint}.
All other components can be gauged away or 
are bosonic derivatives of the fundamental fields. 
By fixing the gauge along the $\Theta$ coordinates
we can restrict to the $\Theta=0$ components 
of $\fldA_M$ and $\Psi^{B}$.
These will become the fundamental fields of 
the field theory 
\[\label{eq:Ten.Super.Fields}
\fldW=(\cder_M,\Psi^A),
\]
which we will collectively refer to as $\fldW$.

The equations of motion \eqref{eq:Ten.Super.EOM}
force the fundamental fields $\fldA_M$
and $\Psi^{B}$ on shell.
These can be encoded into the Lagrangian 
\[\label{eq:Ten.Super.Lagr}
\Lagr_{10}[\fldW]=\sfrac{1}{4}\Tr\fldF^{MN}\fldF_{MN}
+\sfrac{1}{2}\Tr\Psi^A\Sigma^M_{AB}\cder_M\Psi^B.
\]
This is the Lagrangian of a ten-dimensional gauge field $\fldA_M$ coupled to 
a (real) chiral spinor $\Psi^{A}$ in the adjoint representation
of the gauge group. 
%
The covariant derivative $\cder_M \fldW$
of an adjoint field $\fldW$ is defined as the commutator
\[\label{eq:Ten.Super.Gauge}
\cder_M \fldW:=\comm{\cder_M}{\fldW}=\partial_M \fldW-ig\fldA_M \fldW+i g\fldW\fldA_M.
\]
%


Although we have dropped fermionic coordinates $\Theta$, 
translations along them are still possible. 
For that purpose we introduce fermionic translation generators $\algQ_A$ 
which act on fields as though they were derivatives along $\Theta$.
Equivalently we introduce bosonic translation generators $\algP_M$
which act on fields rather than coordinates
%
\[\label{eq:Ten.Super.Vary}
\algQ_A\hateq D_A,\qquad \algP_M\hateq D_M.
\]
The derivatives are taken to be covariant
when acting on gauge invariant states.
Written in terms of variations
$\delta_{\epsilon,e}=\epsilon^A\,\algQ_A+e^M \,\algP_M$
supertranslations of the fundamental fields are given by
\<\label{eq:Ten.Super.Trans}
\delta_{\epsilon,e}\, \cder_M \eq ig\epsilon^A \Sigma_{M,AB}\Psi^B
+ige^N \fldF_{MN},
\nln
\delta_{\epsilon,e}\, \Psi^A\eq\half \Sigma^{M,AB}\Sigma^N_{BC}\epsilon^C\fldF_{MN}
+e^M\cder_M\Psi^A.
\>
These satisfy the usual supersymmetry relations
\eqref{eq:Ten.Super.Algebra} or \eqref{eq:Ten.Super.FieldStrength}.
Note, however, that the algebra closes only on-shell, 
i.e.~up to terms proportional to the equations of
motion.
The Lagrangian \eqref{eq:Ten.Super.Lagr} is invariant under
fermionic translations up to a total derivative.

\section{$\superN=4$ SYM from Ten Dimensions}
\label{app:Ten.Four}

To obtain $\superN=4$ SYM from the ten-dimensional theories,
we split up ten-dimensional spacetime into a four-dimensional 
spacetime and a six-dimensional internal space. The fields
split up according to $\cder_M= \cder_\mu-ig\Phi_m$,
where $\Phi$ is a multiplet of six scalars, and
$\Psi^A=\Psi_{\alpha a}+\dot\Psi_{\dot\alpha}^a$.
The decomposition of the sigma symbols is presented 
in \appref{app:Spinors.Ten}.

\finishchapter 

\chapterbold{The Algebra $\alU(2,2|4)$}
\label{app:U224}

In this appendix we shall present the algebra $\alU(2,2|4)$,
a slightly enlarged version of the superconformal algebra, 
decomposed in terms of spacetime and internal 
$\alSU(2)\times\alSU(2)\times\alSU(4)$ symmetry.

\section{Commutation Relations}
\label{app:U224.Comm}

The (complex) algebra $\alU(2,2|4)=\alGL(4|4)$ is the algebra of 
$(4|4)\times (4|4)$ (complex) supermatrices. 
Using the generators $\algJ$ we can parameterise
an element $j\cdott \algJ$ of the algebra 
by the adjoint vector $j$.
For our purposes it is useful to break up the matrix in
$2|4|2$ (even$|$odd$|$even) rows and columns, 
the supermatrices will split up according to
\[\label{eq:U224.Comm.Matrix}
j\cdott \algJ=\matr{c|c|c}{
l^\beta{}_\alpha+\half \delta_\alpha^\beta(d+b-c)&q^\beta{}_a&p^{\beta\dot\alpha}\\\hline 
s^b{}_\alpha&r^b{}_a-\half \delta^b_a c &\dot q^{b \dot\alpha}\vphantom{\dot{\hat Q}}\\\hline  
-k_{\dot\beta\alpha}&-\dot s_{\dot\beta a}\vphantom{\dot{\hat Q}}&-\dot l_{\dot\beta}{}^{\dot\alpha}+\half \delta_{\dot\beta}^{\dot\alpha}(-d+b-c)}.
\]
The commutation relations of the generators can be read off
from the matrix representation of $\comm{j\cdott \algJ}{j'\cdott \algJ}$. 
Let us now discuss the generators independently.
First of all, there are the $\alSU(2),\alSU(4),\alSU(2)$ rotation generators
$\algL^\alpha{}_\beta,\algR^a{}_b,\algLd^{\dot\beta}{}_{\dot\alpha}$. 
The indices of any generator $\algJ$
transform canonically according to 
\[\label{eq:U224.Comm.Rot}
\arraycolsep0pt
\begin{array}[b]{rclcrcl}
\comm{\algL^\alpha{}_\beta}{\algJ_\gamma}\eq
\delta^\alpha_\gamma \algJ_\beta
-\half \delta^\alpha_\beta \algJ_\gamma,
&\quad&
\comm{\algL^\alpha{}_\beta}{\algJ^\gamma}\eq
-\delta^\gamma_\beta \algJ^\alpha
+\half \delta^\alpha_\beta \algJ^\gamma,
\\[3pt]
\comm{\algR^a{}_b}{\algJ_c}\eq
\delta^a_c \algJ_b
-\quarter \delta^a_b \algJ_c,
&&
\comm{\algR^a{}_b}{\algJ^c}\eq
-\delta_b^c \algJ^a
+\quarter \delta^a_b \algJ^c,
\\[3pt]
\comm{\algLd^{\dot\alpha}{}_{\dot\beta}}{\algJ_{\dot\gamma}}\eq
 \delta^{\dot\alpha}_{\dot\gamma} \algJ_{\dot\beta}
-\half \delta^{\dot\alpha}_{\dot\beta} \algJ_{\dot\gamma},
&&
\comm{\algLd^{\dot\alpha}{}_{\dot\beta}}{\algJ^{\dot\gamma}}\eq
-\delta_{\dot\beta}^{\dot\gamma} \algJ^{\dot\alpha}
+\half \delta^{\dot\alpha}_{\dot\beta} \algJ^{\dot\gamma}.
\end{array}\]
The charges $\algD,\algB,\algC$ (dilatation generator, hypercharge, central charge) 
of the generators are given by
\[\label{eq:U224.Comm.Charge}
\comm{\algD}{\algJ}=\dim(\algJ)\, \algJ,\qquad 
\comm{\algB}{\algJ}= \hyp(\algJ)\, \algJ,\qquad 
\comm{\algC}{\algJ}=0
\]
with non-vanishing dimensions
\[\label{eq:U224.Comm.Dim}
\dim(\algP)=-\dim(\algK)=1,\quad \dim(\algQ)=\dim(\algQd)=-\dim(\algS)=-\dim(\algSd)=\half
\]
and non-vanishing hypercharges
\[\label{eq:U224.Comm.Hyp}
\hyp(\algQ)=-\hyp(\algQd)=-\hyp(\algS)=\hyp(\algSd)=\half.
\]
Finally there are the translations $\algP_{\dot\alpha\beta}$,
boosts $\algK_{\alpha\dot\beta}$ as well as their fermionic partners, the
supertranslations $\algQ^a{}_{\beta},\algQd_{\dot\alpha b}$
and superboosts $\algS^\alpha{}_b,\algSd^{a\dot\beta}$.
The translations and boosts 
commuting into themselves are given by
\[\label{eq:U224.Comm.MomMom}
\arraycolsep0pt
\begin{array}{rclcrcl}
\comm{\algS^\alpha{}_a}{\algP_{\dot\beta\gamma}}\eq 
   \delta^\alpha_\gamma \algQd_{\dot\beta a},
&\qquad&
\comm{\algK^{\alpha\dot\beta}}{\algQd_{\dot\gamma c}}\eq
  \delta^{\dot\beta}_{\dot\gamma} \algS^\alpha{}_c,
\\[3pt]
\comm{\algSd^{a\dot\alpha}}{\algP_{\dot\beta\gamma}}\eq 
  \delta^{\dot\alpha}_{\dot\beta} \algQ^a{}_{\gamma},
&&
\comm{\algK^{\alpha\dot\beta}}{\algQ^c{}_\gamma}\eq
  \delta^\alpha_\gamma \algSd^{c\dot\beta},
\\[3pt]
\acomm{\algQd_{\dot\alpha a}}{\algQ^b{}_\beta}\eq 
  \delta_a^b \algP_{\dot\alpha\beta},
&&
\acomm{\algSd^{a\dot\alpha}}{\algS^\beta{}_b}\eq 
  \delta^a_b \algK^{\beta\dot\alpha},
\end{array}
\]
while the translations and boosts commuting into rotations are given by
\<\label{eq:U224.Comm.MomRot}
\comm{\algK^{\alpha \dot\beta}}{\algP_{\dot\gamma\delta}}\eq 
  \delta_{\dot\gamma}^{\dot\beta} \algL^{\alpha}{}_{\delta}
  +\delta_\gamma^\alpha \algLd^{\dot\beta}{}_{\dot\delta}
  +\delta_\gamma^\alpha\delta_{\dot\delta}^{\dot\beta} \algD
,
\nln
\acomm{\algS^\alpha{}_a}{\algQ^b{}_\beta}\eq
  \delta^b_a \algL^\alpha{}_\beta
  +\delta_\beta^\alpha \algR^b{}_a
  +\half \delta_a^b \delta_\beta^\alpha (\algD-\algC),
\nln
\acomm{\algSd^{a\dot \alpha}}{\algQd_{\dot\beta b}}\eq
  \delta^a_b \algLd^{\dot\alpha}{}_{\dot\beta}
  -\delta_{\dot\beta}^{\dot\alpha} \algR^a{}_b
  +\half \delta^a_b \delta_{\dot\beta}^{\dot\alpha} (\algD+\algC).
\>
As we see, the hypercharge $\algB$ never appears on the right hand side,
it can be dropped, leading to $\alSU(2,2|4)$. 
Conversely, when restricting to representations with 
zero central charge $\algC$, the resulting algebra is $\alPU(2,2|4)$,
which becomes $\alPSU(2,2|4)$ after removing $\algB$ as well.

In this work we deal with two further operators, $\len$ 
and $\delta\algD=g^2\ham$, which are not part of $\alU(2,2|4)$.
The operator $\len$ measures the length, 
i.e.~the number of components fields, of a state.
The anomalous dilatation operator $\delta \algD(g)=\algD(g)-\algD(0)$, 
or equivalently the Hamiltonian $\ham$, commutes with $\alU(2,2|4)$.

\section{Labels}
\label{app:U224.Labels}

\begin{figure}\centering
\includegraphics{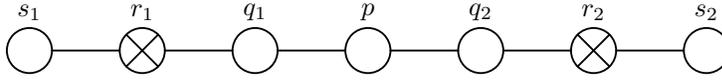}
\caption{Dynkin labels $[s_1;r_1;q_1,r,q_2;r_2;s_2]$ of $\alSU(2,2|4)$. 
Note that in our convention the sign of the odd labels $r_1,r_2$
to be negative for the antisymmetric product
of two fundamental representations.}
\label{fig:U224.Osc.Labels}
\end{figure}

Let us collect some of our notation concerning labels of 
states and multiplets,
see also \secref{sec:N4.Alg}. The Dynkin labels of $\alSU(2,2|4)$ are,
c.f.~\figref{fig:U224.Osc.Labels},
\[\label{eq:U224.Labels.Dynkin}
[s_1;r_1;q_1,p,q_2;r_2;s_2],
\]
where $[q_1,p,q_2]$ and 
$[s_1,r,s_2]$ with $r=-r_1-q_1-p-q_2-r_2$ are the Dynkin labels
of $\alSU(4)$ and $\alSU(2,2)$, respectively. 
Note that $s_1,s_2$ are \emph{twice} the spins of the Lorentz algebra.
These labels are given as the eigenvalues 
$L^\alpha{}_\beta,\dot L^{\dot \alpha}{}_{\dot\beta},R^a{}_b$ 
of Cartan generators
$\algL^\alpha{}_\beta,\algLd^{\dot \alpha}{}_{\dot\beta},\algR^a{}_b$
($\alpha=\beta, \dot\alpha=\dot\beta, a=b$)
\[\label{eq:U224.Labels.DynkinCartan}
\arraycolsep0pt\begin{array}{rclcrcl}
s_1 \eq L^2{}_2-L^1{}_1, 
&\qquad&
s_2 \eq \dot L^2{}_2-\dot L^1{}_1,
\\[5pt]
r_1 \eq \half D-\half C-L^1{}_1+R^1{}_1,
&&
r_2 \eq \half D+\half C-\dot L^1{}_1-R^4{}_4,
\\[5pt]
q_1 \eq R^2{}_2-R^1{}_1,
&&
q_2 \eq R^4{}_4-R^3{}_3,
\\[10pt]
p \eq R^3{}_3-R^2{}_2,
&&
r \eq -D+L^1{}_1+\dot L^1{}_1.
\end{array}
\]
We also use the notation 
\[\label{eq:U224.Labels.Weight}
\weight{D_0,s_1,s_2;q_1,p,q_2;B,L}
\]
to describe states of the classical theory.
The label $B$ is the hypercharge measured by $\algB$.
The length $L$ corresponds to the number of fields within a state,
it is measured by the operator $\len$ which is not part of $\alU(2,2|4)$.
$D_0$ is the classical dimension
and we introduce $E$ as the `energy' or anomalous dimension
measured by $\ham=g^{-2}\delta\algD$.
Together they form the scaling dimension 
$D=D_0+g^2 E$ which is measured by $\algD=\algD_0+g^2\ham$.
Usually we will state only the classical dimension $D_0$ to specify a state; 
the corresponding energy $E$ will be the main result of a our computations.
It is useful to know how to translate between 
the dimension $D$ and Dynkin labels $r_1,r_2,r$,
see \eqref{eq:U224.Labels.DynkinCartan},
\<\label{eq:U224.Labels.FermiLabel}
r_1\eq\half D-\half C-\half p-\sfrac{3}{4}q_1-\sfrac{1}{4}q_2+\half s_1,
\nln
r_2\eq\half D+\half C-\half p-\sfrac{1}{4}q_1-\sfrac{3}{4}q_2+\half s_2,
\nln
r\eq-D-\half s_1-\half s_2,
\nln
r\eq-r_1-q_1-p-q_2-r_2
\nln
D\eq -\half s_1+r_1+q_1+p+q_2+r_2-\half s_2.
\>
%

\section{The Quadratic Casimir}
\label{app:U224.Casimir}

The quadratic Casimir of $\alU(2,2|4)$ is
\<\label{eq:U224.Comm.Casimir}
\algJ^2\eq 
\half \algD^2
+\half \algL^\gamma{}_\delta \algL^\delta{}_\gamma
+\half \algLd^{\dot\gamma}{}_{\dot\delta} \algLd^{\dot\delta}{}_{\dot\gamma}
-\half \algR^c{}_d \algR^d{}_c
\nl
-\half \comm{\algQ^c{}_\gamma}{\algS^\gamma{}_c}
-\half \comm{\algQd_{\dot\gamma c}}{\algSd^{\dot\gamma c}}
-\half \acomm{\algP_{\dot\gamma\delta}}{\algK^{\delta\dot\gamma}}
-\algB\algC.
\>
In $\alPSU(2,2|4)$ the last term $\algB\algC$ is absent.
For highest weight states,
which are annihilated by raising operators $\algJ^+$ \eqref{eq:N4.Alg.PosNeg},
we can conveniently evaluate $\algJ^2$ by using the standard trick of
turning the anticommutators into commutators. We find
\<\label{eq:U224.Comm.CasimirVal}
J^2\eq
\quarter s_1(s_1+2)
+\quarter s_2(s_2+2)
+\half D^2+2D-BC
\nl
-\quarter q_1(q_1+2)
-\quarter q_2(q_2+2)
-\sfrac{1}{8} (2p+q_1+q_2)^2
-(2p+q_1+q_2).
\>
%

\section{The Oscillator Representation}
\label{app:U224.Osc}

Let us explain the use of oscillators
for fields and generators in terms of the algebra $\alGL(M)$:
We write%
\footnote{Strictly speaking, the oscillators $\oscA$ and $\oscA^\dagger$ 
are independent. 
Only in one of the real forms of $\alGL(M)$ they would 
be related by conjugation.}
\[\label{eq:U224.Osc.GenGen}
\algJ^A{}_B=\oscA^\dagger_B\oscA^A,
\qquad
\mbox{with }A,B=1,\ldots,M.
\]
Using the commutators
\[\label{eq:U224.Osc.GenComm}
\comm{\oscA^A}{\oscA^\dagger_B}=\delta^A_B,\qquad
\comm{\oscA^A}{\oscA^B}=\comm{\oscA^\dagger_A}{\oscA^\dagger_B}=0
\]
it is a straightforward exercise to show that the generators
$\algJ$ satisfy the $\alGL(M)$ algebra. 

To construct an oscillator representation for $\alU(2,2|4)$,
c.f.~\cite{Gunaydin:1985fk}, we will consider two sets of bosonic oscillators 
$(\osca^\alpha,\osca^\dagger_\alpha)$, 
$(\oscb^{\dot \alpha},\oscb^\dagger_{\dot\alpha})$ 
with $\alpha,\dot\alpha=1,2$
and one set of fermionic oscillator $(\oscc^a,\oscc^\dagger_a)$
with $a=1,2,3,4$.
The non-vanishing commutators of oscillators are taken to be
\[\label{eq:U224.Osc.Comm}
\comm{\osca^{\alpha}}{\osca^\dagger_{\beta}}=\delta^\alpha_\beta,
\qquad
\comm{\oscb^{\dot\alpha}}{\oscb^\dagger_{\dot\beta}}=\delta^{\dot\alpha}_{\dot\beta},
\qquad
\acomm{\oscc^{a}}{\oscc^\dagger_{b}}=\delta^a_b.
\]
We assume that the oscillators 
$\oscA^A=(\osca,\oscb^\dagger,\oscc)$ and 
$\oscA^\dagger_A=(\osca^\dagger,-\oscb,\oscc^\dagger)$
form a fundamental and 
conjugate fundamental multiplet of $\alU(2,2|4)$.
Then, the bilinears $\oscA^\dagger_A \oscA^B$
generate the algebra $\alU(2,2|4)$ as described above.
By comparing to the matrix form \eqref{eq:U224.Comm.Matrix}
we can read off the generators in $\alSU(2)\times\alSU(2)\times\alSU(4)$ 
notation.
The canonical forms for rotation generators of $\alSU(2)$, $\alSU(2)$
and $\alSU(4)$ are
\<\label{eq:U224.Osc.GenRot}
\algL^{\alpha}{}_{\beta}\eq\osca^\dagger_{\beta}\osca^{\alpha}
-\half \delta^\alpha_\beta\osca^\dagger_{\gamma}\osca^{\gamma},
\nln
\algLd^{\dot\alpha}{}_{\dot\beta}\eq\oscb^\dagger_{\dot\beta}\oscb^{\dot\alpha}
-\half \delta^{\dot\alpha}_{\dot\beta}\oscb^\dagger_{\dot\gamma}\oscb^{\dot\gamma},
\nln
\algR^{a}{}_{b}\eq\oscc^\dagger_{b}\oscc^{a}
-\quarter \delta^{a}_{b}\oscc^\dagger_{c}\oscc^{c}.
\>
Under these the fields \eqref{eq:N4.Fund.Fields} transform canonically.
We write the corresponding three $\alU(1)$ charges as
\<\label{eq:U224.Osc.GenCharge}
\algD\eq 
1+\half \osca^\dagger_{\gamma}\osca^{\gamma}
+\half \oscb^\dagger_{\dot\gamma}\oscb^{\dot\gamma},
\nln
\algC\eq 
1-\half \osca^\dagger_{\gamma}\osca^{\gamma}
+\half \oscb^\dagger_{\dot\gamma}\oscb^{\dot\gamma}
-\half \oscc^\dagger_{c}\oscc^{c},
\nln
\algB\eq \phantom{1+\mathord{}}\half \osca^\dagger_{\gamma}\osca^{\gamma}
-\half \oscb^\dagger_{\dot\gamma}\oscb^{\dot\gamma}.
\>
The remaining off-diagonal generators are
\[\label{eq:U224.Osc.GenMom}
\arraycolsep0pt
\begin{array}[b]{rclcrcl}
\algQ^a{}_{\alpha}\eq\osca^\dagger_\alpha \oscc^{a},&\qquad&
\algS^{\alpha}{}_a\eq\oscc^\dagger_{a} \osca^\alpha , 
\\[3pt]
\algQd_{\dot\alpha a}\eq \oscb^\dagger_{\dot\alpha} \oscc^\dagger_{a} ,&&
\algSd^{\dot\alpha a}\eq \oscb^{\dot\alpha} \oscc^{a} ,
\\[3pt]
\algP_{\alpha \dot \beta}\eq\osca^\dagger_{\alpha}\oscb^\dagger_{\dot \beta} ,&&
\algK^{\alpha \dot \beta}\eq\osca^{\alpha}\oscb^{\dot \beta}.
\end{array}
\]
Quite naturally the algebra $\alU(2,2|4)$ 
is realised by the generators
\eqref{eq:U224.Osc.GenRot,eq:U224.Osc.GenCharge,eq:U224.Osc.GenMom}.%
\footnote{Note that a shift of $\algB$ by a constant ($-1$) does not modify the algebra.
Then the $1$ in $\algD,\algC,\algB$ can be absorbed into 
$1+\half \oscb^\dagger_{\dot\gamma}\oscb^{\dot\gamma}=\half \oscb^{\dot\gamma}\oscb^\dagger_{\dot\gamma}$
to yield a canonical form.}
We have written this in a $\alSU(2)\times\alSU(2)\times\alSU(4)$ 
covariant way. In fact one can combine the indices $a$ and $\alpha$ into
a superindex and obtain a manifest $\alSU(2)\times\alSU(2|4)$
notation. The generators with two lower or two upper indices,
$\algP,\algQd,\algK,\algSd$, 
together with the remaining charges complete the $\alU(2,2|4)$ algebra.

Instead of one fermionic oscillator $(\oscc^a,\oscc^\dagger_a)$
with $a=1,2,3,4$, we can also introduce two
pairs of oscillators $(\oscc^a,\oscc^\dagger_a)$
and $(\oscd^{\dot a},\oscd^\dagger_{\dot a})$ with
$a,\dot a=1,2$. 
These should be grouped as $\oscA^A=(\osca,\oscb^\dagger,\oscc,\oscd^\dagger)$ 
and $\oscA^\dagger_A=(\osca^\dagger,-\oscb,\oscc^\dagger,\oscd)$ 
into fundamental 
representations of $\alU(2,2|4)$.
Despite the fact that only $\alSU(2)^4$ (or $\alPSU(2|2)^2$ when using
superindices)
is manifest in this notation,
it has the added benefit that 
we can define a physical vacuum state $\state{\fldZ}$ by
\[\label{eq:U224.Osc.PhysVac}
\osca^\alpha
\state{\fldZ}=
\oscb^{\dot\alpha}
\state{\fldZ}=
\oscc^a
\state{\fldZ}=
\oscd^{\dot a}
\state{\fldZ}=0.
\]
This is the highest weight state of the field multiplet, 
see \secref{sec:N4.Fund}.

\begin{figure}\centering
\includegraphics{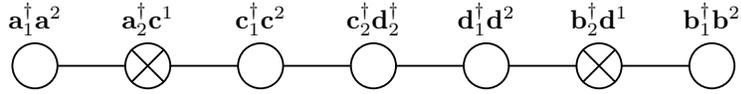}
\caption{Oscillator representation of simple roots.}
\label{fig:U224.Osc.Simple}
\end{figure}

In this context it is useful to know how
to represent a state 
with a given weight 
\[\label{eq:U224.Osc.Weight}
w=\weight{D_0;s_1,s_2;q_1,p,q_2;B,L}\]
in terms of excitations of the oscillators.
We introduce a multi-particle vacuum operator $\state{\fldZ,L}$ which 
is the tensor product of $L$ vacua $\state{\fldZ}$.
The oscillators 
$\osca^\dagger_{p,\alpha},\oscb^\dagger_{p,\dot\alpha},\oscc^\dagger_{p,a},\oscd^\dagger_{p,\dot a}$.
now act on site $p$, where
commutators of two oscillators 
vanish unless the sites agree.
Equivalently, we define the unphysical multi-particle vacuum state
$\state{0,L}$.
A generic state is written as
\[\label{eq:U224.Osc.States}
(\osca^\dagger)^{n_{\osca}}(\oscb^\dagger)^{n_{\oscb}}
(\oscc^\dagger)^{n_{\oscc}}(\oscd^\dagger)^{n_{\oscd}}\state{\fldZ,L}
\qquad\mbox{or}\qquad
(\osca^\dagger)^{n_{\osca}}(\oscb^\dagger)^{n_{\oscb}}
(\oscc^\dagger)^{n_{\oscc}}\state{0,L}.
\]
By considering the weights of the oscillators
as well as the central charge constraint, 
we find the number of excitations as given in \tabref{tab:U224.Osc.Numbers}.%
\footnote{The components of the vectors correspond to 
the numbers of each component of the oscillators.}
\begin{table}\centering
$n_{\osca}=\left(\begin{array}{c}
\half D_0+\half B-\half L-\half s_1\\
\half D_0+\half B-\half L+\half s_1
\end{array}\right),\quad
n_{\oscb}=\left(\begin{array}{c}
\half D_0-\half B-\half L-\half s_2\\
\half D_0-\half B-\half L+\half s_2,
\end{array}\right)$,\bigskip

$n_{\oscc}=\left(\begin{array}{c}
\half L-\half B-\half p-\sfrac{3}{4} q_1-\sfrac{1}{4}q_2\\
\half L-\half B-\half p+\sfrac{1}{4} q_1-\sfrac{1}{4}q_2
\end{array}\right),\quad
n_{\oscd}=\left(\begin{array}{c}
\half L+\half B-\half p-\sfrac{1}{4}q_1-\sfrac{3}{4}q_2\\
\half L+\half B-\half p-\sfrac{1}{4}q_1+\sfrac{1}{4}q_2
\end{array}\right)$,\bigskip

$n_{\oscc}=\left(\begin{array}{c}
\half L-\half B-\half p-\sfrac{3}{4} q_1-\sfrac{1}{4}q_2\\
\half L-\half B-\half p+\sfrac{1}{4} q_1-\sfrac{1}{4}q_2\\
\half L-\half B+\half p+\sfrac{1}{4} q_1-\sfrac{1}{4}q_2\\
\half L-\half B+\half p+\sfrac{1}{4} q_1+\sfrac{3}{4}q_2
\end{array}\right)$.

\caption{Oscillator excitation numbers for a 
state with given charges.}
\label{tab:U224.Osc.Numbers}
\end{table}
It is also useful to know how to represent 
the generators corresponding to the simple roots
in terms of creation and annihilation operators,
c.f.~\figref{fig:U224.Osc.Simple}.

\finishchapter 

\chapterbold{Tools for the $\alSU(2)$ Sector}
\label{app:SU2Tools}

In this appendix we present a basic set of \texttt{Mathematica} 
routines to deal with $\alSU(2)$ spin chains at higher-loops.

\section{States}
\label{app:SU2Tools.States}

One of the basic objects is a \emph{state} $\state{\ldots}$,
in \texttt{Mathematica} it will be represented by the function
\texttt{Chain[...]}, e.g.
\[\label{eq:SU2Tools.States.State}
\verb~Chain[0,0,1,0,1,1,0]~
\hateq\state{0,0,1,0,1,1,0}
=\Tr \fldZ\fldZ\phi\fldZ\phi\phi\fldZ.
\]
This function is undefined and \texttt{Mathematica} will 
leave it as it stands and not try to evaluate it.
For example one can construct linear combinations of 
states, e.g.
\[\label{eq:SU2Tools.States.Ex}
2\state{0,0,1,1}+\state{0,1,0,1}\hateq
\,\verb~2 Chain[0,0,1,1] + Chain[0,1,0,1]~.
\]

We consider spin chain states only 
modulo cyclic permutations. 
The order has to be implemented manually and we need a routine
to shift states into some canonical order
\begin{code}
SortChain[X_] := X /. C_Chain :> 
  Module[{k}, Sort[Table[RotateLeft[C, k], {k, Length[C]}]][[1]]];
\end{code}
This function returns the argument \texttt{X} with 
all chains ordered. It works as follows: First of all, all states 
\texttt{C=Chain[...]} within \texttt{X} are found. 
For each \texttt{C} a list of all possible cyclic permutations is generated
and sorted. The first element is taken as the canonically ordered state
and returned.

We can now define a simple operation on states,
the parity $\gaugepar$ which reverses the spin chain and multiplies
by $(-1)^L$
\begin{code}
ChainParity[X_] := X /. C_Chain :> (-1)^Length[C] SortChain[Reverse[C]];
\end{code}
Conveniently, it shifts the states into a canonical order.

\section{Interactions}
\label{app:SU2Tools.Interact}

The other basic object is an \emph{interaction} 
$\PTerm{\ldots}$ which will 
be represented by \texttt{Perm[...]}
\[\label{app:SU2Tools.Interact.Interact}
\verb~Perm[1,3,2]~
\hateq\PTerm{1,3,2}.
\]
We need a representation for the action
of permutation symbols $\PTerm{\ldots}\state{\ldots}$, this is done
by
\begin{code}
ApplyPerm[P_, C_] := P /. P0_Perm :> (C /. C0_Chain -> ApplyPermPC[P0, C0]);
ApplyPermPC[P_Perm, C_Chain] := 
  Module[{s}, Sum[PermuteList[C, P, s], {s, Length[C]}]];
\end{code}
The function \texttt{ApplyPerm} assumes \texttt{P} and \texttt{C}
are linear combinations of interactions and states. It distributes
the elementary interactions and states and passes on 
to \texttt{ApplyPermPC} for an elementary pair. 
This uses another function \texttt{PermuteList} to 
apply the permutation to each site \texttt{s} of the spin chain
\begin{code}
PermuteList[C_Chain, Perm[], s_] := C;
PermuteList[C_Chain, P_Perm, s_] := 
  PermuteList[PermuteElements[C, Last[P] + s], Drop[P, -1], s];
PermuteElements[C_Chain, p_] := 
  Module[{p0 = Mod[p, Length[C], 1], p1 = Mod[p + 1, Length[C], 1]}, 
    ReplacePart[ReplacePart[C, C[[p0]], p1], C[[p1]], p0]];
\end{code}
The routine \texttt{PermuteList} recursively works on the permutation symbol
\texttt{P} from the right and uses \texttt{PermuteElements} to perform
the pairwise permutations.

\section{Spectrum}
\label{app:SU2Tools.Spec}

To find the spectrum of an operator, we need to find a complete 
basis of states on which the operator closes. This basis is
generated by 
\begin{code}
GenerateChains[L_, K_] := 
  (Chain @@ Join[Array[1 &, K], Array[0 &, L - K]]) 
    // Permutations // SortChain // Union;
\end{code}
The function returns a basis of states of \texttt{L} sites with \texttt{n}
excitations. It is then convenient to have a method
to evaluate the action of an operator on a basis of states
\begin{code}
ActionMatrix[P_, C_] := CoeffList[ApplyPerm[P, C] // SortChain, C];
\end{code}
It returns a matrix that is equivalent to the action of \texttt{P} 
in the basis $\texttt{C}$. It requires the multi-purpose function
\begin{code}
CoeffList[X_, L_] := Map[Coefficient[X, #] &, L];
\end{code}
which expands a linear expression \texttt{X} over a basis 
of atoms \texttt{L}.

\section{An Example}
\label{app:SU2Tools.Example}

We can now find the energy of states with
length \texttt{L=4} and \texttt{K=2} excitations. Let us first 
construct a basis of states
\begin{code}
Ops = GenerateChains[4, 2]
> {Chain[0, 0, 1, 1], Chain[0, 1, 0, 1]}
\end{code}
The one-loop Hamiltonian is given by $H_0=\PTerm{}-\PTerm{1}$
so let us define
\begin{code}
Ham = Perm[] - Perm[1];
\end{code}
and act on the above basis
\begin{code}
MHam = ActionMatrix[Ham, Ops]
> {{2, -4}, {-2, 4}}
\end{code}
The eigenvalues are
\begin{code}
Eigenvalues[MHam]
> {0, 6}
Eigenvectors[MHam]
> {{2, 1}, {-1, 1}}
\end{code}
where the energy $E=0$ belongs to the state 
$2\state{0,0,1,1}+\state{0,1,0,1}$
and $E=6$ to the Konishi state
$-\state{0,0,1,1}+\state{0,1,0,1}$.

\section{Commutators}
\label{app:SU2Tools.Comm}

For investigations of integrability we require methods 
to compute commutators of interactions abstractly.
This is a straightforward implementation
of the commutator of permutation symbols
\begin{code}
CommutePerm[X_, Y_] := 
  X /. P1_Perm :> (Y /. P2_Perm -> CommutePerm12[P1, P2]) // SimplifyPerm;
CommutePerm12[Perm[X___], Perm[Y___]] := 
  Module[{MX = Max[X, 0] + 1, MY = Max[Y, 0] + 1, k}, 
    Sum[Perm @@ Join[{X} + k - 1, {Y} + MX - 1] - 
        Perm @@ Join[{Y} + MX - 1, {X} + k - 1], {k, MX + MY - 1}]];
\end{code}
As above, \texttt{CommutePerm} distributes linear combinations
and calls \texttt{CommutePerm12} for elementary commutators.
Finally, we should simplify the permutation symbols using 
the rules in \secref{sec:HighInt.SU2.Interact}
\begin{code}
SimplifyPerm[YY_] := (YY //. 
  {Perm[X__ /; Min[X] != 1] :> Perm @@ ({X} - Min[X] + 1), 
   Perm[X___, y_, y_, Z___] -> Perm[X, Z], 
   Perm[X___, y_, z_, y_, W___] 
     /; (z == y + 1) || (z == y - 1) 
     -> Perm[X, W] - Perm[X, y, W] - Perm[X, z, W] 
        + Perm[X, y, z, W] + Perm[X, z, y, W], 
   Perm[X___, y_, z_, W___] 
     /; z < y - 1 
     -> Perm[X, z, y, W], 
   Perm[X___, y_, z_, W__, y_, V___] 
     /; (z == y - 1) && (! MemberQ[{W}, k_ /; k > y - 2]) 
     -> Perm[X, y, z, y, W, V], 
   Perm[X___, y_, W__, z_, y_, V___] 
     /; (z == y + 1) && (! MemberQ[{W}, k_ /; k < y + 2]) 
     -> Perm[X, W, y, z, y, V]}) // Expand;
\end{code}

For the construction of higher-loop 
interactions, one can write functions
to construct a basis of interaction symbols
for a given loop order and to 
solve commutators for unknown coefficients.
As the construction still requires a
sufficient amount of manual work, we will not
present these here. 

\finishchapter 

\chapter{The Harmonic Action}
\label{app:Harm}

The Hamiltonian density $\ham_{12}$ is given 
by an $\alPSU(2,2|4)$ invariant function acting on two sites,
\[\label{eq:Harm.Ham}
\ham_{12}=2h(\fldspin_{12}).
\]
We will now describe explicitly how $\ham_{12}$ acts
on a state of two fields in
the oscillator representation, see \secref{sec:N4.Fund}.

\section{Generic Invariant Action}
\label{app:Harm.Inv}

We will investigate the action 
of a generic function $f(\fldspin_{12})$ on two oscillator sites.
Let us introduce a collective oscillator 
$\oscA^\dagger_{A}=(\osca^\dagger_{\alpha},\oscb^\dagger_{\dot\alpha},
\oscc^\dagger_{a},\oscd^\dagger_{\dot a})$.
A general state in $\mdlF\times \mdlF$ can be written as
\[\label{eq:Harm.Inv.State}
\state{p_1,\ldots,p_n;A}=
\oscA^\dagger_{p_1,A_1}
\ldots
\oscA^\dagger_{p_{n},A_{n}}
\state{\fldZ\fldZ},
\]
subject to the central charge constraints $\algC_1\state{X}=\algC_2\state{X}=0$.
The label $p_k=1,2$ determines the site on which the $k$-th oscillator acts.
The action of $\alPSU(2,2|4)$ conserves the number of
each type of oscillator; it can however
move oscillators between both sites.
Therefore the action of $f(\fldspin_{12})$ is 
\[\label{eq:Harm.Inv.Act}
f(\fldspin_{12})\,\state{p_1,\ldots,p_n;A}=
\sum_{p'_1,\ldots p'_{n}}
c_{p,p',A}\,
\delta_{C_1,0}\delta_{C_2,0}\,
\state{p'_1,\ldots,p'_n;A}
\]
with some coefficients $c_{s,s',A}$.
The sums go over the sites $1,2$
and $\delta_{C_1,0}$, $\delta_{C_2,0}$ project 
to states where the central charge at each site is zero.
In view of the fact that oscillators represent
indices of fields, see \eqref{eq:N4.Fund.Example}, a generic 
invariant operator $f(\fldspin_{12})$ acts
on two fields by moving indices between them.

\section{The Harmonic Action}
\label{app:Harm.Harm}

The action of the harmonic numbers
within the Hamiltonian density $\ham_{12}=2h(\fldspin_{12})$
turns out to be particularly simple.
It does not depend on the 
types of oscillators $A_k$,
but only on the number of oscillators which change the site
\[\label{eq:Harm.Harm.Act}
\ham_{12}\state{s_1,\ldots,s_n;A}=
\sum_{s'_1,\ldots s'_{n}}
c_{n,n_{12},n_{21}}\,
\delta_{C_1,0}\delta_{C_2,0}\,
\state{s'_1,\ldots,s'_n;A}.
\]
Here $n_{12},n_{21}$ count the number 
of oscillators hopping 
from site $1$ to $2$ or vice versa.
The coefficients $c_{n,n_{12},n_{21}}$ are given by
\[\label{eq:Harm.Harm.Coeff}
c_{n,n_{12},n_{21}}=
 (-1)^{1+n_{12}n_{21}}
\frac{\Gamma\bigbrk{\half (n_{12}+n_{21})}\Gamma\bigbrk{1+\half (n-n_{12}-n_{21})}}
     {\Gamma\bigbrk{1+\half n}}\,.
\]
In the special case of no oscillator hopping we find
\[\label{eq:Harm.Harm.CoeffZero}
c_{n,0,0}= h(\half n),\]
which can be regarded as a regularisation of \eqref{eq:Harm.Harm.Coeff}.
We will refer to this action given by 
\eqref{eq:Harm.Harm.Act,eq:Harm.Harm.Coeff,eq:Harm.Harm.CoeffZero}
as the `\emph{harmonic action}'.
Interestingly, we find that
the action of the Hamiltonian density using the 
$\alSU(4|2)\times\alSU(2)$ invariant vacuum (c.f.~\secref{sec:N4.Fund})
is given by exactly the same expressions.

\section{Proof}
\label{app:Harm.Proof}

To prove that $\ham_{12}$ is given by 
\eqref{eq:Harm.Harm.Act,eq:Harm.Harm.Coeff,eq:Harm.Harm.CoeffZero}
it suffices to show
\[\label{eq:Harm.Proof.This}
\comm{\algJ_{12}}{\ham_{12}}=0,\qquad
\ham_{12}\,\mdl_{j}=2h(j)\, \mdl_j.
\]
The invariance of $\ham_{12}$ under 
the subalgebra $\alPSU(2|2)\times\alPSU(2|2)$ 
is straightforward:
These generators only change the types of oscillators,
whereas the harmonic action does not depend on that.
In contrast, the remaining generators 
change the number of oscillators by two.

Consider a generator which increases the number of
oscillators by two, e.g.~$\algP_{12,\alpha\dot\beta}$,
and act with $\ham_{12}\algP_{12,\alpha\dot\beta}$ on a generic state. 
First we apply $\algP$ 
\[\label{eq:Harm.Proof.PAct}
\algP_{12,\alpha\dot\beta}
\state{p_1,\ldots,p_n;A}=
\state{1,1,p_1,\ldots,p_n;A'}+
\state{2,2,p_1,\ldots,p_n;A'},
\]
and get a state with two new oscillators, $A'=(\alpha,\dot\beta,A)$.
We apply the Hamiltonian density \eqref{eq:Harm.Harm.Act} to this state
and get eight terms (to be summed over all $p'_k$).
In two of these terms, both new oscillators end up at site $1$
\[\label{eq:Harm.Proof.HPGood}
c_{n+2,n_{12},n_{21}}\state{1,1,p'_1,\ldots,p'_n;A'}+
c_{n+2,n_{12},n_{21}+2}\state{1,1,p'_1,\ldots,p'_n;A'}.
\]
Here, $n_{12},n_{21}$ refer only to the hopping of the old oscillators.
Both coefficients can be combined 
using \eqref{eq:Harm.Harm.Coeff}
\[\label{eq:Harm.Proof.CoeffAdd}
c_{n+2,n_{12},n_{21}}+c_{n+2,n_{12},n_{21}+2}=c_{n,n_{12},n_{21}}.
\]
We pull the additional two oscillators out of the state and get
\[\label{eq:Harm.Proof.PHDone}
\bigbrk{c_{n+2,n_{12},n_{21}}+c_{n+2,n_{12},n_{21}+2}}\state{1,1,p'_1,\ldots,p'_n;A'}=
\algP_{1,\alpha\dot\beta}\, c_{n,n_{12},n_{21}}\state{p'_1,\ldots,p'_n;A}.
\]
Summing over all $p'_k$ therefore yields
$\algP_{1,\alpha\dot\beta}\ham_{12}\state{p_1,\ldots,p_n;A}$. 
If both new oscillators end up at site $2$ we get an equivalent result.
It remains to be shown that the other four terms cancel.
Two of these are
\[\label{eq:Harm.Proof.HPBad}
c_{n+2,n_{12},n_{21}+1}\state{1,2,p'_1,\ldots,p'_n;A'}+
c_{n+2,n_{12}+1,n_{21}}\state{1,2,p'_1,\ldots,p'_n;A'}.
\]
The absolute values in \eqref{eq:Harm.Harm.Coeff}
match for $c_{n+2,n_{12},n_{21}+1}$ and $c_{n+2,n_{12}+1,n_{21}}$
and we sum up the signs
\[\label{eq:Harm.Proof.BadCancel}
(-1)^{1+n_{12}n_{21}+n_{12}}+(-1)^{1+n_{12}n_{21}+n_{21}}=
(-1)^{1+n_{12}n_{21}}\bigbrk{(-1)^{n_{12}}+(-1)^{n_{21}}}.
\]
Now, oscillators always hop in pairs due to the central charge constraint.
One of the new oscillators has changed the site, so 
the number of old oscillators changing site must be odd. 
The above two signs must be opposite and cancel in the sum.
The same is true for the remaining two terms.
This concludes the proof for $\comm{\algP_{12,\alpha\dot\beta}}{\ham_{12}}=0$.
The proof for the other generators which increase
the number of oscillators is equivalent.
To prove invariance under the remaining generators, we note that
these remove two oscillators from one of the two sites.
Assume it will remove the first two oscillators from a state
(for each two oscillators that are removed, the argument will be the same).
Now, the argument is essentially the
same as the proof for $\algP_{12,\alpha\dot\beta}$ 
read in the opposite direction.

To prove that the eigenvalues of $\ham_{12}$ are given by $2h(j)$,
we act on a state $\state{j}$ of $\mdl_j$
within the bosonic $\alSU(1,1)$ subsector, c.f.~\secref{sec:One.Baby}
\[\label{eq:Harm.Proof.SpinJ}
\state{j}=\sum_{k=0}^j \frac{(-1)^k j!}{k!(j-k)!}\,\state{k,j-k}
\]
with a single spin given by 
\[\label{eq:Harm.Proof.Field}
\state{k}=
\frac{1}{k!}\,
(\osca^\dagger_{2})^{k}(\oscb^\dagger_{2})^k
\,\state{\fldZ}.
\]
The state $\state{j}$ has a definite spin and is therefore 
an eigenstate of $\ham_{12}$. We can choose to calculate only the
coefficient of $\state{j,0}$ in $\ham_{12}\state{j}$.
It is given by (see \secref{sec:One.Magic.Eigen})
\[\label{eq:Harm.Proof.Eigen}
h(j)+
\sum_{k=1}^{j}
\frac{(-1)^{1+k} j!}{k\,k!(j-k)!}
=2h(j),
\]
which proves that $\ham_{12}=2h(\fldspin_{12})$. 
This concludes the proof of \eqref{eq:Harm.Proof.This}.

\section{An Example}
\label{app:Harm.Example}

We will now determine the planar anomalous dimensions 
of some states with weight $\weight{2;0,0;0,0,0;0,2}$
to demonstrate how to apply the above Hamiltonian. Using 
\tabref{tab:U224.Osc.Numbers} we find that we have to excite 
each of the four oscillators $\oscc,\oscd$ once.
There must be an equal number of $\oscc$ and $\oscd$ oscillators on each site 
due to the central charge constraint
and the three distinct configurations are
\<\label{eq:Harm.Example.States}
\state{1212}\eq
\oscc^\dagger_{1,1}\oscc^\dagger_{2,2}
\oscd^\dagger_{1,1}\oscd^\dagger_{2,2}
\state{\fldZ\fldZ},
\nln
\state{1221}\eq
\oscc^\dagger_{1,1}\oscc^\dagger_{2,2}
\oscd^\dagger_{2,1}\oscd^\dagger_{1,2}
\state{\fldZ\fldZ},
\nln
\state{1111}\eq \oscc^\dagger_{1,1}\oscc^\dagger_{1,2}
\oscd^\dagger_{1,1}\oscd^\dagger_{1,2}
\state{\fldZ\fldZ}.
\>
Let us now act with $\ham_{12}$ on these states, we find
\<\label{eq:Harm.Example.Act}
\ham_{12}\state{1212}\eq
c_{4,0,0}\state{1212}
+c_{4,0,2}\state{1111}
+c_{4,2,0}\state{2222}
\nl
+c_{4,1,1}\state{2112}
+c_{4,1,1}\state{1221}
+c_{4,2,2}\state{2121}
\nln\eq
\sfrac{3}{2}\state{1212}
-\sfrac{1}{2}\state{1111}
-\sfrac{1}{2}\state{2222}
+\sfrac{1}{2}\state{2112}
+\sfrac{1}{2}\state{1221}
-\sfrac{1}{2}\state{2121}
\nln\eq
\state{1212}+\state{1221}-\state{1111}
\>
using \eqref{eq:Harm.Harm.Act,eq:Harm.Harm.Coeff,eq:Harm.Harm.CoeffZero}
and cyclicity of the trace.
Evaluating the Hamiltonian for the
remaining two states $\state{1212}$ and $\state{1221}$
we find the energy matrix
\[\label{eq:Harm.Example.Matrix}
H=\matr{rrr}{2&2&-2\\2&2&-2\\-2&-2&2}.
\]
the factor of $2$ is due to $\ham=\ham_{12}+\ham_{21}$.
One eigenstate is
\[\label{eq:Harm.Example.Eigen}
\state{\OpK}=\state{1212}+\state{1221}-\state{1111}.\]
with energy $E=6$; it is clearly part of the Konishi multiplet.
The other two, $\state{1212}+\state{1111}$ and $\state{1221}+\state{1111}$,
have vanishing energy and belong to the half-BPS current multiplet.

\finishchapter

\docphysrept[\backmatter]{}
\docphysrept[]{}
\bibliography{doc}
\docphysrept[\bibliographystyle{nb}]{\bibliographystyle{elsart-num}}

\docphysrept{\newpage\listoffigures\newpage\listoftables}

\dochumboldt[\clearpage\thispagestyle{empty}\cleardoublepage]{}

\finishchapter 

\dochumboldt{\include{sechu}}

\end{document}